# HERA AND THE LHC

A workshop on the implications of HERA for LHC physics

March 2004 – March 2005

Parton density functions
Multijet final states
and energy flow
Heavy quarks
Diffraction
Monte Carlo tools

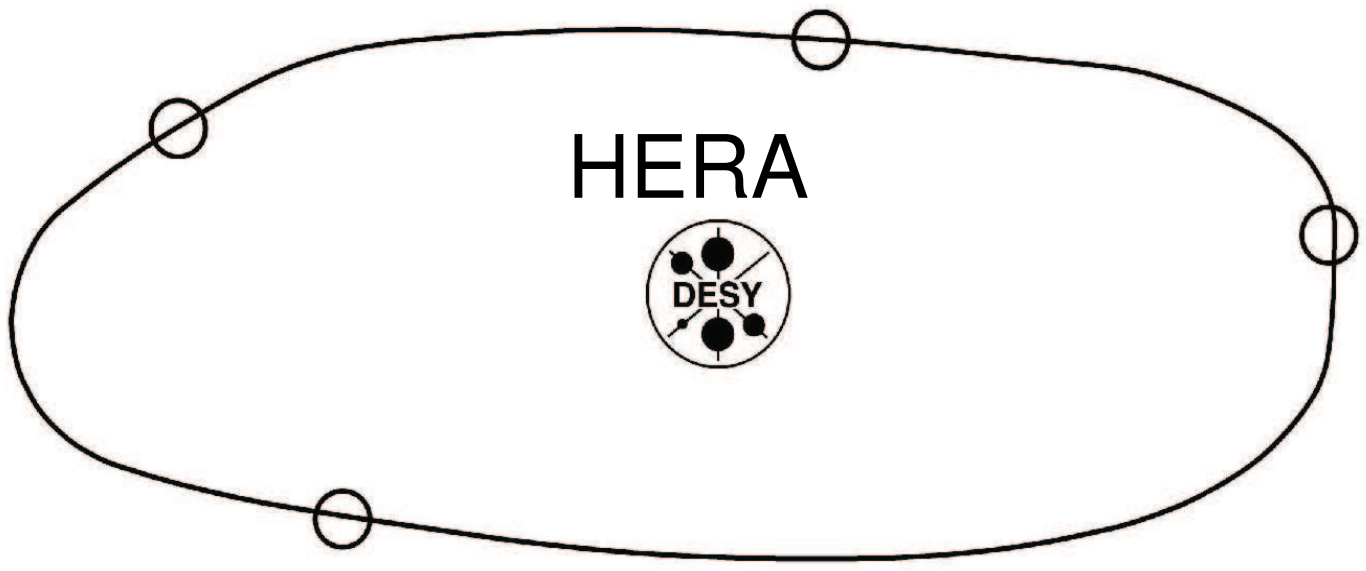

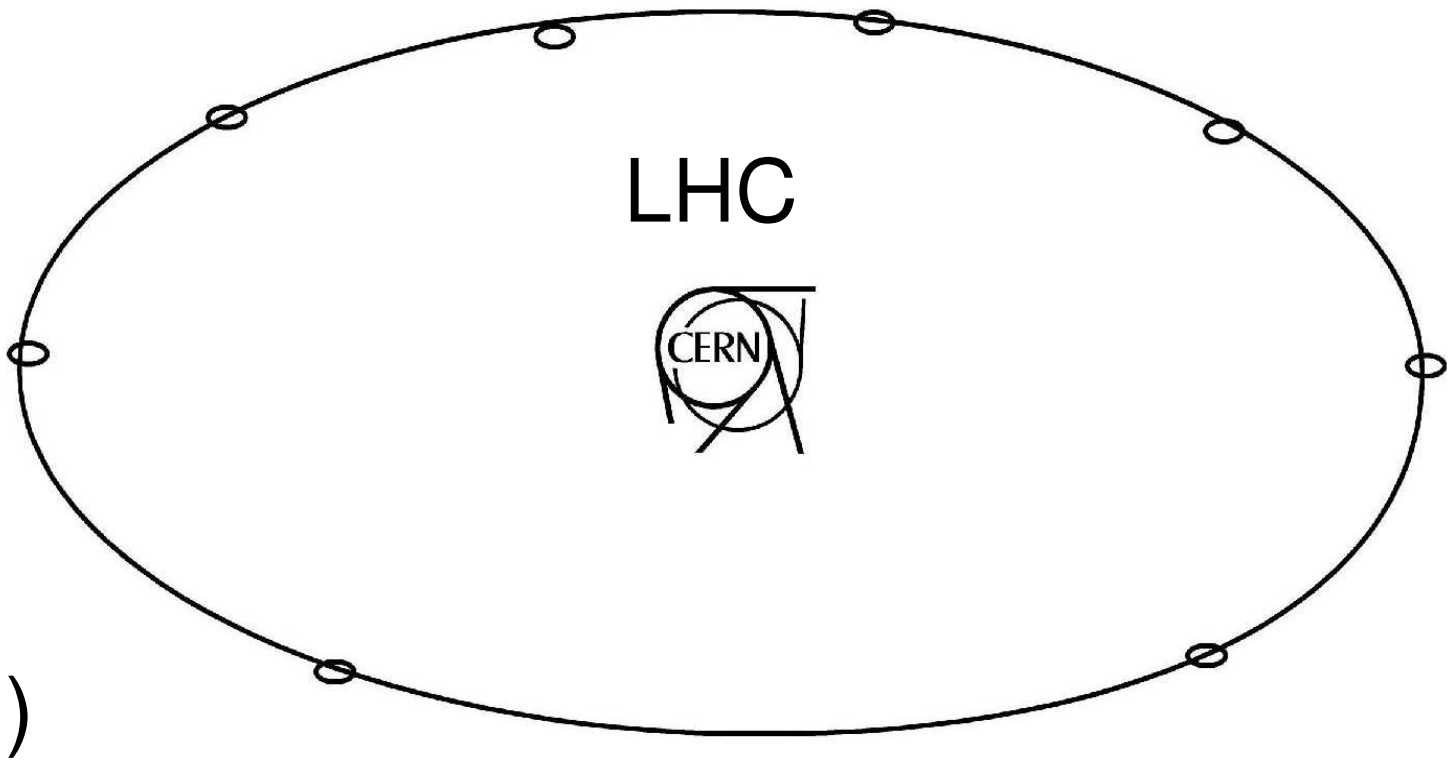

Proceedings
Editors:
A. De Roeck (CERN)
H. Jung (DESY)

**Part B**

**Organizing Committee:**

G. Altarelli (CERN), J. Blümlein (DESY), M. Botje (NIKHEF),
J. Butterworth (UCL), A. De Roeck (CERN) (chair), K. Eggert (CERN),
H. Jung (DESY) (chair), M. Mangano (CERN), A. Morsch (CERN),
P. Newman (Birmingham), G. Polesello (INFN), O. Schneider (EPFL),
R. Yoshida (ANL)

**Advisory Committee:**

J. Bartels (Hamburg), M. Della Negra (CERN), J. Ellis (CERN),
J. Engelen (CERN), G. Gustafson (Lund), G. Ingelman (Uppsala), P. Jenni (CERN),
R. Klanner (DESY), M. Klein (DESY), L. McLerran (BNL), T. Nakada (CERN),
D. Schlatter (CERN), F. Schrempp (DESY), J. Schukraft (CERN),
J. Stirling (Durham), W.K. Tung (Michigan State), A. Wagner (DESY),
R. Yoshida (ANL)

## Abstract

The HERA electron–proton collider has collected 100 $pb^{-1}$ of data since its start-up in 1992, and recently moved into a high-luminosity operation mode, with upgraded detectors, aiming to increase the total integrated luminosity per experiment to more than 500 $pb^{-1}$. HERA has been a machine of excellence for the study of QCD and the structure of the proton. The Large Hadron Collider (LHC), which will collide protons with a centre-of-mass energy of 14 TeV, will be completed at CERN in 2007. The main mission of the LHC is to discover and study the mechanisms of electroweak symmetry breaking, possibly via the discovery of the Higgs particle, and search for new physics in the TeV energy scale, such as supersymmetry or extra dimensions. Besides these goals, the LHC will also make a substantial number of precision measurements and will offer a new regime to study the strong force via perturbative QCD processes and diffraction. For the full LHC physics programme a good understanding of QCD phenomena and the structure function of the proton is essential. Therefore, in March 2004, a one-year-long workshop started to study the implications of HERA on LHC physics. This included proposing new measurements to be made at HERA, extracting the maximum information from the available data, and developing/improving the theoretical and experimental tools. This report summarizes the results achieved during this workshop.

# List of Authors

S. Alekhin [1], G. Altarelli [2,3], N. Amapane [4], J. Andersen [5], V. Andreev [6], M. Arneodo [7], V. Avati [8], J. Baines [9], R.D. Ball [10], A. Banfi [5], S.P. Baranov [6], J. Bartels [11], O. Behnke [12], R. Bellan [4], J. Blümlein [13], H. Böttcher [13], S. Bolognesi [4], M. Boonekamp [14], D. Bourilkov [15], J. Braciník [16], A. Bruni [17], G. Bruni[18], A. Buckley [19], A. Bunyatyan [20], C.M. Buttar [21], J.M. Butterworth [22], S. Butterworth [22], M. Cacciari [23], T. Carli [24], G. Cerminara [4], S. Chekanov [25], M. Ciafaloni [26], D. Colferai [26], J. Collins [27], A. Cooper-Sarkar [28], G. Corcella [2], M. Corradi [29], B.E. Cox [30], R. Croft [31], Z. Czyczula [32], A. Dainese [33], M. Dasgupta [2], G. Davatz [34], L. Del Debbio [2,10], Y. Delenda [30], A. De Roeck [24], M. Diehl [35], S. Diglio [3], G. Dissertori [34], M. Dittmar [34], J. Ellis [2], K.J. Eskola [36], T.O. Eynck [37], J. Feltesse [38], F. Ferro [39], R.D. Field [40], J. Forshaw [30], S. Forte [41], A. Geiser [35], S. Gieseke [42], A. Glazov [35], T. Gleisberg [43], P. Golonka [44], E. Gotsman [45], G. Grindhammer [16], M. Grothe [46], C. Group [40], M. Groys [45], A. Guffanti [13], G. Gustafson [47], C. Gwenlan [28], S. Höche [43], C. Hogg [48], J. Huston [49], G. Iacobucci [18], G. Ingelman [50], S. Jadach [51], H. Jung [35], J. Kalliopuska [52], M. Kapishin [53], B. Kersevan [54], V. Khoze [19], M. Klasen [11,55], M. Klein [13], B.A. Kniehl [11], V.J. Kolhinen [36], H. Kowalski [35], G. Kramer [13], F. Krauss [43], S. Kretzer [56], K. Kutak [11], J.W. Lämsä [52], L. Lönnblad [47], T. Laštovička [24], G. Laštovička-Medin [57], E. Laenen [37], Th. Lagouri [58], J.I. Latorre [59], N. Lavesson [47], V. Lendermann [60], E. Levin [45], A. Levy [45], A.V. Lipatov [61], M. Lublinsky [62], L. Lytkin [63], T. Mäki [52], L. Magnea [64], F. Maltoni [65], M. Mangano [2], U. Maor [45], C. Mariotti [66], N. Marola [52], A.D. Martin [19], A. Meyer [35], S. Moch [13], J. Monk [30], A. Moraes [21], A. Morsch [24], L. Motyka [11], E. Naftali [45], P. Newman [67], A. Nikitenko [68], F. Oljemark [52], R. Orava [52], M. Ottela [52], K. Österberg [52], K. Peters [30,35], F. Petrucci [44], A. Piccione [64], A. Pilkington [30], K. Piotrzkowski [69], O.I. Piskounova [6], A. Proskuryakov [45], A. Prygarin [45], J. Pumplin [70], K. Rabbertz [71], R. Ranieri [26], V. Ravindran [72], B. Reisert [73], E. Richter-Was [74], L. Rinaldi [18], P. Robbe [75], E. Rodrigues [76], J. Rojo [60], H. Ruiz [24], M. Ruspa [18], M.G. Ryskin [77], A. Sabio Vera [11], G.P. Salam [78], A. Schälicke [43], S. Schätzel [35], T. Schörner-Sadenius [79], I. Schienbein [80], F-P. Schilling [24], F. Schrempp [35], S. Schumann [43], M.H. Seymour [81], F. Siegert [24], T. Sjöstrand [2,47], M. Skrzypek [51], J. Smith [37,82], M. Smizanska [83], H. Spiesberger [84], A. Staśto [85], H. Stenzel [86], W.J. Stirling [19], P. Szczypka [31], S. Tapprogge [52,84], C. Targett-Adams [22], M. Tasevsky [87], T. Teubner [88], R.S. Thorne [5], A. Tonazzo [3], A. Tricoli [28], N. Tuning [37], J. Turnau [85], U. Uwer [12], P. Van Mechelen [87], R. Venugopalan [56], M. Verducci [24], J.A.M. Vermaseren [37], A. Vogt [19], R. Vogt [89], B.F.L. Ward [90], Z. Was [44], G. Watt [35], B.M. Waugh [22], C. Weiser [91], M.R. Whalley [19], M. Wing [22], J. Winter [43], S.A. Yost [90], G. Zanderighi [73], N.P. Zotov [61]

[1] Institute for High Energy Physics, 142284 Protvino, Russia
[2] CERN, Department of Physics, Theory Unit, CH 1211 Geneva 23, Switzerland
[3] Dipartimento di Fisica “E.Amaldi”, Università Roma Tre and INFN, Sezione di Roma Tre, via della Vasca Navale 84, I 00146 Roma, Italy
[4] Torino University and INFN Torino, Italy
[5] Cavendish Laboratory, University of Cambridge, Madingley Road, Cambridge, CB3 0HE, UK
[6] P.N. Lebedev Physical Institute of the Russian Academy of Science, Moscow, Russia
[7] Università del Piemonte Orientale, Novara, and INFN-Torino, Italy
[8] CERN, Geneva, Switzerland, and Case Western Reserve University, Cleveland, OH, USA
[9] Rutherford Laboratory, UK
[10] School of Physics, University of Edinburgh, Edinburgh EH9 3JZ, UK
[11] II. Institut für Theoretische Physik, Universität Hamburg Luruper Chaussee 149, D-22761 Hamburg, Germany
[12] Universität Heidelberg, Philosophenweg 12 69120 Heidelberg, Germany
[13] DESY, Platanenallee 6, D 15738 Zeuthen, Germany
[14] Service de physique des particules, CEA/Saclay, 91191 Gif-sur-Yvette CEDEX, France
[15] University of Florida, Gainesville, FL 32611, USA
[16] Max-Planck-Institut für Physik, München, Germany
[17] INFN Bologna, Via Irnerio 46, 40156 Bologna, Italy
[18] INFN Bologna and University of Eastern Piedmont, Italy
[19] Institute for Particle Physics Phenomenology, University of Durham, DH1 3LE, UK
[20] Yerevan Physics Institute, Armenia and MPI-K Heidelberg, Germany
[21] Dept. of Physics and Astronomy, University of Glasgow, UK
[22] Dept. of Physics and Astronomy, University College London, UK
[23] LPTHE - Université P. et M. Curie (Paris 6), Paris, France
[24] CERN, Department of Physics, CH 1211 Geneva 23, Switzerland
[25] HEP Division, Argonne National Laboratory, 9700 S. Cass Avenue, Argonne, IL 60439, USA
[26] Dipartimento di Fisica, Università di Firenze and INFN, Sezione di Firenze, I 50019 Sesto Fiorentino, Italy
[27] Physics Department, Penn State University, USA
[28] Department of Physics, Nuclear and Astrophysics Lab., Keble Road, Oxford, OX1 3RH, UK
[29] INFN Bologna, via Irnerio 46, Bologna, Italy
[30] School of Physics and Astronomy, The University of Manchester, Manchester M13 9PL, UK
[31] University of Bristol, Bristol, UK
[32] Institute of Physics, Jagiellonian University, Krakow, Poland and Niels Bohr Institute, University of Copenhagen, Copenhagen, Denmark
[33] University and INFN, Padova, Italy
[34] Institute for Particle Physics, ETH-Zürich Hönggerberg, CH 8093 Zürich, Switzerland
[35] DESY, Notkestrasse 85, D 22603 Hamburg, Germany
[36] Department of Physics, University of Jyväskylä, Jyväskylä, Finland
[37] NIKHEF Theory Group, Kruislaan 409, 1098 SJ Amsterdam, The Netherlands
[38] DSM/DAPNIA, CEA, Centre d’Etudes de Saclay, F 91191 Gif-sur-Yvette, France
[39] University of Genova and INFN-Genova, Italy
[40] Dept. of Physics, University of Florida, USA
[41] Dipartimento di Fisica, Universitá di Milano, INFN Sezione di Milano, Via Celoria 16, I 20133 Milan, Italy
[42] Institut für Theoretische Physik, Universität Karlsruhe, 76128 Karlsruhe, Germany
[43] Institut für Theoretische Physik, TU Dresden, D-01062 Dresden, Germany
[44] CERN, 1211 Geneva 23, Switzerland, and Institute of Nuclear Physics, ul. Radzikowskiego 152, 31-342 Kraków, Poland
[45] HEP Department, School of Physics and Astronomy, Raymond and Beverly Sackler Faculty of Exact Science, Tel Aviv University, Tel Aviv, 69978, Israel

[46] University of Torino and INFN-Torino, Italy; also at University of Wisconsin, Madison, WI, USA
[47] Dept. of Theoretical Physics, Lund University, Sweden
[48] University of Wisconsin, Madison, WI, USA
[49] Department of Physics and Astronomy, Michigan State University, E. Lansing, MI 48824, USA
[50] High Energy Physics, Uppsala University, Box 535, SE 75121 Uppsala, Sweden
[51] Institute of Nuclear Physics, Academy of Sciences, ul. Radzikowskiego 152, 31-342 Cracow, Poland and CERN, Department of Physics, Theory Unit, CH 1211 Geneva 23, Switzerland
[52] High Energy Physics Division, Department of Physical Sciences, University of Helsinki and Helsinki Institute of Physics, P.O. Box 64, FIN-00014, Finland
[53] JINR Dubna, Russia
[54] Faculty of Mathematics and Physics, University of Ljubljana, Jadranska 19, SI-1000, Slovenia Experimental Particle Physics Department, Jozef Stefan Institute, P.P. 3000, Jamova 39, SI-1000 Ljubljana, Slovenia
[55] Laboratoire de Physique Subatomique et de Cosmologie, Université Joseph Fourier/CNRS-IN2P3, 53 Avenue des Martyrs, 38026 Grenoble, France
[56] Brookhaven National Laboratory, Upton, NY 11973, USA
[57] University of Podgorica, Cetinjski put bb, CNG 81000 Podgorica, Serbia and Montenegro
[58] Institute of Nuclear and Particle Physics, Charles University, Prague, Czech Republic
[59] Departament d’Estructura i Constituents de la Matèria, Universitat de Barcelona, Diagonal 647, E 08028 Barcelona, Spain
[60] Kirchhoff-Institut für Physik, Universität Heidelberg, Im Neuenheimer Feld 227, 69120 Heidelberg, Germany
[61] D.V. Skobeltsyn Institute of Nuclear Physics, Moscow, Russia
[62] University of Connecticut, USA
[63] MPI-K Heidelberg, Germany and JINR Dubna, Russia
[64] Dipartimento di Fisica Teorica, Università di Torino and INFN Sezione di Torino, Via P. Giuria 1, I 10125 Torino, Italy
[65] Institut de Physique Théorique, Université Catholique de Louvain, Chemin du Cyclotron, 2, B-1348, Louvain-la-Neuve, Belgium
[66] INFN Torino, Italy
[67] School of Physics and Astronomy, University of Birmingham, B15 2TT, UK
[68] Imperial College, London, UK
[69] Institut de Physique Nucléaire, Université Catholique de Louvain, Louvain-la-Neuve, Belgium
[70] Department of Physics and Astronomy, Michigan State University, E. Lansing, MI 48824, USA
[71] University of Karlsruhe, EKP, Postfach 6980, D-76128 Karlsruhe, Germany
[72] Harish-Chandra Research Institute, Chhatnag Road, Jhunsi, Allahabad, India
[73] Fermi National Accelerator Laboratory, Batavia, IL 60126, USA
[74] Institute of Physics, Jagiellonian University, 30-059 Krakow, ul. Reymonta 4, Poland. Institute of Nuclear Physics PAN, 31-342 Krakow, ul. Radzikowskiego 152, Poland.
[75] Laboratoire de l’Accélérateur Linéaire, Université Paris-Sud, 91898 Orsay, France
[76] NIKHEF, Amsterdam, The Netherlands
[77] Petersburg Nuclear Physics Institute, Gatchina, St. Petersburg, Russia
[78] LPTHE, Universities of Paris VI and VII and CNRS, F 75005, Paris, France
[79] University of Hamburg, IExpPh, Luruper Chaussee 149, D-22761 Hamburg, Germany
[80] Southern Methodist University Dallas, 3215 Daniel Avenue, Dallas, TX 75275-0175, USA
[81] School of Physics & Astronomy, University of Manchester, UK and Theoretical Physics Group, CERN, Geneva, Switzerland
[82] C.N. Yang Institute for Theoretical Physics, Stony Brook University, Stony Brook, NY 11794, USA
[83] Lancaster University, Lancaster, UK
[84] Johannes-Gutenberg-Universität Mainz, D-55099 Mainz, Germany
[85] H. Niewodniczański Institute of Nuclear Physics, PL 31-342 Kraków, Poland

[86] II. Physikalisches Institut, Universität Giessen, Heinrich-Buff-Ring 16, D 35392 Giessen, Germany
[87] Universiteit Antwerpen, Antwerpen, Belgium
[88] University of Liverpool, UK
[89] Lawrence Berkeley National Laboratory, Berkeley, CA, USA and Physics Department, University of California, Davis, CA, USA
[90] Department of Physics, Baylor University, Waco, TX, USA
[91] Institut für Experimentelle Kernphysik, Universität Karlsruhe, Karlsruhe, Germany

# Contents

# Part IV

# Working Group 3: Heavy Quarks (Charm and Beauty)

# List of participants in the working group

J. Baines, S. Baranov, G. Barbagli, O. Behnke, A. Bertolin, J. Bluemlein, G. Bruno, O. Buchmueller, J. Butterworth, M. Cacciari, T. Carli, J. Catmore, V. Chiochia, A. Cholewa, J. Cole, G. Corcella, M. Corradi, A. Dainese, K. Ellis, K. Eskola, D. Fabris, G. Flucke, A. Geiser, C. Grab, G. Grindhammer, R. Guernane, O. Gutsche, H. Jung, C. Kiesling, R. Klanner, V. Kolhinen, G. Kramer, S. Kretzer, T. Kuhr, A. Kurepin, K. Kutak, E. Laenen, T. Lagouri, E. Levin, A. Likhoded, A. Lipatov, M. Lunardon, F. Maltoni, M. Mangano, L. Marti-Magro, A. Meyer, M. Morando, A. Morsch, N. Panikashvili, N. Pavel, K. Peters, O. Piskounova, R. Ranieri, H. Ruiz, I. Schienbein, R. Sharafiddinov, S. Shulha, M. Smizanska, K. Sridhar, Z. Staykova, P. Thompson, R. Thorne, A. Tonazzo, R. Turrisi, U. Uwre, M. Villa, R. Vogt, B. Vulpescu, Z. Was, C. Weiser, M. Wing, A. Zoccoli, N. Zotov, M. zur Nedden

# Introduction to heavy quarks (charm and beauty)

*O. Behnke, M Cacciari, M. Corradi, A. Dainese, A. Geiser, A. Meyer, M. Smizanska, U. Uwer, C. Weiser*

Perturbative QCD is expected to provide reliable predictions for the production of bottom and (to a lesser extent) charm quarks since their masses are large enough to assure the applicability of perturbative calculations. A direct comparison of perturbative QCD predictions to heavy-flavour production data is not straightforward. Difficulties arise from the presence of scales very different from the quark masses that reduce the predictivity of fixed-order theory, from the non-perturbative ingredients needed to parametrize the fragmentation of the heavy quarks into the observed heavy hadrons, and from the limited phase space accessible to present detectors. Moreover, a breakdown of the standard collinear factorization approach can be expected at low-$x$. The study of heavy-quark production in hadronic interactions and in e–p collisions at HERA has been therefore an active field in the effort to overcome these difficulties and to get a deeper understanding of hard interactions.

Besides its intrinsic interest, a precise understanding of heavy-quark production is important at the LHC because charm and beauty from QCD processes are relevant backgrounds to other interesting processes from the Standard Model (e.g., Higgs to $b\bar{b}$) or beyond. Theoretical and experimental techniques developed at HERA in the heavy-quark field, such as heavy-quark parton densities or $b$-tagging, are also of great value for future measurements at the LHC.

The present status of heavy-quark production theory is critically reviewed in the first contribution. The second contribution summarizes the present heavy-flavour data from HERA and gives an outlook of what can be expected from HERA-II. The potential of the LHC experiments for charm and beauty physics is reviewed in the third contribution. Then the relevance of saturation and low-$x$ effects to heavy-quark production at HERA and at the LHC are discussed. The non-perturbative aspects of heavy-quark fragmentation and their relevance to HERA and LHC are discussed in the next contribution. Finally, a comparison of different theoretical predictions for HERA and the LHC based on different approaches is presented.

# Theoretical review of various approaches in heavy quark production

Coordinators: *M. Cacciari*[1], *E. Laenen*[2]
Contributing authors: *S.P. Baranov*[3], *M. Cacciari*[1], *S. Diglio*[4], *T.O. Eynck*[2], *H. Jung*[5], *B.A. Kniehl*[6], *S. Kretzer*[7], *E. Laenen*[2], *A.V. Lipatov*[8], *F. Maltoni*[9], *F. Petrucci*[4], *O.I. Piskounova*[3], *I. Schienbein*[6], *J. Smith*[2,10], *A. Tonazzo*[4], *M. Verducci*[11], *N.P. Zotov*[8]
[1]LPTHE - Université P. et M. Curie (Paris 6), France
[2]NIKHEF Theory Group, Kruislaan 409, 1098 SJ Amsterdam, The Netherlands
[3]P.N. Lebedev Physical Institute of Russian Academy of Science, Moscow, Russia
[4]Università Roma Tre, Dipartimento di Fisica "E.Amaldi" and
INFN Sezione Roma III, Via della Vasca Navale 84, 00146 Rome, Italy
[5]Deutsches Elektronen-Synchroton Hamburg, FRG
[6]II. Institut für Theoretische Physik, Universität Hamburg, Luruper Chaussee 149, 22761, Hamburg, Germany
[7] Brookhaven National Laboratory, Upton, NY 11973, USA
[8]D.V. Skobeltsyn Institute of Nuclear Physics, Moscow, Russia
[9]Institut de Physique Théorique, Université Catholique de Louvain,Chemin du Cyclotron, 2, B-1348, Louvain-la-Neuve, Belgium
[10]C.N. Yang Institute for Theoretical Physics, Stony Brook University, Stony Brook, NY 11794, USA
[11]CERN, CH-1211 Genève 23, Switzerland

**Abstract**
We review some of the main theoretical aspects of heavy quark production at HERA that will be important for understanding similar processes at the LHC.

## 1 Introduction

The value for the LHC physics program of heavy quark production studies at HERA consists not only of measured quantities such as parton distributions, heavy quark masses etc. but at least as much of the theoretical ideas on heavy quark production that were developed and refined in the course of these studies. The strong experimental interest in heavy quark observables at HERA has led to a significantly increased understanding of the benefits and limitations of finite order calculations. It has stimulated theorists to deepen their insight into the issue of when a heavy quark should be treated as a parton, and it has provoked novel proposals to explain the hadronization of heavy quarks. In what follows we review and critically assess some of these ideas.

## 2 Heavy quark production

The study of heavy quarks, historically plagued by low production rates and large uncertainties, has now entered the regime of 'precision physics'. On the one hand, the larger centre-of-mass energies of the colliders running now (Tevatron, HERA) and in the near future (LHC) lead to a much more copious production yield. On the other hand, technological advances such as the introduction of microvertex detectors based on semiconductor devices allow for much better tagging of the produced heavy flavours, and hence better measurements. Needless to say, an equally substantial improvement of the theoretical calculations has been needed in order to match this progress and therefore deliver predictions with an accuracy at least as good as that of the experimental measurements. Properly testing and constraining the theoretical calculations will in turn help in refining the predictions for the LHC.

One example for which a good theoretical accuracy at the LHC is desirable is in calculating the total $Z$ boson production rate, a process which can be used as a luminosity candle and which we would like to have under control at the one per cent level. One channel contributing to this process is gluon-gluon fusion followed by bottom-antibottom annihilation, $gg \to b\bar{b} \to Z$. This channel provides about 5% of the total $Z$ yield [1]: hence, it must be under control at the 20% level in order to achieve the sought-for final 1% accuracy.

As it turns out, it is more efficient and more reliable to rewrite this in terms of a perturbatively calculated parton distribution function (PDF) for the bottom quark, i.e. as the effective process $b\bar{b} \to Z$. The theoretical tools that we use to construct such heavy quark parton distribution functions must therefore be tested by employing them in other theoretical predictions, to be compared to the available experimental data. In the following section we shall list a number of examples where this is done.

From the point of view of 'standard' perturbative QCD calculations, the situation has not changed since the beginning of the '90s: fully massive next-to-leading order (NLO) calculations were made available for hadron-hadron [2–6], photon-hadron [7–9] (i.e. photoproduction) and electron-hadron [10–13] (i.e. Deep Inelastic Scattering, DIS) collisions. These calculations still constitute the state of the art as far as fixed order results are concerned, and they form the basis for all modern phenomenological predictions.

Over the years, and with increasing experimental accuracies, it however became evident that perturbative QCD alone did not suffice. In fact, real particles - hadrons and leptons - are observed in the detectors, not the unphysical quarks and gluons of perturbative QCD. A proper comparison between theory and experiment requires that this gap be bridged by a description of the transition. Of course, the accuracy of such a description will reflect on the overall accuracy of the comparison. When the precision requirements were not too tight, one usually employed a Monte Carlo description to 'correct' the data, deconvoluting hadronization effects and extrapolating to the full phase space. The final 'experimental' result could then easily be compared to the perturbative calculation. This procedure has the inherent drawback of including the bias of our theoretical understanding (as implemented in the Monte Carlo) into an experimental measurement. This bias is of course likely to be more important when the correction to be performed is very large. It can sometimes become almost unacceptable, for instance when exclusive measurements are extrapolated by a factor of ten or so in order to produce an experimental result for a total photoproduction cross section or a heavy quark structure function.

The alternative approach is to present (multi)differential experimental measurements, with cuts as close as possible to the real ones, which is to say with as little theoretical correction/extrapolation as possible. The theoretical prediction must then be refined in order to compare with the real data that it must describe. This has two consequences. First, one has to deal with differential distributions which, in certain regions of phase space, display a bad convergence in perturbation theory. All-order resummations must then be performed in order to produce reliable predictions. Second, differential distributions of real hadrons depend unavoidably on some non-perturbative phenomenological inputs, fragmentation functions. Such inputs must be extracted from data and matched to the perturbative theory in a proper way, pretty much like parton distribution functions of light quarks and gluons are.

In the following sections we review the state of the art of theoretical calculations of heavy quark production in a number of high energy processes, pointing out similarities and differences. In particolar, resummations aimed at improving the theoretical description of heavy quark production at large transverse momentum or large photon virtuality in DIS (Section 3), small centre-of-mass energy (Section 5) and large centre of-mass energy (Section 6) are described in some detail.

## 3 Collinear resummations and heavy quark PDFs

Perturbative calculations of heavy quark production contain badly converging logarithmic terms of quasi-collinear origin in higher orders when a second energy scale is present and it is much larger than the heavy quark mass $m$. Examples are the (square root of the) photon virtuality $Q^2$ in DIS and the transverse momentum $p_T$ in either hadroproduction or photoproduction. Naming generically $E$ the large scale, we can write schematically the cross section for the production of the heavy quark $Q$ as

$$\sigma_Q(E,m) = \sigma_0 \left(1 + \sum_{n=1} \alpha_s^n \sum_{k=0}^{n} c_{nk} \ln^k \left[\frac{E^2}{m^2} + \mathcal{O}\left(\frac{m}{E}\right)\right]\right) , \tag{1}$$

where $\sigma_0$ stands for the Born cross section, and the coefficients $c_{nk}$ can contain constants as well as functions of $m$ and $E$, vanishing as powers of $m/E$ when $E \gg m$.

Resummation approaches bear many different names, (ZM-VFNS, ACOT, FONLL, BSMN to name but a few) but they all share the goal of resumming leading ($\alpha_s^n \ln^n(E^2/m^2)$, LL) and sometimes also next-to-leading ($\alpha_s^n \ln^{n-1}(E^2/m^2)$, NLL) logarithmic terms to all orders in the cross section above. This is achieved by discarding power suppressed $m/E$ terms, and factoring all the logarithms into a resummation factor, to be obtained via Altarelli-Parisi evolution of an initial condition set at the heavy quark mass scale,

$$\sigma_Q^{res}(E,m) = \sigma_0 C(E,\mu) f(\mu,m) = \sigma_0 C(E,\mu) E(\mu,\mu_0) f(\mu_0,m) , \tag{2}$$

where $\mu$ and $\mu_0$ represent artificial factorization scales, to be taken of order $E$ and $m$ respectively. The 'products' between the various functions actually hide convolution operations with respect to momentum fractions, not explicitly shown as arguments. $C(E,\mu)$ is a perturbatively calculable coefficient function, which does not contain large logarithms thanks to the choice $\mu \simeq E$. The function $f(\mu,m)$ can represent either a parton distribution or a fragmentation function for a heavy quark, and contains the resummation of the collinear logarithms. Due to the large heavy quark mass, its initial condition $f(\mu_0,m)$ can be calculated in perturbation theory [14, 15]: this is the distinctive feature that sets heavy quark parton and fragmentation functions apart from light flavour ones, whose initial conditions are instead entirely non-perturbative and must be fitted to data.

Once a massless but resummed result, valid in the $E \gg m$ region, is obtained, one would like to interpolate it with a fixed order cross section, valid instead in the $E \simeq m$ region, so as to retain predictivity over the whole $E$ range.

The differences between the various approaches are then to be found essentially in two points:

– the perturbative order to which the initial condition $f(\mu_0,m)$ is evaluated, and the perturbative accuracy of the evolution;
– the way the matching with the fixed order calculation is performed.

We summarize below the features of the most commonly used implementations.

### 3.1 ACOT - Aivazis, Collins, Olness, Tung

This approach was the first to try to improve the prediction of the heavy quark structure functions $F_2^c(Q^2,m_c^2)$ and $F_2^b(Q^2,m_b^2)$ at large $Q^2 \gg m_c^2, m_b^2$, by moving potentially large logarithms $\ln(Q^2/m^2)$ into heavy quark parton densities [16, 17]. A general all-order analysis of factorization for the total inclusive $F_2(Q^2)$ in this context was presented in [18].

### 3.2 Simplified ACOT and ACOT($\chi$)[1]

The original ACOT prescription [16, 17] has been simplified in [19] along lines suggested in [18, 20]. In a nutshell, diagrams with initial state heavy quark legs can be treated as if they represented massless quarks. More generally, the diagrams can be manipulated by power suppressed terms provided that higher order diagrams are regularized consistently. ACOT($\chi$) [21, 22] explores this freedom to improve on the threshold behaviour of partonic heavy quark schemes by enforcing the physical pair-production threshold on a term-by-term basis. Heuristically, it comes down to a simple re-scaling of Bjorken-$x$, i.e. in LO

$$F_2^{c\bar{c}} \propto c(\chi)|_{\chi = x_{\rm Bj}(1+4m^2/Q^2)} \quad . \tag{3}$$

[1]Contributed by S. Kretzer

Physical arguments –mostly kinematic– have been given in [21–23], here we will establish the connection with the FONLL terminology of Section 1.3.3 while focusing on the inclusive DIS process. Much of the following has appeared before, in one form or another, in the literature [16–19, 24–28].

We formulate ACOT($\chi$) as an explicit manipulation of resummed terms of the perturbative series. We follow [24] in notation and add an $\mathcal{O}\left(\alpha_s^1\right)$ fixed order (FO) calculation to an all order collinearly resummed (RS) result. In RS heavy quark mass dependence other than logarithmic is neglected. When we remove double-counting terms from FO + RS the zero mass limit (FOM0) of the FO calculation will be required as an auxiliary quantity. Just as in RS, only asymptotic mass logarithms are retained in FOM0. We write therefore, as usual,

$$\sigma^{ACOT}(Q,m) = \mathrm{FO} + (\mathrm{RS} - \mathrm{FOM0}) \times G \tag{4}$$

where $G$ is an arbitrary operation which behaves like $G = 1 + \mathcal{O}\left(\frac{m^2}{Q^2}\right)$. In [24] $G$ was chosen to be an overall multiplicative factor. More generally, it can be seen as an operation which only modifies, with $\mathcal{O}(m^2/Q^2)$ power-suppressed terms, perturbative coefficients beyond those which have been explicitly calculated, and which are therefore unknown anyway. Any choice for $G$ with this behaviour is therefore legitimate.

To motivate the ACOT($\chi$) choice for $G$ we first re-write more explicitly the three terms given above in the case of inclusive DIS:

$$\begin{aligned} \mathrm{FO} &= \alpha_s\, g\, \tilde{\otimes} H(Q,m) && (5)\\ \mathrm{FOM0} &= \alpha_s \left( g \otimes P_{qg}^{(0)} \ln \frac{\mu^2}{m^2} + g \otimes C_g \right) && (6)\\ \mathrm{RS} &= c(x) + \alpha_s \left( g \otimes C_g + c \otimes C_q \right) && (7) \end{aligned}$$

where $H(Q,m)$ is the massive coefficient function for the $FO$ gluon fusion process, $C_g$ and $C_q$ are the gluon and light quark coefficient functions (the $\overline{\rm MS}$ scheme is implied), and $g$ and $c$ are the gluon and charm (i.e. heavy quark) parton distribution functions (both the coefficient functions and the PDFs depend, of course, on the factorization scale $\mu \simeq Q$). $P_{qg}^{(0)}$ is the leading order Altarelli-Parisi splitting vertex. The symbol $\tilde{\otimes} \equiv \int_\chi^1 d\xi/\xi$ ... denotes a threshold-respecting convolution integral. One can convince oneself that the standard convolution $\otimes$, with $x \to \chi$ in the lower limit of integration, only differs by $\tilde{\otimes}$ by power-suppressed terms, $\tilde{\otimes} = \otimes + \mathcal{O}(m^2/Q^2)$.

The combined result (4) reads now

$$\begin{aligned} \sigma^{ACOT}(Q,m) &= \mathrm{FO} + (\mathrm{RS} - \mathrm{FOM0}) \times G \\ &= \alpha_s\, g\, \tilde{\otimes} H + \left[ c(x) - \alpha_s\, g \otimes P_{qg}^{(0)} \ln \frac{\mu^2}{m^2} + \alpha_s\, c \otimes C_q \right] \times G \,, \end{aligned} \tag{8}$$

and we recognize the Krämer-Olness-Soper simplified ACOT framework of [19][2] if we set $G = 1$. Different choices for $G$ can still be made, but natural demands are that:

- In kinematic regions where FO represents the relevant physics (i.e. $Q \sim m$), $G$ should efficiently suppress uncontrolled spurious higher order terms in the square bracket of eq.(8).
- For computational efficiency, the simple $c(x)$ term alone should provide an optimized effective $\mathcal{O}\left(\alpha_s^0\right)$ approximation.

[2] See Eqs. (7), (8) there. General choices for $G$ correspond to the discussion above these equations.

The ACOT($\chi$) scheme implements these requests by making an implicit choice for $G$ which corresponds to writing

$$\begin{aligned} \sigma^{ACOT(\chi)}(Q,m) &= \text{FO} + (\text{RS} - \text{FOM0}) \times G \\ &= \alpha_s\, g\, \tilde{\otimes} H + \left[ c(\chi) - \alpha_s\, g\, \tilde{\otimes} P_{qg}^{(0)} \ln \frac{\mu^2}{m^2} + \alpha_s\, c\, \tilde{\otimes} C_q \right] . \end{aligned} \tag{9}$$

Further details on ACOT($\chi$) can be found in [21–23]. These articles also contain a more intuitive perspective of ACOT($\chi$). Moreover, [22] describes a PDF set that is consistent with ACOT($\chi$) applications.

### 3.3 BSMN - Buza, Smith, Matiounine, van Neerven

In Refs. [29–33] the treatment of heavy quarks as a parton was fully explored through next-to-next-leading order (NNLO), based on a precise two-loop analysis of the heavy quark structure functions from an operator point of view. This analysis yielded a number of results. One result is important beyond the observable at hand: the authors obtained the complete set of NNLO matching conditions for parton evolution across flavor thresholds. They found that, unlike at NLO, the matching conditions are *discontinuous* at the flavor thresholds. These conditions are necessary for any NNLO calculation at the LHC, and have already been implemented in a number of evolution packages [34, 35].

Furthermore, their two-loop calculations explicitly showed that the heavy *quark* structure functions in such a variable flavor approach are not infrared safe: one needs to either define a heavy *quark-jet* structure function, or introduce a fragmentation function to absorb the uncancelled divergence. In either case, a set of contributions to the inclusive light parton structure functions must be included at NNLO.

A dedicated analysis [36] for charm electroproduction showed that even at very large $Q^2$ one could not distinguish the fixed order NLO calculation of [10] and the NNLO VFNS calculations of [31], given the experimental data available in the year 2000. This demonstrates the possibility that the large logarithms $\ln(Q^2/m^2)$ together with small coefficients can in the end have little weight in the overall hadronic cross section.

### 3.4 FONLL - Fixed Order plus Next-to-Leading Log resummation

This approach was developed for improving the large-$p_T$ differential cross section for heavy quark production in hadron-hadron collisions [37]. It was successively extended to photoproduction [38], and in a second phase a matching to the fixed order NLO calculations was performed [24, 39]. The FONLL acronym refers specifically to the matched version.

From the point of view of perturbative logarithms, it contains a NLO-accurate initial condition and full NLL evolution. It therefore reproduces the full NLL structure of the NLO calculation, and resums to all orders the large logarithms with NLL accuracy.

The matching with the fixed order result is performed according to the following master formula (see eq.(16) of [24]):

$$\sigma_Q^{\text{FONLL}}(p_T, m) = \text{FO} + (\text{RS} - \text{FOM0}) G(m, p_T) \;, \tag{10}$$

where FO stands for the NLO fixed order massive calculation, FOM0 for its $m/p_T \to 0$ limit (where however $\ln p_T/m$ terms and non-vanishing terms are kept), and RS for the massless, resummed calculation[3]. The RS$-$FOM0 subtraction is meant to cancel the terms which are present in both RS and FO. This difference starts therefore at order $\alpha_s^2$ with respect to the Born cross section: at large $p_T$ it resums correctly the NLL terms, at small $p_T$ it only contains spurious terms, which are suppressed by the function $G(m, p_T) = p_T^2/(p_T^2 + c^2m^2)$, with $c = 5$, in order to ensure a physically correct behaviour. The choice of the suppression factor was motivated in [24] by the observation that the massless limit starts to approach the massive hadroproduction calculation at $\mathcal{O}(\alpha_s^3)$ only for $p_T > 5m$. Below this value the massless limit returns unreliable results, and its contribution must therefore be suppressed. It is important to realize that $G(m, p_T)$ only affects terms which are beyond the control of perturbation theory, and therefore it does not spoil the NLO+NLL accuracy. The choice to control such terms by means of an ad-hoc function might seem a somewhat unpleasant characteristic of this approach. However, it simply portraits the freedom one has in performing the matching, and does not represent a shortcoming of the approach: different matching procedures will simply make other implicit or explicit choices for $G(m, p_T)$.

[3]This term might also be referred to as a 'zero-mass variable flavour number scheme' (ZM-VFNS) contribution. However this name, while by itself completely general, has been used in the past for specific approaches with different overall perturbative accuracies. We shall therefore avoid its use. It will be understood that 'RS' in this approach has full NLL accuracy.

For the sake of making comparisons with other approaches easier, the formula (10) can be rewritten with some more details as follows:

$$\sigma_Q^{\rm FONLL}(p_T, m) = \sum_{ij\in\mathcal{L}} F_i F_j\ \sigma_{ij\to QX}(p_T, m)$$
$$+ \left( \sum_{ijk\in\mathcal{L}+\mathcal{H}} F_i F_j\ \hat{\sigma}^{\overline{MS}}_{ij\to kX}(p_T) D_{k\to Q} - \sum_{ij\in\mathcal{L}} F_i F_j\ \sigma_{ij\to QX}(p_T, m; m\to 0) \right) G(m, p_T) \quad (11)$$

A few ingredients needing definition have been introduced. The kernel cross sections $\sigma_{ij\to QX}(p_T, m)$ are the massive NLO calculations for heavy quark production of Refs. [2–6]. When convoluted with the PDFs for light flavours $F_i$ $(i \in \mathcal{L})$ they yield the FO term in eq. (10). The $\sigma_{ij\to QX}(p_T, m; m \to 0)$ terms represent the $m \to 0$ limit of the massive NLO cross sections, performed by sending to zero $m/p_T$ terms while preserving $\ln(p_T/m)$ contributions and non-vanishing constants. When convoluted with light flavour PDFs they give FOM0. Finally, $\hat{\sigma}^{\overline{MS}}_{ij\to kX}(p_T)$ are the massless $\overline{MS}$-subtracted NLO cross section kernels given in [40]. In addition to the light flavour PDFs, they are also convoluted with the perturbatively-calculated parton distribution functions for the heavy quarks $(i \in \mathcal{H})$ and with the fragmentation functions describing the transformation of a parton into a heavy quark, $D_{k\to Q}$ [15], to give the term RS.

The formula given above returns the differential cross section for heavy *quark* production, evaluated with NLO + NLL accuracy. In order to obtain the corresponding cross section for an observable heavy meson it must still be convoluted with the proper scale-independent non-perturbative fragmentation function, extracted from experimental data, describing the heavy quark $\to$ heavy hadron transition:

$$\sigma_H^{\rm FONLL}(p_T, m) = \sigma_Q^{\rm FONLL}(p_T, m) D^{NP}_{Q\to H}\ . \quad (12)$$

Phenomenological analyses of charm- and bottom-flavoured hadrons production within the FONLL approach have been given in [41–45].

### 3.5 GM-VFNS - General mass variable flavour number scheme

This approach also combines a massless resummed calculation with a massive fixed order one, for predicting $p_T$ distributions in hadron-hadron collisions. One difference with respect to FONLL is that this approach does not include the perturbative NLO parton-to-heavy-quark fragmentation functions $D_{k\to Q}$. Rather, it directly convolutes a properly $\overline{MS}$ subtracted cross section (with mass terms also included, hence the 'general mass' name) with non-perturbative fragmentation functions for heavy mesons $D^{NP,\overline{MS}}_{Q\to H}$, fitted at LEP in a pure $\overline{MS}$ scheme. The cross section can be schematically written as

$$\sigma_H^{\rm GM-VFNS}(p_T, m) = \sum_{ij\in\mathcal{L}} F_i F_j\ \hat{\sigma}_{ij\to QX}(p_T, m) D^{NP,\overline{MS}}_{Q\to H} + \sum_{ijk\in\mathcal{L}+\mathcal{H}} F_i F_j\ \hat{\sigma}^{\overline{MS}}_{ij\to kX}(p_T) D^{NP,\overline{MS}}_{k\to H}\ , \quad (13)$$

where the 'massive-but-subtracted' cross section kernels $\hat{\sigma}_{ij\to QX}(p_T, m)$ are defined by

$$\hat{\sigma}_{ij\to QX}(p_T, m) \equiv \sigma_{ij\to Q}(p_T, m) - \sigma_{ij\to QX}(p_T, m; m \to 0) + \hat{\sigma}^{\overline{MS}}_{ij\to QX}(p_T) \ . \tag{14}$$

The new kernels $\hat{\sigma}_{ij\to QX}(p_T, m)$ defined by this operation (of the form FO-FOM0+RS) can be convoluted with an evolved $\overline{MS}$-subtracted fragmentation function, but they also retain power suppressed $m/p_T$ terms. It should also be noted that the sum in the second term of (13) only runs over contributions not already included in the first.

Recalling the way the perturbative parton-to-heavy-quark $D_{k\to Q}$ fragmentation functions are defined in [15], setting

$$D^{NP,\overline{MS}}_{k\to H} = D_{k\to Q} D^{NP}_{Q\to H}, \qquad k \in \mathcal{L} + \mathcal{H} \ , \tag{15}$$

and comparing eqs.(13) and (11), it can be seen that the GM-VFNS master formula is a reshuffling of the FONLL one, up to higher-orders terms.

Two comments are worth making. The first is that due to the absence of the perturbative $D_{k\to Q}$ terms, eq. (13) cannot reproduce the NLO heavy *quark* production cross section: even the normalization must be extracted from the experimental data. Eq. (11), on the other hand, can reproduce the heavy quark spectrum, and only the heavy quark $\to$ heavy meson transition is fitted to data. The second remark concerns the higher order power suppressed terms: since GM-VNFS implicitly makes a different choice for the $G(m, p_T)$ function, the results from the two approaches might differ considerably in the $p_T \sim m$ region since, while formally suppressed, such terms can be numerically important.

An example of a phenomenological application of the GM-VFNS scheme is given below.

### 3.6 Hadroproduction of heavy mesons in a massive VFNS[4]

Various approaches for next-to-leading-order (NLO) calculations in perturbative QCD have been applied to one-particle-inclusive hadroproduction of heavy mesons. The general-mass variable-flavor-number scheme (GM-VFNS) devised by us in Ref. [46, 47] is closely related to the conventional massless variable-flavor-number scheme (ZM-VFNS), but keeps all $m^2/p_T^2$ terms in the hard-scattering cross sections, where $m$ is the mass of the heavy quark and $p_T$ the transverse momentum of the observed meson, in order to achieve better accuracy in the intermediate region $p_T \geq m$. The massive hard-scattering cross sections have been constructed in such a way that the conventional hard-scattering cross sections in the $\overline{\mathrm{MS}}$ scheme are recovered in the limit $p_T \to \infty$ (or $m \to 0$). The requirement to adjust the massive theory to the ZM-VFNS with $\overline{\mathrm{MS}}$ subtraction is necessary, since all commonly used PDFs and FFs for heavy flavors are defined in this particular scheme. In this sense, this subtraction scheme is a consistent extension of the conventional ZM-VFNS for including charm-quark mass effects. It should be noted that our implementation of a GM-VFNS is similar to the ACOT scheme [16,17], which has been extended to one-particle-inclusive production of $B$ mesons a few years ago [48]. As explained in the second paper of Ref. [46,47], there are small differences concerning the collinear subtraction terms. Furthermore, in Ref. [48], the resummation of the final-state collinear logarithms has been performed only to leading logarithmic accuracy. The field-theoretical foundation of a GM-VFNS has been provided a few years ago by a factorization proof including heavy-quark masses [18]. Therefore, it is possible to extract improved universal parton distribution functions (PDFs) [22] and fragmentation functions (FFs) [49] from fits employing massive hard-scattering cross sections. From this perspective, it is important to compute massive hard-scattering cross sections in a given massive scheme for all relevant processes. Explicit calculations in the original ACOT scheme have been performed in Ref. [50,51] for inclusive and semi-inclusive deep-inelastic scattering (DIS). Furthermore, our calculation in Ref. [46,47] for hadronic collisions completes earlier work in the GM-VFNS on $D$-meson production in $\gamma\gamma$ and $\gamma p$ collisions [52–54], and it is planned to extend our analysis to the case of heavy-meson production in DIS.

[4]Contributed by B.A. Kniehl and I. Schienbein.

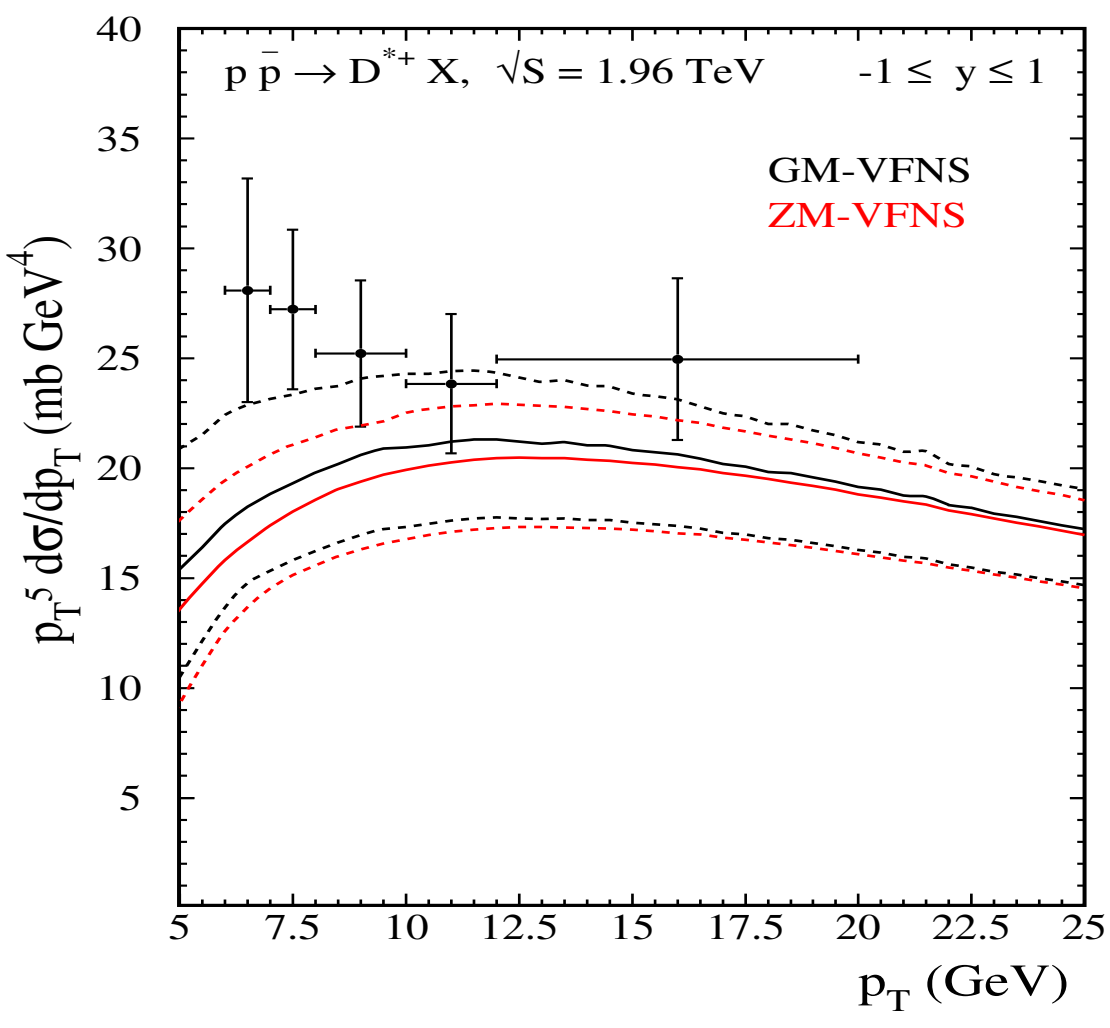

**Fig. 1:** QCD predictions for one-particle-inclusive production of $D^\star$ mesons at the Tevatron Run II in comparison with CDF data [57]. The results are shown for the average $D^\star = (D^{\star+} + D^{\star-})/2$. The solid lines have been obtained with $\mu_\mathrm{R} = \mu_\mathrm{F} = \mu'_\mathrm{F} = m_T$. The upper and lower dashed curves represent the maximum and minimum cross sections found by varying $\mu_\mathrm{R}$, $\mu_\mathrm{F}$, and $\mu'_\mathrm{F}$ independently within a factor of 2 up and down relative to the central values while keeping their ratios in the range $0.5 \le \mu_\mathrm{F}/\mu_\mathrm{R}, \mu'_\mathrm{F}/\mu_\mathrm{R}, \mu_\mathrm{F}/\mu'_\mathrm{F} \le 2$.

Next, we show predictions for the cross section $d\sigma/dp_T$ of $D^\star$-meson production obtained in the GM-VFNS and the ZM-VFNS. The cross section has been scaled with $p_T^5$ in order to arrive at a flat $p_T$ distribution, which is useful for visualizing the heavy-quark mass effects. The hard-scattering cross sections are convoluted with the (anti-)proton PDFs and FFs for the transition of the final-state parton into the observed $D^\star$ meson. We use the CTEQ6M PDFs [55] and the FFs for $D^\star$ mesons from Ref. [56]. As in the experimental analysis, the theoretical results are presented for the average $(D^{\star+}+D^{\star-})/2$. We consider $d\sigma/dp_T$ at $\sqrt{S} = 1.96$ TeV as a function of $p_T$ with $y$ integrated over the range $-1.0 < y < 1.0$. We take the charm mass to be $m = 1.5$ GeV and evaluate $\alpha_s^{(n_f)}(\mu_R)$ with $n_f = 4$ and scale parameter $\Lambda^{(4)}_{\overline{\mathrm{MS}}} = 328$ MeV, corresponding to $\alpha_s^{(5)}(m_Z) = 0.1181$. The results are presented in Fig. 1 for the GM-VFNS (black lines) and the ZM-VFNS (red lines) in comparison with CDF data [57]. The solid lines have been obtained with $\mu_\mathrm{R} = \mu_\mathrm{F} = \mu'_\mathrm{F} = m_T$. The upper and lower dashed curves represent the maximum and minimum cross sections found by varying $\mu_\mathrm{R}$, $\mu_\mathrm{F}$, and $\mu'_\mathrm{F}$ independently within a factor of 2 up and down relative to the central values requiring for their ratios to satisfy the inequalities $0.5 \le \mu_\mathrm{F}/\mu_\mathrm{R}, \mu'_\mathrm{F}/\mu_\mathrm{R}, \mu_\mathrm{F}/\mu'_\mathrm{F} \le 2$. As can be seen, for large values of $p_T$, the predictions of the GM-VFNS nicely converge to the corresponding results in the ZM-VFNS. Both approaches lead to reasonable descriptions of the data, but the inclusion of the positive mass effects clearly improves the agreement with the data. It should be noted that the mass effects are largest for the upper curves of the uncertainty band, which have been obtained with the smaller value of the renormalization scale implying a larger $\alpha_s(\mu_R)$. At $p_T = 5$ GeV, one observes an increase of the massless cross section by about 35%. A more detailed comparison of the GM-VFNS with CDF data [57] including $D^0$, $D^+$, and $D_s^+$ mesons can be found in Refs. [58, 59].

Residual sources of theoretical uncertainty include the variations of the charm mass and the employed PDF and FF sets. A variation of the value of the charm mass does not contribute much to the theoretical uncertainty. Also, the use of other up-to-date NLO proton PDF sets produces only minor differences. Concerning the choice of the NLO FF sets, we obtain results reduced by a factor of 1.2–1.3 when we use the NLO sets from Ref. [60], which is mainly caused by a considerably different gluon FF. A more detailed discussion can be found in Ref. [56].

**Table 1:** Process relevant for SM measurements and SUSY discoveries at the LHC which entail the use of bottom in the initial state. All of them are known at least at NLO accuracy.

| Name | LO Process | Interest | Accuracy |
|---|---|---|---|
| single-top t-channel | $qb \to qt$ | top EW couplings | NLO |
| single-top tW-associated | $gb \to tW^-$ | Higgs bckg, top EW couplings | NLO |
| Vector boson + 1 b-jet | $gb \to (\gamma, Z)b$ | b-pdf, SUSY Higgs benchmark | NLO |
| Vector boson + 1 b-jet +1 jet | $qb \to (\gamma, Z, W)bq$ | single-top and Higgs bckgs | NLO |
| Higgs inclusive | $b\bar{b} \to (h, H, A)$ | SUSY Higgs discovery at large $\tan\beta$ | NNLO |
| Higgs + 1 b-jet | $gb \to (h, H, A)b$ | SUSY Higgs discovery at large $\tan\beta$ | NLO |
| Charged Higgs | $gb \to tH^-$ | SUSY Higgs discovery | NLO |

## 4 A case study in collinear resummation: $b$-quark PDF from $Z + b$ production at LHC[5]

### 4.1 Introduction

The discovery of new physics at LHC will probably rely on the detailed understanding of standard-model background processes. Outstanding among these is the production of weak bosons ($W, Z$) in association with jets, one or more of which contains a heavy quark ($Q = c, b$). The prime example is the discovery of the top quark at the Fermilab Tevatron, which required a thorough understanding of the $W$+jets background, with one or more heavy-quark jets. The discovery of single-top-quark production via the weak interaction will require an even more sophisticated understanding of this background [61, 62].

For many processes involving production of heavy quarks, there are two ways (schemes) to perform the calculation in QCD: the fixed-flavor-scheme (FFS) and variable-flavor-scheme (VFS). The main practical difference between the two approaches is simple: in the VFS the heavy-quark can also be in the initial state, and in that case is assumed to be massless, while in the FFS it appears only as a final state (massive) particle. QCD factorisation tells us that if calculations could be performed at arbitrary high order, the two schemes would be equivalent. At fixed order, on the other hand, differences arise and one should choose that describing more effectively the kinematics of the process of interest. This freedom has sometimes created intense and fruitful debates among the QCD practitioners (see, *e.g.*, Ref. [63] for a detailed comparison of Higgs boson production in association with bottom quarks). Here we just recall the main two reasons for using a heavy-quark distribution function. First, it resums collinear logarithms of the form $\ln Q/m_Q$ to all orders, where $Q$ is the scale of the hard scattering and $m_Q$ is the mass of the heavy quark. Second, it simplifies the leading-order process, which often makes a higher-order calculation feasible. There are many processes in the standard model and in models beyond it, such as SUSY, that are better described using a bottom in the initial state. In Table 1, we give a non-exhaustive list of processes that will be relevant for QCD, EW and SUSY studies at the LHC, and the QCD order at which they are known.

At present the $b$ distribution function is derived perturbatively from the gluon distribution function [17, 18, 34, 55]. Recently, direct, albeit not very precise, measurements of $F_2^b$ have become available that are compatible with the perturbative determination [64, 65]. In the light of its phenomenological importance, a better direct determination of the $b$ distribution function is certainly desirable.

To this aim it has been proposed to use the associated production of a photon and a $b$-jet via $gb \to \gamma b$ at the LHC [66]. This measurement suffers from two main limitations. The first is the large contamination from charm which has a much larger cross section due to both the pdf and the electromag-netic coupling. The second is that the theoretical prediction at NLO for an isolated photon is uncertain, due to necessity of introducing a photon fragmentation function, which is at present poorly known.

[5] Contributed by S. Diglio, F. Maltoni, F. Petrucci, A. Tonazzo and M. Verducci

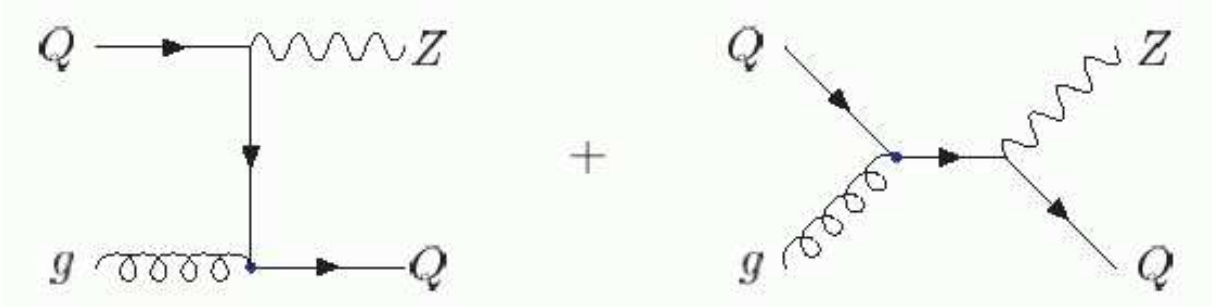

**Fig. 2:** Leading Order Feynman diagrams for associated production of a $Z$ boson and a single high-$p_T$ heavy quark ($Q = c, b$).

In this note we follow the suggestion of Ref. [67] to use $Z$ production in association with a $b$-jet to extract information on the $b$-pdf. At leading order, it proceeds via $gb \to Zb$, as shown in Fig. 2. This process is known at NLO, including $\gamma/Z$ interference effects. The advantages of using a $\gamma/Z$ decaying into leptons with respect to a real photon are noticeable. The NLO cross section is theoretically very well known and, apart from the PDF's, free of non-perturbative inputs. In addition, the competing process $gc \to Zc$ is suppressed by the ratio of the couplings of the charm and the bottom to the $Z$, and makes the $b$-pdf determination much cleaner.

The D0 Experiment at Tevatron has recently measured the cross-section ratio $\sigma(Z+b)/\sigma(Z+jet)$ [68], and their result is consistent with the NLO calculation.

As pointed out in [67], the measurement of this process at the LHC should be even more interesting because the contribution of the leading order process, sensitive to the $b$ content of the proton, is more relevant than at the Tevatron. In addition, the total cross-section is larger by a factor 50, and the relative contribution of background processes, mainly $Z+c$, is smaller. These features are summarised in Table 2, taken from Ref. [67].

**Table 2:** Next-to-leading-order inclusive cross sections (pb) for $Z$-boson production in association with heavy-quark jets at the Tevatron ($\sqrt{s}$ = 1.96 TeV $p\bar{p}$) and the LHC ($\sqrt{s}$ = 14 TeV $pp$). A jet lies in the range $p_T > 15$ GeV/c and $|\eta| < 2$ (Tevatron) or $|\eta| < 2.5$ (LHC). $ZQ$ indicates events containing a heavy quark, $Zj$ events which do not contain a heavy quark.

| Cross sections (pb) | Tevatron | LHC |
|---|---|---|
| Process | $ZQ$ inclusive | |
| $gb \to Zb$ | $13.4 \pm 0.9 \pm 0.8 \pm 0.8$ | $1040^{+70+70\ +30}_{-60-100-50}$ |
| $gb \to Zb\bar{b}$ | 6.83 | 49.2 |
| $gc \to Zc$ | $20.3^{+1.8}_{-1.5} \pm 0.1^{+1.3}_{-1.2}$ | $1390 \pm 100^{+60+40}_{-70-80}$ |
| $gc \to Zc\bar{c}$ | 13.8 | 89.7 |
| | $Zj$ inclusive | |
| $q\bar{q} \to Zg, gq \to Zq$ | $1010^{+44+9+7}_{-40-2-12}$ | $15870^{+900+60\ +300}_{-600-300-500}$ |

Besides the possibility of extracting the $b$-pdf, $Z + b$ represents also a benchmark and in some cases a background to the search of the Higgs boson, when it is produced in association with a single high-$p_T$ $b$ quark [63]: the dominant leading-order subprocess for the production of a Higgs boson via its coupling to the $b$ is $b\bar{b} \to h$; however, if the presence of a single $b$ with high $p_T$ is demanded, the dominant process becomes $gb \to hb$, with cross-sections of the order of tens of fb. The $h$ can then decay to the same final states as the $Z$; in particular, the decay $h \to \mu^+\mu^-$ is enhanced in some models [69–71].

A preliminary analysis on the potential of the ATLAS experiment to measure the $Z$+$b$-jet production at the LHC is presented in the following.

### 4.2 A study of LHC measurement potential

A sample of $Z$+jet events generated using the PYTHIA Monte Carlo [72] was processed with a fast simulation of the ATLAS detector, the ATLFAST package [73]. Only decays of the $Z$ boson to $\mu^+\mu^-$ were taken into account. The signal was defined as the sample events containing a $b$ quark with $p_T > 15$ GeV/c and $|\eta| < 2.5$. The background samples containing respectively a $c$ quark within the same cuts, or a jet originating from a light quark or a gluon in the same range, were considered separately. The NLO cross-sections computed in [67] were used for the signal and for these two classes of background, while the cross-section given by PYTHIA was taken for the other types of events.

The experimental selection of $Z$+jet events with $Z \to \mu^+\mu^-$ required the detection of two muons of opposite charge with $p_T > 20$ GeV/c and $|\eta| < 2.5$ and one hadronic jet. The presence of two high-$p_T$ muons ensures the possibility to have high trigger efficiency on this type of events. In addition, to reject the contribution from virtual photons, the invariant mass $M_{\mu\mu}$ of the muon pair was required to be close to the $Z$ mass (80 GeV/c$^2$ $< M_{\mu\mu} <$ 105 GeV/c$^2$). About 50% of signal events are retained after applying these cuts, the loss being equally due to the $\eta$ acceptance and to the $p_T$ cut.

The selection of events where the jet originates from a $b$ quark was based on two different tagging methods, as described in the following. Their complementarity is still to be studied in detail, however the comparison of two independent selections will be important to control the systematic uncertainties.

The first method to select $Z + b$ events was based solely on the presence of a third muon. Hadrons containing a $b$ quark give origin to prompt muon decays in about 12% of the cases. The efficiency of this method, therefore, cannot exceed this value, however the background is also expected to be small. The "third muon", considered to be the muon from the $b$ hadron decay, will in general be softer and closer to a jet than the muons from the $Z$ decay. The distribution of the transverse momentum of the third muon in $Z + j$ events is shown in Fig. 3. Different thresholds on the third muon $p_T$ were considered for the final selection.

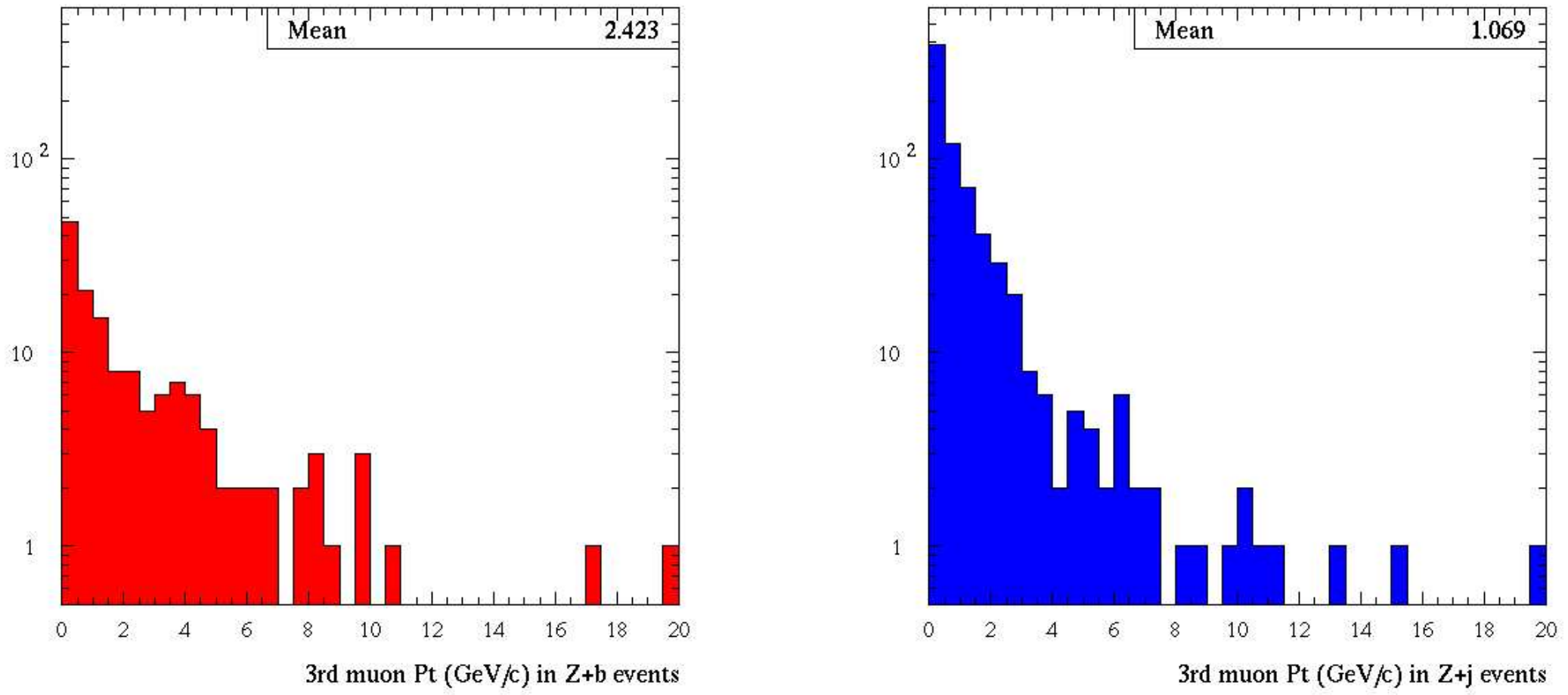

**Fig. 3:** Distribution of the transverse momentum of the third muon in a $Z$+jet sample, for signal events (left) and for events with no $b$ quark (right).

The second analysis used an inclusive method for $b$-tagging, based on the presence of secondary vertices and of tracks with high impact parameter with respect to the primary vertex, originated from the decay of the long-lived $b$ hadrons. The ATLFAST package reproduces the ATLAS $b$-tagging capabilities by applying the tagging efficiency on $b$ jets and a mis-tag rate on non-$b$ jets on a statistical basis, according to the values set by the user to reproduce the actual detector performance. The efficiency of the inclusive $b$-tagging on signal events, after the selection described above, is about 30%. The mistagging probability is about 4% on $c$-quark jets and 0.5% on light jets.

**Table 3:** Expected efficiency, statistics and purity in a data sample corresponding to an integrated luminosity of 10 fb$^{-1}$, using the soft muon tagging with different thresholds on the muon transverse momentum and the inclusive $b$-tagging. $N_b$ denotes the number of expected signal events as defined in the text, $N_c$ the number of selected events with a $c$ jet with $p_T > 15$ GeV and $|\eta| < 2.5$, $N_{other}$ the selected events from other processes. The statistical error on efficiencies and purities, due to the limited size of the simulated sample, is at the level of 1-2%.

| Cut | Efficiency | $N_b^{p_T>15\text{ GeV}, \|\eta\|\leq 2.5}$ | $N_c^{p_T>15\text{ GeV}, \|\eta\|\leq 2.5}$ | $N_{other}$ | Purity |
|---|---|---|---|---|---|
| $p_T^\mu > 4$ GeV/c | 4% | 13990 | 6270 | 0 | 69% |
| $p_T^\mu > 5$ GeV/c | 3% | 11090 | 5210 | 0 | 69% |
| $p_T^\mu > 6$ GeV/c | 2.5% | 8430 | 4180 | 0 | 67% |
| incl. $b$-tag | 14% | 49500 | 17400 | 49600 | 43% |

The overall efficiency on signal events, the expected number of signal and background events with an integrated luminosity of 10 fb$^{-1}$ and the expected purity of the selected samples are reported in table 3. With the fast simulation, the soft muon tagging capabilities are optimistic, in that full efficiency and no mis-tag are assumed for the lepton identification; more realistic assumptions will be made when the study is carried on with the full detector simulation. The efficiency on signal events achieved with the inclusive $b$-tagging method, where the results of the fast simulation are more realistic, is higher than with the soft muon tagging, while the purity of the selected sample is still quite good. Consistent results were obtained with a full simulation of the ATLAS experiment, on a small statistics sample.

A better determination of the signal component in the selected sample will eventually be achieved by exploiting the information on the transverse momentum of the $b$-jet or of the third muon.

Given the large statistics of the available data samples, the measurement will be limited by systematic effects.

The possibility to control the systematic effects directly from data samples has been explored, in particular the evaluation of $b$-tagging performance and of the residual background.

The $b$-tagging efficiency can be checked using $b$-enriched samples. Based on previous experience at Tevatron and LEP, we can expect a relative uncertainty of about 5%.

The background in the selected sample is mainly due to mis-tagged jets from $c$ and light quarks. This can be controlled by looking at the number of $b$-tagged jets in data samples that in principle should contain no $b$-jets at first order. $W$+jet events, for example, will be available with large statistics and with jets covering the full $p_T$ range of the signal. It can therefore be expected to estimate the background from mis-tagging with a relative uncertainty at the level of few percent, as shown by the plots in figure 4.

### 4.3 Conclusions and outlook

$Z$ boson production in association with a $b$-jet can provide information on the $b$-pdf.

A preliminary study of the $Z + b$ channel using a fast simulation of the ATLAS detector has shown that this type of event will be observed with very high statistics and good purity at the LHC. Given the large statistics of the samples, the precision of the $Z + b$ cross-section measurement will be limited by systematic effects. Some possibilities to evaluate systematic uncertainties directly form the data have been considered. An overall accuracy on the measurement at the level of 5% can be expected.

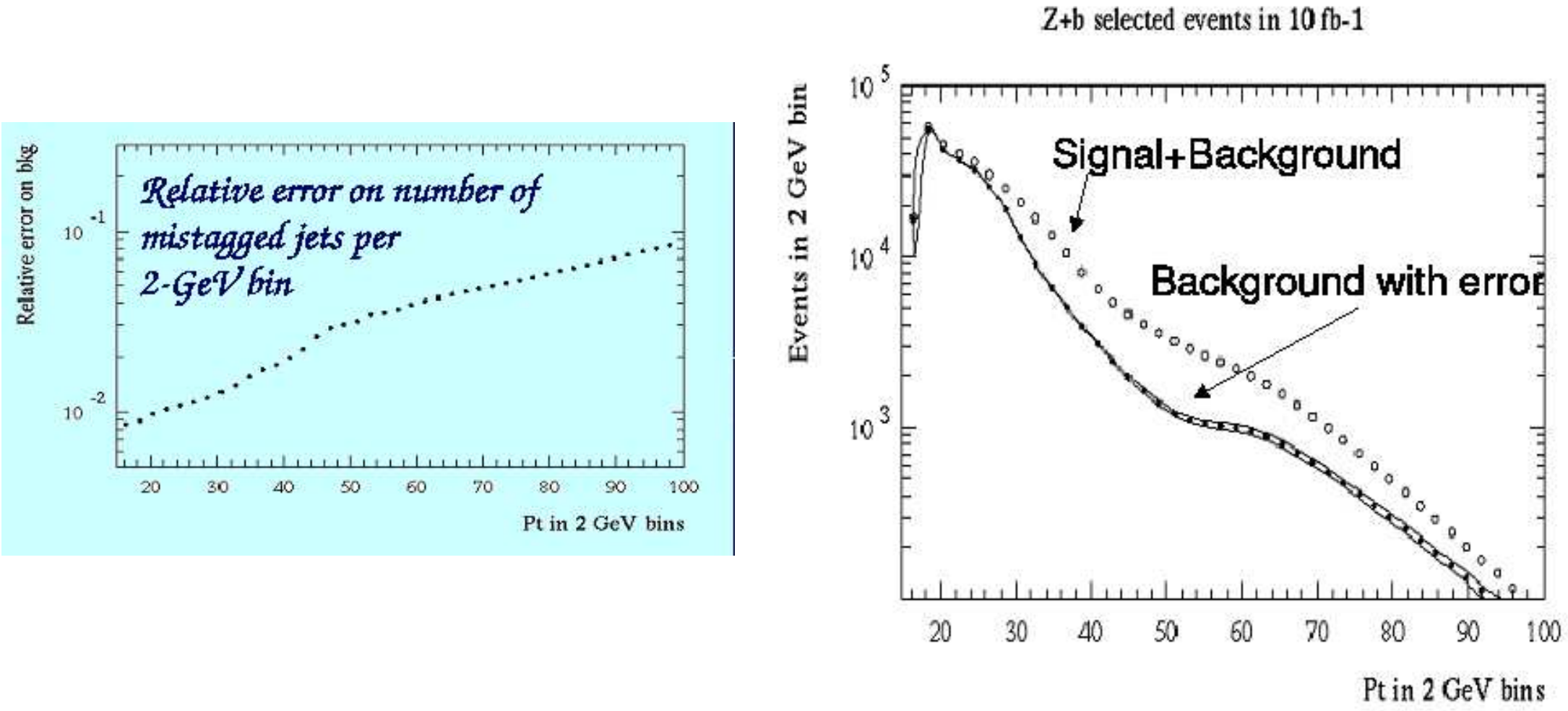

**Fig. 4:** Systematics due to mis-tagging of $b$-jets as evaluated from $W$+jet events. Left: relative error on background level per jet $p_T$ bin. Right: $p_T$ distribution of jets in event selected as $Z + b$; the error band on the background contribution represents the systematic uncertainty, as derived from the previous plot.

The availability of large samples opens interesting possibilities for the study of differential distributions: for instance, measuring the cross-section as a function of the $\eta$ and $p_T$ of the $Z$ boson would allow for the measurement of the $b$ PDF as a function of the momentum fraction carried by the quark inside the proton. These items are an additional topic for further studies.

## 5 Soft-gluon resummation[6]

QCD factorizes long- and short distance dynamics in inclusive cross sections with initial state hadrons into non-perturbative, but universal parton distribution functions, and perturbatively calculable hard scattering functions. Large remnants of the long-distance dynamics occur near the threshold edge of phase space in the form of logarithmic distributions that are singular at the edge. Resummation [74,75] of these effects organizes them to all orders in perturbation theory, and thereby extends the predictive power of QCD.

Threshold resummation is now a well-established calculational scheme with systematically improvable accuracy. It allows organization of all subleading powers of the logarithmic enhancements, and can be consistently matched to finite order perturbation theory. Resummed expressions, which take the form of exponentiated integrals over functions of the running coupling, require however a prescription for their numerical evaluation to handle a Landau pole singularity of the coupling. But for this intrinsic ambiguity (which must cancel against ambiguities in power corrections), threshold resummation is just as systematically improvable as the standard coupling constant expansion.

As stated earlier, the more differential a cross section, the better suited it is for phenomenology, because one may incorporate detector-specific acceptance cuts and thereby reduce the need for extrapolation. Therefore we should like to better understand the behavior of threshold-resummed expressions for double-differential cross sections. A study for the inclusive threshold-resummed heavy quark structure function can be found in Ref. [76]. Here we examine the differential structure function for the reaction

$$\gamma^*(q) + \mathrm{P}(p) \to \mathrm{Q}(p_1) + X'(p_2') \tag{16}$$

[6]Contributed by T.O. Eynck and E. Laenen.

which we write as

$$\frac{d^2F_2^Q(S,T_1,U_1)}{dT_1\,dU_1} \tag{17}$$

We define the invariants

$$\begin{aligned} S &= (p+q)^2 \equiv S' - Q^2, \qquad T_1 = (p-p_1)^2 - m^2, \\ U_1 &= (q-p_1)^2 - m^2, \qquad S_4 = S' + T_1 + U_1\,. \end{aligned} \tag{18}$$

The invariant mass squared of the final state $X'$ is given by

$$M_{X'}^2 = m^2 + S_4 \tag{19}$$

so that the elastic (threshold) limit for the subprocess (16) is approached by $S_4 \to 0$. It may be converted to the double-differential structure function in terms of the heavy quark transverse momentum and rapidity, e.g.

$$\frac{d^2F_k^Q}{d(p_T^Q)^2\,dy^Q} = S'\frac{d^2F_k^Q}{dT_1\,dU_1}\,, \tag{20}$$

where e.g. [11]

$$p_T^Q = \left[\frac{S'T_1U_1 + Q^2T_1^2 + Q^2S'T_1}{S'^2} - m^2\right]^{(1/2)}\,. \tag{21}$$

At the parton level one may define invariants equivalent to those in (18), which we will denote by using lower case. The order-by-order perturbation theory expansion for the partonic version of this distribution $\omega(s_4,t_1,u_1)$ and its all-order resummation have the following schematic forms

$$\begin{aligned} \omega &= 1 + \alpha_s(L^2+L+1) + \alpha_s^2(L^4+L^3+L^2+L+1) + \ldots \\ &= \exp\left(\underbrace{\underbrace{Lg_1(\alpha_sL)}_{LL} + g_2(\alpha_sL) + \ldots}_{NLL}\right)\underbrace{C(\alpha_s)}_{\text{constants}} \\ &\quad + \text{ suppressed terms} \end{aligned} \tag{22}$$

with

$$g_1(\lambda) = \frac{C_F}{\pi b_0\lambda}\Big[\lambda + (1-\lambda)\ln(1-\lambda)\Big]\,, \quad \lambda = b_0\alpha_s\ln N\,. \tag{23}$$

(We have also computed $g_2(\lambda)$; by including ever more $g_i$ functions in the exponent in Eq. (22) we can increase the parametric accuracy of the resummation.) The symbol $L^i$ represents, in this case, the logarithmically singular plus-distributions

$$\left[\frac{\ln^{i-1}(\rho)}{\rho}\right]_+ \tag{24}$$

with $\rho = s_4/m^2$, or, after a Laplace transform $\int d\rho\exp(-N\rho)$ by $\ln^i N$. The conversion to momentum space then requires a numerical inverse Laplace transform. For the case at hand one needs to compute

$$S'^2\,\frac{d^2F_2^Q(S_4,T_1,U_1)}{dT_1\,dU_1} = \int\limits_{c-i\infty}^{c+i\infty}\frac{dN}{2\pi i}e^{NS_4/m^2}\,\bar{\phi}_g\left(N\frac{S'+T_1}{m^2}\right)\,\omega\Big(N,T_1,U_1\Big)\,, \tag{25}$$

with $c$ the intercept of the contour with the real $N$ axis, and $\phi_g(N)$ the gluon density in moment space. We chose a toy density for the gluon PDF, and the minimal prescription [77] to perform the $N$ integral.

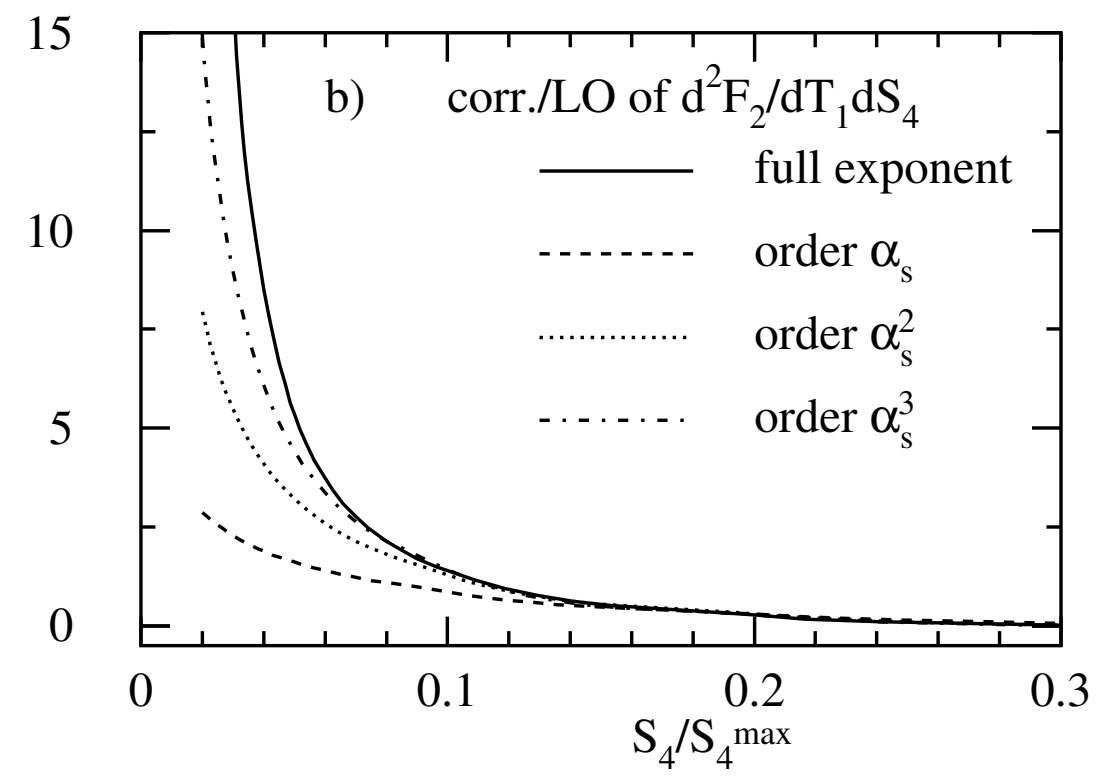

**Fig. 5:** Expandability of the resummed expressions for $d^2F_2^c/dT_1dS_4$ with NLL exponent (ratio to LO)

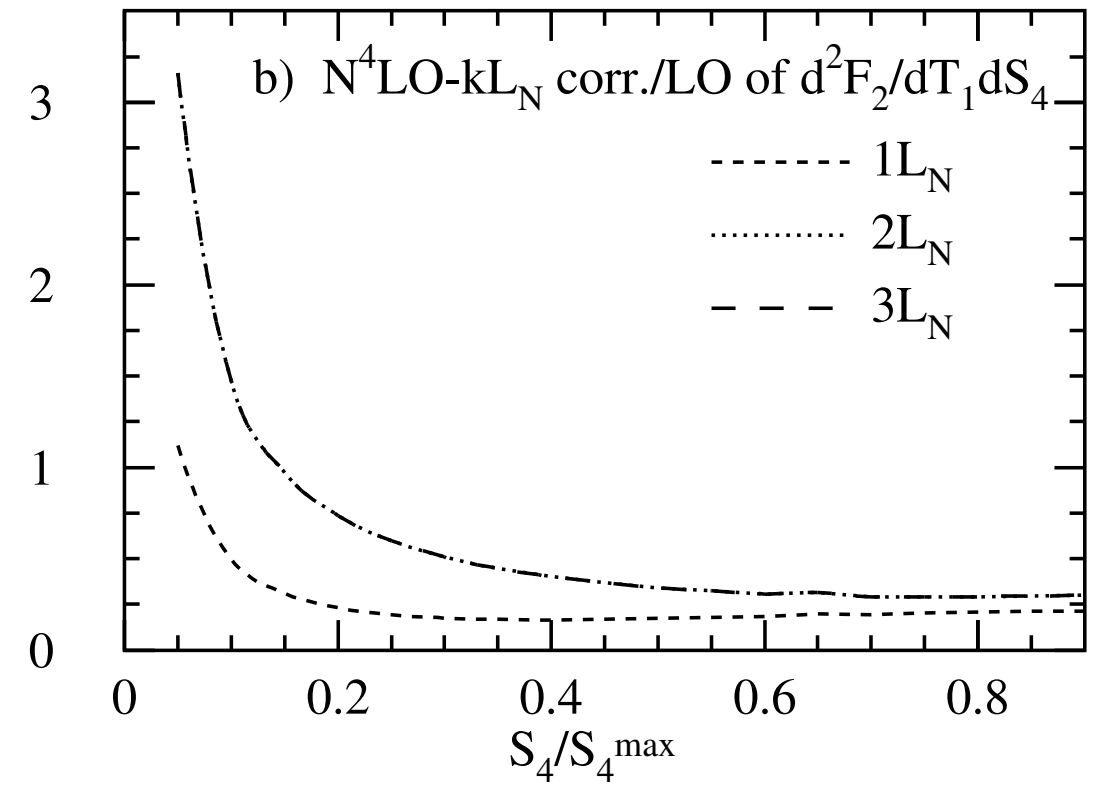

**Fig. 6:** Tower resummation at $\mathrm{N^4LO} - k\mathrm{L}_N$, $k \in \{1, 2, 3\}$ ($\mathrm{N^4LO} - 2\mathrm{L}_N$ and $\mathrm{N^4LO} - 3\mathrm{L}_N$ almost coincide).

In Fig. 5 we evaluate this expansion as a function of the recoil mass $S_4$, and compare it to its finite order expansions. We keep the variable $T_1$ fixed at the average of its minimum and maximum allowed value. Clearly, for reasonable values of $S_4$ the resummed result is already well-approximated by its 2nd and 3rd order expansions.

Another way to evaluate the resummed expression is in terms of towers of [78] $L = \ln N$.

$$\omega = h_{00}(\alpha_s) \left[ 1 + \sum_{k=1}^{\infty} \left( \frac{\alpha_s}{\pi} \right)^k \left( c_{k1}\, L^{2k} + c_{k2}\, L^{2k-1} + c_{k3}\, L^{2k-2} + \dots \right) \right] . \tag{26}$$

where the indicated coefficients $c_{kj}$ can be determined exactly. More accuracy here means including more subleading towers. This method is equivalent, but not identical to the minimal prescription method. In practice, one need only include the first 4 terms in each tower, the higher terms are vanishingly small. The ambiguities mentioned earlier are shifted to far-subleading towers in this approach. To exhibit the convergence of terms in the towers, it will be useful and illustrative to exhibit contributions of a particular order in the strong coupling and the large logarithms. We will employ the notation

$$\mathrm{N}^k\mathrm{LO} - l\mathrm{L}_N \tag{27}$$

for finite order results, where $k$ indicates the order in the strong coupling, the subscript $N$ denotes moments, and $l$ expresses if only the leading term ($l = 1$, $\mathrm{L}^2k_N$), or also the next-to-leading term ($l = 2$, $\mathrm{L}_N^{2k-1}$) is included, etc. In Fig. 6 we see also in this approach a rapid convergence toward the tower-resummed result.

A more complete study of the relevance of threshold resummation for electroproduction of heavy quarks at HERA still awaits. We note that even if the size of the corrections does not cause much concern for the perturbative analysis of an observable, threshold resummation or its finite order approximations, often lead to a reduction of scale dependence [79], indeed also seen in Ref. [76].

## 6 $k_t$ - factorization[7]

### 6.1 Introduction

The transverse momenta of the partons initiating a hard scattering process, like heavy quark production via $\gamma g \to Q\bar{Q}$ or $gg \to Q\bar{Q}$ in lepto- (hadro-) production, respectively, is mainly generated by the QCD evolution, which can reach large values, in DGLAP up to the factorization scale, in BFKL/CCFM/LDC even larger.

The typical transverse momenta of the gluons in a process $gg \to X$ for different masses $M$ of the system $X$ are shown in Fig. 7 as a function of the momentum fraction $x$ of one of the gluons for LHC energies. The transverse momenta can become large, so that they cannot be neglected. A theoretical approach, formulated for small $x$, which takes into account the tranverse momenta is the $k_t$-factorization [80, 81] or semi-hard [82] approach.

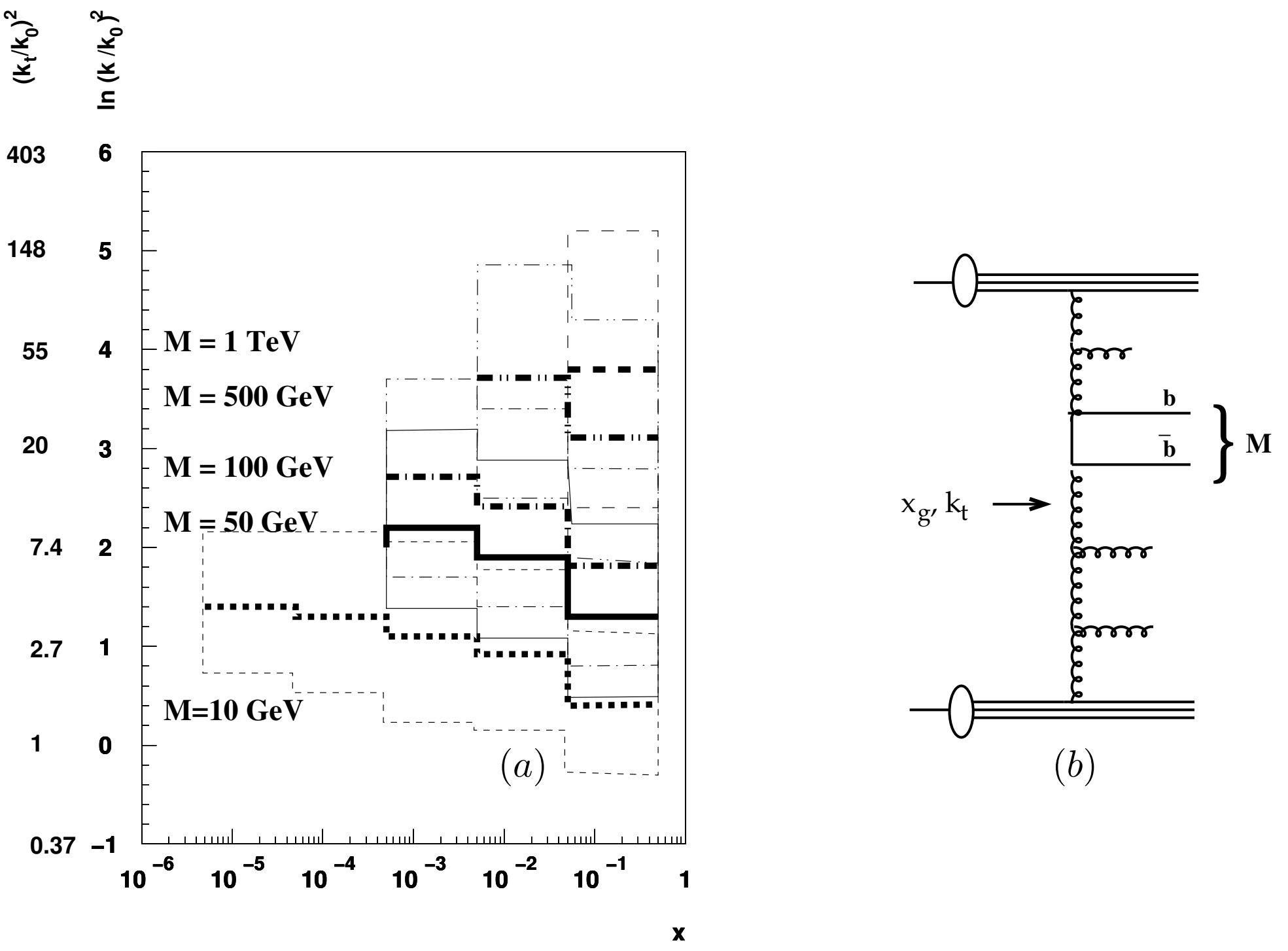

**Fig. 7:** Average tranverse gluon momenta $k_t$ in processes $gg \to X$ for different masses $M$ of the system $X$ as a function of the momentum fraction of one of the gluons $x$. The thin lines indicate the RMS spread of the distributions. In (b) is shown the definition of $x, k_t$ and $M$ for a gluon induced process.

In $k_t$-factorization the cross section for any process $pp \to X$ can be written as:

$$\sigma = \int dx_1 dx_2 \int dk_{t\,1} dk_{t\,2} \mathcal{A}(x_1, k_{t\,1}, q) \mathcal{A}(x_2, k_{t\,2}, q) \hat{\sigma}(x_1, x_2, k_{t\,1} k_{t\,2}, q) \tag{28}$$

with $\mathcal{A}(x, k_t, q)$ being the un-integrated ($k_t$-dependent) parton density function uPDF, $q$ defines the factorization scale and $\hat{\sigma}$ is the partonic cross section. The off-shell matrix-elements $\hat{\sigma}$ are calculated in [80, 81].

[7]Contributed by S.P. Baranov, H. Jung, A.V. Lipatov and N.P. Zotov

The effects of finite transverse momenta are present independent of the evolution scheme: uPDFs can be defined also for the DGLAP evolution. A more detailed discussion on these effects can be found in [83, 84].

It is interesting to note, that the $k_t$-factorization approach (in LO $\alpha_s$) agrees very well with calculations performed in the collinear approach in NLO $\alpha_s$, which is shown in [85]. The main effect comes from a more realistic treatment of the kinematics already in LO, which otherwise has to be covered in NLO. The $k_t$ factorization approach, however, is strictly valid only at small $x$, where the virtuality of the exchanged gluons can be identified with its tranverse momentum $k^2 \sim -k_t^2$. The full expression for the virtuality is [86]:

$$k^2 = \frac{-k_t^2}{1-x} - \frac{x \cdot m^2}{1-x} \tag{29}$$

with $m$ being the recoiling mass of the hadronic system except the hard scattering process, taking into account the history of the evolution process. For finite $x$ the mass effects can be substantial.

### 6.2 Open $b\bar{b}$ production and correlations at the LHC

Heavy quark production in the $k_t$-factorization approach at HERA and the Tevatron was considered already in many papers (see, for example, [82, 87–90]). In Ref. [91] the $k_t$-factorization approach was used for a more detailed analysis of the D0 and CDF experimental data. The effect of the initial gluon tranverse momenta on the kinematics of the $b\bar{b}$ production at the LHC were investigated [92]. The renormalization and factorization scales were set equal to either the initial gluon virtualities, $\mu_R^2 = \mu_F^2 = q_{T1,2}^2$, or $\mu_F^2 = m_{bT}^2$, as is in the standard collinear QCD, and the quark mass of $m_b = 4.75$ GeV was used.

In Fig. 8$a$ we show the transverse momentum distributions of $B$ mesons at LHC energies. The calculation was performed in the range $|\eta^B| < 1$ and the Peterson fragmentation with $\epsilon = 0.006$ using the KMS [93] parameterization for the un-integrated gluon density (see [83, 84]). The prediction for the azimuthal correlations between the muons coming from $B$ meson decays are shown in Fig. 8$b$ with $p_t^\mu > 6$ GeV and $|\eta^\mu| < 2.5$. The azimuthal correlations indicate an important theoretical difference between the collinear and $k_t$-factorization approaches. In the collinear approximation at parton level and leading order, the $b$ quarks are be produced exactly back-to-back, which is clearly unphysical when the gluon is evolved up to a large enough scale. Only starting with NLO a significant deviation from the back-to-back scenario is found. Thus the NLO calculation has to correct for the wrong kinematics in LO together with higher order corrections, leading to large $K$ factors. In the $k_t$-factorization, the transverse momenta of the gluons are correctly treated already in LO. In the $k_t$ - factorization approach the NLO corrections are therefore expected to be much smaller, since here only the purely dynamical corrections have to be applied, whereas the kinematics are already correctly treated in LO.

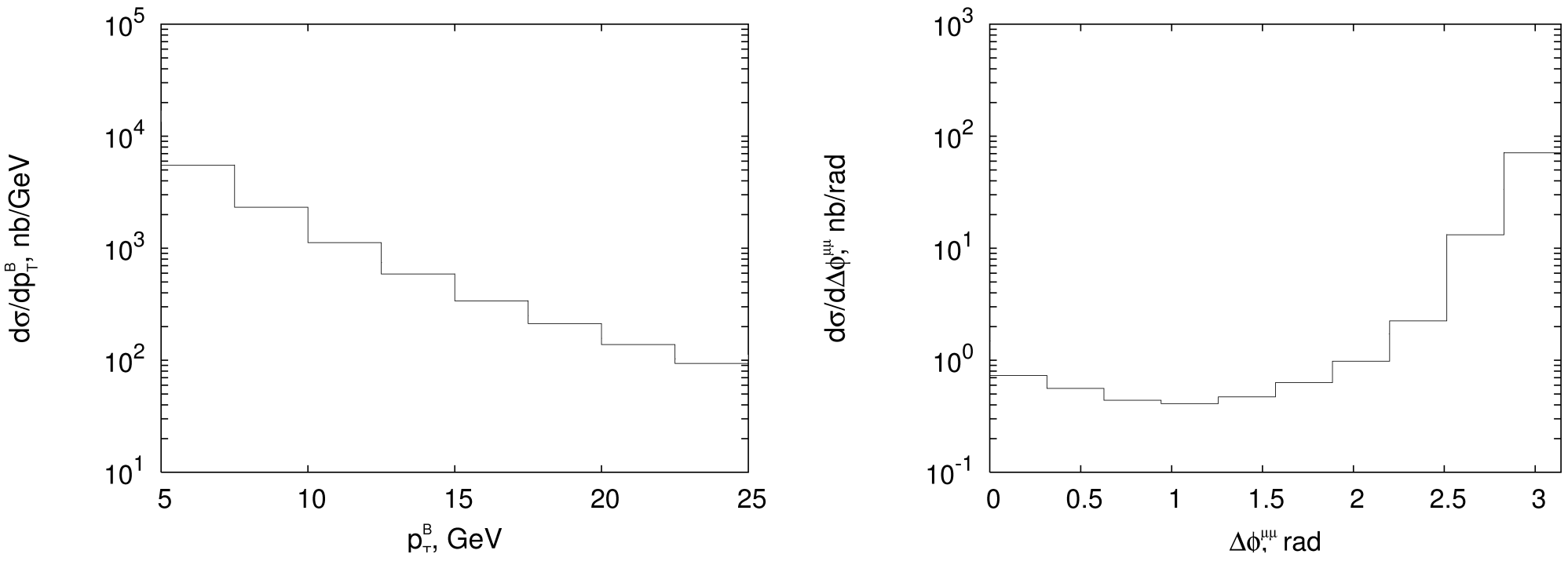

**Fig. 8:** Prediction for $B$-meson production at the LHC using the KMS un-integrated gluon density. In $a$ the $p_T$ distribution of $B$-mesons is shown. In $b$ the azimuthal $\mu\mu$ correlation coming from the $B$ decays is shown.

### 6.3 Quarkonium production and polarization at the LHC

Since the initial gluons have non-zero transverse momenta, they are off-shell, and they have a longitudinal component in their polarization vector. Typically, the $k_t$ values of the two colliding gluons are much different, as the parton evolution is equivalent to the random walk in the $\ln|k_t|$ plane, not in $k_t$ plane. Roughly speaking, the $k_t$ of one of the gluons can be neglected in comparison with that of the other. So, in the initial state we have one nearly on-shell (transversely polarized) gluon and one off-shell (longitudinally polarized) gluon. After the interaction, they convert into one on-shell gluon and a heavy vector meson. Simple helicity conservation arguments show that the polarization of vector meson must be longitudinal, in contrast with the ordinary parton model, where the initial gluons are both on-shell. This effect has been already studied for the HERA [94] and Tevatron [95] conditions. Fig.9$a$ shows the predictions for the LHC energy obtained with KMS [93] parameterization for un-integrated gluon densities. The calculations are restricted to the pseudorapidity interval $-2.5 < \eta_\Upsilon < 2.5$ and assume ATLAS "$\mu 6 \mu 3$" trigger cut, which means one muon with $p_t > 6$ GeV and another muon with $p_t > 3$ GeV.

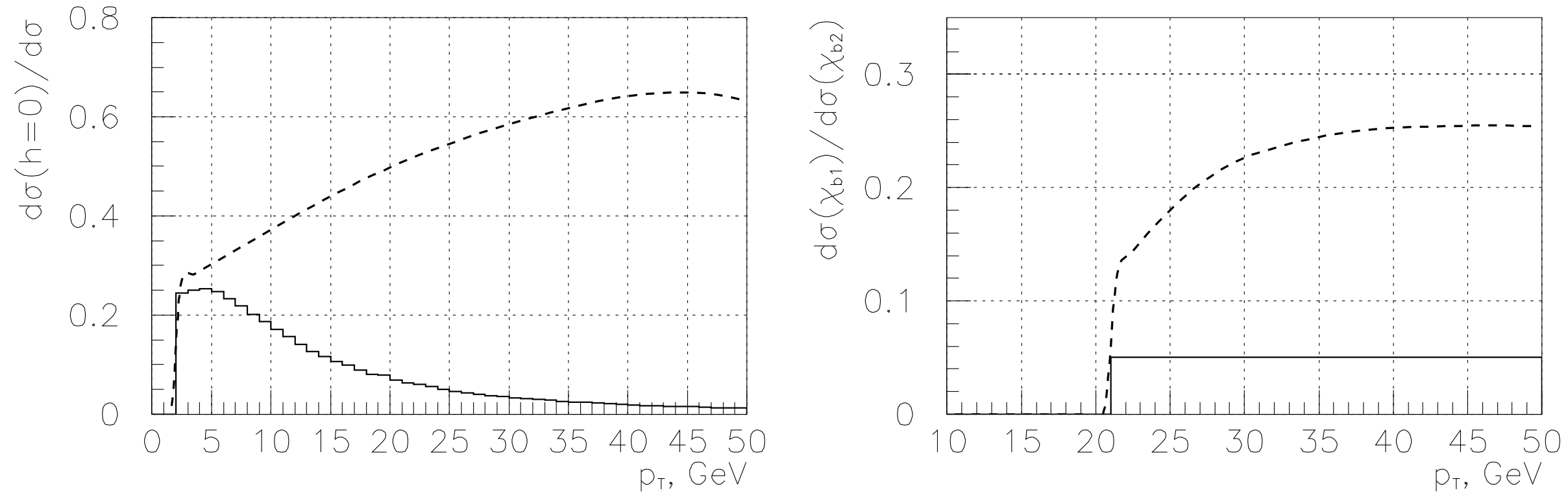

(a) (b)

**Fig. 9:** Predictions of different theoretical approaches for quarkonium production. In ($a$) the fraction of longitudinally polarized $\Upsilon$ mesons is shown: solid histogram – collinear parton model, singlet + octet; dashed – $k_t$-factorization with KMS u.g.d.. In ($b$) the ratio of the production rates $\chi_{b1}/\chi_{b2}$ is shown: solid histogram – collinear parton model, singlet + octet; dashed – $k_t$ factorization with KMS u.g.d.

Important effects are also seen in the production of $P$-wave bottomium states with different spins $\chi_{b1}$ and $\chi_{b2}$. At the Tevatron energies, this process has been considered in Ref. [96], and the predictions for the LHC are presented in Fig.9$b$. The $P$-wave states are assumed to be detected via the decay $\chi_b \to \Upsilon + \gamma$, with an additional requirement that the energy of the decay photon be greater than 2 GeV. The ratio of the production rates $\sigma(\chi_1)/\sigma(\chi_2)$ is qualitatively different in the $k_t$-factorization and the collinear parton model. The underlying physics is clearly connected with gluon off-shellness. In the collinear parton model, the relative suppression of $\chi_1$ states becomes stronger with increasing $p_T$ because of the increasing role of the color-octet contribution: in this approach, the leading-order fragmentation of an on-shell transversely polarized gluon into a vector meson is forbidden. In contrast with that, in the $k_t$-factorization approach, the increase in the final state $p_T$ is only due to the increasing transverse momenta (and virtualities) of the initial gluons, and, consequently, the suppression motivated by the Landau-Yang theorem becomes weaker at large $p_T$.

### 6.4 Associated Higgs + jets production at the LHC

The dominant mechanism for Higgs production at the LHC is gluon-gluon fusion, and the calculations can be significantly simplified in the large top mass limit ($M_H \leq 2M_{top}$) [97].

The differential cross section of the inclusive Higgs production $p\bar{p} \to H + X$ in the $k_t$-factorization approach has been calculated in [98, 99] and can be written as:

$$\frac{d\sigma(p\bar{p} \to H + X)}{dy_H} = \int \frac{\alpha_s^2(\mu^2)}{288\pi} \frac{G_F\sqrt{2}}{x_1 x_2 m_H^2 s} \left[m_H^2 + \mathbf{p}_T^2\right]^2 \cos^2(\Delta\varphi) \times$$
$$\times \mathcal{A}(x_1, \mathbf{k}_{1T}^2, \mu^2) \mathcal{A}(x_2, \mathbf{k}_{2T}^2, \mu^2) d\mathbf{k}_{1T}^2 d\mathbf{k}_{2T}^2 \frac{d(\Delta\varphi)}{2\pi}, \quad (30)$$

where $G_F$ is the Fermi coupling constant, $\mathcal{A}(x, \mathbf{k}_T^2, \mu^2)$ is the un-integrated gluon distribution, $\Delta\varphi$ the azimuthal angle between the momenta $\mathbf{k}_{1T}$ and $\mathbf{k}_{2T}$, and the transverse momentum of the produced Higgs boson is $\mathbf{p}_T = \mathbf{k}_{1T} + \mathbf{k}_{2T}$. It should be noted, that this process is particularly interesting in $k_t$-factorization, as the transverse momenta of the gluons are in the same order as their longitudinal momenta ($\sim \mathcal{O}(10$ GeV$)$) [100].

The total inclusive Higgs production cross section at the LHC energies ($\sqrt{s} = 14$ TeV) is plotted in Fig. 10($a$) as a function of the Higgs mass in the mass range $m_H = 100 - 200$ GeV. The solid line is obtained by fixing both the factorization and renormalization scales at the default value $\mu = m_H$ with the J2003 (set 1) CCFM un-integrated gluon distribution [101]. In order to estimate the theoretical uncertainties, we take $\mu = \xi m_H$ and vary the scale parameter $\xi$ between $1/2$ and 2 about the default value $\xi = 1$. The uncertainty band is presented by the upper and lower dashed lines. We find that our central values agree very well with recent NNLO results [102].

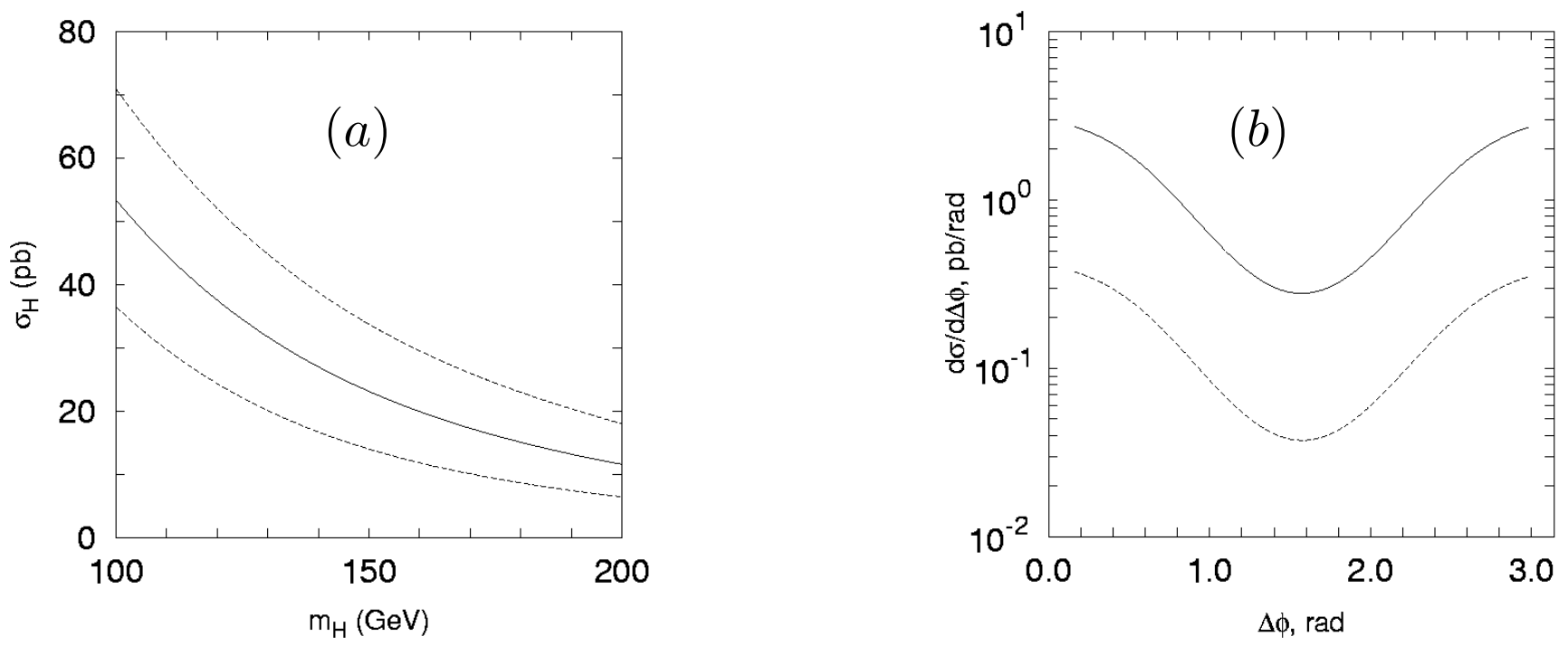

**Fig. 10:** Higg production at the LHC. In $(a)$ the total cross section for Higgs boson production as a function of Higgs mass is shown: the solid curve corresponds to the default scale $\mu = m_H$, upper and lower dashed curves - $\mu = m_H/2$ and $\mu = 2m_H/2$ respectively. In $(b)$ the jet-jet azimuthal angle distribution in the Higgs+jet+jet production at $\sqrt{s} = 14$ TeV. The kinematical cut $|\mathbf{p}_{\text{jet}\,T}| > 20$ GeV was applied for both jets. Solid and dashed lines correspond to the J2003 (set 1) and J2003 (set 2) [101] u.g.d. respectively.

To demonstrate the capabilities of the $k_t$-factorization approach, we calculate the azimuthal angle $\Delta\phi$ distribution between the two final jets transverse momenta in the Higgs+jet+jet production process. Our results are shown in Fig. 10($b$). The dip at $\Delta\phi = \pi/2$ comes from the $\cos(\Delta\varphi)$ in eq.(30). In the approach presented here, the $k_t$ of the initial gluons is approximately compensated by the transverse momenta of the jets [103]: $\mathbf{k}_T \simeq -\mathbf{p}_{T,jet}$, and, consequently, $\Delta\phi \simeq \Delta\varphi$ applying a cut-off $|\mathbf{p}_{\text{jet}\,T}| > 20$ GeV. This dip is characteristic for the loop-induced Higgs coupling to gluons in the framework of fixed-order perturbative QCD calculations [102]. Thus, we illustrate that the features usually interpreted as NNLO effects are reproduced in the $k_t$-factorization with LO matrix elements.

However, we see a very large difference (about one order of magnitude) between the predictions based on the J2003 gluon densities set 1 and set 2 [101], showing the sensitivity to the shape of the un-integrated gluon density.

### 6.5 Conclusions

The finite $k_t$ of the initial state gluons significantly modifies the kinematics of the gluon-gluon fusion process and leads to nontrivial angular correlations between the final state heavy quarks. The longitudinal polarization of the initial off-shell gluons manifests in the longitudinal polarization of $J/\psi$ and $\Upsilon$ mesons at moderate $p_T$ and, also, affects the production rates of $P$-wave quarkonia.

The predictions in $k_t$-factorization are very close to NNLO pQCD results for the inclusive Higgs production at the LHC, since the main part of high-order collinear pQCD corrections is already included in the $k_t$-factorization. In the $k_t$-factorization approach the calculation of associated Higgs+jets production is much simpler than in the collinear factorization approach. However, the large scale dependence of our calculations (of the order of $20 - 50\%$) probably indicates the sensitivity to the unintegrated gluon distributions.

## 7 Baryon charge transfer and production asymmetry of $\Lambda^0/\bar{\Lambda}^0$ in hadron interactions[8]

### 7.1 Introduction to the QGSM

The phenomenon of nonzero asymmetry of baryon production with nonbaryonic beams ($\pi$,$\mu$,$e$) was mentioned and explained in a few theoretical papers. Baryon charge can be transferred from proton or nucleus targets through the large rapidity gap with the string junction. In baryonic beam reactions ($p$,$A$, etc.) this effect is displayed in the valuable baryon/antibaryon spectrum asymmetry at $y = 0$. Every theoretical discourse on baryon charge transfer appeals to the value of the intercept, $\alpha_{SJ}(0)$, that is an intercept of the Regge-trajectory of imaginary particles which consists only of string junctions from baryon and antibaryon. Practically, the models that can account for this effect are only non perturbative QCD phenomenological models: the Dual Parton Model (DPM) [104] and the Quark Gluon String Model (QGSM) [105] as well as the DPMJET Monte Carlo expansion of these two models. Both analytical models are similar and they were based on the common Regge asymptotic presentation of constituent quark structure functions and string (quark) fragmentation functions. Here we are considering QGSM. In the comparison to the other models, QGSM accounts for many Pomeron exchanges. This approach works very well to give us the correct description of particle production cross sections at very high energies. The QGSM procedures of constructing of quark/diquark structure functions and fragmentation functions were presented in many previous publications. We take into consideration the $\pi$-$p$ reaction that gives similar asymmetries as the $\gamma$-$p$ reaction. The spectra in this reaction are more sensitive to the baryon excess in the region of positive $x_F$ than the spectra of baryons in $p$-$p$ collisions.

### 7.2 Comparison with Experimental Data

The asymmetry $A(y)$ between the spectra of baryons and antibaryons is defined as:

$$A(x) = \frac{dN^{\Lambda^0}/dx - dN^{\bar{\Lambda}^0}/dx}{dN^{\Lambda^0}/dx + dN^{\bar{\Lambda}^0}/dx}, \tag{31}$$

The EHS and the NA49 experiments have presented evidence for a nonzero baryon production asymmetry in proton-proton fixed target interactions, measuring at $y = 0$ values of the order of 0.5 - 0.3. In pion-proton interactions (E769) we can see the $y$ dependence of the asymmetry and the measured asymmetry, which was multiplied by a factor of 2 in order to be compared with the $pp$ asymmetry.

The data from these experiments can be presented in one plot for $A(\Delta y)$, where $\Delta y$ is the rapidity distance from interacting target-proton (see Fig. 11). It is seen that the points are situated on the same line. If we add the data from proton-nucleus experiments (HERA-B and RHIC) they are still approximately on this line. Only the STAR asymmetry point at $\sqrt{s} = 130$ GeV can be interpreted as a sign that the curve is bent. And the result of the H1 experiment at HERA [106] calls certainly for a steeper curve.

[8]Contributed by O.I. Piskounova.

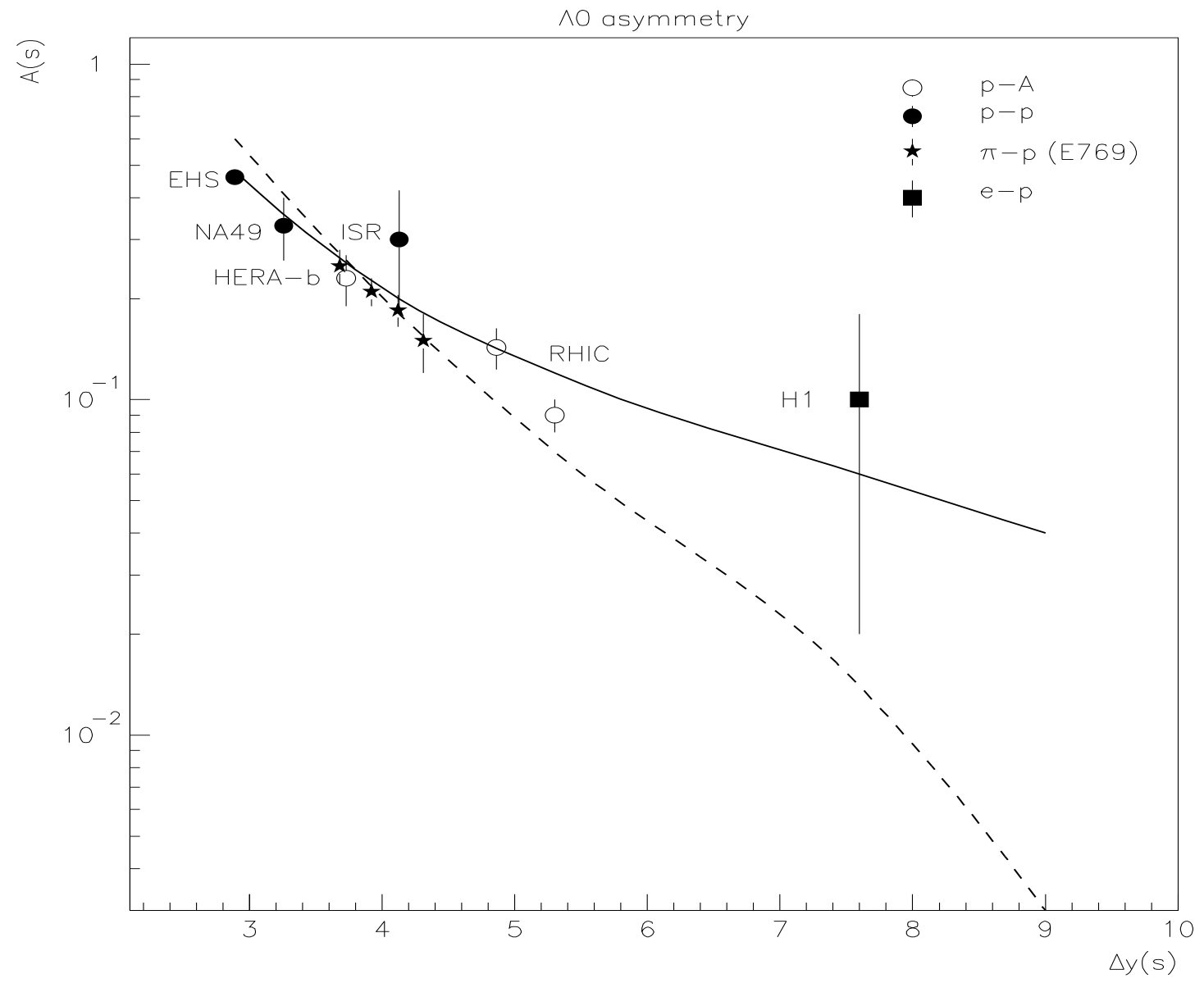

**Fig. 11:** Asymmetry in $\Lambda^0$ and $\bar{\Lambda}^0$ production and QGSM curves: $\alpha_{SJ}$(0)=0.5(dashed line) and $\alpha_{SJ}$(0)=0.9 (solid line).

What means do we have in QGSM to describe this dependence? In Ref. [107] it was shown that the data of the E769 and H1 experiments can not be described with the same value of $\alpha_{SJ}(0)$: the points at lower energy correspond to $\alpha_{SJ}$(0)=0.5, while the H1 point requires $\alpha_{SJ}$(0)=0.9.

### 7.3 Summary

The purpose of this contribution is to show the band of asymmetries that can be predicted for the LHC experiments between the two possibilities given above for $\alpha_{SJ}$(0). The results are shown in Fig. 11. The solid line represents the case of $\alpha_{SJ}$(0)=0.9. This curve fits the data at low energies (small $\Delta y$) due to varying the energy splitting between string junction and diquark configuration: 0.1*SJ+0.9*DQ. What we had to tune also was the fragmentation parameter af=0.15 instead of 0.2 accepted in previous papers. Also the curve for $\alpha_{SJ}$(0)=0.5 is shown in Fig. 11 with a dashed line. This line certainly doesn't fit the H1 point and gives a negligible asymmetry at the energy of the LHC experiments. Finally, we have the prediction for strange baryon asymmetries at the LHC within the range: $0.003 < A < 0.04$. The same procedure has to be applied to charmed baryon asymmetry to get the predictions at LHC energy.

# Experimental overview of heavy quark measurements at HERA

*O. Behnke*[1], *A. Geiser*[2], *A. Meyer*[2]
[1] Universität Heidelberg, Germany
[2] DESY Hamburg, Germany

**Abstract**
Experimental results on heavy flavour production at HERA are reviewed in the context of their relevance for future measurements at the LHC.

## 1 Introduction

Measurements of heavy flavour production at HERA can have significant impacts on the preparation and understanding of heavy flavour measurements at the LHC, and on the understanding of background processes to new physics discoveries [1]. The purpose of this contribution is to summarize the current status of heavy flavour measurements at HERA, and provide an outlook on how they might improve in the near future. The relation of these measurements to measurements at the LHC will be covered in more detail in subsequent contributions [2–4]. Since the top quark is too heavy to be produced at HERA with a significant rate, the term “heavy flavour” refers to $b$ and $c$ quarks only. The dominant diagram for heavy flavour production at HERA is shown in Fig. 1. The theory of heavy quark production at HERA is covered in the theoretical review section [5].

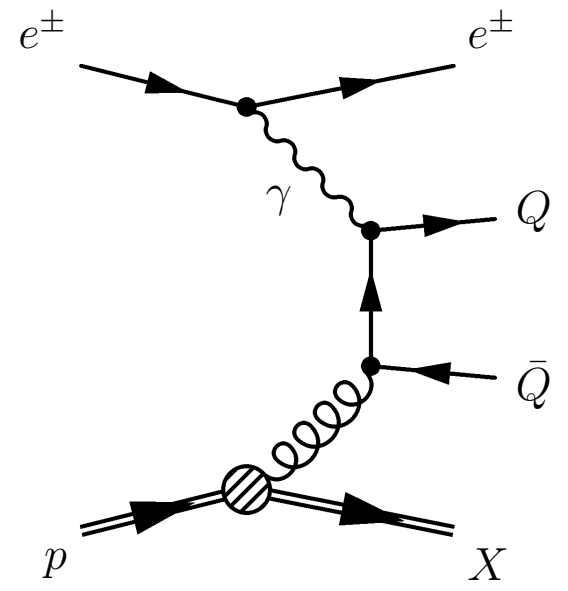

**Fig. 1:** Feynman graph for the production of a heavy quark pair via the leading order boson-gluon-fusion (BGF) process.

The interest in heavy flavour production arises from several aspects.

- Tagging a heavy flavour particle, e.g. inside a jet, establishes that this jet arises from a quark rather than a gluon. The number of possible QCD diagrams is thus reduced, and specific QCD final states can be studied more precisely than in inclusive measurements. This is even more true when *both* quarks of a heavy flavour quark pair can be tagged.
- The charm and beauty masses ($m_b, m_c \gg \Lambda_{QCD}$) provide energy scales which are large enough to allow perturbative calculations using a “massive” scheme [6, 7]. *All* QCD-processes involving heavy quarks should thus be reliably calculable. However, these mass scales often compete with other scales occurring in the same process, such as the transverse momentum ($p_T$) of the heavy quarks, or the virtuality of the exchanged photon, $Q^2$. Since the perturbative expansion can not be optimized for all scales at once, additional theoretical uncertainties enter which reduce the reliability of the predictions. If one of the competing scales ($p_T, Q^2$) is much larger than the quark mass, approximations in which the heavy quarks are treated as massless [8–14] can improve the reliability. Mixed schemes [15–17] are also possible. Understanding and resolving these difficulties should contribute to the understanding of multi-scale problems in general.

– Tagging the final state also constrains the initial state. Therefore, heavy flavour measurements can be used to measure or constrain parton density functions. In particular, Fig. 1 illustrates the direct sensitivity to the gluon density in the proton. Alternatively, in appropriate kinematic ranges, the initial state splitting of the gluon or photon into a heavy quark pair can be absorbed into the parton density definition. If the mass can be neglected, the same diagram (or higher order variants of it) can be reinterpreted as a way to measure the heavy flavour content of the proton or of the photon. These can in turn be used to calculate cross sections for other processes, such as Higgs production at the LHC.
– The production of "hidden" heavy flavour states (quarkonia) yields further insights into the interplay of (perturbative) heavy quark production and (non-perturbative) binding effects.

At HERA, the fraction of charm production vs. inclusive QCD processes is of order 10% in the perturbative QCD regime. Reasonably large samples can therefore be obtained despite the partially rather low tagging efficiency. Beauty production is suppressed with respect to charm production by the larger $b$ mass, and by the smaller coupling to the photon. The resulting total cross section is two orders of magnitude smaller than the one for charm. Beauty studies at HERA are thus often limited by statistics. Kinematically, beauty production at HERA is similar to top production at LHC ($m_b/\sqrt{s_{HERA}} \sim m_t/\sqrt{s_{LHC}}$). On the other hand, in the "interesting" physics region beauty is produced as copiously at the LHC as charm is at HERA.

## 2 Open charm production

Charmed mesons are tagged at HERA in different ways. A typical mass distribution for the "golden" channel $D^{*+} \to D^0\pi^+, D^0 \to K^-\pi^+$ (+ c.c.) is shown in Fig. 2 [20]. Despite the low branching ratio, this channel yields large statistics charm samples of high purity. Fig. 3 [21] shows a corresponding D* production cross section in photoproduction for different kinematic variables. In general, D* production is well described by next-to-leading order QCD predictions, although the data often prefer the upper edge of the theoretical error band. Some deviations are observed in particular regions of phase space. For instance, there are indications that forward (i.e. in the direction of the proton) charm production might be slightly larger than theoretical expectations (Fig. 3b). Also, there are regions of phase space which effectively require four-body final states which are not covered by NLO calculations (see Fig. 5 in [1]). In order to describe such phase space regions, either NNLO calculations, or parton shower extensions to NLO calculations such as MC@NLO [18, 19] will be needed.

Other ways to tag charm include the reconstruction of a secondary vertex in a high resolution Micro-Vertex-Detector (MVD) in addition to the reconstruction of a charmed meson mass (Fig. 4) [22], or the reconstruction of inclusive multiple impact parameters resulting from the finite charm meson lifetime. A resulting cross section for $D^+$ production is shown in Fig. 5.

Since the charm mass of approximately 1.5 GeV is not very much above the threshold at which perturbative calculations are believed to produce reliable results, the generally good agreement of perturbative QCD predictions with the data is highly nontrivial, and encouraging concerning the validity of corresponding predictions for the LHC.

## 3 Open beauty production

Open beauty production is detected at HERA using essentially three different methods, related to the large $b$ mass or long $b$ lifetime.

– If a jet is built out of the fragmentation and decay products of a $b$ meson/quark, the transverse momentum of the decay products with respect to the jet axis will be of order half the $b$ mass. This is significantly larger than for decay products of charm particles, or the transverse momenta induced by non-perturbative fragmentation effects, which are both of order 1 GeV or less. This distribution of this transverse momentum, called $p_T^{rel}$, can thus be used to measure the beauty contribution to a given jet sample.

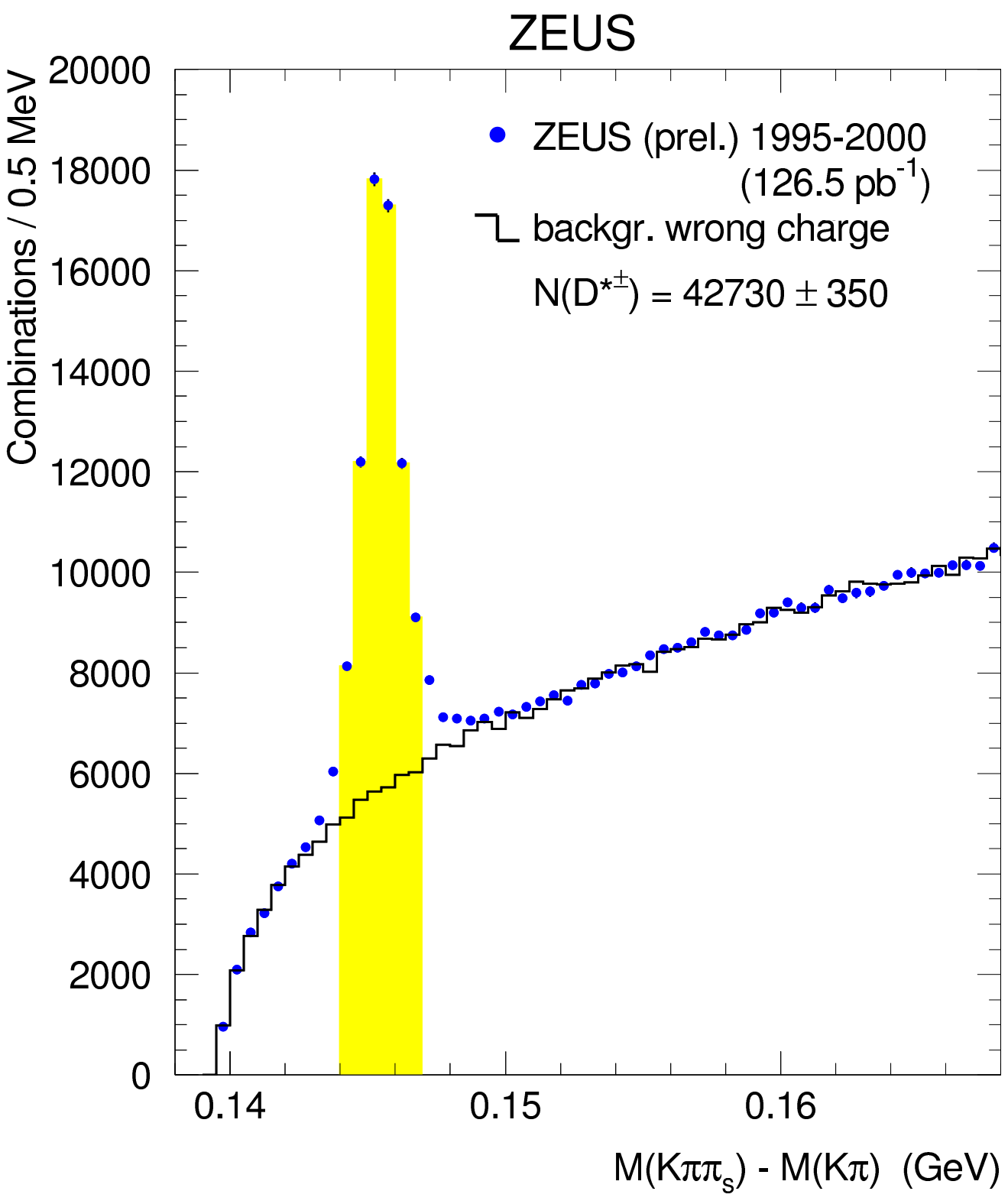

**Fig. 2:** Total inclusive $D^{*\pm}$ sample obtained by ZEUS for the HERA I data period in the golden decay channel $D^{*+} \rightarrow D^0\pi_s^+ \rightarrow (K^-\pi^+)\pi_s^+$.

- Due to the CKM-suppressed weak deacy of the $b$ quark, the lifetime of $b$ hadrons is longer than that of charmed particles. Furthermore, the larger decay angle due to the larger mass results in a higher significance of the decay signature. Detectors with a resolution in the $100\ \mu$ region or better can thus separate the beauty contribution from charm and light flavour contributions.
- Again due to their mass, $b$ hadrons take a larger fraction of the available energy in the fragmentation process. Furthermore, they produce decay products with sizeable transverse momentum even when produced close to the kinematic threshold. Simple lower cuts on the transverse momentum of such decay products therefore enrich the beauty content of a sample. Applying such cuts on two different decay products (double tagging) often sufficiently enriches the beauty content such that the remaining background can be eliminated or measured by studying the correlation between these decay products.

An example for an analysis using the first two methods with muons from semileptonic $b$ decays is shown in Fig. 6 [23]. Some cross sections resulting from this type of analysis are shown in Fig. 3 of [1]. In general, reasonable agreement is observed between the data and corresponding NLO QCD predictions, although, as in the charm case, the data tend to prefer the upper edge of the theoretical error band. In some regions of phase space, e.g. at low $p_T^\mu$ or high $\eta^\mu$ differences of up to two standard deviations are observed. More precise measurements (section 8) will be needed to decide whether these deviations are really significant.

An example for an analysis using the 2nd method only is shown in Fig. 7 [24], while an example for an analysis using the third method is shown in Fig. 8 [25].

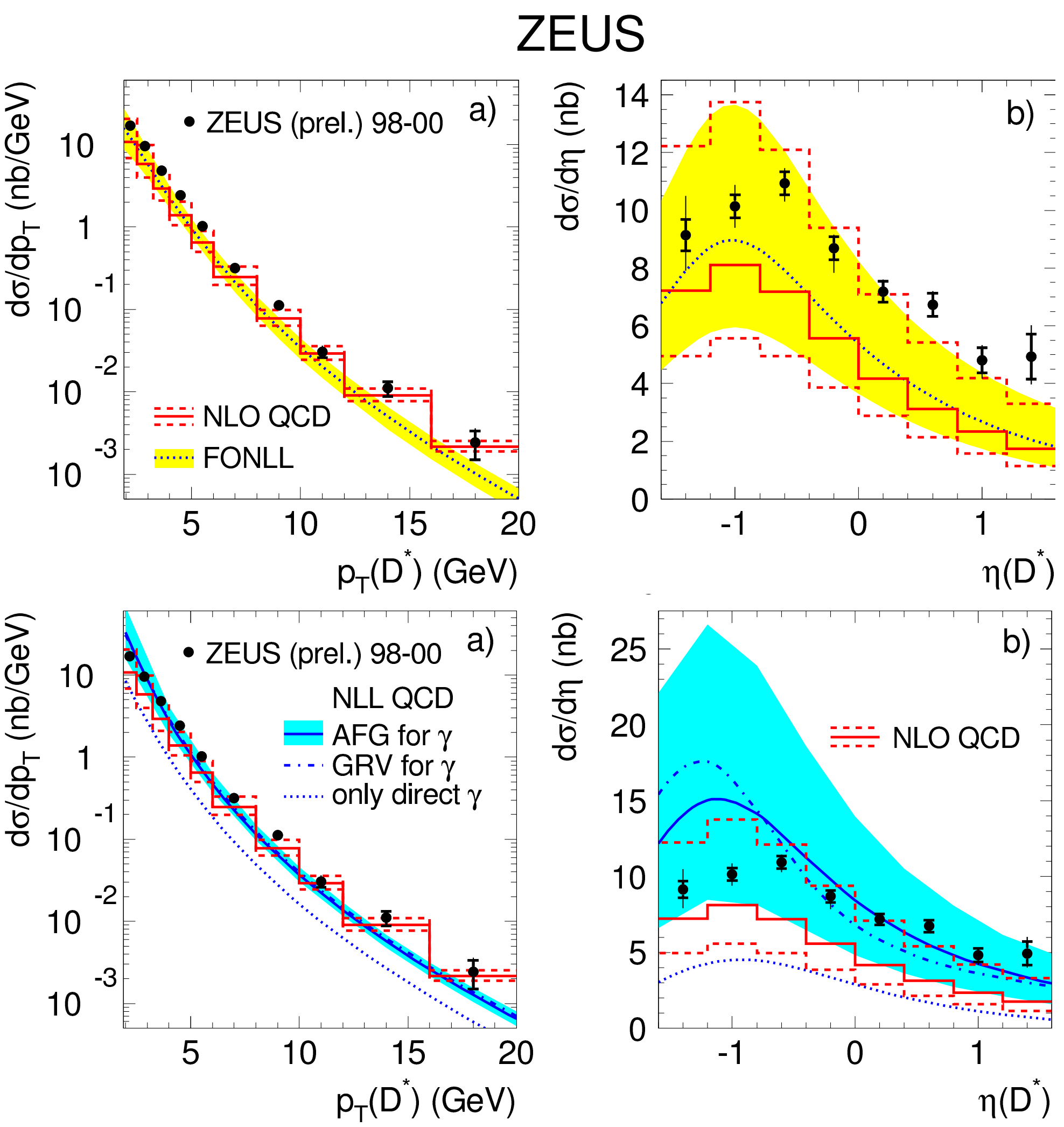

**Fig. 3:** $D^{*\pm}$ single differential cross sections in photoproduction as function of the $D^{*\pm}$ transverse momentum and pseudorapidity. The measurements are compared to NLO calculations in the massive (NLO), massless (NLL) and mixed (FONLL) scheme.

Figure 9 shows a summary of the data/theory comparison for all HERA beauty results as a function of $Q^2$. For the measurements sensitive to $b$ quarks with $p_T^b \sim m_b$ or lower (black points) there is a trend that the "massive" NLO QCD predictions [7] tend to underestimate the $b$ production rate at very low $Q^2$. Depending on the chosen set of structure functions and parameters, a "mixed" prediction (VFNS) [16,17] might describe the data better. For the higher $p_T$ measurements (red/grey points), no clear trend is observed. Note that theoretical errors, which are typically of order 30%, are not shown. Fig. 10 shows a similar compilation for all HERA measurements in photoproduction ($Q^2 < 1$ GeV), now as a function of the $b$ quark $p_T$. A similar trend is observed towards low $p_T$ (but note that several measurements appear in both figures). Again, more precise measurements are needed to determine whether these trends can be confirmed.

## 4 Quarkonium production

Inelastic heavy quarkonia, like open charm and beauty production, are produced at HERA via the process of photon-gluon fusion. The two charm or beauty quarks hadronize to form a charmonium or bottomonium state.

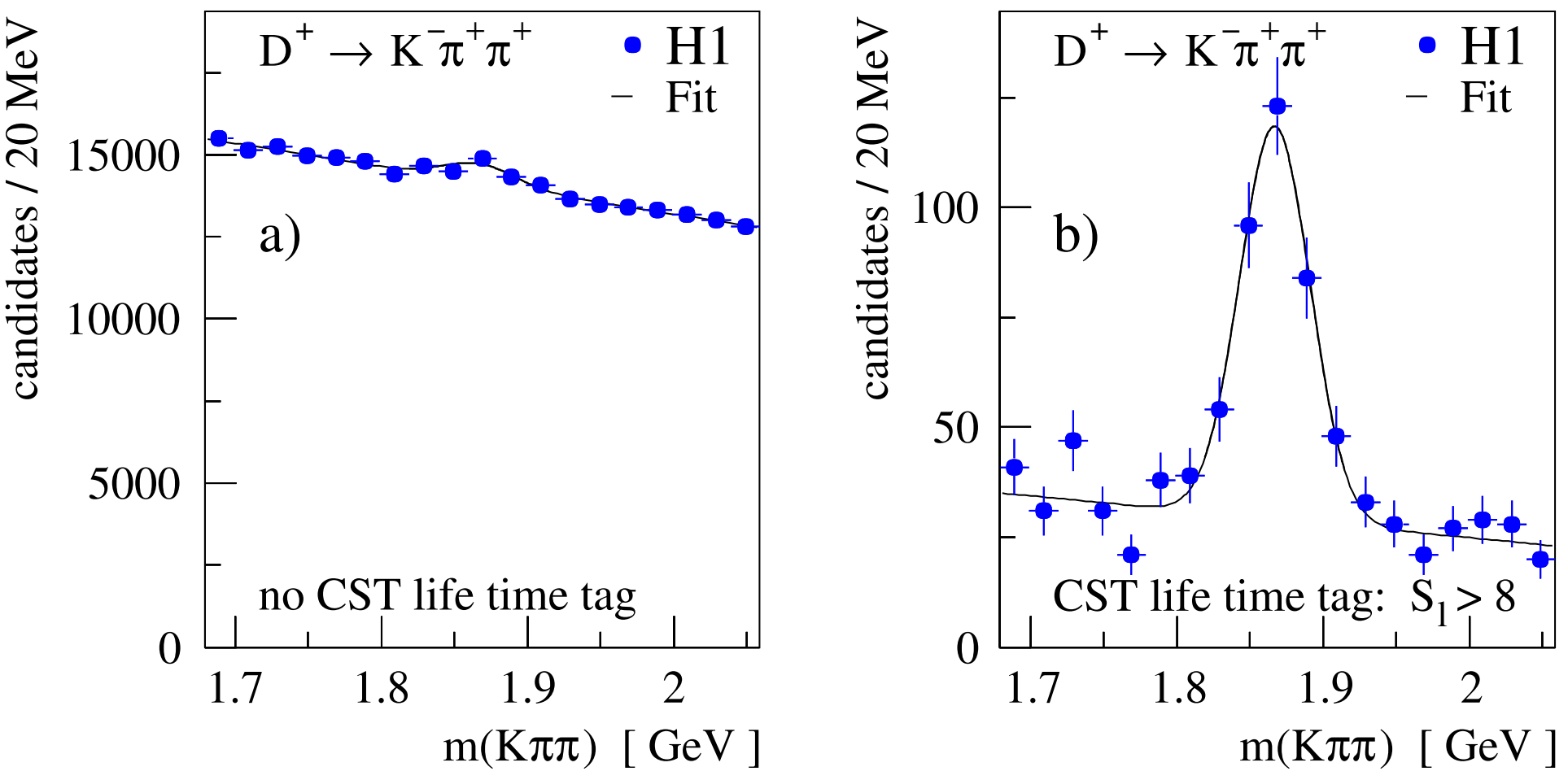

**Fig. 4:** $D^+$ mass peak in H1 before (left) and after (right) a cut on the decay length significance.

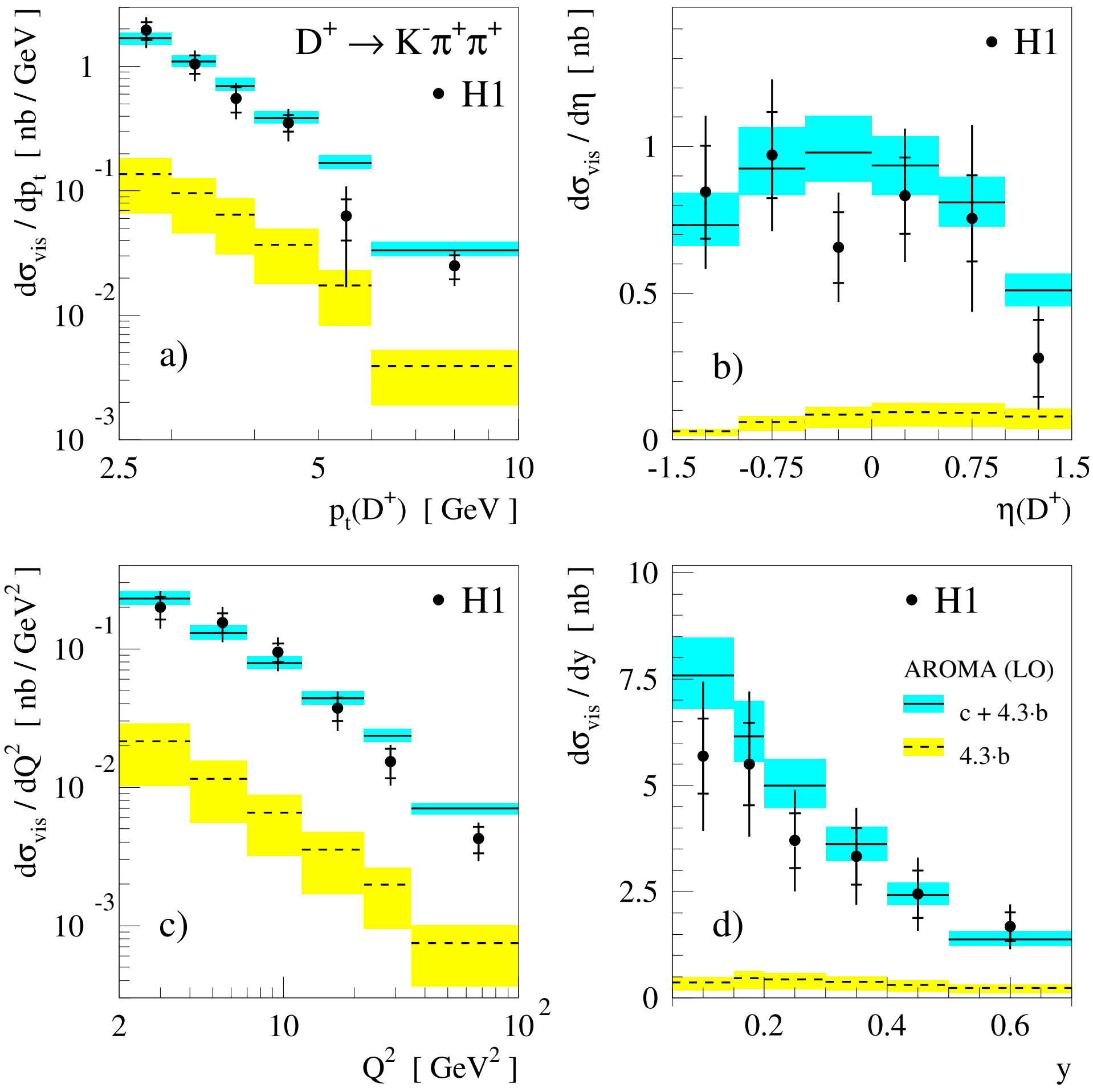

**Fig. 5:** Cross sections for $D^+$ production in DIS. A leading order + parton shower QCD prediction (AROMA) is shown for comparison.

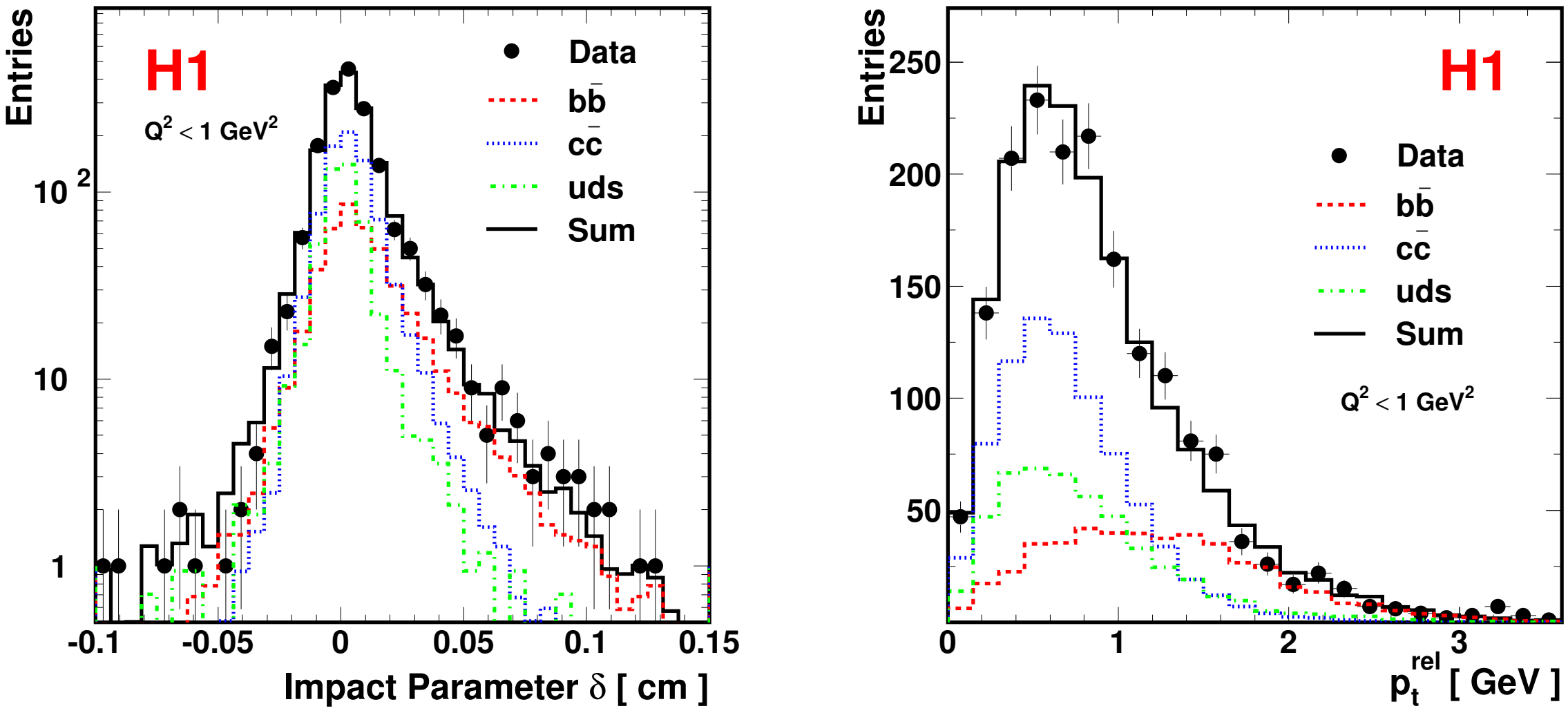

**Fig. 6:** Distributions of the impact parameter $\delta$ of the muon track (left) and the transverse muon momentum $p_t^{rel}$ relative to the axis of the associated jet (right) in H1. Also shown are the estimated contributions of events arising from $b$ quarks (dashed line), $c$ quarks (dotted line) and the light quarks (dash-dotted line).

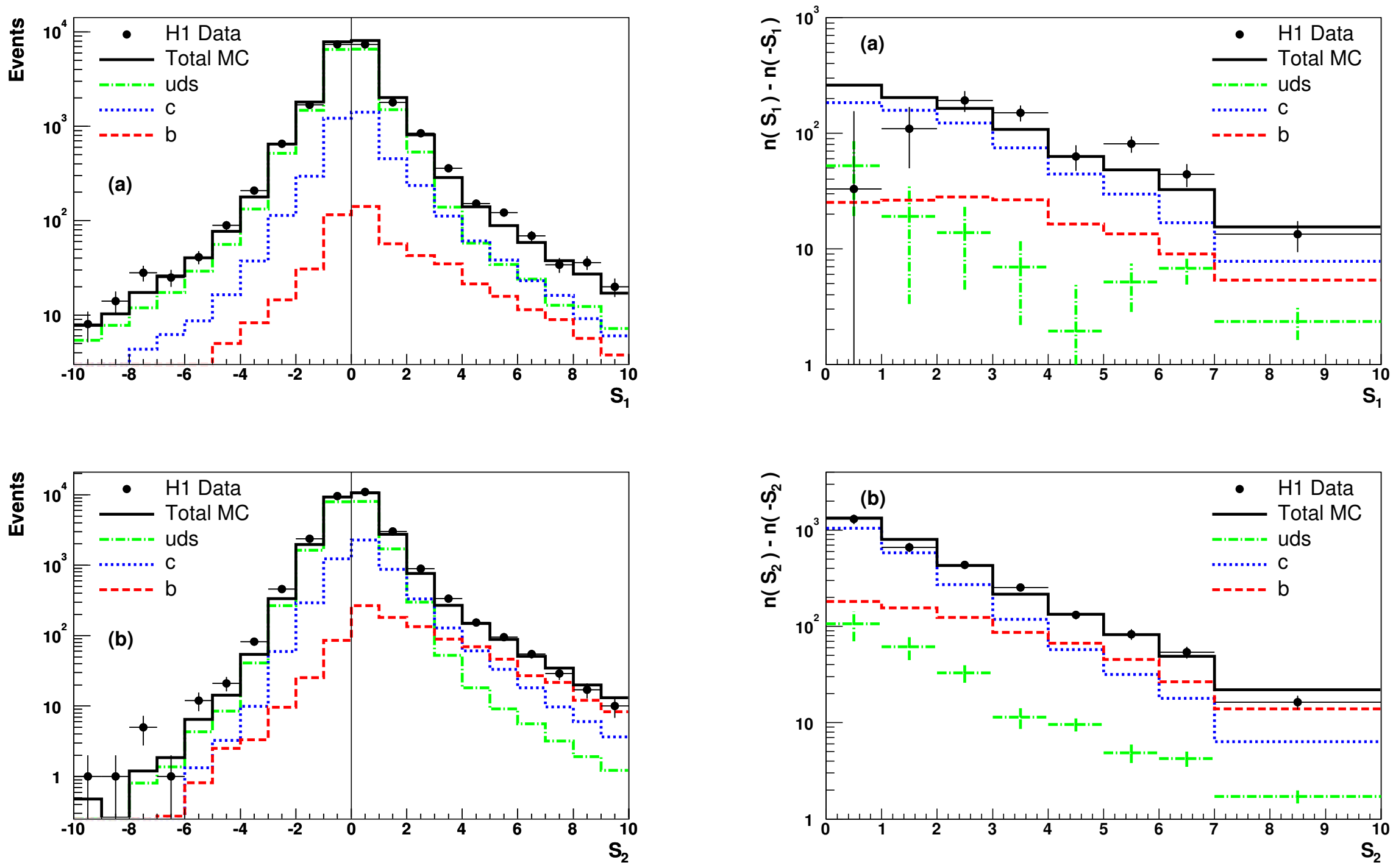

**Fig. 7:** Upper left: significance $S_1 = \delta/\sigma(\delta)$ distribution per event for events that contain one selected track associated to the jet axis. Lower left: significance $S_2 = \delta/\sigma(\delta)$ distribution per event of the track with the second highest absolute significance for events with $\geq 2$ selected tracks associated to the jet. Right: $S_1$ and $S_2$ distributions after subtracting the negatvie bins in the $S_1$ and $S_2$ distributions from the positive.

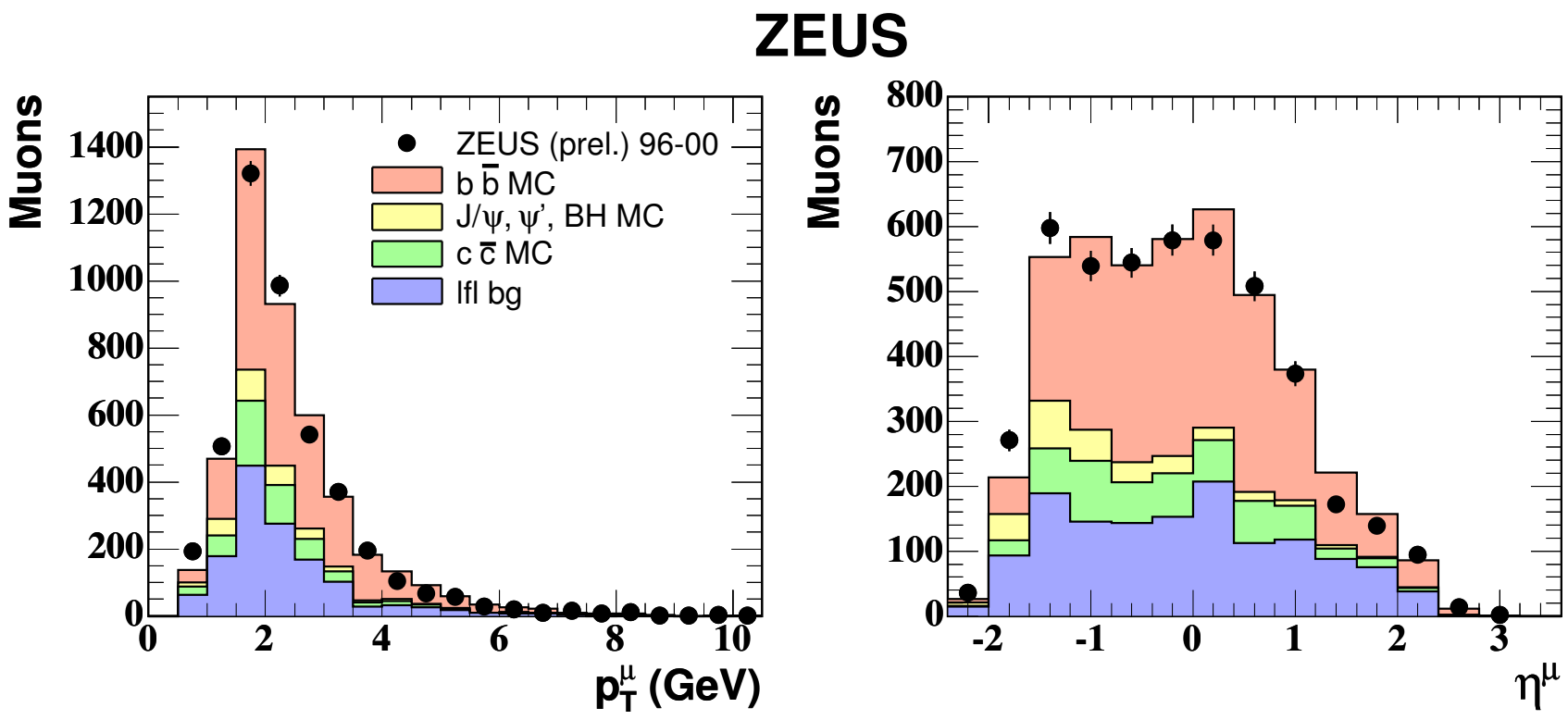

**Fig. 8:** Muon transverse momentum (a) and pseudorapity (b) distributions for nonisolated low transverse momentum muon pairs in ZEUS (two entries per event). The beauty contribution is dominant.

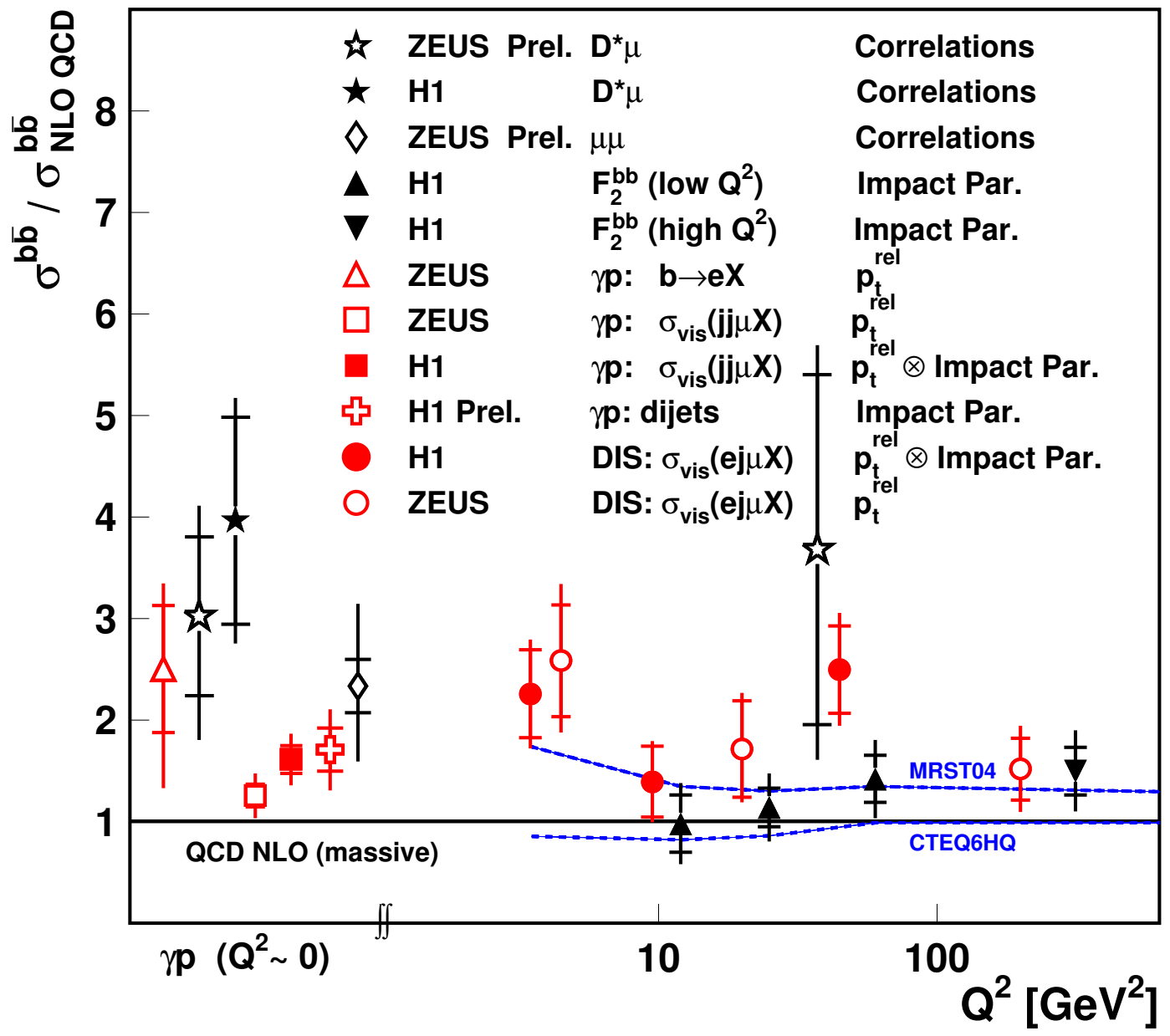

| Key | Ref. | Signature | $p_T$ cuts |
|---|---|---|---|
| Photoproduction | | | |
| ◇ | [25] | $\mu\mu$ | low |
| ☆ | [26] | $D^*\mu$ | low |
| ★ | [27] | $D^*\mu$ | low |
| △ | [28] | 2 jets+$e$ | medium |
| □ | [29] | 2 jets+$\mu$ | medium |
| ■ | [30] | 2 jets+$\mu$ | medium |
| ✚ | [23] | tracks | high |
| DIS | | | |
| ☆ | [26] | $D^*\mu$ | low |
| ▲ | [31] | tracks | low |
| ▼ | [24] | tracks | low |
| ● | [30] | 1 jet+$\mu$ | medium |
| ○ | [32] | 1 jet+$\mu$ | medium |

**Fig. 9:** Ratio of beauty production cross section measurements at HERA to NLO QCD predictions in the massive scheme as function of the photon virtuality $Q^2$. Measurements with low $p_T$ cuts are shown in black, while measurements with medium or high $p_T$ cuts are shown in red/grey. For more details see Table. The predictions from the VFNS NLO calculations by MRST and CTEQ for the DIS kinematic regime $Q^2 > 2$ GeV$^2$ are also shown (valid for comparison with the black low threshold points). Since theoretical errors are different for each point, they are not included in this plot.

A number of models have been suggested to describe inelastic quarkonium production in the framework of perturbative QCD, such as the color-singlet model (CSM) [33, 34], the color-evaporation model [35, 36] and soft color interactions [37]. Comprehensive reports on the physics of charmonium production are available [38, 39].

Recently the ansatz of non-relativistic quantum chromodynamics (NRQCD) factorization was introduced. In the NRQCD approach non-perturbative effects associated with the binding of a $q\bar{q}$ pair into a quarkonium are factored into NRQCD matrix elements that scale in a definite manner with the typical relative velocity $v$ of the heavy quark in the quarkonium. This way, colour octet quark anti-quark states, carrying different angular momenta and color charges than the quarkonium, can contribute to the charmonium production cross section. Theoretical calculations based on the NRQCD factorization approach [40–42] are available in leading order [43–48]. In the NRQCD factorization approach the size of the color octet contributions, which are described by long distance matrix elements (LDME), are additional free parameters and have been determined in fits to the Tevatron data [49]. The NRQCD factorization approach incorporates the color singlet model i.e. the state $q\bar{q}[1,^3S_1]$ which is recovered in the limit in which the long distance matrix elements for other $q\bar{q}$ states tend to zero.

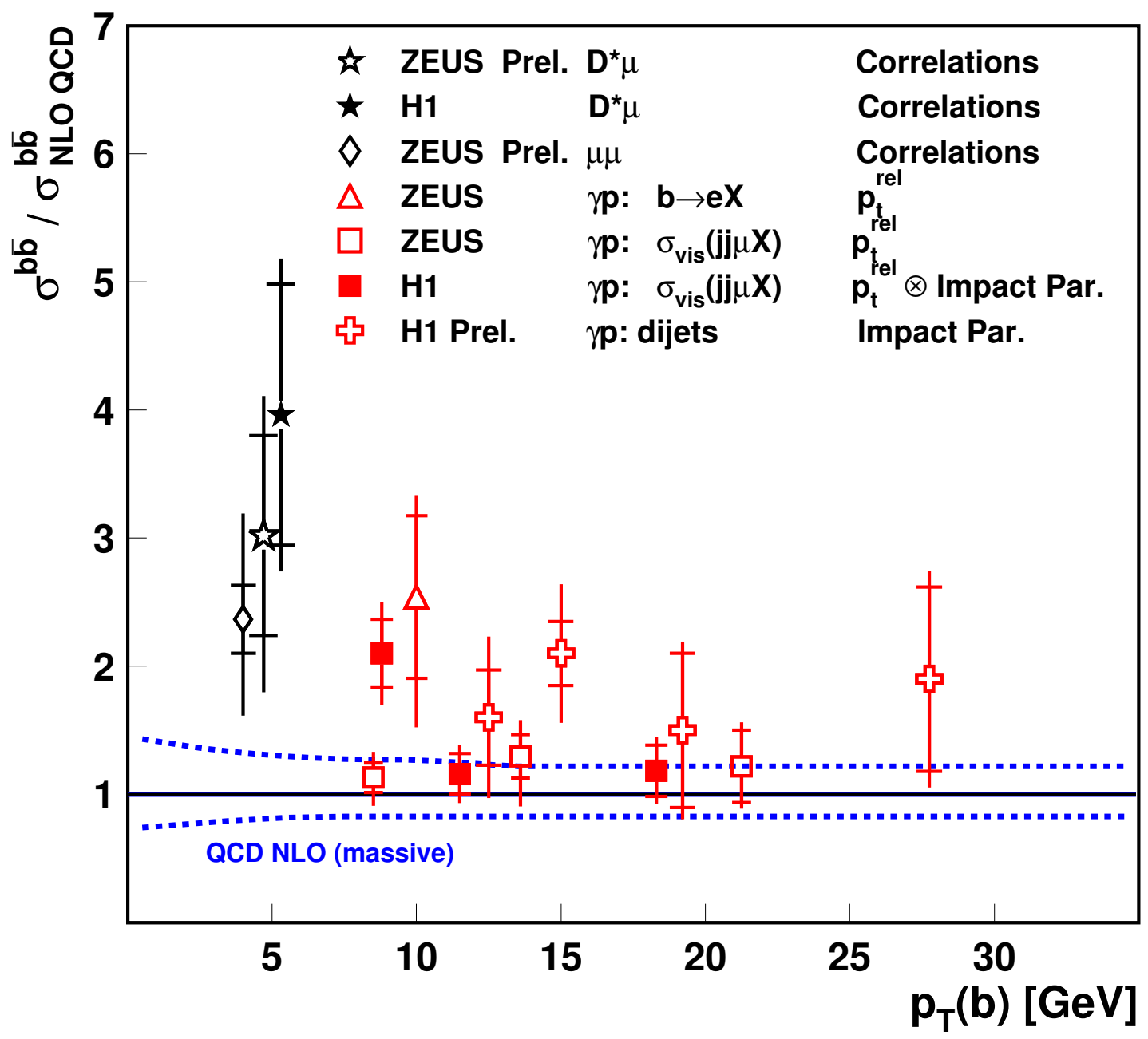

**Fig. 10:** Ratio of beauty production cross section measurements in photoproduction at HERA to NLO QCD predictions in the massive scheme as function of the transverse momentum of the $b$ quark $p_{Tb}$. The dashed line gives an indication of the size of the theoretical uncertianties.

At HERA, cross sections measurements for photoproduction of $J/\psi$ and $\psi(2S)$ and for electroproduction of $J/\psi$ mesons have been performed [52–55]. Bottomonium data are not available due to statistical limitations of the data.

For $J/\psi$ and $\psi(2S)$ photoproduction, calculations of the color-singlet contribution are available to next-to-leading order perturbation theory [50, 51]. Calculations which include the color octet contributions as predicted by NRQCD are available in leading order.

Figure 11 shows the measurements of the $J/\psi$ photoproduction cross section by the H1 collaboration [52] and the ZEUS collaboration [53] which are in good agreement with each other. The variable $z$ (left figure) denotes the fraction of the photon energy in the proton rest frame that is transferred to the $J/\psi$. Reasonable agreement is found between the HERA data and the NRQCD factorization ansatz in leading order (LO, CS+CO). The uncertainty indicated by the open band is due to the uncertainty in the color-octet NRQCD matrix elements. In contrast, the shaded band shows the calculation of the color-singlet contribution (NLO, CS) which is performed to next-to-leading order in $\alpha_s$ [50, 51]. This NLO, CS contribution alone describes the data quite well without inclusion of color-octet contributions. Comparison between the NLO,CS prediction (shaded band) and the LO,CS prediction (dotted line) shows that the NLO corrections are crucial for the description of the HERA photoproduction data.

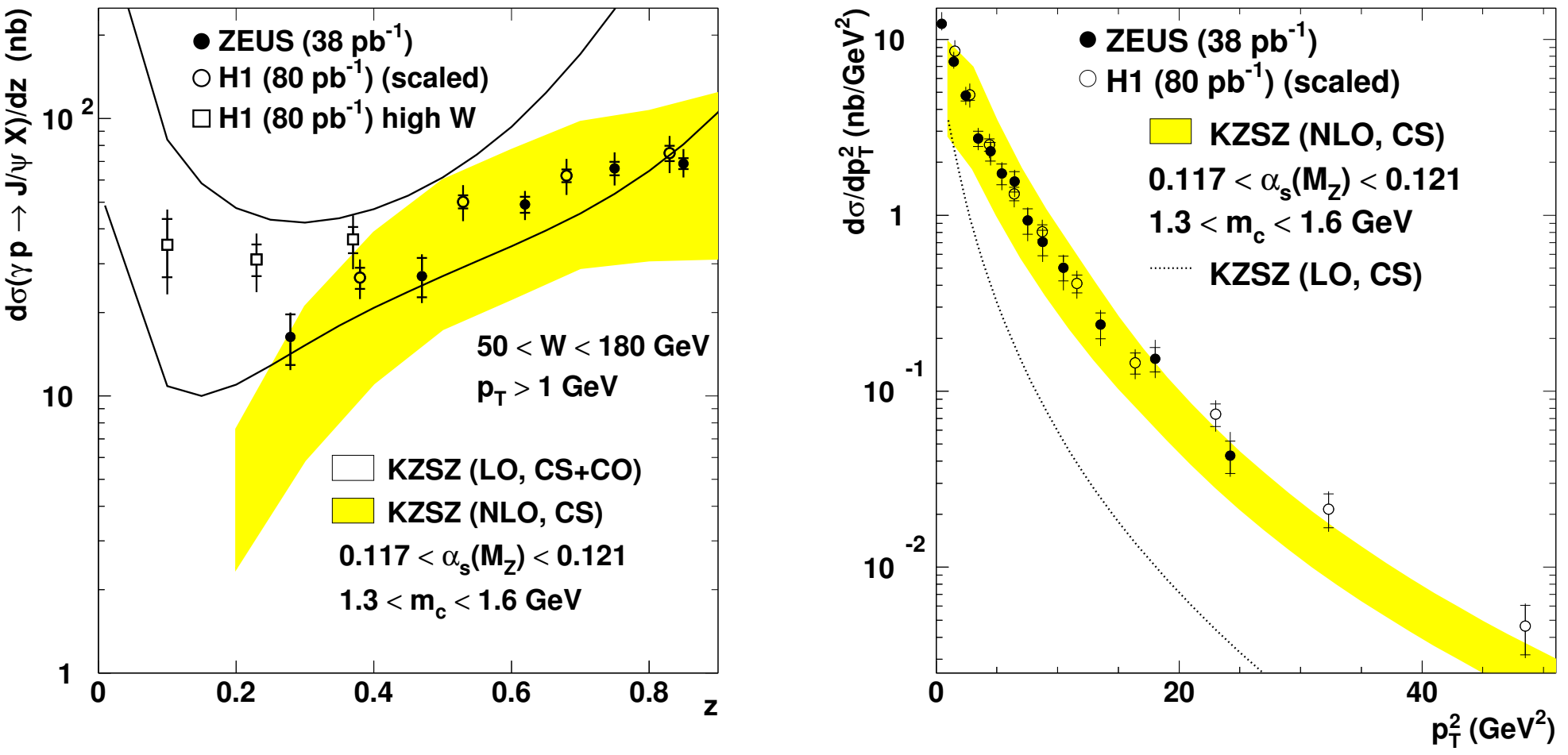

**Fig. 11:** Differential charmonium photoproduction cross sections as measured by H1 and ZEUS in comparison to calculations from LO NRQCD factorization (open band), NLO Color singlet contribution (shaded band) and LO color-singlet contribution (dotted line).

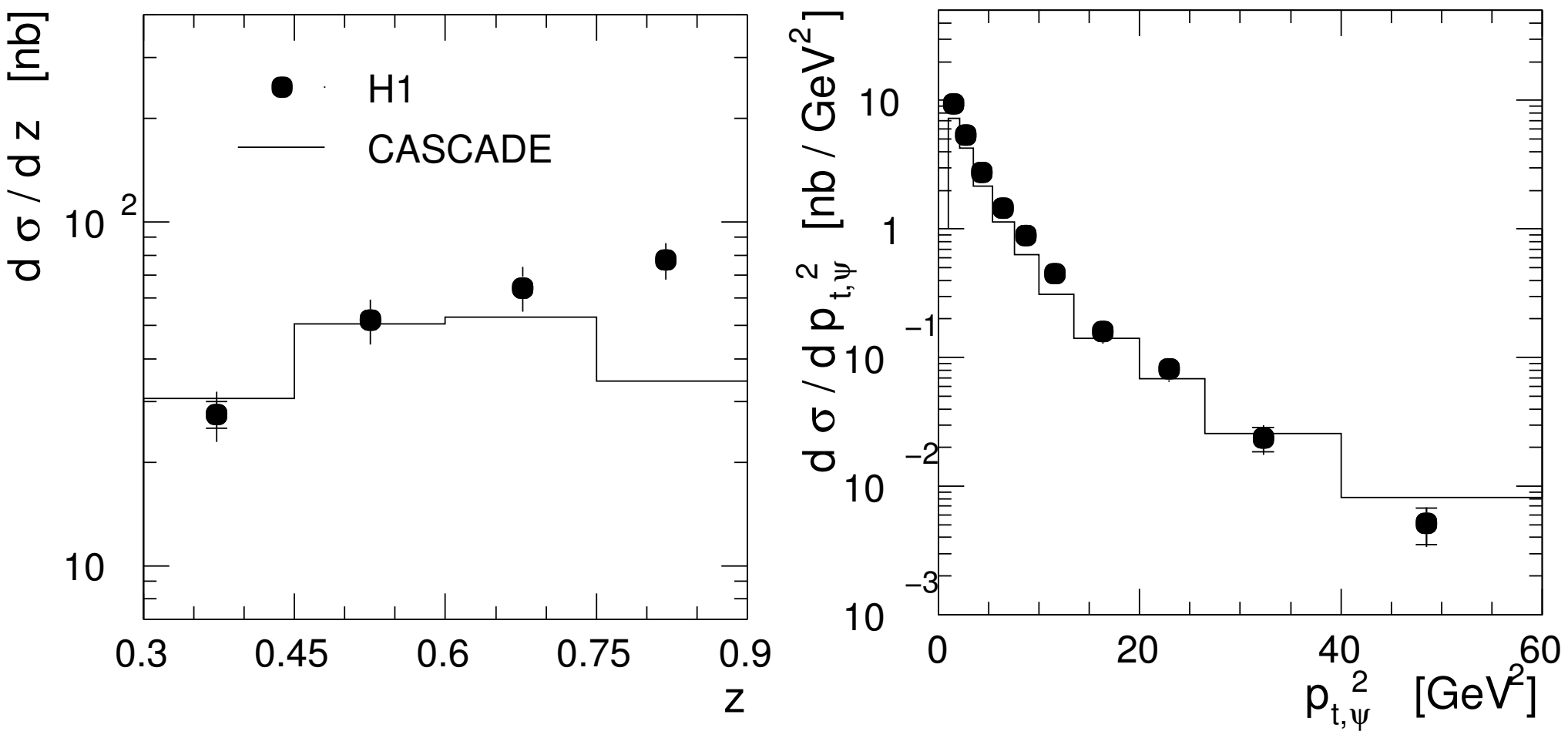

**Fig. 12:** Differential Charmonium photoproduction in comparison with a prediction using the color singlet-model and $k_t$ factorization as implemented in the Monte Carlo generator CASCADE.

Charmonium production cross sections have also been calculated in the $k_t$ factorization approach (see Refs. [56–58]). In these calculations the color-singlet model is used to describe the formation of the charmonium state. Figure 12 shows a comparison of the H1 data with the predictions from the $k_t$ factorization approach as implemented in the Monte Carlo generator CASCADE [59]. Good agreement is observed between data and predictions for $z < 0.8$. At high $z$ values, the CASCADE calculation underestimates the cross section. The CASCADE predictions for the the $p_{t,\psi}^2$ dependence of the cross section fit the data considerably better than the LO,CS calculation in the collinear factorization approach (dotted curve in Fig. 11).

In fig 13 the differential cross sections for electroproduction of $J/\psi$ mesons as measured by H1 [54] and ZEUS [55] are shown as a function of $z$ and compared with predictions from the color singlet model (shaded band), with the NRQCD calculation [60] (CS+CO, open band), and also with calculations in the $k_t$ factorization approach (dotted line) as provided by [58] and implemented in the Monte Carlo generator CASCADE (dash-dotted line).

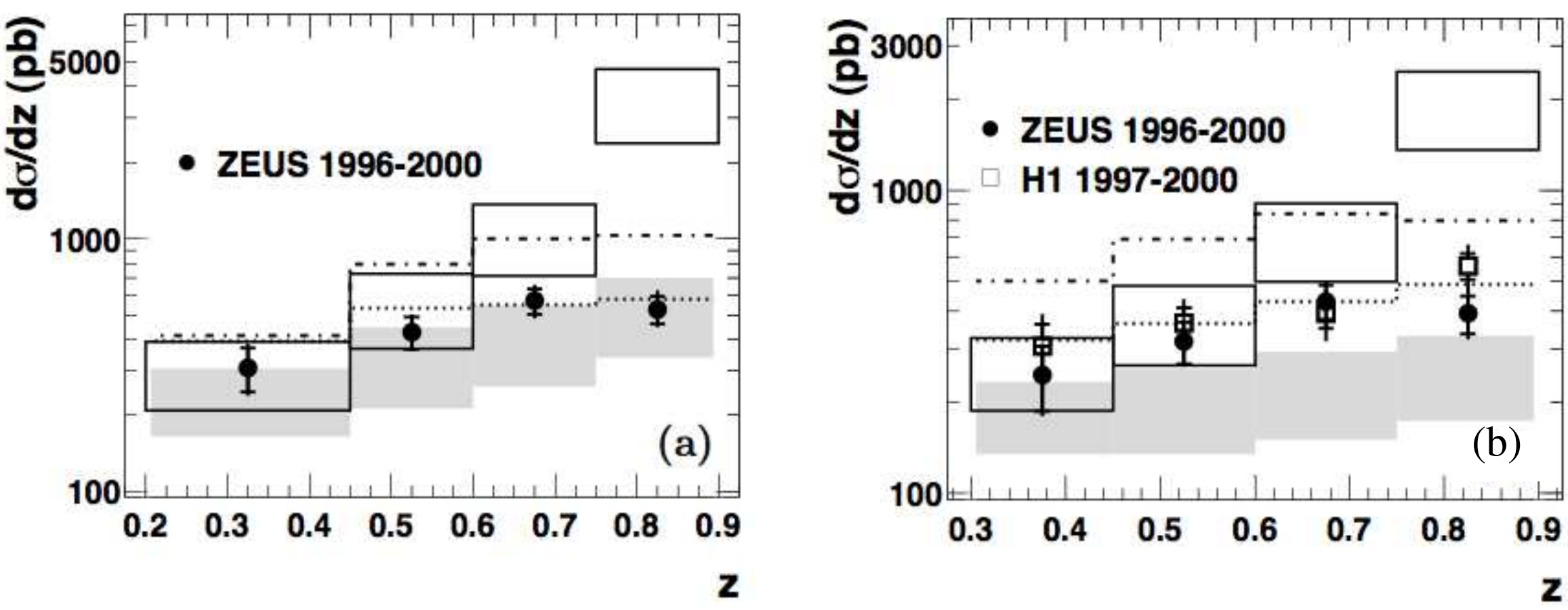

**Fig. 13:** Differential cross sections $d\sigma/dz$ a) without and b) with a cut on $p^{*2}_{t,\psi} > 1$ GeV. The data are compared to the NRQCD calculation (CS+CO, open band), the color-singlet contribution (CS, shaded band), with a prediction in the $k_t$ factorization approach assuming the CSM (dotted line) and with the Monte Carlo generator CASCADE (dash-dotted line).

In the left figure the data are seen to agree well with the predictions using the color singlet model (shaded band and lines) while the full NRQCD calculation (open band), including color-octet contributions is wrong in shape and normalization. The agreement deteriorates when the cut $p^{*2}_{t,\psi} > 1$ GeV is applied (right Fig. 13). This cut is justified, however, as towards small $p^{*2}_{t,\psi}$ perturbation theory becomes increasingly unreliable due to collinear singularities for the contributions $e + g \to e + c\bar{c}[n] + g$ with $n{=}^1S_0^{(8)}$ and $^3P_J^{(8)}$ [60].

In conclusion, NRQCD, as presently available in leading order, does not give a satisfactory description of the HERA data. In contrast, the color singlet model shows a reasonable description of the HERA data, when implemented in calculations to next-to-leading order perturbation theory or in calculations in which the $k_t$-factorization approach is used.

## 5 Charm and Beauty contributions to structure functions

To a good approximation, except at very high $Q^2$, the cross section for inclusive deep inelastic electron scattering off the proton at HERA can be described in terms of a single proton structure function $F_2$ (for formula see [1]). This structure function only depends on the photon virtuality, $Q^2$, and on the Bjorken scaling variable $x$. Assuming that the electron scatters off a single quark in the proton (0th order QCD, quark-parton model) $x$ can be reinterpreted as the fraction of the proton momentum carried by the struck quark. This is a reasonable approximation for the light quark content of the proton.

For heavy quarks, the situation is a bit more complicated. Due to the heavy quark mass, on-shell heavy quarks can not exist within the proton. Rather, the dominant process for heavy quark production is the 1st order QCD BGF process depicted in Fig. 1. However, this process (and other higher order processes) still contributes to electron scattering, and hence to $F_2$. This can be interpreted in two ways.

In the massive approximation, heavy quarks are treated as being produced dynamically in the scattering process. The heavy quark contribution to $F_2$, frequently denoted as $F_2^{c\bar{c}}$ and $F_2^{b\bar{b}}$, therefore indirectly measures the *gluon* content of the proton. If $Q^2$ is large enough such that the quark mass can be neglected ($Q^2 \gg (2m_Q)^2$), the splitting of the gluon into a heavy quark pair can be reinterpreted to occur *within* the proton. $F_2^{c\bar{c}}$ and $F_2^{b\bar{b}}$ then measure the occurrence of *virtual* heavy quark pairs in the proton, or the "heavy quark structure" of the proton.

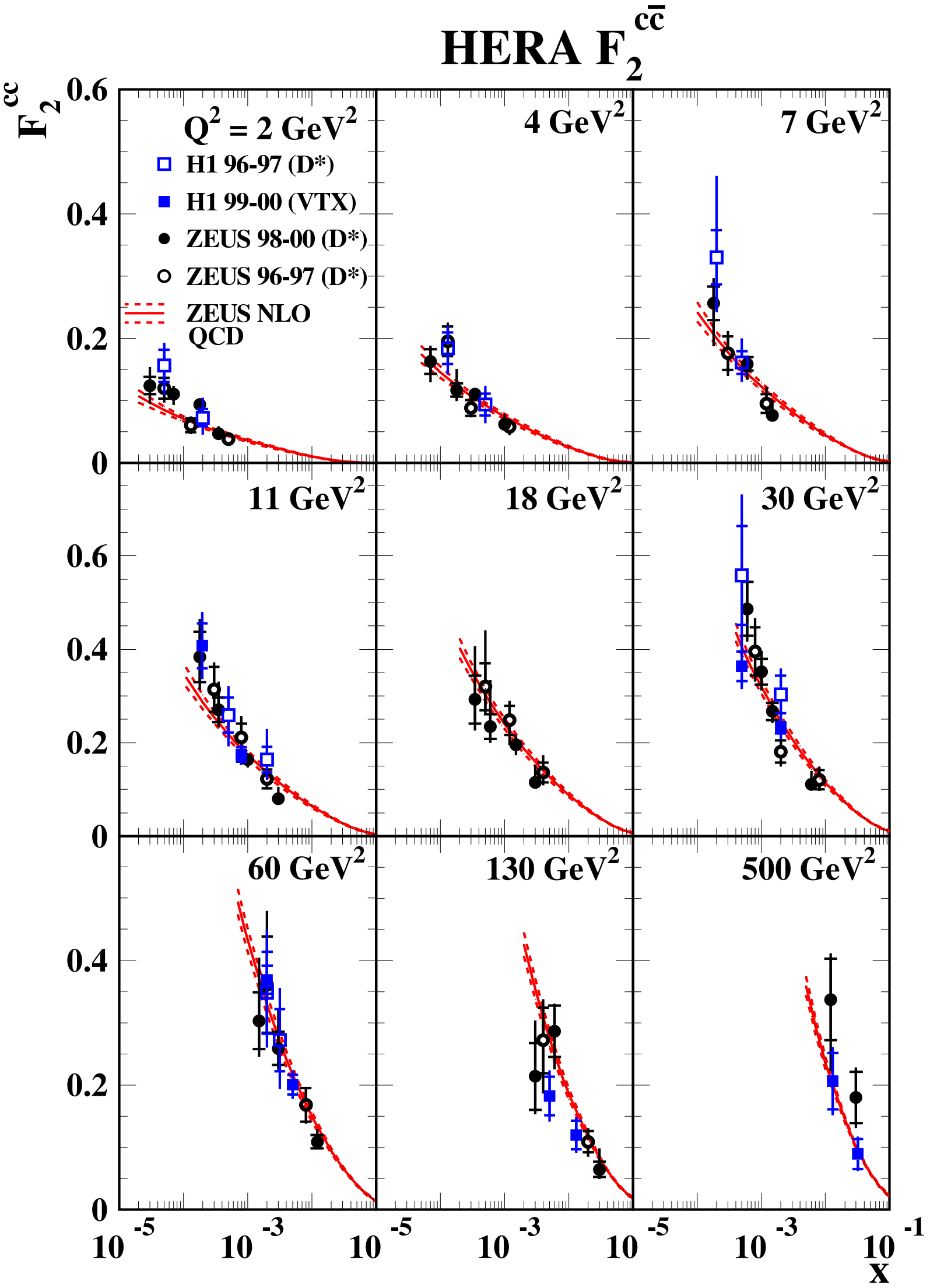

**Fig. 14:** $F_2^{c\bar{c}}$ results as a function of $x$ in bins of $Q^2$, from the H1 and ZEUS $D^{*\pm}$ analyses and from the H1 inclusive lifetime tagging measurements The data are compared to a NLO prediction using the ZEUS NLO fit results for the proton parton densities.

For charm production, the condition $Q^2 \gg (2m_Q)^2$ is valid for a large part of the HERA phase space. For beauty, it is only satisfied at very large $Q^2$. This is also the region most interesting for physics at the LHC.

Similar arguments hold for the heavy quark structure of the photon.

As an example, Fig. 14 [24,31,61–63] shows $F_2^{c\bar{c}}$ as measured by the ZEUS and H1 collaborations. A different representation of these results is shown in Fig. 6 of [1]. There, also $F_2^{b\bar{b}}$ is shown. Good agreement is observed with QCD predictions. Parametrizations of heavy quark densities of the proton at LHC energies should therefore be valid within their respective errors.

## 6 Charm fragmentation

The large cross section for charm production at HERA allows measurements of charm fragmentation which are very competitive with $e^+e^-$ measurements. As this topic is covered very nicely in [1] and [2] it is not treated further here.

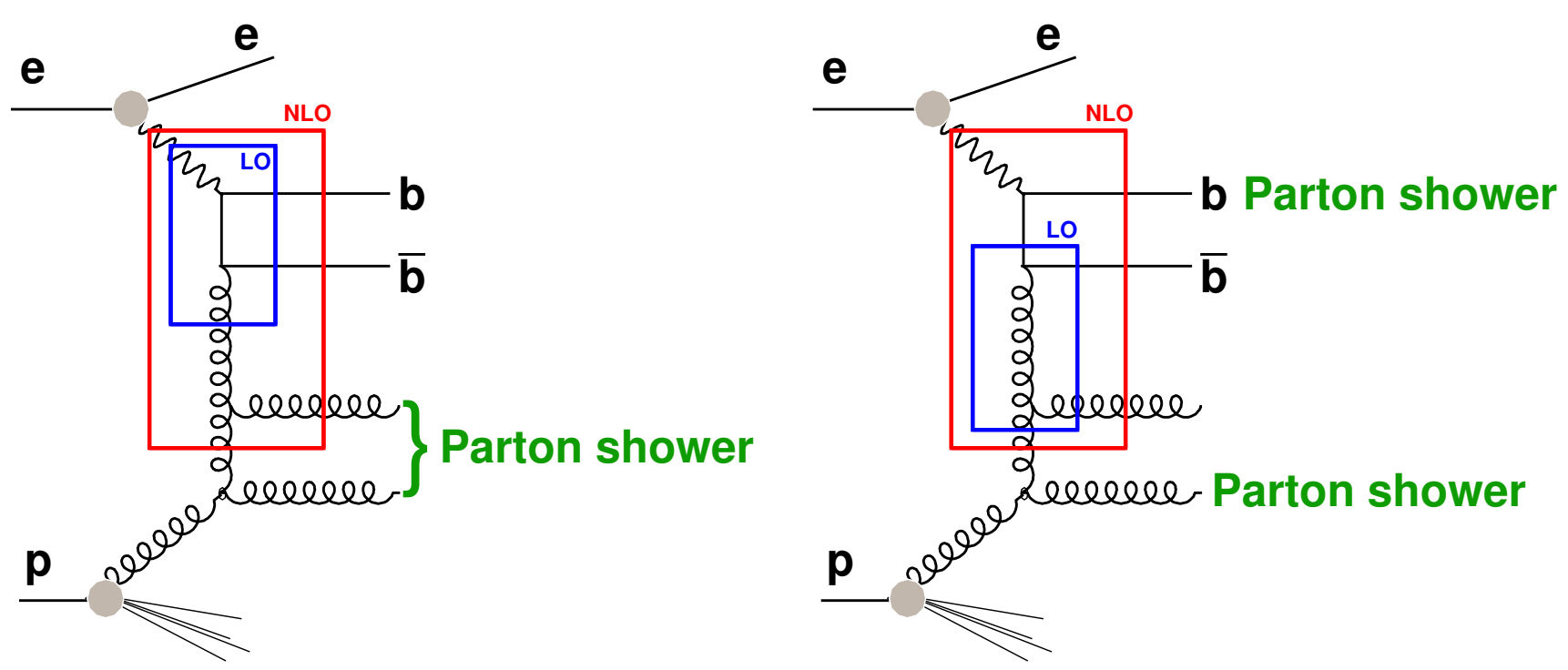

**Fig. 15:** Example for higher order Feynman graph for beauty production. Different interpretations of the graph in terms of NLO or LO matrix elements plus parton showers are highlighted. Depending on the kinematics and the scheme chosen, part of the gluons could also be reabsorbed into the proton structure function definition, and/or the $\gamma b\bar{b}$ vertex could be interpreted as part of the photon structure.

## 7 Quark-antiquark correlations

Heavy quarks are always produced in pairs. An interesting way to check QCD is thus to verify whether the kinematic correlations between the quark pair are correctly described by QCD.

Figure 15 shows different interpretations of the same higher order beauty production process. These different interpretations partially manifest themselves in different kinematic regions of beauty production phase space. If the highest virtuality part of the process occurs in the leading order BGF-like subprocess (left), the two $b$ quarks will be almost back-to-back in the detector transverse plane. The two extra gluons can either be reabsorbed into the proton structure, recovering the original BGF graph, or manifest themselves as visible "parton shower" activity in the direction of the proton. Alternatively, if the dominant leading order subprocess is gluon exchange with one of the $b$ quarks (right, "flavour excitation in the photon"), this $b$ quark will recoil against a gluon jet. At sufficiantly large momentum transfer (rare at HERA), the second $b$ quark can be treated as a "spectator", and will approximately follow the initial photon direction. At next-to-leading order, contributions to both processes are described by the same Feynman graph, but the two extreme kinematic cases (and all variants in between) are still included. If both heavy quarks are tagged, these different kinematic regions can be distinguished by measuring the momentum and angular correlations between the two quarks.

Figure 16 [25] shows the angular correlations between the two muons originating from different $b$ quarks of a $b\bar{b}$ pair. Reasonable agreement is observed with QCD predictions. The predominantly back-to-back topology comfirms the dominance of the BGF-like contribution.

## 8 HERA II prospects

Both the HERA collider and its detectors have been upgraded in 2001/2 to provide more luminosity with polarized electron beams, and improve heavy flavour detection. This program is called HERA II. The luminosity accumulated so far already exceeds the HERA I luminosity. An integrated luminosity up to 700 $pb^{-1}$ is expected at the end of the HERA program in 2007. This enhances the statistics for many studies by almost an order of magnitude with respect to HERA I. The improved detectors offer further handles for improved heavy flavour measurements. H1 has improved the forward coverage of its Micro-Vertex-Detector [68], and added a Fast Track Trigger [67]. ZEUS has implemented a Micro-Vertex-Detector (MVD) [65] for the first time for HERA II, and has added an upgraded forward tracking detector [66]. These improvements allow the application of measurement techniques which could not be used at HERA I, and can be used to improve the data quality, add additional statistics, and/or cover new phase space regions.

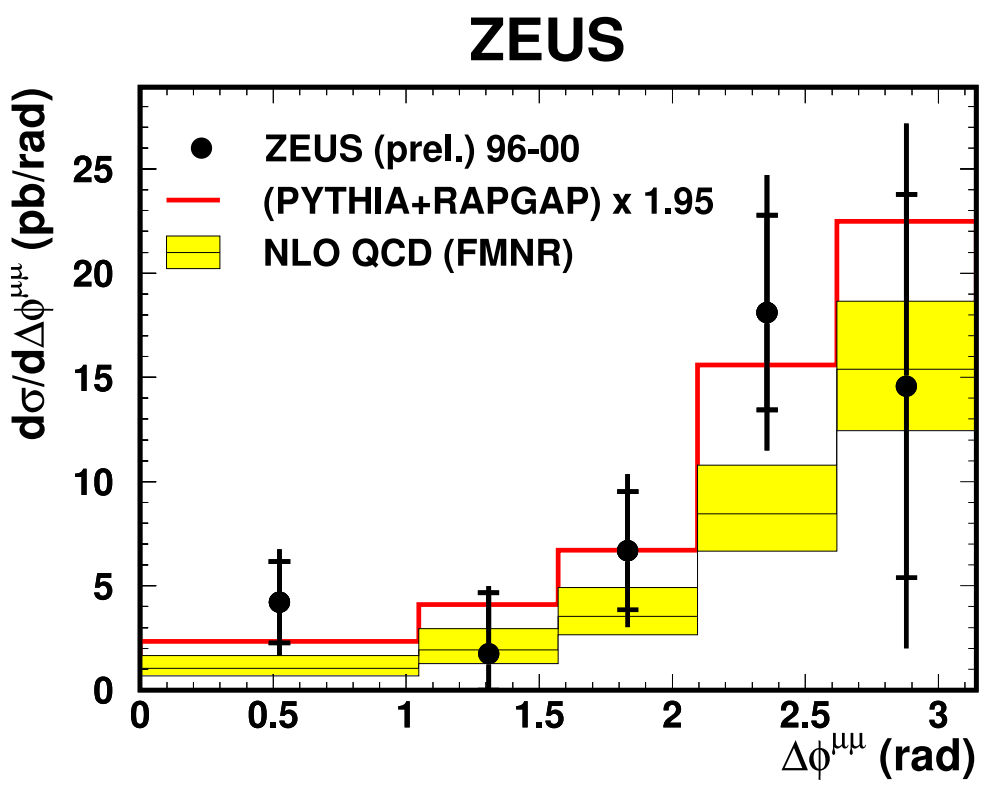

**Fig. 16:** Differential cross section $d\sigma/d\Delta\phi^{\mu\mu}$ for dimuon events from $b\bar{b}$ decays in which each muon originates from a different $b(\bar{b})$ quark. The data (solid dots) are compared to the leading order + parton shower generators PYTHIA and RAPGAP (histogram) and to massive NLO QCD predictions (shaded band).

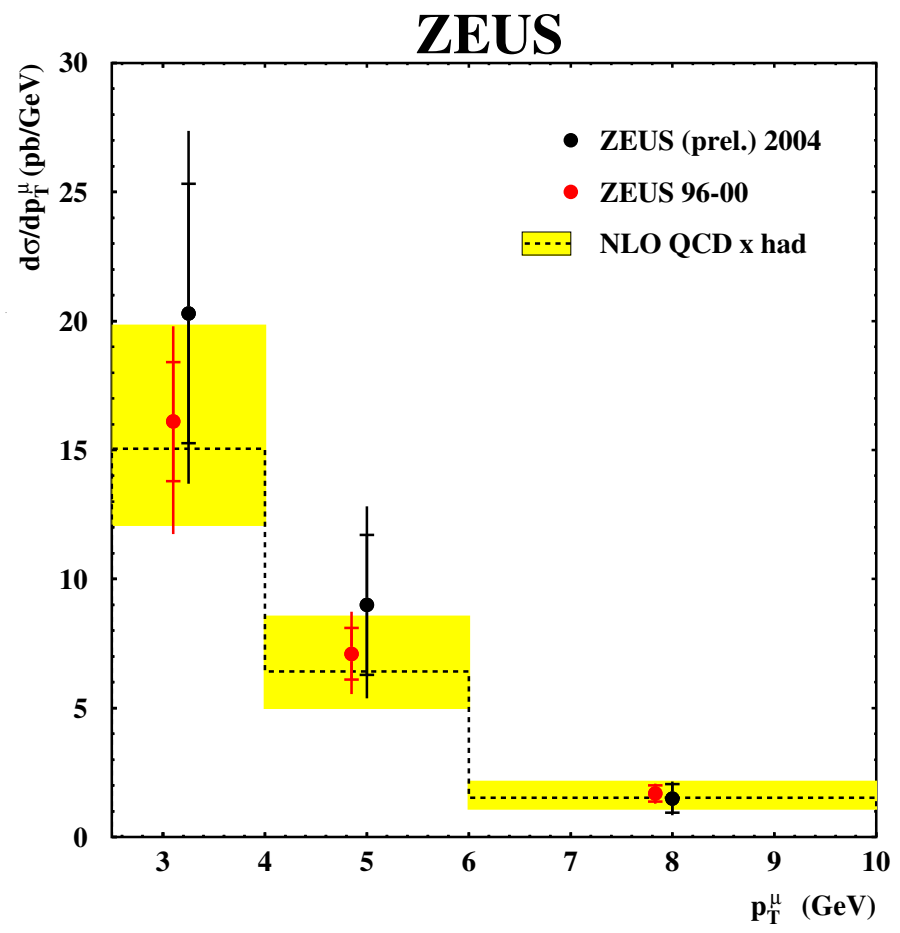

**Fig. 17:** Differential cross section as function of muon $p_T$ for dimuon + jet events in photoproduction. Preliminary results from the first 33 $pb^{-1}$ of HERA II data are compared to HERA I results and QCD predictions.

New detectors require time to fully understand their systematics, but first preliminary results have already been obtained. Figure 17 [64] shows the cross section for beauty production obtained using the new ZEUS MVD with the first 33 $pb^{-1}$ of HERA II data, compared to the HERA I result. Good agreement is observed.

The measurements which will profit most from the improved HERA II data sets include double differential measurements such as the beauty and charm contributions to the proton structure function $F_2$, and multi-tag measurements to explicitly study quark-quark correlations. Statistical improvements of at least one order of magnitude can be expected when the increased luminosity and improved measurement techniques are combined.

## 9 Conclusions

Heavy flavour production at HERA is a very active field of research yielding multiple insights into the applicability of perturbative QCD. The problem of multiple scales complicates the perturbative expansions and limits the achievable theoretical precision. In general, QCD predictions agree well with the data, although indications for deviations persist in specific regions of phase space. Some of these might be attributable to missing NNLO or even higher order contributions.

The overall reasonable agreement, as well as the self-consistency of the structure functions tested by or derived from heavy flavour production at HERA, enhances confidence in corresponding cross-section predictions at LHC, within their respective theoretical uncertainties.

# Experimental aspects of heavy flavour production at the LHC

*J. Baines[a], A. Dainese[b], Th. Lagouri[c], A. Morsch[d], R. Ranieri[e], H. Ruiz[d], M. Smizanska[f], and C. Weiser[g]*
[a] Rutherford Laboratory, UK
[b] University and INFN, Padova, Italy
[c] Institute of Nuclear and Particle Physics, Charles University, Prague, Czech Republic
[d] CERN, Geneva, Switzerland
[e] University and INFN, Firenze, Italy
[f] Lancaster University, Lancaster, UK
[g] Institut für Experimentelle Kernphysik, Universität Karlsruhe, Karlsruhe, Germany

**Abstract**
We review selected aspects of the experimental techniques being prepared to study heavy flavour production in the four LHC experiments (ALICE, ATLAS, CMS and LHCb) and we present the expected performance for some of the most significative measurements.

*Coordinators: A. Dainese, M. Smizanska, and C. Weiser*

## 1 Introduction[1]

Unprecedently large cross sections are expected for heavy-flavour production in proton–proton collisions at the LHC energy of $\sqrt{s} = 14$ TeV. Next-to-leading order perturbative QCD calculations predict values of about 10 mb for charm and 0.5 mb for beauty, with a theoretical uncertainty of a factor 2–3. Despite these large cross sections, the LHC experiments, ALICE [1,2], ATLAS [3], CMS [4], and LHCb [5], will have to deal with rejection of background coming from non-heavy flavour inelastic interactions for which the predicted cross is about 70 mb. The four experiments will work at different luminosity conditions. ATLAS and CMS are designed to work in a wide range of luminosities up to nominal $10^{34}$ $\mathrm{cm}^{-2}\mathrm{s}^{-1}$, while the LHCb optimal luminosity will vary in the range $(2–5) \times 10^{32}$ $\mathrm{cm}^{-2}\mathrm{s}^{-1}$ and ALICE is designed to work at $3 \times 10^{30}$ $\mathrm{cm}^{-2}\mathrm{s}^{-1}$ in proton–proton collisions. Luminosity conditions in ATLAS, CMS and LHCb allow multiple interactions per bunch crossing, thus leading to requirements of even stronger identification and selection of heavy-flavour events already at trigger level. The first task will be the measurement of integrated and differential charm and beauty production cross sections in the new energy domain covered at the LHC. ALICE will play an important role, having acceptance down to very low transverse momentum, as we discuss in Section 4. These measurements can be performed within a relatively short period of running. Afterwards, the heavy-flavour studies will focus on less inclusive measurements addressing specific production mechanisms that allow to test higher order perturbative QCD predictions, as well as on rare decays of heavy-flavour hadrons, that may carry information on New Physics beyond the Standard Model. In order to meet these requirements dedicated and sophisticated trigger strategies have been prepared by the LHC experiments.

## 2 Heavy flavour detection in the LHC experiments[2]

The four detectors that will take data at the LHC have different features and design requirements, but all of them are expected to have excellent capabilities for heavy-flavour measurements. Their complementarity will provide a very broad coverage in terms of phase-space, decay channels and observables.

[1] Author: M. Smizanska
[2] Author: A. Dainese

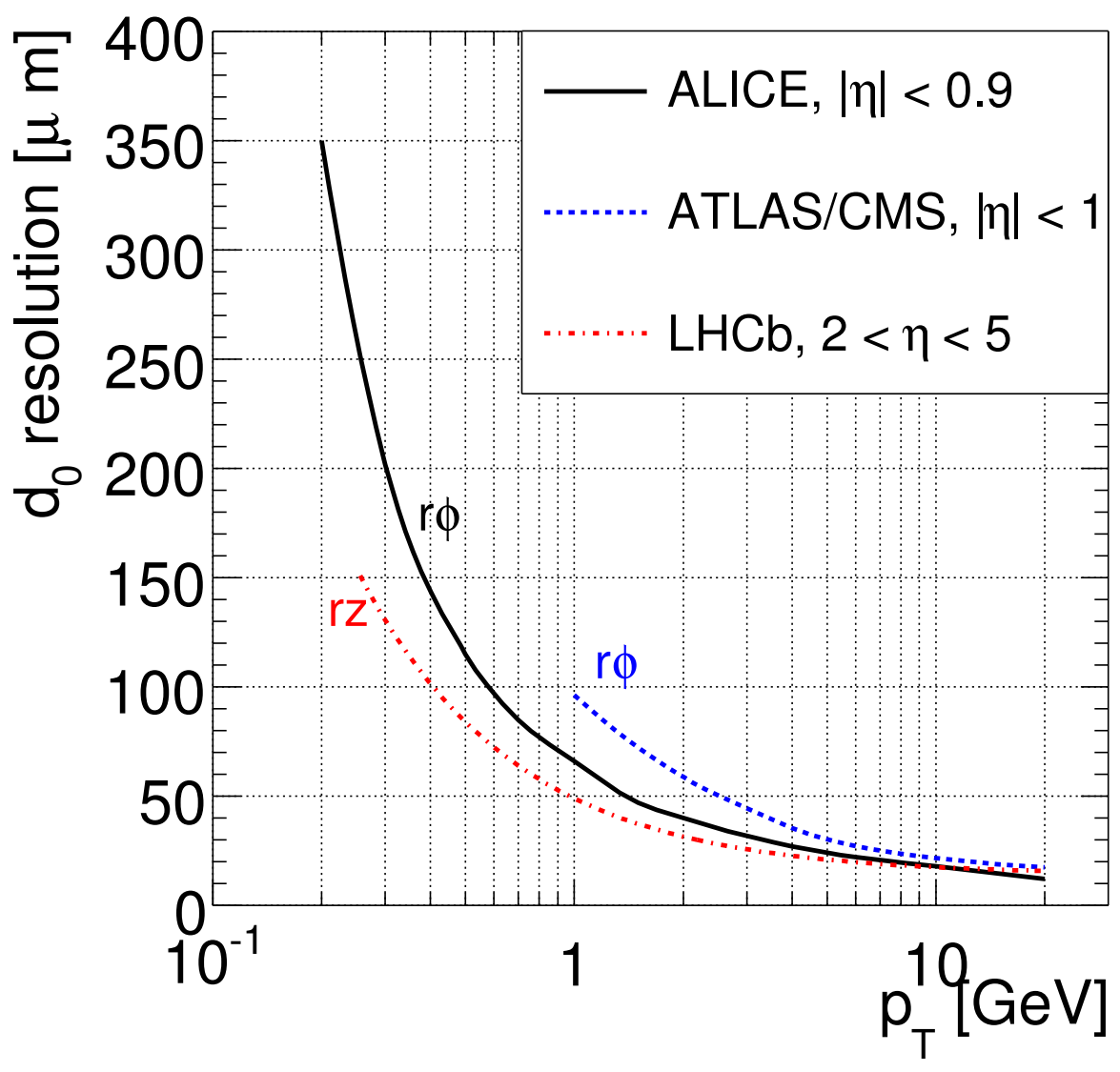

**Fig. 1:** Track impact parameter resolutions for the four LHC experiments. Note that for ALICE, ATLAS and CMS the impact parameter is defined in the $r\phi$ plane, while for LHCb it is defined in the $rz$ plane.

Experimentally, the key elements for a rich heavy-flavour program are track and vertex reconstruction and particle identification (PID).

Open charm and beauty mesons have typical life-times in the order of a $ps$ ($c\tau$ values are about $125$–$300$ $\mu$m for D mesons and $500$ $\mu$m for B mesons) and the most direct detection strategy is the identification of single tracks or vertices that are displaced from the interaction vertex. The detector capability to perform this task can be characterized by the transverse impact parameter[3] ($d_0$) resolution. All experiments will be equipped with high position-resolution silicon-detector layers, including pixel detector for the innermost layers, for precise tracking and impact parameter measurement. Tracking is done in the central (pseudo)rapidity region for ALICE ($|\eta| < 0.9$), ATLAS and CMS ($|\eta| \lesssim 2.5$), and in the forward region for LHCb ($2 \lesssim \eta \lesssim 5$). In Fig. 1 we show the $d_0$ resolution, which is similar for the different experiments, and better than $50$ $\mu$m for $p_T \gtrsim 1.5$–$3$ GeV. The inner detector systems of ATLAS, CMS and ALICE will operate in different magnetic fields: The ALICE magnetic field will vary within low values (0.2–0.5 T) leading to a very low $p_T$ cutoff of 0.1–0.2 GeV, while ATLAS (2 T) and CMS (4 T) have higher cutoffs of 0.5 and 1 GeV, respectively, but better $p_T$ resolution at high $p_T$ (e.g., at $p_T = 100$ GeV, $\delta p_T / p_T \approx 1$–$2\%$ for ATLAS/CMS at central rapidity and $\approx 9\%$ for ALICE).

Both lepton and hadron identification are important for heavy-flavour detection. D and B mesons have relatively large branching ratios (BR) in the semi-leptonic channels, $\simeq 10\%$ to electrons and $\simeq 10\%$ to muons, and inclusive cross-section measurements can be performed via single leptons or di-leptons. Alternatively, high-$p_T$ leptons can be used as trigger-level tags to select $\mathrm{B} \to \mathrm{J}/\psi + X$ candidate events, that provide more accurate cross section measurements. ALICE can identify electrons with $p_T > 1$ GeV and $|\eta| < 0.9$, via transition radiation and $\mathrm{d}E/\mathrm{d}x$ measurements, and muons in the forward region, $2.5 < \eta < 4$, which allows a very low $p_T$ cutoff of $1$ GeV. CMS and ATLAS have a broad pseudorapidity coverage for muons, $|\eta| < 2.4$ and $|\eta| < 2.7$, respectively, but they have a higher $p_T$ cutoff varying between 4 and 6 GeV, depending on $\eta$. Both CMS and ATLAS have high-resolution electro-magnetic calorimeters that will be used to identify electrons. Semi-leptonic inclusive measurements do not provide direct information on the D(B)-meson $p_T$ distribution, especially at low $p_T$, because of the weak corre-lation between the lepton and the meson momenta. Therefore, for charm in particular, the reconstruction of exclusive (hadronic) decays is preferable. In this case, hadron identification allows a more effective rejection of the combinatorial background in the low-$p_T$ region. ALICE disposes of $\pi/\mathrm{K}/\mathrm{p}$ separation via $\mathrm{d}E/\mathrm{d}x$ and time-of-flight measurements for $p < 3$–$4$ GeV and $|\eta| < 0.9$.

[3]We define as impact parameter the distance of closest approach to the interaction vertex of the track projection in the plane transverse to the beam direction.

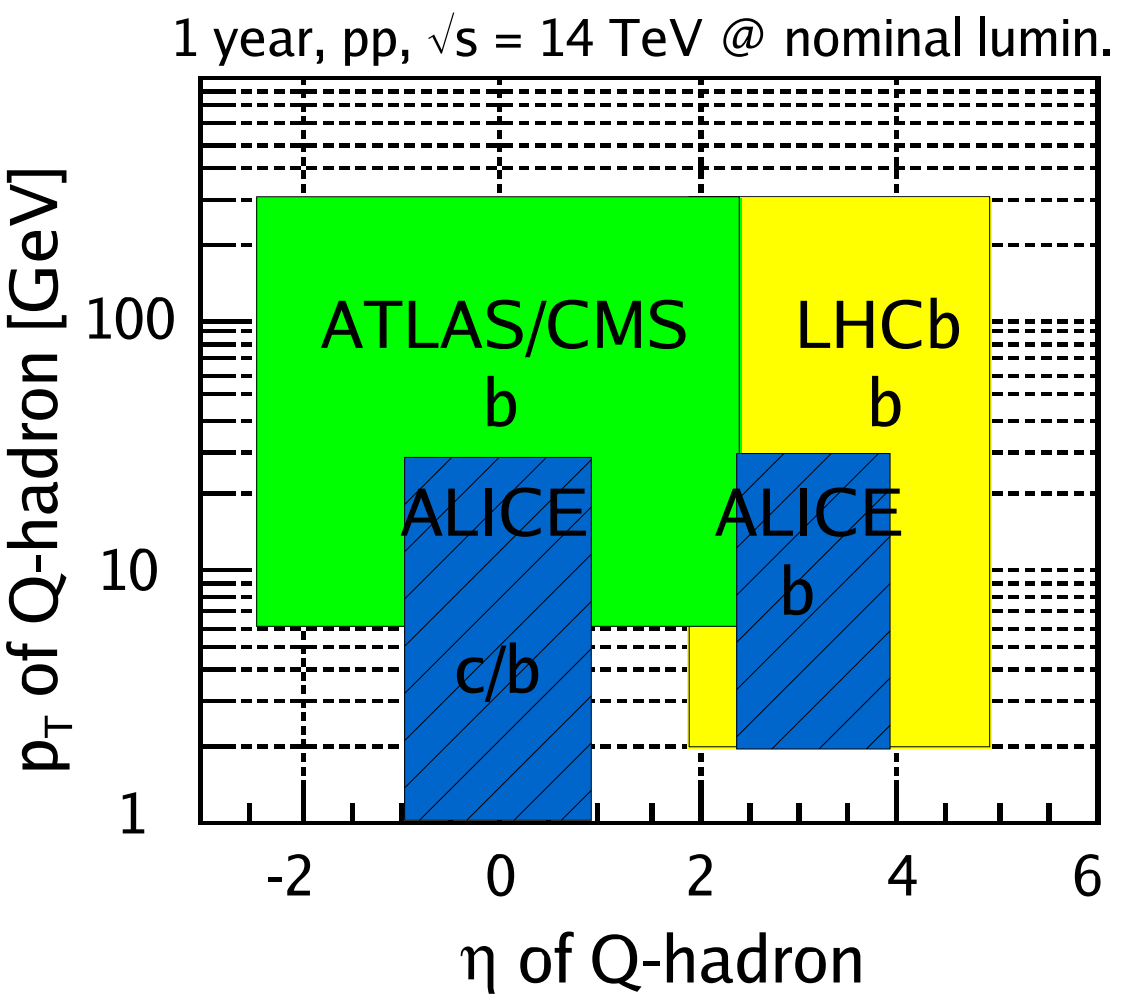

**Fig. 2:** Schematic acceptances in transverse momentum and pseudorapidity for open heavy flavour hadrons (indicated as 'Q-hadrons') in the four LHC experiments. The high-$p_T$ coverages correspond to one year (i.e. 7 months) of running at nominal luminosity (see beginning of this section).

Figure 2 shows schematically the $p_T$ vs. $\eta$ acceptances for charm (c) and beauty (b) hadrons in the four experiments, as expected for one year of running at nominal luminosity (note that the value of the luminosity is different for each experiment, as previously discussed). ATLAS and CMS have similar acceptances for beauty measurements; the minimum accessible $p_T$ is relatively large because of the strong magnetic fields, which in turn, together with the high luminosity, allow to cover transverse momenta up to 200–300 GeV. The acceptance of LHCb, although centred at forward rapidity, has a significant overlap, with those of ATLAS and CMS. The acceptance of ALICE for beauty overlaps with ATLAS and CMS at central rapidity and with LHCb at forward rapidity. The moderate magnetic field allows measurements down to transverse momenta of about 2 GeV for B mesons in the forward muon arm and in the barrel, and down to about 1 GeV for D mesons in the barrel.

## 3 Beauty triggers at the LHC

### 3.1 ATLAS beauty trigger[4]

The ATLAS trigger consists of three levels [6]. Level-1 is implemented in hardware, whilst the higher level triggers (level-2 and the Event Filter, EF) are based on general-purpose processors. The level-1 triggers are based on information from the calorimeter and muon trigger chambers. At higher trigger levels, information from the Inner Detector (ID) and precision muon detector is included. The size of the level-2 and EF processor farms is limited, which in turn limits the amount of data processing that can be performed in the trigger. The B-trigger must, therefore, have the flexibility to adapt selections both as the luminosity falls during a beam-coast and, over a longer time-scale, as the peak luminosity of the LHC increases. This is achieved by using a di-muon trigger at the start of higher luminosity fills and introducing additional triggers for lower luminosity fills or as the luminosity falls during a beam coast [7].

[4]Author: J. Baines

A di-muon trigger provides a very effective selection for a range of important channels, e.g. $\mathrm{B_d^0} \to \mathrm{J}/\psi(\mu^+\mu^-)\mathrm{K_s^0}$, $\mathrm{B} \to \mathrm{K^{0\star}}\mu\mu$ and $\mathrm{B} \to \rho^0\mu\mu$. The Level-1 muon trigger is efficient down to a $p_T$ of about 5 GeV in the barrel region and about 3 GeV in the end-caps. However the actual thresholds used for the di-muon trigger will be determined by rate limitations. For example, a $p_T$ threshold of 6 GeV would give a di-muon trigger rate of about 600 Hz after level-1 at a luminosity of $2 \times 10^{33}$ $\mathrm{cm^{-2}s^{-1}}$. These triggers are mostly due to muons from heavy flavour decays plus some single muons which are doubly counted due to overlaps in the end-cap trigger chambers. The later are removed when the muons are confirmed at level-2 using muon precision chambers and ID information from inside the level-1 Region of Interest (RoI). At the EF tracks are refit, inside regions identified by level-2, and specific selections made on the basis of mass and decay length cuts. These consist of semi-inclusive selections, for example to select $\mathrm{J}/\psi(\mu^+\mu^-)$ decays with a displaced vertex, and in some cases exclusive selections such as for $\mathrm{B} \to \mu^+\mu^-$. The final trigger rate, after the EF, is about 20 Hz at a luminosity of $2 \times 10^{33}$ $\mathrm{cm^{-2}s^{-1}}$.

At lower luminosities, additional triggers are introduced which are based on a single muon trigger ($p_T \gtrsim 8$ GeV) together with a calorimeter trigger. The calorimeter trigger identifies clusters of energy deposition in the electromagnetic and hadronic calorimeter consistent with an electron or photon (EM RoI) or a jet (Jet RoI). For hadronic final states, such as $\mathrm{B_s^0} \to \mathrm{D_s^-}\pi^+$ and $\mathrm{B_s^0} \to \mathrm{D_s^-}\mathrm{a_1^+}$ track are reconstructed in the Inner Detector in RoI of about $\Delta\eta \times \Delta\phi = 1.0 \times 1.5$. By limiting track reconstruction to the part of the ID lying within the RoI, about 10% on average, there is potential for up to a factor of ten saving in execution time compared to reconstruction in the full Inner Detector. Preliminary studies of efficiency and jet-cluster multiplicity have been made using a fast simulation which includes a detailed parameterization of the calorimeter. These studies indicate that a threshold on the jet cluster energy of $E_T > 5$ GeV gives a reasonable multiplicity, i.e. a mean of about two RoI per event for events containing a muon trigger. This threshold would give a trigger that is efficient for $\mathrm{B_s^0} \to \mathrm{D_s^-}\pi^+$ events with a $B$-hadron $p_T$ above about 15 GeV.

Track reconstruction inside e/gamma RoI can be used to select channels such as $\mathrm{B_d} \to \mathrm{K^{0\star}}\gamma$, $\mathrm{B_d^0} \to \mathrm{J}/\psi(e^+e^-)\mathrm{K_s^0}$, and $\mathrm{B_s} \to \phi\gamma$. Preliminary studies show that a reasonable compromise between RoI multiplicity and electron efficiency might be obtained with a cluster energy threshold of $E_T > 2$ GeV. This gives a mean RoI multiplicity of about one for events containing a muon trigger and is efficient for channels containing an electron with $p_T > 5$ GeV. Following the ID track reconstruction further selections are made for specific channels of interest. These are kept as inclusive as possible at level-2 with some more exclusive selections at the EF.

In LHC running, there will be competing demands for resources in the level-2 and EF trigger farms and for trigger band-width. By adopting a flexible strategy and making the maximum use of RoI information to guide reconstruction at level-2 and the EF, the B-physics coverage of ATLAS can be maximized.

### 3.2 CMS beauty trigger[5]

The Large Hadron Collider (LHC) will provide 40 MHz proton-proton collisions at the centre of mass energy of 14 TeV. At the beginning a luminosity of $2{\times}10^{33}$ $\mathrm{cm^{-2}s^{-1}}$ is expected, corresponding to 20 $\mathrm{fb^{-1}}$ collected per year. Assuming the $\mathrm{b\overline{b}}$ production cross section to be 0.5 mb, $10^{13}$ b-physics events per year are foreseen: all kind of b-particles will be produced and studies will be performed not only in $\mathrm{B_d^0}$, but also in $\mathrm{B_s^0}$ meson system. A wide b-physics programme, including CP violation, $\mathrm{B_s^0} - \mathrm{\overline{B}_s^0}$ mixing and rare decays can therefore be covered by the CMS experiment. The apparatus will be equipped with a very precise tracking system made with silicon microstrip and pixel detectors [8, 9].

The rate at which events can be archived for offline analyses is 100 Hz [10, 11]. The trigger thresholds are optimized for a wide physics discovery program with selection of high transverse momentum

[5] Author: R. Ranieri

($p_T$) processes. Low-$p_T$ events, as required for b-physics , are selected mainly by the first level muon trigger, then an exclusive reconstruction of few relevant *benchmark* channels can separate interesting events from the background. The b-physics programme could evolve with time following both the theoretical developments and the results which will be obtained in the next years by b-factories and Tevatron experiments.

The lowest trigger level (Level-1) is based on the fast response of calorimeters and muon stations with coarse granularity. No information on secondary vertices is available, hence the Level-1 selection of b-physics events exploits the leptonic signatures from beauty hadron decays, therefore a single muon or a di-muon pair is required. The Level-1 output at start-up will be 50 kHz. Several studies have been done to optimize the trigger thresholds in order to have the possibility of selecting most of the interesting physics signatures. A total of 3.6 kHz rate is dedicated to the Level-1 muon selection. It is obtained by requiring a single muon with $p_T > 14$ GeV or at least two muons with $p_T > 3$ GeV.

A further selection is made during the High-Level trigger (HLT) by using also the information from the tracking system. The CMS High-Level trigger is entirely based on a CPU farm with some thousand CPUs. Each processor analyses a single event; in principle offline event reconstruction can be performed, but in order to reduce the processing time fast track reconstruction has to be done. Some algorithms will be dedicated to the fast reconstruction and identification of physics processes, thus allowing to start the offline analysis directly from the online selection. They have to fulfill the HLT time constraint, hence they have to be able to analyze and accept (or reject) data within the time limits imposed by the HLT latency. To lower the execution time, which is due mainly to the processing of tracking system signals, track reconstruction is preferably performed only in limited regions of the space (*regional track reconstruction*) and stopped when a certain precision is reached in the measurement of some track parameters, such as transverse momentum and impact parameter (*conditional track finding*). Invariant mass of b-hadrons can thus be measured online with good resolution, allowing to select the searched event topologies.

An additional trigger strategy, which relies on the possibility of lowering the trigger thresholds during the LHC beam coast or lower luminosity fills to collect more b-physics events is under study.

The rare decay $\mathrm{B}^0_{\mathrm{s,d}} \to \mu^+\mu^-$ is triggered at Level-1 with 15.2% efficiency. At HLT, the two muons are required to be opposite charged and isolated, to come from a displaced common vertex and have an invariant mass within 150 MeV from the $\mathrm{B}^0_{\mathrm{s}}$ mass. The estimated background rate is below 2 Hz and nearly 50 signal events are expected with 10 fb$^{-1}$.

The determination of $\Delta m_s$ and $\Delta\Gamma_s$ will be a valuable input for flavour dynamics in the Standard Model and its possible extensions. The measurement of $\Delta m_s$ is allowed by the $\mathrm{B}^0_{\mathrm{s}} \to \mathrm{D_s}^-\pi^+$ decay followed by $\mathrm{D_s}^- \to \phi\pi^-$ and $\phi \to \mathrm{K}^+\mathrm{K}^-$. The $\mathrm{B}^0_{\mathrm{s}}$ CP state at decay time is tagged by the charge of the pion associated to the $\mathrm{D_s}$ (in this case the $\pi^+$). The only way to trigger on these hadronic events is to search for the muon coming from the decay of the other b quark in the event. In addition to the single muon Level-1 trigger, it was studied the possibility of a combined trigger with a low-$p_T$ muon and a soft jet. The CMS High-Level trigger algorithm reconstructs the charged particle tracks with only three points by using the precise pixel detector. Topological and kinematical cuts are applied to reconstruct the three resonances $\phi$, $\mathrm{D_s}$ and $\mathrm{B}^0_{\mathrm{s}}$. A 20 Hz output rate is achieved with about 1000 signal events in 20 fb$^{-1}$. Since the overall possible rate on tape is 100 Hz, the bandwidth allocated to this channel probably could not exceed 5 Hz. If the fraction of events written to tape is scaled accordingly, more than 300 signal events are expected for 20 fb$^{-1}$. In order to fully cover the range allowed by the Standard Model, about 1000 events are needed.

The decay channel $\mathrm{B}^0_{\mathrm{s}} \to \mathrm{J}/\psi\phi$ is very important because it can not be studied with large accuracy before LHC and can reveal hints for physics beyond the Standard Model. Events with a couple of muons are passed to the HLT. The inclusive selection of $\mathrm{J}/\psi \to \mu^+\mu^-$ decays, obtained with mass requirements on the di-muon system, leads to a total of 15 Hz rate, 90% of which is made of $\mathrm{J}/\psi$ from b quarks. With an additional amount of CPU time, perhaps sustainable by the HLT computing power, about 170 000 events are expected in 20 fb$^{-1}$ with less than 2 Hz rate.

### 3.3 LHCb beauty trigger[6]

The LHCb detector [5] is optimized for exploiting the B-physics potential of LHC. Together with excellent vertexing and particle identification, an efficient trigger on a wide variety of B decays is one of the main design requirements of the experiment.

The LHCb trigger system [12] is organized in three levels. The first one (L0) runs on custom electronics and operates synchronously at 40 MHz, with a 4 $\mu$s latency. The remaining two trigger levels (L1 and HLT) run on a shared farm of 1400 commercial CPUs. A brief description of the three trigger levels and their performance follows.

L0 exploits the relatively high $p_T$ of B decay products. High $p_T$ candidates are identified both in the calorimeter and in the muon system, with $p_T$ thresholds of about 3 and 1 GeV respectively. Complicated events that would consume unreasonable time at higher levels are promptly vetoed in two different ways. First, multiple primary vertex topologies are rejected by using two dedicated silicon layers of the vertex detector. Secondly, events with large multiplicity, measured at a scintillating pad layer, are vetoed. The input rate of events visible in the detector is about 10 MHz, with a $b\bar{b}$ content of $1.5\%$. L0 reduces this rate by a factor of 10 while increasing the $b\bar{b}$ content to $3\%$. The typical efficiency of L0 is $90\%$ for channels with dimuons, $70\%$ for radiative decays and $50\%$ for purely hadronic decays.

At the 1 MHz input rate of L1 it becomes feasible to use tracking information, allowing for the search for B vertex displacement signatures. Tracks are first searched at the vertex detector, and then confirmed in two dedicated tracking layers (trigger tracker or TT) which provide a rough estimation of the momentum of the tracks ($\delta p_T/p_T \sim 25\%$). The generic L1 decision is based on the presence of two tracks with an impact parameter higher than 0.15 mm with a sufficiently high value of $log(p_{T1} + p_{T2})$. Alternative selection criteria are applied, based on the presence of tracks matched to L0 neutral calorimeter objects and muon candidates. The output rate of L1 is 40 kHz with a $b\bar{b}$ content of $15\%$. The efficiencies are at the level of 90, 80 and $70\%$ for channels with di-muons, only hadrons and radiative decays respectively.

The HLT [13] consists of two sequential *layers*. The first one refines the L1 decision with the all the tracking information from the detector, improving the $p_T$ measurement to the level of $\delta p_T/p_T \sim 1\%$. The rate is reduced to 13 kHz and the $b\bar{b}$ content is enriched to $30\%$. The second layer consists on a series of alternative selections. A first group of them aims for maximal efficiency on the base-line physics channels and the corresponding control samples, by making use of the complete reconstruction of the decay vertex and its kinematical properties. These selections fill 200 Hz of bandwidth, while providing efficiencies typically higher than $90\%$. The rest of selections aim for more generic signatures that will provide robustness and flexibility to the trigger system. In addition, the samples selected will be useful for calibration and systematic studies. The selections aim for generic J/$\psi$ and $D^\star$ (600 Hz and 300 Hz respectively), and generic B decays (900 Hz). The latter is based on the detection of single muons with high $p_T$ and impact parameter.

In total, 2 kHz of events will be written on tape, with an expected overall efficiency ranging between $75\%$ for channels with di-muons to $35\%$ for purely hadronic final states.

## 4 Measurements in preparation at the LHC and expected performance

In the following we present, as examples, the expected performance for the detection of D and B mesons in ALICE[7], and for the study of $\mathrm{b\bar{b}}$ azimuthal correlation in ATLAS. We also include a summary of the capability of ALICE of quarkonia measurements ($\psi$ family and $\Upsilon$ family).

[6] Author: H. Ruiz

[7] Given that ALICE is dedicated to the study of nucleus–nucleus collisions at the LHC, some of the presented results are relative to Pb–Pb collisions at $\sqrt{s} = 5.5$ TeV per nucleon–nucleon collisions. These results can be taken as lower limits for the performance in pp collisions, where the background contributions are much lower.

### 4.1 Charm reconstruction in ALICE[8]

One of the most promising channels for open charm detection is the $D^0 \to K^-\pi^+$ decay (and charge conjugate) that has a BR of $3.8\%$. The expected yields (BR $\times$ $\mathrm{d}N/\mathrm{d}y$ at $y = 0$), in pp collisions at $\sqrt{s} = 14$ TeV and in central Pb–Pb (0–5% $\sigma^{\rm tot}$) at $\sqrt{s_{\rm NN}} = 5.5$ TeV are $7.5 \times 10^{-4}$ and $5.3 \times 10^{-1}$ per event, respectively [14].

The main feature of this decay topology is the presence of two tracks with impact parameters $d_0 \sim 100\ \mu\mathrm{m}$. The detection strategy [15] to cope with the large combinatorial background from the underlying event is based on the selection of displaced-vertex topologies, i.e. two tracks with large impact parameters and good alignment between the $D^0$ momentum and flight-line, and on invariant-mass analysis to extract the signal yield. This strategy was optimized separately for pp and Pb–Pb collisions, as a function of the $D^0$ transverse momentum, and statistical and systematic errors were estimated [16, 17]. The results, in terms of $p_T$ coverage and statistical precision, are found to be similar for the two colliding systems [16, 17].

Figure 3 (left) shows the expected sensitivity of ALICE for the measurement of the $D^0$ $p_T$-differential cross section in pp collisions, along with NLO pQCD [18] calculation results corresponding to different choices of the charm quark mass and of renormalization and factorization scales. In the right-hand panel of the figure we present the ratio 'data/theory' ('default parameters/theory parameters') which better allows to compare the different $p_T$-shapes obtained by changing the input 'theory parameters' and to illustrate the expected sensitivity of the ALICE measurement. The estimated experimental errors are much smaller than the theoretical uncertainty band. We note that the data cover the region at low transverse momentum where the accuracy of the pQCD calculation becomes poorer and where novel effects, determined by the high partonic density of the initial state at LHC energies, may play an important role (see "Small-$x$ effects in heavy quark production" section of this report).

### 4.2 Beauty production measurements in ALICE[9]

The expected yields (BR $\times$ $\mathrm{d}N/\mathrm{d}y$ at $y = 0$) for $B \to e^\pm + X$ plus $B \to D\,(\to e^\pm + X) + X'$ in pp collisions at $\sqrt{s} = 14$ TeV and in central Pb–Pb (0–5% $\sigma^{\rm tot}$) at $\sqrt{s_{\rm NN}} = 5.5$ TeV are $2.8 \times 10^{-4}$ and $1.8 \times 10^{-1}$ per event, respectively [14].

The main sources of background electrons are: (a) decays of D mesons; (b) neutral pion Dalitz decays $\pi^0 \to \gamma e^+ e^-$ and decays of light mesons (e.g. $\rho$ and $\omega$); (c) conversions of photons in the beam pipe or in the inner detector layers and (d) pions misidentified as electrons. Given that electrons from beauty have average impact parameter $d_0 \simeq 500\ \mu\mathrm{m}$ and a hard momentum spectrum, it is possible to obtain a high-purity sample with a strategy that relies on: electron identification with a combined $\mathrm{d}E/\mathrm{d}x$ and transition radiation selection, which allows to reduce the pion contamination by a factor $10^4$; impact parameter cut to reject misidentified pions and electrons from sources (b) and (c); transverse momentum cut to reject electrons from charm decays. As an example, with $d_0 > 200\ \mu\mathrm{m}$ and $p_T > 2$ GeV, the expected statistics of electrons from B decays is $8 \times 10^4$ for $10^7$ central Pb–Pb events, allowing the measurement of electron-level $p_T$-differential cross section in the range $2 < p_T < 18$ GeV. The residual contamination of about $10\%$, acculated in the low-$p_T$ region, of electrons from prompt charm decays and from misidentified charged pions can be evaluated and subtracted using a Monte Carlo simulation tuned to reproduce the measured cross sections for pions and $D^0$ mesons. A Monte-Carlo-based procedure can then be used to compute, from the electron-level cross section, the B-level cross section $\mathrm{d}\sigma^{\rm B}(p_T > p_T^{\rm min})/\mathrm{d}y$ [17]. In the left-hand panel of Fig. 4 we show this cross section for central Pb–Pb collisions with the estimated uncertainties. The covered range is $2 < p_T^{\rm min} < 30$ GeV.

B production can be measured also in the ALICE forward muon spectrometer, $2.5 < \eta < 4$, analyzing the single-muon $p_T$ distribution and the opposite-sign di-muons invariant mass distribution [17].

[8] Author: A. Dainese

[9] Author: A. Dainese

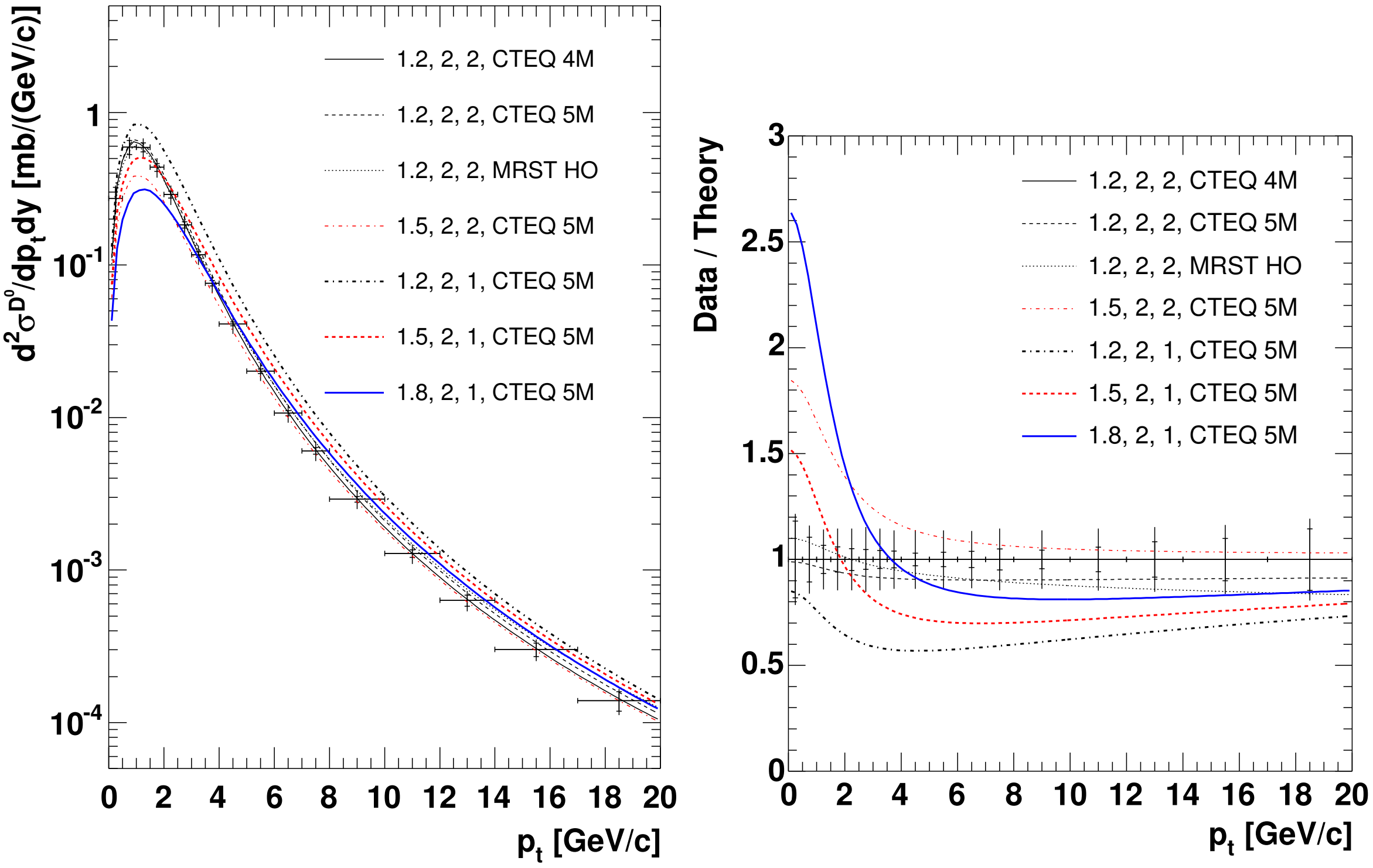

**Fig. 3:** ALICE sensitivity on $\mathrm{d}^2\sigma^{\mathrm{D}^0}/\mathrm{d}p_T\mathrm{d}y$, in pp at 14 TeV, compared to the pQCD predictions obtained with different sets of the input parameters $m_{\mathrm{c}}$ [GeV], $\mu_F/\mu_0$, $\mu_R/\mu_0$ and PDF set ($\mu_0$ is defined in the text). The inner bars represent the statistical error, the outer bars the quadratic sum of statistical and $p_T$-dependent systematic errors. A normalization error of 5% is not shown. The panel on the right shows the corresponding 'data/theory' plot.

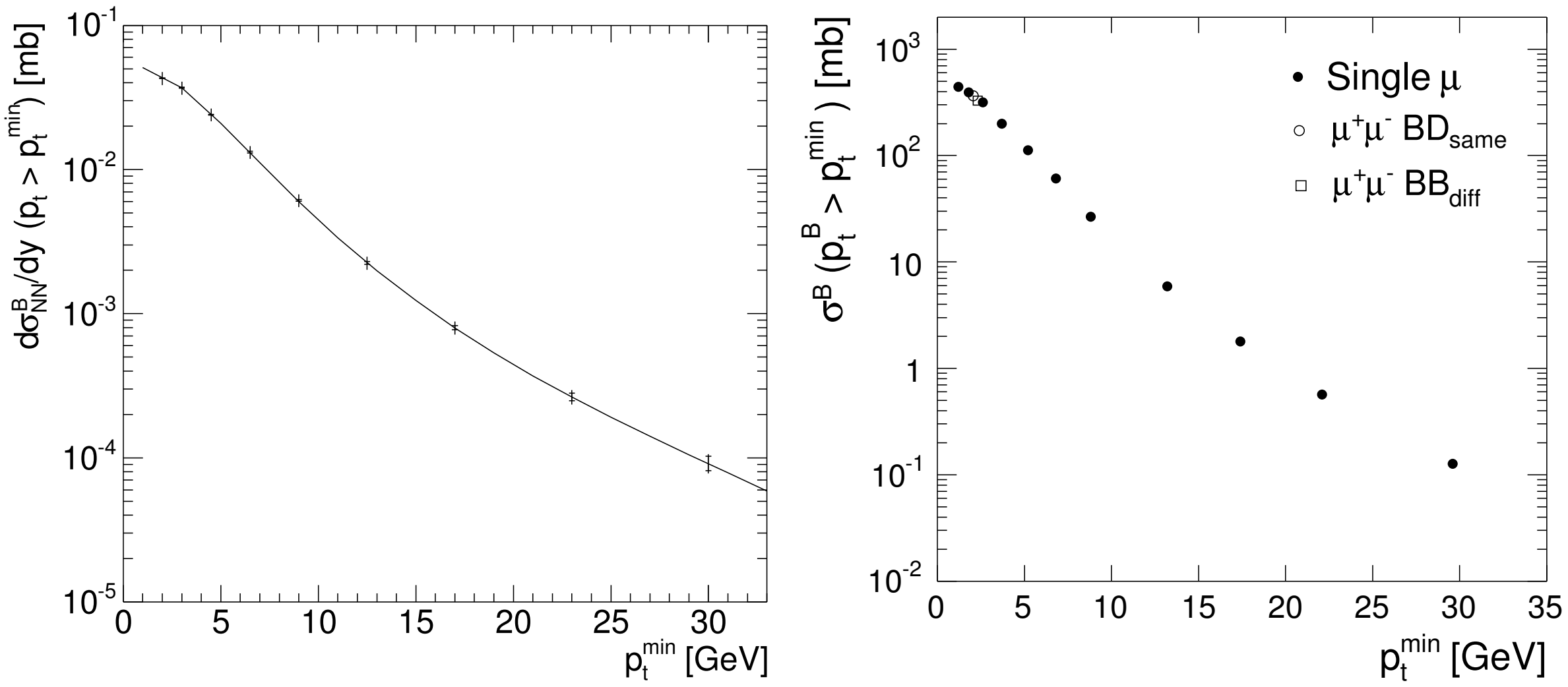

**Fig. 4:** B production cross section vs. $p_T^{\min}$ reconstructed by ALICE in $10^7$ central Pb–Pb events. Left: $\mathrm{d}\sigma^{\mathrm{B}}_{\mathrm{NN}}/\mathrm{d}y$ at $y = 0$ (normalized to one nucleon–nucleon collision) from single electrons in $|\eta| < 0.9$; statistical (inner bars) and quadratic sum of statistical and $p_T$-dependent systematic errors (outer bars) are shown; a 9% normalization error is not shown. Right: $\sigma^{\mathrm{B}}$ integrated in $2.5 < y^{\mathrm{B}} < 4$ (not normalized to one nucleon–nucleon collision) from single muons and di-muons in $2.5 < \eta < 4$; only (very small) statistical errors shown.

The main backgrounds to the 'beauty muon' signal are $\pi^{\pm}$, $\mathrm{K}^{\pm}$ and charm decays. The cut $p_T > 1.5$ GeV is applied to all reconstructed muons in order to increase the signal-to-background ratio. For the opposite-sign di-muons, the residual combinatorial background is subtracted using the technique of event-mixing and the resulting distribution is subdivided into two samples: the low-mass region, $M_{\mu^+\mu^-} < 5$ GeV, dominated by muons originating from a single b quark decay through $\mathrm{b} \to \mathrm{c}(\to \mu^+)\mu^-$ ($\mathrm{BD_{same}}$), and the high-mass region, $5 < M_{\mu^+\mu^-} < 20$ GeV, dominated by $\mathrm{b\overline{b}} \to \mu^+\mu^-$, with each muon coming from a different quark in the pair ($\mathrm{BB_{diff}}$). Both samples have a background from $\mathrm{c\overline{c}} \to \mu^+\mu^-$ and a fit is done to extract the charm- and beauty-component yields. The single-muon $p_T$ distribution has three components with different slopes: K and $\pi$, charm, and beauty decays. Also in this case a fit technique allows to extract a $p_T$ distribution of muons from B decays. A Monte Carlo procedure, similar to that used for semi-electronic decays, allows to extract B-level cross sections for the data sets (low-mass $\mu^+\mu^-$, high-mass $\mu^+\mu^-$, and $p_T$-binned single-muon distribution), each set covering a specific B-meson $p_T > p_T^{\min}$ region, as preliminarily shown in Fig. 4 (right). Since only minimal cuts are applied, the reported statistical errors are very small and high-$p_T$ reach is excellent. Systematic errors are currently under study.

### 4.3 Study of $\mathrm{b\overline{b}}$ correlations in ATLAS[10]

The ATLAS detector [19] is well engineered for studies of b-production, and together with the huge rate of b-quark production that will be seen at LHC, offers great potential for the making of novel precise b production measurements. Correlations between b and $\overline{\mathrm{b}}$ quarks and events with more than one heavy-quark pair, $\mathrm{b\overline{b}b\overline{b}}$, $\mathrm{b\overline{b}c\overline{c}}$, $\mathrm{b\overline{b}s\overline{s}}$, that were difficult to access in previous experiments due to limited statistics, will be investigated in detail. A new technique has been developed in ATLAS for measuring correlations, and this will yield results that will shed new light on our understanding of the QCD cross-section for $\mathrm{b\overline{b}}$-production.

A detailed study investigated a possibility of $\mathrm{b\overline{b}}$ correlations measurement using the $\Delta\phi$(J/$\psi$-$\mu$) distribution, the azimuthal separation of a J/$\psi$ and a muon [20–22]. This technique is expected to be superior to earlier methods used at the Tevatron Run-1 based on muon–muon or muon–b-jet correlations. The new method does not require separation cuts between the two objects. Such cuts were necessary to control the background, but they required a model-dependent extrapolation of the results to full azimuthal space [23]. Using a full simulation of the Inner Detector and the Muon Spectrometer of the ATLAS detector [19] it is shown that such a distribution can be extracted from heavy flavour events at LHC.

ATLAS studies were done for two channels selected to measure the azimuthal angle difference $\Delta\phi(\mathrm{b\overline{b}})$ between b and $\overline{\mathrm{b}}$ quarks:

$$\overline{\mathrm{b}} \to \mathrm{B_d} \to \mathrm{J}/\psi(\to \mu\mu)\mathrm{K}^0\,, \quad \mathrm{b} \to \mu + \mathrm{X} \quad \text{and} \quad \overline{\mathrm{b}} \to \mathrm{B_s} \to \mathrm{J}/\psi(\to \mu\mu)\phi\,, \quad \mathrm{b} \to \mu + \mathrm{X}\,.$$

The numbers of events expected for 30 $\mathrm{fb}^{-1}$ as might be achieved after 3 years of running at a luminosity of $10^{33}$ $\mathrm{cm}^2\mathrm{s}^{-1}$ are $4.8\times10^4$ and $3.2\times10^4$ respectively for these channels. No isolation cuts are needed to separate exclusively reconstructed B-decays from the muon produced in the semi-leptonic decay of the other B-particle in the event. The reconstruction efficiency remains high in topologies where the azimuthal angle difference $\Delta\phi$(J/$\psi$-$\mu$) between J/$\psi$ and the muon is small.

Special attention was devoted to background events in which the muon is produced from the decays $\mathrm{K}^{\pm}$, $\pi^{\pm} \to \mu^{\pm} + X$ instead of b$\to$ $\mu + X$. The study showed that this background is not problematic in $\mathrm{B_d}$ decays, however it is important in the case of $\mathrm{B_s^0}$ meson.

In summary, the results of the analysis suggest that backgrounds from K/$\pi$ decays are small, and that backgrounds from events containing 4 b quarks are relatively flat in $\Delta\phi$(J/$\psi$-$\mu$). The efficiency of the reconstruction of muons with this technique is also relatively flat in $\Delta\phi$(J/$\psi$-$\mu$) and so we conclude that corrections to the measured $\Delta\phi$(J/$\psi$-$\mu$) distribution are likely to be small.

[10] Author: Th. Lagouri

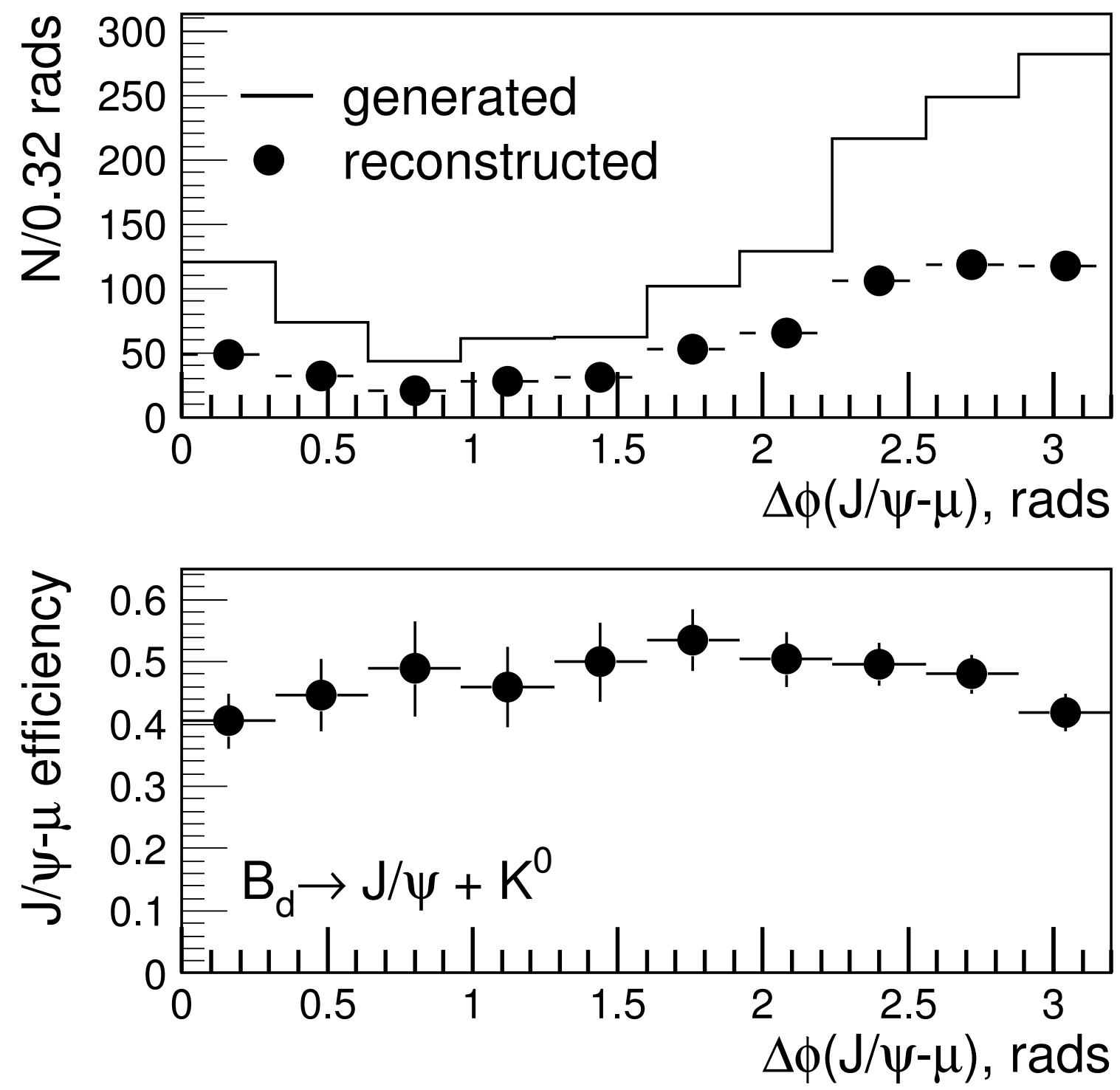

**Fig. 5:** Distribution of the opening angle $\Delta\phi$(J/$\psi$-$\mu$) between the J/$\psi$ from the decay $\mathrm{B_d} \to \mathrm{J}/\psi\mathrm{K}^0$ and the muon coming from the associated B hadron decay to muon, both direct $\mathrm{b} \to \mu$ and indirect $\mathrm{b} \to \mathrm{c} \to \mu$.

### 4.4 Quarkonia measurements in ALICE[11]

Heavy quarkonia states are hard, penetrating probes which provide an essential tool to study the earliest and hottest stages of heavy-ion collisions [24]. They can probe the strongly interacting matter created in these reactions on short distance scales and are expected to be sensitive to the nature of the medium, i.e. confined or de-confined [25, 26]. The suppression (dissociation) of the heavy-quark resonances is considered as one of the most important observables for the study of the QGP at the LHC (see Ref. [2] for a recent review).

In ALICE, quarkonia will be measured in the di-electron channel using a barrel ($|\eta| < 0.9$) Transition Radiation Detector (TRD) [2] and in the di-muon channel using a forward Muon Spectrometer ($2.5 < \eta < 4$) [2]. The complete spectrum of heavy-quark vector mesons (J/$\psi$, $\psi'$, $\Upsilon$, $\Upsilon'$, $\Upsilon''$) can be measured down to zero $p_T$. In particular the good mass resolution allows to resolve the Upsilon family.

The Muon Spectrometer uses a low-$p_T$ trigger threshold, $p_T > 1$ GeV, on single muons for charmonia and a high-$p_T$ trigger, $p_T > 2$ GeV, for bottonia detection. The TRD can trigger on single electrons with $p_T > 3$ GeV, which results in a minimum transverse momentum of triggered charmonia of 5.2 GeV. Electron identification combined with the excellent vertexing capabilities of the inner tracking system allows ALICE to distinguish direct charmonium production from secondary charmonium production through B decays.

The energy density dependence will be studied by varying the impact parameters and by studying in addition to the heaviest collision system (Pb–Pb) also intermediate mass and low mass A–A systems. To determine the primary production cross-section of the resonances and the amount of pre-resonance absorption, corresponding measurement have to be performed for pA and pp collisions.

[11]Author: A. Morsch

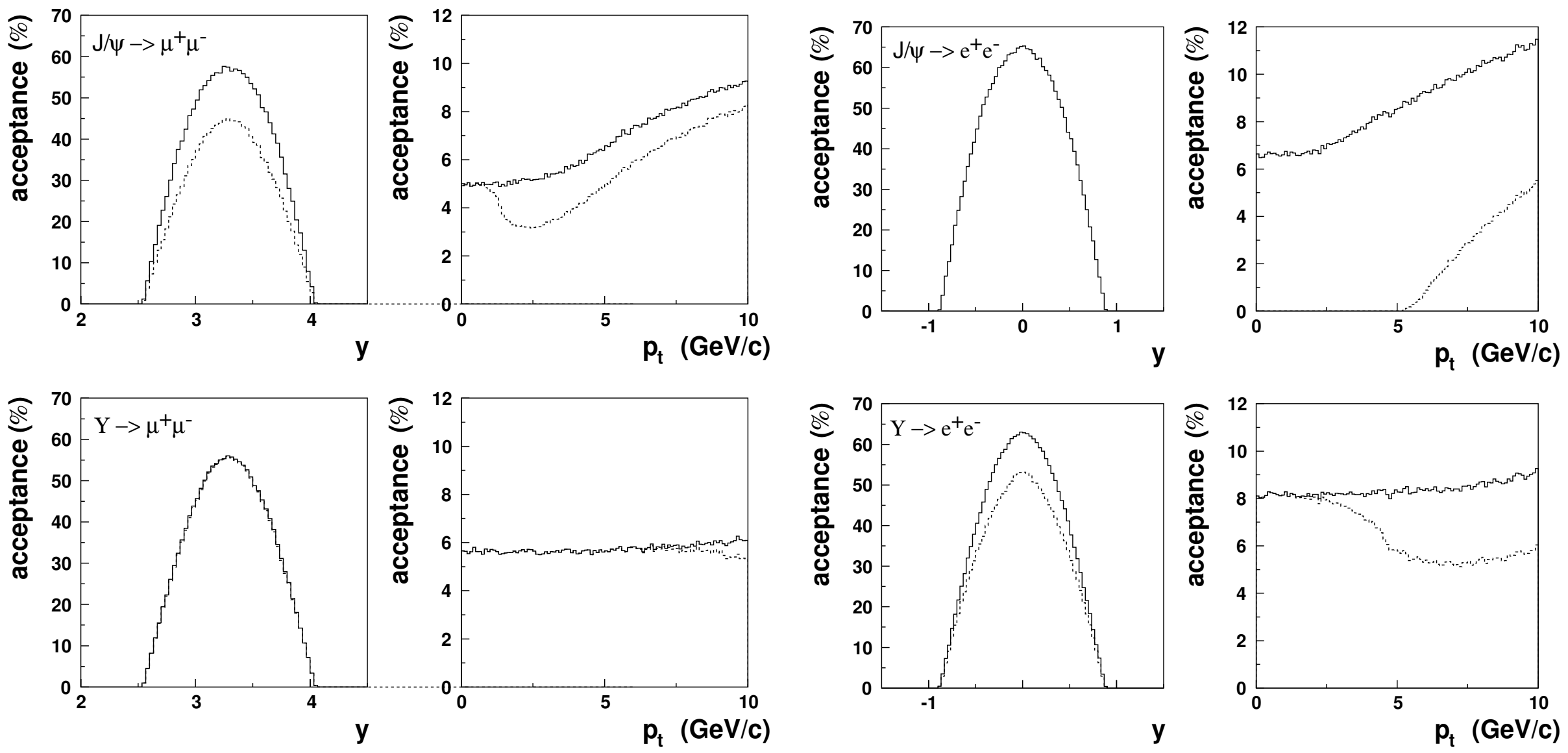

**Fig. 6:** Acceptance for J/$\psi$ and $\Upsilon$ as a function of $y$ and $p_T$ for measurements in the di-muon channel and di-electron channels. To give an idea of the effect of the trigger, the acceptances are shown without (solid) and with (dashed) a sharp cut on the transverse momentum of single muons of 1 GeV/$c$ (2 GeV/$c$) for J/$\psi$ ($\Upsilon$ ) and for single electrons of 3 GeV.

Table 1 shows the main quarkonia detection characteristics of the TRD and the Muon Spectrometer and the acceptances for J/$\psi$ and $\Upsilon$ as a function of $y$ and $p_T$ are shown in Fig. 6.

**Table 1:** Main characteristics of quarkonia detection with the TRD and the Muon Spectrometer in ALICE.

| | Muon Spectrometer | TRD |
|---|---|---|
| Acceptance | $2.5 < \eta < 4$ | $\|\eta\| < 0.9$ |
| Mass Resolution J/$\psi$ | 72 MeV | 34 MeV |
| Mass Resolution $\Upsilon$ | 99 MeV | 93 MeV |

In one year of pp running at $\langle L \rangle = 3 \times 10^{30}$ cm$^{-2}$s$^{-1}$ ALICE will detect several $10^6$ J/$\psi$s and several $10^4$ $\Upsilon$s in the di-muon channel. A $\Upsilon$ statistics of $10^2$–$10^3$ can be obtained in the di-electron channel. For the J/$\psi$ we expect $\approx 10^4$ untriggered low-$p_T$ and $\approx 10^4$ high-$p_T$ triggered events.

# Small-$x$ effects in heavy quark production

*A. Dainese[a], K. J. Eskola[b], H. Jung[c], V. J. Kolhinen[b], K. Kutak[d,e], A. D. Martin[f], L. Motyka[c], K. Peters[c,g], M. G. Ryskin[h], and R. Vogt[i,j]*
[a] University of Padova and INFN, Padova, Italy
[b] Department of Physics, University of Jyväskylä, Jyväskylä, Finland
[c] Deutsches Elektronen-Synchroton DESY, Hamburg, FRG
[d] University of Hamburg, Hamburg, FRG
[e] Henryk Niewodniczanski Institute of Nuclear Physics, Polish Academy of Sciences, Poland
[f] Institute for Particle Physics Phenomenology, University of Durham, Durham, UK
[g] Department of Physics and Astronomy, University of Manchester, UK
[h] Petersburg Nuclear Physics Institute, Gatchina, St. Petersburg, Russia
[i] Lawrence Berkeley National Laboratory, Berkeley, CA, USA
[j] Physics Department, University of California, Davis, CA, USA

**Abstract**
We study small-$x$ effects on heavy flavor production at the LHC in two approaches including nonlinear, saturation-motivated, terms in the parton evolution. One approach is based on collinear factorization, the other on $k_T$ factorization. The prospects for direct experimental study of these effects in $pp$ collisions at the LHC are discussed.

*Coordinators: A. Dainese, H. Jung, and R. Vogt*

## 1 Introduction

HERA data are used to constrain the small $x$, moderate $Q^2$ parton densities in two approaches. In the first, HERA $F_2$ data are refit using DGLAP evolution with the first nonlinear recombination terms. Recombination slows the evolution so that, after refitting the data, the gluon distribution is enhanced relative to that obtained by DGLAP alone. The resulting set of parton densities produces charm enhancement in $pp$ collisions at the LHC. On the other hand, assuming $k_T$ factorization, the unintegrated gluon distribution is determined from the HERA $F_2^c$ data, the only inclusive HERA measurements which directly accesses the gluon density. Saturation effects are then included, reducing the small $x$ gluon densities with little distinguishable effect on $F_2$. This approach leads instead to heavy flavor suppression at the LHC. After a short general introduction, both approaches and their predicted effects on heavy quark production are discussed in detail. Direct experimental study of these effects in $pp$ collisions at the LHC may be able to differentiate between the two approaches.

## 2 Small-$x$ partons, absorption and the LHC[1]

### 2.1 Partons densities at low $x$?

Almost nothing is known about the behaviour of partons at low $x$. There are essentially no data available for $x < 10^{-4}$ with $Q^2$ in the perturbative region and there is no reliable theory to extrapolate down in $x$.

In the Dokshitzer-Gribov-Lipatov-Altarelli-Parisi (DGLAP)-based [1–4] global analyses, small-$x$ behaviour is driven by input distributions at a starting scale $Q = Q_0$. Usually these 'input' distributions are written in the form $x^{-\lambda}(1-x)^\eta$ where $\lambda$ and $\eta$ are free parameters fit to the data. So one can say nothing without data in the $x$ region of interest. Moreover, there may be large low-$x$ contributions to the gluon of the form $(\alpha_s \ln(1/x))^n$ – the so-called Balitsky-Fadin-Kuraev-Lipatov (BFKL) effects [5–8], beyond the DGLAP approximation.

[1] Authors: A.D. Martin and M.G. Ryskin

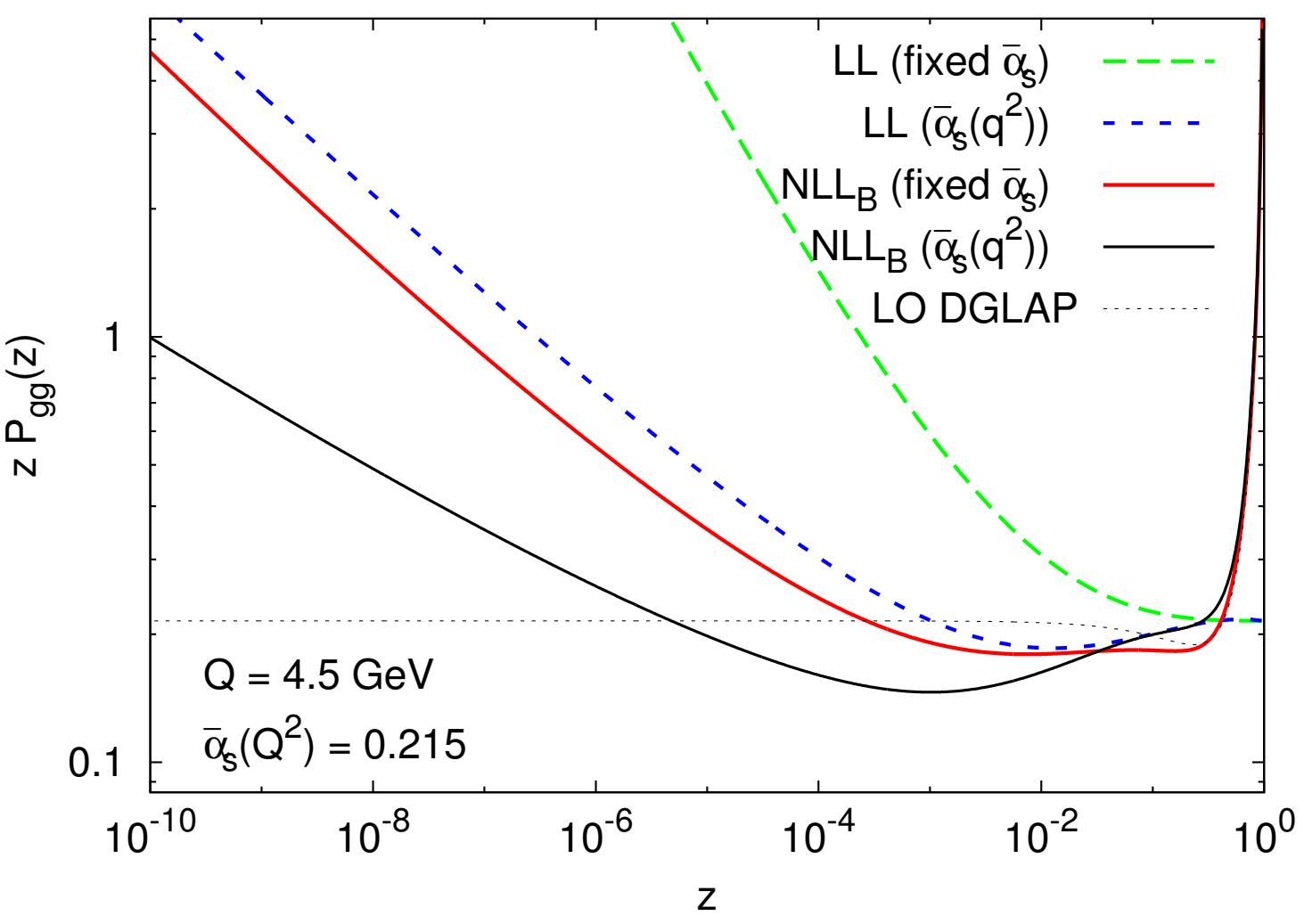

**Fig. 1:** The gluon-gluon splitting function, $P_{gg}$, with fixed and running coupling in the LL and resummed NLL BFKL approximations, compared with the LO DGLAP behaviour. The figure is taken from Ciafaloni et al. [9–14]. The subscript B refers to scheme B which ensures energy-momentum conservation in the splitting.

Thus it seems better to discuss low-$x$ behaviour in terms of BFKL-evolution. However there are also problems here. The next-to-leading logarithm (NLL) corrections to the leading order (LO) BFKL (CCFM) amplitude are known to be very large and one needs to resum such corrections to obtain a relatively stable result. We cannot justify the perturbative QCD approach at low $Q^2$ so that the solution of the BFKL equation need to be matched to some non-perturbative amplitude at $Q = Q_0$. This non-perturbative distribution (analogous to the 'input' in the DGLAP case) is not known theoretically. Either it has to be fit to low $x$ data or some phenomenological model such as a Regge-based parametrization has to be used.

After a reasonable resummation of the NLL corrections in the region where the starting virtuality $Q_0$ is not close to the final value of $Q$, $Q > Q_0$, the resummed BFKL amplitude turns out to be similar to that resulting from DGLAP evolution [9–14]. For example, the preasymptotic effects on the resummation of the gluon-gluon splitting function are so large that the NLL BFKL power growth only sets in for $z < 10^{-5}$, as can be seen from Fig. 1. Moreover, the behaviour of the convolution $P_{gg} \otimes g/g$, normalized to $g$, in the NNLO DGLAP and NLL approximations is exactly the same down to $z \sim 10^{-4}$ [15].

Thus, in practice, the DGLAP and BFKL based approaches are rather close to each other in the HERA kinematic regime. In both cases, the main problem is the low-$x$ behaviour of the amplitude at $Q = Q_0$ where we need to phenomenologically determine possible non-perturbative contributions, power corrections and so on.

## 2.2 The puzzle of the $x^{-\lambda}$ behaviour

Since the BFKL amplitude grows as a power of $x$, $A \propto x^{-\lambda}$, it will violate unitarity as $x \to 0$. Indeed, even after the NLL resummation, the expected power, $\lambda \simeq 0.3$, is rather large. Thus, we first discuss absorption effects which tame the violation of unitarity. The upper limit of the small $x$ behaviour of the parton distributions $a = g, q$ is given by the extrapolation

$$xa(x, q^2) = \left(\frac{x_0}{x}\right)^{0.3} x_0 a(x_0, q^2) \tag{1}$$

below $x_0 = 0.001$. The distributions are reliably determined from global parton analyses at $x > x_0$.

On the other hand, it is reasonable to expect that at $Q \lesssim Q_0 \sim 1$ GeV the behaviour will reflect that of hadron-hadron interactions: $\lambda = 0.08$ [16]. Most likely the lower value of $\lambda$ is explained by absorptive/screening effects in hadron-hadron collisions. So, for extrapolation down to $x \sim 10^{-7} - 10^{-6}$ we may regard $\lambda = 0.08$ as a lower limit since, in DIS, we expect smaller absorptive effects than those in hadron-hadron interactions.

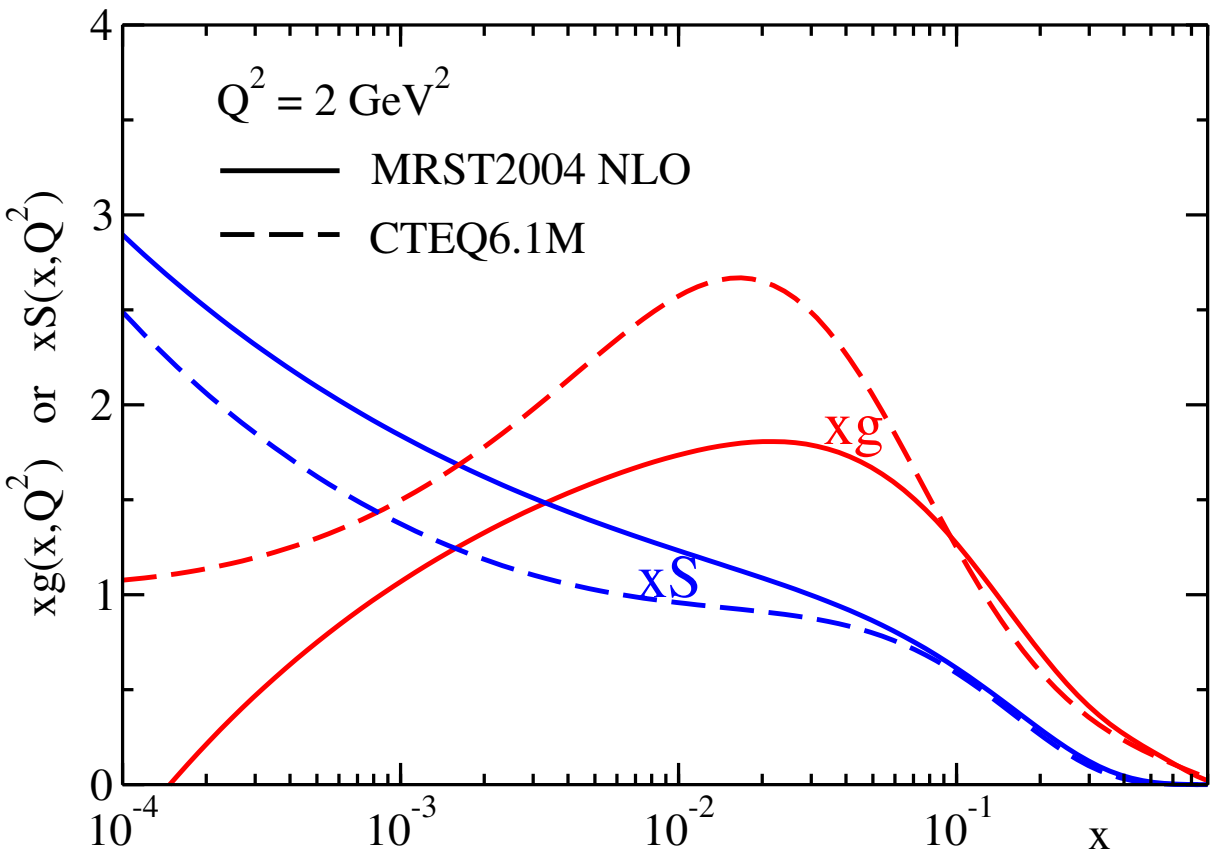

**Fig. 2:** The behaviour of the gluon and sea quark distributions at $Q^2 = 2$ GeV$^2$ found in the CTEQ6.1M [17] and MRST2004 NLO [18] global analyses. The valence-like behaviour of the gluon is evident.

However, present global analyses, which do not allow for absorption effects, reveal that at $Q \sim 1 - 1.5$ GeV and low $x$, the sea quarks have a Pomeron-like behaviour, $xq \sim x^{-0.2}$, whereas the gluon distribution has a valence-like behaviour, $xg \sim \sqrt{x}$. This different behaviour is evident from Fig. 2, which shows the behaviour of the gluon and sea quark distributions, $xS = 2x(\bar{u} + \bar{d} + \bar{s})$ for $Q^2 = 2$ GeV$^2$. Such a result looks strange from the Regge viewpoint where the same vacuum singularity (Pomeron) should drive both the sea quarks and the gluons since the same power is expected for sea quarks and gluons, $\lambda_g = \lambda_q$.

This difference demonstrates that the actual situation is even more complicated. It is worth noting that a simultaneous analysis of inclusive and diffractive DIS data indicates that, after accounting for screening effects and allowing for some power corrections, it may be possible to describe the HERA data with $\lambda_g = \lambda_q = 0$ [19]. The absorptive effects, estimated from fitting the diffractive DIS data, enlarge the input gluon distribution at low $x$.

It may initially seem strange that accounting for absorptive effects gives a *larger* gluon density[2] at low $x$ and $Q^2$. The point is that the only way to describe the data, *which are sensitive to absorptive effects*, within the framework of DGLAP evolution without absorption, is to choose a very low 'input' gluon density in order to *mimic* the screening corrections 'hidden' in the data. Indeed, there is a tendency for the gluon distribution to even become negative at low $x$ and $Q^2$. On the other hand, allowing for absorption during DGLAP evolution (with the help of the Gribov-Levin-Ryskin (GLR) and Mueller-Qiu (MQ), GLRMQ, equations [22,23]) the same data may be described with a larger and definitely positive input gluon density at $Q = Q_0$.

### 2.3 Estimates of absorptive effects: GLRMQ to BK

The saturation of parton densities ($\lambda = 0$) may be obtained using the Balitski-Kovchegov (BK) [24,25] equation, based on the BFKL equation, as well as the aforementioned GLRMQ equations. The latter equation is based on DGLAP evolution. These equations sum the set of fan diagrams which describe the rescattering of intermediate partons on the target nucleon. The screening caused by these rescatterings prohibits the power growth of the parton densities.

[2]The same result was obtained in Ref. [20,21] – note, however, it was based on LO evolution and the large NLO correction to $P_{qg}$ will change the $q, g$ relationship.

The GLR equation for the gluon may be written symbolically as

$$\frac{\partial xg}{\partial \ln Q^2} = P_{gg} \otimes g + P_{gq} \otimes q - \frac{81\alpha_s^2}{16R^2Q^2} \int \frac{dy}{y} [y\, g(y, Q^2)]^2 \,. \tag{2}$$

The nonlinear shadowing term, $\propto -[g]^2$, arises from perturbative QCD diagrams which couple four gluons to two gluons so that two gluon ladders recombine into a single gluon ladder. The minus sign occurs because the scattering amplitude corresponding to a gluon ladder is predominantly imaginary. The parameter $R$ is a measure of the transverse area $\pi R^2$ where the gluon density is sufficiently dense for recombination to occur.

The BK equation is an improved version of the GLR equation. It accounts for the more precise triple-pomeron vertex (first calculated in Ref. [26–28]) and can be used for the non-forward amplitude. The GLR equation was in momentum space, whereas the BK equation is written in coordinate space in terms of the dipole scattering amplitude $N(\mathbf{x}, \mathbf{y}, Y) \equiv N_{\mathbf{xy}}(Y)$. Here $\mathbf{x}$ and $\mathbf{y}$ are the transverse coordinates of the two $t$-channel gluons which form the colour-singlet dipole and $Y = \ln(1/x)$ is the rapidity. The BK equation reads

$$\frac{\partial N_{\mathbf{xy}}}{\partial Y} = \frac{\alpha_s N_c}{\pi} \int \frac{d^2\mathbf{z}}{2\pi} \frac{(\mathbf{x}-\mathbf{y})^2}{(\mathbf{x}-\mathbf{z})^2(\mathbf{y}-\mathbf{z})^2} \left\{ N_{\mathbf{xz}} + N_{\mathbf{yz}} - N_{\mathbf{xy}} - N_{\mathbf{xz}} N_{\mathbf{yz}} \right\} \,. \tag{3}$$

For small dipole densities, $N$, the quadratic term in the brackets may be neglected and Eq. (3) reproduces the conventional BFKL equation. However for large $N$, that is $N \to 1$, the right-hand side of Eq. (3) vanishes and we reach saturation when $N = 1$. The equation sums up the set of fan diagrams where at small $Y$ the target emits any number of pomerons (i.e. linear BFKL amplitudes) while at large $Y$ we have only one BFKL dipole.

Starting from the same initial conditions, the solution of the BK equation gives *fewer* small-$x$ partons than that predicted by its *parent* linear BFKL/CCFM equation[3].

In principle, it would appear more appropriate to use the BFKL-based BK equation to describe the parton densities at low $x$. Unfortunately, however, the BK equation is only a model. It cannot be used for numerical predictions. We discuss the reasons below.

### 2.4 Status of the BK equation

The Balitski-Kovchegov (BK) equation [24,25] is an attempt to describe saturation phenomena. However it is just a 'toy model' and cannot, at present, be used to reliably estimate absorptive effects at small $x$. The reasons are as follows:

- The BK equation is based on the LO BFKL/CCFM equation. We know that the NLL corrections are large. We need to know the NLL corrections not only for the linear part of the evolution, but also for the nonlinear term.
- Even neglecting the NLL corrections, we need to match the solution to a boundary condition at rather low $Q^2$. This boundary condition is not theoretically known.
- It sums a limited set of diagrams. The selection of diagrams (the fan graphs) was justified in the region where absorptive effects are relatively small. When these corrections become important, as in the saturation region, one has to allow for many other graphs whose contributions become comparable to the fan diagram contributions[4].

[3] Analogously, starting from the same input (and not fitting the input to the data) the GLR equation gives fewer small-$x$ partons than that predicted by the parent linear DGLAP equation.

[4] Unfortunately the problem of summing all relevant diagrams has not been solved, even in the simpler case of Reggeon field theory.

– To solve the BK equation we need an initial condition at fixed $x$ and all $Q^2$. These conditions are not well enough known. In particular, the maximum (saturation) value of the gluon density depends on the radius: $xg(x, q^2) \propto R^2 q^2$. At the moment, the radius $R$ is a free parameter. It may be small — the so-called 'hot spot' scenario. Moreover, we should account for the possibility of dissociation of the target proton[5]. The contribution coming from the dissociation is expected to have a smaller $R$.

### 2.5 Relevance to, and of, the LHC

How do the uncertainties at low $x$ affect the predictions for the LHC? Fortunately for inclusive production of possible massive new particles with $M \gtrsim 100$ GeV, the partons are sampled at $x$ values and scales $M$ reliably determined from NLO and NNLO global analyses. For illustration, we discuss $W$ production which has been studied in detail [29–31]. Central $W$ production ($y_W = 0$) at the LHC samples partons at $x = M_W/\sqrt{s} = 0.006$. However to predict the total cross section, $\sigma_W$, we need to integrate over rapidity, important for $|y_W| \lesssim 4$ so that $\sigma_W$ has some sensitivity to partons as low as $x \sim 10^{-4}$. The total uncertainty on the NNLO prediction of $\sigma_W$ has been estimated to be $\pm 4\%$ [29]. Therefore $W$ production at the LHC can serve as a good luminosity monitor. To reduce the uncertainty in the prediction of $\sigma_W$ will require a better theoretical understanding of low $x$ partons.

Of course, if the new particles are sufficiently massive, $M \gtrsim 1$ TeV, and produced by gluon fusion, then the uncertainties due to the PDFs will be larger. However, there are situations where the scale is considerably lower such as exclusive double-diffractive Higgs production which depends on the unintegrated gluon at $Q^2 \approx 5$ GeV$^2$ with $x \sim M_H/\sqrt{s} \sim 0.01$. The absorptive effects are also expected to be small here.

Turning the discussion around, is it possible for the LHC experiments to determine the behaviour of partons in the $x$ region below $10^{-4}$ at low scales? One possibility is $\mu^+\mu^-$ Drell-Yan production in which events are observed with the $\mu^+\mu^-$ invariant mass as low as possible and the rapidity as large as possible. For example, for $M_{\mu\mu} = 4$ GeV and $y_{\mu\mu} = 3$, we sample quarks at $x = 1.4 \times 10^{-5}$. This process predominantly samples the sea quark distributions. To study the small $x$ behaviour of the gluon at low scales we may consider $\chi_c$ production, or prompt photon production driven by the subprocess $gq \to \gamma q$.

In practice, rather than $\chi_c$, it may be better to study $pp \to J/\psi\ X$ as a function of $y_{J/\psi}$. This process is also sensitive to the gluon distribution through the subprocesses $gg \to J/\psi\ g,\ \ gg \to \chi \to J/\psi\ \gamma$. There are also contributions from the subprocesses $gg \to b\bar{b}$ with $b \to J/\psi$, and $q\bar{q} \to J/\psi$. The analysis of such data will be considerably helped by the detailed observations of prompt $J/\psi$ and $J/\psi$ from $b$ in central production at the Tevatron [32]. In fact, the first ever NLO global parton analysis [33] used $J/\psi$ data as a function of rapidity to constrain the gluon distribution.

The LHCb detector covers the rapidity region of $2 < \eta < 5$ [34], and may be able to perform some of the above measurements. There is another possibility. Since LHCb will operate at a luminosity of $2 \times 10^{32}$ cm$^{-2}$s$^{-1}$, there will usually be a single collision per bunch crossing and thus practically no 'pile-up' problems. Installing a forward detector at LHCb would offer the possibility of observing asymmetric events with one very large rapidity gap to probe the region of $x_{I\!P} \leq 10^{-5}$.

## 3 Including nonlinear terms in gluon evolution: the GLRMQ and BK approaches

### 3.1 GLRMQ approach[6]

The DGLAP [1–4] evolution equations describe the scale evolution of the parton distribution functions (PDFs) well in the region of large interaction scale, $Q^2 \gtrsim 4$ GeV$^2$ [17, 35, 36]. However, toward small values of $x$ and $Q^2$, the gluon recombination terms start to play an increasingly important role. The inclusion of correction terms which arise from fusion of two gluon ladders leads to nonlinear power corrections to the DGLAP evolution equations. The first of these nonlinear corrections are the GLRMQ terms.

[5]We know that these channels provide more than $30 - 40\%$ of $F_2^D$ measured at HERA.

[6]Authors: K.J. Eskola and V.J. Kolhinen

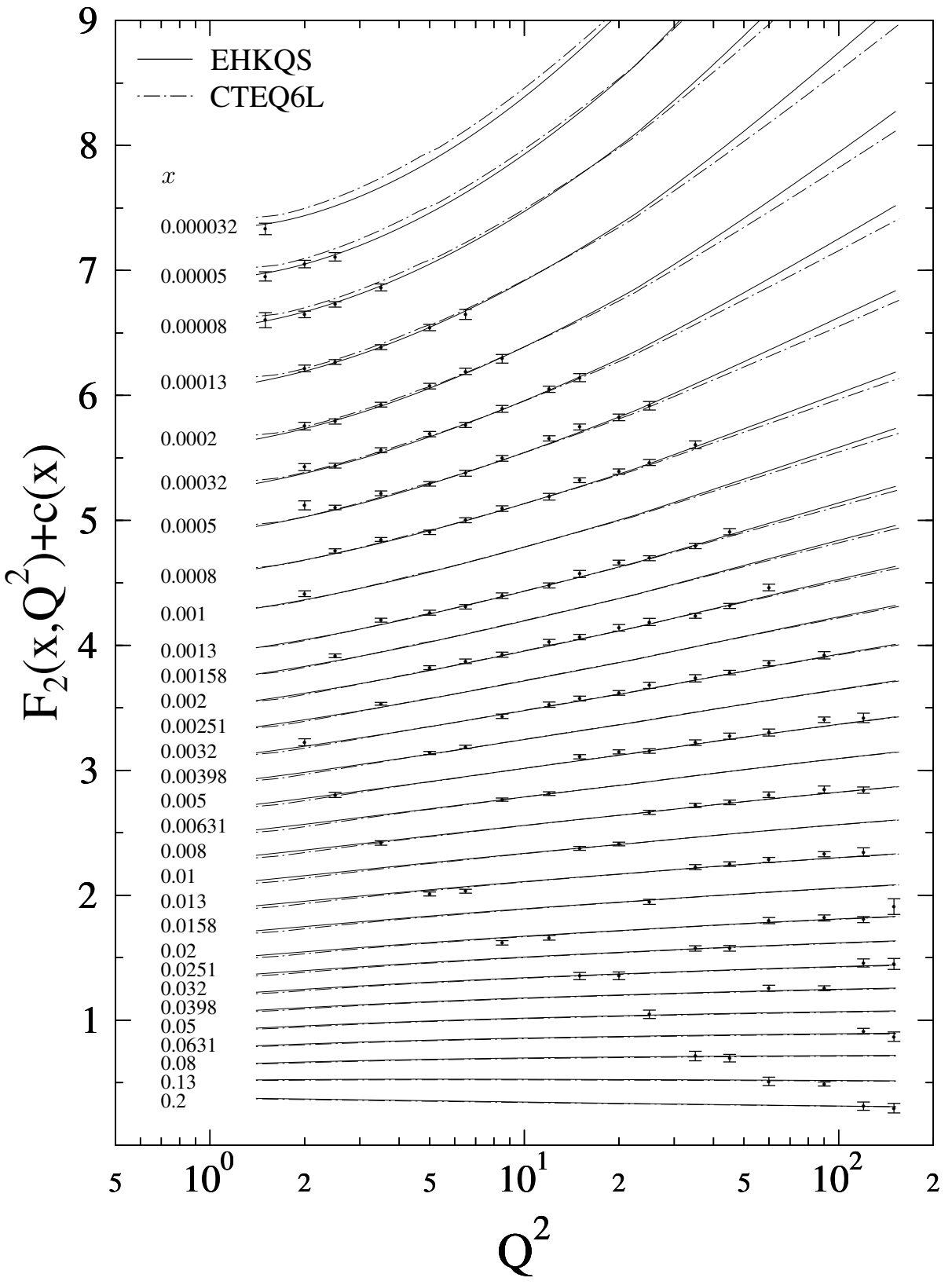

**Fig. 3:** Calculated $F_2(x, Q^2)$ values compared with the H1 data.

With the GLRMQ corrections, the gluon evolution equation becomes

$$\frac{\partial xg(x,Q^2)}{\partial \ln Q^2} = \frac{\partial xg(x,Q^2)}{\partial \ln Q^2}\bigg|_{\text{DGLAP}} - \frac{9\pi}{2}\frac{\alpha_s^2}{Q^2}\int_x^1 \frac{dy}{y} y^2 G^{(2)}(y,Q^2). \tag{4}$$

We model the two-gluon density in the latter term on the right-hand side as

$$x^2 G^{(2)}(x,Q^2) = \frac{1}{\pi R^2}[xg(x,Q^2)]^2, \tag{5}$$

where $R = 1$ fm is the radius of the proton (we comment further on this later). The corrections to the sea quark distributions are

$$\frac{\partial xq(x,Q^2)}{\partial \ln Q^2} \approx \frac{\partial xq(x,Q^2)}{\partial \ln Q^2}\bigg|_{\text{DGLAP}} - \frac{3\pi}{20}\frac{\alpha_s^2}{Q^2} x^2 G^{(2)}(x,Q^2).$$

We have assumed that the higher-twist gluon term, $G_{HT}$ [23], is negligible.

Since these correction terms are negative, they slow down the evolution of the PDFs. Due to the $1/Q^2$ dependence, they also die out in the evolution so that at large scales Eqs. (4) and (6) relax into the linear DGLAP equations.

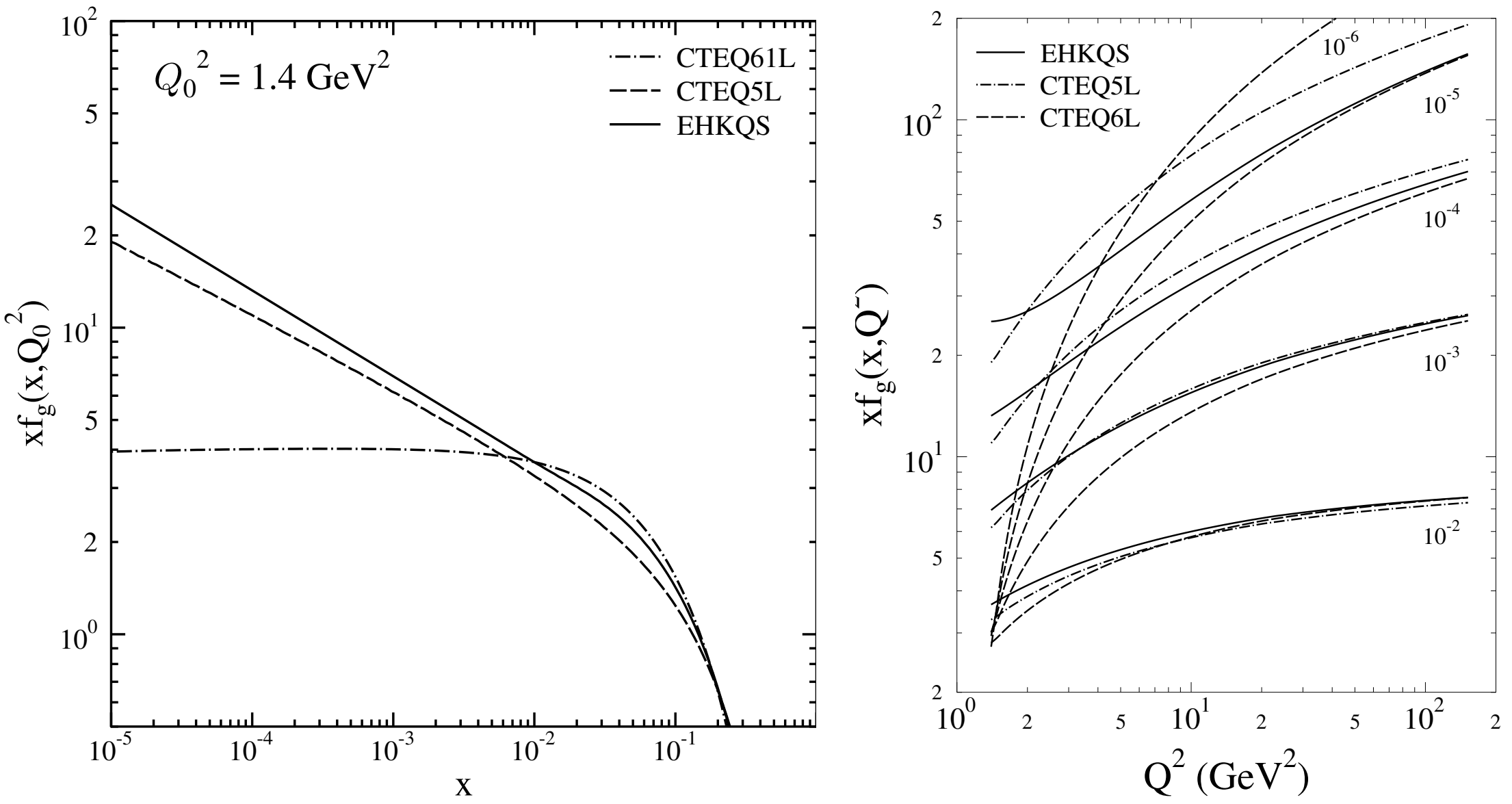

**Fig. 4:** Left: initial gluon distributions at $Q_0^2 = 1.4$ GeV$^2$. Right: evolution of gluon distributions for several fixed values of $x$ shows that the effect of the nonlinear terms vanishes as $Q^2$ increases.

In order to study the interplay between the nonlinear corrections and the initial PDFs and observe the nonlinear effects in fits to the DIS data, in Ref. [37] we compared the structure function $F_2(x, Q^2) = \sum_q e_q^2[xq(x, Q^2)+x\bar{q}(x, Q^2)]$, calculated with the nonlinearly-evolved PDFs, to the HERA H1 data [38]. As reference distributions we used the CTEQ5L and CTEQ6L PDF sets at large scales. We chose these sets because the CTEQ collaboration uses only the large scale, $Q^2 > 4$ GeV$^2$, data in their fits, thus avoiding some of the possible nonlinear effects appearing in the small $x$, $Q^2 < 4$ GeV$^2$ region in their initial distributions.

At small $x$, sea quarks dominate $F_2$ and the gluon distribution dictates its scale evolution. At leading order (LO), the DGLAP contribution can be approximated as [39]:

$$\partial F_2(x, Q^2)/\partial \ln Q^2 \approx (10\alpha_s/27\pi)xg(2x, Q^2) \, .$$

Larger $xg(x, Q^2)$ values correspond to faster $F_2(x, Q^2)$ evolution. The scale evolution of $F_2(x, Q^2)$ at small $x$ computed with CTEQ5L is too fast due to a large small-$x$ small-$Q^2$ gluon distribution. The newer CTEQ6L set has much smaller gluon distribution in this region (see Fig. 4 (left)), giving a slower evolution and hence a good fit to the H1 data.

Our goal in Ref. [37] was to determine whether the good fit to the data could be maintained using the GLRMQ-corrected DGLAP scale evolution together with initial scale PDFs differing from CTEQ6L. We constructed a new set of PDFs using the CTEQ5L and CTEQ6L distributions piecewise as baselines at scales $Q^2 \sim 3 - 10$ GeV$^2$ where the linear terms dominate the evolution and evolved them nonlinearly to lower $Q^2$. We then interpolated between the sets in $x$ and assumed a power-like dependence at small-$x$ for gluons and sea quarks. These initial PDF candidates were then evolved to higher scales and compared to the data. This iterative procedure was repeated until a sufficiently good fit to the data was found.

As a result, we obtained a new set of initial PDFs[7], called EHKQS, which, when evolved using the nonlinear DGLAP+GLRMQ evolution equations, produced an equally good or even better fit to the H1 data relative to CTEQ6L, shown in Fig. 3. At $Q^2 \sim 1.4$ GeV$^2$ and $x \sim 10^{-5}$, a good fit to the HERA data requires the nonlinear evolution to be compensated by a larger gluon distribution than obtained with

[7] In fact, we produced three new sets of initial distributions, differing by the charm quark mass and parton distribution at the initial scale, see Ref. [37] for more details. All sets produced equally good fits to the HERA data.

DGLAP alone. The enhancement is a factor of $\sim 6$ relative to CTEQ6L, as shown in Fig. 4 (left). The $Q^2$ dependence of EHKQS is compared to CTEQ6L and CTEQ5L in Fig. 4 (right) for several different values of $x$.

We used $R = 1$ fm as the free proton radius in the two-gluon density term. We did not repeat the calculations with different $R$ but, depending on the transverse matter density of the free proton, some $\sim 20\%$ uncertainty in $R$ can be expected. Since the nonlinear contributions decrease as $R$ increases, a larger $R$ would lead to reduced enhancement of the small $x$ gluons below $Q^2 \sim 10$ GeV$^2$. Thus, minimizing the $\chi^2$ of the fit with respect to $R$ is a future task.

### 3.2 BK approach[8]

A theoretical framework capable of describing the QCD evolution of parton densities taking gluon rescattering (corresponding to nonlinear effects) into account is the Balitsky-Kovchegov (BK) equation [24, 25, 40–42]. The equation, based on the BFKL approach [6, 7, 43], may be used to determine the unintegrated gluon density. The BK equation resums the BFKL pomeron fan diagrams with the triple pomeron vertex derived in the high energy limit of QCD. In the doubly logarithmic limit, the BK equation reduces [25] to the collinear Gribov-Levin-Ryskin (GLR) equation [22]. It is the non-collinear limit, however, which gives the dominant contribution to the triple pomeron vertex [44, 45]. We conclude that GLR approach misses an essential part of the nonlinear gluon dynamics.

The solution to the BK equation, constrained by the low-$x$ HERA data will be used to extrapolate the parton densities to the LHC kinematical domain. Extrapolation is necessary as the LHC may probe very low values of $x$, down to $10^{-7}$ for $M = 10$ GeV and $\eta \sim 9$, where unitarity corrections may be important even at relatively large scales of a few GeV$^2$. Last but not least, unitarity corrections may also break $k_T$ factorization. We will also discuss which processes may be affected.

This section is organized as follows. First we give a brief description of the formalism used to determine the gluon evolution. Within this formalism, we fit the HERA charm structure function, $F_2^c$, data, the most relevant inclusive measurement directly sensitive to the gluon density. Using further assumptions about the sea quarks, $F_2$ can also be described well. The resulting gluon density is then used to compute heavy quark production and to investigate the nonlinear effects. First we estimate $b\bar{b}$ production at CDF and D0. Then, cross sections for heavy quark production at various LHC experiments are estimated, tracing the impact of the unitarity corrections. Finally, conclusions are given.

The standard framework to determine parton evolution is the collinear DGLAP formalism. It works rather well for inclusive quantities but, for more exclusive processes, the $k_T$-factorization scheme is more appropriate because both the longitudinal and transverse components of the gluon momenta are considered. In this framework, the process-independent quantity is the unintegrated gluon distribution, connected to the process-dependent hard matrix element via the $k_T$-factorization theorem. Linear evolution of the unintegrated gluon distribution may be described by one of the small $x$ evolution equations using the $k_T$-factorization scheme, the BFKL and CCFM [46–49] equations. These equations are based on resummation of large logarithmic pQCD corrections, $\alpha_s^n \ln^m(1/x)$, and are equivalent at the leading logarithmic level.

The very small $x$ kinematic region is also the regime where the growth of the gluon density must be tamed in order to preserve unitarity. Recently, a successful description of unitarity corrections to DIS was derived within the color dipole formulation of QCD. This is the Balitsky-Kovchegov (BK) equation which describes the BFKL evolution of the gluon in a large target, including a nonlinear term corresponding to gluon recombination at high density.

In our analysis, we determine the unintegrated gluon distribution from the BK equation unified with the DGLAP equation following KMS (Kwieciński, Martin and Staśto) [50–53]. We use the abbreviation KKMS (Kutak, Kwieciński, Martin and Staśto) [52, 53] for the unified nonlinear equation.

[8] Authors: H. Jung, K. Kutak, K. Peters, L. Motyka

The linear part of this equation is given by the BFKL kernel with subleading $\ln(1/x)$ corrections, supplemented by the non-singular parts of the DGLAP splitting functions. Thus resummation of both the leading $\ln Q^2$ and $\ln(1/x)$ terms are achieved. The subleading terms in $\ln(1/x)$ are approximated by the so-called consistency constraint and the running coupling constant. The nonlinear part is taken directly from the BK equation, ensuring that the unitarity constraints are preserved. One expects that this framework provides a more reliable description of the gluon evolution at extremely small $x$, where $\ln(1/x) \gg 1$ and the unitarity corrections are important, than does DGLAP.

We give a short review of the KKMS equation, starting from the impact parameter dependent BK equation. The equation for the unintegrated gluon density, $h(x, k^2, b)$, at impact parameter $b$ from the center of the target, becomes

$$\frac{\partial h(x,k^2,b)}{\partial \ln 1/x} = \frac{\alpha_s N_c}{\pi} k^2 \int_{k_0^2} \frac{dk'^2}{k'^2} \left\{ \frac{h\left(x,k'^2,b\right) - h\left(x,k^2,b\right)}{|k'^2 - k^2|} + \frac{h\left(x,k^2,b\right)}{[4k'^4 + k^4]^{\frac{1}{2}}} \right\}$$
$$-\pi\alpha_s \left(1 - k^2 ddk^2\right)^2 k^2 \left[\int_{k^2}^{\infty} \frac{dk'^2}{k'^4} \ln\left(\frac{k'^2}{k^2}\right) h(x,k'^2,b)\right]^2 , \quad (6)$$

the BFKL equation at LL$x$ accuracy, extended by the negative recombination term. The (dimensionless) unintegrated gluon distribution is obtained from $h(x, k^2, b)$ by integration over $b$,

$$f(x,k^2) = \int d^2b\, h(x,k^2,b). \quad (7)$$

A comment about the impact parameter treatment is in order. In Eq. (7), we assume that the evolution is local in $b$. However, the complete BK equation results in some diffusion in the impact parameter plane. This diffusion effect may be neglected if the target is much larger than the inverse of the saturation scale. In this scheme, the impact parameter dependence enters through the initial condition at large $x_0$, $h(x_0, k^2, b) = f(x_0, k^2)S(b)$ where $f(x_0, k^2)$ is the unintegrated gluon distribution. Note that, due to nonlinearities, the $b$ dependence of $h(x, k^2, b)$ does not factorize from $x$ and $k$ at low $x$.

The input profile function is assumed to be Gaussian, $S(b) = \exp(-b^2/R^2)/\pi R^2$, where $R^2$ corresponds to the square of the average nucleon radius. Since the size of the target, $R$, sets the magnitude of the initial parton density in the impact parameter plane, $h(x_0, k^2, b)$, the unitarity corrections depend on $R$. At smaller $R$, gluons are more densely packed in the target and the nonlinear effects are stronger.

References [52, 53] proposed to combine Eq. (6) with the unified BFKL-DGLAP framework developed in Ref. [50]. In this scheme, the (linear) BFKL part is modified by the consistency constraint [54, 55], resulting in the resummation of most of the subleading corrections in $\ln(1/x)$ which arise from imposing energy-momentum conservation on the BFKL kernel [56–59]. In addition, we assume that the strong coupling constant runs with scale $k^2$, another source of important NLL$x$ corrections. Finally, the non-singular part of the leading order DGLAP splitting function and quark singlet distribution were included in the $x$ evolution. The final improved nonlinear equation for the unintegrated gluon density is

$$\begin{aligned} h(x,k^2,b) = {} & \tilde{h}^{(0)}(x,k^2,b) + \\ & + \frac{\alpha_s(k^2) N_c}{\pi} k^2 \int_x^1 \frac{dz}{z} \int_{k_0^2} \frac{dk'^2}{k'^2} \left\{ \frac{h(\frac{x}{z},k'^2,b)\,\Theta(\frac{k^2}{z}-k'^2) - h(\frac{x}{z},k^2,b)}{|k'^2-k^2|} + \frac{h(\frac{x}{z},k^2,b)}{|4k'^4+k^4|^{\frac{1}{2}}} \right\} + \\ & + \frac{\alpha_s(k^2)}{2\pi} \int_x^1 dz \left[ \left(P_{gg}(z) - \frac{2N_c}{z}\right) \int_{k_0^2}^{k^2} \frac{dk'^2}{k'^2} h(\tfrac{x}{z},k'^2,b) + P_{gq}(z)\Sigma\left(\tfrac{x}{z},k'^2,b\right) \right] + \\ & - \pi \left(1 - k^2 \frac{d}{dk^2}\right)^2 k^2 \int_x^1 \frac{dz}{z} \left[ \int_{k^2}^{\infty} \frac{dk'^2}{k'^4} \alpha_s(k'^2) \ln\left(\frac{k'^2}{k^2}\right) h(z,k'^2,b) \right]^2 . \end{aligned} \quad (8)$$

The second line of the equation corresponds to the BFKL evolution. The theta function, $\Theta(\frac{k^2}{z} - k'^2)$, reflects the consistency constraint that generates the dominant part of the subleading BFKL corrections.

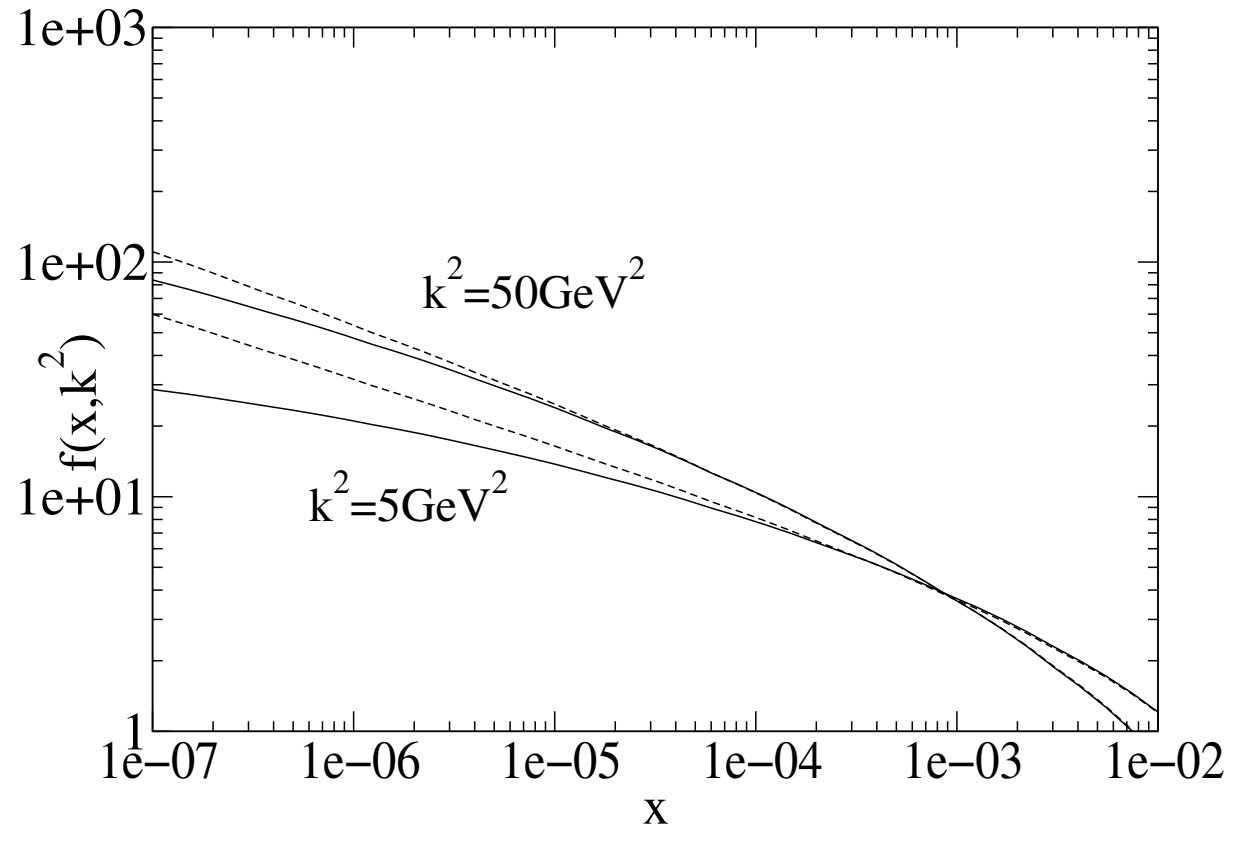

**Fig. 5:** The unintegrated gluon distribution obtained from Eq. (8) as a function of $x$ for different values of $k_T^2$. The solid lines correspond to the solution of the nonlinear equation with $R = 2.8$ GeV$^{-1}$ while the dashed lines correspond to the linear part.

The third line corresponds to the DGLAP effects generated by the part of the splitting function, $P_{gg}(z)$, that is not singular in the limit $z \to 0$ and also by the quarks where $\Sigma(x, k^2, b^2)$ corresponds to the impact-parameter dependent singlet quark distribution. The nonlinear screening contribution following from the BK equation is given in the last term. The inhomogeneous contribution, defined in terms of the integrated gluon distribution, carries information about the transverse profile of the proton,

$$\tilde{h}^{(0)}(x, k^2, b) = \frac{\alpha_s(k^2)}{2\pi} S(b) \int_x^1 dz P_{gg}(z) \frac{x}{z} g\left(\frac{x}{z}, k_0^2\right) , \tag{9}$$

at $k_0^2 = 1$ GeV$^2$. The initial integrated density at $k_0^2$ is parameterized as

$$xg(x, k_0^2) = N(1-x)^\rho \tag{10}$$

where $\rho = 2.5$. The size of the dense gluon system inside the proton is assumed to be $R = 2.8$ GeV$^{-1}$, in accord with the diffractive slope, $B_d \simeq 4$ GeV$^{-2}$, of the elastic $J/\psi$ photoproduction cross section at HERA. In this process, the impact parameter profile of the proton defines the $t$ dependence of the elastic cross section, $B_d \simeq R^2/2$, by Fourier transform. In the 'hot-spot' scenario, the radius can be smaller, $R = 1.5$ GeV$^{-1}$. We also use the hot spot value to compare with measurements and make predictions for the LHC.

Equation (8) was solved numerically both in the linear approximation and in full. The method for solving Eq. (8) was developed in Refs. [50, 52]. In Fig. 5, the effects of linear and nonlinear evolution on the unintegrated gluon distribution are given as a function of $x$ for $k^2 = 5$ and 50 GeV$^2$. Nonlinear evolution leads to sizeable suppression at the smallest $x$ values. While the nonlinear effects are small in the HERA $x$ range, they may be important at the LHC. In the following sections, we address the importance of these nonlinear effects.

The initial distribution in Eq. (10) was obtained by fitting the HERA $F_2^c$ measurements [60,61] using the Monte Carlo CASCADE [62,63] for evolution and convolution with the off-shell matrix elements. We find $\chi^2$ per degree of freedom of 0.46 (1.17) for H1 (ZEUS). The fits were repeated both with the standard KMS evolution without the nonlinear contribution and with extended KMS evolution including the nonlinear part. The predicted $F_2^c$ is equivalent for both linear and nonlinear evolution, independent of $R$. Thus nonlinear evolution is only a small effect at HERA, even in the hot-spot scenario with $R = 1.5$ GeV$^{-1}$.

In Fig. 6(a) we compare the measured $F_2^c$ [61] to our prediction at $Q^2 = 4$ GeV$^2$. We have determined our initial distribution from $F_2^c$ since it is the only inclusive measurement at HERA directly sensitive to the gluon distribution. However, we can also describe $F_2$ [64] by making further assumptions about the sea quark distribution, following the KMS approach. The agreement with the data, shown in Fig. 6(b), is also quite good. There is only a small effect for $Q^2 > 5$ GeV$^2$, even in the hot-spot scenario with $R = 1.5$ GeV$^{-1}$.

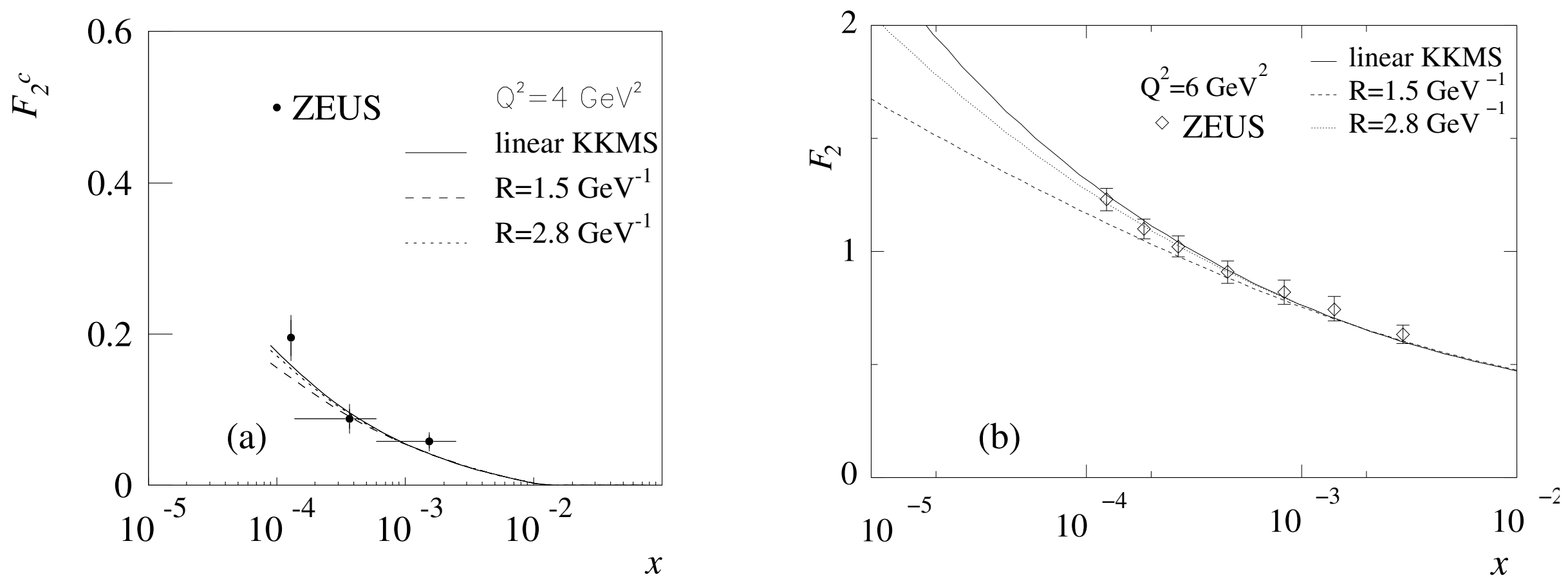

**Fig. 6:** The charm structure function, $F_2^c$, [61] at $Q^2 = 4$ GeV$^2$ (a) and $F_2$ [64] at $Q^2 = 6$ GeV$^2$ (b) obtained for KKMS evolution with different values of $R$.

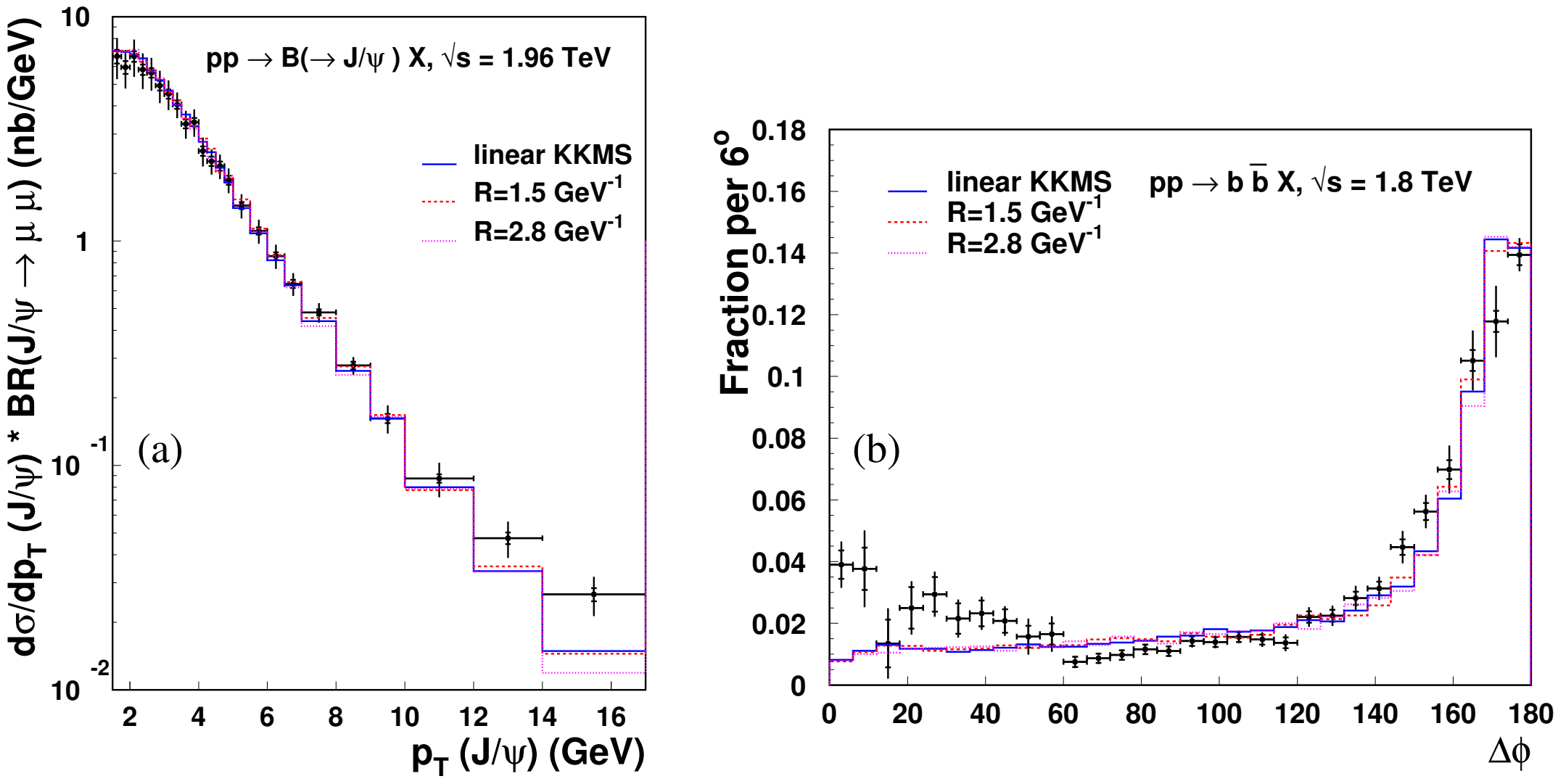

**Fig. 7:** Bottom production, measured by CDF, is compared to predictions using CASCADE with linear and non-linear KKMS evolution, including two values of $R$. (a) The $p_T$ distribution of $B$ meson decays to $J/\psi$. (b) The azimuthal angle, $\Delta\phi$, distribution of $b\bar{b}$ pair production smeared by the experimental resolution.

Next, this constrained gluon density was used to calculate $gg \to b\bar{b}$ production at the Tevatron as a cross check of the fit and the evolution formalism. We use $m_b = 4.75$ GeV and a renormalization scale in $\alpha_s$ of $Q^2 = 4m_b^2 + p_T^2$. The predicted cross section was then compared to both CDF [65, 66] and D0 [67] measurements. The predictions agree well with the data.

In Fig. 7(a) the cross section for $B$ decays to $J/\psi$ is shown as a function the $J/\psi$ $p_T$ [65,66]. The KKMS gluon density fits the data well in all three scenarios with deviations only appearing for $p_T > 12$ GeV. It is interesting to note that the approach described here gives even better agreement than the NLO collinear approach [68].

In Fig. 7(b), the azimuthal angle distribution between the $b$ and $\bar{b}$ quarks, $\Delta\phi$, is given. The $\Delta\phi$ and $b\bar{b}$ $p_T$ distributions are correlated since $\Delta\phi < 180^\circ$ corresponds to higher pair $p_T$. Since the $k_T$-factorization formula allows the incoming gluons to have sizable transverse momenta, the calculated $\Delta\phi$ distribution agrees very well with the data for $\Delta\phi > 60^\circ$ with only smearing due to the experimental resolution. The enhancement of the data relative to the calculations at low $\Delta\phi$ requires further study.

Both plots compare linear (solid histograms) and nonlinear KKMS evolution (dotted and dashed histograms) for $R = 1.5$ GeV$^{-1}$ and 2.8 GeV$^{-1}$ respectively. The nonlinear part of the evolution also has no impact in this kinematic region.

## 4 Phenomenological applications: heavy quark production at the LHC

### 4.1 GLRMQ approach[9]

Since the HERA $F_2$ data can be described by both linear DGLAP and nonlinear DGLAP+GLRMQ evolution, as shown in Fig. 3 of Section 3.1, additional independent probes are needed. Here, we discuss how charm quark production in $pp$ collisions could probe the gluon enhancement predicted in Section 3.1 and described in detail in Ref. [20,21]. Charm production is an ideal choice since the charm mass is low and its production is dominated by gluons. Assuming factorization, the inclusive differential charm cross section is

$$d\sigma_{pp\to c\bar{c}X}(Q^2,\sqrt{s}) = \sum_{i,j,k=q,\bar{q},g} f_i(x_1,Q^2)\otimes f_j(x_2,Q^2)\otimes d\hat{\sigma}_{ij\to c\bar{c}\{k\}}(Q^2,x_1,x_2) \tag{11}$$

where $\hat{\sigma}_{ij\to c\bar{c}\{k\}}(Q^2,x_1,x_2)$ are the perturbatively calculable partonic cross sections for charm production at scales $Q^2 \sim m_T^2 \gg \Lambda^2_{\rm QCD}$, $x_1$ and $x_2$ are the parton momentum fractions and $f_i(x,Q^2)$ are the proton parton densities. We assume that the renormalization and factorization scales are equal. Only the leading order $gg$ and $q\overline{q}$ channels are considered here.

The values of the charm quark mass and scale used in the calculations are chosen to give good agreement with the total cross section data at NLO: $m = 1.2$ GeV and $Q^2 = 4m^2$ for standard DGLAP-evolved NLO PDFs such as CTEQ6M [69] and MRST [70]. Nearly equivalent agreement may be obtained with $m = 1.3$ GeV and $Q^2 = m^2$ [71,72]. Both choices assure that the PDFs are evaluated above the minimum scales. While scales proportional to $m$ are used in the total cross section, inclusive calculations of distributions also depend on the transverse momentum scale, $p_T$, so that $m_T = \sqrt{m^2 + p_T^2}$ is used instead [73].

To illustrate the effects of the nonlinear EHKQS distributions [37] of Section 3.1 on charm production at the LHC, we show

$$R(y) \equiv \frac{d\sigma(\text{EHKQS})/dy}{d\sigma(\text{CTEQ61L})/dy} \quad \text{and} \quad R(p_T) \equiv \frac{d\sigma(\text{EHKQS})/dp_T}{d\sigma(\text{CTEQ61L})/dp_T} \tag{12}$$

in Fig. 8 where $y$ is the charm quark rapidity. The results are calculated for the maximum LHC $pp$, $p$Pb and Pb+Pb energies, $\sqrt{S} = 14$ (solid), 8.8 (dashed) and 5.5 (dot-dashed) TeV respectively. The results for $m = 1.2$ GeV and $Q^2 = 4m_T^2$ are on the left-hand side while those with $m = 1.3$ GeV and $Q^2 = m_T^2$ are on the right-hand side.

The change in the slope of $R(y)$ occurs when one $x$ drops below the minimum value of the EHKQS set where further nonlinearities become important, $x_{\rm min}^{\rm EHKQS} = 10^{-5}$, and enters the unconstrained $x$ region. The minimum $x$ of CTEQ61L is lower, $x_{\rm min}^{\rm CTEQ61L} = 10^{-6}$. While the EHKQS gluon distribution is fixed at its minimum for $x < x_{\rm min}^{\rm EHKQS}$, the CTEQ61L distribution continues to change until $x_{\rm min}^{\rm CTEQ61L}$.

[9]Author: R. Vogt

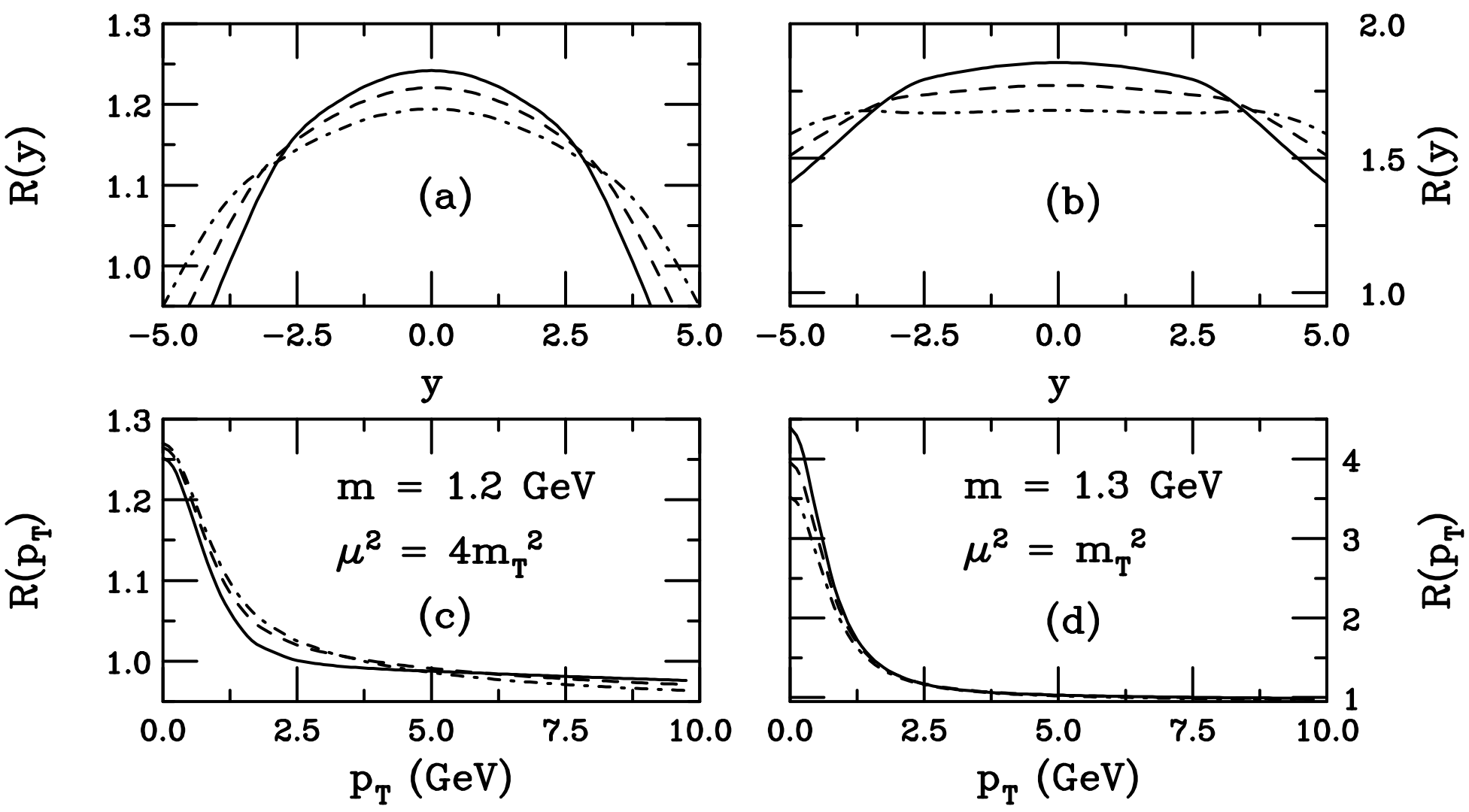

**Fig. 8:** We present $R(y)$, (a) and (c), and $R(p_T)$, (b) and (d), in $pp$ collisions at $\sqrt{S} = 14$ (solid), 8.8 (dashed) and 5.5 (dot-dashed) TeV. The left-hand side shows $m = 1.2$ GeV and $Q^2 = 4m_T^2$, the right-hand side $m = 1.3$ GeV and $Q^2 = m_T^2$.

In inclusive kinematics with an identified charm quark and fixed $x_T = 2m_T/\sqrt{S}$, the unconstrained $x$-region contributes to charm production in the region

$$y_l \equiv \ln\left(1/x_T - \sqrt{1/x_T^2 - 1/x_{\rm min}}\right) \leq |y| \leq \ln\left(1/x_T + \sqrt{1/x_T^2 - 1/x_{\rm min}}\right) . \quad (13)$$

The upper limit is close to the phase space boundary. Expanding the lower limit, $y_l$, in powers of $x_T^2/x_{\rm min} \ll 1$, $y_l \approx \ln[m_T/(x_{\rm min}\sqrt{S})] \geq \ln[m/(x_{\rm min}\sqrt{S})]$. If $m = 1.2$ GeV, the small $x$ region contributes to charm production at $|y| \geq y_l = 2.2$, 2.6 and 3.1 for $\sqrt{S} = 14$, 8.8 and 5.5 TeV, respectively. The rather sharp turnover in $R(y)$ indicates where the $x < 10^{-5}$ region begins to contribute. For $|y| > y_l$ and $Q^2 > 4$ GeV$^2$, as $x$ decreases, the CTEQ61L gluon distribution increases considerably above that of the EHKQS distribution. Thus $R(y) < 1$ at large rapidities when $Q^2 = 4m_T^2$. At midrapidity $R(y)$ is insensitive to the EHKQS extrapolation region, $x < x_{\rm min}^{\rm EHKQS}$. Since $R(y)$ is integrated over $p_T$, it not only reflects the enhancement at $m_T = m$ because $Q^2 \propto m_T^2$ and the $p_T$ distribution peaks around $p_T \approx 1$ GeV. When $Q^2 = m_T^2$, the ratios are broad because the CTEQ61L gluon distribution is relatively flat as a function of $x$ for $Q^2 \sim 2 - 3$ GeV$^2$. The enhancement decreases and broadens with decreasing energy.

Since the rapidity distributions are rather flat, there are still important contributions to the $p_T$ distributions from the extrapolation region, up to $\sim 30\%$ at $\sqrt{S} = 14$ TeV for $m = 1.2$ GeV and $Q^2 = 4m^2$. Thus the sensitivity of $R(p_T)$ to the unconstrained region should be kept in mind. At the largest $\sqrt{S}$, the contribution from the $x < 10^{-5}$ region is greatest and if $Q^2 \geq 4m^2$, $xg^{\rm CTEQ61L}(x, Q^2) > xg^{\rm EHKQS}(x, Q^2)$. Because the contribution from the region $x < 10^{-5}$ decreases with $\sqrt{S}$, at low $p_T$ $R(p_T)$ decreases with energy. In contrast, for $Q^2 = m_T^2$, $xg^{\rm EHKQS}(x, Q^2) > xg^{\rm CTEQ61L}(x, Q^2)$ and the enhancement decreases with energy.

Because the DGLAP gluon distributions are already well constrained by HERA data, they cannot absorb additional large effects. Therefore we conclude that, if a low-$p_T$ enhancement in the charm cross section relative to the DGLAP-based result is observed in future experiments, it is a signal of nonlinear effects on the PDF evolution.

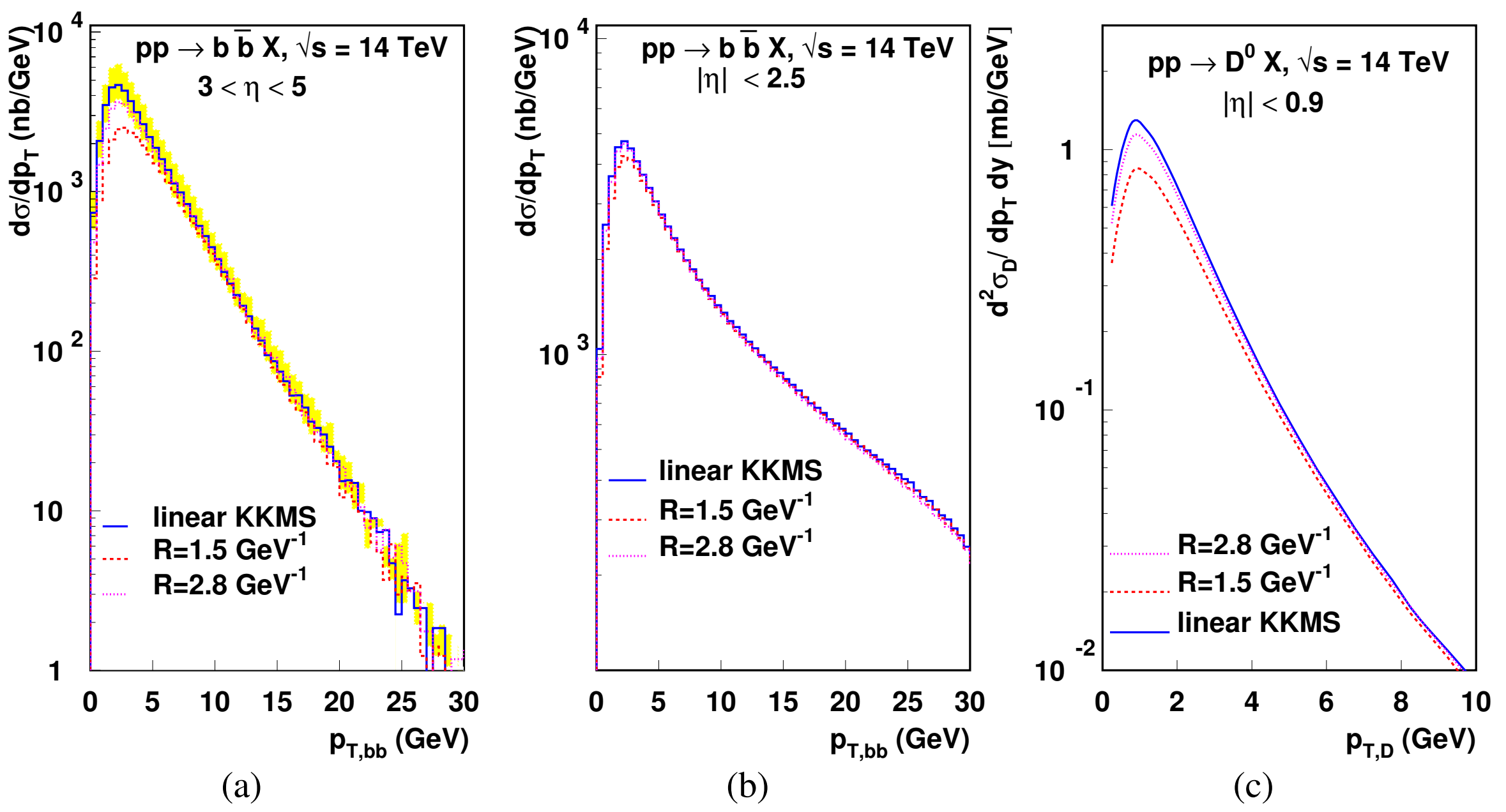

**Fig. 9:** (a) and (b) show $b\bar{b}$ production as a function of pair $p_T$ without cuts in $3 < |\eta| < 5$ (a) and in the ATLAS/CMS acceptance (b). The $D^0$ meson $p_T$ distribution in the ALICE acceptance is shown in (c).

### 4.2 BK approach[10]

Since the Tevatron measurements are well described using the unintegrated parton densities constrained by HERA and convoluted with the off-shell matrix elements, the same approach may be used for heavy quark production at the LHC at *e.g.* $\sqrt{s} = 14$ TeV. As discussed previously, see also Fig. 5, heavy quark production at this energy is already in the region where saturation effects may be relevant. In the kinematic regions, such as at the LHC, where nonlinear evolution may become important, the cross section will be suppressed due to the negative sign of the nonlinear term in Eq. (8).

First, we compute the $b\bar{b}$ production cross section at 14 TeV without any experimental cuts. In Fig. 9(a) the $b\bar{b}$ differential cross section is shown as a function of pair $p_T$ in the forward region, $3 < |\eta| < 5$. We compare linear evolution (solid histogram), nonlinear evolution with $R = 1.5$ GeV$^{-1}$ (dashed histogram) and $R = 2.8$ GeV$^{-1}$ (dotted histogram). The grey band shows the uncertainty in the linear result due to the $b$ quark mass. We take a central value of 4.75 GeV (the solid histogram) and vary $m_b$ from 4.5 to 5 GeV. The $b\bar{b}$ pair results are shown since the pair $p_T$ is most sensitive to the gluon $k_T$ and thus to the saturation effects. In the hot-spot scenario, saturation effects are visible for $p_{T_{b\bar{b}}} < 5$ GeV. These saturation effects grow with rapidity, increasing the suppression to a factor of $3 - 4$ in the higher rapidity regions. For larger $R$, the saturation effects are not very significant.

In Fig. 9(b), the $b\bar{b}$ production cross section is computed within the ATLAS and CMS acceptance ($p_T > 10$ GeV and $|\eta| < 2.5$ for both the $b$ and $\bar{b}$ quarks, see Ref. [74]). With these cuts, the observed suppression due to nonlinear effects nearly vanishes. This result suggests that $k_T$ factorization can safely be applied in the central $\eta$ region. Thus saturation effects due to nonlinear gluon evolution are seen only for $p_{T_{b\bar{b}}} < 10$ GeV and at high $\eta$. This regime is accessible with upgraded ATLAS/CMS detectors or in LHCb where the $b$ quark $p_T$ can be measured to 2 GeV for $1.9 < \eta < 4.9$. In this kinematic regime, the hot-spot scenario predicts a factor of two suppression of the $b\bar{b}$ cross section.

Similarly, we investigated $c\bar{c}$ production at ALICE. In ALICE, it will be possible to measure the $D^0$ down to $p_T \sim 0.5$ GeV in $|\eta| < 0.9$. The result is shown in Fig. 9(c) with $m_c = 1.5$ GeV. In the hot-spot scenario (dashed curve), a factor of two suppression occurs at $p_T \sim 1$ GeV.

[10] Authors: H. Jung, K. Kutak, K. Peters, L. Motyka

## 5 Perspectives for experimental observation at LHC[11]

### 5.1 Introduction

In Section 4.1, charm production in $pp$ collisions at the LHC was suggested as a promising way to study the effects of nonlinear evolution on the parton densities. Due to gluon dominance of charm production and the small values of $x$ and $Q^2$ probed, $x \approx 2 \times 10^{-4}$ and $Q^2 \approx 1.69 - 6$ GeV$^2$ at midrapidity and transverse momentum[12] $p_T \approx 0$, charm production at the LHC is sensitive to the gluon enhancement arising from nonlinear evolution. The resulting charm enhancement was quantified in Ref. [20,21] by the LO ratios of the differential cross sections computed with the nonlinearly-evolved EHKQS PDFs [37], obtained from DGLAP+GLRMQ evolution, relative to the DGLAP-evolved CTEQ61L PDFs.

The enhancement of the nonlinearly-evolved gluons increases as $x$ and $Q^2$ decrease. Consequently, the charm enhancement increases with center of mass energy, $\sqrt{S}$. Thus the maximum enhancement at the LHC will be at $\sqrt{S} = 14$ TeV and small charm quark transverse momentum. The sensitivity of the charm enhancement to the value of the charm quark mass, $m$, as well as to the choice of the factorization, $Q_F^2$, and renormalization, $Q_R^2$, scales was studied in Ref. [20, 21] assuming $Q^2 = Q_F^2 = Q_R^2 \propto m_T^2$ where $m_T^2 = p_T^2 + m^2$. The most significant charm enhancement occurs when $m$ and $Q^2/m_T^2$ are both small. A comparison of the NLO total cross sections with low energy data shows that the data prefer such small $m$ and $Q^2$ combinations [71, 72]. The largest enhancement is obtained with $m = 1.3$ GeV and $Q^2 = m_T^2$, see Fig. 8 in Section 4.1.

In Section 4.1, only charm enhancement was described. Neither its subsequent hadronization to $D$ mesons nor its decay and detection were considered. In this section, we address these issues to determine whether the charm enhancement survives hadronization and decay to be measured in the ALICE detector [75]. The calculation described in that section was to leading order since the EHKQS sets are evolved according to the LO DGLAP+GLRMQ equations using a one-loop evaluation of $\alpha_s$. Thus these LO distributions should generally not be mixed with NLO matrix elements and the two-loop $\alpha_s$. However, the charm quark total cross section is increased and the $p_T$ distribution is broadened at NLO relative to LO [76]. Thus, to determine whether or not the enhancement is experimentally measurable, we assume that the enhancement is the same at NLO and LO and employ a NLO cross section closest to the calculation of the enhancement in Ref. [20, 21].

As described in Ref. [76], the theoretical $K$ factor may be defined in more than one way, depending on how the LO contribution to the cross section is calculated. In all cases, the $\mathcal{O}(\alpha_s^3)$ contribution to cross section is calculated using NLO PDFs and the two-loop evaluation of $\alpha_s$. If the LO contribution is also calculated using NLO PDFs and a two-loop $\alpha_s$, this is the "standard NLO" cross section. It is used in most NLO codes, both in the global analyses of the NLO PDFs and in evaluations of cross sections and rates [76]. The $K$ factor formed when taking the ratio of the "standard NLO" cross section to the LO cross section with the NLO PDFs [76], $K_0^{(1)}$, indicates the convergence of terms in a fixed-order calculation [77]. On the other hand, if the LO contribution to the total NLO cross section employs LO PDFs and the one-loop $\alpha_s$, we have a cross section which we refer to here as "alternative NLO". The $K$ factor calculated taking the ratio of the "alternative NLO" cross section to the LO cross section with LO PDFs [76], $K_2^{(1)}$, indicates the convergence of the hadronic cross section toward a result. If $K_0^{(1)} > K_2^{(1)}$, convergence of the hadronic cross section is more likely [77]. This is indeed the case for charm production [76]. We also note that $K_2^{(1)}$ is a much weaker function of energy than $K_0^{(1)}$. Since, in the absence of nonlinear NLO PDFs, the "alternative NLO" cross section is more consistent with the calculated enhancement, we use this cross section to calculate the NLO $D$ meson rates and $p_T$ spectra. In both cases, the $p_T$ distributions have the same slope even though $K_2^{(1)}$, for the alternative NLO cross section, is somewhat smaller. Thus, using a non-standard NLO calculation will not change the slope of the $p_T$ distributions, distorting the result.

[11] Authors: A. Dainese and R. Vogt

[12] Here we use $p_T$ for the transverse momentum of the charm quark and $p_T^D$ for the transverse momentum of the $D$ meson.

The LO and NLO calculations used to obtain the full NLO result in both cases can be defined by modification of Eq. (11) in Section 4.1. We define the full LO charm production cross section as

$$d\sigma_{\rm LO}^{\rm 1L} = \sum_{i,j=q,\overline{q},g} f_i^{\rm LO}(x_1,Q_F^2) \otimes f_j^{\rm LO}(x_2,Q_F^2) \otimes d\hat{\sigma}_{ij\to c\overline{c}}^{\rm LO}(\alpha_s^{\rm 1L}(Q_R^2),x_1,x_2) \tag{14}$$

where the superscript "LO" on $d\hat{\sigma}_{ij\to c\overline{c}}$ indicates the use of the LO matrix elements while the superscript "1L" indicates that the one-loop expression of $\alpha_s$ is used. The LO cross section typically used in NLO codes employs the NLO PDFs and the two-loop (2L) $\alpha_s$ so that

$$d\sigma_{\rm LO}^{\rm 2L} = \sum_{i,j=q,\overline{q},g} f_i^{\rm NLO}(x_1,Q_F^2) \otimes f_j^{\rm NLO}(x_2,Q_F^2) \otimes d\hat{\sigma}_{ij\to c\overline{c}}^{\rm LO}(\alpha_s^{\rm 2L}(Q_R^2),x_1,x_2)\ . \tag{15}$$

In either case, the NLO contribution, $\mathcal{O}(\alpha_s^3)$ for heavy quark production, is

$$d\sigma_{\mathcal{O}(\alpha_s^3)} = \sum_{i,j=q,\overline{q},g} f_i^{\rm NLO}(x_1,Q_F^2) \otimes f_j^{\rm NLO}(x_2,Q_F^2) \otimes \sum_{k=0,q,\overline{q},g} d\hat{\sigma}_{ij\to c\overline{c}k}^{\rm NLO}(\alpha_s^{\rm 2L}(Q_R^2),Q_F^2,x_1,x_2) \tag{16}$$

where the superscript "NLO" on $d\hat{\sigma}_{ij\to c\overline{c}k}$ indicates the use of the NLO matrix elements. The additional sum over $k$ in Eq. (16) includes the virtual ($k=0$) and real ($k=q$, $\overline{q}$ or $g$ depending on $i$ and $j$) NLO corrections. In the calculations of $d\sigma_{\rm LO}^{\rm 2L}$ and $d\sigma_{\mathcal{O}(\alpha_s^3)}$, we use the value of $\Lambda_{\rm QCD}^{(4)}$ given for the NLO PDFs and work in the $\overline{\rm MS}$ scheme. The standard NLO cross section is then

$$d\sigma_{\rm NLO}^{\rm std} = d\sigma_{\rm LO}^{\rm 2L} + d\sigma_{\mathcal{O}(\alpha_s^3)} \tag{17}$$

while our "alternative NLO" cross section is defined as

$$d\sigma_{\rm NLO}^{\rm alt} = d\sigma_{\rm LO}^{\rm 1L} + d\sigma_{\mathcal{O}(\alpha_s^3)}\ . \tag{18}$$

Since the enhancement in Ref. [20, 21] was defined using $d\sigma_{\rm LO}^{\rm 1L}$ only, the best we can do is to use the alternative NLO cross section in our analysis, as described below.

We now discuss how the enhancement is taken into account in the context of the NLO computation. We calculate the LO inclusive charm $p_T$ distribution, $d^2\sigma/dp_Tdy$, with the detected charm (anticharm) quark in the rapidity interval $\Delta y$ with $|y|<1$, motivated by the pseudorapidity acceptance of the ALICE tracking barrel, $|\eta|<0.9$. The rapidity, $y_2$, of the undetected anticharm (charm) quark is integrated over. The charm enhancement factor $R(p_T,\Delta y)$ is then

$$R(p_T,\Delta y) = \frac{\displaystyle\int_{\Delta y} dy \int dy_2 \frac{d^3\sigma({\rm EHKQS})}{dp_Tdydy_2}}{\displaystyle\int_{\Delta y} dy \int dy_2 \frac{d^3\sigma({\rm CTEQ61L})}{dp_Tdydy_2}}\ . \tag{19}$$

Next, we assume that the enhancement calculated at LO is the same when calculated at NLO. This is the only reasonable assumption we can make to test whether the enhancement can be detected with ALICE which will measure the physical $p_T^D$ distribution. The alternative NLO cross section is therefore the closest in spirit to the LO computation in Ref. [20,21]. Thus, the enhanced NLO charm $p_T$ distribution is

$$R(p_T,\Delta y)\ d\sigma_{\rm NLO}^{\rm alt}(\Delta y)/dp_T\ . \tag{20}$$

Our results are obtained with the same parameters used in Section 4.1, $m=1.2$ GeV and $Q^2=4m_T^2$ as well as $m=1.3$ GeV and $Q^2=m_T^2$. These two choices are the baseline results against which other parameter choices will be compared to see if the enhancement is detectable.

### 5.2 From charm to $D$ enhancement

To make a more realistic $D$ meson distribution, we have modified the charm $p_T$ distribution by the heavy quark string fragmentation in PYTHIA [78]. Charm events in $pp$ collisions at $\sqrt{S} = 14$ TeV are generated using PYTHIA (default settings) with the requirement that one of the quarks is in the interval $|y| < 1$. The charm quarks are hadronized using the default string model. Since $c$ and $\bar{c}$ quarks fragment to $D$ and $\overline{D}$ mesons[13], respectively, in each event related $(c, D)$ and $(\bar{c}, \overline{D})$ pairs can easily be identified[14]. These pairs are reweighted to match an arbitrary NLO charm quark $p_T$ distribution, $dN^c_{\rm NLO}/dp_T$. If $dN^c_{\rm PYTHIA}/dp_T$ is the charm $p_T$ distribution given by PYTHIA, each $(c, D)$ pair is assigned the weight

$$\mathcal{W}(p_T) = \frac{dN^c_{\rm NLO}/dp_T}{dN^c_{\rm PYTHIA}/dp_T} \qquad (21)$$

where $p_T$ is the transverse momentum of the charm quark of the pair. Therefore, the reweighted final-state $D$ distribution corresponds to the one that would be obtained by applying string fragmentation to the NLO $c$-quark distribution. The resulting $D$ distribution is significantly harder than that obtained using the Peterson fragmentation function [79]. The enhancement survives after fragmentation although the $D$ enhancement is somewhat lower than that of the charm because for a given $p_T^D$, the $D$ spectrum receives contributions from charm quarks with $p_T \gtrsim p_T^D$, where the charm enhancement is smaller.

### 5.3 Sensitivity to the enhancement

Figure 10 shows the double-differential $D^0$ cross section, $d^2\sigma_D/dp_T^D dy$, in $|y| < 1$ as a function of the transverse momentum. The points represent the expected "data" measured by ALICE, obtained from the alternative NLO cross section scaled by the enhancement factor $R(p_T, \Delta y)$ defined in Eq. (19), and modified by string fragmentation. The solid and dashed curves are obtained by applying string fragmentation to the alternative NLO and standard NLO $c\bar{c}$ cross sections, respectively. Thus, the "data" points include the enhancement while the curves do not. The horizontal error bars indicate the bin width, the vertical error bars represent the statistical error and the shaded band gives the $p_T$-dependent systematic error. The 5% $p_T$-independent systematic error on the normalization is not shown. (See Ref. [80] for a discussion of the error analysis. The standard NLO cross section, Eq. (17), and the $\mathcal{O}(\alpha_s^3)$ contribution to the alternative NLO cross section, Eq. (16), were calculated using the HVQMNR code [81] with CTEQ6M and $\Lambda^{(4)}_{\rm QCD} = 0.326$ GeV. The LO contribution to the alternative NLO cross section, Eq. (14), was calculated using the CTEQ61L PDFs. Fragmentation was included as described in Section 5.2. The enhancement, the difference between the data and the solid curves for $p_T^D \lesssim 3$ GeV, is more pronounced for the larger mass and lower scale, on the right-hand side of Fig. 10.

There is a significant difference between the alternative and standard NLO distributions. Part of the difference is due to the one- and two-loop evaluations of $\alpha_s$ since $\alpha_s^{\rm 2L} < \alpha_s^{\rm 1L}$. However, the most important contribution is the large differences between the LO and NLO gluon distributions, especially at low scales [80].

In order to address the question of the experimental sensitivity to the effect of nonlinear gluon evolution on low-$p_T$ charm production, we consider, as a function of $p_T^D$, the ratio of the simulated data, including the enhancement, to alternative NLO calculations using a range of $m$ and $Q^2$ with PYTHIA string fragmentation. We denote this ratio as "Data/Theory" and try to reproduce this ratio with NLO calculations employing recent linearly-evolved PDFs and tuning $m$ and $Q^2$.

Since the enhancement has disappeared for $p_T^D \gtrsim 5$ GeV, we refer to this unenhanced region as high $p_T^D$. The $p_T^D$ region below 5 GeV, where the enhancement is important, is referred to as low $p_T^D$. If no set of parameters can describe both the high- and low-$p_T^D$ components of the distribution equally well, and, if the set that best reproduces the high-$p_T^D$ part underestimates the low-$p_T^D$ part, this would be a strong indication of the presence of nonlinear effects.

[13] Here $D \equiv D^+, D^0$.

[14] Events containing charm baryons were rejected.

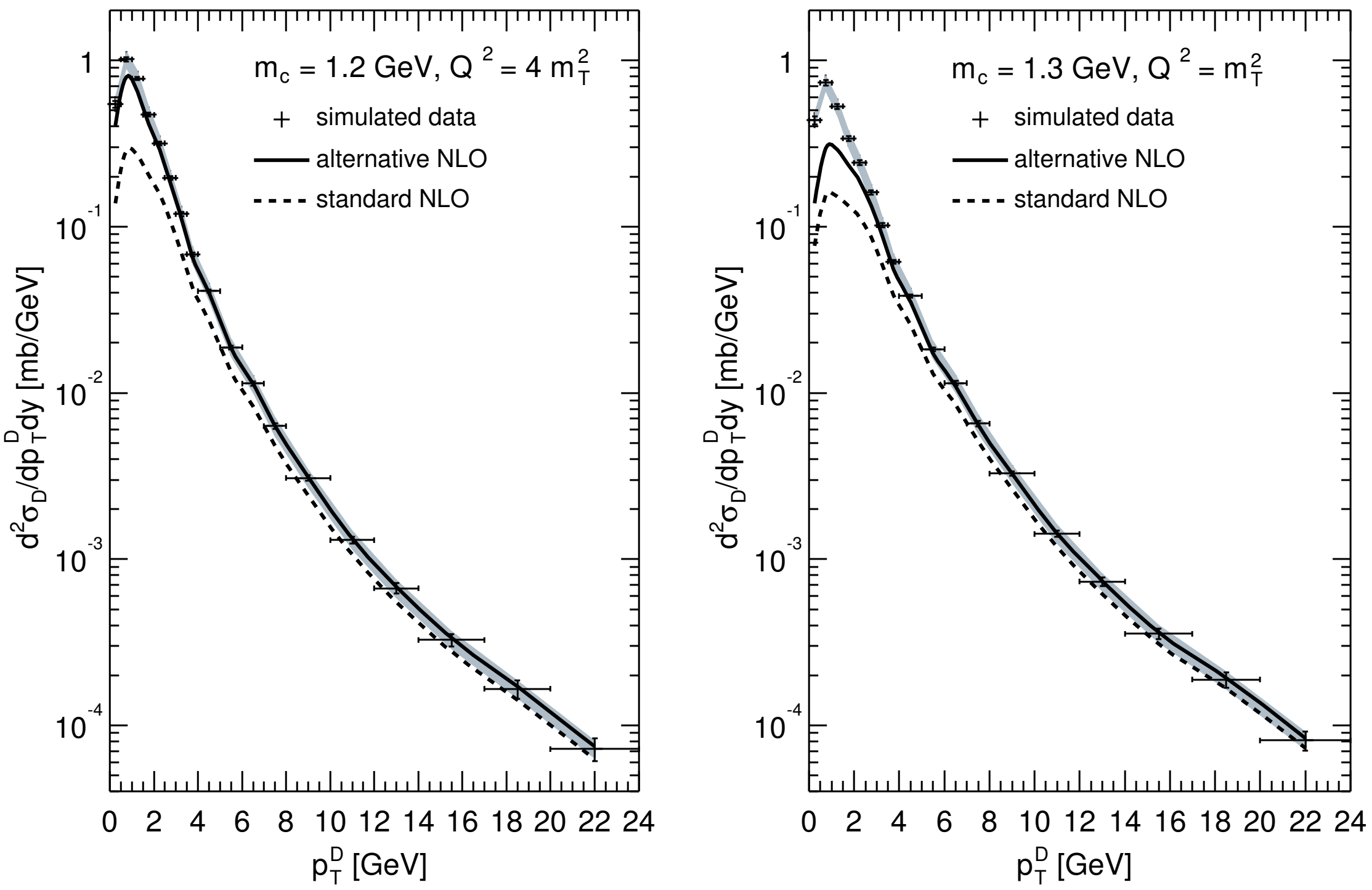

**Fig. 10:** Comparison of the simulated ALICE data generated from $R(p_T, \Delta y) d\sigma_{\rm NLO}^{\rm alt}$ with the alternative (solid) and standard (dashed) NLO calculations. The effect of string fragmentation is included in the "data" points as well as in the curves. The left-hand side shows the result for $m = 1.2$ GeV and $Q^2 = 4m_T^2$ while the right-hand side is the result for $m = 1.3$ GeV and $Q^2 = m_T^2$. The error bars on the data represent the statistical error and the shaded band represents the $p_T$-dependent systematic error. The 5% normalization error is not shown.

The Data/Theory plots are shown in Fig. 11. The points with the statistical (vertical bars) and $p_T$-dependent systematic (shaded region) error correspond to the data of Fig. 10, including the enhancement, divided by themselves, depicting the sensitivity to the theory calculations. The black squares on the right-hand sides of the lines $\rm Data/Theory = 1$ represent the 5% $p_T$-independent error on the ratio coming from the cross section normalization. This error is negligible relative to present estimates of other systematic uncertainties ($\simeq 13\%$).

On the left-hand side, the thick solid curve with $m = 1.2$ GeV and $Q^2 = 4m_T^2$ best agrees with the high-$p_T^D$ ratio by construction since $R \approx 1$ at large $p_T^D$. It also shows the effect of the enhancement well beyond the error band for $p_T^D \lesssim 2$ GeV. Better agreement with the data over the entire $p_T^D$ range can be achieved only by choosing a charm quark mass lower than 1.2 GeV, below the nominal range of charm masses, as illustrated by the dashed curve for $m = 1.1$ GeV. Higher masses with $Q^2 = 4m_T^2$ produce much larger Data/Theory ratios than the input distribution. The ratio with $m = 1.3$ GeV and $Q^2 = m_T^2$ (dot-dot-dashed curve) gives a much larger ratio at low $p_T^D$ and drops below $\approx 1$ for $p_T^D > 8$ GeV.

We also present the ratio using the MRST parton densities (MRST2001 LO [36] in Eq. (14) and MRST2002 NLO [82] in Eq. (16)) with $m = 1.2$ GeV and $Q^2 = 4m_T^2$. We find that this result also agrees reasonably well with the CTEQ6 results for the same $m$ and $Q^2$. Thus, the enhancement seems to be rather independent of the PDF. The CTEQ61L and the MRST2001 LO distributions are similar at low $x$, suggesting that non-linearly evolved PDFs based on MRST2001 LO would produce an enhancement like that of Ref. [20, 21].

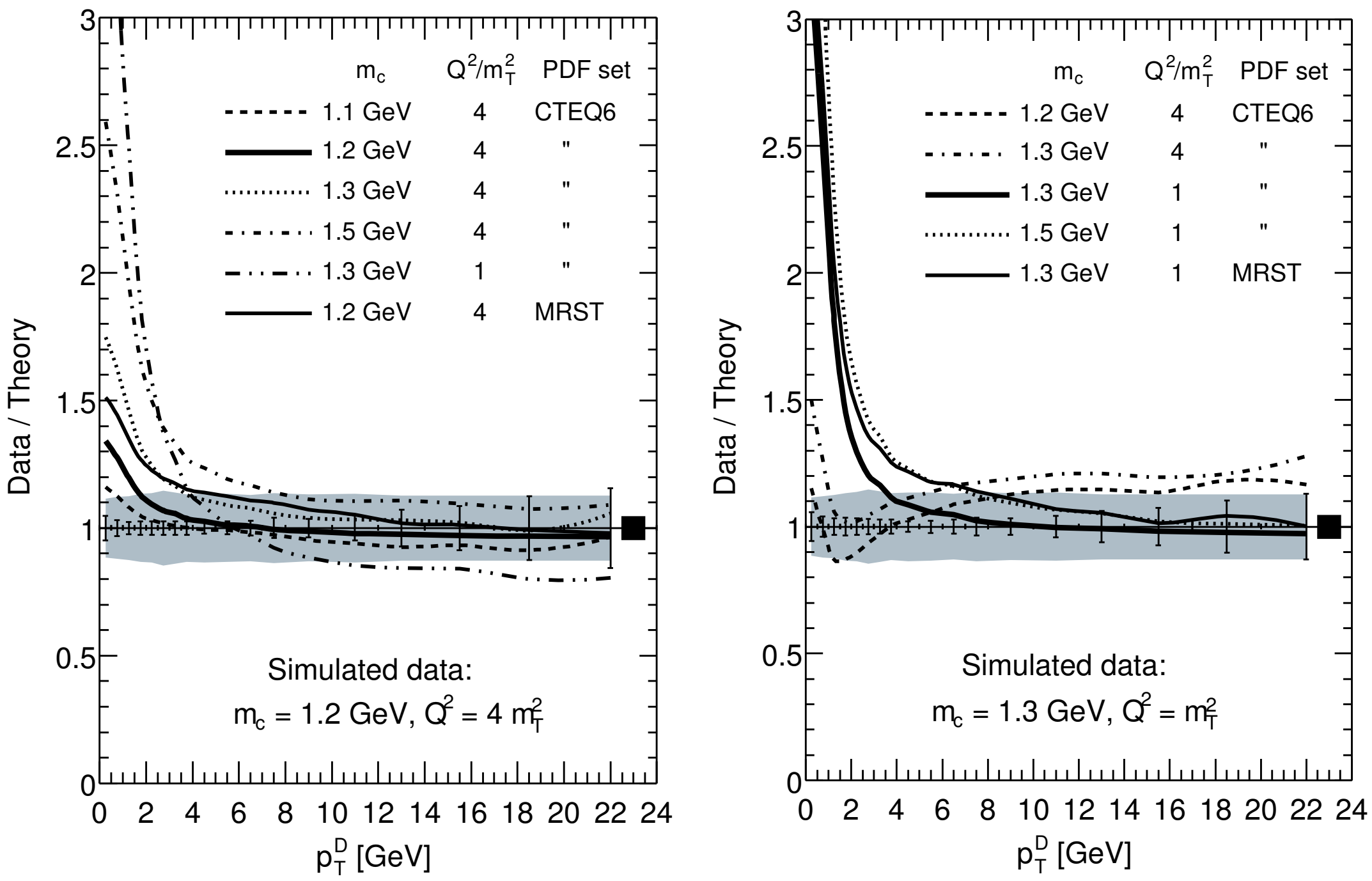

**Fig. 11:** Ratio of the generated ALICE data relative to calculations of the alternative NLO cross sections with several sets of parameters and PYTHIA string fragmentation. The left-hand side shows the result for $m = 1.2$ GeV and $Q^2 = 4m_T^2$ while the right-hand side is the result for $m = 1.3$ GeV and $Q^2 = m_T^2$.

On the right-hand side of Fig. 11 the thick solid curve, employing the same parameters as the data, gives the best agreement at high $p_T^D$. We note that even though the results with $Q^2 = 4m_T^2$ and $m \leq 1.3$ GeV lie closer to the data at low $p_T^D$ and within the combined statistical and systematic error at higher $p_T^D$, the curves with these parameters have the wrong slopes for $p_T^D \lesssim 8$ GeV. The statistical sensitivity is expected to be good enough to distinguish the difference in curvature. The results with the MRST PDFs do not alter the conclusions.

### 5.4 Conclusions

We have studied whether the EHKQS gluon distributions [37] could generate an observable $D$ meson enhancement in $pp$ collisions at the LHC. Using the EHKQS LO PDFs and LO matrix elements for charm quark production and PYTHIA string fragmentation for $D$ meson hadronization, the enhancement indeed survives to the $D$ mesons.

The $D$ meson enhancement, however, drops rapidly with transverse momentum. Therefore, $D$ measurement capability at small $p_T^D$ is necessary to verify the effect experimentally. The ALICE detector can directly reconstruct $D^0 \to K^-\pi^+$. We have demonstrated that, in the most optimistic case, the enhancement can be detected above the experimental statistical and systematic errors. When the charm mass is somewhat smaller, $m = 1.2$ GeV, but the scale is larger, $Q^2 = 4m_T^2$, it is more difficult to detect the enhancement over the experimental uncertainties.

## Acknowledgements

The work of K. Kutak was supported in part by the Graduiertenkolleg Zukünftige Entwicklungen in der Teilchenphysik. M.G. Ryskin would like to thank the IPPP at the University of Durham for hospitality, and A.D. Martin thanks the Leverhulme Trust for an Emeritus Fellowship. Their work was supported by the Royal Society, by grants INTAS 00-00366, RFBR 04-02-16073, and by the Federal Program of the Russian Ministry of Industry, Science and Technology SS-1124.2003.2. The work of RV was supported in part by the Director, Office of Energy Research, Division of Nuclear Physics of the Office of High Energy and Nuclear Physics of the U. S. Department of Energy under Contract Number DE-AC02-05CH11231.

# Heavy quark fragmentation

*J. Bracinı́k*[1], *M. Cacciari*[2], *M. Corradi*[3], *G. Grindhammer*[1]
[1]Max-Planck-Institut für Physik, München, Germany
[2]LPTHE - Université P. et M. Curie (Paris 6), Paris, France
[3]INFN Bologna, via Irnerio 46, Bologna, Italy

**Abstract**
The fragmentation of heavy quarks into hadrons is a key non-perturbative ingredient for the heavy quark production calculations. The formalism is reviewed, and the extraction of non-perturbative parameters from $e^+e^-$ and from $ep$ data is discussed.

Coordinator: *M. Corradi*

## 1 Introduction[1]

When we try to describe in QCD the production of a hadron we are always faced with the necessity to take into account the non-perturbative hadronization phase, i.e. the processes which transform perturbative objects (quarks and gluons) into real particles. In the case of light hadrons the QCD factorization theorem [1–6] allows to factorize these non-perturbative effects into universal (but factorization-scheme dependent) *fragmentation functions* (FF):

$$\frac{d\sigma_h}{dp_T}(p_T) = \sum_i \int \frac{dx}{x} \frac{d\sigma_i}{dp_T}\left(\frac{p_T}{x};\mu\right) D_{i\to h}(x;\mu) + \mathcal{O}\left(\frac{\Lambda}{p_T}\right) . \tag{1}$$

In this equation, valid up to higher twist corrections of order $\Lambda/p_T$ ($\Lambda$ being a hadronic scale of the order of a few hundred MeV and $p_T$ for instance a transverse momentum), the partonic cross sections $d\sigma_i/dp_T$ for production of the parton $i$ are calculated in perturbative QCD, while the fragmentation functions $D_{i\to h}(x;\mu)$ are usually extracted from fits to experimental data. Thanks to their universality they can be used for predictions in different processes. The artificial factorization scale $\mu$ is a reminder of the non-physical character of both the partonic cross sections and the fragmentation functions: it is usually taken of the order of the hard scale $p_T$ of the process, and the fragmentation functions are evolved from a low scale up to $\mu$ by means of the DGLAP evolution equations.

This general picture becomes somewhat different when we want to calculate the production of heavy-flavoured mesons. In fact, thanks to the large mass of the charm and the bottom quark, acting as a cutoff for the collinear singularities which appear in higher orders in perturbative calculations, one can calculate the perturbative prediction for heavy *quark* production. Still, of course, the quark $\to$ hadron transition must be described. Mimicking the factorization theorem given above, it has become customary to complement the perturbative calculation for heavy quark production with a non-perturbative fragmentation function accounting for its hadronization into a meson:

$$\frac{d\sigma_H}{dp_T}(p_T) = \int \frac{dx}{x} \frac{d\sigma_Q^{pert}}{dp_T}\left(\frac{p_T}{x}, m\right) D^{np}_{Q\to H}(x) . \tag{2}$$

It is worth noting that at this stage this formula is not given by a rigorous theorem, but rather by some sensible assumptions. Moreover, it will in general fail (or at least be subject to large uncertainties) in the region where the mass $m$ of the heavy quark is not much larger than its transverse momentum $p_T$, since the choice of the scaling variable is not unique any more, and $\mathcal{O}(m/p_T)$ corrections cannot be neglected.

[1]Author: M. Cacciari

Basic arguments in QCD allow to identify the main characteristics of the non-perturbative fragmentation function $D^{np}_{Q\to H}(x)$. In 1977 J.D. Bjorken [7] and M. Suzuki [8] independently argued that the average fraction of momentum lost by the heavy quark when hadronizing into a heavy-flavoured hadron is given by

$$\langle x\rangle^{np} \simeq 1 - \frac{\Lambda}{m} \ . \tag{3}$$

Since (by definition) the mass of a heavy quark is much larger than a hadronic scale $\Lambda$, this amounts to saying that the non-perturbative FF for a heavy quark is very hard, i.e. the quark loses very little momentum when hadronizing. This can also be seen with a very simplistic argument: a fast-traveling massive quark will lose very little speed (and hence momentum) when picking up from the vacuum a light quark of mass $\Lambda$ to form a heavy meson[2].

This basic behaviour is to be found as a common trait in all the non-perturbative heavy quark FFs derived from various phenomenological models. Among the most commonly used ones we can cite the Kartvelishvili-Likhoded-Petrov [12], Bowler [13], Peterson-Schlatter-Schmitt-Zerwas [14] and Collins-Spiller [15] fragmentation functions. These models all provide some functional form for the $D^{np}_{Q\to H}(x)$ function, and one or more free parameters which control its hardness. Such parameters are usually not predicted by the models (except perhaps on an order-of-magnitude basis), and must be fitted to the experimental data.

During the '80s many such fits were performed, and these and similar functions were also included in many Monte Carlo event generators. Eventually, some 'best' set of parameter values (for instance for the PSSZ form) was determined [16] and subsequently widely used.

These first applications, given the limited accuracy of the available data, tended to overlook two aspects which have become more important in recent years, when the accuracy of the data has vastly improved:

- A non-perturbative FF is designed to describe the heavy quark $\to$ hadron transition, dealing with events mainly populated by soft gluons of energies of a few hundred MeV. However, if a heavy quark is produced in a high energy event it will initially be far off shell: perturbative hard gluons will be emitted to bring it on-shell, reducing the heavy quark momentum and yielding in the process large collinear logarithms (for instance of the form $\alpha_s^n \log^n(p_T/m)$ in a transverse momentum differential cross section). Of course, the amount of gluon radiation is related to the distance between the heavy quark mass scale and the hard scale of the interaction, and is therefore process-dependent. One can (and it was indeed done) either fit different free parameters at different centre-of-mass energies (or transverse momenta), or try to evolve directly the non-perturbative FF by means of the DGLAP equations, hence including into it the perturbative collinear logarithms. However, this is not what non-perturbative fragmentation functions are meant for, and doing so spoils the validity of the relation in Eq. (3).
- Since only the final heavy hadron is observed, both the non-perturbative FF and the perturbative cross section for producing the heavy quark must be regarded as non-physical objects. The details of the fitted non-perturbative FF (e.g. the precise value(s) of its free parameter(s)) will depend on those of the perturbative cross sections: different perturbative calculations (leading order, next-to-leading order, Monte Carlo, ...) and different perturbative parameters (heavy quark masses, strong coupling, ...) will lead to different non-perturbative FFs. These in turn will have to be used only with a perturbative description similar to the one they have been determined with. Hence the limited accuracy (and hence usefulness) of a 'standard' determination of the parameters [17].

The first point was addressed by Mele and Nason in a paper [18] which deeply changed the field of heavy quark fragmentation, and essentially propelled it into the modern era. Mele and Nason observed that, in the limit where one neglects heavy quark mass terms suppressed by a large energy scale, a heavy quark cross section can be factored into a massless, $\overline{MS}$-subtracted cross section for producing a light parton, and a process-independent[3], *perturbative* heavy quark fragmentation function describing the transition of the massless parton into the heavy quark:

$$\frac{d\sigma_Q^{pert,res}}{dp_T}(p_T, m) = \sum_i \int \frac{dx}{x} \frac{d\sigma_i}{dp_T}\left(\frac{p_T}{x}; \mu\right) D_{i\to Q}(x; \mu, m) + \mathcal{O}\left(\frac{m}{p_T}\right) . \tag{4}$$

[2] More modern and more rigorous derivations of this result can be found in [9–11].

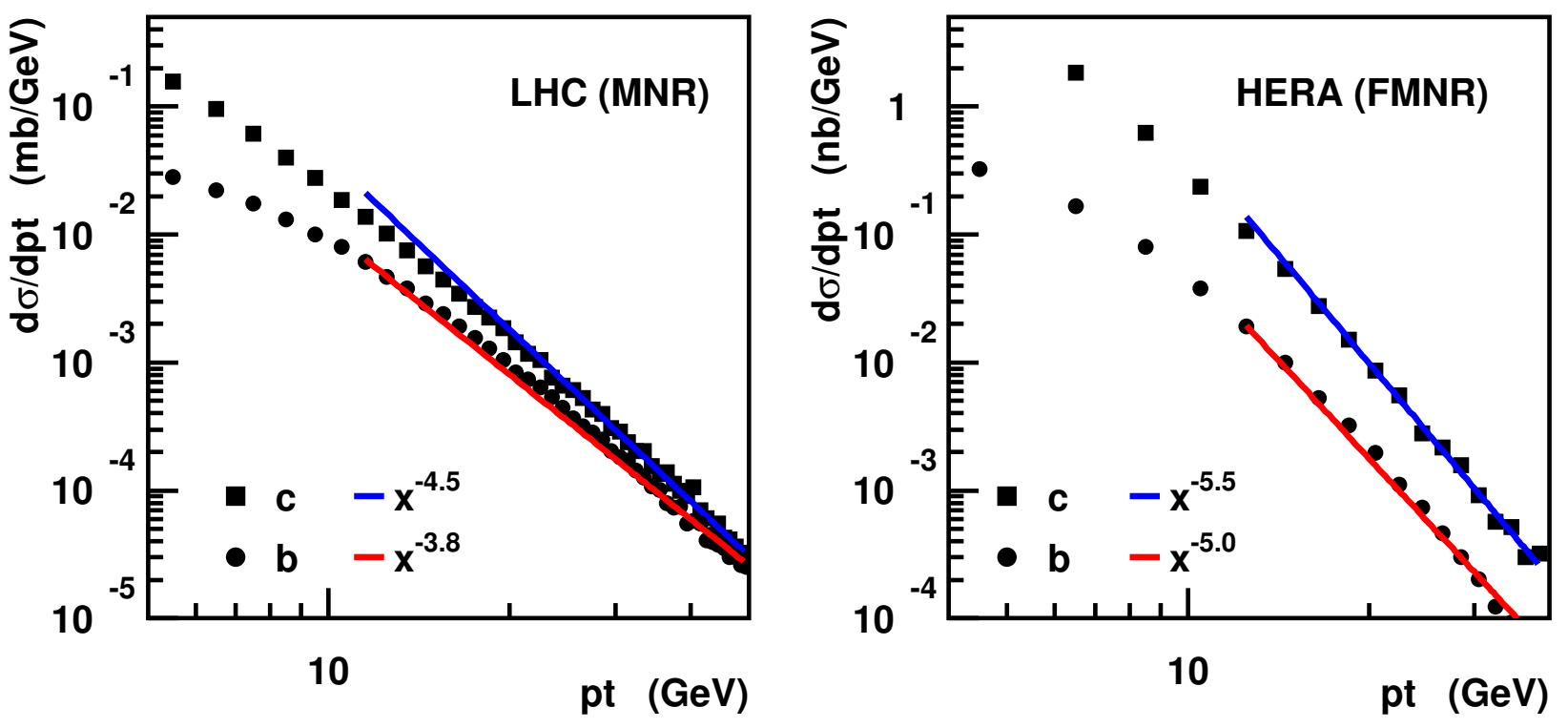

**Fig. 1:** Power-law fits to the heavy quark $p_T$ distributions at LHC (left) and HERA (right) obtained with the NLO programs MNR and FMNR. The resulting exponents are $N = 4.5/3.8$ for charm/beauty at LHC and $N = 5.5/5.0$ for *c*/*b* at HERA.

The key feature of this equation is that it is entirely perturbative: every term can be calculated in perturbative QCD. The perturbative fragmentation functions $D_{i\to Q}(x;\mu,m)$ (not to be confused with the non-perturbative one $D^{np}_{Q\to H}(x)$) can be evolved via DGLAP equations from an initial scale of the order of the heavy quark mass up to the large scale of the order of $p_T$. This resums to all orders in the strong coupling the collinear logarithms generated by the gluon emissions which bring the heavy quark on its mass shell, leading to a more accurate theoretical prediction for $d\sigma_Q/dp_T$.

Once a reliable perturbative cross section for the production of a heavy quark is established, one is simply left with the need to account for its hadronization. For this purpose one of the functional forms listed above can be used for the non-perturbative FF, and implemented as in Eq. (2), but using the improved, resummed cross section given by Eq. (4). Since most of the the scaling-violation logarithms are accounted for by the evolution of the perturbative FF, the non-perturbative one can now be scale-independent and only contain the physics related to the hadronization of the heavy quark. It will always, however, depend on the details of the perturbative picture used.

## 2 Extraction of heavy quark fragmentation parameters from $e^+e^-$ and their impact on HERA and LHC[4]

### 2.1 Importance of $\langle x\rangle^{\mathrm{np}}$

According to the factorization of the fragmentation functions (FF), the differential cross section $d\sigma/dp_T$ for the production of a heavy hadron $H$ can be written as the convolution of the perturbative heavy quark differential cross section $d\sigma^{\mathrm{pert}}/dp_T$ and the non-perturbative fragmentation function $D^{\mathrm{np}}(x)$:

$$\frac{d\sigma}{dp_T}(p_T) = \int \frac{dx}{x} D^{\mathrm{np}}(x) \frac{d\sigma^{\mathrm{pert}}}{dp_T}\left(\frac{p_T}{x}\right) . \tag{5}$$

[3]Mele and Nason extracted this function from the $e^+e^-$ cross section, convincingly conjecturing its process independence, which was successively established on more general grounds in [19]

[4]Author: M. Corradi

**Table 1:** Test functions used in Fig. 2. The functions assume a value different from zero in the range given by the third column.

| Function | $D(x)$ | parameters | $x$ range |
|---|---|---|---|
| Kartvelishvili | $(1-x)x^{\alpha}$ | $\alpha = 2/\delta - 3$ | $[0,1]$ |
| Peterson | $\frac{1}{x}\left(1-\frac{1}{x}-\frac{\epsilon}{(1-x)}\right)^{-2}$ | $\epsilon$ | $[0,1]$ |
| Gauss | $\exp(-(\frac{x-\mu}{2\sigma})^2)$ | $\mu=1-\delta$ $\sigma=\delta/2$ | $[-\infty,\infty]$ |
| Box | const. | – | $[1-2\delta,1]$ |
| Triangular: | $x-x_0$ | $x_0=1-3\delta$ | $[1-3\delta,1]$ |

This convolution neglects mass terms $\mathcal{O}(m_Q/p_T)$ and non-perturbative terms $\mathcal{O}(\Lambda_{\rm qcd}/m_Q)$.

The heavy quark $p_T$ distribution behaves at large $p_T$ like a power law $d\sigma^{\rm pert}/dp_T = Cp_T^{-N}$. Figure 1 shows power-law fits to the $p_T$ distributions of heavy quarks at LHC and in photoproduction at HERA as obtained from the NLO programs MNR [20] and FMNR [21]. For $p_T > 10$ GeV $N$ was found to range from 3.8 ($b$ at LHC) to 5.5 ($c$ at HERA). Combining this power-law behavior with Eq. (5), the hadron $p_T$ distribution is given by

$$\frac{d\sigma}{dp_T}(p_T) = \int dx\, x^{N-1}\, D^{\rm np}(x)\, Cp_T^{-N} = \frac{d\sigma^{\rm pert}}{dp_T}\hat{D}_N^{\rm np}, \tag{6}$$

where $\hat{D}_N^{\rm np} = \int dx\, x^{N-1}\, D^{\rm np}(x)$ is the $N^{th}$ Mellin moment of the non-perturbative FF.

The hadron distribution is therefore governed by the $4^{th}$ - $5^{th}$ Mellin moments of $D^{\rm np}(x)$. It is interesting to translate the Mellin moments into more intuitive central moments

$$\mu_n = \int dx\, (x-\langle x\rangle)^n\, D^{\rm np}(x) \quad \text{for} \quad n \geq 2 \tag{7}$$

where $\langle x\rangle = \int dx\, xD^{\rm np}(x)$ is the mean value. The first Mellin moments, written in terms of $\langle x\rangle$ and $\mu_n$, are: $\hat{D}_1 = 1$ , $\hat{D}_2 = \langle x\rangle$ , $\hat{D}_3 = \langle x\rangle^2 + \mu_2$ , $\hat{D}_4 = \langle x\rangle^3 + 3\mu_2\langle x\rangle + \mu_3$.

In heavy quark fragmentation, the mean value of $D^{\rm np}(x)$ can be written as $\langle x\rangle = 1-\delta$ where $\delta = \mathcal{O}(\Lambda_{\rm qcd}/m_Q)$ is small [11]. For any positive function with $\langle x\rangle = 1-\delta$, defined in the interval $[0,1]$, the central moments are limited by $\delta$, $|\mu_n| \leq \delta$. In practice, reasonable heavy quark fragmentation functions are concentrated in a small region around $1-\delta$ and therefore the higher central moments are small. To be specific, if the function is different from zero in a region of size $\pm K\delta$ (with $K = \mathcal{O}(1)$) around $1-\delta$ then $|\mu_n| \leq (K\delta)^n$. This means that the Mellin moments of reasonable FFs are given, to a good approximation, by the mean value to the $N-1$ power:

$$\hat{D}_N = \langle x\rangle^{N-1} + \mathcal{O}(\delta^2). \tag{8}$$

The expansion to $\delta^2$ involves the second central moment $\mu_2$: $\hat{D}_N = \langle x\rangle^{N-1} + \frac{(N-1)!}{2(N-3)!}\mu_2\langle x\rangle^{N-3} + \mathcal{O}(\delta^3)$. For a reasonable FF and a perturbative distribution falling with the power $-N$, Eq. 6 and 8 give

$$\frac{d\sigma}{dp_T}(p_T) = \frac{d\sigma^{\rm pert}}{dp_T^Q}(p_T)\ (\langle x\rangle^{\rm np})^{N-1} + \mathcal{O}(\delta^2). \tag{9}$$

Therefore the effect of the non-perturbative FF is to introduce a shift in the normalisation that depends on the average $x$, while the details of the shape of $D(x)$ have negligible effect. To check that this reasoning works with realistic fragmentation functions and realistic perturbative $p_T$ distributions, various functions with the same $\langle x \rangle$ but different shapes have been tested in convolution with the perturbative $p_T$ spectrum for $b$ production at LHC obtained with the NLO program MNR. The test functions considered are the Peterson [14] and Kartvelishvili [12] fragmentation functions, a Gaussian distribution with $\sigma = 1 - \mu$, a flat and a triangular distribution. Table 1 gives more detail about these functions. Three average values were chosen: $\langle x \rangle = 0.9, 0.8, 0.666$. Figure 2 shows the result of this test. For each average value, the convolutions are very similar, even if the test functions are very different. For $\langle x \rangle = 0.9, 0.8$, which are typical beauty or charm values, the hadron spectra agree within few %. For the extreme value of $\langle x \rangle = 0.666$, the results for the Peterson and Kartvelishvili functions give very similar hadron spectra while the less realistic Gaussian and Box shapes differ at most 10% from Peterson at large $p_T$ and the extreme Triangular function shows deviations up to $\sim 20\%$.

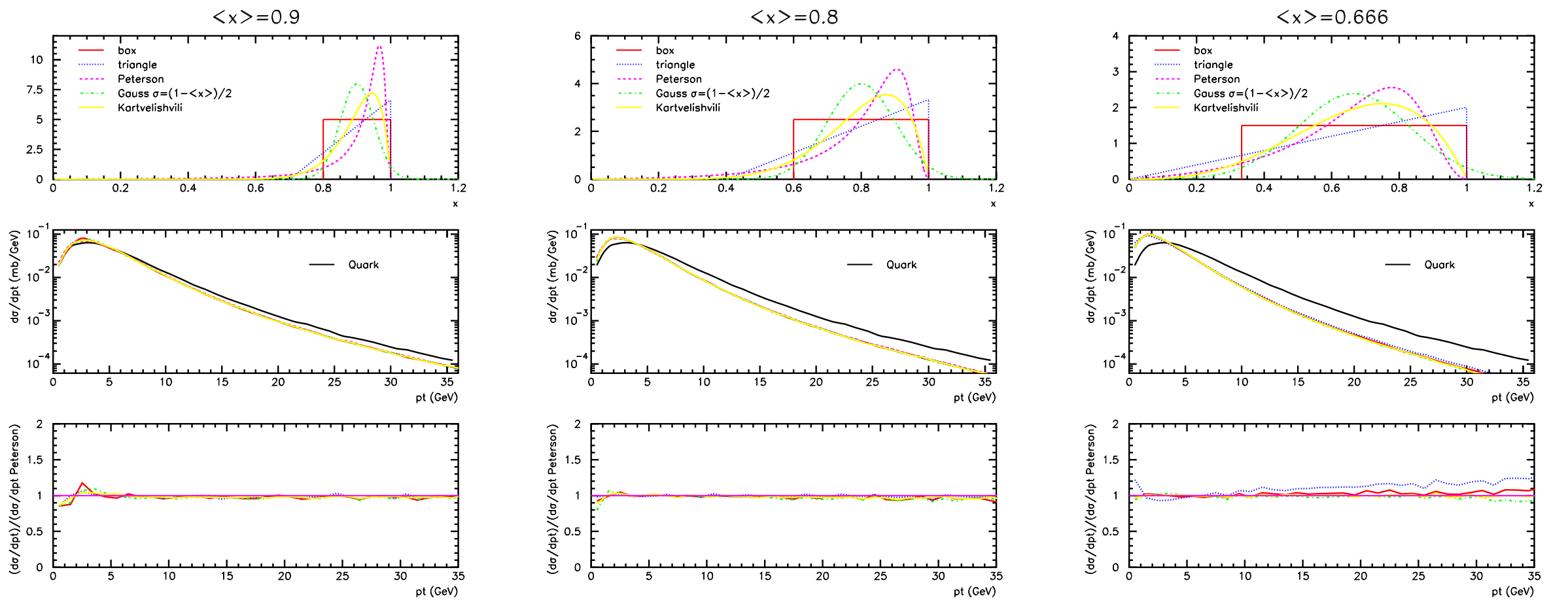

**Fig. 2:** Effect of the convolution of the heavy quark transverse momentum distribution with different test functions for different values of the non-perturbative FF $\langle x \rangle = 0.9$ (left), 0.8 (center), 0.666 (right). For each $\langle x \rangle$, the upper plot shows the test functions, the middle plot shows the perturbative $p_T$ distribution obtained with the MNR program for beauty at LHC and the hadron $p_T$ distributions after the convolution with the test functions. The lower plot shows the ratio of the different hadron $p_T$ distributions to the result obtained with the Peterson one.

In conclusion the relevant fragmentation parameter for the inclusive hadron spectra at $pp$ and $ep$ colliders is the mean value $\langle x \rangle^{\rm np}$ of the non-perturbative FF. The next part will discuss, on the basis of $e^+e^-$ data, what values of $\langle x \rangle^{\rm np}$ are relevant for different calculations.

### 2.2 Extraction of $\langle x \rangle^{\rm np}$ from $e^+e^-$ data

In $e^+e^-$ interactions it is convenient to express the factorization ansatz, given for the heavy-hadron $p_T$ in Eq. (5), in terms of the heavy-hadron momentum normalized to the maximum available momentum: $x_p = p^H/p^H_{\rm max}$, where $p^H_{\rm max} = \sqrt{(\frac{1}{2}E_{\rm cms})^2 - m_H^2}$:

$$\frac{d\sigma}{dx_p}(x_p) = \int \frac{dx}{x}\, D^{\rm np}(x)\, \frac{d\sigma^{\rm pert}}{dx_p}(\frac{x_p}{x})$$

which corresponds to the following relation for the mean values: $\langle x_p \rangle = \langle x \rangle^{\rm np} \langle x \rangle^{\rm pert}$ where $\langle x_p \rangle$ is the mean hadron $x_p$, $\langle x \rangle^{\rm np}$ is the mean value of the non-perturbative FF and $\langle x \rangle^{\rm pert} = \int dx\, x\, \frac{d\sigma^{\rm pert}}{dx_p}$ is the mean value of the perturbative distribution. Then, taking $\langle x_p \rangle$ from experimental data and $\langle x \rangle^{\rm pert}$ from a particular perturbative calculation, it is possible to extract the value of $\langle x \rangle^{\rm np}$ valid for that calculation as

$$\langle x \rangle^{\rm np} = \langle x_p \rangle / \langle x \rangle^{\rm pert}. \qquad (10)$$

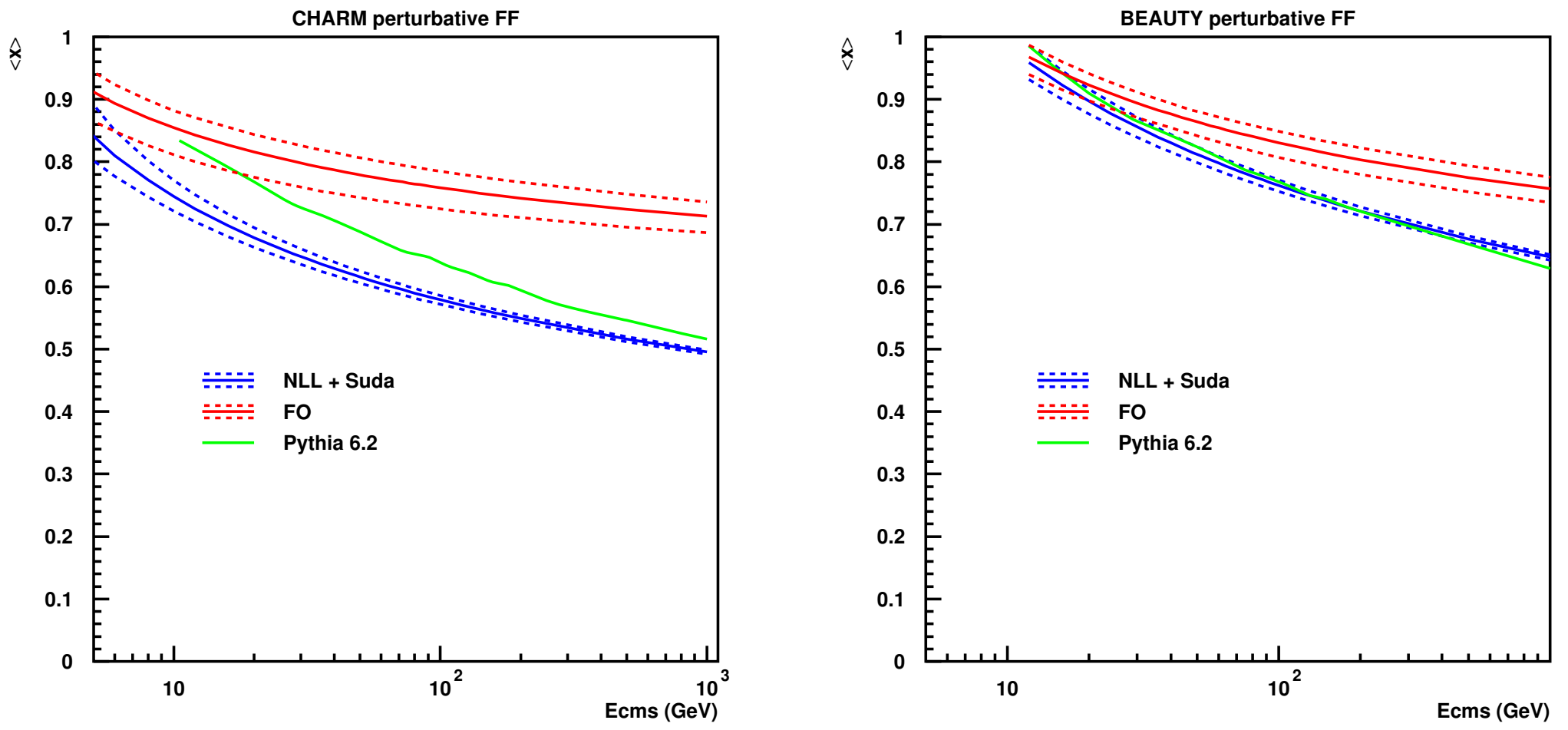

**Fig. 3:** Average fragmentation function from the perturbative calculations for charm (left) and beauty (right) as a function of the $e^+e^-$ center of mass energy.

Two perturbative calculations will be considered to extract $\langle x \rangle^{\text{pert}}$: a fixed-order (FO) next-to-leading order (i.e. $\mathcal{O}(\alpha_S)$) calculation and a calculation that includes also the resummation of next-to-leading logarithms (NLL) and Sudakov resummation, both obtained with the HVQF program [19]. From the point of view of fragmentation, the FO calculation only considers the emission of a gluon from one of the two heavy quarks generated in the $e^+e^-$ collision while the NLL calculation includes the evolution of the FF from the hard interaction scale down to the scale given by the heavy quark mass. The parameters used for the FO and NLL models are $m_c = 1.5$ GeV, $m_b = 4.75$ GeV, $\Lambda_{\text{QCD}} = 0.226$ GeV and the renormalisation and factorization scales $\mu_R = \mu_F = E_{\text{cms}}$. The starting scale for FF evolution in the NLL model was chosen to be $m_Q$. The theoretical uncertainty was obtained by varying independently the normalisation and factorization scales by a factor 2 and $1/2$ and taking the largest positive and negative variations as the uncertainty.

The experimental data are also compared to the PYTHIA 6.2 Monte Carlo program [37] which contains an effective resummation of leading-logarithms based on a parton-shower algorithm and which is interfaced to the Lund fragmentation model. In this case the MC model gives directly $\langle x_p \rangle$, while $\langle x \rangle^{\text{pert}}$ has been obtained taking the heavy quark at the end of the parton shower phase. The quark masses have been set to $m_c = 1.5$ GeV and $m_b = 4.75$ GeV, and all the parameters were set to the default values except for specific fragmentation parameters explained below. Three sets of fragmentation parameters were chosen for charm: the default fragmentation (Lund-Bowler), a longitudinal string fragmentation of the Peterson form with $\epsilon = 0.06$ (`MSTJ(11)=3, PARJ(54)=-0.06`) and the Lund-Bowler fragmentation with parameters re-tuned by the CLEO collaboration [28] (`PARJ(41)=0.178, PARJ(42)=0.393, PARJ(13)=0.627`). The two sets chosen for beauty are the default Lund-Bowler fragmentation and the Peterson fragmentation with $\epsilon = 0.002$ (`MSTJ(11)=3, PARJ(55)=-0.002`). Figure 3 shows $\langle x \rangle^{\text{pert}}$ from the perturbative calculations as a function of the centre of mass energy $E_{\text{cms}}$ for charm and bottom.

## 2.3 Charm

Charm fragmentation data are available from various $e^+e^-$ experiments. The most precise are those at the $Z^0$ pole at LEP (ALEPH [23], OPAL [22], DELPHI [24]) and near the $\Upsilon(4s)$ (ARGUS [27], CLEO [28], BELLE [29]). Less precise data are available in the intermediate continuum region from DELCO [26] at PEP and TASSO [25] at PETRA. Measurements in which the beauty component was not subtracted have been discarded [38–40]. The experimental data are reported in Table 2. Only measurements relative to the $D^{*\pm}(2010)$ meson are considered, to avoid the complications due to cascade decays that are present for ground state mesons. Charm quarks originating from gluon splitting rather than from the virtual boson from $e^+e^-$ annihilation may be relevant at LEP energies. This contribution is anyway already subtracted in the published data considered here, and it is consistently not considered in the perturbative calculations. Most of the experiments published the mean value of the $x$ distribution. The only exception is ARGUS, for which the mean value was computed from the published distribution. Some of the experiments give the results directly in terms of $x_p$, others in terms of the energy fraction $x_E = 2E_H/E_{\rm cms}$. The latter has been corrected to $x_p$ using the PYTHIA MC. The difference between $\langle x_p \rangle$ and $\langle x_E \rangle$ can be as large as $12\%$ at $E_{\rm cms} = 10.5$ GeV and reduces to less than $1\%$ at $E_{\rm cms} = 92$ GeV. Since the low-mass measurements are already given in terms of $x_p$, the applied corrections from $x_E$ to $x_p$ was always small. QED corrections are also needed to compare the experimental data to the QCD predictions. The initial state radiation (ISR) from the electrons has the effect of reducing the energy available for the $e^+e^-$ annihilation and therefore to reduce the observed value of $\langle x_p \rangle$. A correction, obtained by comparing the PYTHIA MC with and without ISR, was applied to the data to obtain $\langle x_p \rangle^{\rm corr}$. The correction is $\sim 4\%$ at $E_{\rm cms} = 10.5$ GeV, is largest in the intermediate region and is negligible at $E_{\rm cms} = 92$ GeV.

**Table 2:** Experimental results on the average fragmentation function in $e^+e^-$ collisions for $D^*$ mesons and weakly decaying beauty hadrons. The table reports, for each experiment, the published variable and the corrections applied to obtain $\langle x_p \rangle^{\rm corr}$. All the measurement have been corrected for initial state radiation (ISR). Measurements reported in terms of $\langle x_E \rangle$ have been corrected to $\langle x_p \rangle$. In the case of ARGUS the average has been calculated from the full distribution. In the case of TASSO the error on the average was re-evaluated using the full distribution since the published error seems incompatible with the data. DELCO reports a fit with a Peterson distribution that has been translated into $\langle x_p \rangle^{\rm corr}$. Systematical and statistical uncertainties, where reported separately, have been added in quadrature. The ALEPH beauty measurement refers to $B^+$ and $B^0$ mesons only (i.e. excluding $B_s$ and $\Lambda_b$), a MC study shows that this correspond to underestimating $\langle x_p \rangle^{\rm corr}$ by $\sim 0.1\%$ only, which is negligible.

| Charm ($D^*$) measurement | $E_{\rm cms}$ (GeV) | Measured variable | Value | ISR corr. (%) | $x_E \to x_p$ (%) | $\langle x_p \rangle^{\rm corr}$ |
|---|---|---|---|---|---|---|
| OPAL [22] | 92 | $\langle x_E \rangle$ | $0.516^{+0.008}_{-0.005} \pm 0.010$ | $+0.4$ | $-0.4$ | $0.516 \pm 0.012$ |
| ALEPH [23] | 92 | $\langle x_E \rangle$ | $0.4878 \pm 0.0046 \pm 0.0061$ | $+0.4$ | $-0.4$ | $0.488 \pm 0.008$ |
| DELPHI [24] | 92 | $\langle x_E \rangle$ | $0.487 \pm 0.015 \pm 0.005$ | $+0.4$ | $-0.4$ | $0.487 \pm 0.016$ |
| TASSO [25] | 36.2 | $\langle x_E \rangle$ | $0.58 \pm 0.02$ | $+6.7$ | $-1.8$ | $0.61 \pm 0.02$ |
| DELCO [26] | 29 | $\epsilon^*_{\rm Pet.}$ | $0.31^{+0.10}_{-0.08}$ | $+6.3$ | - | $0.55 \pm 0.03$ |
| ARGUS [27] | 10.5 | $\langle x_p \rangle$ | $0.64 \pm 0.03$ | $+4.2$ | - | $0.67 \pm 0.03$ |
| CLEO [28] | 10.5 | $\langle x_p \rangle$ | $0.611 \pm 0.007 \pm 0.004$ | $+4.2$ | - | $0.637 \pm 0.008$ |
| BELLE [29] | 10.58 | $\langle x_p \rangle$ | $0.61217 \pm 0.00036 \pm 0.00143$ | $+4.2$ | - | $0.6379 \pm 0.0016$ |
| **Beauty ($B^{wd}$) measurement** | **$E_{\rm cms}$ (GeV)** | **Measured variable** | **Value (%)** | **ISR corr. (%)** | **$x_E \to x_p$** | **$\langle x_p \rangle^{\rm corr}$** |
| OPAL [30] | 92 | $\langle x_E \rangle$ | $0.7193 \pm 0.0016 \pm 0.0038$ | $+0.3$ | $-0.9$ | $0.715 \pm 0.004$ |
| SLD [31] | 92 | $\langle x_E \rangle$ | $0.709 \pm 0.003 \pm 0.005$ | $+0.3$ | $-0.9$ | $0.705 \pm 0.006$ |
| ALEPH [32] | 92 | $\langle x_E \rangle$ | $0.716 \pm 0.006 \pm 0.006$ | $+0.3$ | $-0.9$ | $0.712 \pm 0.008$ |
| DELPHI [33] | 92 | $\langle x_E \rangle$ | $0.7153 \pm 0.0007 \pm 0.0050$ | $+0.3$ | $-0.9$ | $0.711 \pm 0.005$ |
| JADE [34] | 36.2 | $\langle x_E \rangle$ | $0.76 \pm 0.03 \pm 0.04$ | $+5.4$ | $-3.5$ | $0.77 \pm 0.06$ |
| DELCO [35] | 29 | $\langle x_E \rangle$ | $0.72 \pm 0.05$ | $+4.8$ | $-4.7$ | $0.72 \pm 0.05$ |
| PEP4-TPC [36] | 29 | $\langle x_E \rangle$ | $0.77 \pm 0.04 \pm 0.03$ | $+4.8$ | $-4.7$ | $0.77 \pm 0.07$ |

Only LEP data at $E_{\rm cms} = 92$ GeV were used to extract $\langle x \rangle^{\rm np}$ since the factorization of the non-perturbative FF could be spoiled by large $\mathcal{O}(m_Q/E_{\rm cms})$ terms at lower energies. Table 3 reports the LEP average $\langle x_p \rangle^{\rm corr}$, the perturbative results at 92 GeV and the resulting $\langle x \rangle^{\rm np}$ for NLL and FO calculations as well as $\langle x \rangle$ and $\langle x \rangle^{\rm pert}$ from PYTHIA. Figure 4 (left) shows $\langle x_p \rangle$ obtained by multiplying the perturbative calculations with the corresponding $\langle x \rangle^{\rm np}$, compared to the experimental data and to the PYTHIA MC with different fragmentation parameters.

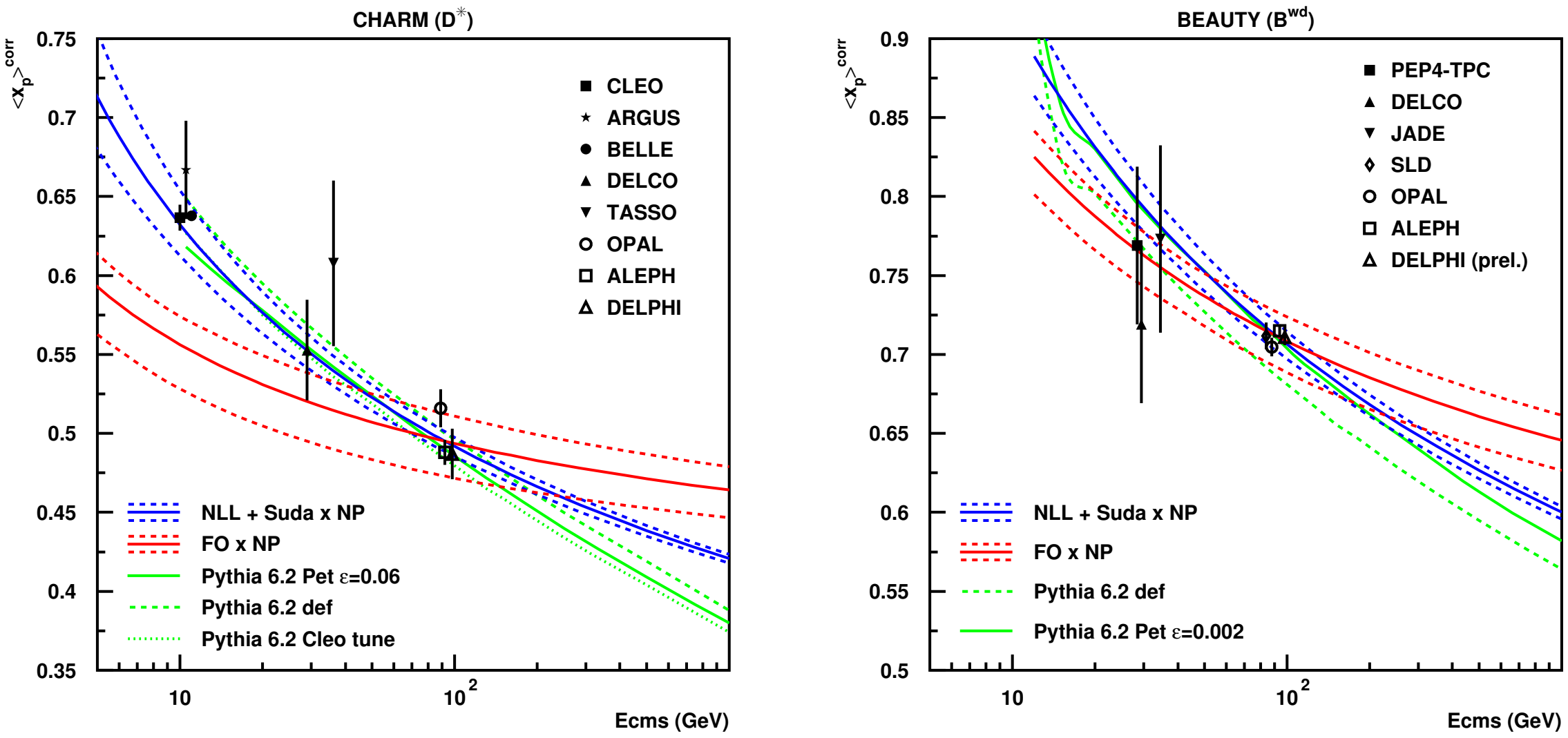

**Fig. 4:** Average fragmentation function as a function of the center of mass energy for charm (left) and beauty (right). The plots show the experimental results and the curves from NLL and FO theory with a non-perturbative fragmentation obtained using the data at the $Z^0$ energy. The curves from PYTHIA 6.2 with different fragmentation choices are also shown. The experimental points at the $\Upsilon(4s)$ and $Z^0$ resonances are shown slightly displaced in the horizontal axis for better legibility.

With the non-perturbative $\langle x\rangle^{\rm np} = 0.849 \pm 0.018$ obtained at LEP energies, the NLL calculation can reproduce all the data within a quite small theoretical uncertainty. The FO calculation is instead too flat to reproduce the data even considering its large theoretical uncertainty band. The non-perturbative fragmentation $\langle x\rangle^{\rm np}$ obtained at LEP energy for the FO calculation is quite small $(0.65 \pm 0.04)$ since it compensates the effect of the FF evolution that is missing in the perturbative part. Therefore FO calculations with $\langle x\rangle^{\rm np}$ extracted at LEP energy undershoot drastically the data at the $\Upsilon(4s)$.

The PYTHIA MC with the Lund-Bowler fragmentation reproduces the data reasonably well. The result with default parameters is slightly above the data while the result with the parameters tuned by the CLEO collaboration is slightly below. Both are compatible within the experimental uncertainty with all the experimental values with the exception of the very precise measurement from Belle from which they differ anyway by less than 2%. PYTHIA with the Peterson fragmentation with $\epsilon = 0.06$ reproduces well the LEP data but is too low at lower energies.

### 2.4 Beauty

In the case of beauty we consider fragmentation measurements for the mix of weakly decaying hadrons $B^{\rm wd}$. Precise measurements are available only at the $Z^0$ peak (SLD [31], ALEPH [32], OPAL [30], DELPHI [33]). Lower energy measurements from PEP (PEP4-TPC [36], DELCO [35]) and PETRA (JADE [34]) have larger uncertainties. As for charm, corrections have been applied for ISR and to convert $\langle x_E\rangle$ to $\langle x_p\rangle$. The data are shown in Table 2 and the results in Table 3 and Figure 4 (right). Since precise data are available only at a single energy, it is impossible to test the energy beaviour of the theoretical predictions. As in the charm case, the energy dependence of PYTHIA and NLL theory are similar, while the FO prediction is much more flat, suggesting that also for beauty the non-perturbative fragmentation obtained for FO at the $Z^0$ could not be applied at lower energy. PYTHIA with Peterson fragmentation with $\epsilon = 0.002$ reproduces the data, while the default Lund-Bowler fragmentation is too soft.

**Table 3:** Average fragmentation functions at the $Z^0$ resonance for charm (top) and beauty. The table shows the average of the experimental data, the results from the NLL and FO calculations and from the PYTHIA MC with different fragmentation parameters. For the NLL and FO calculations $\langle x_p \rangle^{\rm np}$ is obtained by dividing the average from the experimental data by the perturbative result $\langle x_p \rangle^{\rm np} = \langle x_p \rangle^{\rm corr} / \langle x_p \rangle^{\rm pert}$.

| Charm ($D^*$) @ 92 GeV | $\langle x_p \rangle^{\rm corr}$ | $\langle x_p \rangle^{\rm pert}$ | $\langle x_p \rangle^{\rm np}$ |
|---|---|---|---|
| Data | $0.495 \pm 0.006$ | – | – |
| NLL | – | $0.583 \pm 0.007$ | $0.849 \pm 0.018$ |
| FO | – | $0.76 \pm 0.03$ | $0.65 \pm 0.04$ |
| PYTHIA Def. | 0.500 | 0.640 | – |
| PYTHIA CLEO | 0.484 | 0.640 | – |
| PYTHIA Pet. $\epsilon = 0.06$ | 0.490 | 0.640 | – |
| **Beauty ($B^{wd}$) @ 92 GeV** | $\langle x_p \rangle^{\rm corr}$ | $\langle x_p \rangle^{\rm pert}$ | $\langle x_p \rangle^{\rm np}$ |
| Data | $0.7114 \pm 0.0026$ | – | – |
| NLL | – | $0.768 \pm 0.010$ | $0.927 \pm 0.013$ |
| FO | – | $0.83 \pm 0.02$ | $0.85 \pm 0.02$ |
| PYTHIA Def. | 0.686 | 0.773 | – |
| PYTHIA Pet. $\epsilon = 0.002$ | 0.710 | 0.773 | – |

### 2.5 Effect on predictions for heavy quark production at HERA and LHC

Going back to the heavy-hadron production in $ep$ and $pp$ collisions, Eq. 9 shows that the uncertainty on the differential heavy-hadron cross section $d\sigma/dp_T$ is related to the uncertainty on the average non-perturbative fragmentation by

$$\Delta(d\sigma/dp_T) = N\Delta(\langle x \rangle^{\rm np}),$$

where $-N$ is the exponent of the differential cross section.

The state of the art calculations for photo- and hadro-production (FONLL [41,42]) include NLO matrix elements and the resummations of next-to-leading logarithms. The appropriate non-perturbative fragmentation for FONLL is therefore obtained with the NLL theory which has the same kind of perturbative accuracy [43]. Since the NLL calculation gives a good description of $e^+e^-$ data, it seems appropriate to use the value and the uncertainty of $\langle x \rangle^{\rm np}$ as obtained from $e^+e^-$ data at the $Z^0$ peak. The relative error for the $D^*$ fragmentation is $\Delta\langle x \rangle^{\rm np}/\langle x \rangle^{\rm np} = 2\%$ which translates into an uncertainty of 9% on charm production at large $p_T$ at LHC ($N = 4.5$) of 9%. For beauty, the relative uncertainty $\Delta\langle x \rangle^{\rm np}/\langle x \rangle^{\rm np} = 1.4\%$ translates into an uncertainty on large-$p_T$ $B$-hadron production at LHC ($N = 3.8$) of 5.3%. These uncertainty are smaller or of the order of the perturbative uncertainties of the calculation. Nevertheless, it should be noted that this approach is only valid for large transverse momenta. At small transverse momenta the factorization ansatz breaks down and large corrections of order $m_Q/p_T$ may appear. Therefore, for the low-$p_T$ region, the uncertainty on the $p_T$ distribution is large and difficult to evaluate.

For processes such as DIS and for particular observables FONLL calculations are not available. The best theory available in this case is the fixed order NLO theory. In this case the situation is complex since the equivalent FO calculation for $e^+e^-$ does not reproduce the experimental data. The proposed solution is to vary $\langle x \rangle^{\rm np}$ from the same value obtained in the NLL case (that would be correct at low $p_T$, where the FF evolution is irrelevant) to the value obtained at the $Z^0$ energy (that would be valid at $p_T \sim 100$ GeV). Therefore we consider for charm $\langle x \rangle^{\rm np} = 0.075 \pm 0.010$ and for beauty $\langle x \rangle^{\rm np} = 0.089 \pm 0.004$. When these values are transported to heavy-hadron production at LHC, the corresponding uncertainties on $d\sigma/dp_T$ at large $p_T$ are 60% for charm and 20% for beauty. Therefore the NLO fixed order calculations cannot be used for precise predictions of the charm (and to a lesser extent beauty) production at $pp$ and $ep$ colliders.

**Table 4:** Proposed value and uncertainty on $\langle x\rangle^{\rm np}$ to be used with FO-NLO and FONLL programs for photo- and hadro-production of $D^*$ mesons and weakly decaying $B$ hadrons. The corresponding value and range for the Peterson $\epsilon$ and for the Kartvelishvili $\alpha$ parameters are also reported. The last columns show the corresponding relative uncertainty on $d\sigma/dp_T$ at LHC (assuming a negative power $N = 4.5/3.8$ for charm/beauty) and HERA ($N = 5.5/5.0$ for c/b).

| | $\langle x^{\rm np}\rangle$ | $\epsilon$(min : max) | $\alpha$(min : max) | $\Delta\langle x^{\rm np}\rangle/\langle x^{\rm np}\rangle$ | $\Delta\sigma/\sigma$ (LHC) | $\Delta\sigma/\sigma$ (HERA) |
|---|---|---|---|---|---|---|
| FONLL $D^*$ | $0.849 \pm 0.018$ | 0.0040(0.0027 : 0.0057) | 10(9 : 12) | 2.1% | 9% | 12% |
| FONLL $B^{wd}$ | $0.927 \pm 0.013$ | 0.00045(0.00026 : 0.00072) | 24(20 : 30) | 1.4% | 5% | 7% |
| FO-NLO $D^*$ | $0.75 \pm 0.10$ | 0.02(0.004 : 0.08) | 5(3 : 10) | 13% | 60% | 70% |
| FO-NLO $B^{wd}$ | $0.89 \pm 0.04$ | 0.0015(0.0004 : 0.004) | 15(10 : 25) | 4.5% | 20% | 22% |

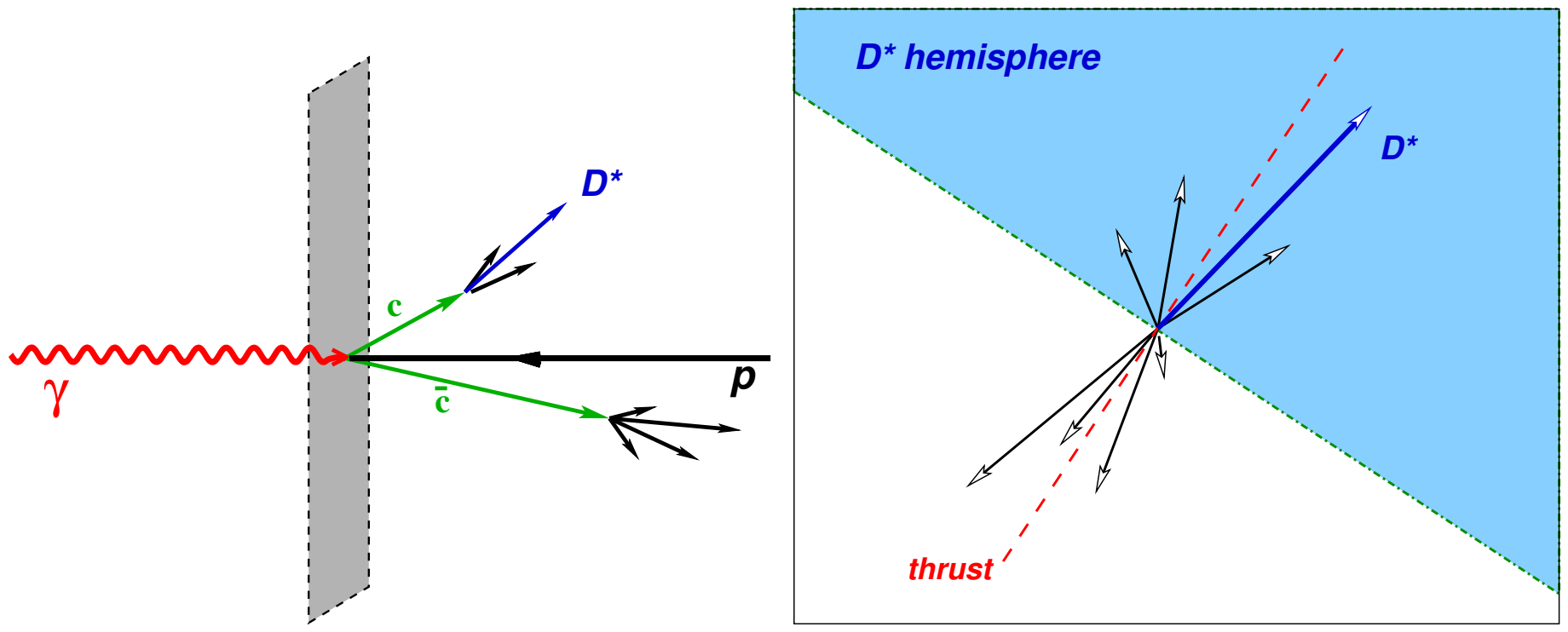

**Fig. 5:** Hemisphere method

In the FO-NLO and FONLL programs the hadron distributions are obtained by reducing the quark momenta according to a given fragmentation functions. Typical fragmentation functions used in these programs are the Peterson and Kartvelishvili forms. Table 4 summarises the proposed values and uncertainties for $\langle x\rangle^{\rm np}$ to be used with FO-NLO and FONLL calculations and reports the corresponding values and ranges for the Peterson and Kartvelishvili parameters. Similar ranges are used in the calculations presented in the section on "Benchmark cross sections" in these proceedings.

## 3 Measurements of the charm quark fragmentation function at HERA[5]

The differential cross section for the inclusive production of a heavy hadron $H$ from a heavy quark $h$ can be computed in perturbative QCD (pQCD) as a convolution of a short-distance cross section $\hat{\sigma}(p_h)$ with a fragmentation function $D_H^h(z)$:

$$d\sigma(p_H) = \int dz dp_h d\hat{\sigma}(p_h) D_H^h(z)\delta(p - zp_h) \tag{11}$$

The quantity $z$ is the fractional momentum of the heavy quark $h$ which is transferred to the heavy hadron $H$, and the normalized fragmentation function $D_H^h(z)$ gives the probability to observe the hadron $H$ with a momentum fraction $z$.

The precise definition of $D_{\rm H}^{(h)}(z)$ is in some sense arbitrary. Due to the short and long-distance processes involved, the fragmentation function contains a perturbative and a non-perturbative component.

[5]Authors: J. Bracinı́k and G. Grindhammer

Since the former can be calculated only up to some order in the strong coupling, the non-perturbative component in practice will have to absorb some of the missing higher order corrections. The calculable perturbative part can be absorbed into the definition of $\hat{\sigma}(p_h)$. Since for heavy quarks perturbative gluon emission do not lead to collinear divergencies, the perturbative evolution is well defined, and it is possible to absorb them into $\hat{\sigma}(p_h)$ and to perform perturbative evolution down to a scale of the heavy quark mass $m_h$. In this case the non-perturbative fragmentation function $D_H^h(z)$ accounts for the transition of an almost on-shell quark $h$ into a heavy hadron $H$.

According to the QCD factorization theorem, the non-perturbative fragmentation functions (FF) depend neither on the type of the hard process nor on the scale at which the heavy quark $h$ is originally produced. This implies universality of FF and allows - if valid - to extract fragmentation functions from data for one particular reaction (usually $e^+e^-$ annihilation) and to use them to predict cross sections for other reactions (e.g. in $pp$ and $ep$-collisions). In order to be able to check the reliability of pQCD predictions, it is necessary to check the universality of FF.

In practice, different theoretically motivated functional forms for $D_H^h(z)$ are used, depending on one more free parameters which are fitted to data. Among frequently used expressions are those by Peterson et al. [14] and by Kartvelishvili et al. [12].

From Equation 11 it is clear that $D_H^h(z)$ cannot be measured directly, since all observables are convoluted with the perturbative cross section. In case of $ep$ and $pp$ scattering there are additional convolutions with the parton density functions of one or two interacting hadrons. However, there are some observables which are more sensitive to $D_H^h(z)$ then others.

In $e^+e^-$, a convenient way to study fragmentation is to study the differential cross section of a heavy meson as a function of a scaled momentum or energy z. A customary experimental definition[6] of z is $\mathrm{z} = E_H/E_{\mathrm{beam}}$, where $E_{\mathrm{beam}}$ is the energy of the beams in the center-of-mass system. In leading order, i.e. without gluon emissions, it is also the energy of the charm and anti–charm quark and is equal to $D_H^h(z)$. In contrast to $e^+e^-$ annihilation the choice of a fragmentation observable in $ep$ collisions is more difficult. Two different observables have been used so far, both of them having the feature that in leading order QCD, the z-distributions are equal to $D_H^h(z)$.

In the case of what is called here the jet method, the energy of the charm quark is approximated by the energy of the charm-jet, tagged by a $D^*$-meson, which is considered to be part of the jet. The scaling variable is then defined as $\mathrm{z}_{\mathrm{jet}} = (E + p_{\mathrm{L}})_{D^*}/(E + p)_{\mathrm{jet}}$.

The idea of the so called hemisphere method (see Figure 5) is to exploit the special kinematics of charm events in $ep$ collisions. The dominant charm production process has been shown to be boson-gluon fusion. If such an event is viewed in the photon-proton center-of-mass frame, the photon puts its full energy into the hard subprocess, while the proton interacts via a gluon, which typically carries only a small fraction of the proton momentum. As the result, both quarks produced, $c$ and $\bar{c}$, move in the direction of the photon. Assuming no initial gluon $k_{\mathrm{T}}$ and no gluon radiation, their transverse momenta are balanced (see Fig. 5, left).

This can be seen best by projecting the quark momenta onto the plane perpendicular to the $\gamma$-p axis. In this plane it is possible to distinguish rather efficiently between the products of the fragmentation of the charm quark and its antiquark. The momenta of all particles are projected onto the plane and the thrust axis in this plane is found (see Fig. 5, right). The plane is then divided into two hemispheres by the line perpendicular to the thrust axis. All particles, lying in the hemisphere containing the $D^*$-meson are marked and their three-momenta and energy are summed-up to give the hemisphere's momentum and energy, which is used to approximate the momentum and energy of the respective charm/anti-charm quark. The scaling variable $\mathrm{z}_{\mathrm{hem}}$ is then defined as $\mathrm{z}_{\mathrm{hem}} = (E + p_{\mathrm{L}})_{D^*}/\sum_{\mathrm{hem}}(E + p)$.

The ZEUS collaboration has provided preliminary results [44] on a measurement of normalized differential cross sections of $D^*$-mesons as a function of $\mathrm{z}_{\mathrm{jet}}$. The measurement was done in photopro-duction, in the kinematic range $Q^2 < 1$ GeV$^2$ and $130 < W < 280$ GeV. The $D^*$-mesons were reconstructed using the 'golden channel' $D^* \rightarrow D^0\pi_s \rightarrow K\pi\pi_s$ and were required to be in the central rapidity region $|\eta| < 1.5$ and to have $p_{\rm T} > 2$ GeV. Jets were reconstructed using the inclusive $k_\perp$ algorithm. They fulfill the conditions $|\eta_{\rm jet}| < 2.4$ and $E_{\rm T,jet} > 9$ GeV. The jets were reconstructed as massless jets. The beauty contribution to the $D^*$-meson cross section, which amounts to about 9%, was subtracted using the prediction of PYTHIA. The scaling variable was calculated as $z_{\rm jet} = (E + p_{\rm L})_{D^*}/(2E_{\rm jet})$. The cross section as a function of $z_{\rm jet}$ is shown in Fig. 6. The uncertainties due to choice of the model used to correct for detector effects, and the subtraction of the beauty component were the largest contributions to the total uncertainty.

[6] Sometime there are slightly different definitions of z [28] in case of heavy meson production close to threshold.

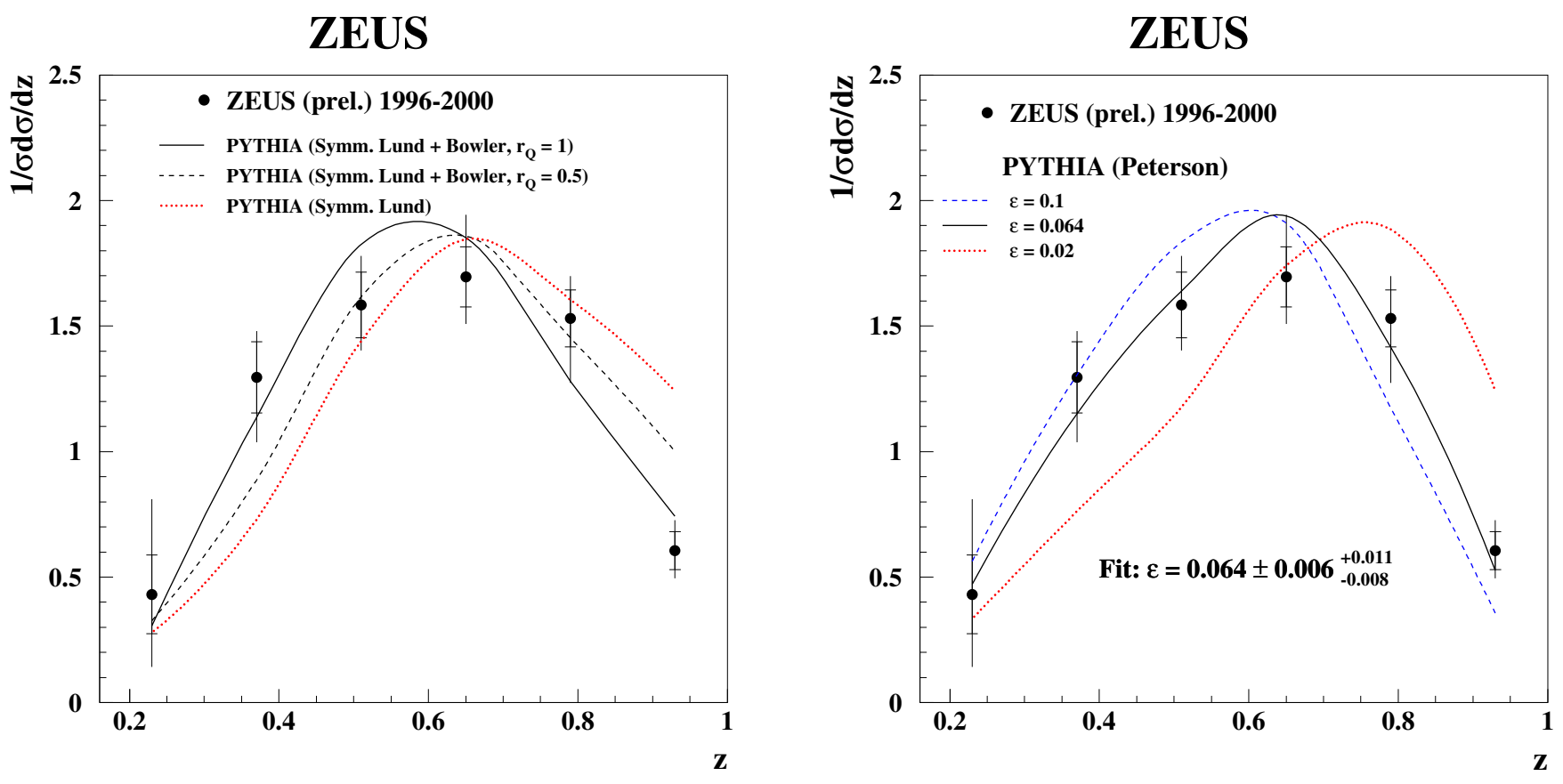

**Fig. 6:** Normalized differential cross section as a function of $z_{\rm jet}$ as measured by ZEUS in photoproduction for jets with an associated $D^*$-meson with $|\eta_{\rm jet}| < 2.4$ and $E_{\rm T,jet} > 9$ GeV.

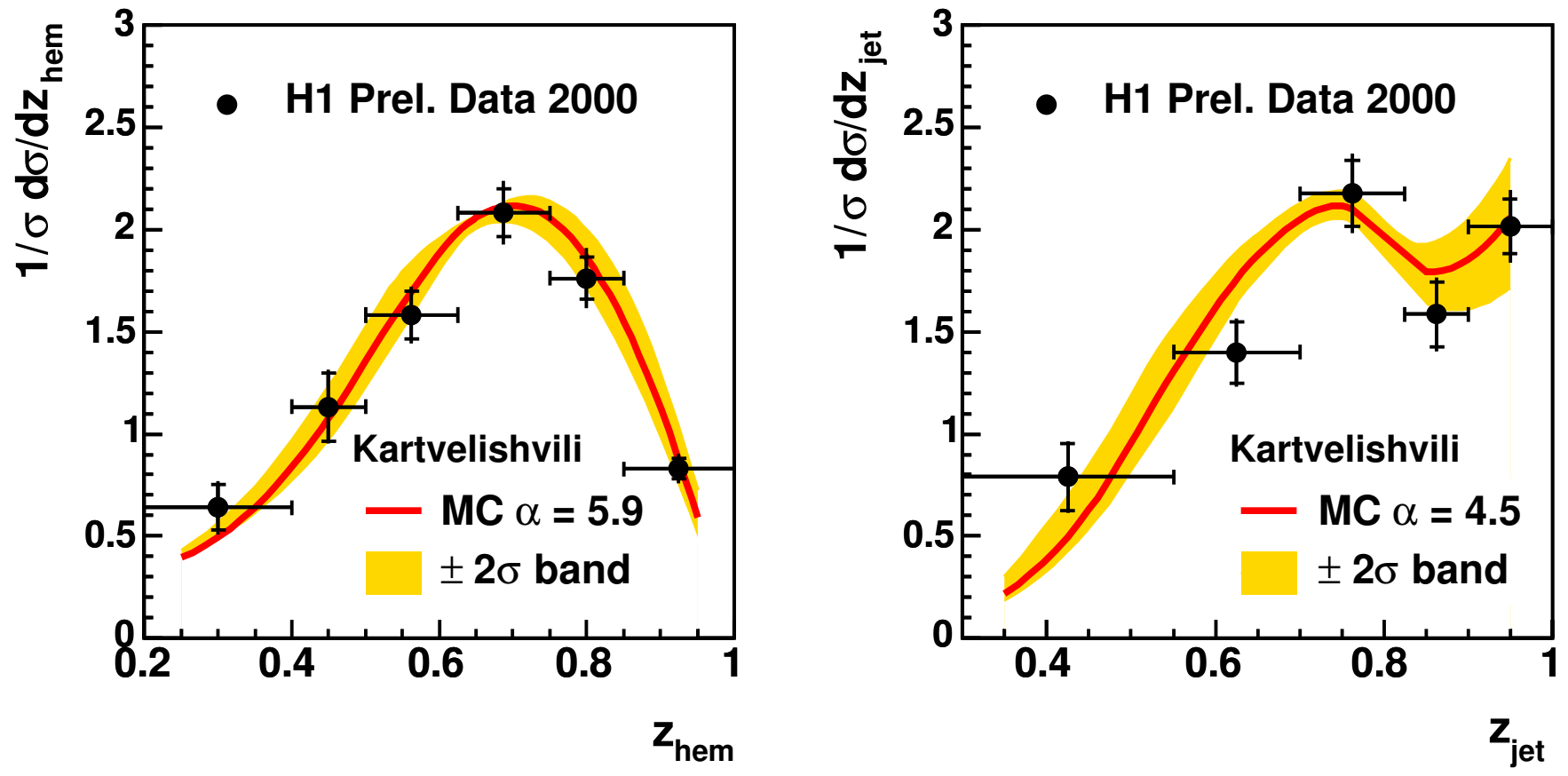

**Fig. 7:** Normalized differential cross section of $D^*$-meson as a function of $z_{\rm jet}$ and $z_{\rm hem}$ in DIS as measured by H1.

The H1 collaboration has recently presented preliminary results [45] on the normalized differential cross section also of $D^*$-mesons as a function of both $z_{\rm hem}$ and $z_{\rm jet}$. Their measurement was performed in the kinematic range $2 < Q^2 < 100$ GeV$^2$ and $0.05 < y < 0.7$. The $D^*$-mesons were reconstructed using the 'golden channel' with $|\eta| < 1.5$ in the central rapidity region and $p_{\rm T} > 1.5$ GeV. The jets were reconstructed using the inclusive $k_\perp$ algorithm in the photon-proton center of mass frame, using the massive recombination scheme. The jets were required to have $E_{\mathrm{T,jet}} > 3$ GeV. The scaling variables were calculated as $\mathrm{z_{jet}} = (E + p_{\mathrm{L}})_{D^*}/(E + p)_{\mathrm{jet}}$ and $\mathrm{z_{hem}} = (E + p_{\mathrm{L}})_{D^*}/\sum_{\mathrm{hem}}(E + p)$ and are shown in Fig. 7. The resolved contribution was varied between 10 and 50% and the beauty contribution as predicted by the model was varied by a factor of two. The resulting uncertainties are part of the systematic error of the data points. For these distributions, the contribution of $D^*$-mesons coming from the fragmentation of beauty, as predicted by RAPGAP, was subtracted. It amounts to about $1.3\%$ for the hemisphere method and $1.8\%$ for the jet method. The dominant systematic errors are due to the model uncertainty and the signal extraction procedure.

Both collaborations used the normalized z-distributions to extract the best fragmentation parameters for a given QCD model.

In case of ZEUS, PYTHIA was used together with the Peterson fragmentation function. The MC was fit to the data using a $\chi^2$-minimization procedure to determine the best value of $\epsilon$. The result of the fit is $\epsilon = 0.064 \pm 0.006^{+0.011}_{-0.008}$.

The H1 collaboration used RAPGAP 3.1 interfaced with PYTHIA 6.2. The contribution due to $D^*$-mesons produced in resolved photon processes (in DIS), which amounts to $33\%$ as predicted by the model, has been included in addition to the dominant direct photon contribution. The Peterson and Kartvelishvili parametrizations were both fitted to the data. The results are shown in Table 5.

**Table 5:** Extracted fragmentation parameters for $\mathrm{z_{jet}}$ and $\mathrm{z_{hem}}$ from H1.

| Parametrization | | Hemisphere Method | Jet Method | Suggested range |
|---|---|---|---|---|
| Peterson | $\varepsilon$ | $0.018^{+0.004}_{-0.004}$ | $0.030^{+0.006}_{-0.005}$ | $0.014 < \varepsilon < 0.036$ |
| Kartvelishvili | $\alpha$ | $5.9^{+0.9}_{-0.6}$ | $4.5^{+0.5}_{-0.5}$ | $4 < \alpha < 6.8$ |

The parameter of the Peterson fragmentation function as measured by ZEUS and H1 do not agree with each other. This may be due to the different phase-space regions covered by the two measurements (photoproduction versus DIS, $E_{\mathrm{T,jet}} > 9$ GeV versus $E_{\mathrm{T,jet}} > 3$ GeV ) and most importantly, the parameters were extracted for two different models[7]. More detailed investigations are needed to resolve this question.

The fragmentation function parameters extracted by H1 with the hemisphere and the jet method differ by less than 3 $\sigma$. At the present level of statistical and systematic errors it is not possible to exclude a statistical fluctuation. On the other hand, the potential discrepancy may be a sign of deficiencies in the modelling of the hadronic final state in RAPGAP.

The measured $\mathrm{z_{hem}}$ distribution of H1 is compared to data from the ALEPH [23], OPAL [22] and CLEO [28] collaborations in Fig. 8 (left) and to ZEUS [44] and Belle [29] in Figure 8 (right)[8]. The results of H1 are in rough agreement with recent data from CLEO and Belle, taken at at 10.5 and 10.6 GeV, corresponding roughly to the average energy of the system at H1. Differences beyond the measurement errors can be observed. However, this may be due to the somewhat different definitions used for the fragmentation observable z, different kinematics, different processes, or it may be a sign of the violation of universality.

While the z distributions don't need to agree, the fragmentation parameters, which are extracted from them, should agree. This can be expected only, if a model with consistent parameter settings is used which provides an equally good description of the different processes at their respective scales.

[7] While ZEUS has used the default parameters for PYTHIA, H1 has taken the tuned parameter values of the ALEPH collaboration [46]

[8] Data points were taken from the figure in [29] and [44].

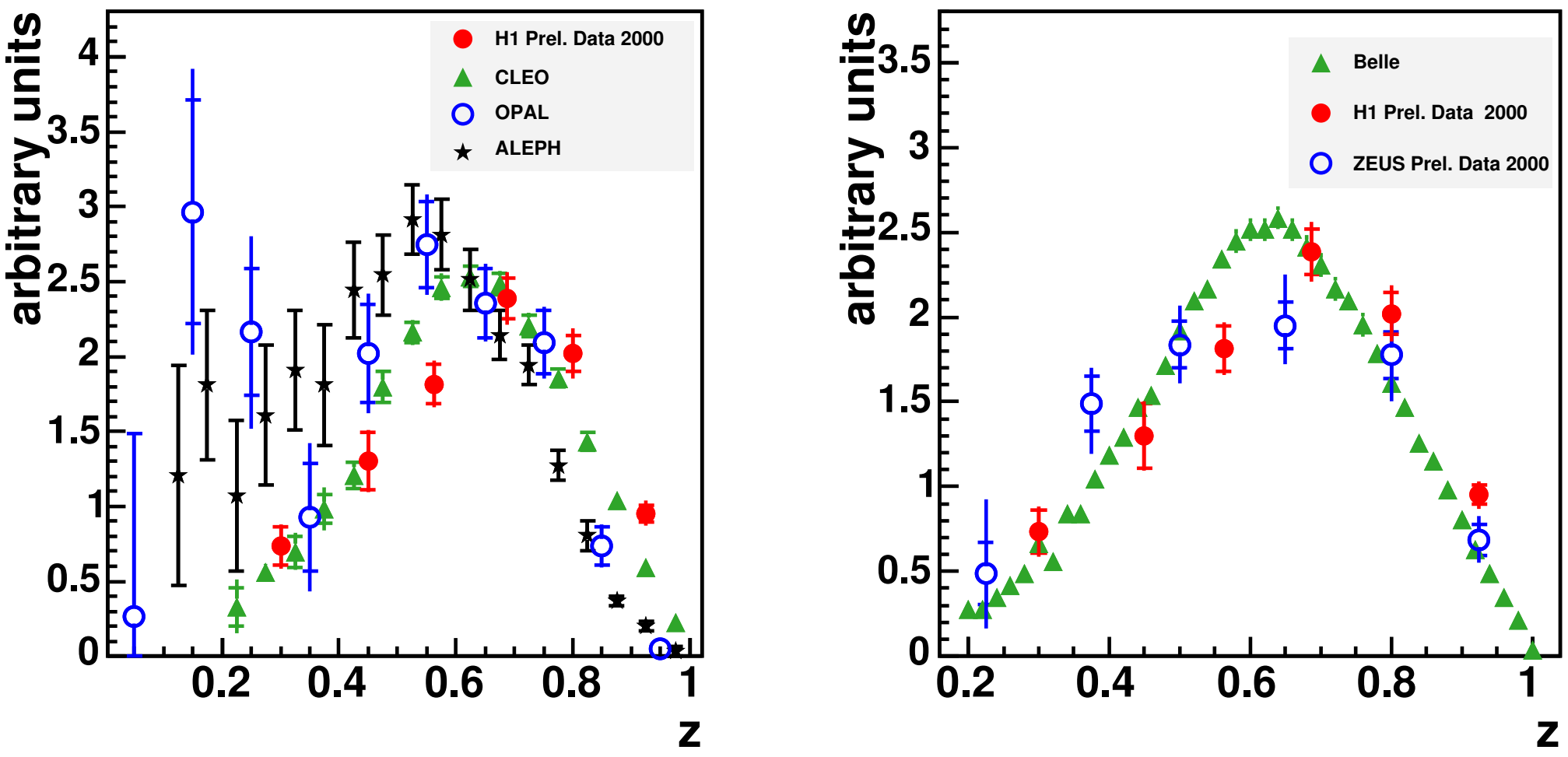

**Fig. 8:** Comparison of the z-distributions from CLEO, OPAL and ALEPH (left) and Belle and ZEUS (right) with the one from the hemisphere method from H1. All distributions are normalized to unit area from z = 0.4 to z = 1.

The values of the Peterson fragmentation parameter, as extracted by different experiments within the PYTHIA/JETSET models, are summarized in Table 6.

**Table 6:** Extracted fragmentation parameters from $e^+e^-$ annihilation data by ALEPH [23], OPAL [22] and BELLE [29] and from $ep$ data by ZEUS [44] and H1 [45].

| PARAMETRIZATION | | ALEPH | OPAL | BELLE | ZEUS | H1: $z_{\rm hem}$ | H1: $z_{\rm jet}$ |
|---|---|---|---|---|---|---|---|
| Peterson | $\varepsilon$ | $0.034 \pm 0.0037$ | $0.034 \pm 0.009$ | 0.054 | $0.064^{+0.013}_{-0.010}$ | $0.018^{+0.004}_{-0.004}$ | $0.030^{+0.006}_{-0.005}$ |
| Kartvelishvili | $\alpha$ | — | $4.2 \pm 0.6$ | 5.6 | — | $5.9^{+0.9}_{-0.6}$ | $4.5^{+0.5}_{-0.5}$ |

Contrary to expectations, discrepancies between various experiments can be seen. A consistent phenomenological analysis of these data is therefore needed in order to resolve the reasons for the discrepancies.

The measurement of the charm fragmentation function at HERA provides an important test of our understanding of heavy quark production. We may hope that HERA II data and a phenomenological analysis of existing data will bring new insights in this area.

# Benchmark cross sections for heavy-flavour production

*O. Behnke*[1], *M. Cacciari*[2], *M. Corradi*[3], *A. Dainese*[4], *H. Jung*[5], *E. Laenen*[6], *I. Schienbein*[7], *H. Spiesberger*[8]
[1]Universität Heidelberg, Philosophenweg 12 69120 Heidelberg, FRG;
[2]LPTHE - Université P. et M. Curie (Paris 6), Paris, France;
[3]INFN Bologna, via Irnerio 46, Bologna, Italy;
[4]Università di Padova and INFN, Padova, Italy;
[5]Deutsches Elektronen-Synchroton DESY, Hamburg, FRG;
[6]NIKHEF,Theory Group, Kruislaan 409, 1098 SJ Amsterdam, The Netherlands;
[7]Southern Methodist University Dallas, 3215 Daniel Avenue, Dallas, TX 75275-0175, USA;
[8]Johannes-Gutenberg-Universität Mainz, D-55099 Mainz, FRG

**Abstract**
Reference heavy-flavour cross sections at HERA and LHC have been computed following different theoretical approaches and the results have been compared.

Coordinators: *M. Corradi, A. Dainese*

## 1 Introduction

This section presents a comparison of cross sections for HERA and LHC calculated according to different theoretical approaches. Different programs were used to calculate the same reference cross sections, using, as far as possible, the same input parameters and a consistent method to evaluate uncertainties. In this way it is possible to identify processes and kinematical regions in which different approaches give the same answer and regions where they differ. Unified criteria to evaluate the theoretical uncertainty should also allow to understand what approach is expected to be more precise. Moreover these calculations, which incorporate up-to-date parameters and PDF parametrisations, can be used as a reference for experiments and for further theoretical predictions. The cross sections presented here, are available in computer-readable format from the web page `http://www-zeus.desy.de/~corradi/benchmarks`, where figures in color can also be found.

## 2 Programs

A list of the programs used for the cross section calculations is given below. For further details see the references and the theoretical review on heavy quark production in these proceedings.

- MNR [1] is a fixed-order (FO) NLO program for heavy-flavour hadro-production, it was used for LHC cross sections;
- FMNR [2, 3] is an extension of the previous program to photoproduction, it was used for photoproduction at HERA;
- HVQDIS [4, 5] is a FO-NLO program for heavy-flavour production in deep-inelastic scattering (DIS), it has been used for DIS at HERA;
- FONLL [6, 7] provides matched massive-massless calculations with NLO accuracy and resummation of large $p_T$ logarithms. It is available for hadro- and photo-production and was used for HERA photoproduction and LHC cross sections;
- GM-VFNS [8–11] is a calculation in the generalised massive variable flavour number scheme. It has been used for charmed hadron $p_T$ spectra at LHC and in photoproduction at HERA;

**Table 1:** The table shows input parameter to the different programs with the corresponding lower and upper values used for the uncertainty: $\Lambda_{\rm QCD}$, the quark masses, the proton and photon parton densities, the fraction of c quarks decaying into a $D^*$ meson, and the parameters used for fragmentation. The fragmentation form are abbreviated to Pet. for Peterson, Kart. for Kartvelishvili, Def. for the default PYTHIA fragmentation

| Parameter | program | central value | lower/upper |
|---|---|---|---|
| $\Lambda^5_{\rm QCD}$ | all | 0.226 GeV | fix |
| $m_c$ | all | 1.5 GeV | 1.3/1.7 GeV |
| $m_b$ | all | 4.75 GeV | 4.5/5.0 GeV |
| p-PDF | all-CASCADE | CTEQ6.1 [15] | MRST2002 [16]/Alekhin [17] |
| | CASCADE | CCFM A0 | – |
| $\gamma$-PDF | FMNR, FONLL | AGF [18] | GRV [19] |
| $f(c \to D^*)$ | all | 0.235 | fix |
| $c$ fragmentation: | (F)MNR,HVQDIS | Pet. [20] $\epsilon_c = 0.021$ | 0.002/0.11 |
| | FONLL | BCFY $r = 0.1$ | $0.06/0.135$ |
| | GM-VFNS | [9] | - |
| | CASCADE, RAPGAP | Pet. $\epsilon_c = 0.075$ | Def./$\epsilon_c = 0.05$ |
| $b$ fragmentation: | (F)MNR,HVQDIS | Pet. $\epsilon_b = 0.001$ | 0.0002/0.004 |
| | FONLL | Kart. $\alpha = 29.1$ | $25.6/34.0$ |
| | CASCADE, RAPGAP | Pet. $\epsilon_b = 0.002$ | Def./$\epsilon_b = 0.005$ |

- CASCADE 1.2009 [12] is a full Monte Carlo program based on unintegrated parton densities and $K_T$ factorisation. It has been used to calculate cross sections for Photoproduction and DIS at HERA and for LHC;
- RAPGAP 3 [13] is a multi-purpose MC program for $ep$ collisions, it implements heavy-flavour production through the boson-gluon-fusion process $\gamma^* g \to Q\bar{Q}$ at leading order. It has been used for DIS at HERA. Both CASCADE and RAPGAP use PYTHIA [14] routines for fragmentation.

## 3 Parameters and uncertainties

The different calculations were compared using the same input parameters and, where possible, with total uncertainty bands computed in a consistent way. The total uncertainty band includes the effect of the uncertainty on the input parameters and on the missing higher orders in the perturbative expansion.

### 3.1 Perturbative uncertainty

The perturbative uncertainty was obtained by varying the renormalisation and factorisation scales independently in the range $0.5\mu_0 < \mu_F, \mu_R < 2\mu_0$, while keeping $1/2 < \mu_R/\mu_F < 2$, were $\mu_0$ is the nominal value, typically set to the transverse mass $p_T^2 + m_Q^2$ or to $4m^2 + Q^2$ in the DIS case. The largest positive and negative variations were taken as the perturbative uncertainty band.

### 3.2 Input parameters

The uncertainty from the input parameters was obtained by varying each parameter the central value. An effort was made within the working group to find the best central value and uncertainty for the input parameters. The values used for the perturbative parameters $\Lambda^5_{\rm QCD}$, $m_c$, $m_b$ as well as the parton distribution functions (PDF) for the proton and for the photon are reported in Table 1.

For practical reasons, rather than using the full treatment of the PDF uncertainty, few different parametrisations were tried and it was checked that the choice of the PDF set always gives a small contribution to the total uncertainty band. In the case of CASCADE, the CCFM A0 parametrisation was used as the central value while the PDF parametrisations A0+ and A0-, obtained from fits to DIS data with different renormalisation scales, were used in conjunction with the variation of the renormalisation scale.

Since the different programs have different perturbative contents, different parameters for the non-perturbative fragmentation function were used. The values were chosen in order to correspond to the same average fragmentation in $e^+e^-$ collisions as explained in the section on heavy quark fragmentation in these proceedings. Table 1 reports the fragmentation form and the corresponding parameters used in the different programs.

In the FONLL calculation for charm, the BCFY [21] fragmentation parameter $r$ was varied in conjunction with the variation of the charm mass since different values of $r$ are obtained from $e^+e^-$ data for different $m_c$ [22]. Similarly for beauty, the Kartvelishvili [23] parameter $\alpha$ was varied in conjunction with the variation of the $b$ mass [23]. For GM-VFNS, the fragmentation functions and fractions were taken from [9].

The total uncertainty band was obtained from the sum of the uncertainties added in quadrature coming from the parameter variations and the perturbative uncertainty.

## 4 Results

### 4.1 HERA Photoproduction

The results for HERA Photoproduction are given as $ep$ cross-sections for $0.2 < y < 0.8$ ($y$ is the Bjorken variable while $Y$ is the rapidity in the laboratory frame) and $Q^2 < 1$ GeV$^2$. The beam energies have been set to $E_e = 27.52$ GeV, $E_p = 920$ GeV with the proton beam going in the positive rapidity direction.

Figure 1 shows the differential cross sections as a function of the charm quark transverse momentum (a) and pseudorapidity (b). In (c) and (d) the same cross sections are given for the charmed $D^*$ meson. A meaningful comparison can be performed only for the hadron variables, which are the real physical observables, since the quark level may be defined differently in different approaches. The FO calculation (FMNR) shows a large uncertainty ($\sim 60\%$) at the hadron level due to the related uncertainty on the fragmentation parameters. The resummed programs FONLL and GM-VFNS have much smaller uncertainty and are within the FMNR uncertainty band. The central values from FMNR and FONLL coincide at low transverse momenta. GM-VFNS, instead, tends to grow unphysically at low $p_T(D^*)$. As can be seen in (c), the quark-level disagreement between FO (FMNR) and FONLL calculations is consistently removed at the hadron-level. The unintegrated-PDF Monte Carlo CASCADE tends to be above the other calculations, in particular at large $p_T$. In the case of beauty (Fig. 2) the uncertainty bands are smaller ($\sim 20\%$ for FMNR), CASCADE and FMNR are in good agreement. Due to the large $b$ mass, the resummed calculation FONLL (not shown) is expected to be similar to the fixed-order one (FMNR). For both beauty and charm, FMNR and FONLL show a shoulder at positive rapidities (b, d) due to the "hadron-like" component of the photon that is not present in CASCADE.

Figure 3 shows the different components of the FMNR uncertainty band for charm and beauty. The uncertainties for quark production are typically dominated by the perturbative scale uncertainty with the exception of the low transverse momentum region ($p_T \sim m_Q$) where the uncertainty from the quark-mass can dominate. For hadron production, the fragmentation dominates the FMNR uncertainty at large $p_T$. The PDF uncertainty was found to be small. Resummed calculation have smaller uncertainty bands due to the smaller perturbative and fragmentation contributions at large $p_T$.

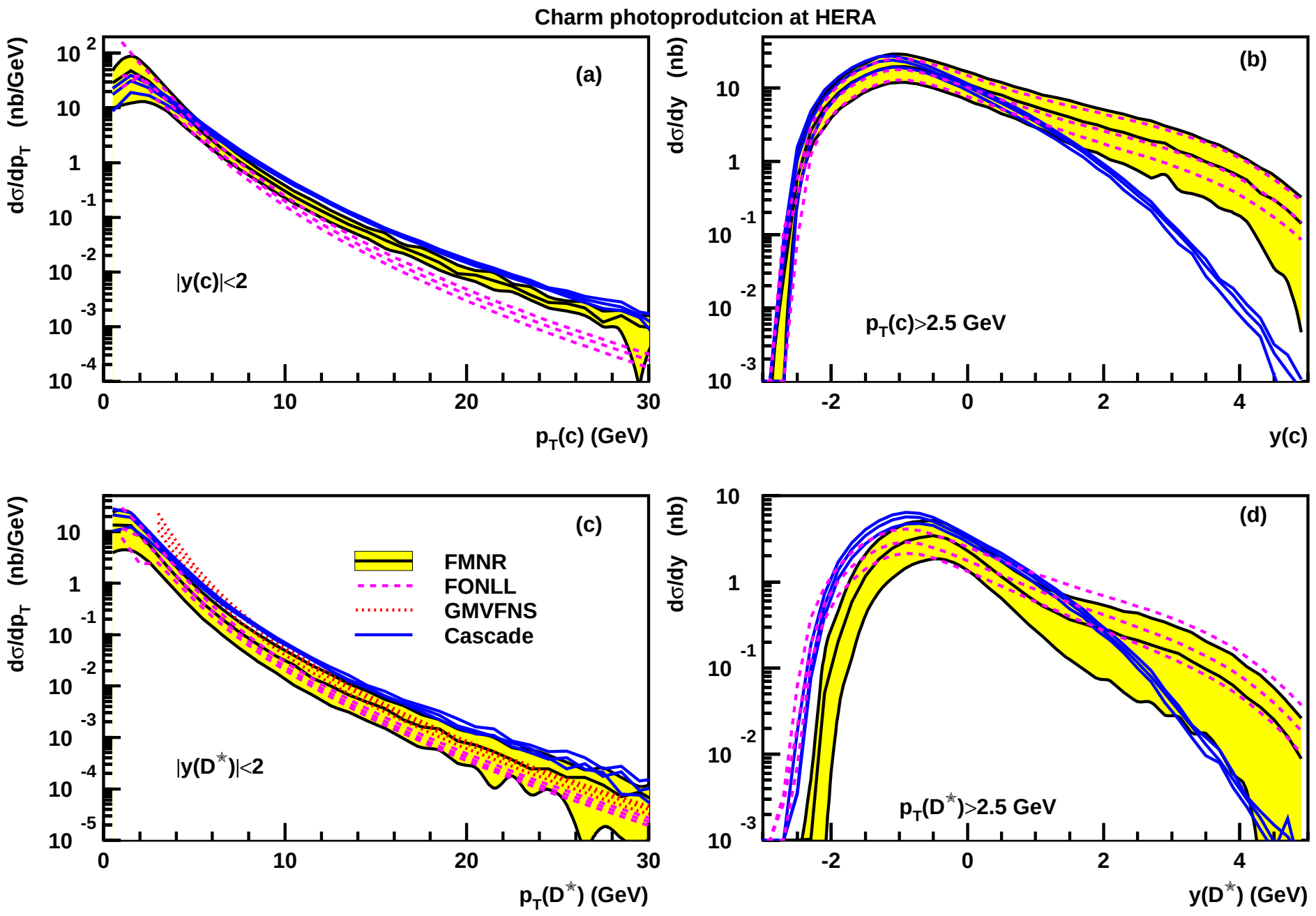

**Fig. 1:** Cross sections for charm photoproduction at HERA ($Q^2 < 1$ GeV$^2$, $0.2 < y < 0.8$). The differential cross sections as a function of the $p_T$ of the $c$ quark for rapidity $|Y| < 2$ and as a function of the rapidity of the $c$ quark for $p_T > 2.5$ GeV are shown in (a), (b). Plots (c) and (d) show similar cross sections for the production of a $D^*$ meson. The cross sections are shown for FMNR (shaded band), FONLL (empty band with dashed lines), GM-VFNS (empty band with dotted lines) and CASCADE (empty band with full lines).

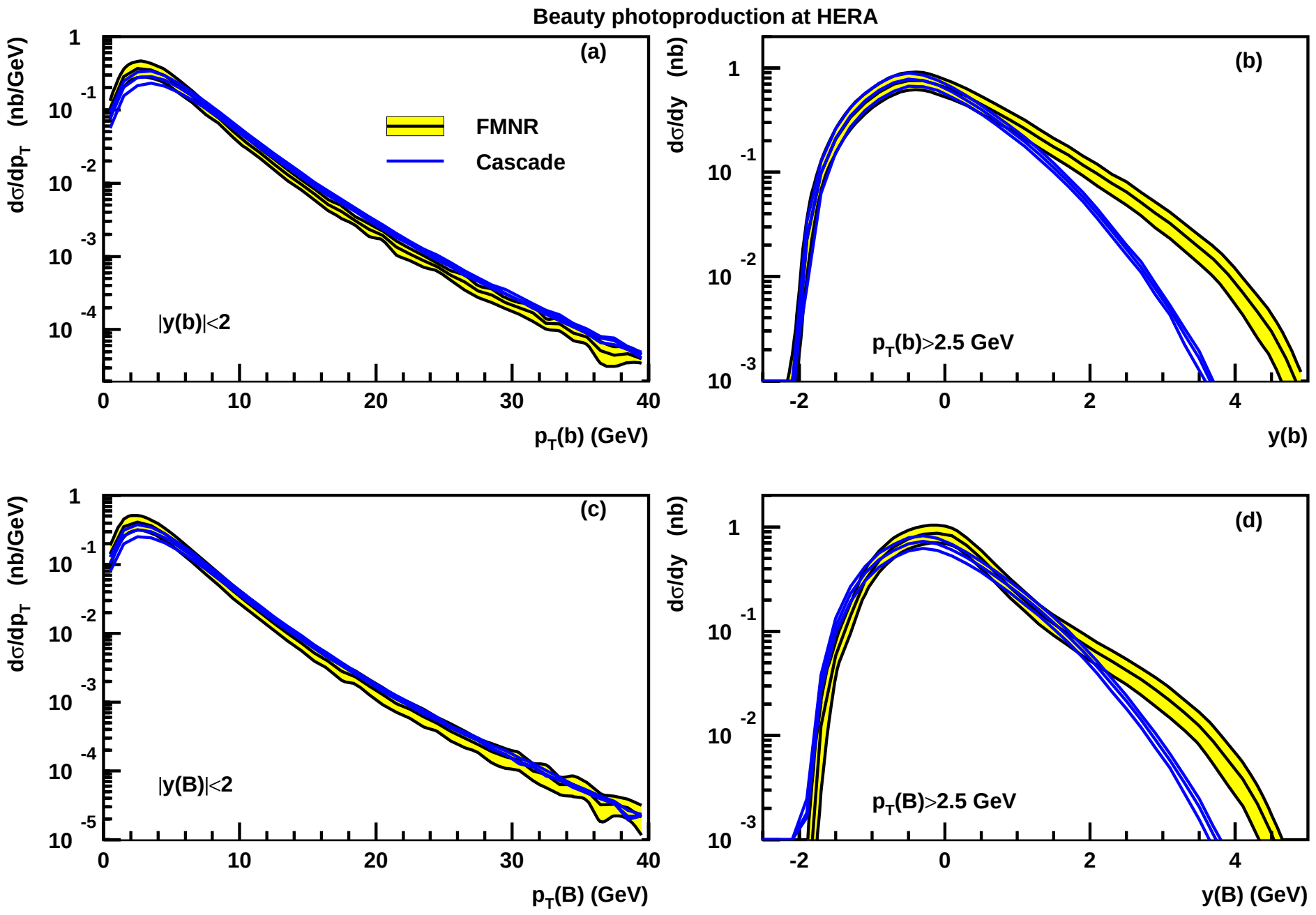

**Fig. 2:** Cross sections for beauty photoproduction at HERA ($Q^2 < 1$ GeV$^2$, $0.2 < y < 0.8$). The differential cross sections in $p_T$ and rapidity of the $b$ quark are shown in (a), (b). Plots (c) and (d) show the cross sections for the production of a weakly-decaying $B$ hadron as a function of $p_T(B)$ and $Y(B)$. The cross sections are shown for FMNR (shaded band) and CASCADE (empty band with full lines).

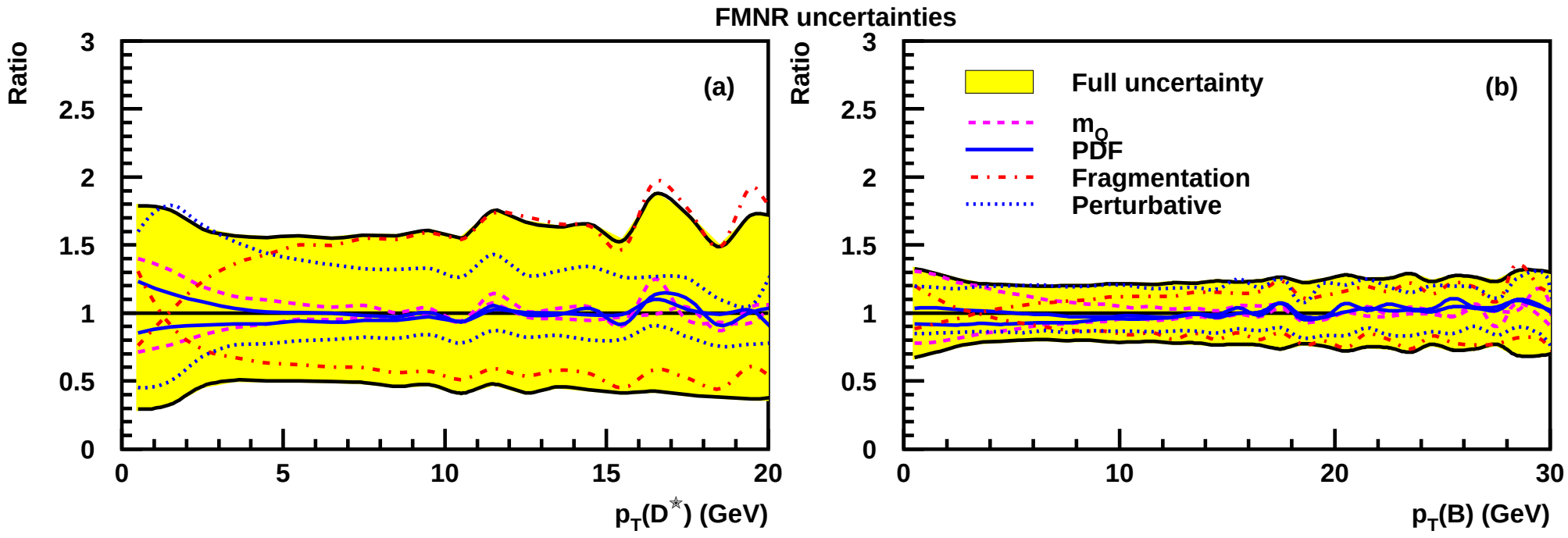

**Fig. 3:** Breakdown of the different components of the FMNR uncertainty for $d\sigma/dp_T$ for charmed (a) and beauty (b) hadrons in photoproduction at HERA. The plots show the ratio of the upper/lower side of each uncertainty to the nominal value. The following sources of uncertainty are shown: quark mass ($m_Q$), parton density parametrisation (PDF), fragmentation parameter and the perturbative uncertainty from scale variations.

### 4.2 HERA DIS

Heavy quark production in DIS is not available in the matched massive-massless approach (except for total cross sections). Therefore the DIS comparison was limited to the FO-NLO program HVQDIS, the unintegrated-PDF MC CASCADE and the RAPGAP Monte Carlo. The DIS cross sections at HERA are reported as $d\sigma/d\log_{10}(x)$ for different bins of $Q^2$ and are intended at the Born level, without electroweak corrections. Figure 4 shows, for each $Q^2$ bin, the inclusive charm cross-section, the cross section for observing a $D^*$ meson in the "visible" range $p_T(D^*) > 1.5$ GeV, $|Y(D^*)| < 1.5$ and for observing a muon in the range $p_T(\mu) > 3$ GeV, $|Y(\mu)| < 2$. The three calculations are compatible at intermediate values of $x$ ($\sim 10^{-3}$). At large $x$ and low $Q^2$, CASCADE and RAPGAP drop to zero much faster than HVQDIS. At low $x$ RAPGAP is significantly larger than HVQDIS while both are within the uncertainty band given by CASCADE. A similar behavior is seen for beauty (Fig. 5). The uncertainty on HVQDIS, not given here, is expected to be small ($\sim 10-20\%$ for beauty [24]). The high-$x$ discrepancy between HVQDIS and the other two calculations seems therfore to be beyond the program uncertainties and deserves further investigations.

### 4.3 LHC

For LHC, we computed the cross sections in $pp$ collisions at $\sqrt{s} = 14$ TeV.

Figures 6 and 7 show the single inclusive cross sections as a function of $p_T$, at quark (upper panels) and hadron (lower panels) level, for charm and beauty, respectively. Two rapidity intervals are considered: $|Y| < 2.5$, approximately covering the acceptance of the barrel detectors of ATLAS ($|\eta| < 2.5$), CMS ($|\eta| < 2.5$), and ALICE ($|\eta| < 0.9$); $2.5 < |Y| < 4$, approximately covering the acceptance of LHCb ($2 < \eta < 5$) and of the ALICE muon spectrometer ($2.5 < \eta < 4$).

For charm, we compare the fixed-order NLO results from MNR to the results from the CASCADE event generator, from the GM-VFNS calculation and from the FONLL calculation. The agreement is in general good, in particular in the low-$p_T$ region; at high-$p_T$ CASCADE predicts a larger cross section than the other calculations, especially at forward rapidities. The FONLL central prediction is in agreement with that of the FO NLO calculation at low $p_T$, while deviating from it at high $p_T$, where it gives a smaller cross section.

For beauty, we compare FO NLO (MNR), FONLL and CASCADE. Again, there is agreement at low $p_T$, where, as expected, the FONLL result coincides with the MNR result. At high $p_T$, both CASCADE and FONLL predict a larger cross section than the MNR central values, but all models remain compatible within the theoretical uncertainties. At forward rapidities, for beauty as for charm, CASCADE gives a significantly larger cross section than MNR.

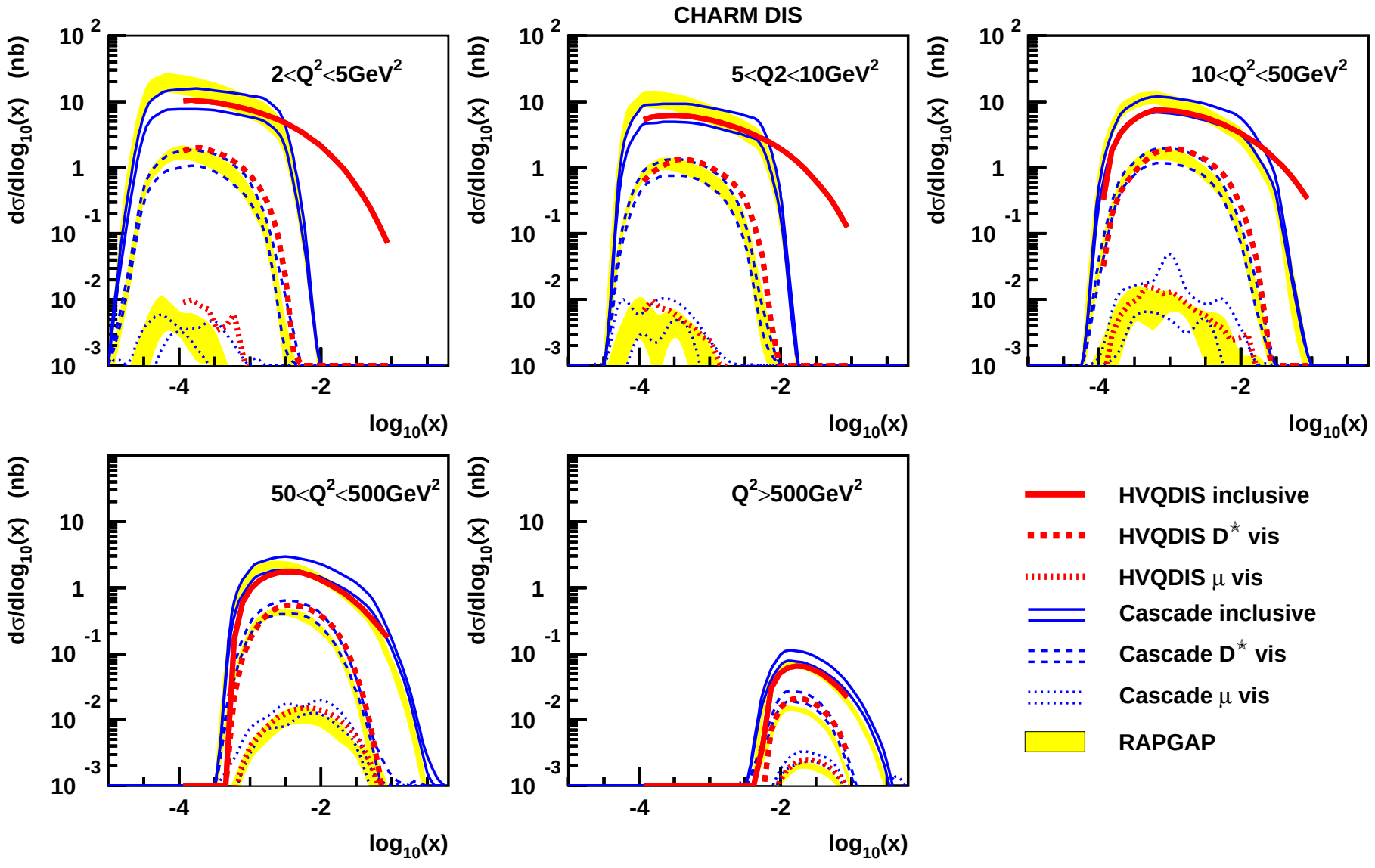

**Fig. 4:** Charm cross sections in DIS at HERA. Each plot shows the distribution of $\log_{10}(x)$ in a different $Q^2$ range for the inclusive cross-section, the cross-section for a $D^*$ meson in the "visible" range $p_T(D^*) > 1.5$ GeV, $|Y(D^*)| < 1.5$ and the cross-section for a muon from charm decay in the range $p_T(\mu) > 3$ GeV, $|Y(\mu)| < 2$. The thick curves show the central value from HVQDIS, the thin curves represent the uncertainty band from CASCADE and the shaded area shows the uncertainty band from RAPGAP. The fluctuations in the muon cross sections are due to the limited statistics.

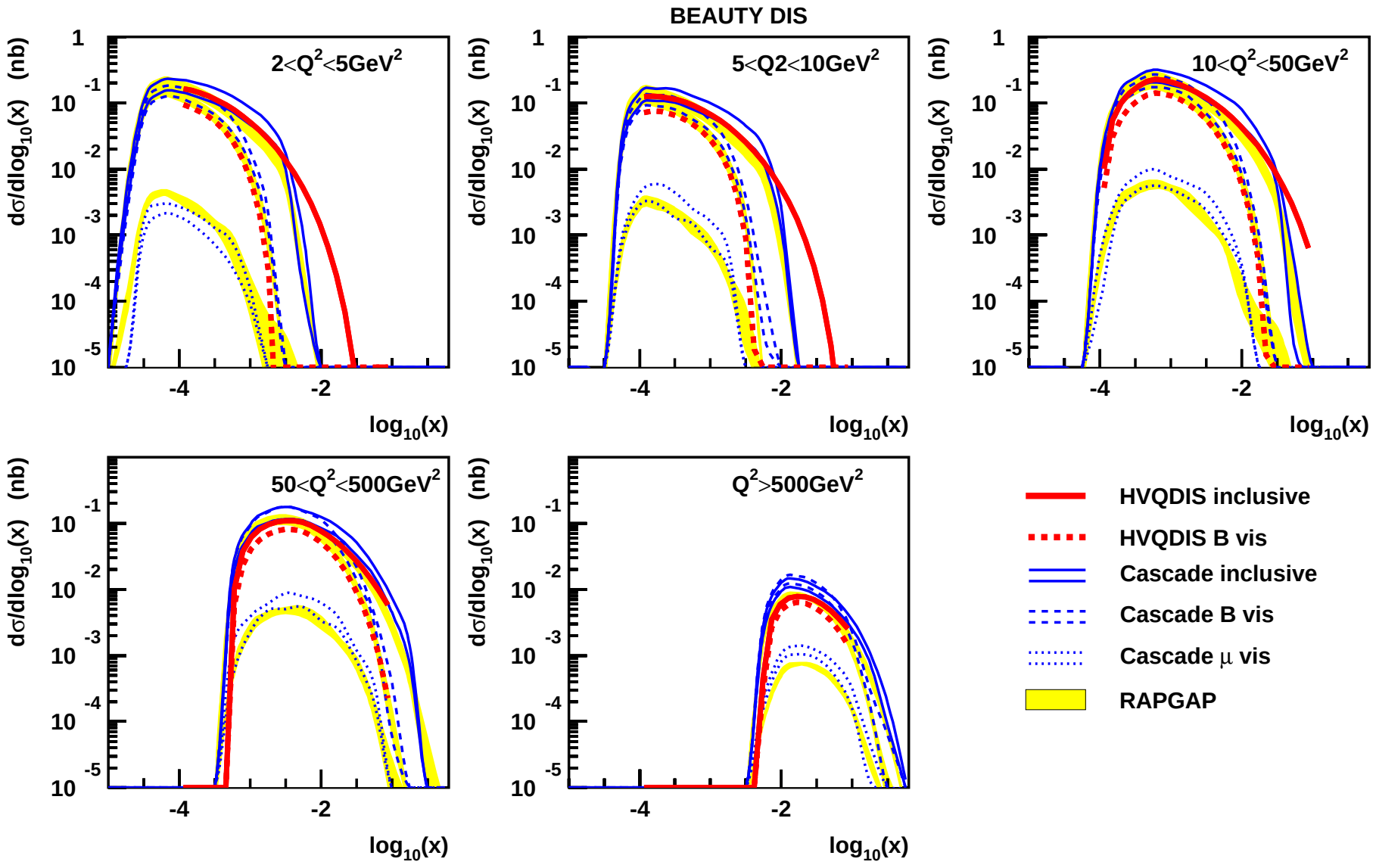

**Fig. 5:** Beauty cross sections in DIS at HERA. Each plot shows the distribution of $\log_{10}(x)$ in a different $Q^2$ range for the inclusive cross-section, the cross section for a hadron containing a $b$ quark in the "visible" range $p_T(B) > 3$ GeV, $|Y(B)| < 2$ and the cross section for a muon from beauty decay in the range $p_T(\mu) > 3$ GeV, $|Y(\mu)| < 2$. The thick curves show the central value from HVQDIS, the thin curves represent the uncertainty band from CASCADE and the shaded area shows the uncertainty band from RAPGAP. The muon distributions are not given for HVQDIS.

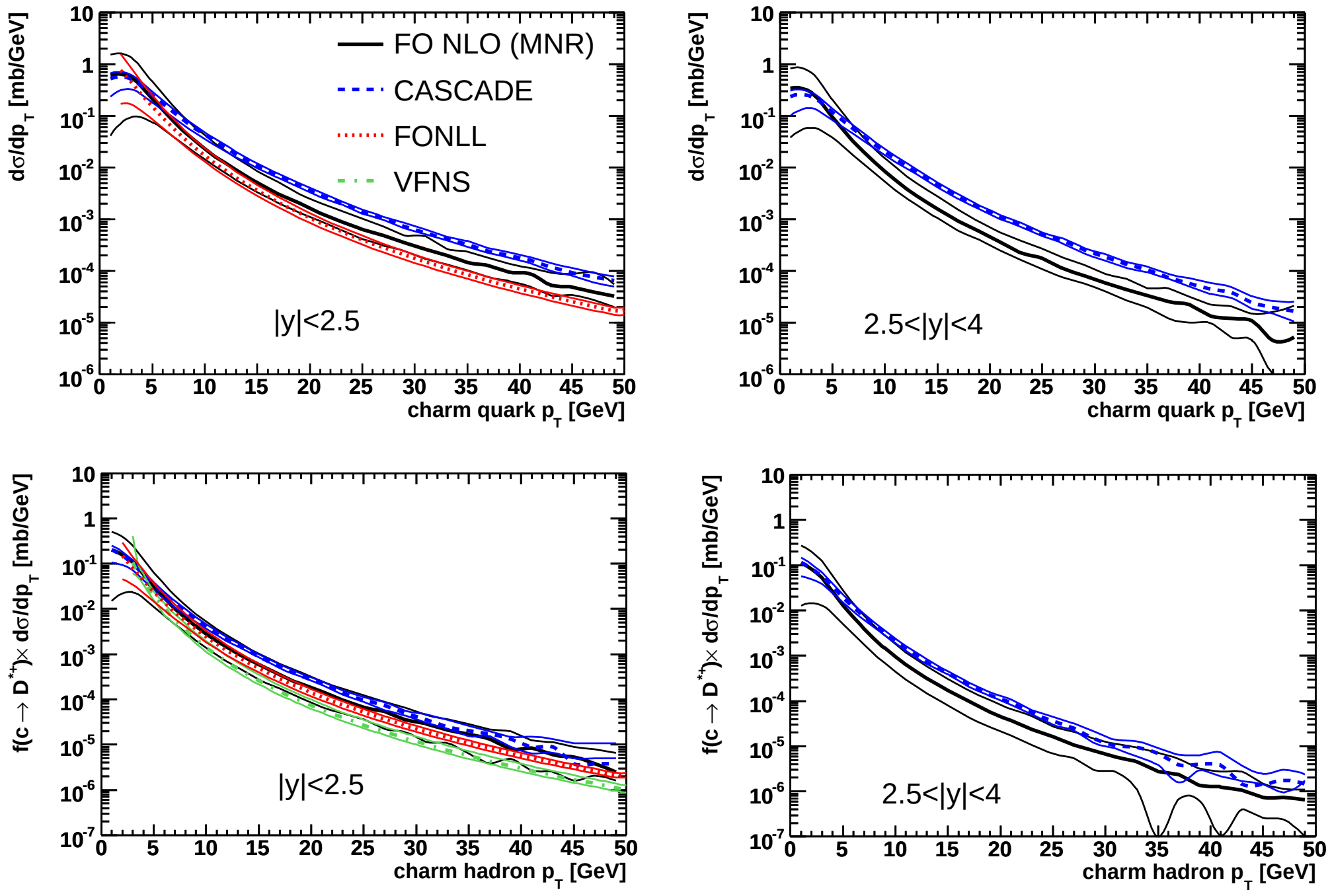

**Fig. 6:** Cross sections for charm production in $pp$ collisions at the LHC with $\sqrt{s} = 14$ TeV. The differential cross sections in $p_T$ for $c$ quark in the two rapidity ranges $|Y| < 2.5$ and $2.5 < |Y| < 4$ are shown in the upper panels. The lower panels show the cross sections for the production of a $D^*$ meson as a function of $p_T(D^*)$ in the same rapidity ranges.

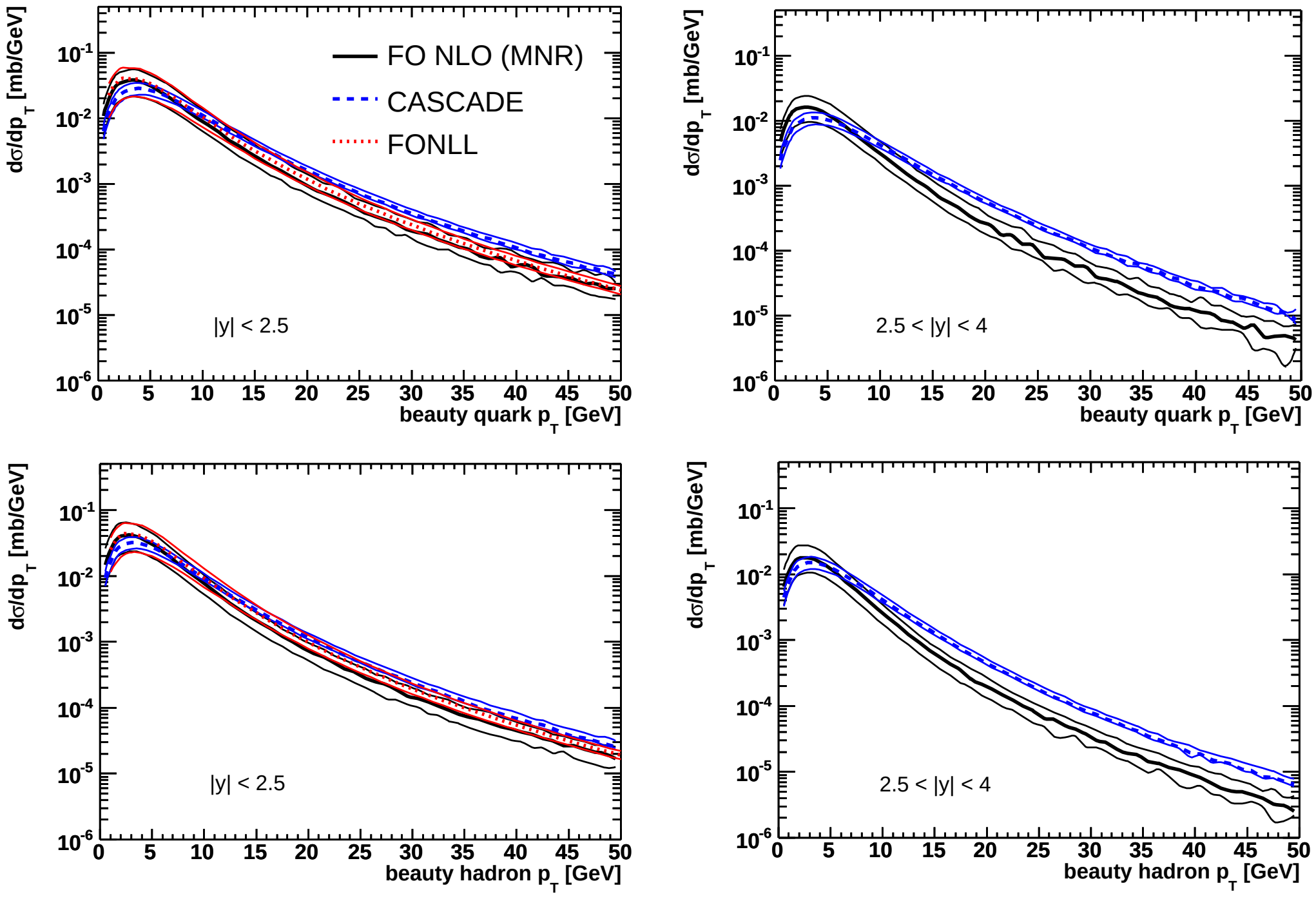

**Fig. 7:** Cross sections for beauty production in $pp$ collisions at the LHC with $\sqrt{s} = 14$ TeV. The differential cross sections in $p_T$ for $b$ quark in the two rapidity ranges $|Y| < 2.5$ and $2.5 < |Y| < 4$ are shown in the upper panels. The lower panels show the cross sections for the production of a beauty hadron as a function of $p_T$ in the same rapidity ranges.

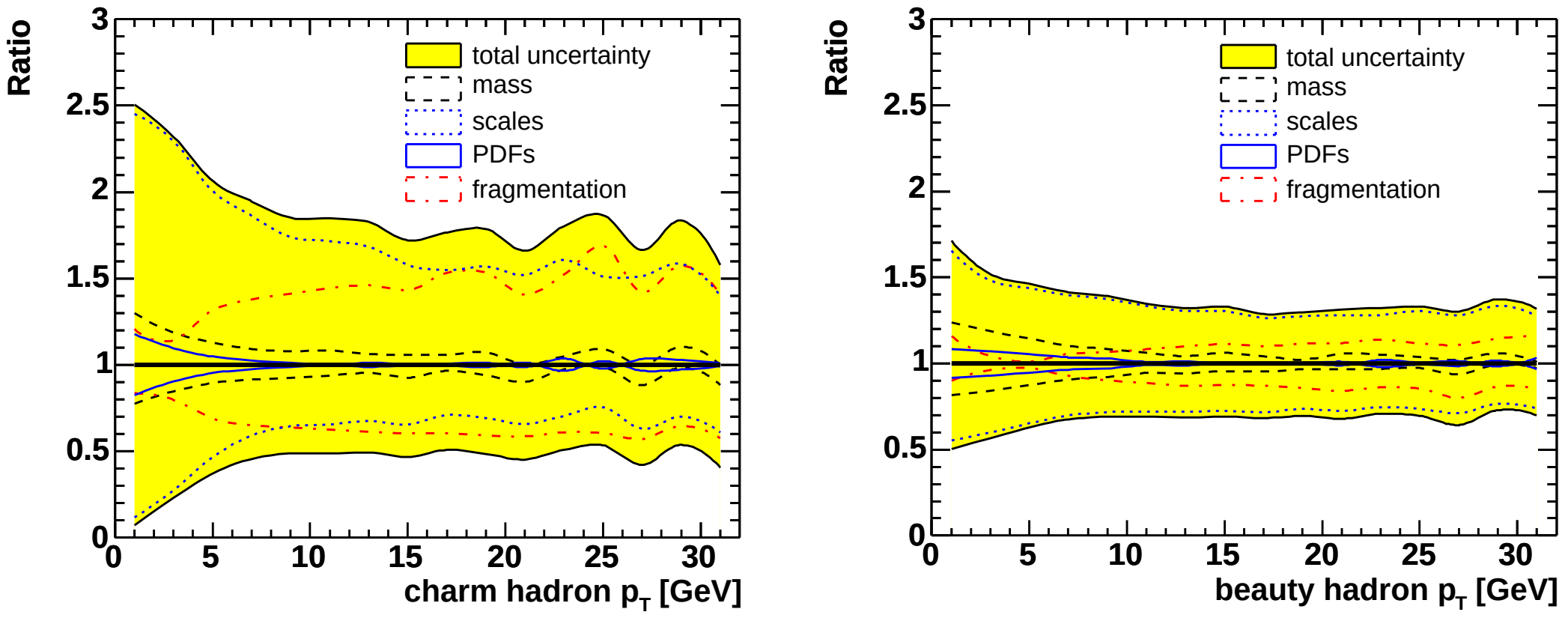

**Fig. 8:** Breakdown of the different components of the uncertainty on $d\sigma/dp_T$ for charmed (a) and beauty (b) hadrons at LHC as obtained from MNR. The plots show the ratio of the upper/lower side of each uncertainty to the nominal value. The following sources of uncertainty are shown: quark mass ($m_Q$), parton density parametrisation (PDF), fragmentation parameter and the perturbative uncertainty from scale variations.

Figure 8 shows the breakdown of the uncertainties for hadron production as obtained with MNR. The perturbative component dominates at LHC. Only the fragmentation component for charm hadron production becomes comparable in size to the perturbative one at large $p_T$.

### 4.4 $Q$-$\bar{Q}$ correlations

The azimuthal separation between the two heavy quarks $\Delta\phi(Q\bar{Q})$ and the transverse momentum of the quark-antiquark system $p_T(Q\bar{Q})$ are particularly sensitive to higher-order effects since at leading order their distributions are delta functions peaked at $\Delta\phi(Q\bar{Q}) = \pi$ and $p_T(Q\bar{Q}) = 0$. The distribution of these variables is therefore a direct probe of QCD radiation and is well suited for comparing different calculations.

Figures 9 and 10 show the heavy-quark pair $p_T$ distribution and the quark-antiquark relative azimuthal angle distribution for charm and beauty at LHC, respectively. For both distributions, the two quarks of the pair are required to have $|Y| < 2.5$; also minimum $p_T$ selections are applied to mimic the effect of realistic experimental cuts ($p_T^Q > 3$ GeV and $p_T^{\bar{Q}} > 6$ GeV). In the region near $\Delta\phi(Q\bar{Q}) = \pi$ and $p_T(Q\bar{Q}) = 0$, where the cancellation of soft and collinear divergencies occur, the fixed-order NLO calculation gives an unphysical negative cross section with next to a large positive peak. A larger binning would be needed to average this behavior and produce a more physical results. The CASCADE MC, has a more realistic behavior. Both calculations have a non-zero value at $\Delta\phi(Q\bar{Q}) = 0$ related to "gluon-splitting" events. A similar result was found for HERA as shown in Figure 11. This kind of distribution is expected to be well described by programs that merge NLO matrix elements to the parton-shower MC approach such as MC@NLO [25].

## 5 Conclusions

Heavy-flavour cross sections for HERA and LHC, obtained with fixed-order NLO programs, with matched massive/massless calculations and within the $K_T$-factorisation approach have been compared. Similar results are found for photoproduction at HERA and for the LHC. As expected the resummed calculations were found to be compatible with the fixed-order results but have smaller uncertainties at large $p_T$. Resummed calculations for charm in two different schemes (GM-VFNS and FONLL) are anyway somewhat incompatible both at HERA and LHC, suggesting that their uncertainty may be underestimated.

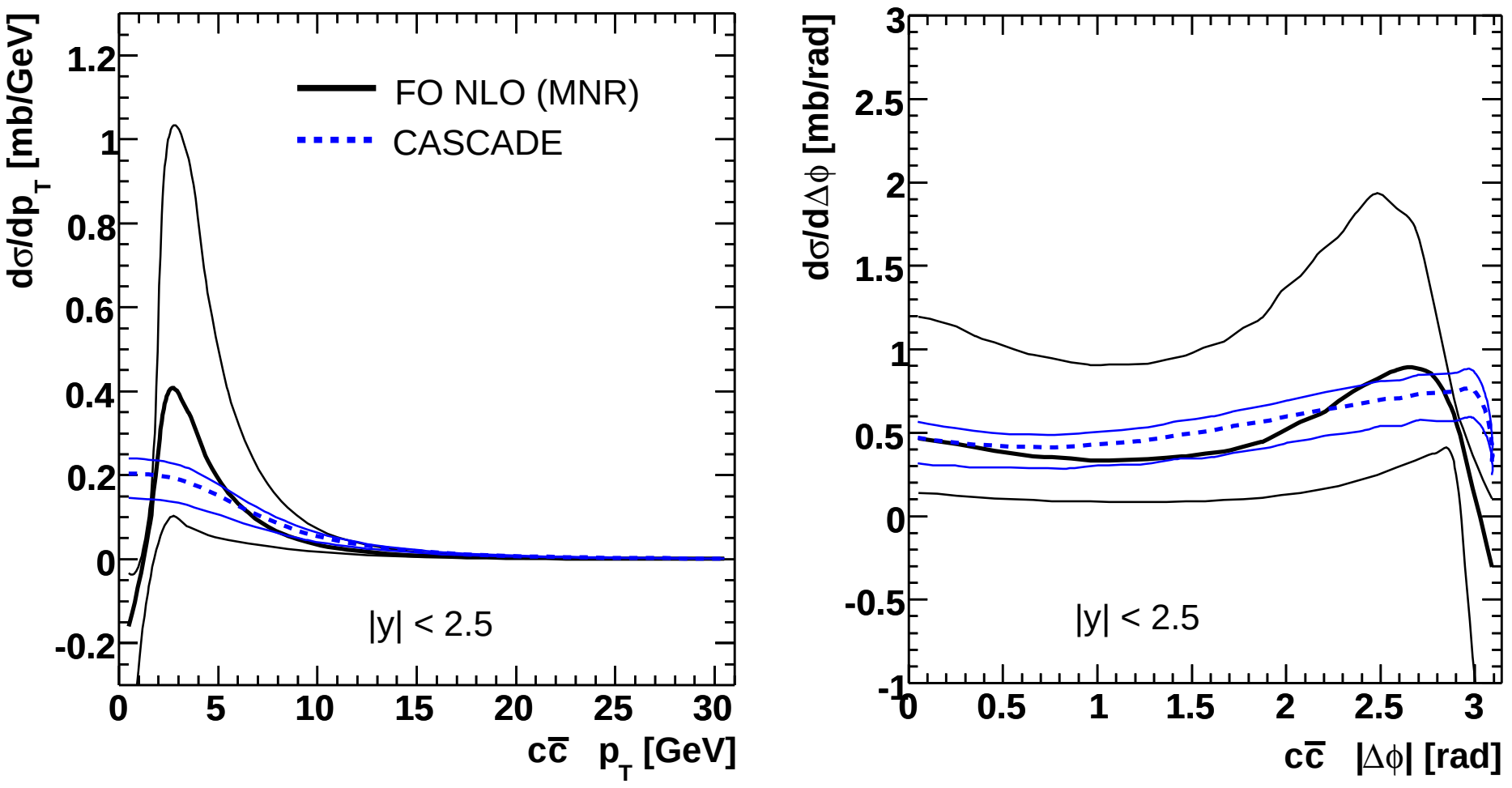

**Fig. 9:** $Q$-$\bar{Q}$ correlations for charm at LHC: $p_T$ of the $c\bar{c}$ pair (left) and azimuthal angle $\Delta\phi$ between the $c$ and the $\bar{c}$ (right). For both cross sections, the following kinematic cuts are applied: $|Y^c| < 2.5$, $|Y^{\bar{c}}| < 2.5$, $p_T^c > 3$ GeV, $p_T^{\bar{c}} > 6$ GeV.

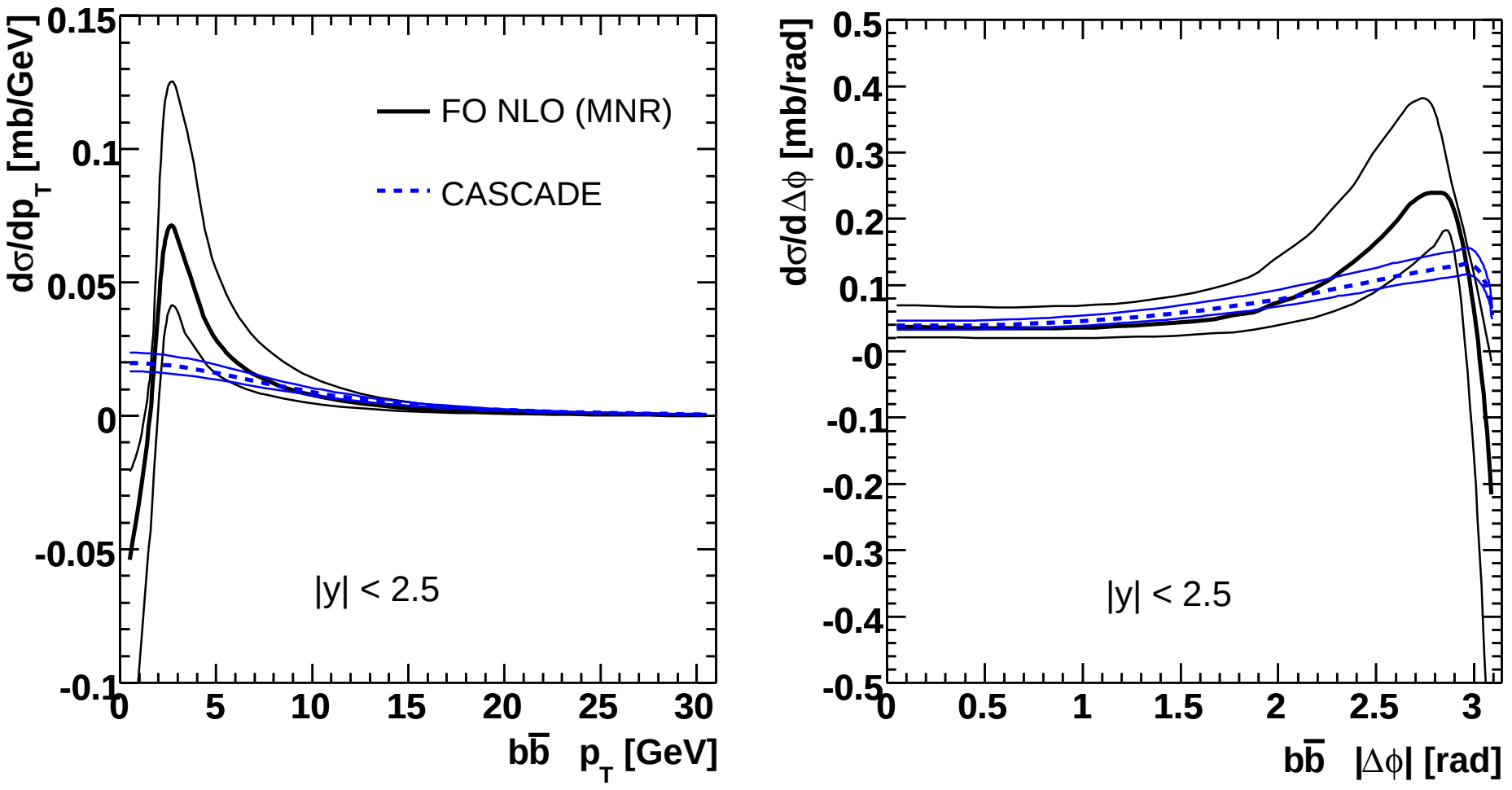

**Fig. 10:** $Q$-$\bar{Q}$ correlations for charm at LHC: $p_T$ of the $b\bar{b}$ pair (left) and azimuthal angle $\Delta\phi$ between the $b$ and the $\bar{b}$ (right). For both cross sections, the following kinematic cuts are applied: $|Y^b| < 2.5$, $|Y^{\bar{b}}| < 2.5$, $p_T^b > 3$ GeV, $p_T^{\bar{b}} > 6$ GeV.

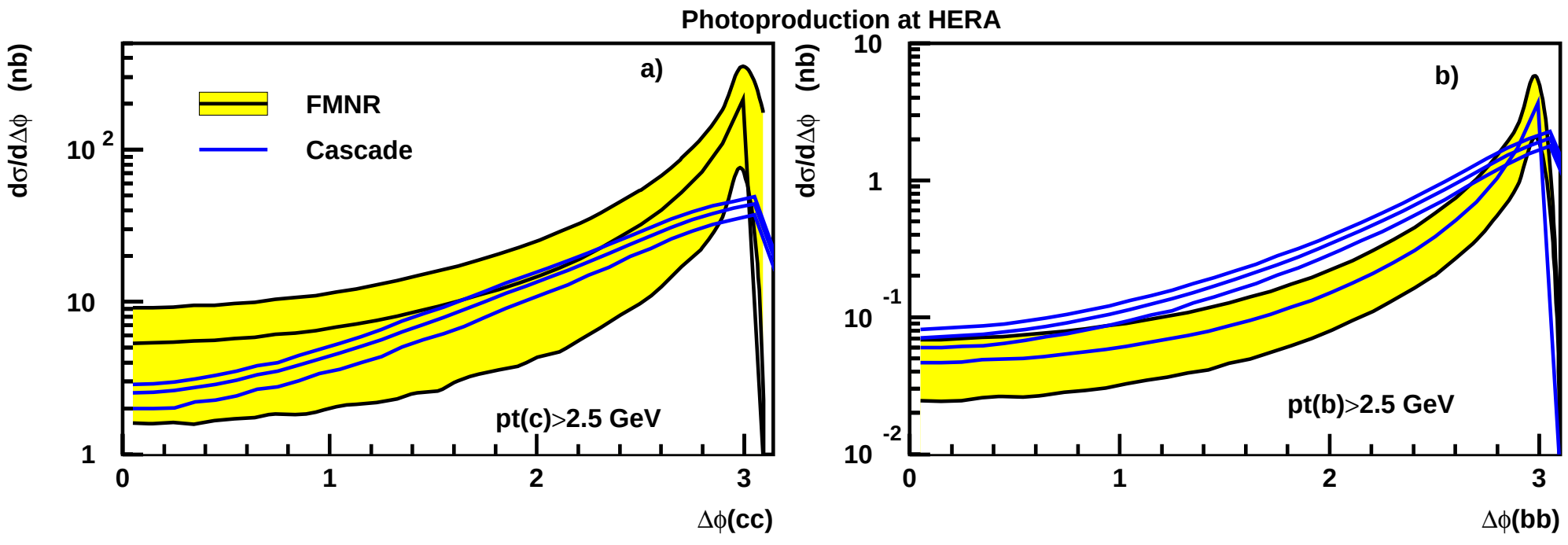

**Fig. 11:** Azimuthal $Q$-$\bar{Q}$ correlations in photoproduction at HERA for charm (a) and beauty (b). One of the two quarks was required to be in the "visible" region $p_T(Q) > 2.5$ GeV and $|Y(Q)| < 2$.

The $K_T$-factorisation program CASCADE predicts larger cross sections than the other approaches at large $p_T$ at LHC and for charm at HERA. The comparison for DIS was limited to FO-NLO and a MC program with leading order matrix elements. Large discrepancies, which deserve further investigations, were found in this case. A comparison with experimental data would be needed for further understanding of the quality of the available calculations.

# Part V

# Working Group 4: Diffraction

# List of participants

L. Adamczyk, N. Amapane, V. Andreev, M. Arneodo, V. Avati, C. Avila, J. Bartels, R. Bellan, S. Bolognesi, M. Boonekamp, A. Bruni, A. Bunyatyan, G. Cerminara, B. Cox, R. Croft, A. De Roeck, M. Diehl, I. Efthymiopoulos, K. Eggert, F. Ferro, J. Forshaw, E. Gotsman, D. Goulianos, P. Grafström, M. Grothe, M. Groys, C. Hogg, G. Ingelman, X. Janssen, H. Jung, J. Kalliopuska, M. Kapishin, D. Kharzeev, V. Khoze, M. Klasen, H. Kowalski, G. Kramer, J. W. Lämsä, P. V. Landshoff, P. Laycock, E. Levin, A. Levy, L. Lönnblad, M. Lublinsky, L. Lytkin, D. Macina, T. Mäki, U. Maor, C. Mariotti, N. Marola, C. Marquet, A. D. Martin, V. Monaco, J. Monk, L. Motyka, S. Munier, M. Murray, E. Naftali, P. Newman, J. Nystrand, F. Oljemark, R. Orava, K. Österberg, M. Ottela, A. Panagiotou, R. Peschanski, A. Pilkington, K. Piotrzkowski, A. Proskuryakov, A. Prygarin, J. Raufeisen, M. Rijssenbeek, M. Ruspa, M. G. Ryskin, M. Ryynänen, A. Sabio-Vera, R. Sacchi, S. Schätzel, F.-P. Schilling, A. Sobol, M. Strikman, S. Tapprogge, M. Taševský, P. Van Mechelen, R. Venugopalan, K. Vervink, G. Watt, K. Wichmann, G. Wolf.

# Introduction to Diffraction

*M. Arneodo* [a], *J. Bartels* [b], *A. Bruni* [c], *B. E. Cox* [d], *M. Diehl* [e], *J. Forshaw* [d], *M. Grothe* [f], *H. Kowalski* [e], *R. Orava* [g], *K. Piotrzkowski* [h] *and P. Van Mechelen* [i]
[a]Università del Piemonte Orientale, Novara, and INFN-Torino, Italy
[b]II. Institut für Theoretische Physik, Universität Hamburg, Luruper Chaussee 149, 22761 Hamburg, Germany
[c]INFN Bologna, Via Irnerio 46, 40156 Bologna, Italy
[d]School of Physics and Astronomy, The University of Manchester, Manchester M139PL, United Kingdom
[e]Deutsches Elektronen-Synchroton DESY, 22603 Hamburg, Germany
[f]University of Torino and INFN-Torino, Italy; also at University of Wisconsin, Madison, WI, USA
[g]High Energy Physics Division, Department of Physical Sciences, University of Helsinki and Helsinki Institute of Physics, P. O. Box 64, 00014 University of Helsinki, Finland
[h]Institut de Physique Nucléaire, Université Catholique de Louvain, Louvain-la-Neuve, Belgium
[i]Universiteit Antwerpen, Antwerpen, Belgium

**Abstract**

We summarize the main activities of the Working Group on Diffraction in this workshop, which cover a wide range of experimental, phenomenological and theoretical studies. Central themes are exclusive and inclusive diffraction at HERA and the LHC, multiple interactions and rapidity gap survival, and parton saturation.

## 1 Forward proton tagging at the LHC as a means to discover new physics

The use of forward proton tagging detectors at CMS and ATLAS as a means to search for and identify the nature of new physics at the LHC was one of the major topics of discussion at the workshop. The process of interest is the so-called 'central exclusive' production process $pp \to p \oplus \phi \oplus p$, where $\oplus$ denotes the absence of hadronic activity (a 'gap') between the outgoing intact protons and the decay products of the central system $\phi$. The final state therefore consists of *only* the decay products of the system $\phi$, which can be seen in the central detectors, and the two outgoing protons, which must be detected at some point downstream of the interaction point where they emerge far enough from the LHC beams. To this end, the feasibility of installing proton tagging detectors at 420 m from the interaction points of ATLAS and/or CMS, at a suitable time after the initial start-up of the LHC, is currently being assessed [1]. These would complement and increase the acceptance of the detectors already planned in the 220 m/240 m region by CMS / TOTEM and ATLAS. The choice of the 420 m region is set by the central system masses of interest; protons which lose approximately 60 GeV of their longitudinal momentum — the interesting range from the point of view of Higgs boson searches — emerge from the beam in this region.

The motivation for these studies stems from the unique properties of central exclusive production. Firstly, the mass of the central system $\phi$ can be measured to high accuracy by measuring the four-momenta of the outgoing protons alone, without reference to the central system (the so-called 'missing mass method' [2]). The achievable mass resolution and the acceptance as a function of mass of the 420 m detectors (in combination with the already planned 220 m proton detectors) are discussed in detail in these proceedings [3,4]. The resolution can be as good as 1 GeV for a Higgs boson of mass 140 GeV. As an example, in the case of a 140 GeV Standard Model Higgs decaying to two W bosons, and the subsequent leptonic decays of one or both of the $W$'s to leptons plus neutrinos, six events are expected with no modification of the level-1 trigger thresholds of ATLAS and CMS for 30 fb$^{-1}$ of delivered luminosity. We discuss the trigger issues in more detail below. This number is expected to double if realistic changes are made to the leptonic trigger thresholds [5].

A second crucial advantage is that, to a good approximation, the central system $\phi$ is produced in the $J_z = 0$, $C$ and $P$ even state, and an absolute determination of the quantum numbers of any resonance is possible by measuring correlations between the outgoing proton momenta. Observation of any resonance production with associated proton tags, therefore, allows a determination of its quantum numbers, something that is difficult to do in any other process at the LHC. Such a determination could be made with only a few 'gold-plated' events.

Thirdly, states which would otherwise be very difficult to detect in conventional channels can be detected in the central exclusive channel. Perhaps the best-studied example is the high $\tan\beta$ region of the MSSM, where over 100 signal events can be detected with backgrounds lower by an order of magnitude or more, within 30 fb$^{-1}$ of delivered luminosity at the LHC [6]. There are extensions to the MSSM in which central exclusive production becomes in all likelihood the only method at the LHC of isolating the underlying physics. One example [7] is the case where there are non-vanishing $CP$ phases in the gaugino masses and squark couplings. In such scenarios, the neutral Higgs bosons are naturally nearly degenerate for large values of $\tan\beta$ and charged Higgs masses around 150 GeV. In such scenarios, observing the mass spectrum using forward proton tagging may well be the only way to explore such a Higgs sector at the LHC. Explicit $CP$-violation in the Higgs sector can be observed as an asymmetry in the azimuthal distributions of the tagged protons [8].

From an experimental perspective, the key issue along with the mass resolution and acceptance is the level-1 (L1) trigger efficiency. The problem is that detectors at 420 m from the interaction points of ATLAS or CMS are too far away to participate in a L1 trigger decision without an increase in the trigger latency. This means that the central detectors, or forward detectors up to 220 m, must be relied upon to keep candidate events until the signals from 420 m can be used in higher level trigger decisions. A full description of the work done at the workshop is presented in Refs. [9, 10] in these proceedings. The most difficult case is that of a low-mass (120 GeV) Higgs boson decaying in the $b$-quark channel (a decay mode that will not be observed in any other measurement at the LHC). The relatively low transverse momenta of the $b$-jets necessitate L1 jet $E_T$ thresholds as low as 40 GeV. Thresholds that low would result in a L1 trigger rate of more than 50 kHz, because of the QCD background, and thus would essentially saturate the available output bandwidth. The output rate of a 2-jet L1 trigger condition with thresholds of 40 GeV per jet can be kept at an acceptable level of order 1 kHz in the absence of pile-up (i.e. for a single proton–proton interaction per bunch crossing) by either using the TOTEM T1 and T2 detectors (or the ATLAS forward detectors) as vetoes — central exclusive events have no energy in these regions — or by requiring that a proton be seen in the TOTEM (or ATLAS) detectors at 220 m on one side of the interaction point. This gives a sufficient reduction of the QCD background event rate. At higher luminosities, up to $2 \times 10^{33}\,\mathrm{cm}^{-2}\,\mathrm{s}^{-1}$, where pile-up is present, it is necessary to combine a 220 m tag with additional conditions based on event topology and on $H_T$, the scalar sum of all L1 jet $E_T$ values. These L1 trigger conditions result in signal efficiencies between 15% and 20%. A further 10% of the Higgs events can be retained by exploiting the muon-rich final state in the $H \to b\bar{b}$ mode, with no requirements on the forward detectors. Other interesting decay channels, such as $WW$ and $\tau\tau$, should be possible at the highest luminosities ($1 \times 10^{34}\,\mathrm{cm}^{-2}\,\mathrm{s}^{-1}$) since both ATLAS and CMS will trigger on such events routinely using only the central detectors.

As well as upgrading the proton tagging capabilities of ATLAS and CMS, there was also discussion of upgrading the very forward region of CMS to extend the pseudo-rapidity coverage up to $|\eta| \sim 11$. This would allow proton $x$ values down to $10^{-8}$ to be probed, opening up an unexplored region of small-$x$ parton dynamics [11].

In summary, central exclusive production provides an excellent means of measuring the masses of new particles with a precision at the 1 GeV level, irrespective of the decay mode of the particles. It also provides a clean way of unambiguously determining the quantum numbers of any resonances produced in the central exclusive process (including Standard Model and MSSM Higgs bosons) at the LHC.

In certain regions of the MSSM, and indeed for any scenarios in which the new particles couple strongly to gluons, central exclusive production may be the discovery channel[1]. The challenge is to design and build proton tagging detectors with the capability to measure the momentum loss of the outgoing protons at the 1 GeV level.

## 2 Theory of diffractive Higgs production

It is a fact that the theoretical predictions for central exclusive production suffer from several sources of uncertainty. The theoretical framework is presented and critically assessed in the contribution by Forshaw [13]. The emphasis is on the calculations of the Durham group, which are performed within perturbative QCD. The use of perturbative QCD is shown to be justified, with around 90% of the contribution to the Standard Model Higgs production cross-section ($m_H = 120$ GeV) coming from the region where the gluon virtualities are all above 1 GeV.

One of the main sources of uncertainty in the perturbative calculation arises from a lack of knowledge of the proton's generalized, unintegrated gluon distribution function, and so far estimates are based upon theoretically motivated corrections to the more familiar gluon distribution function. It is hard to make an accurate assessment of the uncertainty arising from this source, but currently a factor of 2 uncertainty on the Higgs production cross-section is probably not unrealistic. Measurements of exclusive diffraction at HERA can help constrain the generalized gluon distribution in kinematics similar to the one relevant for exclusive Higgs production at the LHC [14]. High-quality data are now available for $ep \to e\, J/\Psi\, p$. Exclusive production of $\Upsilon$ mesons and deeply virtual Compton scattering $ep \to e\gamma p$ involve smaller theoretical uncertainties, but are experimentally more demanding and should be explored in more detail with HERA II data.

Since the focus is on exclusive final states such as $p \oplus H \oplus p$, it is necessary to sum the Sudakov logarithms which arise in perturbation theory. One must go beyond summing the leading double logarithms and sum also the single logarithms. Without the single logs, one vastly underestimates the production rate. Unfortunately, perturbative emissions are not the only way to spoil the exclusive nature of the final state: extra particles can also be produced as a result of soft interactions between the colliding protons. To account for such soft interactions is clearly outside of the scope of perturbation theory and one is forced to resort to non-perturbative models. It is universally assumed that one can estimate the effect of forbidding additional particle production by simply multiplying the perturbative cross-section by an overall 'gap survival' factor [15]. The two most sophisticated models of this factor are discussed in some detail and compared with each other in the contribution of Gotsman et al. [16]. It turns out that, although the approaches are different in many respects, they tend to predict very similar values for the gap survival factor. Nevertheless, both models are essentially multi-channel eikonal models and one would like to test them against data. Fortunately that is possible: data from HERA and the Tevatron already tend to support the theoretical models and future measurements at the LHC will allow one to further constrain them.

Uncertainties in the gluon densities and in our knowledge of gap survival can be reduced as we test our ideas against data, both at present colliders and at the LHC itself. Fortunately, these uncertainties essentially factorize (from the hard subprocess which produces the central system) into a universal 'effective gluon luminosity' function. Thus one can hope to extract the important physics associated with the production of the central system by first measuring the luminosity function in a 'standard candle' process. The ideal candidate is $pp \to p + \gamma\gamma + p$ [17] since the hard subprocess is well known ($gg \to \gamma\gamma$) and the effective gluon luminosity can be extracted over a wide kinematic range. In this way one might hope to extract the effective coupling of any centrally produced new physics to two gluons.

[1]For a recent review of the physics case for FP420, see [12] and references therein.

During the period of the workshop, Monte Carlo codes have been developed which simulate the theoretical predictions for both interesting signal processes and also the associated backgrounds. These codes are now routinely used, for example, to help develop the case for the installation of low-angle proton detectors at the LHC, and new processes are being added with time. A review and comparison of the various Monte Carlos is to be found in the contribution of Boonekamp et al. [18].

## 3 Diffractive structure functions and diffractive parton distributions

The cross-section for the reaction $ep \to eXp$ can be expressed in terms of the diffractive structure functions $F_2^D$ and $F_L^D$, in analogy to the way in which $d\sigma/dx\, dQ^2$ is related to the structure functions $F_2$ and $F_L$ for inclusive DIS, $ep \to eX$. The function $F_2^D$ describes the proton structure in processes in which a fast proton is present in the final state; $F_L^D$ corresponds to longitudinal polarization of the virtual photon. Since in diffractive events the proton typically loses a fraction of less than 0.02–0.03 of its initial momentum, the parton participating in a diffractive interaction has a fractional momentum which is also less than 0.02–0.03. Diffractive DIS thus probes the low-$x$ structure of the proton, in a way complementary to that provided by non-diffractive DIS.

Diffractive structure functions, like the usual ones, can be expressed as the convolution of universal partonic cross-sections and a specific type of parton distribution functions, the diffractive PDFs. This is the so-called diffractive factorization theorem. Diffractive PDFs can be determined by means of QCD fits similar to those used for extracting the standard PDFs from the $F_2$ data.

Several measurements of $F_2^D$ are available from the H1 and ZEUS collaborations. Three alternative approaches have been used to select diffractive events:

1. a fast proton is required in the final state; this can be done only be means of a proton spectrometer able to detect scattered protons which do not leave the beam pipe (see e.g. [19]);
2. a rapidity gap in the forward direction is required;
3. the different shape of the $M_X$ distribution for diffractive and non-diffractive events is exploited.

Method 1 selects the reaction $ep \to eXp$ with a high degree of purity; the acceptance of proton spectrometers is, however, small, yielding comparatively small samples. Methods 2 and 3 select the reaction $ep \to eXY$, as opposed to $ep \to eXp$, with $Y$ a proton or a low-mass system. Samples selected with these two methods may include some contamination from non-diffractive processes. Method 3 suppresses the contribution of subleading exchanges (i.e. Reggeon and pion exchanges, as opposed to Pomeron exchange), which is instead present in the samples selected with methods 1 and 2.

Results obtained with the three methods are presented and compared in these proceedings [20]. Methods 2 and 3 yield results for $F_2^D$ which are higher than those obtained with the LPS by factors as large as 1.4, depending on the degree of forward coverage. This normalization difference is due to the proton-dissociative background (from $ep \to eXY$) and is relatively well understood. Having corrected for this effect, the results of the three methods exhibit, at present, a fair degree of agreement. However, differences in the shapes of the $Q^2$, $\beta$ and $x_{I\!P}$ dependences become apparent especially when comparing the results obtained with method 3 and those obtained with methods 1 and 2. The origin of these differences is at present not clear. An urgent task for the HERA community is to understand these discrepancies and provide a consistent set of measurements of $F_2^D$.

Several NLO fits of the $F_2^D$ data were discussed at the workshop [20–22]. The corresponding parametrizations are available in Ref. [23]. The diffractive PDFs are dominated by gluons, as expected given the low-$x$ region probed, with the density of gluons larger than that of quarks by a factor 5–10. There are significant discrepancies between the results of the fits, reflecting, at least in part, the differences in the fitted data. In addition, Martin, Ryskin and Watt [22] argue that the leading-twist formula used in Refs. [20,21] is inadequate in large parts of the measured kinematics, and use a modified expression which includes an estimate of power-suppressed effects.

The discrepancies between the various diffractive PDFs, while not fully understood, are at the moment the best estimate of their uncertainties. Here as well, it is imperative that the HERA community provide a consistent set of diffractive PDFs. Not only are they important for our understanding of the proton structure, but they are also an essential input for any calculation of the cross-sections for *inclusive* diffractive reactions at the LHC — which are interesting in themselves in addition to being a potentially dangerous background to the central *exclusive* production processes discussed in Sections 1 and 2.

No direct measurement exists of $F_L^D$. The dominant role played by gluons in the diffractive parton densities implies that the leading-twist $F_L^D$ must also be relatively large. A measurement of $F_L^D$ to even modest precision would provide an independent and theoretically very clean tool to verify our understanding of the underlying dynamics and to test the gluon density extracted indirectly in QCD fits from the scaling violations of $F_2^D$. This is discussed in Ref. [24].

## 4 Diffractive charm and dijet production at HERA

As mentioned in Section 2, the possibility to observe central exclusive processes depends critically on the survival probability of large rapidity gaps. This probability is not unity as a consequence of the rescattering between the spectator partons in the colliding hadrons: these interactions generate final-state particles which fill the would-be rapidity gap and slow down the outgoing proton or antiproton [16]. This is why the diffractive factorization theorem [25] is expected to fail for hadron–hadron scattering — and therefore also for resolved photoproduction, where the photon acts as a hadron.

In $p\bar{p}$ collisions at the Tevatron, breaking of diffractive factorization was indeed observed. The fraction of diffractive dijet events at CDF is a factor 3 to 10 smaller than that predicted on the basis of the diffractive parton densities obtained at HERA. Similar suppression factors were observed in all hard diffractive processes in proton–antiproton collisions.

In photoproduction processes, however, the situation is far from clear at the moment. A recent ZEUS result [26] indicates that the cross-section for diffractive photoproduction of $D^*$ mesons, a process dominated by the *direct* photon component, is well described by NLO QCD predictions based on the diffractive PDFs. This lends support to the idea that in direct processes the photon is pointlike and that the diffractive factorization theorem holds in this case. Conversely, diffractive dijet data from H1 and ZEUS are better described by a global suppression of *both* the direct and resolved contribution. A discussion of how this might be understood is given in Refs. [27, 28], where a critical study of the factorization scheme and scale dependence of resolved and direct contributions is presented.

## 5 Multiple scattering at HERA and the LHC

A thorough analysis of the event structure at the LHC will have to take into account contributions from multiple-parton interactions, i.e. from interactions involving more than one parton in each of the colliding protons. Such multiple interactions are expected to be particularly important in the region of small longitudinal momentum fractions and not too high momentum scales. At HERA there are several pieces of evidence that multiple interactions are present; the strongest one comes from the observation of diffractive final states in deep-inelastic electron–proton scattering. A useful tool for analysing these multiple interactions are the so-called AGK cutting rules. During this workshop several groups have studied their application to HERA and to future LHC scattering processes.

The theoretical basis of the AGK rules in perturbative QCD has been outlined in Ref. [29], and a few first applications to HERA and to LHC scattering processes have been addressed. The contribution by Watt et al. [30] uses the AGK rules for deriving, from the measured *diffractive* structure function, absorptive corrections to the *inclusive* structure function $F_2$. An iterative scheme is then set up which leads to corrected parton densities: at low $Q^2$ and small $x$, they tend to be higher than those without absorptive corrections. In particular, they seem to weaken the trend of the gluon density becoming negative, which has been seen in the global parton analyses of both MRST2004 and CTEQ6.

The study presented in Ref. [31] is based upon a specific saturation model that has been successfully applied both to the total $\gamma^* p$ cross-section and to the diffractive process $\gamma^* p \to J/\Psi\, p$. An analysis of this model, based upon the AGK rules, leads to the conclusion that contributions of multiple interactions to $F_2$ are quite sizeable, even for $Q^2$ as large as $40$ GeV$^2$.

## 6 Parton saturation: from HERA to the LHC

A key experimental finding of HERA is the strong rise of structure functions at small $x$, which implies a high density of small-$x$ gluons in the proton. From theoretical considerations, it is clear that for sufficiently large parton densities, dynamics beyond what can be described by leading-twist factorization and linear DGLAP evolution must become important. If the associated momentum scale is high enough, the strong coupling is still small enough to serve as an expansion parameter, but at very high gluon densities the gluon potential can be so strong that the non-linear term $g_s f^{abc} A^b_\mu A^c_\nu$ in the gluon field strength is as large as the linear term $\partial_\mu A^a_\nu - \partial_\nu A^a_\mu$. High parton densities thus offer the possibility to study QCD in a strongly non-linear regime, and the effective theory of such a 'colour glass condensate' is reviewed in Ref. [32]. A possible link between the strong gluon fields in this description and QCD instantons is discussed in Ref. [33].

The theory and phenomenology of parton saturation are in rapid development, of which the workshop could only provide a snapshot. Data on both inclusive and diffractive deeply inclusive scattering, in particular their very similar energy dependence at given $Q^2$, suggest that saturation effects are relevant in HERA kinematics, see Ref. [34] and references therein. When saturation is important, the usual parton densities cease to be the key input quantities for describing physical processes. For many reactions a suitable quantity is instead the colour-dipole cross-section — a concept that has been successfully applied in HERA phenomenology. An important theoretical laboratory to study saturation effects is provided by the non-linear Balitsky–Kovchegov equation. In a contribution to the workshop, this equation has been applied to the colour-dipole cross-section for the proton [35]. To describe saturation in $pp$ collisions in general requires non-perturbative functions that can be written as matrix elements of Wilson line operators; *one* of these functions is the colour-dipole cross-section just mentioned [32]. The formulation of suitable evolution equations for $pp$ scattering is an active area of research [36].

## 7 Rapidity gaps in electroweak processes

Diffractive processes are characterized by rapidity gaps. Such gaps can also originate from the exchange of a photon, a $W$ or a $Z$ boson (see for example Ref. [15]). Selecting events with large rapidity gaps filters out specific final states and, at the same time, leads to better-constrained event kinematics. However, the event rate is lowered by the gap survival probability, as discussed in the previous sections.

The contribution by Amapane et al. [37] discusses the possibility to study the scattering of longitudinally polarized vector bosons ($V_L$) in $pp$ collisions with the CMS detector at the LHC. $V_L V_L$ fusion may lead to Higgs production; should the Higgs boson not exist, the cross-section for $V_L V_L$ scattering will deviate from the Standard Model prediction at high invariant masses of the $V_L V_L$ system. In all cases, $V_L V_L$ scattering should shed light on the mechanism behind the electroweak symmetry breaking. Preliminary studies based on Pythia and a fast simulation of the CMS detector are encouraging. It will be interesting to investigate in more detail the potential of the rapidity-gap signature for improved signal extraction and background control.

Large rapidity gaps at hadron colliders can also be due to photon exchange. In this case, a direct tagging of high-energy photon interactions can be achieved by using forward proton detectors [38]. Both photon–photon and photon–proton interactions at the LHC have been studied [39]. Some of these events can be used to scan the gap survival probability in impact parameter space, which would help to constrain models for gap survival. A reference point is given by single $W$ boson photoproduction, which has been studied theoretically in this context [40] and is being investigated at HERA.

Finally, diffractive photoproduction of $\Upsilon$ mesons, currently being studied at HERA, can be accessed at the LHC in an extended range of small $x$. This will provide a very clean channel to study the generalized gluon distribution (see Section 2) and can be seen as a complement to measurements of the usual gluon distribution at very small $x$, for instance in forward jet production at the LHC.

## Acknowledgements

We wish to thank all participants of the Working Group on Diffraction for their valuable contributions to this Workshop. This work has been supported by PPARC and the Royal Society in the UK, and the Italian Ministry for Education, University and Scientific Research under the programme 'Incentivazione alla mobilità di studiosi stranieri e italiani residenti all'estero'.

# Diffraction for non-believers

*Michele Arneodo*[a] *and Markus Diehl*[b]
[a]Università del Piemonte Orientale, 28100 Novara, Italy
[b]Deutsches Elektronen-Synchroton DESY, 22603 Hamburg, Germany

**Abstract**
Diffractive reactions involving a hard scale can be understood in terms of quarks and gluons. These reactions have become a valuable tool for investigating the low-$x$ structure of the proton and the behavior of QCD in the high-density regime, and they may provide a clean environment to study or even discover the Higgs boson at the LHC. In this paper we give a brief introduction to the description of diffraction in QCD. We focus on key features studied in $ep$ collisions at HERA and outline challenges for understanding diffractive interactions at the LHC.

## 1 Introduction

In hadron-hadron scattering a substantial fraction of the total cross section is due to diffractive reactions. Figure 1 shows the different types of diffractive processes in the collision of two hadrons: in elastic scattering both projectiles emerge intact in the final state, whereas single or double diffractive dissociation corresponds to one or both of them being scattered into a low-mass state; the latter has the same quantum numbers as the initial hadron and may be a resonance or continuum state. In all cases, the energy of the outgoing hadrons $a, b$ or the states $X$, $Y$ is approximately equal to that of the incoming beam particles, to within a few per cent. The two (groups of) final-state particles are well separated in phase space and in particular have a large gap in rapidity between them.

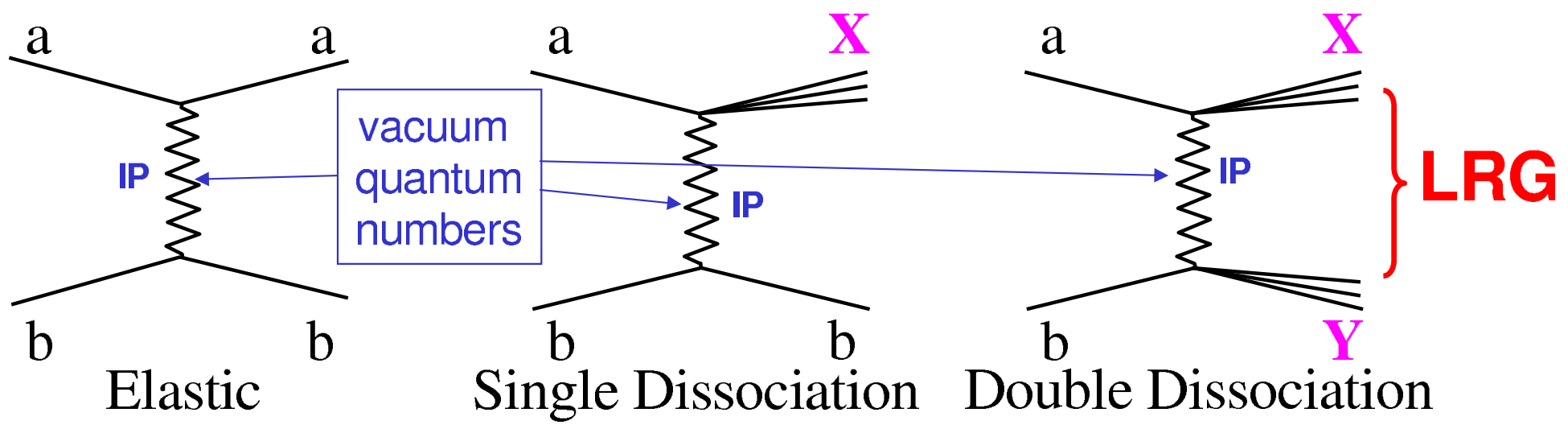

**Fig. 1:** Elastic scattering, single diffractive dissociation and double diffractive dissociation in the collision of two hadrons $a$ and $b$. The two (groups of) final-state hadrons are separated by a large rapidity gap (LRG). The zigzag lines denote the exchange of a Pomeron ($I\!P$) in the $t$-channel. There are further graphs, not shown, with multiple Pomeron exchange.

Diffractive hadron-hadron scattering can be described within Regge theory (see e.g. [1]). In this framework, the exchange of particles in the $t$-channel is summed coherently to give the exchange of so-called “Regge trajectories”. Diffraction is characterized by the exchange of a specific trajectory, the “Pomeron”, which has the quantum numbers of the vacuum. Regge theory has spawned a successful phenomenology of soft hadron-hadron scattering at high energies. Developed in the 1960s, it predates the theory of the strong interactions, QCD, and is based on general concepts such as dispersion relations. Subsequently it was found that QCD perturbation theory in the high-energy limit can be organized following the general concepts of Regge theory; this framework is often referred to as BFKL after the authors of the seminal papers [2].

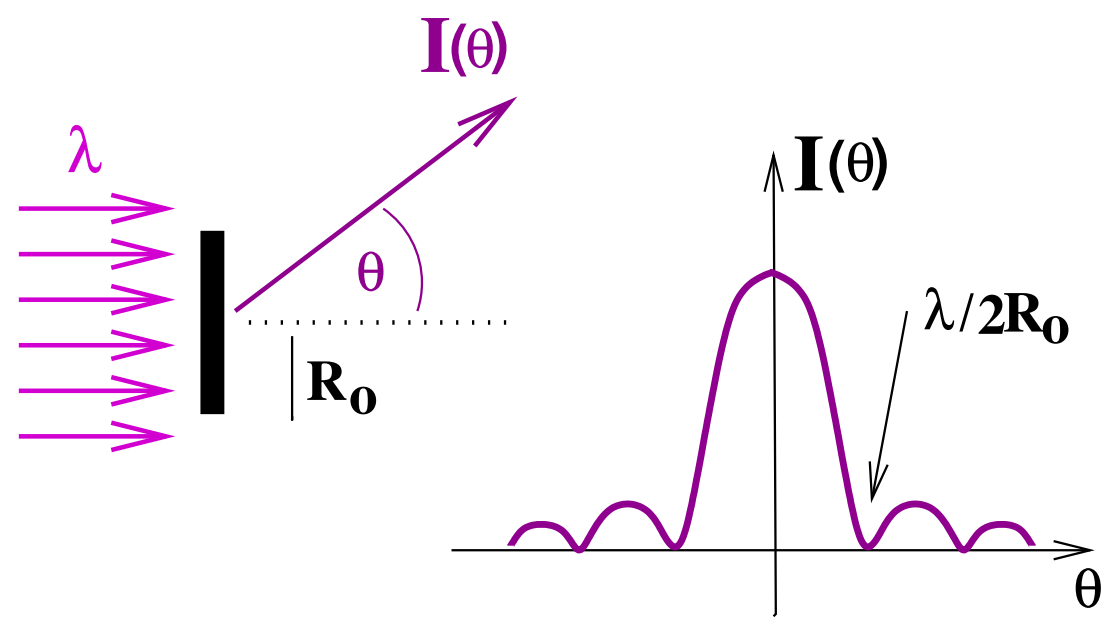

**Fig. 2:** Distribution of the intensity $I$ in the diffraction of light of wavelength $\lambda$ from a circular target of size $R_0$.

It is clear that a $t$-channel exchange leading to a large rapidity gap in the final state must carry zero net color: if color were exchanged, the color field would lead to the production of further particles filling any would-be rapidity gap. In QCD, Pomeron exchange is described by the exchange of two interacting gluons with the vacuum quantum numbers.

The effort to understand diffraction in QCD has received a great boost from studies of diffractive events in $ep$ collisions at HERA (see e.g. [3] for further reading and references). The essential results of these studies are discussed in the present paper and can be summarized as follows:

- Many aspects of diffraction are well understood in QCD when a hard scale is present, which allows one to use perturbative techniques and thus to formulate the dynamics in terms of quarks and gluons. By studying what happens when the hard scale is reduced towards the non-perturbative region, it may also be possible to shed light on soft diffractive processes.
- Diffraction has become a tool to investigate low-momentum partons in the proton, notably through the study of diffractive parton densities in inclusive processes and of generalized parton distributions in exclusive ones. Diffractive parton densities can be interpreted as conditional probabilities to find a parton in the proton when the final state of the process contains a fast proton of given four-momentum. Generalized parton distributions, through their dependence on both longitudinal and transverse variables, provide a three-dimensional picture of the proton in high-energy reactions.
- A fascinating link has emerged between diffraction and the physics of heavy-ion collisions through the concept of saturation, which offers a new window on QCD dynamics in the regime of high parton densities.

Perhaps unexpectedly, the production of the Higgs boson in diffractive $pp$ collisions is drawing more and more attention as a clean channel to study the properties of a light Higgs boson or even discover it. This is an example of a new theoretical challenge: to adapt and apply the techniques for the QCD description of diffraction in $ep$ collisions to the more complex case of $pp$ scattering at the LHC. A first glimpse of phenomena to be expected there is provided by the studies of hard diffraction in $p\bar{p}$ collisions at the Tevatron.

### 1.1 A digression on the nomenclature: why "diffraction" ?

Physics students first encounter the term "diffraction" in optics. Light of wavelength $\lambda$ impinging on a black disk of radius $R_0$ produces on a distant screen a diffraction pattern, characterized by a large forward peak for scattering angle $\theta = 0$ (the "diffraction peak") and a series of symmetric minima and maxima, with the first minimum at $\theta_{\min} \simeq \pm\lambda/(2R_0)$ (Fig. 2). The intensity $I$ as a function of the scattering angle $\theta$ is given by

$$\frac{I(\theta)}{I(\theta = 0)} = \frac{[2J_1(x)]^2}{x^2} \simeq 1 - \frac{R_0^2}{4}(k\theta)^2, \tag{1}$$

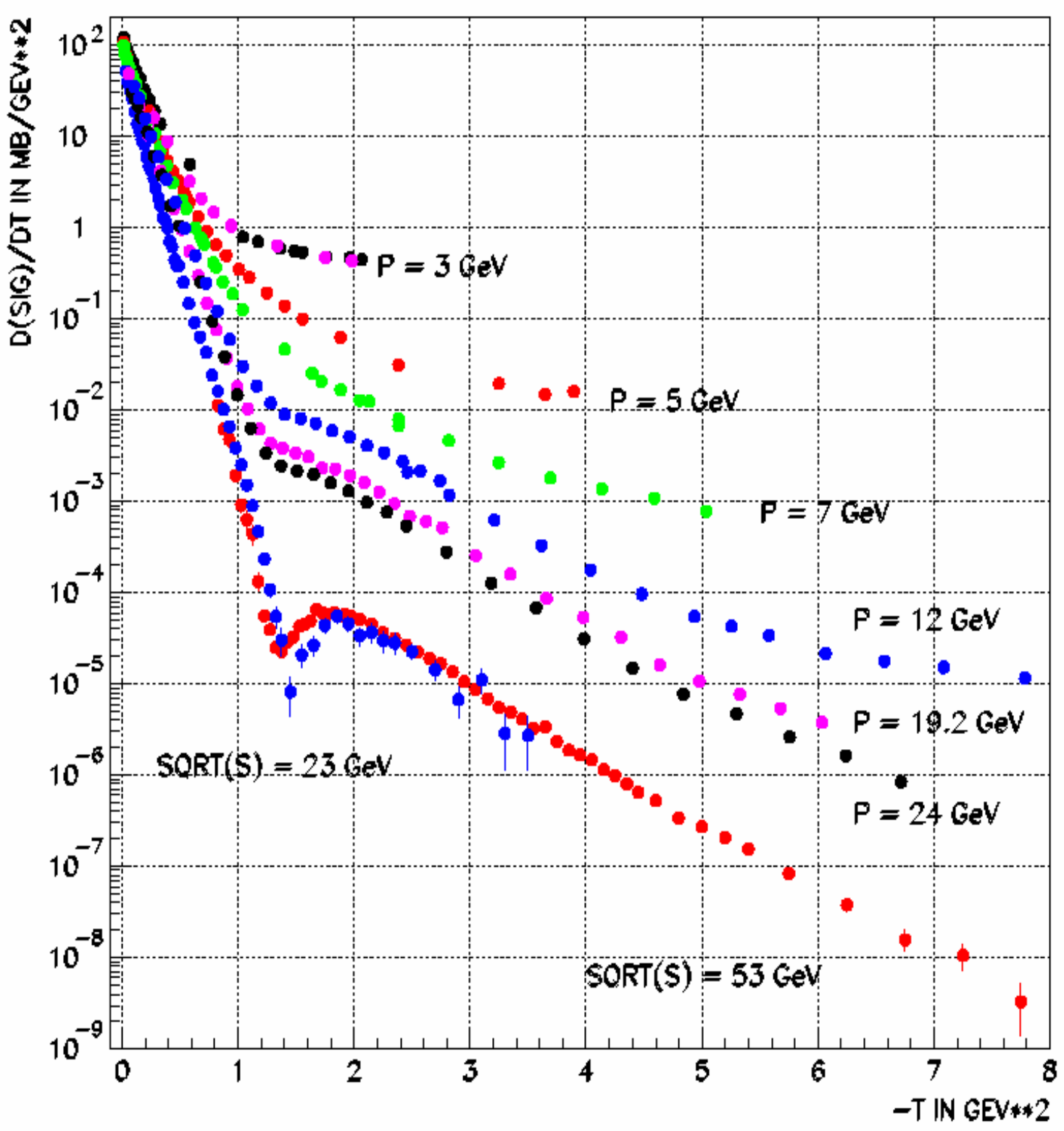

**Fig. 3:** Compilation of proton-proton elastic cross section data as a function of $t$. The symbol $P$ indicates the momentum of the incoming proton in a fixed target experiment and $\sqrt{s}$ the center-of-mass energy in a $pp$ collider setup.

where $J_1$ is the Bessel function of the first order and $x = kR_0 \sin\theta \simeq kR_0\,\theta$ with $k = 2\pi/\lambda$. The diffraction pattern is thus related to the size of the target and to the wavelength of the light beam.

As shown in Fig. 3, the differential cross section $d\sigma/dt$ for elastic proton-proton scattering, $pp \to pp$, bears a remarkable resemblance to the diffraction pattern just described (see e.g. [4]). At low values of $|t|$ one has

$$\frac{\frac{d\sigma}{dt}(t)}{\frac{d\sigma}{dt}(t=0)} \simeq e^{-b|t|} \simeq 1 - b\,(P\theta)^2, \tag{2}$$

where $|t| \simeq (P\theta)^2$ is the absolute value of the squared four-momentum transfer, $P$ is the incident proton momentum and $\theta$ is the scattering angle. The $t$-slope $b$ can be written as $b = R^2/4$, where once again $R$ is related to the target size (or more precisely to the transverse distance between projectile and target). A dip followed by a secondary maximum has also been observed, with the value of $|t|$ at which the dip appears decreasing with increasing proton momentum. It is hence not surprising that the term diffraction is used for elastic $pp$ scattering. Similar $t$ distributions have been observed for the other diffractive reactions mentioned above, leading to the use of the term diffraction for all such processes.

### 1.2 Diffraction at HERA ?!

Significant progress in understanding diffraction has been made at the $ep$ collider HERA, where 27.5 GeV electrons or positrons collide with 820 or 920 GeV protons. This may sound peculiar: diffraction is a typical hadronic process while $ep$ scattering at HERA is an electro-weak reaction, where the electron radiates a virtual photon (or a $Z$ or $W$ boson), which then interacts with the proton.[1] To understand this, it is useful to look at $ep$ scattering in a frame where the virtual photon moves very fast (for instance in the proton rest frame, where the $\gamma^*$ has a momentum of up to about 50 TeV at HERA). The virtual photon can fluctuate into a quark-antiquark pair. Because of its large Lorentz boost, this virtual pair has a lifetime much longer than a typical strong interaction time. In other words, the photon fluctuates into a pair long before the collision, and it is the pair that interacts with the proton. This pair is a small color dipole. Since the interaction between the pair and the proton is mediated by the strong interaction, diffractive events are possible.

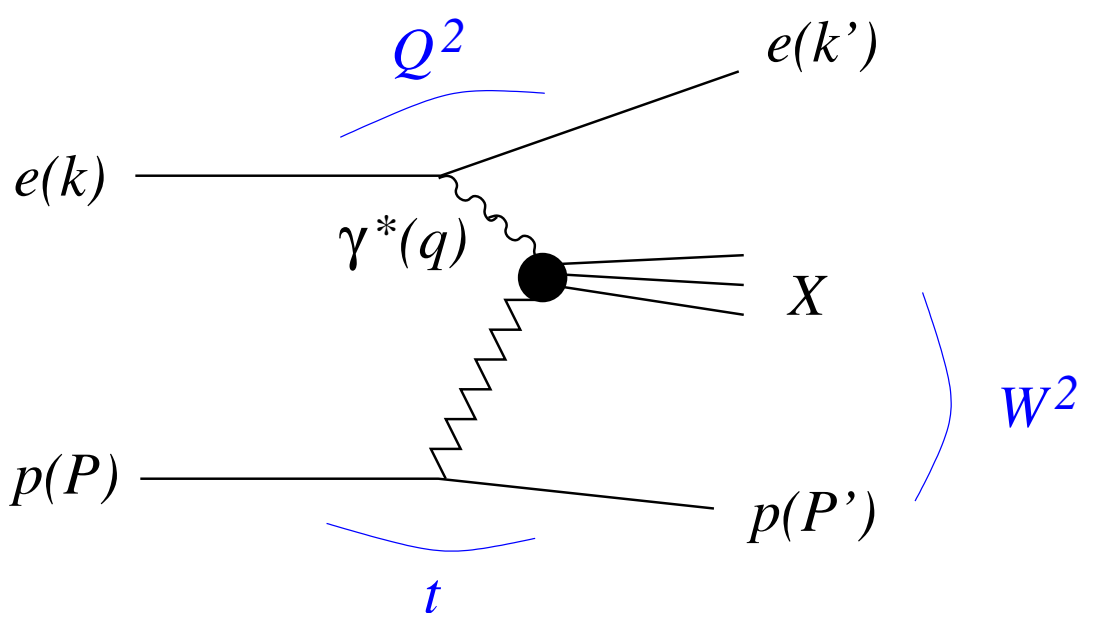

**Fig. 4:** Schematic diagram of inclusive diffractive DIS, $ep \to eXp$. Four-momenta are indicated in parentheses.

An advantage of studying diffraction in $ep$ collisions is that, for sufficiently large photon virtuality $Q^2$, the typical transverse dimensions of the dipole are small compared to the size of a hadron. Then the interaction between the quark and the antiquark, as well as the interaction of the pair with the proton, can be treated perturbatively. With decreasing $Q^2$ the color dipole becomes larger, and at very low $Q^2$ these interactions become so strong that a description in terms of quarks and gluons is no longer justified. We may then regard the photon as fluctuating into a vector meson – this is the basis of the well-known vector meson dominance model – and can therefore expect to see diffractive reactions very similar to those in hadron-hadron scattering.

A different physical picture is obtained in a frame where the incident proton is very fast. Here, the diffractive reaction can be seen as the deep inelastic scattering (DIS) of a virtual photon on the proton target, with a very fast proton in the final state. One can thus expect to probe partons in the proton in a very specific way. For suitable diffractive processes there are in fact different types of QCD factorization theorems, which bear out this expectation (see Sects. 2 and 3).

## 2 Inclusive diffractive scattering in *ep* collisions

Figure 4 shows a schematic diagram of inclusive diffractive DIS. The following features are important:

- The proton emerges from the interaction carrying a large fraction $x_L$ of the incoming proton momentum. Diffractive events thus appear as a peak at $x_L \approx 1$, the diffractive peak, which at HERA approximately covers the region $0.98 < x_L < 1$ (see the left panel of Fig. 5). The right panel of Fig. 5 shows that large values of $|t|$ are exponentially suppressed, similarly to the case of elastic $pp$ scattering we discussed in Sect. 1.1. These protons remain in the beam-pipe and can only be measured with detectors located inside the beam-pipe.
- The collision of the virtual photon with the proton produces a hadronic final state $X$ with the photon quantum numbers and invariant mass $M_X$. A large gap in rapidity (or pseudorapidity) is present between $X$ and the final-state proton. Figure 6 shows a typical diffractive event at HERA.

[1]For simplicity we will speak of a virtual photon in the following, keeping in mind that one can have a weak gauge boson instead.

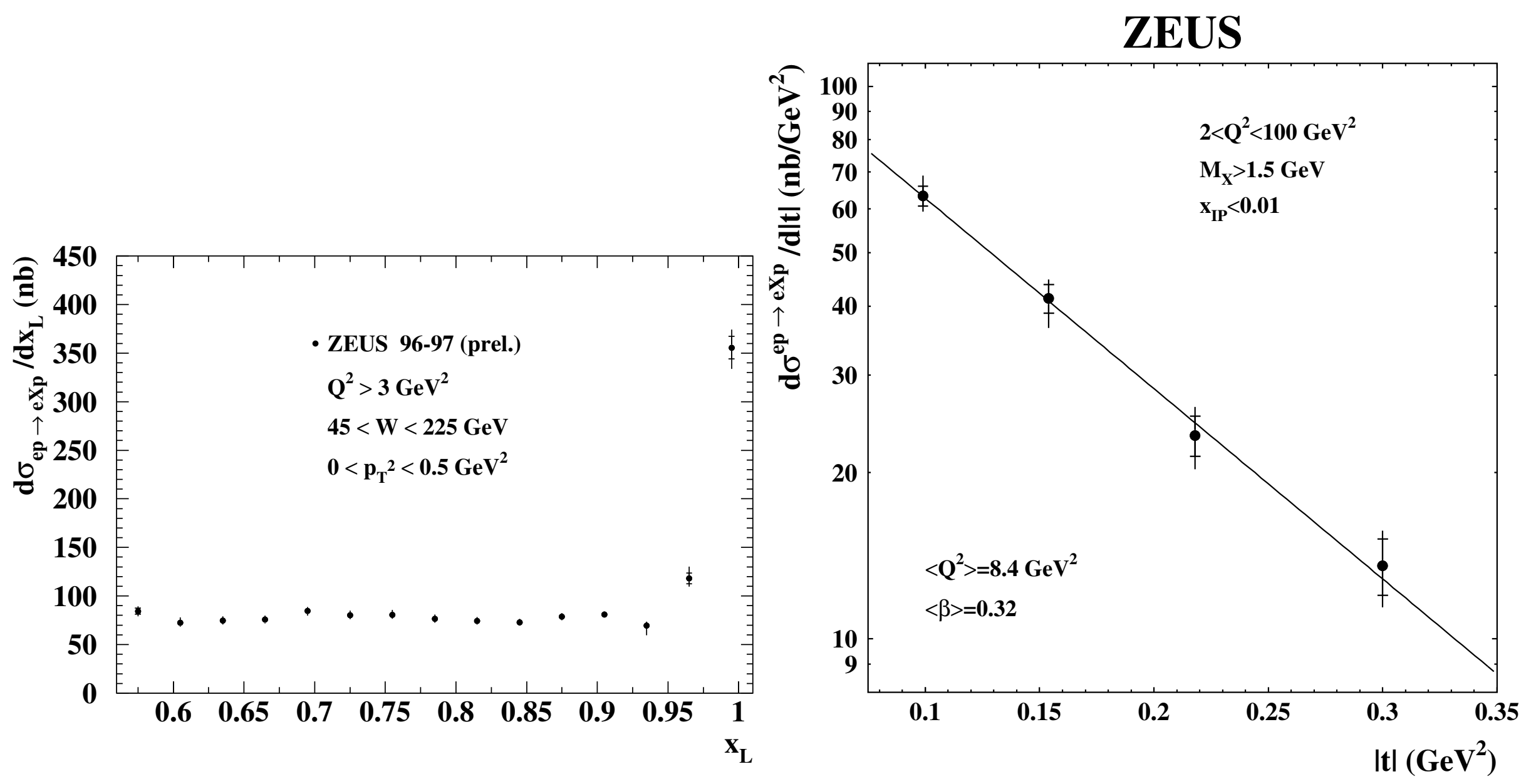

**Fig. 5:** Left: Differential cross section $d\sigma/dx_L$ for the process $ep \to eXp$ (from [5]). The diffractive peak at $x_L \approx 1$ is clearly visible. Right: Differential cross section $d\sigma/dt$ for the same process for $x_L > 0.99$ (from [6]). The average $|t|$ of this spectrum is $\langle |t| \rangle \approx 0.15$ GeV$^2$.

Diffractive $ep$ scattering thus combines features of hard and soft scattering. The electron receives a large momentum transfer; in fact $Q^2$ can be in the hundreds of GeV$^2$. In contrast, the proton emerges with its momentum barely changed.

### 2.1 Diffractive structure functions

The kinematics of $\gamma^* p \to Xp$ can be described by the invariants $Q^2 = -q^2$ and $t = (P - P')^2$, and by the scaling variables $x_{I\!P}$ and $\beta$ given by

$$x_{I\!P} = \frac{(P-P')\cdot q}{P\cdot q} = \frac{Q^2 + M_X^2 - t}{W^2 + Q^2 - M_p^2}, \qquad \beta = \frac{Q^2}{2(P-P')\cdot q} = \frac{Q^2}{Q^2 + M_X^2 - t}, \tag{3}$$

where $W^2 = (P+q)^2$ and the four-momenta are defined in Fig. 4. The variable $x_{I\!P}$ is the fractional momentum loss of the incident proton, related as $x_{I\!P} \simeq 1 - x_L$ to the variable $x_L$ introduced above. The quantity $\beta$ has the form of a Bjorken variable defined with respect to the momentum $P - P'$ lost by the initial proton instead of the initial proton momentum $P$. The usual Bjorken variable $x_B = Q^2/(2P\cdot q)$ is related to $\beta$ and $x_{I\!P}$ as $\beta x_{I\!P} = x_B$.

The cross section for $ep \to eXp$ in the one-photon exchange approximation can be written in terms of diffractive structure functions $F_2^{D(4)}$ and $F_L^{D(4)}$ as

$$\frac{d\sigma^{ep\to eXp}}{d\beta\, dQ^2\, dx_{I\!P}\, dt} = \frac{4\pi\alpha_{\text{em}}^2}{\beta Q^4}\left[\left(1 - y + \frac{y^2}{2}\right)F_2^{D(4)}(\beta, Q^2, x_{I\!P}, t) - \frac{y^2}{2}F_L^{D(4)}(\beta, Q^2, x_{I\!P}, t)\right], \tag{4}$$

in analogy with the way $d\sigma^{ep\to eX}/(dx_B\, dQ^2)$ is related to the structure functions $F_2$ and $F_L$ for inclusive DIS, $ep \to eX$. Here $y = (P\cdot q)/(P\cdot k)$ is the fraction of energy lost by the incident lepton in the proton rest frame. The structure function $F_L^{D(4)}$ corresponds to longitudinal polarization of the virtual photon;

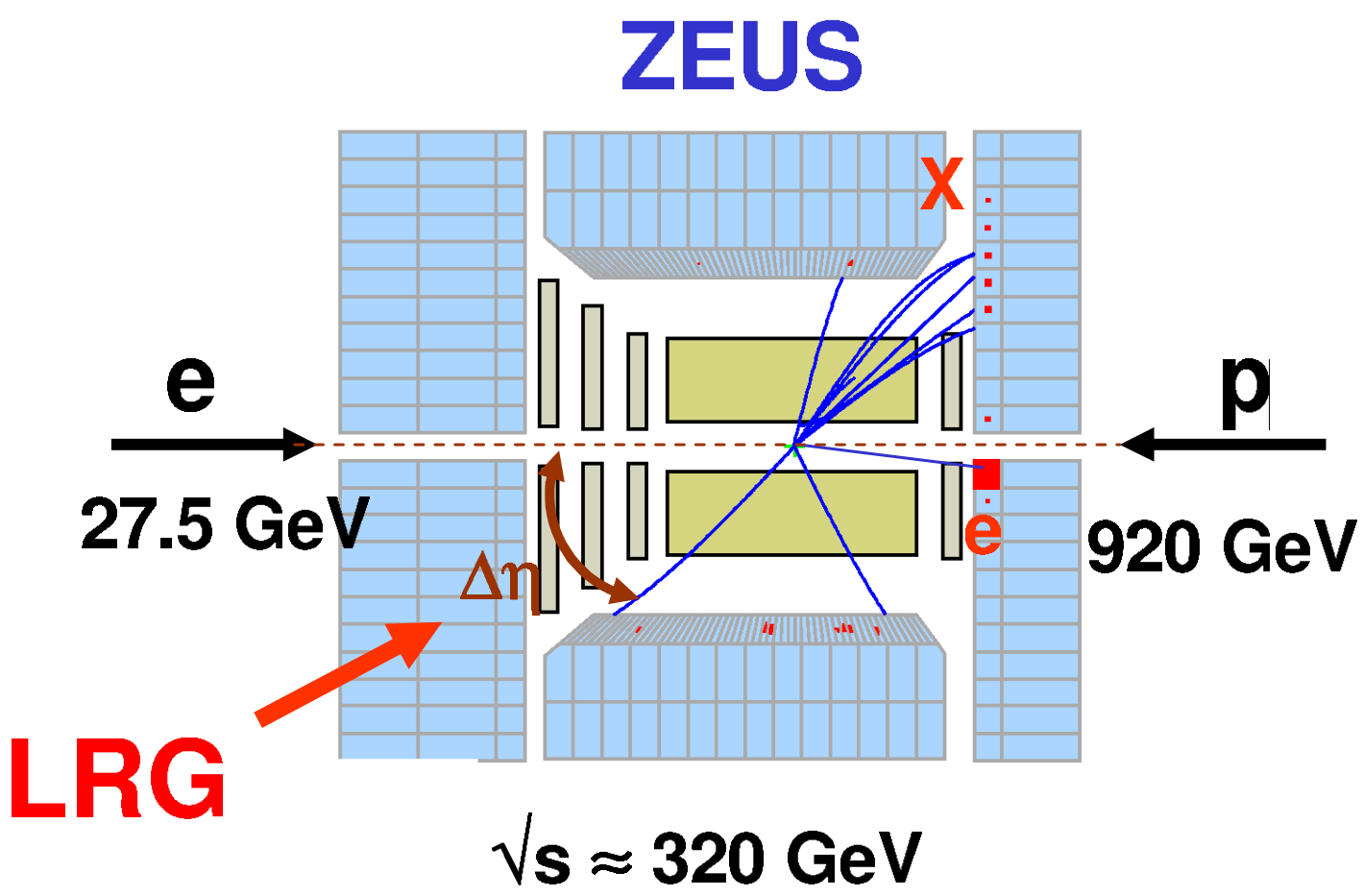

**Fig. 6:** A DIS event with a large rapidity gap (LRG) observed with the ZEUS detector at HERA. The scattered proton escapes into the beam-pipe. The symbol $\Delta\eta$ denotes the difference in pseudorapidity between the scattered proton and the most forward particle of the observed hadronic system $X$. Pseudorapidity is defined as $\eta = -\ln\tan(\theta/2)$ in terms of the polar angle $\theta$ measured with respect to the incoming proton direction, which is defined as "forward".

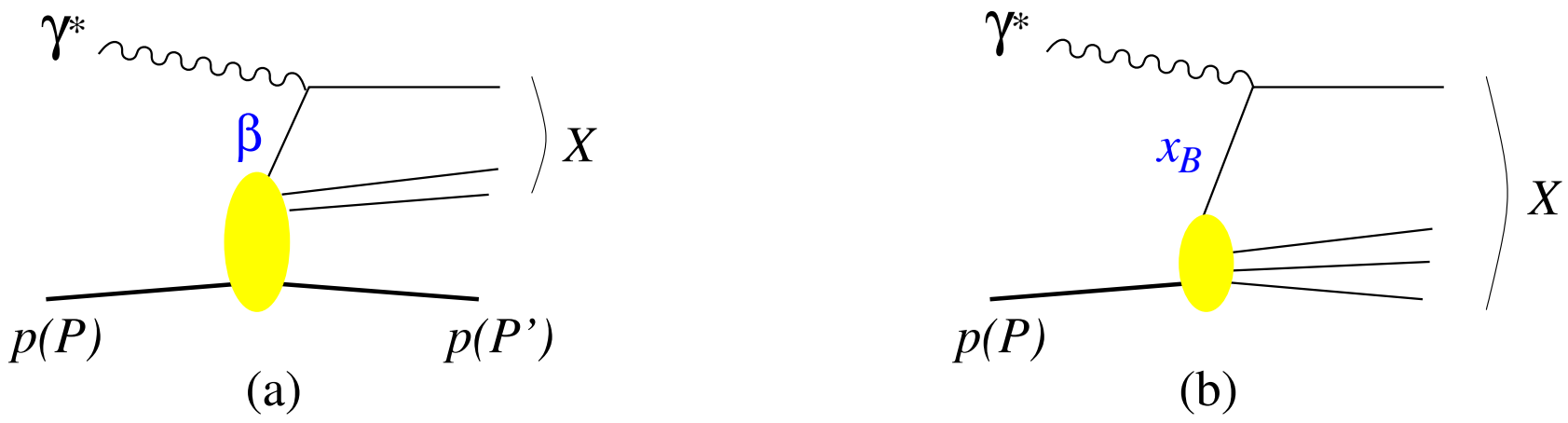

**Fig. 7:** Parton model diagrams for deep inelastic diffractive (a) and inclusive (b) scattering. The variable $\beta$ is the momentum fraction of the struck quark with respect to $P - P'$, and $x_B$ its momentum fraction with respect to $P$.

its contribution to the cross section is small in a wide range of the experimentally accessible kinematic region (in particular at low $y$). The structure function $F_2^{D(3)}$ is obtained from $F_2^{D(4)}$ by integrating over $t$:

$$F_2^{D(3)}(\beta, Q^2, x_{I\!P}) = \int dt\, F_2^{D(4)}(\beta, Q^2, x_{I\!P}, t). \tag{5}$$

In a parton model picture, inclusive diffraction $\gamma^* p \to Xp$ proceeds by the virtual photon scattering on a quark, in analogy to inclusive scattering (see Fig. 7). In this picture, $\beta$ is the momentum fraction of the struck quark with respect to the exchanged momentum $P - P'$ (indeed the allowed kinematical range of $\beta$ is between 0 and 1). The diffractive structure function describes the proton structure in these specific processes with a fast proton in the final state. $F_2^D$ may also be viewed as describing the structure of whatever is exchanged in the $t$-channel in diffraction, i.e. of the Pomeron (if multiple Pomeron exchange can be neglected). It is however important to bear in mind that the Pomeron in QCD cannot be interpreted as a particle on which the virtual photon scatters, as we will see in Sect. 2.5.

Figures 8 and 9 show recent H1 data [7] on $F_2^{D(3)}$ at fixed $x_{I\!P}$ as a function of $\beta$ for different $Q^2$ bins, and as a function of $Q^2$ for different bins of $\beta$.[2] The data have two remarkable features:

[2] To be precise, the H1 data are for the so-called reduced diffractive cross section, which equals $F_2^{D(3)}$ if $F_L^D$ can be neglected.

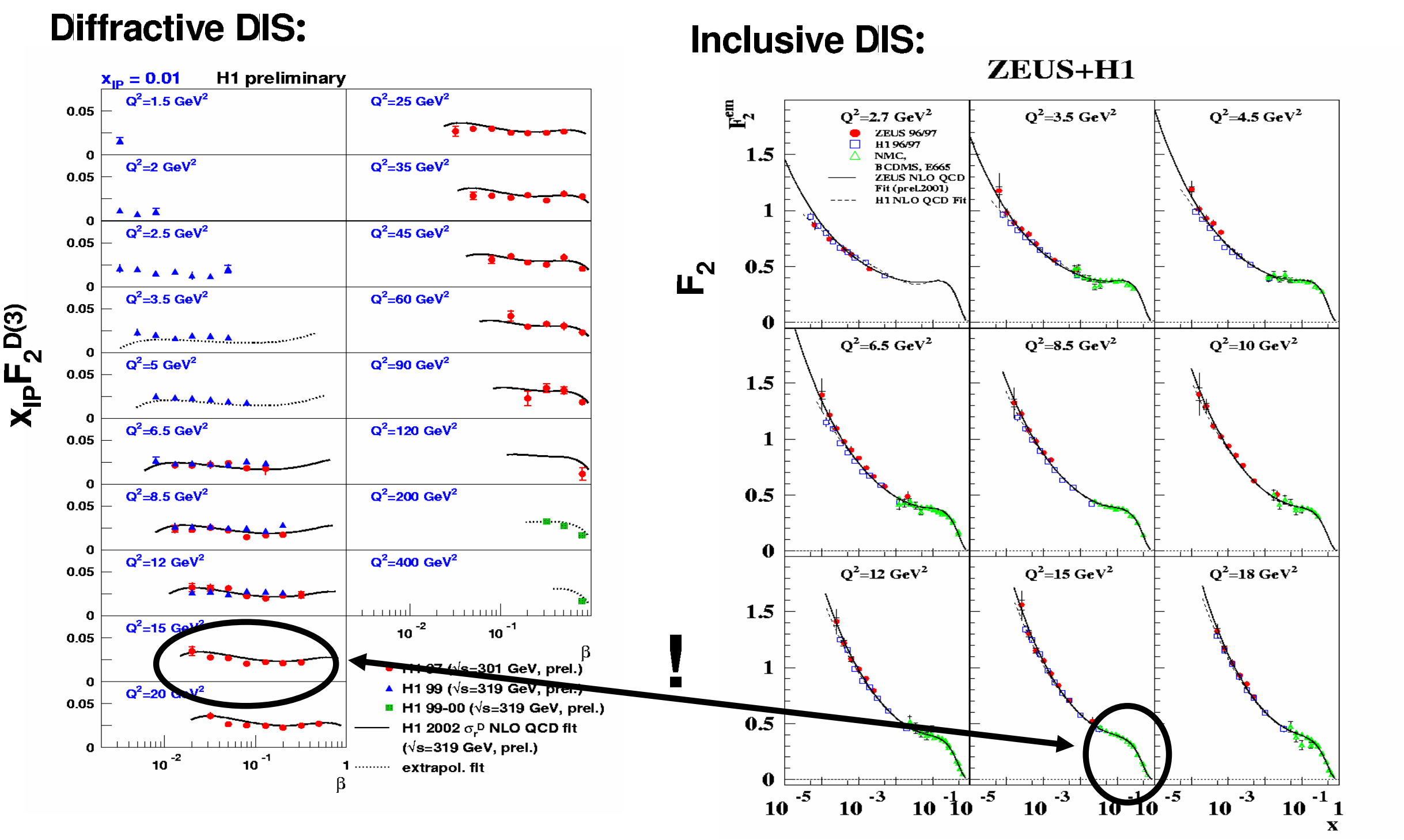

**Fig. 8:** Left: the diffractive structure function of the proton as a function of $\beta$ (from [7]). Right: the structure function of the proton as a function of $x_B$ (from [8]). The two highlighted bins show the different shapes of $F_2^D$ and $F_2$ in corresponding ranges of $\beta$ and $x_B$ at equal $Q^2$.

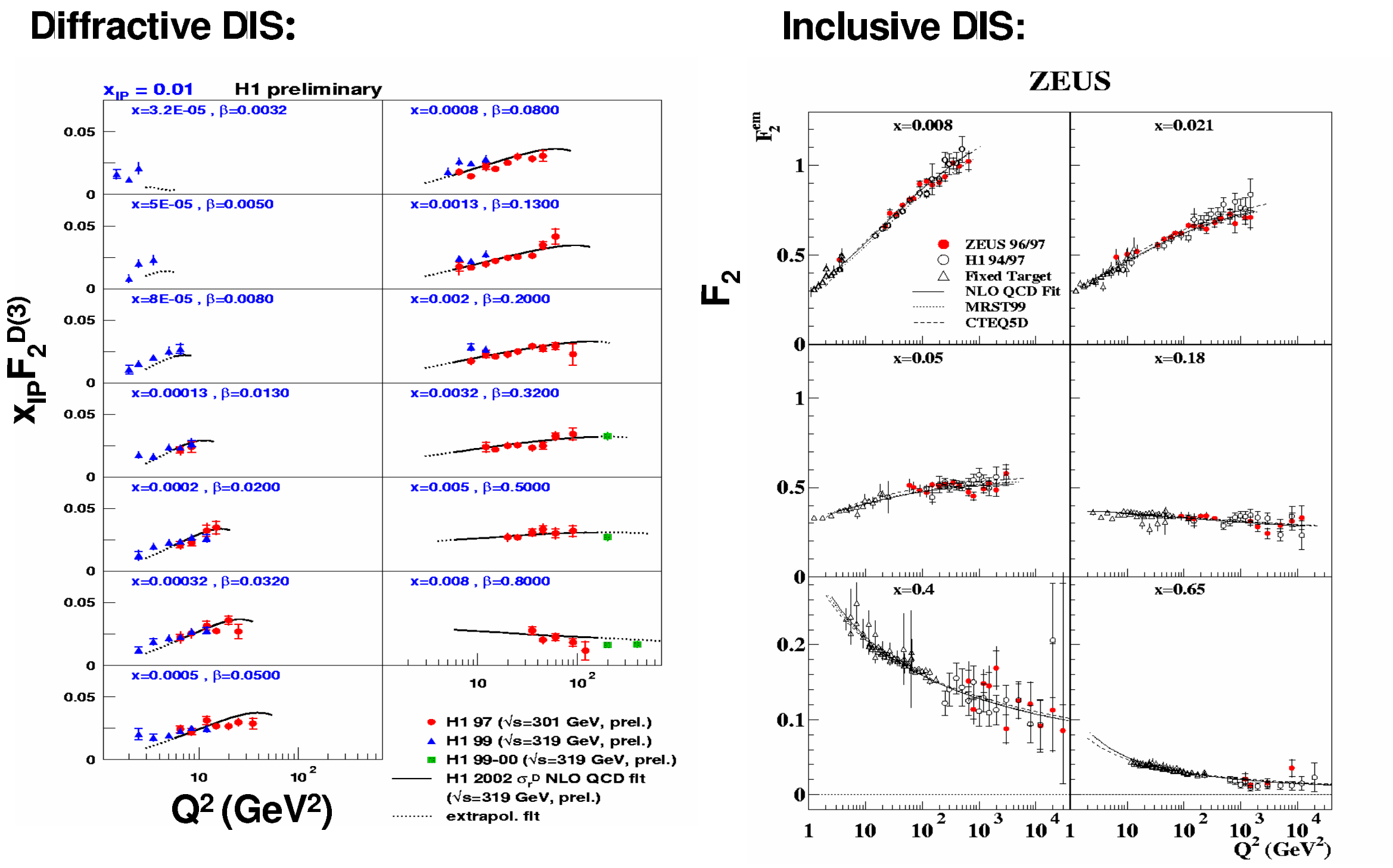

**Fig. 9:** Left: the diffractive structure function of the proton as a function of $Q^2$ (from [7]). Right: the structure function of the proton as a function of $Q^2$ (from [9]).

- $F_2^D$ is largely flat in the measured $\beta$ range. Keeping in mind the analogy between $\beta$ in diffractive DIS and $x_B$ in inclusive DIS, this is very different from the behavior of the "usual" structure function $F_2$, which strongly decreases for $x_B \gtrsim 0.2$ (see Fig. 8).
- The dependence on $Q^2$ is logarithmic, i.e. one observes approximate Bjorken scaling. This indicates the applicability of the parton model picture to inclusive $\gamma^* p$ diffraction. The structure function $F_2^D$ increases with $Q^2$ for all $\beta$ values except the highest. This is reminiscent of the scaling violations of $F_2$, except that $F_2$ rises with $Q^2$ only for $x_B \lesssim 0.2$ and that the scaling violations become negative at higher $x_B$ (see Fig. 9). In the proton, negative scaling violations reflect the presence of the valence quarks radiating gluons, while positive scaling violations are due to the increase of the sea quark and gluon densities as the proton is probed with higher resolution. The $F_2^D$ data thus suggest that the partons resolved in diffractive events are predominantly gluons. This is not too surprising if one bears in mind that these partons carry only a small part of the proton momentum: the struck quark in the diagram of Fig. 7a has a momentum fraction $\beta x_{I\!P} = x_B$ with respect to the incident proton, and $x_{I\!P} \lesssim 0.02 - 0.03$ in diffractive events.

## 2.2 Diffractive parton distributions

The conclusion just reached can be made quantitative by using the QCD factorization theorem for inclusive diffraction, $\gamma^* p \to Xp$, which formalizes the parton model picture we have already invoked in our discussion. According to this theorem, the diffractive structure function, in the limit of large $Q^2$ at fixed $\beta$, $x_{I\!P}$ and $t$, can be written as [10–12]

$$F_2^{D(4)}(\beta, Q^2, x_{I\!P}, t) = \sum_i \int_\beta^1 \frac{dz}{z}\, C_i\Big(\frac{\beta}{z}\Big)\, f_i^D(z, x_{I\!P}, t; Q^2), \tag{6}$$

where the sum is over partons of type $i$. The coefficient functions $C_i$ describe the scattering of the virtual photon on the parton and are exactly the same as in inclusive DIS. In analogy to the usual parton distribution functions (PDFs), the diffractive PDFs $f_i^D(z, x_{I\!P}, t; Q^2)$ can be defined as operator matrix elements in a proton state, and their dependence on the scale $Q^2$ is given by the DGLAP evolution equations. In parton model language, they can be interpreted as conditional probabilities to find a parton $i$ with fractional momentum $z x_{I\!P}$ in a proton, probed with resolution $Q^2$ in a process with a fast proton in the final state (whose momentum is specified by $x_{I\!P}$ and $t$).

During the workshop, several fits of the available $F_2^D$ data were discussed which are based on the factorization formula (6) at next-to-leading order (NLO) in $\alpha_s$ [13,14]. Figure 10 compares the diffractive PDFs from an earlier H1 fit [7] to those from the fit of the ZEUS data [15] by Schilling and Newman [13]. As expected the density of gluons is larger than that of quarks, by about a factor 5–10. Discrepancies between the two sets are evident, and it remains to be clarified to which extent they reflect differences in the fitted data. Martin, Ryskin and Watt [16] have argued that the leading-twist formula (6) is inadequate in large parts of the measured kinematics, and performed a fit to a modified expression which includes an estimate of power-suppressed effects. The discrepancies between the various diffractive PDFs, while not fully understood, may be taken as an estimate of the uncertainties on these functions at this point in time. A precise and consistent determination of the diffractive PDFs and their uncertainties is one of the main tasks the HERA community has to face in the near future. They are a crucial input for predicting cross sections of inclusive diffractive processes at the LHC.

## 2.3 Diffractive hard-scattering factorization

Like usual parton densities, diffractive PDFs are process-independent functions. They appear not only in inclusive diffraction but also in other processes where diffractive hard-scattering factorization holds. In analogy with Eq. (6), the cross section of such a process can be evaluated as the convolution of the relevant parton-level cross section with the diffractive PDFs. For instance, the cross section for charm production in diffractive DIS can be calculated at leading order in $\alpha_s$ from the $\gamma^* g \to c\bar{c}$ cross section and the diffractive gluon distribution. An analogous statement holds for jet production in diffractive DIS. Both processes have been analyzed at next-to-leading order in $\alpha_s$.

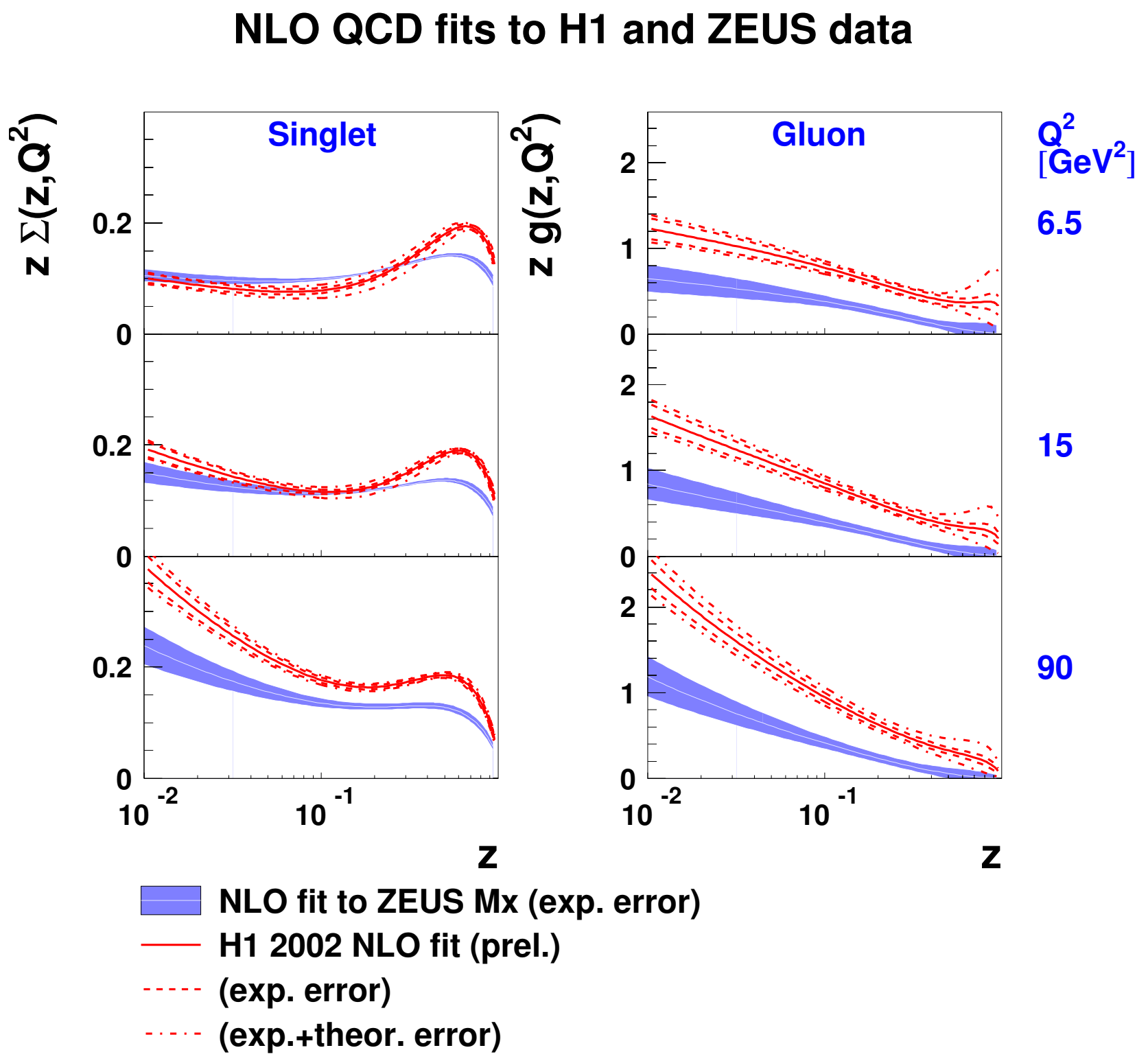

**Fig. 10:** Diffractive quark singlet and gluon distributions obtained from fits to H1 [7] and ZEUS [15] data (from [13]).

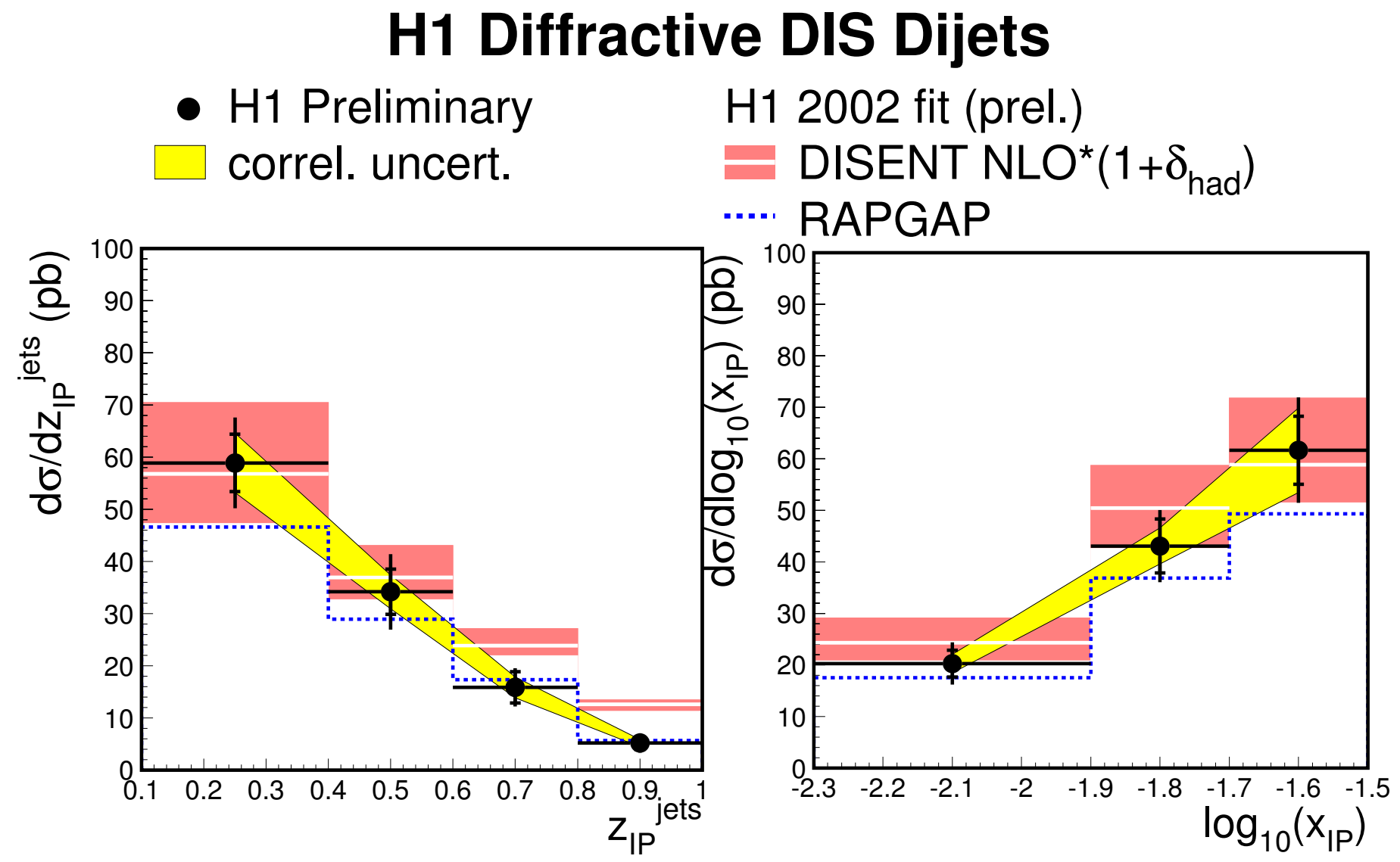

(a) (b)

**Fig. 11:** Cross section for dijet production in diffractive DIS, compared with the expectations based on the diffractive PDFs [7] (from [17]). The variable $z_{I\!P}^{\rm jets}$ estimates the fractional momentum of the parton entering the hard subprocess.

As an example, Fig. 11 shows a comparison between the measured cross sections for diffractive dijet production and the expectations based on diffractive PDFs extracted from a fit to $F_2^D$. These data lend support to the validity of hard-scattering factorization in diffractive $\gamma^* p$ interactions. For further discussion we refer the reader to [18].

### 2.4 Limits of diffractive hard-scattering factorization: hadron-hadron collisions

A natural question to ask is whether one can use the diffractive PDFs extracted at HERA to describe hard diffractive processes such as the production of jets, heavy quarks or weak gauge bosons in $p\bar{p}$ collisions at the Tevatron. Figure 12 shows results on diffractive dijet production from the CDF collaboration [19] compared to the expectations based on the diffractive PDFs [6, 7] from HERA. The discrepancy is spectacular: the fraction of diffractive dijet events at CDF is a factor 3 to 10 smaller than would be expected on the basis of the HERA data. The same type of discrepancy is consistently observed in all hard diffractive processes in $p\bar{p}$ events, see e.g. [20]. In general, while at HERA hard diffraction contributes a fraction of order 10% to the total cross section, it contributes only about 1% at the Tevatron.

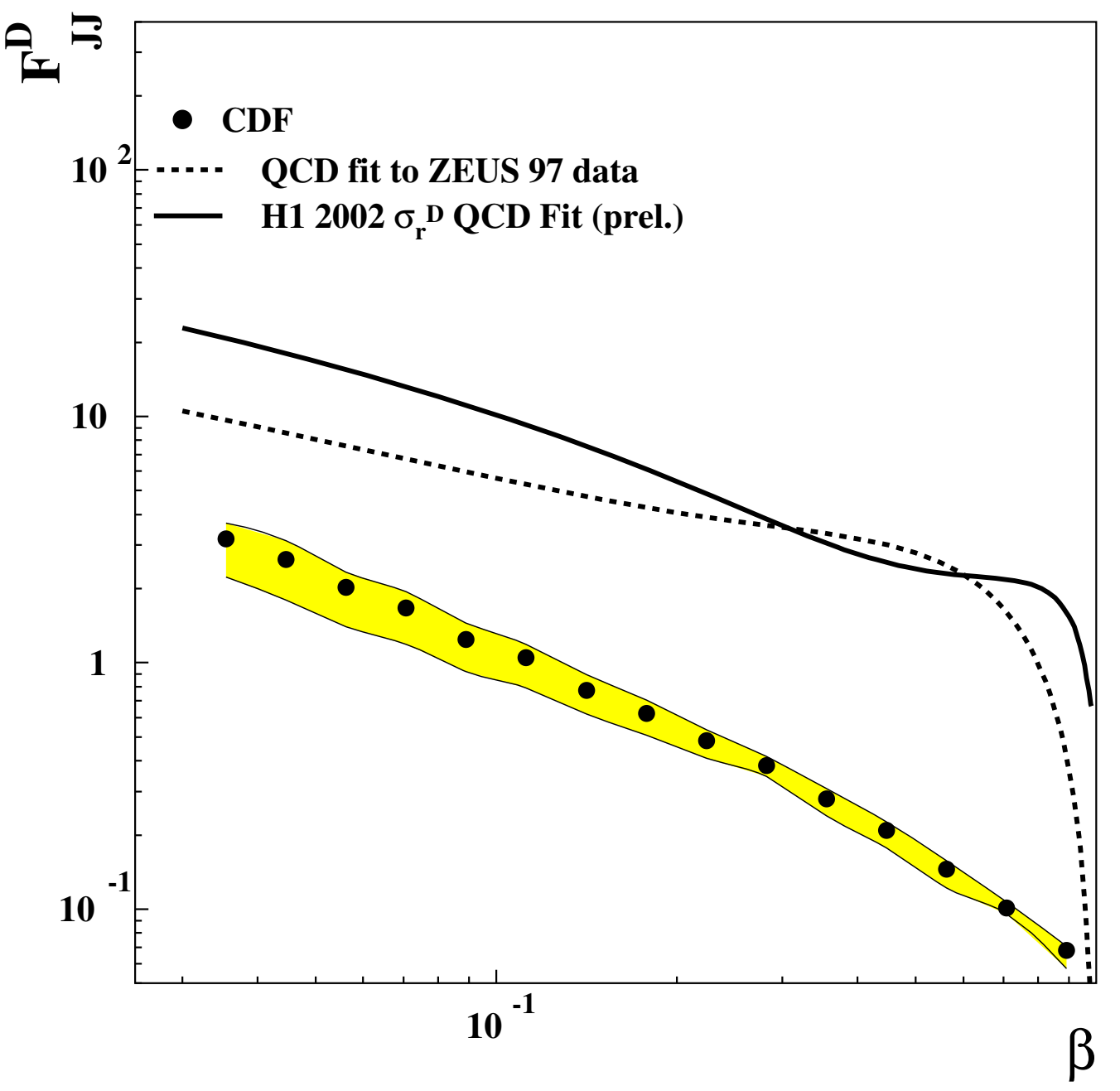

**Fig. 12:** CDF results for the cross section of diffractive dijet production with a leading antiproton in $p\bar{p}$ collisions (expressed in terms of a structure function $F_{JJ}^D$), compared with the predictions obtained from the diffractive PDFs [6] and [7] extracted at HERA (from [21]). See also the analogous plot in the original CDF publication [19].

In fact, diffractive hard-scattering factorization does not apply to hadron-hadron collisions [11,12]. Attempts to establish corresponding factorization theorems fail because of interactions between spectator partons of the colliding hadrons. The contribution of these interactions to the cross section does not decrease with the hard scale. Since they are not associated with the hard-scattering subprocess (see Fig. 13), we no longer have factorization into a parton-level cross section and the parton densities of one of the colliding hadrons. These interactions are generally soft, and we have at present to rely on phenomenological models to quantify their effects [22].

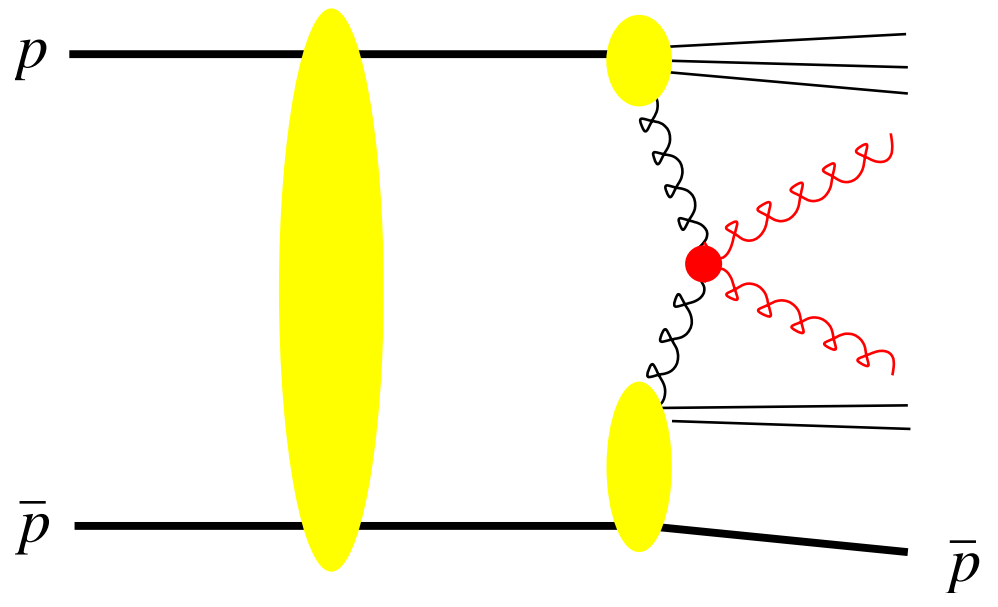

**Fig. 13:** Example graph for diffractive dijet production with a leading antiproton in a $p\bar{p}$ collision. The interaction indicated by the large vertical blob breaks hard diffractive factorization. It reduces the diffractive cross section, as explained in the text.

The yield of diffractive events in hadron-hadron collisions is lowered precisely because of these soft interactions between spectator partons (often referred to as "reinteractions" or "multiple scatterings"). They can produce additional final-state particles which fill the would-be rapidity gap (hence the often-used term "rapidity gap survival"). When such additional particles are produced, a very fast proton can no longer appear in the final state because of energy conservation. Diffractive factorization breaking is thus intimately related to multiple scattering in hadron-hadron collisions; understanding and describing this phenomenon is a challenge in the high-energy regime that will be reached at the LHC [23].

In $pp$ or $p\bar{p}$ reactions, the collision partners are both composite systems of large transverse size, and it is not too surprising that multiple interactions between their constituents can be substantial. In contrast, the virtual photon in $\gamma^* p$ collisions has small transverse size, which disfavors multiple interactions and enables diffractive factorization to hold. According to our discussion in Sect. 1.2, we may expect that for decreasing virtuality $Q^2$ the photon behaves more and more like a hadron, and diffractive factorization may again be broken. This aspect of diffractive processes in photoproduction at HERA was intensively discussed during the workshop, and findings are reported in [18].

## 2.5 Space-time structure: the Pomeron is not a particle

It is tempting to interpret diffractive $\gamma^* p$ processes as the scattering of a virtual photon on a Pomeron which has been radiated off the initial proton. Diffractive DIS would then probe the distribution of partons in a "Pomeron target". This is indeed the picture proposed by Ingelman and Schlein long ago [24].

This picture is however not supported by an analysis in QCD (see e.g. [25]). There, high-energy scattering is dominated by the exchange of two gluons, whose interaction is (in an appropriate gauge) described by ladder diagrams, as shown in Fig. 14. By analyzing these diagrams in time-ordered perturbation theory, one can obtain the dominant space-time ordering in the high-energy limit. The result depends on the reference frame, as illustrated in the figure. In the Breit frame, which is natural for a parton-model interpretation, the photon does *not* scatter off a parton in a pre-existing two-gluon system; in fact some of the interactions in the gluon ladder building up the Pomeron exchange take place long after the virtual photon has been absorbed. The picture in the Breit frame is however compatible with the interpretation of diffractive parton densities given in Sect. 2.2, namely the probability to find a parton under the condition that subsequent interactions will produce a fast proton in the final state.

We note that the Ingelman-Schlein picture suggests that the diffractive structure function takes a factorized form $F_2^{D(4)} = f_{I\!P}(x_{I\!P}, t)\, F_2^{I\!P}(\beta, Q^2)$, where $f_{I\!P}$ is the "Pomeron flux" describing the emission of the Pomeron from the proton and its subsequent propagation, and where $F_2^{I\!P}$ is the "structure function of the Pomeron". Phenomenologically, such a factorizing ansatz works not too badly and is often used, but recent high-precision data have shown its breakdown at small $x_{I\!P}$ [15].

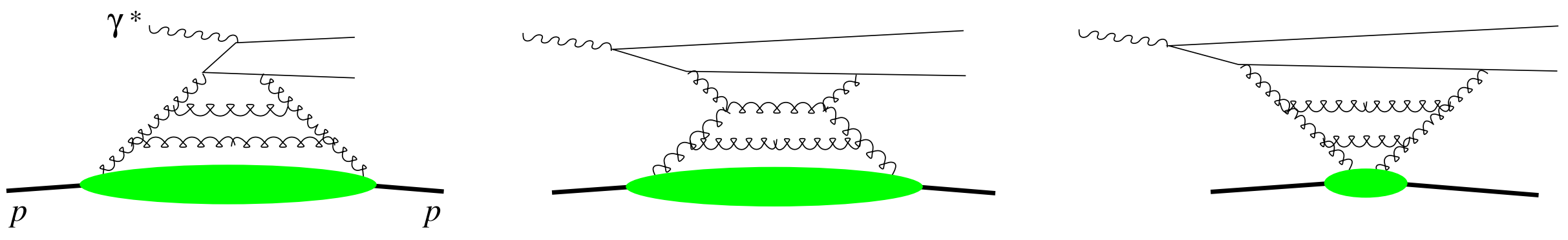

**Fig. 14:** Dominant time ordering for diffractive dissociation of a virtual photon in (a) the Breit frame, (b) the photon-proton center-of-mass, (c) the proton rest frame. The physical picture in (a) corresponds to the parton-model description of diffraction, and the one in (b) and (c) to the picture of the photon splitting into a quark-antiquark dipole which subsequently interacts with the proton.

## 3 Exclusive diffractive processes

Let us now discuss diffractive processes where a real or virtual photon dissociates into a single particle. Since diffraction involves the exchange of vacuum quantum numbers, this particle can in particular be a vector meson (which has the same $J^{PC}$ quantum numbers as the photon) – in this case the process is sometimes referred to as "elastic" vector meson production. Another important case is deeply virtual Compton scattering (DVCS), $\gamma^* p \to \gamma p$.[3] A striking feature of the data taken at HERA (Figs. 15 and 16) is that the energy dependence of these processes becomes steep in the presence of a hard scale, which can be either the photon virtuality $Q^2$ or the mass of the meson in the case of $J/\Psi$ or $\Upsilon$ production. This is similar to the energy dependence of the $\gamma^* p$ total cross section (related by the optical theorem to forward Compton scattering, $\gamma^* p \to \gamma^* p$), which changes from flat to steep when going from real photons to $Q^2$ of a few GeV$^2$.

To understand this similarity, let us recall that in perturbative QCD diffraction proceeds by two-gluon exchange. The transition from a virtual photon to a real photon or to a quark-antiquark pair subsequently hadronizing into a meson is a short-distance process involving these gluons, provided that either $Q^2$ or the quark mass is large. In fact, in an approximation discussed below, the cross sections for DVCS and vector meson production are proportional to the square of the gluon distribution in the proton, evaluated at a scale of order $Q^2 + M_V^2$ and at a momentum fraction $x_{I\!P} = (Q^2 + M_V^2)/(W^2 + Q^2)$, where the vector meson mass $M_V$ now takes the role of $M_X$ in inclusive diffraction [28]. In analogy to the case of the total $\gamma^* p$ cross section, the energy dependence of the cross sections shown in Figs. 15 and 16 thus reflects the $x$ and scale dependence of the gluon density in the proton, which grows with decreasing $x$ with a slope becoming steeper as the scale increases.

There is however an important difference in how the gluon distribution enters the descriptions of inclusive DIS and of exclusive diffractive processes. The inclusive DIS cross section is related via the optical theorem to the imaginary part of the forward virtual Compton amplitude, so that the graphs in Fig. 17 represent the *cross section* of the inclusive process. Hence, the gluon distribution in Fig. 17a gives the *probability* to find *one* gluon in the proton (with any number of unobserved spectator partons going into the final state). In contrast, the corresponding graphs for DVCS and exclusive meson production in Fig. 18 represent the *amplitudes* of exclusive processes, which are proportional to the *probability amplitude* for first extracting a gluon from the initial proton and then returning it to form the proton in the final state. In the approximation discussed below, this probability amplitude is given by the gluon distribution. The cross sections of DVCS and exclusive meson production are then proportional to the *square* of the gluon distribution.

A detailed theoretical analysis of DVCS and exclusive meson production at large $Q^2$ shows that short-distance factorization holds, in analogy to the case of inclusive DIS. QCD factorization theorems [29] state that in the limit of large $Q^2$ (at fixed Bjorken variable $x_B$ and fixed $t$) the Compton amplitude factorizes into a hard-scattering subprocess and a hadronic matrix element describing the emission and reabsorption of a parton by the proton target (see Fig. 18a). As shown in Fig. 18b, the analogous result for exclusive meson production involves in addition the quark-antiquark distribution amplitude of the meson (often termed the meson wave function) and thus a further piece of non-perturbative input.

[3] We do not discuss processes with diffractive dissociation of the proton in this paper, but wish to mention interesting studies of vector meson or real photon production at large $|t|$, where the proton predominantly dissociates, see e.g. [26].

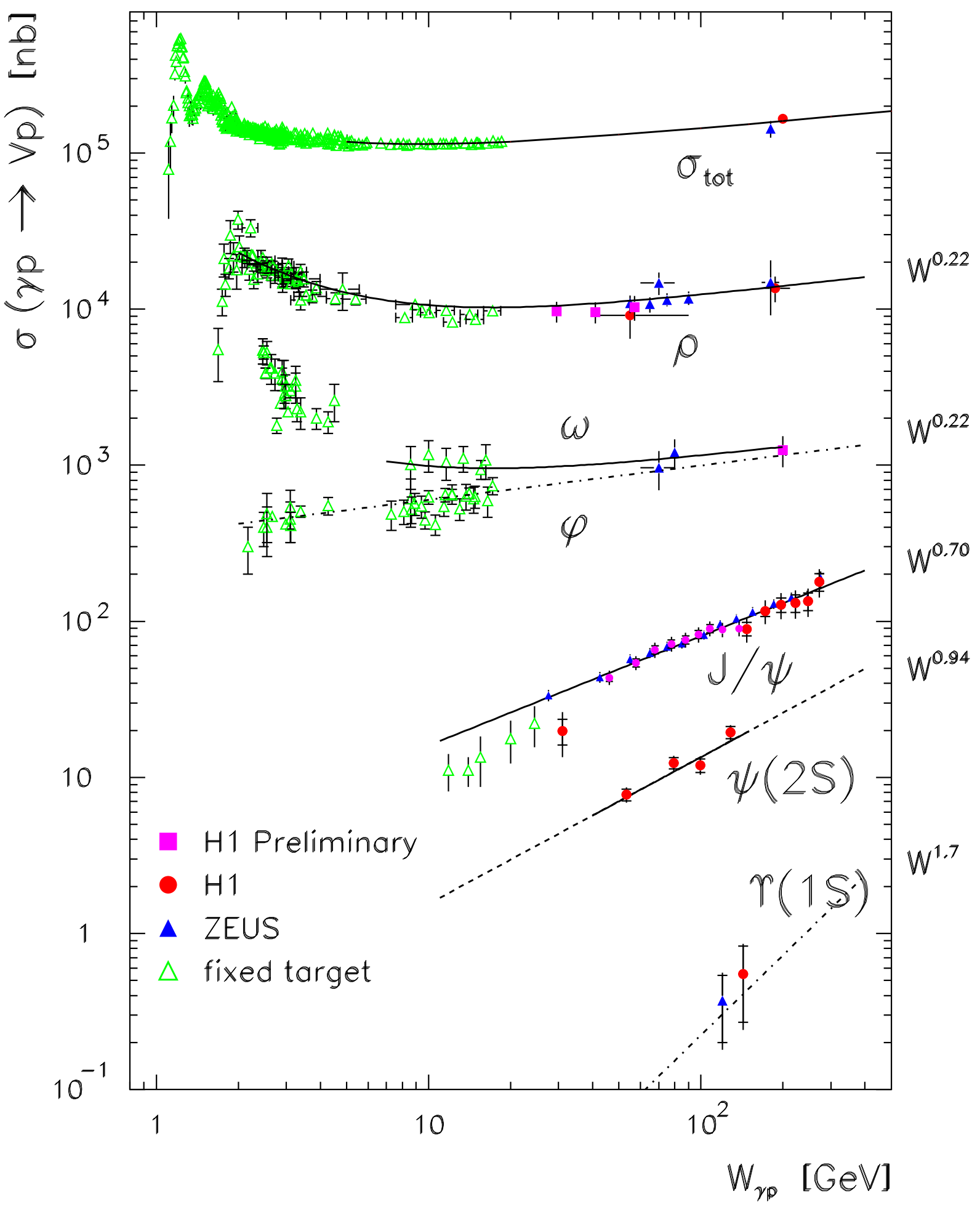

**Fig. 15:** Compilation of results on the cross section for vector meson photoproduction, $\gamma p \to Vp$, with $V = \rho, \omega, \phi, J/\Psi, \psi', \Upsilon$, as a function $W$. The total $\gamma p$ cross section $\sigma_{\mathrm{tot}}$ is also shown.

The hadronic matrix elements appearing in the factorization formulae for exclusive processes would be the usual PDFs if the proton had the same momentum in the initial and final state. Since this is not the case, they are more general functions taking into account the momentum difference between the initial and final state proton (or, equivalently, between the emitted and reabsorbed parton). These "generalized parton distributions" (GPDs) depend on two independent longitudinal momentum fractions instead of a single one (compare Figs. 17a and 18a), on the transverse momentum transferred to the proton (whose square is $-t$ to a good approximation at high energy), and on the scale at which the partons are probed. The scale dependence of the GPDs is governed by a generalization of the DGLAP equations. The dependence on the difference of the longitudinal momenta (often called "skewness") contains information on correlations between parton momenta in the proton wave function. It can be neglected in the approximation of leading $\log x$ (then the GPDs at $t = 0$ reduce to the usual PDFs as anticipated above), but it is numerically important in typical HERA kinematics. The dependence on $t$ allows for a very intuitive interpretation if a Fourier transformation is performed with respect to the transverse momentum transfer. We then obtain distributions depending on the impact parameter of the partons, which describe the two-dimensional distribution of the struck parton in the transverse plane, and on its longitudinal momentum fraction in the proton. The $t$ dependence of exclusive diffractive processes thus provides unique information beyond the longitudinal momentum spectra encoded in the usual parton densities. The study of the generalized parton distributions is a prime reason to measure DVCS and exclusive meson production in $ep$ scattering. Detailed discussions and references can be found in the recent reviews [30, 31].

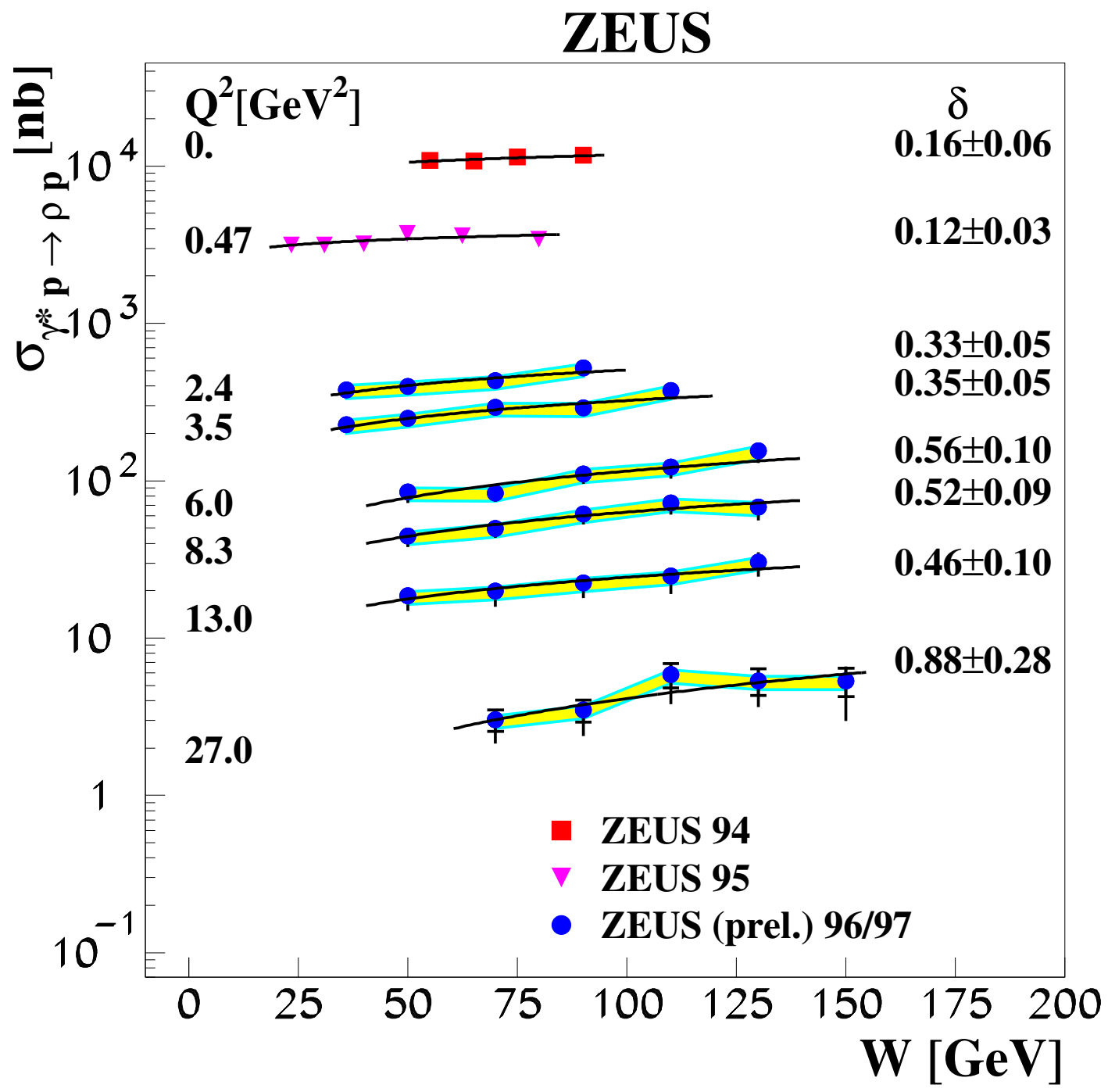

**Fig. 16:** Cross section for exclusive $\rho$ production as a function of $W$ (from [27]). The lines represent the result of fits to the data with the form $\sigma(\gamma^* p \to \rho p) \propto W^\delta$, yielding the exponents given in the figure.

An observable illustrating the short-distance factorization in meson production at high $Q^2$ is the ratio of the $\phi$ and $\rho$ production cross sections, shown in Fig. 19. At large $Q^2$ the process is described in terms of a light quark coupling to the photon and of the generalized gluon distribution. Using approximate flavor SU(3) symmetry between the $\rho$ and $\phi$ wave functions, the only difference between the two channels is then due to different quark charge and isospin factors, which result in a cross section ratio of $2/9$.

### 3.1 High-energy factorization and the dipole picture

So far we have discussed the description of hard exclusive diffraction within short-distance, or collinear factorization. A different type of factorization is high-energy, or $k_t$ factorization, which is based on the BFKL formalism. Here the usual or generalized gluon distribution appearing in the factorization formulae depends explicitly on the transverse momentum $k_t$ of the emitted gluon. In collinear factorization, this $k_t$ is integrated over in the parton distributions and set to zero when calculating the hard-scattering process (the partons are thus approximated as "collinear" with their parent hadron). Likewise, the meson wave functions appearing in $k_t$ factorization explicitly depend on the relative transverse momentum between the quark and antiquark in the meson, whereas this is integrated over in the quark-antiquark distribution amplitudes (cf. Sect. 3) of the collinear factorization formalism. Only gluon distributions appear in $k_t$ factorization, whereas collinear factorization formulae involve both quark and gluon distributions (see e.g. Sects. 8.1 and 8.2 in [30] for a discussion of this difference). We note that other

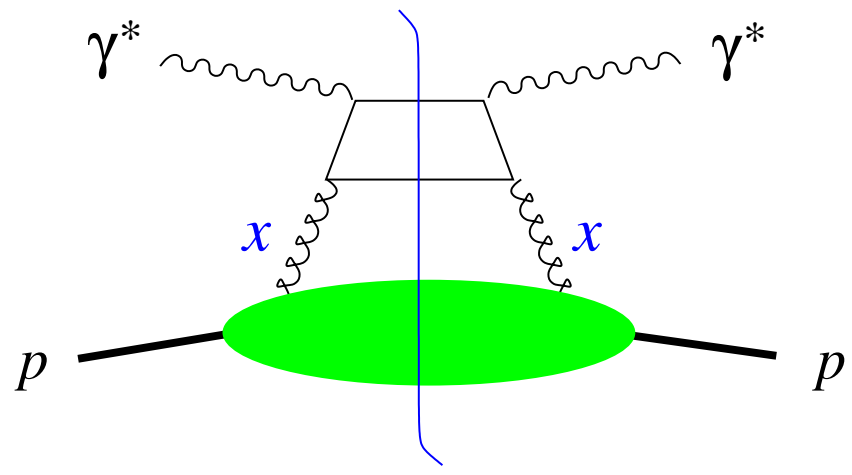

(a)

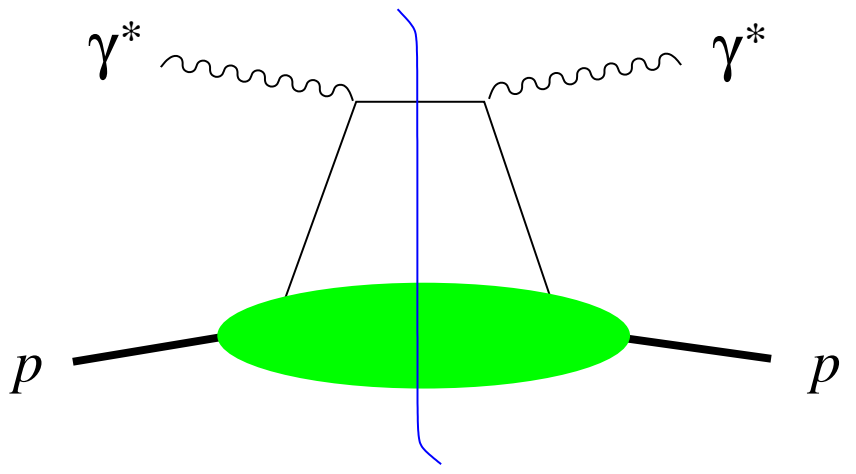

(b)

**Fig. 17:** Factorization of forward Compton scattering, which is related to the total inclusive structure function via the optical theorem, $\mathrm{Im}\,\mathcal{A}(\gamma^* p \to \gamma^* p) = \frac{1}{2}\sum_X |\mathcal{A}(\gamma^* p \to X)|^2 \propto \sigma(\gamma^* p \to X)$. The final state of the inclusive process is obtained by cutting the diagrams along the vertical line. The blobs represent the gluon or quark distribution in the proton. Graph (b) is absent in the $k_t$ factorization formalism (see Sect. 3.1): its role is taken by graph (a) in the "aligned jet configuration", where the quark line joining the two photons carries almost the entire photon momentum.

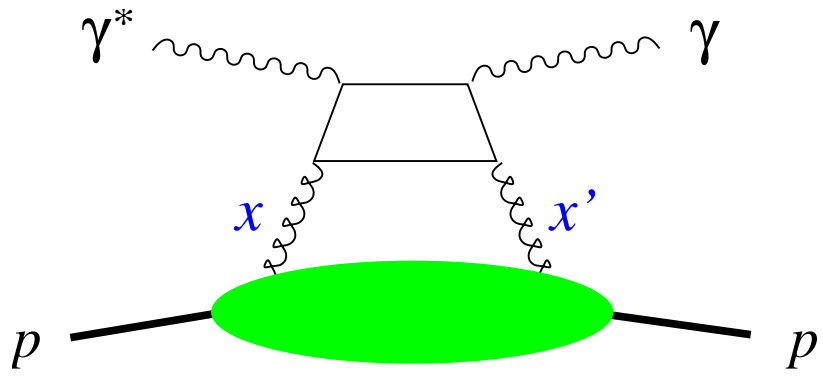

(a)

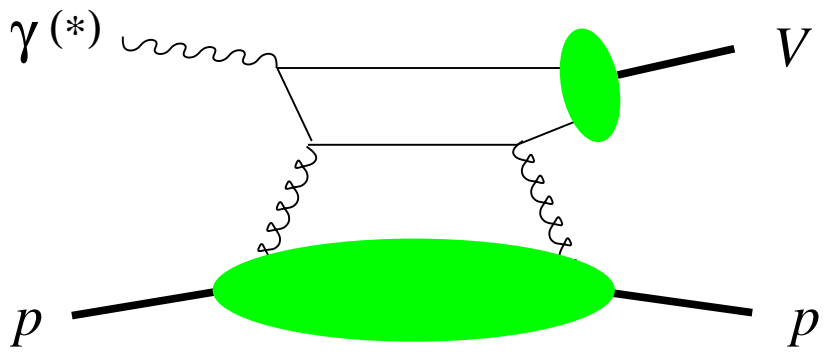

(b)

**Fig. 18:** (a) Factorization of deeply virtual Compton scattering, $\gamma^* p \to \gamma p$, which can be measured in the exclusive process $ep \to ep\gamma$. The blob represents the generalized gluon distribution, with $x$ and $x'$ denoting the momentum fractions of the gluons. (b) Factorization of exclusive meson production. The small blob represents the vector meson wave function. In the collinear factorization formalism, there are further graphs (not shown) involving quark instead of gluon exchange.

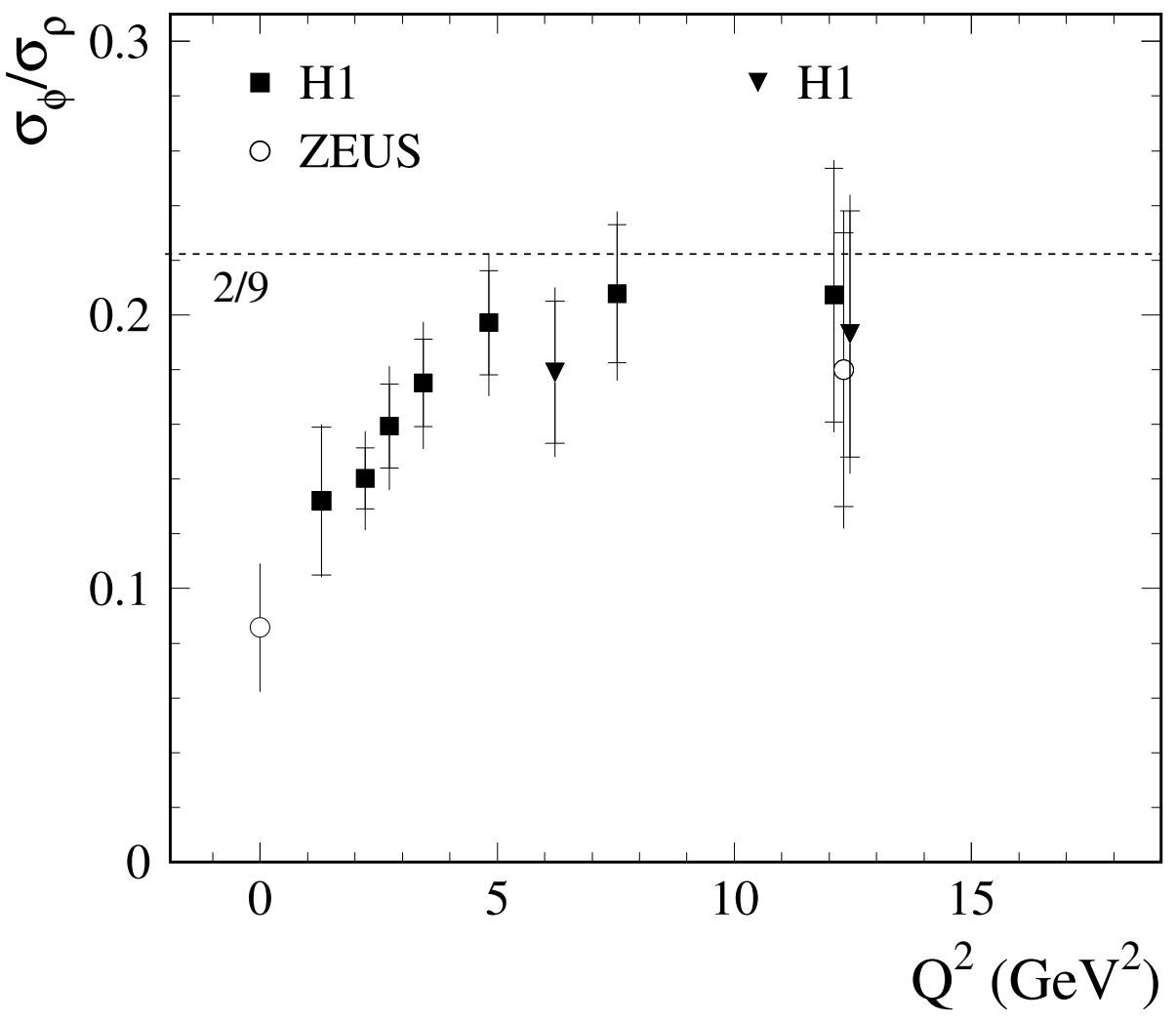

**Fig. 19:** The ratio of cross sections for $\gamma^* p \to \phi p$ and $\gamma^* p \to \rho p$ as a function of the photon virtuality (from [32]).

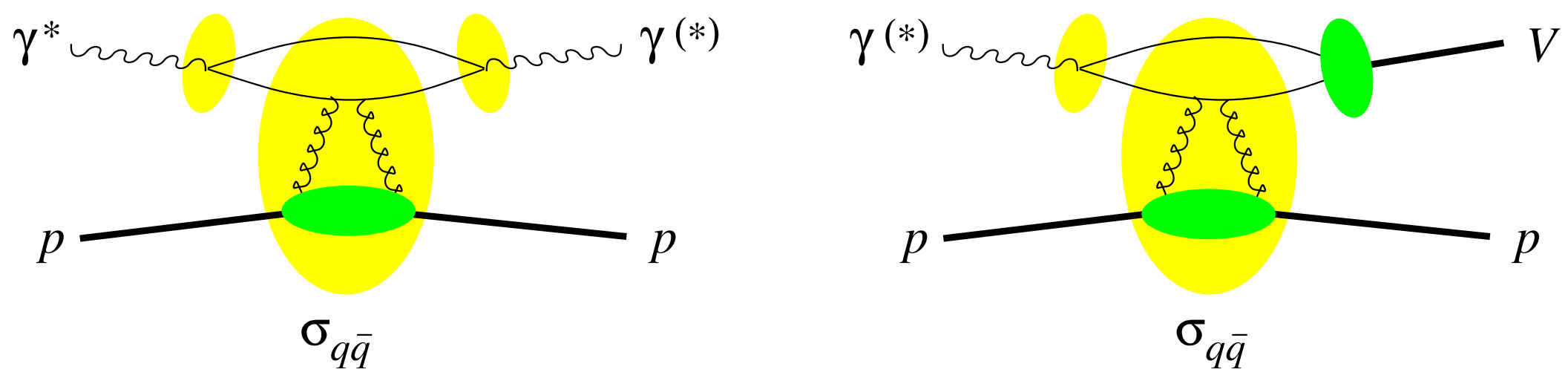

**Fig. 20:** The dipole representation of the amplitudes for Compton scattering (a) and for meson production (b), corresponding to the graphs in Figs. 17a and 18.

factorization schemes have been developed, which combine features of the collinear and $k_t$ factorization formalisms.

The two different types of factorization implement different ways of separating different parts of the dynamics in a scattering process. The building blocks in a short-distance factorization formula correspond to either small or large particle virtuality (or equivalently to small or large transverse momentum), whereas the separation criterion in high-energy factorization is the particle rapidity. Collinear and $k_t$ factorization are based on taking different limits: in the former case the limit of large $Q^2$ at fixed $x_B$ and in the latter case the limit of small $x_B$ at fixed $Q^2$ (which must however be large enough to justify the use of QCD perturbation theory). In the common limit of large $Q^2$ and small $x_B$ the two schemes give coinciding results. Instead of large $Q^2$ one can also take a large quark mass in the limits just discussed.

A far-reaching representation of high-energy dynamics can be obtained by casting the results of $k_t$ factorization into a particular form. The different building blocks in the graphs for Compton scattering and meson production in Figs. 17a and 18 can be rearranged as shown in Fig. 20. The result admits a very intuitive interpretation in a reference frame where the photon carries large momentum (this may be the proton rest frame but also a frame where the proton moves fast, see Fig. 14): the initial photon splits into a quark-antiquark pair, which scatters on the proton and finally forms a photon or meson again. This is the picture we have already appealed to in Sect. 1.2.

In addition, one can perform a Fourier transformation and trade the relative transverse momentum between quark and antiquark for their transverse distance $r$, which is conserved in the scattering on the target. The quark-antiquark pair acts as a color dipole, and its scattering on the proton is described by a "dipole cross section" $\sigma_{q\bar{q}}$ depending on $r$ and on $x_{I\!P}$ (or on $x_B$ in the case of inclusive DIS). The wave functions of the photon and the meson depend on $r$ after Fourier transformation, and at small $r$ the photon wave function is perturbatively calculable. Typical values of $r$ in a scattering process are determined by the inverse of the hard momentum scale, i.e. $r \sim (Q^2 + M_V^2)^{-1/2}$. An important result of high-energy factorization is the relation

$$\sigma_{q\bar{q}}(r, x) \propto r^2 x g(x) \tag{7}$$

at small $r$, where we have replaced the generalized gluon distribution by the usual one in the spirit of the leading $\log x$ approximation. A more precise version of the relation (7) involves the $k_t$ dependent gluon distribution. The dipole cross section vanishes at $r = 0$ in accordance with the phenomenon of "color transparency": a hadron becomes more and more transparent for a color dipole of decreasing size.

The scope of the dipole picture is wider than we have presented so far. It is tempting to apply it outside the region where it can be derived in perturbation theory, by modeling the dipole cross section and the photon wave function at large distance $r$. This has been very been fruitful in phenomenology, as we will see in the next section.

The dipole picture is well suited to understand the $t$ dependence of exclusive processes, parameterized as $d\sigma/dt \propto \exp(-b|t|)$ at small $t$. Figure 21 shows that $b$ decreases with increasing scale $Q^2 + M_V^2$ and at high scales becomes independent of the produced meson. A Fourier transform from momentum to impact parameter space readily shows that $b$ is related to the typical transverse distance between the colliding objects, as anticipated by the analogy with optical diffraction in Sect. 1.1. At high scale, the $q\bar{q}$ dipole is almost pointlike, and the $t$ dependence of the cross section is controlled by the $t$ dependence of the generalized gluon distribution, or in physical terms, by the transverse extension of the proton. As the scale decreases, the dipole acquires a size of its own, and in the case of $\rho$ or $\phi$ photoproduction, the values of $b$ reflect the fact that the two colliding objects are of typical hadronic dimensions; similar values would be obtained in elastic meson-proton scattering.

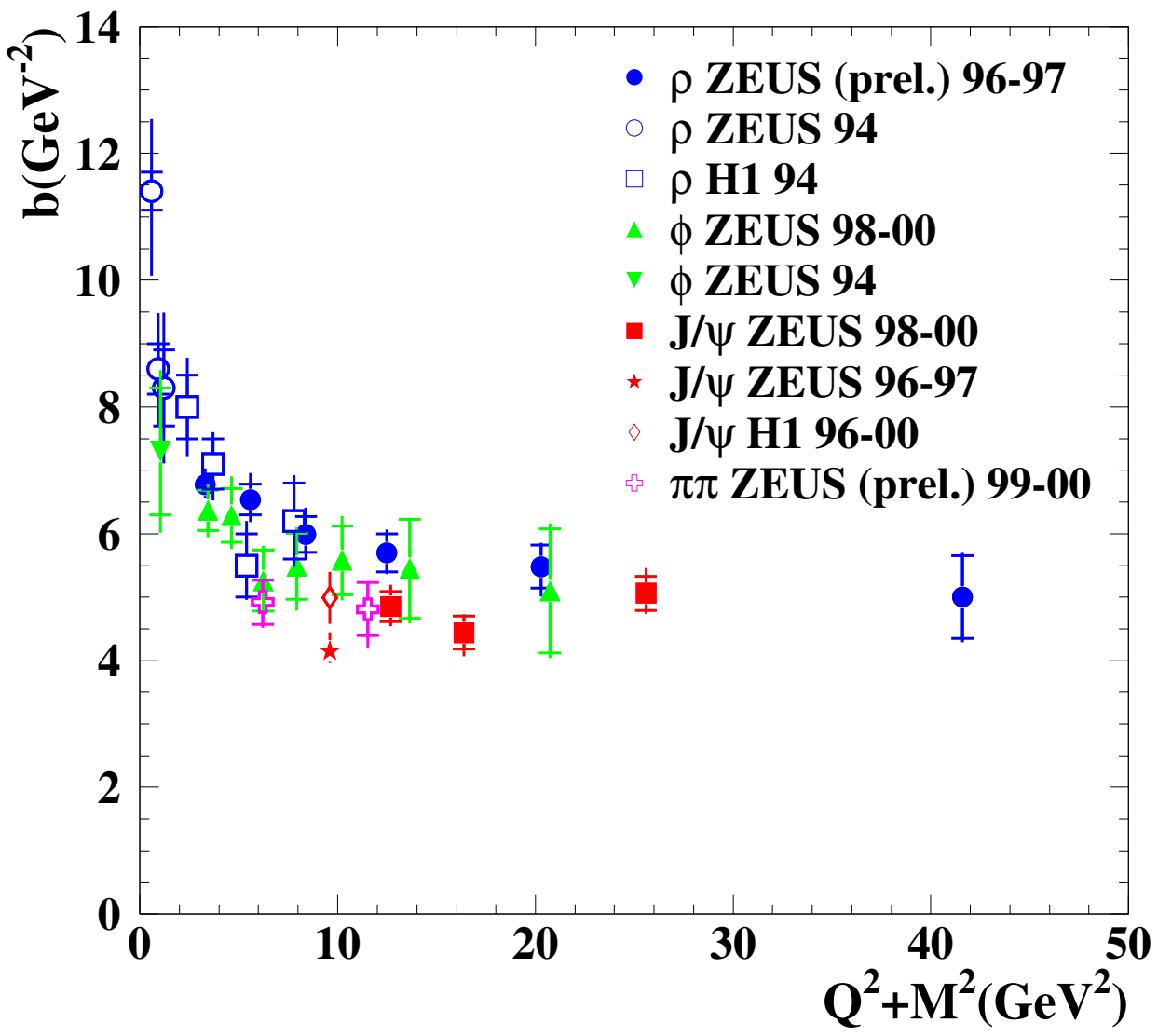

**Fig. 21:** The logarithmic slope of the $t$ dependence at $t = 0$ for different meson production channels, as well as for non-resonant dipion production.

### 3.2 Exclusive diffraction in hadron-hadron collisions

The concepts we have introduced to describe exclusive diffraction can be taken over to $pp$ or $p\bar{p}$ scattering, although further complications appear in these processes. A most notable reaction is exclusive production of a Higgs boson, $pp \to pHp$, sketched in Fig. 22. The generalized gluon distribution is a central input in this description. The physics interest, theory description, and prospects to measure this process at the LHC have been discussed in detail at this workshop [33, 34]. A major challenge in the description of this process is to account for secondary interactions between spectator partons of the two projectiles, which can produce extra particles in the final state and hence destroy the rapidity gaps between the Higgs and final-state protons – the very same mechanism we discussed in Sect. 2.4.

## 4 Parton saturation

We have seen that diffraction involves scattering on small-$x$ gluons in the proton. Consider the density in the transverse plane of gluons with longitudinal momentum fraction $x$ that are resolved in a process with hard scale $Q^2$. One can think of $1/Q$ as the "transverse size" of these gluons as seen by the probe. The number density of gluons at given $x$ increases with increasing $Q^2$, as described by DGLAP evolution (see Fig. 23). According to the BFKL evolution equation it also increases at given $Q^2$ when $x$ becomes smaller, so that the gluons become more and more densely packed. At some point, they will start to overlap and thus reinteract and screen each other. One then enters a regime where the density of partons saturates and where the linear DGLAP and BFKL evolution equations cease to be valid. If $Q^2$ is large enough to have a small coupling $\alpha_s$, we have a theory of this non-linear regime called "color glass condensate", see e.g. [35]. To quantify the onset of non-linear effects, one introduces a saturation scale $Q_s^2$ depending on $x$, such that for $Q^2 < Q_s^2(x)$ these effects become important. For smaller values of $x$, the parton density in the target proton is higher, and saturation sets in at larger values of $Q^2$ as illustrated in Fig. 23.

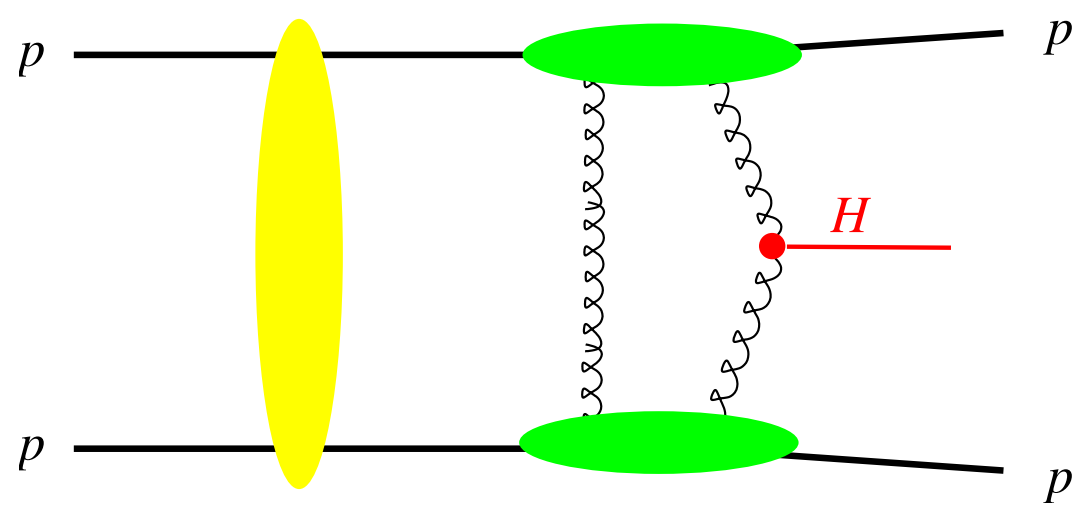

**Fig. 22:** Graph for the exclusive production of a Higgs boson in $pp$ scattering. The horizontal blobs indicate generalized gluon distributions, and the vertical blob represents secondary interactions between the projectiles (cf. Fig. 13).

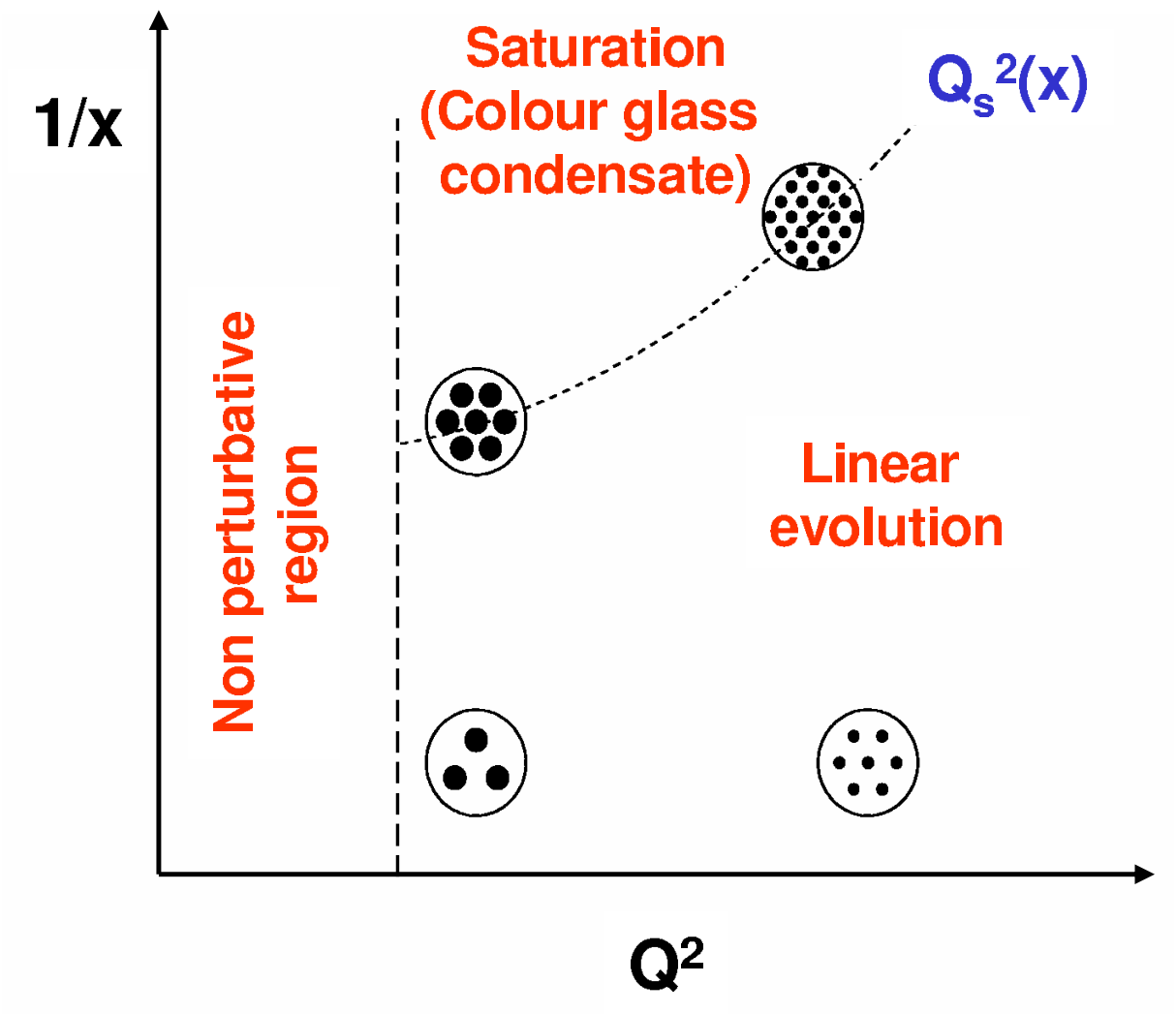

**Fig. 23:** Schematic view of the density of gluons in the transverse plane, as a function of the momentum fraction $x$ and the resolution scale $Q^2$. Above the line given by $Q_s^2(x)$, saturation effects set in.

The dipole picture we introduced in Sect. 3.1 is well suited for the theoretical description of saturation effects. When such effects are important, the relation (7) between dipole cross section and gluon distribution ceases to be valid; in fact the gluon distribution itself is then no longer an adequate quantity to describe the dynamics of a scattering process. In a certain approximation, the evolution of the dipole cross section with $x$ is described by the Balitsky-Kovchegov equation [36], which supplements the BFKL equation with a non-linear term taming the growth of the dipole cross section with decreasing $x$.

Essential features of the saturation phenomenon are captured in a phenomenological model for the dipole cross section, originally proposed by Golec-Biernat and Wüsthoff, see [37, 38]. Figure 24 shows $\sigma_{q\bar{q}}$ as a function of $r$ at given $x$ in this model. The dipole size $r$ now plays the role of $1/Q$ in our discussion above. At small $r$ the cross section rises following the relation $\sigma_{q\bar{q}}(r,x) \propto r^2 xg(x)$. At some value $R_s(x)$ of $r$, the dipole cross section is so large that this relation ceases to be valid, and $\sigma_{q\bar{q}}$ starts to deviate from the quadratic behavior in $r$. As $r$ continues to increase, $\sigma_{q\bar{q}}$ eventually saturates at a value typical of a meson-proton cross section. In terms of the saturation scale introduced above, $R_s(x) = 1/Q_s(x)$. For smaller values of $x$, the initial growth of $\sigma_{q\bar{q}}$ with $r$ is stronger because the gluon distribution is larger. The target is thus more opaque and as a consequence saturation sets in at lower $r$.

A striking feature found both in this phenomenological model [39] and in the solutions of the Balitsky-Kovchegov equation (see e.g. [40]) is that the total $\gamma^* p$ cross section only depends on $Q^2$ and $x_B$ through a single variable $\tau = Q^2/Q_s^2(x_B)$. This property, referred to as geometric scaling, is well satisfied by the data at small $x_B$ (see Fig. 25) and is an important piece of evidence that saturation effects are visible in these data. Phenomenological estimates find $Q_s^2$ of the order 1 GeV$^2$ for $x_B$ around $10^{-3}$ to $10^{-4}$.

The dipole formulation is suitable to describe not only exclusive processes and inclusive DIS, but also inclusive diffraction $\gamma^* p \to Xp$. For a diffractive final state $X = q\bar{q}$ at parton level, the theory description is very similar to the one for deeply virtual Compton scattering, with the wave function for the final state photon replaced by plane waves for the produced $q\bar{q}$ pair. The inclusion of the case $X = q\bar{q}g$ requires further approximations [37] but is phenomenologically indispensable for moderate to small $\beta$. Experimentally, one observes a very similar energy dependence of the inclusive diffractive and the total cross section in $\gamma^* p$ collisions at given $Q^2$ (see Fig. 26). The saturation mechanism implemented in the Golec-Biernat Wüsthoff model provides a simple explanation of this finding. To explain this aspect of the data is non-trivial. For instance, in the description based on collinear factorization, the energy dependence of the inclusive and diffractive cross sections is controlled by the $x$ dependence of the ordinary and the diffractive parton densities. This $x$ dependence is not predicted by the theory.

The description of saturation effects in $pp$, $pA$ and $AA$ collisions requires the full theory of the color glass condensate, which contains concepts going beyond the dipole formulation discussed here and is e.g. presented in [35]. We remark however that estimates of the saturation scale $Q_s^2(x)$ from HERA data can be used to describe features of the recent data from RHIC [41].

## 5 A short summary

Many aspects of diffraction in $ep$ collisions can be successfully described in QCD if a hard scale is present. A key to this success are factorization theorems, which render parts of the dynamics accessible to calculation in perturbation theory. The remaining non-perturbative quantities, namely diffractive PDFs and generalized parton distributions, can be extracted from measurements and contain specific information about small-$x$ partons in the proton that can only be obtained in diffractive processes. To describe hard diffractive hadron-hadron collisions is more challenging since factorization is broken by rescattering between spectator partons. These rescattering effects are of interest in their own right because of their intimate relation with multiple scattering effects, which at LHC energies are expected to be crucial for understanding the structure of events in hard collisions. A combination of data on inclusive and diffractive $ep$ scattering hints at the onset of parton saturation at HERA, and the phenomenology developed there is a helpful step towards understanding high-density effects in hadron-hadron collisions.

## Acknowledgements

It is a pleasure to thank our co-convenors and all participants for the fruitful atmosphere in the working group on diffraction, and A. De Roeck and H. Jung for their efforts in organizing this workshop. We are indebted to A. Proskuryakov for Figs. 5 and 12, to P. Fleischmann for Fig. 15, and to A. Bonato, K. Borras, A. Bruni, J. Forshaw, M. Grothe, H. Jung, L. Motyka, M. Ruspa and M. Wing for valuable comments on the manuscript.

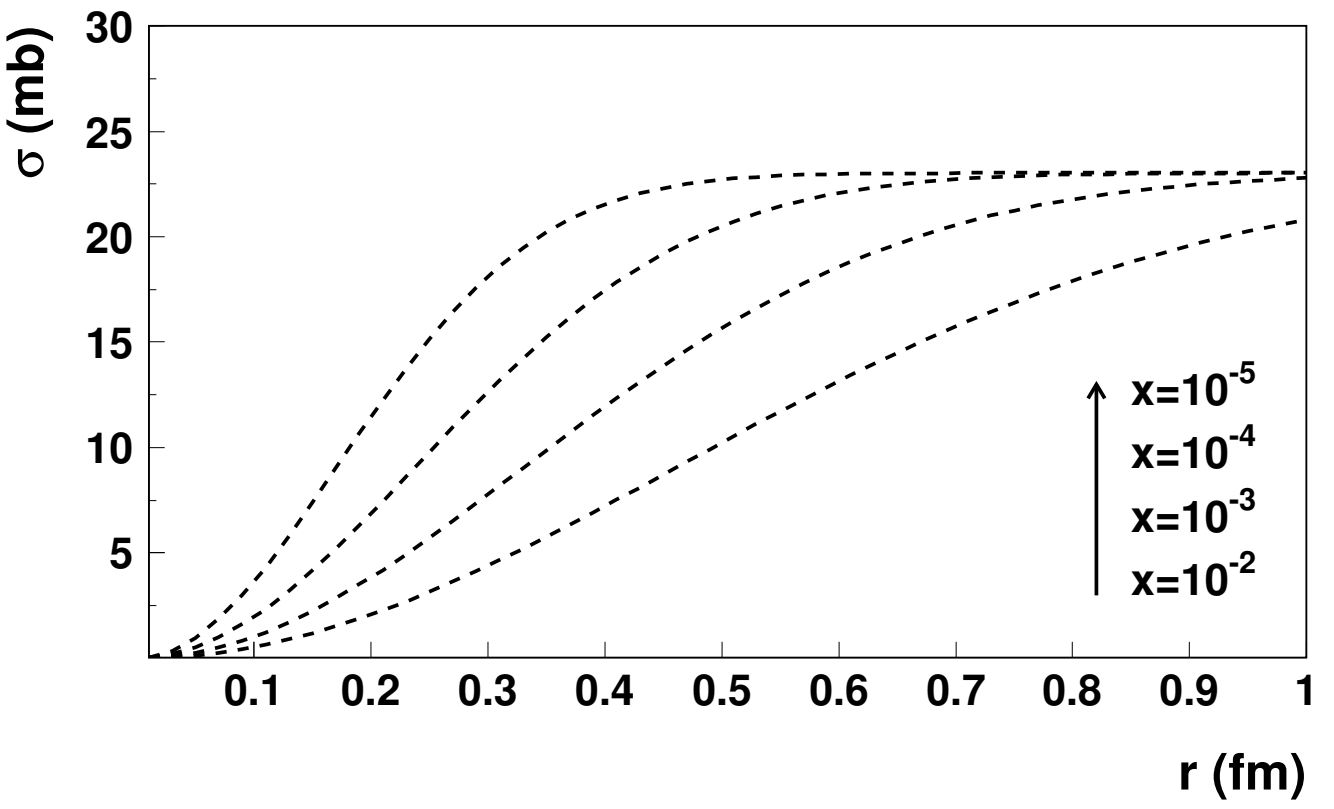

**Fig. 24:** The dipole cross section $\sigma_{q\bar{q}}$ in the Golec-Biernat Wüsthoff model as a function of dipole size $r$ for different $x$ (from [38]).

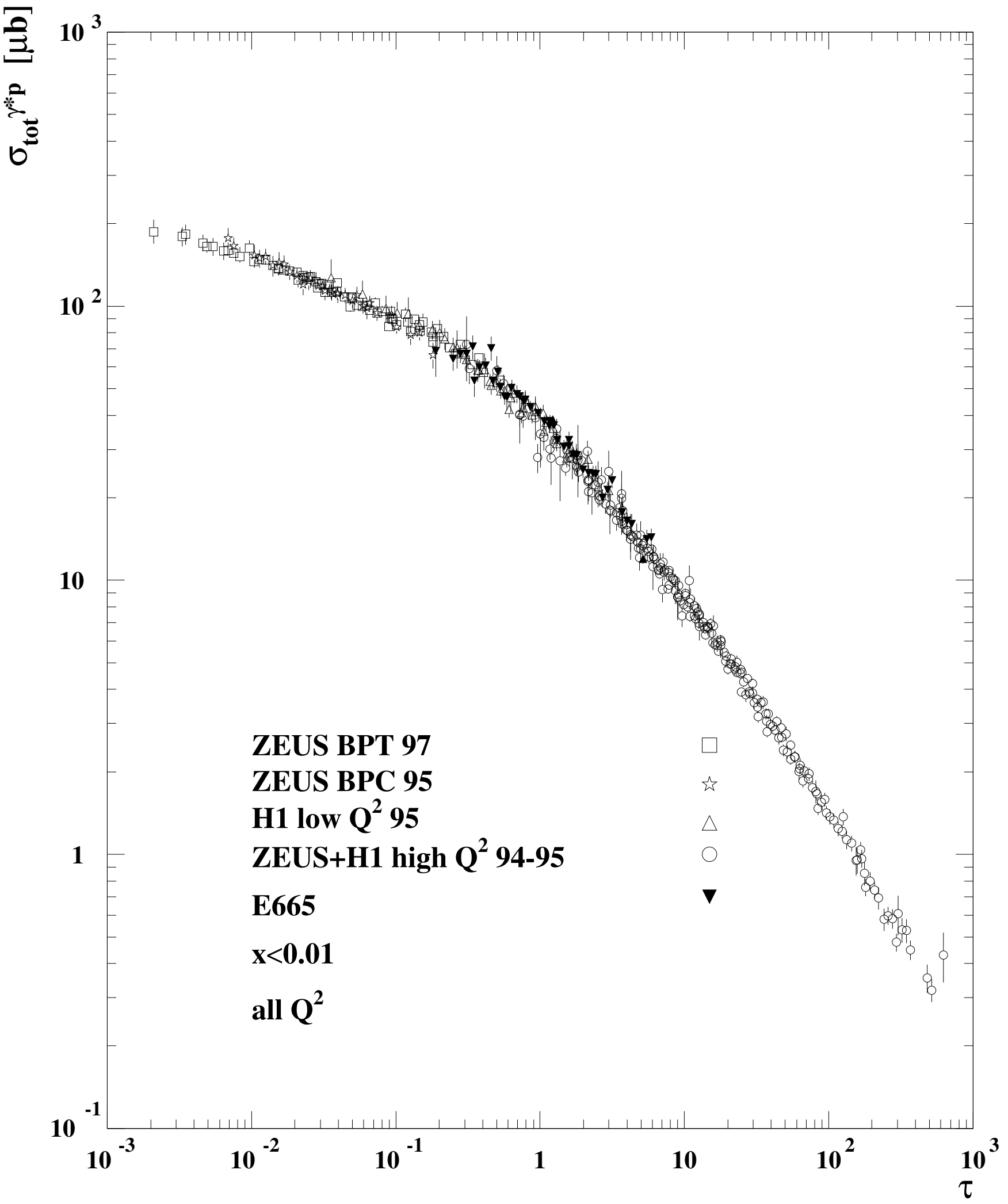

**Fig. 25:** Geometric scaling of the $\gamma^* p$ cross section in a single variable $\tau = Q^2/Q_s^2(x_B)$, as determined in [39]. The $Q^2$ of these data ranges from $0.045$ to $450$ GeV$^2$.

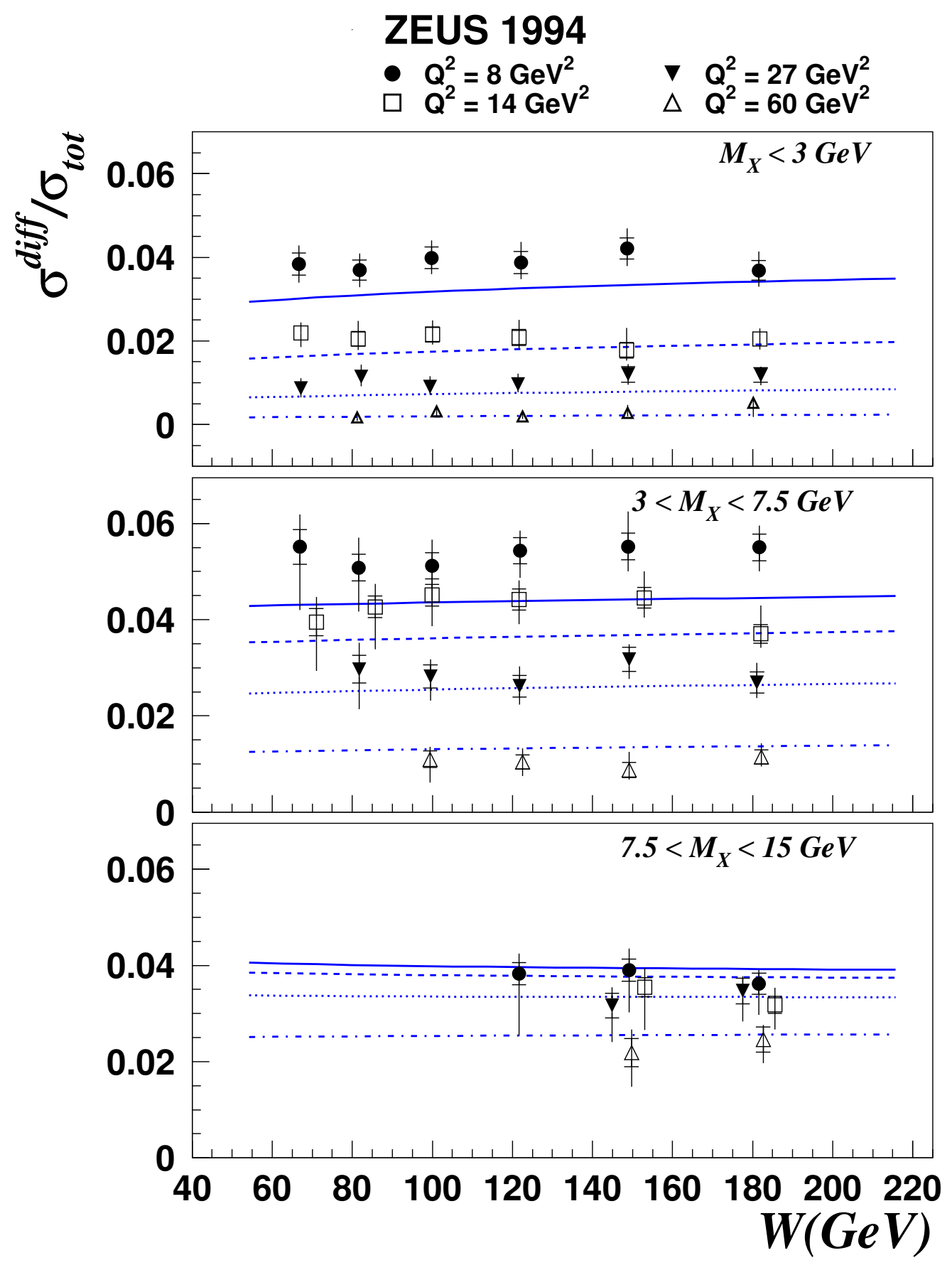

**Fig. 26:** Data on the ratio of diffractive and total $\gamma^* p$ cross sections compared with the result of the Golec-Biernat Wüsthoff model (from [37]).

# Diffractive Higgs Production: Experiment

*Editors: B. E. Cox[a] and M. Grothe[b]*
[a] School of Physics and Astronomy, The University of Manchester, Manchester M139PL, UK
[b] University of Torino and INFN-Torino, Italy; also at University of Wisconsin, Madison, WI, USA

## TOTEM forward measurements: exclusive central diffraction

*J. Kalliopuska, J.W. Lämsä, T. Mäki, N. Marola, R. Orava, K. Österberg, M. Ottela and S. Tapprogge*

**Abstract**

In this contribution, we present a first systematic study of the precision of the momentum measurement of protons produced in the central exclusive diffractive processes, $pp \rightarrow p + X + p$, as well as the accuracy of the reconstructed mass for particle state $X$ based on these proton measurements. The scattered protons are traced along the LHC beam line using the nominal LHC optics, accounting for uncertainties related to beam transport and proton detection.

To search for and precisely measure new particle states $X$ with masses below 200 GeV, additional leading proton detectors are required at about 420 m from the interaction point in addition to the already approved detectors. Using these additional detectors, a mass resolution of the order of 1 GeV can be achieved for masses beyond $\sim$120 GeV.

## TOTEM forward measurements: leading proton acceptance

*V. Avati and K. Österberg*

**Abstract**

We report about the acceptance of forward leading protons in Roman Pot stations placed along the LHC beam line. The TOTEM stations plus additional detectors at 420 m from the interaction point have been considered using the low–$\beta^*$ optics V6.5 for LHC physics runs.

## Diffractive Higgs: CMS/TOTEM Level-1 Trigger Studies

*M. Arneodo, V. Avati, R. Croft, F. Ferro, M. Grothe, C. Hogg, F. Oljemark, K. Österberg and M. Ruspa*

**Abstract**

Retaining events containing a Higgs Boson with mass around 120 GeV poses a special challenge to triggering at the LHC due to the relatively low transverse momenta of the decay products. We discuss the potential of including into the CMS trigger the TOTEM forward detectors and possible additional detectors at a distance of 420 m from the CMS interaction point. We find that the output rate of a 2-jet Level-1 trigger condition with thresholds sufficiently low for the decay products of a 120 GeV Higgs Boson can be limited to $\mathcal{O}(1)$ kHz for luminosities of up to $2 \times 10^{33} \mathrm{cm}^{-2}\mathrm{s}^{-1}$ by including the TOTEM forward detectors in the Level-1 trigger.

## Proposal to upgrade the very forward region at CMS

*V. Andreev, A.Bunyatyan, H. Jung, M. Kapishin and L. Lytkin*

**Abstract**

The possibilities of extending the acceptance of LHC experiments beyond 7 units of pseudorapidity are investigated. With additional detectors it would be possible to measure the particles with energies above 2 TeV in the pseudorapidity range between 7 and 11.

# TOTEM forward measurements: exclusive central diffraction

*J. Kalliopuska, J.W. Lämsä*, T. Mäki, N. Marola, R. Orava, K. Österberg†, M. Ottela and S. Tapprogge‡*
High Energy Physics Division, Department of Physical Sciences, University of Helsinki and
Helsinki Institute of Physics, P.O. Box 64, FIN-00014 University of Helsinki, Finland

**Abstract**
In this contribution, we present a first systematic study of the precision of the momentum measurement of protons produced in the central exclusive diffractive processes, $pp \rightarrow p + X + p$, as well as the accuracy of the reconstructed mass for particle state $X$ based on these proton measurements. The scattered protons are traced along the LHC beam line using the nominal LHC optics, accounting for uncertainties related to beam transport and proton detection.

To search for and precisely measure new particle states $X$ with masses below 200 GeV, additional leading proton detectors are required at about 420 m from the interaction point in addition to the already approved detectors. Using these additional detectors, a mass resolution of the order of 1 GeV can be achieved for masses beyond $\sim$120 GeV.

## 1 Introduction

It has been recently suggested that the Higgs boson mass could be measured to an accuracy of $\mathcal{O}$(1 GeV) in the central exclusive diffractive process (CED) [1,2]:

$$pp \rightarrow p + H + p \tag{1}$$

In contrast to this, the direct measurement of the Higgs boson mass, based on the two final state $b$-jets in $H \rightarrow b\bar{b}$, is estimated to yield a precision of $\mathcal{O}$(10 GeV). The precise reconstruction of the centrally produced system $X$, i.e. the Higgs mass in Eq. 1, is based on the four-momenta of the incoming ($p_{1,2}$) and scattered ($p'_{1,2}$) protons and since the two scattered protons are expected to have small transverse momenta, the following approximation for the mass of the centrally produced system can be made:

$$M^2 = (p_1 + p_2 - p'_1 - p'_2)^2 \approx \xi_1 \xi_2 s\,, \tag{2}$$

where $\xi_{1,2} = 1 - |\vec{p}\,'_{1,2}|/|\vec{p}_{1,2}|$ denote the momentum loss fractions of the two scattered protons.

The acceptance for forward leading protons for nominal LHC runs ($\beta^* \sim$ 0.5 m) is described in detail elsewhere (see [3]). This contribution focuses on the CED process and the precision with which the proton momenta and the mass of the centrally produced system can be reconstructed.

## 2 Leading proton uncertainties and transport

The study is done in multiple steps, which include the event generation (ExHuME [4] or PHOJET [5]), simulation of the interaction point (IP) region, tracking of the protons through the LHC beam line, a detector simulation and a proton momentum reconstruction algorithm using the detector information [6]. The following beam related uncertainties are inputs to the study[1]:

- $pp$ interaction region width: $\sigma_{x,y}$ = 16 $\mu$m, $\sigma_z$ = 5 cm,
- beam angular divergence: $\Theta_{x,y}$ = 30 $\mu$rad

* also Physics Department, Iowa State University, Ames, USA
† corresponding author: kenneth.osterberg@helsinki.fi
‡ currently at Institute of Physics, Johannes-Gutenberg Universität Mainz, Germany
[1]The reference system $(x, y, z)$ used in the study corresponds to the reference orbit in the accelerator; the $z$-axis is tangent to the orbit and positive in the beam direction; the $x$-axis (horizontal) is negative toward the center of the ring.

– beam energy spread: $1.1 \cdot 10^{-4}$.

Concerning the detector response, only the horizontal plane is considered with the following inputs:

– The detector is assumed to be fully efficient at a distance $10\sigma_x(z)$ + 0.5 mm from the beam center[2], where $\sigma_x(z)$ is the horizontal beam width at distance $z$. The second term takes into account the distance from the bottom of the vacuum window to the edge of the fully sensitive detector area.
– For the protons within the fully sensitive detector area, a position reconstruction uncertainty is introduced by smearing the hit coordinates according to a Gaussian distribution with a $\sigma$ of 10 $\mu$m.
– The uncertainty due to the beam position knowledge at each detector location is accounted for by smearing the hit coordinates by a correlated Gaussian distribution with a $\sigma$ of 5 $\mu$m.

The transverse displacement $(x(z), y(z))$ of a scattered proton at a distance $z$ from the IP is determined by tracing the proton along the LHC beam line using the MAD program [7]. The optics layout version 6.2 for nominal LHC runs ($\beta^*$ = 0.5 m) with a 150 $\mu$rad horizontal crossing angle is used [8]. Although the study was carried out for CMS/TOTEM (IP5), the results should be equally valid for ATLAS (IP1).

## 3 Proton momentum reconstruction

The $x$-coordinate of the proton observed at any given location along the beam line, depends on three initial parameters of the scattered proton: its fractional momentum loss, $\xi$, its initial horizontal scattering angle, $\Theta_x^*$, and its horizontal position of origin, $x^*$, at the IP. Consequently, more than one $x$-measurement of a particular proton is needed to constrain its parameters. In the procedure chosen, two $x$-measurements from a detector doublet are used to determine $\xi$ and $\Theta_x^*$, neglecting the $x^*$ dependence. The effect of the $x^*$ on the reconstructed proton momentum will be treated as an independent source of uncertainty.

To obtain a large acceptance in $\xi$, the following two detector locations, each consisting of a doublet of proton detectors, are chosen based on the LHC optics layout:

– 215 and 225 meters from IP5 ("215 m location"), and
– 420 and 430 meters from IP5 ("420 m location").

The 215 m location corresponds to a TOTEM approved Roman Pot location [9], while the 420 m location in the cryogenic section of the accelerator will require special design and further investigation.

Each detector doublet yields two observables, which are related to the horizontal offset and angle with respect to the beam axis. The $\xi$ dependence of these observables has been derived by fitting a functional form to the simulated average values of $\xi$, as a function of the values of the two observables [6].

## 4 Acceptance and resolution on $\xi$ and mass

The $\xi$ and $t$ acceptance of protons moving in the clockwise and counter-clockwise directions are slightly different due to differences in the optical functions, for details see [3]. As a summary: a proton from the CED process is seen when its $\xi$ is between 0.025 (0.002) and 0.20 (0.015) for the 215 (420) location.

The relative resolution on $\xi$, $\Delta\xi/\xi = (\xi - \xi_{rec})/\xi$ as a function of $\xi$ for protons produced in the CED process and seen in either the 215 m or the 420 m location is shown in Fig. 1 for protons circulating in the LHC both in the clockwise and counter-clockwise direction. Included are the separate effects from the uncertainty of the transverse IP position, the resolution of the proton detector, the beam energy uncertainty, the beam angular divergence at the IP, and the beam position resolution at the proton detector.

At both detector locations, major contributors to the over-all $\xi$ resolution are the uncertainty of the transverse IP position and the resolution of the proton detector. In addition to these two uncertainties, the beam energy uncertainty contributes significantly to the resolution at the 420 m location.

[2]The LHC collimators extend to $6\sigma_x(z)$. The closest safe position can be assumed to lay anywhere between 10 and 15.

The acceptance as function of the mass of the centrally produced system is shown in Fig. 2a. Each leading proton is required to be within the acceptance of either the 215 or 420 m locations. Independently shown is the case (sub-set of above) where both protons are within the acceptance of the 420 m locations. In the mass range shown, there is no acceptance for detecting both protons at the 215 m location. The $\xi_1$-$\xi_2$ combinations result from the gluon density function in the proton and the mass of the centrally produced system (see Eq. 2). The ExHuME generator favours a harder gluon distribution than that of PHOJET. Thus, the Higgses are produced more centrally. This yields a higher acceptance for ExHuME.

The resolution effects of the two scattered protons are, in general, uncorrelated from each other. The only correlation comes from the production point, whose transverse component is determined by the rms spread of the beam at the IP and by an independent measurement using the Higgs decay products [10]. It can be determined to 10 $\mu$m or better, and therefore for the mass resolution of the centrally produced system, a 10 $\mu$m uncertainty on the transverse IP position is used. For the mass resolution, all other uncertainties are assumed to be uncorrelated between the two protons.

The mass resolutions for events with protons within the acceptance of the 420 m location on both sides, and for events with one proton within the acceptance of the 215 m location on one side and the other proton within the acceptance of the 420 m location on the other side (labelled "asym." in the figure) are shown as a function of the mass of the centrally produced system in Fig. 2b. The values quoted in the figure are based on Gaussian fits to the reconstructed mass distributions. The two-proton acceptance requirement imposes a restriction on the allowed $\xi_1$-$\xi_2$ combinations; as a result the mass resolutions obtained with ExHuME and PHOJET are very similar.

## 5 Conclusions

The first comprehensive study of the CED process at the LHC is reported. The study is based on detailed simulations along the LHC beam line of the diffractively scattered protons, accounting for the known sources of uncertainties related to beam transport and proton detection. The feasibility of measuring such events during nominal LHC runs for masses of the central system, $X$, below $\sim$200 GeV is addressed.

On the basis of this study, it is concluded that with an additional pair of leading proton detectors at $\pm$420 m from the interaction point, a Higgs boson with a mass of 120–180 GeV could be measured with a mass resolution of the order of 1 GeV. Such additional proton detectors would also enable large statistics of pure gluon jets to be collected, thereby turning the LHC into a gluon factory.

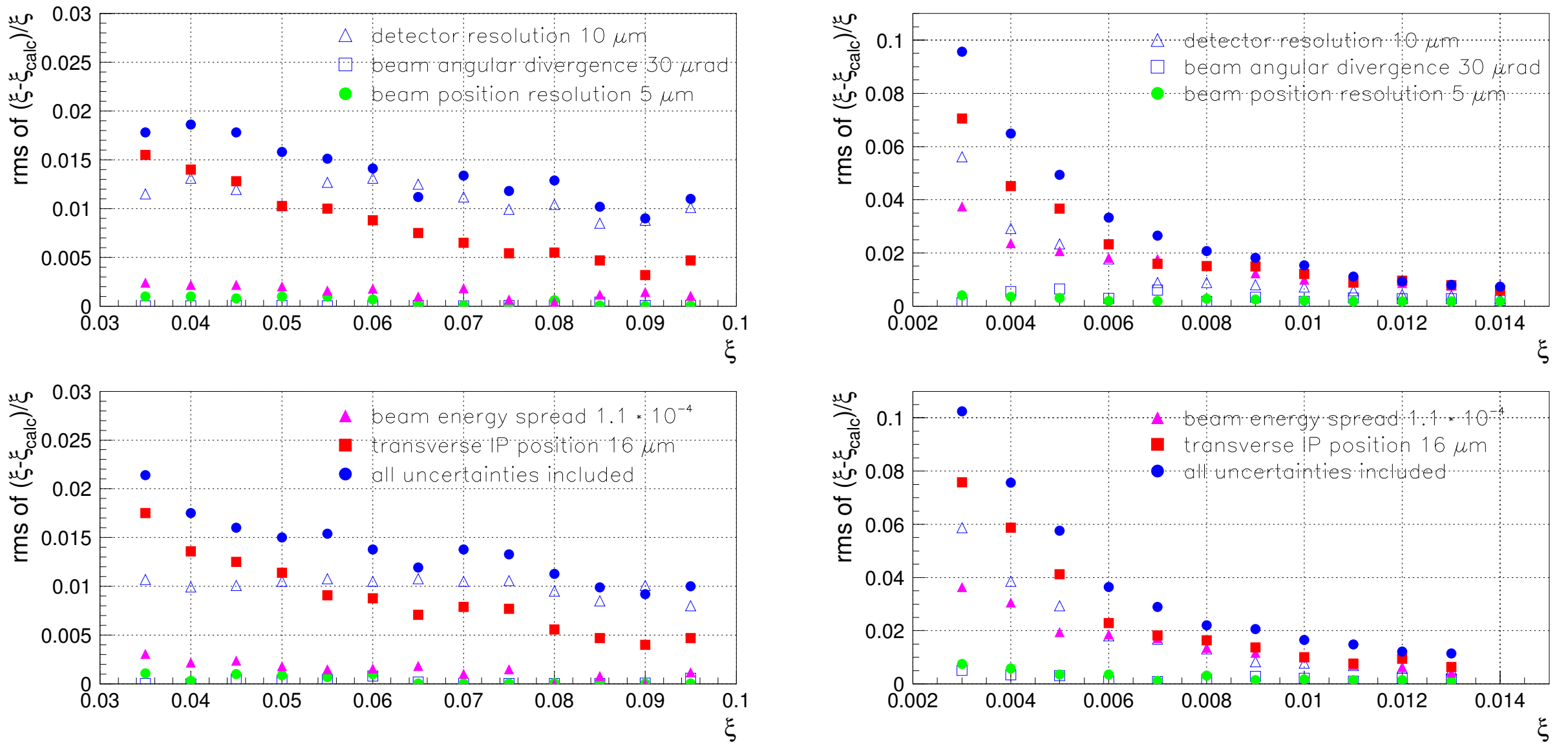

**Fig. 1:** Summary of all effects studied contributing to the over-all $\xi$ resolution for the 215 m (left) and 420 m location (right). The upper and lower plots are for protons circulating clockwise and counter-clockwise along the LHC beam line, respectively. The $t$ values of the protons used for each $\xi$ bin is similar to the $t$ distribution originating from central exclusive diffraction.

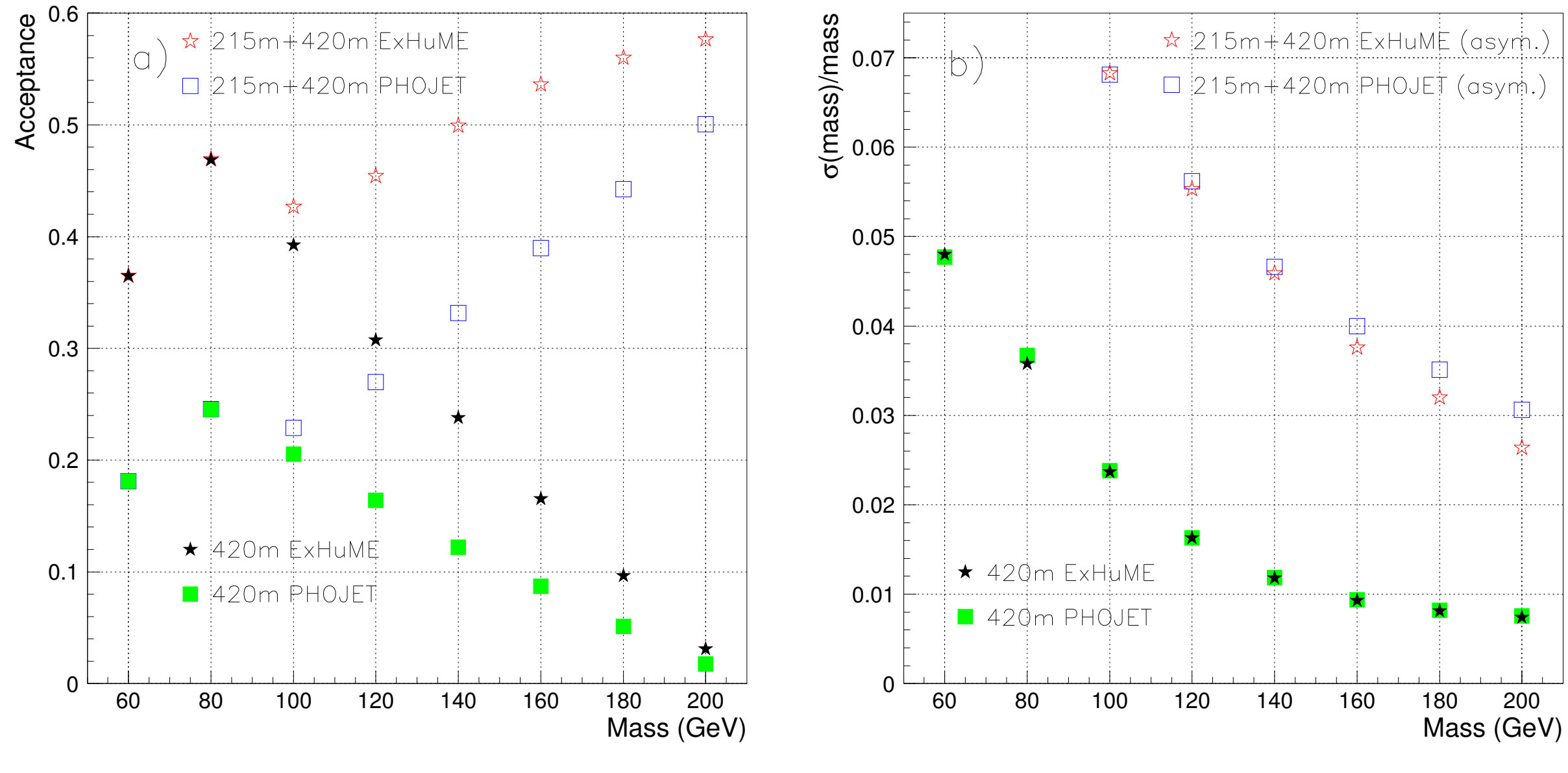

**Fig. 2:** a) Mass acceptance for events with protons within the acceptance of the 420 m location on each side of the interaction point (”420m”); and for events with protons within the combined acceptance of the 215 m and 420 m location on each side of the interaction point (”215m+420m”). b) Mass resolution for events with protons within the acceptance of the 420 m location on each side of the interaction point (”420m”); and for events with one proton within the acceptance of the 215 m location on one side of the interaction point and the other proton within the acceptance of the 420 m location on the other side (”215m+420m (asym.)”). ExHuME or PHOJET denotes the generator used for producing the central exclusive diffractive events.

# TOTEM forward measurements: leading proton acceptance

*V. Avati[a] and K. Österberg[b]*
[a] CERN, Geneva, Switzerland and Case Western Reserve University, Cleveland, OH, USA
[b] Department of Physical Sciences, University of Helsinki and Helsinki Institute of Physics, Finland

**Abstract**
We report about the acceptance of forward leading protons in Roman Pot stations placed along the LHC beam line. The TOTEM stations plus additional detectors at 420 m from the interaction point have been considered using the low–$\beta^*$ optics V6.5 for LHC physics runs.

## 1 Introduction

The TOTEM very forward detectors consist of telescopes of "Roman Pots" (RP) placed symmetrically on both sides of the interaction region IP5. The RP stations will be placed at 147 m and 220 m from IP5: each station is composed of two units separated by 2.5–4 m and each unit is equipped with two vertical and one horizontal silicon detector package. For more details on the TOTEM RPs, please refer to [1]. The possibility to add a detector in the cryogenic sections of the LHC is under investigation, therefore, we have included one more RP station at 420 m in these acceptance studies. This work is an update, due to the release of a new LHC optics, of previous studies done by the TOTEM Collaboration [2].

### 1.1 Low $\beta^*$ optics acceptance study

The transverse displacement $(x(s), y(s))$[1] of a scattered leading proton (with momentum loss $\xi = \Delta P/P < 0$) at distance $s$ from the interaction point (IP) is determined by tracking the proton through the accelerator lattice using the MAD-X program [3].

The new optics version 6.5 for the standard LHC physics runs is used. Notable changes (at IP5) from the previous versions are :

- $\beta^*$ = 0.55 m (previously 0.5 m)
- Beam offset in the horizontal plane = 0.5 mm (previously zero)
- Horizontal crossing angle = 142 $\mu$rad (previously 150 $\mu$rad)

The protons at the IP are generated with flat distributions in the azimuthal angle $\phi$, in Log($-\xi$) and in Log($-t$) in the kinematically allowed region of the $\xi$–$t$ plane, i.e. for physical values of the scattering angle of the proton. The Mandelstam variable $t$ is defined as $t = (p_{\text{orig}} - p_{\text{scatt}})^2$, where $p_{\text{orig(scatt)}}$ is the four-momentum of the incoming (scattered) proton. The scattering angle of the proton is physical when $t \geq t_0(\xi)$, where $t_0(\xi)$ is given by

$$t_0(\xi) = 2\left(P^2_{\text{orig}} + m_p^2\right)\left[\sqrt{1 + \left(P^2_{\text{orig}}\left[\xi^2 + 2\xi\right]\right) / \left(P^2_{\text{orig}} + m_p^2\right)} - 1\right] - 2\xi P^2_{\text{orig}}\,. \quad (1)$$

In Eq. 1, $P_{\text{orig}}$ is the momentum of the incoming proton and $m_p$ is the proton mass.

The transverse vertex position and the scattering angle at the IP are smeared assuming Gaussian distributions with widths given by the transverse beam size (16 $\mu$m) and the beam divergence (30 $\mu$rad).

[1]The reference system (x,y,s) defines the reference orbit in the accelerator; the s-axis is tangent to the orbit and positive in the beam direction; the two other axes are perpendicular to the reference orbit. The x-axis (horizontal, bending plane) is negative toward the center of the ring.

To determine the acceptance of a RP station, the minimum distance of a detector to the beam and constraints imposed by the beam pipe or beam screen size are considered.

The minimum distance of detector approach to the beam is proportional to the beam size:

$$x(y)_{min} = 10{\sigma_{x(y)}}^{beam} + c\,, \tag{2}$$

where $c$ is a constant that takes into account the distance from the edge of the sensitive detector area to the bottom of the RP window ($\sim$ 0.5 mm). For the nominal transverse beam emittance $\epsilon$ = 3.75 $\mu\mathrm{m}\cdot\mathrm{rad}$ typical values of the horizontal detector distance are $\sim$1 mm (at 220 m) and $\sim$4 mm (at 420 m). In the results shown later, the detector shape has not been included. The beam pipe apertures can be found in the LHC–LAYOUT Database [4].

### 1.2 Results

Figures 1–3 show the acceptance in Log($-\xi$), Log($-t$) for the RP stations at 220 and 420 m for the clockwise ("beam1") and counter–clockwise ("beam2") circulating beam.

One should note that these results refer to non-physical distributions in the variables $\xi$ and $t$ in order to have good statistics in each interval and describe all possible processes. To use these results in a general simulation program, the $\phi$ dependence has to be taken into account, since it is not negligible in many kinematical configurations. More detailed analysis such as detector alignment samples, collimator effects, etc. can be found in [5].

These results have been included in FAMOS (FAst MOntecarlo Simulation of the CMS detector) by M. Tasevsky ("Diffractive Higgs production", these proceedings) and they have been used in the CMS/TOTEM studies on triggering a diffractively produced light Higgs boson with the CMS Level-1 trigger ("Diffractive Higgs: CMS/TOTEM Level-1 Trigger Studies", these proceedings).

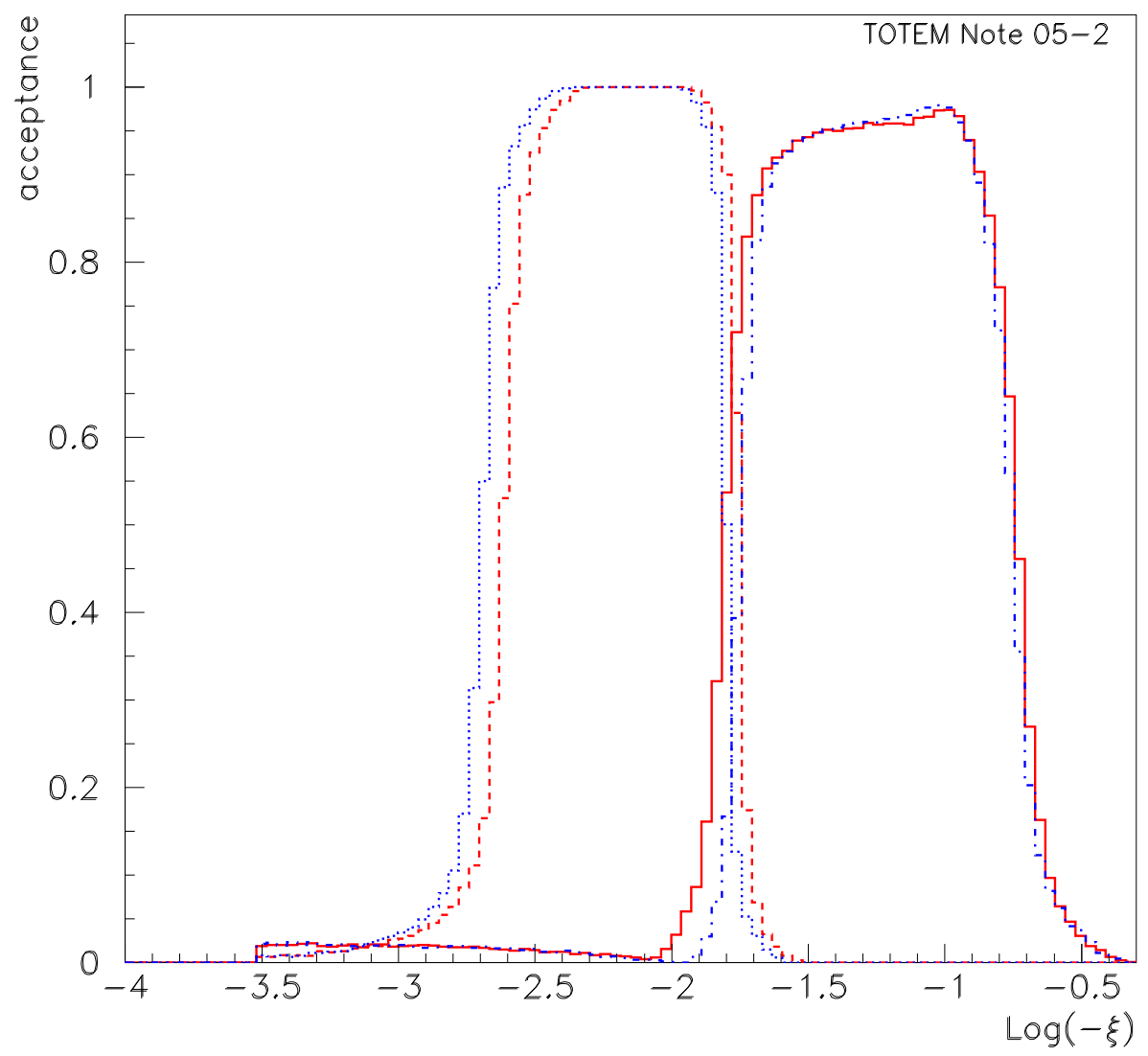

**Fig. 1:** Log($-\xi$) acceptance for: beam1 station at 220 m (solid-red) and 420 m (dashed-red) and beam2 station at 220 m (dashed-dotted-blue) and 420 m (dotted-blue).

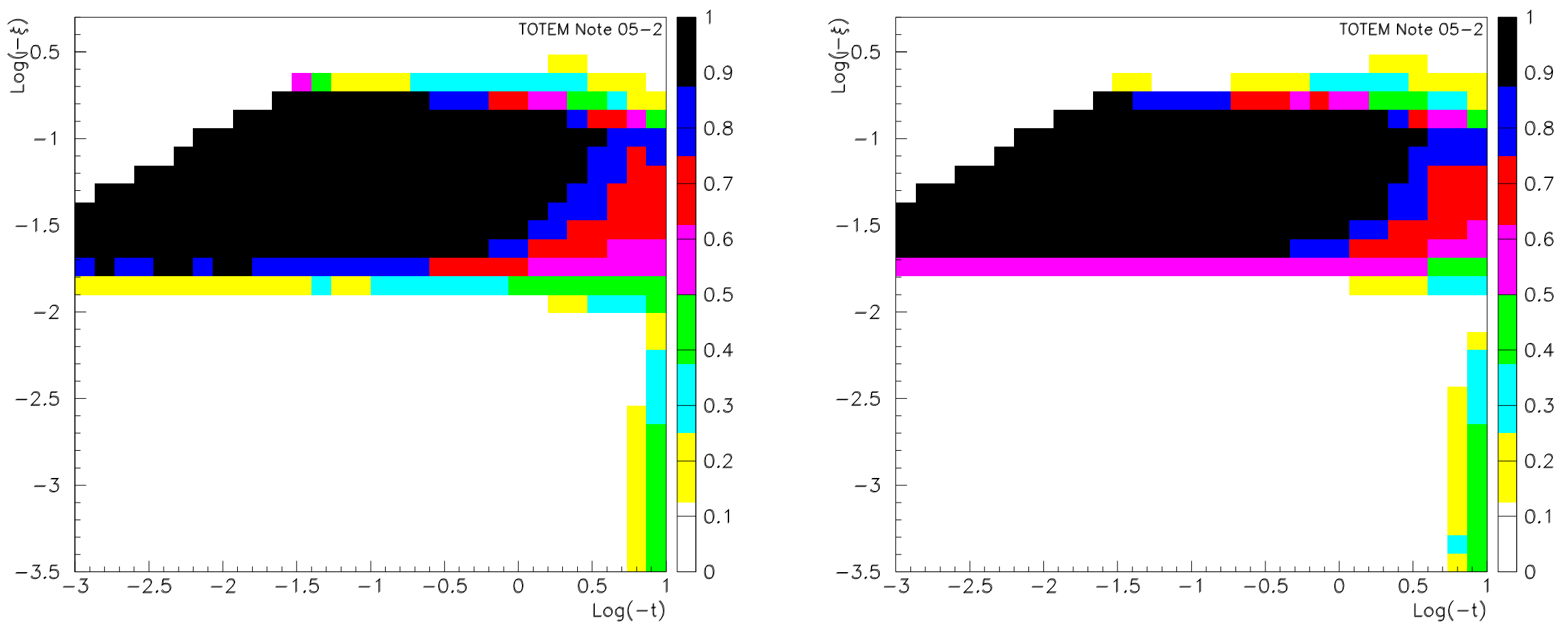

**Fig. 2:** Log($-\xi$) vs Log($-t$) acceptance for beam1 (left) and beam2 (right) stations at 220 m.

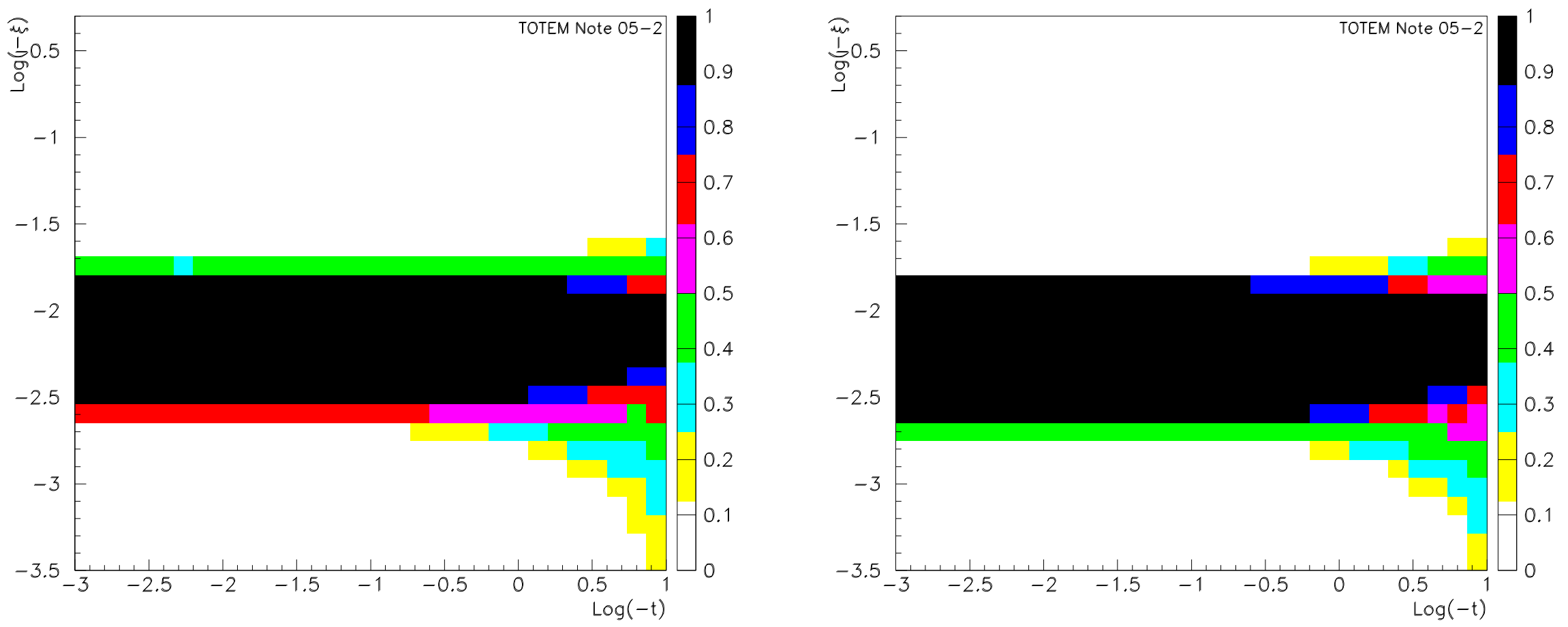

**Fig. 3:** Log($-\xi$) vs Log($-t$) acceptance for beam1 (left) and beam2 (right) stations at 420 m.

# Diffractive Higgs: CMS/TOTEM Level-1 Trigger Studies

*M. Arneodo[a], V. Avati[b], R. Croft[c], F. Ferro[d], M. Grothe[e*†], C. Hogg[f], F. Oljemark[g], K. Österberg[g], M. Ruspa[a]*

[a]Università del Piemonte Orientale, Novara, and INFN-Torino, Italy

[b] European Organisation for Nuclear Research CERN, Geneva, Switzerland, and Case Western Reserve University, Cleveland, OH, USA

[c] University of Bristol, Bristol, UK

[d] University of Genova and INFN-Genova, Italy

[e] University of Torino and INFN-Torino, Italy; also at University of Wisconsin, Madison, WI, USA

[f] University of Wisconsin, Madison, WI, USA

[g] Department of Physical Sciences, University of Helsinki, and Helsinki Institute of Physics, Finland

**Abstract**

Retaining events containing a Higgs Boson with mass around 120 GeV poses a special challenge to triggering at the LHC due to the relatively low transverse momenta of the decay products. We discuss the potential of including into the CMS trigger the TOTEM forward detectors and possible additional detectors at a distance of 420 m from the CMS interaction point. We find that the output rate of a 2-jet Level-1 trigger condition with thresholds sufficiently low for the decay products of a 120 GeV Higgs Boson can be limited to $\mathcal{O}(1)$ kHz for luminosities of up to $2 \times 10^{33}\mathrm{cm}^{-2}\mathrm{s}^{-1}$ by including the TOTEM forward detectors in the Level-1 trigger.

## 1 Introduction

A Higgs Boson with mass close to the current exclusion limit poses a special challenge to triggering at the LHC. The dominant decay of a Standard Model Higgs Boson of mass $\sim$120 GeV is into two $b$-quarks and generates 2 jets with at most 60 GeV transverse momentum, $p_T$, each. The so far considered Level-1 (L1) trigger tables of CMS [1] are optimized for events with high $p_T$; the necessity of keeping the overall L1 rate at acceptable levels requires thresholds in two-jet events above $p_T$ =100 GeV per jet. Conversely, triggering is not a problem should the mass of the Higgs Boson be sufficiently high so that its final states are rich in high $p_T$ leptons, as is the case for $H \rightarrow WW^{\star}$.

In order to retain a potential Higgs signal with mass close to the current exclusion limit, information beyond that from the central CMS detector needs to be included in the L1 trigger. A proton that scatters diffractively at the CMS interaction point (IP) may be detected by Roman Pot (RP) detectors further downstream. Roman Pot detectors up to 220 m downstream of CMS will be part of the TOTEM experiment [2]. Information from TOTEM will be available to the CMS L1 trigger. Furthermore, detectors at up to 420 m distance from the IP are currently discussed as part of the FP420 project [3]. Including information from them into the CMS L1 trigger is however not possible without an increase in the trigger latency.

This article discusses the effect of including the TOTEM forward detectors and/or those planned at 420 m distance on rate and selection efficiency of the CMS L1 trigger. All results reported in the following are preliminary; further studies are still on-going at the time of writing.

* Work supported by the Italian Ministry for Education, University and Scientific Research under the program "Incentivazione alla mobilità di studiosi stranieri e italiani residenti all'estero".

† Corresponding author: Monika.Grothe@cern.ch

## 2 Experimental apparatus

The CMS trigger system is designed to reduce the input rate of $10^9$ interactions per second at the nominal LHC luminosity of $10^{34}\text{cm}^{-2}\text{s}^{-1}$ to an output rate of not more than 100 Hz. This reduction of $10^7$ is achieved in two steps, by the CMS L1 trigger (output rate 100 kHz) and the CMS Higher-Level Trigger (HLT). The L1 trigger carries out its data selection algorithms with the help of three principal components: the Calorimeter Trigger, the Muon Trigger and the Global Trigger. The decision of the Calorimeter Trigger is based on the transverse energy, $E_T$, information of the CMS calorimeters (pseudorapidity coverage $|\eta| < 5$). A L1 jet consists of $3 \times 3$ regions, each with $4 \times 4$ trigger towers, where the $E_T$ in the central region is above the $E_T$ in any of the outer regions. A typical L1 jet has dimensions $\Delta\eta \times \Delta\phi = 1 \times 1$, where $\phi$ is the azimuthal angle. The $E_T$ reconstructed by the L1 trigger for a given jet corresponds on the average only to 60% of its true $E_T$. All studies in this article use calibrated jet $E_T$ values, obtained from the reconstructed value by means of an $\eta$ and $E_T$ dependent correction.

The TOTEM experiment [2] will have two identical arms, one at each side of the CMS IP. Each arm will comprise two forward tracker telescopes, T1 (Cathode Strip Chambers) and T2 (Gas Electron Multipliers), as well as Silicon detectors housed in RP stations along the LHC beam-line. The TOTEM detectors will provide input data to the Global Trigger of the CMS L1 trigger. Track finding in T1 and T2 (combined coverage $3.2 < |\eta| < 6.6$) for triggering purposes is optimized with respect to differentiating between beam-beam events that point back to the IP and beam-gas and beam-halo events that do not. The TOTEM RP stations will be placed at a distance of $\pm 147$ m and $\pm 220$ m from the CMS IP. Each station will consist of two units, 2.5 m and 4 m apart, each with one horizontally and two vertically movable pots equipped with Silicon strip detectors. The possibility of implementing a cut on $\xi$ in the L1 trigger is currently under investigation.

The fractional momentum loss, $\xi$, of diffractively scattered protons peaks at $\xi = 0$ ("diffractive peak"). The combination of CMS and TOTEM will permit to measure protons that have undergone a fractional momentum loss $0.2 > \xi > 0.02$. Detectors at a distance of 420 m, in the cryogenic region of the LHC ring, are currently being considered by the FP420 project [3]. They would provide a coverage of $0.02 > \xi > 0.002$, complementary to that of the TOTEM detectors, but cannot be included in the L1 trigger without an increase in the L1 latency of 3.2 $\mu$s. A special, long latency running mode might be feasible at lower luminosities. This option is currently under investigation. Using detectors at 420 m in the L1 trigger is included as an option in the studies discussed in this article.

The studies discussed in the following assume that the RP detectors are 100% efficient in detecting all particles that emerge at a distance of at least $10\sigma_{beam} + 0.5$ mm from the beam axis. Their acceptance was calculated by way of a simulation program that tracks particles through the accelerator lattice [4]. This has been done for the nominal LHC optics, the so-called low-$\beta^*$ optics, version V6.5. Further details can be found in [5]. All Monte Carlo samples used in the following assume LHC bunches with 25 ns spacing.

## 3 Level-1 trigger rates and signal efficiencies

We consider here perhaps the most challenging case, that of a low-mass (120 GeV) Standard Model Higgs Boson, decaying into two $b$-jets. There, the jets have transverse energies of at most 60 GeV. In order to retain as large a signal fraction as possible, as low an $E_T$ threshold as possible is desirable. In practice, the threshold value cannot be chosen much lower than 40 GeV per jet. The L1 trigger applies cuts on the calibrated $E_T$ value of the jet. Thus, a threshold of 40 GeV corresponds to 20 to 25 GeV in reconstructed $E_T$, i.e. to values where noise effects start becoming sizable.

In the trigger tables forseen for the first LHC running period, a L1 2-jet rate of $\mathcal{O}(1)$ kHz is planned. For luminosities of $10^{32}\text{cm}^{-2}\text{s}^{-1}$ and above, the rate from standard QCD processes for events with at least 2 central jets ($|\eta| < 2.5$) with $E_T > 40$ GeV is above this. Thus additional conditions need to be employed in the L1 trigger to reduce the rate from QCD processes. The efficiency of several conditions was investigated and, in the following, the corresponding rate reduction factors are always quoted with respect to the rate of QCD events that contain at least 2 central jets with $E_T > 40$ GeV per jet. These conditions are:

1) Condition based on additional central detector quantities available to the Calorimeter Trigger.
2) Condition based on T1 and T2 as vetoes.
3) Condition based on the RP detectors at $\pm 220$ m and $\pm 420$ m distance from the CMS IP.
4) Condition based on the Muon Trigger.

The QCD background events were generated with the Pythia Monte Carlo generator.

At higher luminosities more than one interaction takes place per bunch crossing; the central exclusive production of a Higgs boson is overlaid with additional, typically soft events, the so-called pile-up. In order to assess the effect when the signal is overlaid with pile-up, a sample of 500,000 pile-up events was generated with Pythia. This sample includes inelastic as well as elastic and diffractive events. Pythia underestimates the number of final state protons in this sample. The correction to the Pythia leading proton spectrum described in [6] was used to obtain the results discussed in the following.

The effect from beam-halo and beam-gas events on the L1 rate is not yet included in the studies discussed here. Preliminary estimates suggest that the size of their contribution is such that the conclusions of this article are not invalidated.

Table 1 summarizes the situation for luminosities between $10^{32}\mathrm{cm}^{-2}\mathrm{s}^{-1}$ and $10^{34}\mathrm{cm}^{-2}\mathrm{s}^{-1}$. Given a target rate for events with 2 central L1 jets of $\mathcal{O}(1)$ kHz, a total rate reduction between a factor 20 at $1 \times 10^{33}\mathrm{cm}^{-2}\mathrm{s}^{-1}$ and 200 at $1 \times 10^{34}\mathrm{cm}^{-2}\mathrm{s}^{-1}$ is necessary.

**Table 1:** Reduction of the rate from standard QCD processes for events with at least 2 central L1 jets with $E_T > 40$ GeV, achievable with requirements on the tracks seen in the RP detectors. Additional rate reductions can be achieved with the $H_T$ condition and with a topological condition (see text). Each of them yields, for all luminosities listed, an additional reduction by about a factor 2.

| Lumi nosity $[\mathrm{cm}^{-2}\mathrm{s}^{-1}]$ | # Pile-up events per bunch crossing | L1 2-jet rate [kHz] for $E_T > 40$GeV per jet | Total reduc tion needed | Reduction when requiring track in RP detectors | | | | |
|---|---|---|---|---|---|---|---|---|
| | | | | at 220 m | | at 420 m | at 220 m & 420 m (asymmetric) | |
| | | | | | $\xi < 0.1$ | | | $\xi < 0.1$ |
| $1 \times 10^{32}$ | 0 | 2.6 | 2 | 370 | | | | |
| $1 \times 10^{33}$ | 3.5 | 26 | 20 | 7 | 15 | 27 | 160 | 380 |
| $2 \times 10^{33}$ | 7 | 52 | 40 | 4 | 10 | 14 | 80 | 190 |
| $5 \times 10^{33}$ | 17.5 | 130 | 100 | 3 | 5 | 6 | 32 | 75 |
| $1 \times 10^{34}$ | 35 | 260 | 200 | 2 | 3 | 4 | 17 | 39 |

### 3.1 Condition based on central CMS detector quantities

In addition to the $E_T$ values of individual L1 jets, the CMS Calorimeter Trigger has at its disposal the scalar sum, $H_T$, of the $E_T$ values of all jets. Requiring that essentially all the $E_T$ be concentrated in the two central L1 jets with highest $E_T$, i.e. $[E_T^1 + E_T^2]/H_T > 0.9$ ($H_T$ condition), corresponds to imposing a rapidity gap of at least 2.5 units with respect to the beam direction. This condition reduces the rate of QCD events by approximately a factor 2, independent of the presence of pile-up and with only a small effect on the signal efficiency.

### 3.2 Condition based on TOTEM detectors T1 and T2

Using T1 and T2 as vetoes in events with 2 central L1 jets imposes the presence of a rapidity gap of at least 4 units. This condition suppresses QCD background events by several orders of magnitude. At luminosities low enough so that not more than one interaction takes place per bunch crossing, the signal efficiency is very high ($> 90\%$). In the presence of pile-up, the signal efficiency falls rapidly. The non-diffractive component in pile-up events tends quickly to fill in the rapidity gap in the Higgs production process. Only about 20 (5) % of signal events survive in the presence of 1 (2) pile-up event(s).

### 3.3 Condition based on Roman Pot detectors

Demanding that a proton be seen in the RP detectors at 220 m results in excellent suppression of QCD background events in the absence of pile-up. This is demonstrated in Figure 1 for a luminosity of $10^{32}\mathrm{cm}^{-2}\mathrm{s}^{-1}$. There, the rate of QCD background events with at least 2 central L1 jets with $E_T$ above a threshold is shown as function of the threshold value. The two histograms reflect the rate without and with the requirement that a proton be seen in the RP detectors at 220 m. The rate of QCD background events containing at least 2 central L1 jets with $E_T > 40$ GeV each is reduced by a factor $\sim 370$. At $2 \times 10^{33}\mathrm{cm}^{-2}\mathrm{s}^{-1}$, where on the average 7 pile-up events overlay the signal event, the diffractive component in the pile-up causes the reduction to decrease to a factor $\sim 4$, and at $10^{34}\mathrm{cm}^{-2}\mathrm{s}^{-1}$, to a factor $\sim 2$, as can be seen from table 1.

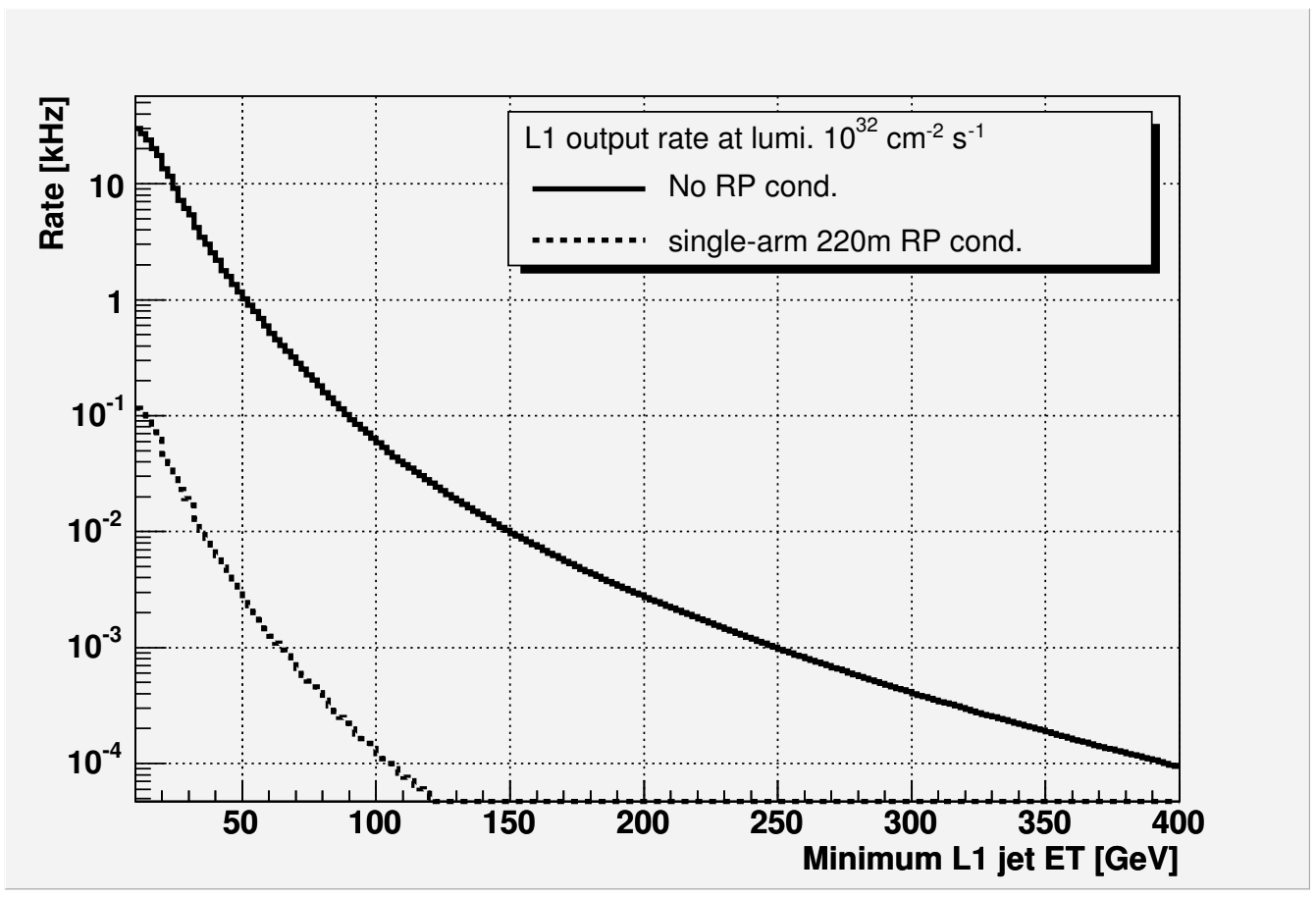

**Fig. 1:** L1 rate for the QCD background at a luminosity of $10^{32}\mathrm{cm}^{-2}\mathrm{s}^{-1}$ as function of the L1 threshold value when requiring at least 2 central L1 jets with $E_T$ above threshold.

Table 1 summarizes the reduction factors achieved with different conditions for tracks in the RP detectors: a track in the RP detectors at 220 m distance on one side of the IP (single-arm 220 m), without and with a cut on $\xi$, a track in the RP detectors at 420 m distance on one side of the IP (single-arm 420 m), a track in the RP detectors at 220 m and 420 m distance (asymmetric). Because the detectors at 220 m and 420 m have complementary coverage in $\xi$, the last condition in effect selects events with two tracks of very different $\xi$ value, in which one track is seen at 220 m distance on one side of the IP and a second track is seen on the other side at 420 m. If not by the L1 trigger, these asymmetric events can be selected by the HLT and are thus of highest interest. The effect on the acceptance of the RP detectors of a collimator located in front of the LHC magnet Q5, which will be operative at higher luminosities, has not been taken into account in table 1.

A further reduction of the QCD rate could be achieved with the help of a topological condition. The 2-jet system has to balance the total momentum component of the two protons along the beam axis. In signal events with asymmetric $\xi$ values, the proton seen on one side in the RP detectors at 220 m distance is the one with the larger $\xi$ and thus has lost more of its initial momentum component along the beam axis. Hence the jets tend to be located in the same $\eta$-hemisphere as the RP detectors that detect this proton. A trigger condition requiring that $[\eta^{jet1} + \eta^{jet2}] \times \mathrm{sign}(\eta^{220m\ RP}) > 0$ would reduce the QCD background by a factor 2, independent of pile-up, and with no loss in signal efficiency.

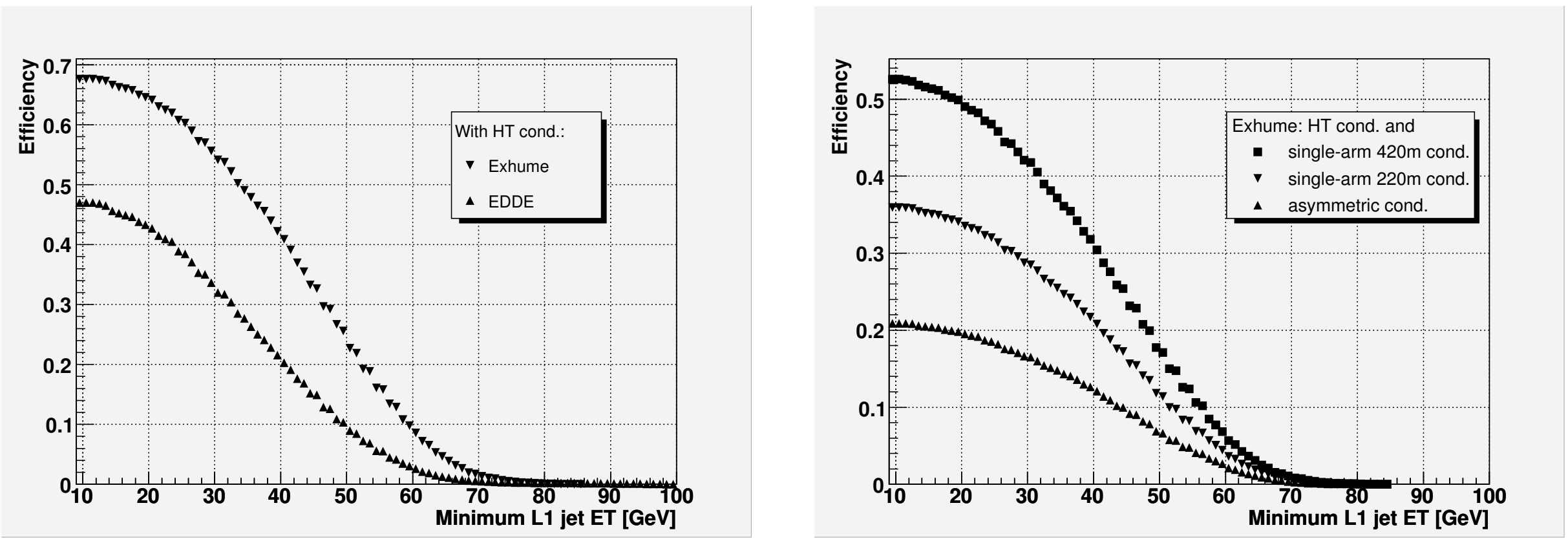

**Fig. 2:** L1 selection efficiency as function of the $E_T$ threshold value when requiring at least 2 central L1 jets with $E_T$ above threshold. All plots are for the non-pile-up case and the $H_T$ condition (see text) has been applied. Left: Comparison between the EDDE and Exhume Monte Carlo generators, without applying any additional RP conditions. Right: Comparison of the effect of different RP conditions on the efficiency in the Exhume Monte Carlo sample.

A reduction of the QCD rate to levels compatible with the trigger bandwidth requirements by including RP detectors at a distance of 220 m from the CMS IP thus appears feasible for luminosities up to $2 \times 10^{33}\mathrm{cm}^{-2}\mathrm{s}^{-1}$, as long as a $\xi$ cut can be administered in the L1 trigger such that the accepted events can be restricted to the diffractive peak region around $\xi = 0$. Higher luminosities would necessitate inclusion of the RP detectors at 420 m distance in the L1 trigger.

In order to study the effect of the L1 trigger selection on the Higgs signal, signal samples of 20,000 events with central exclusive production of a Higgs Boson were generated with the Monte-Carlo programs EDDE [7] (version 1.1) and Exhume [8] (version 0.9). Figure 2 shows the L1 selection efficiency as a function of the $E_T$ threshold values when requiring at least 2 central L1 jets with $E_T$ above threshold. The histograms show the case when no pile-up is present. The presence of pile-up has only a small effect on the efficiency curves. The plot on the left-hand side compares the efficiency curves obtained for EDDE and Exhume. For a threshold of 40 GeV per jet, Exhume yields an efficiency of about 40%. As a consequence of its less central jet $\eta$ distribution (see [9]), the efficiency for EDDE is about 20% lower than the one of Exhume. The plot on the right-hand side overlays the efficiency curves obtained with Exhume when including three different RP detector conditions in the L1 2-jet trigger: single-arm 220 m, single-arm 420 m and the asymmetric 220 & 420 m condition. At an $E_T$ threshold of 40 GeV per jet, the single-arm 220 m (420 m) condition results in an efficiency of the order 20% (30%), the asymmetric condition in one of 15%. This also means that even without the possibility of including the RP detectors at 420 m distance from the CMS IP in the L1 trigger, 15% of the signal events can be triggered with the single-arm 220 m condition, but will have a track also in the 420 m detectors which can be used in the HLT.

### 3.4 Condition based on the Muon Trigger

An alternative trigger strategy may be to exploit the relatively muon-rich final state from $B$-decays. We estimate that up to 10% of the signal events could be retained using this technique. Further investigations are underway at the time of writing.

## 4 Conclusions

Retaining a Higgs Boson with mass around 120 GeV poses a special challenge to triggering at the LHC. The relatively low transverse momenta of its decay products necessitate L1 jet $E_T$ thresholds as low as 40 GeV. Thresholds that low would result in a L1 trigger rate of more than 50 kHz, essentially saturating the available output bandwidth.

The results we presented in this article are preliminary and should be taken as a snapshot of our present understanding. They can be summarized as follows: The output rate of a 2-jet L1 trigger condition with thresholds of 40 GeV per jet can be kept at an acceptable $\mathcal{O}(1)$ kHz by including the TOTEM forward detectors in the CMS L1 trigger. In the absence of pile-up, either using the TOTEM T1 and T2 detectors as vetoes or requiring that a proton be seen in the TOTEM RP detectors at 220 m on one side of the CMS IP (single-sided 220 m condition) results in a sufficient reduction of the QCD event rate that dominates the L1 trigger output rate. At higher luminosities, up to $2 \times 10^{33}\mathrm{cm}^{-2}\mathrm{s}^{-1}$, where pile-up is present, it is necessary to combine the single-sided 220 m condition with conditions based on event topology and on $H_T$, the scalar sum of all L1 jet $E_T$ values. Going to even higher luminosities, up to $1 \times 10^{34}\mathrm{cm}^{-2}\mathrm{s}^{-1}$, would necessitate additional L1 trigger conditions, such as inclusion of RP detectors at 420 m distance from the CMS IP, which, however, would require an increase in the L1 trigger latency. These L1 trigger conditions result in signal efficiencies between 15% and 20%.

We expect no trigger problems for final states rich in high $p_T$ leptons, such as the $WW$ decay modes of the Standard Model Higgs Boson.

## Acknowledgement

We are grateful to Sridhara Dasu for sharing with us his expertise on the CMS L1 trigger simulation and for providing considerable pratical help with it. We would like to thank him and Dan Bradley for their invaluable help with producing the Monte Carlo event samples used in this article.

# Proposal to upgrade the very forward region at CMS

*V. Andreev*[1], *A.Bunyatyan*[2,3], *H. Jung*[4], *M. Kapishin*[5], *L. Lytkin*[3,5]
[1] Lebedev Physics Institute, Moscow, [2] Yerevan Physics Institute,
[3] MPI-K Heidelberg, [4] DESY Hamburg, [5] JINR Dubna

**Abstract**
The possibilities of extending the acceptance of LHC experiments beyond 7 units of pseudorapidity are investigated. With additional detectors it would be possible to measure the particles with energies above 2 TeV in the pseudorapidity range between 7 and 11.

## 1 Introduction

At the LHC experiments, CMS and ATLAS, the acceptance for forward energy measurements is limited to about 5 units of pseudorapidity. The acceptance of CMS detector will be extended by proposed CASTOR calorimeter, which will cover the angular range $5.4 < \eta < 6.7$. Already with this device small-$x$ parton dynamics can be studied down to very small $x$-values of $10^{-6} - 10^{-7}$ with Drell-Yan, prompt photon and jet events at small invariant masses of the order of $M \sim 10$ GeV.

In the present work we investigate the technical possibilities of extending the angular acceptance for forward energy measurement beyond 7 units of pseudorapidity. Extending the acceptance down to $\eta \sim 11$, $x$-values down to $10^{-8}$ can be reached, which is a completely unexplored region of phase space. In this region, effects coming from new parton dynamics are expected to show up, as well as effects coming from very high density gluonic systems, where saturation and recombination effects will occur. In this region of phase space, a breakdown of the usual factorization formalism is expected, and multiple interactions will be dominant (see Fig 1) [1]. The full angular coverage from the central to the most forward region allows a systematic study of the transition from single particle exchange processes to complex systems and a systematic understanding of non-linear and collective phenomena.

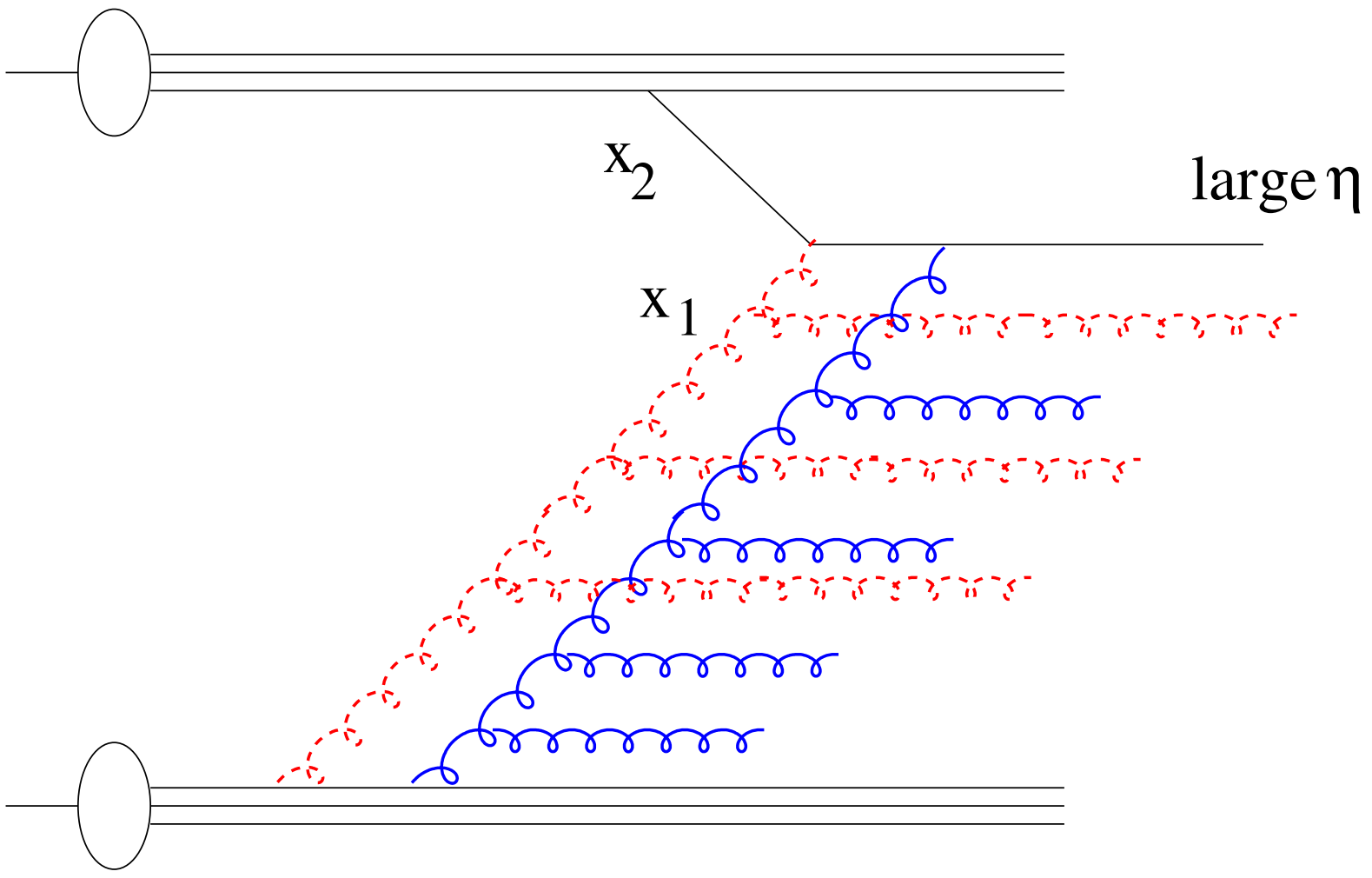

**Fig. 1:** Schematic picture of multiple interactions at small $x$

The interest in the very forward region of phase space is not only motivated by the fundamental understanding of QCD in a new phase of matter, but is also important for the further understanding of high energetic cosmic rays [2].

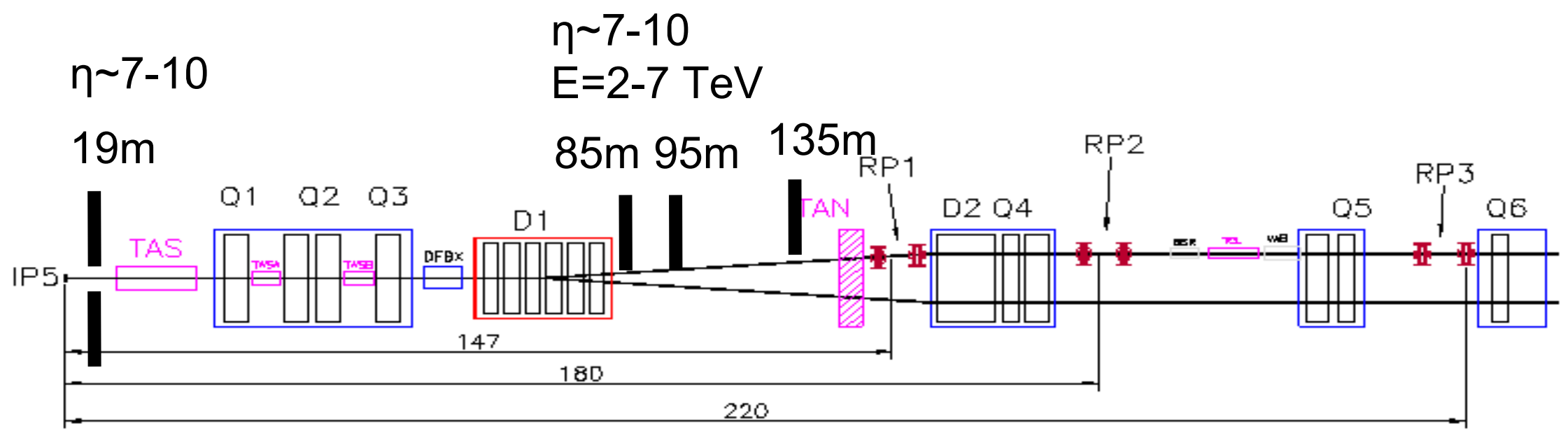

**Fig. 2:** Schematic view of forward beamline at CMS detector up to 220m. The positions of proposed tracking devices at 19m, 85m and 95m and calorimeter at 135m are indicated.

## 2 Tracking and energy measurement in the very forward region

For this study the geometry of the beam-line around the CMS detector up to 150m from interaction point has been implemented in the GEANT-3 [3] simulation program. The PYTHIA Monte-Carlo generator program [4] was used to generate charged particles produced in the interaction point, which were subsequently fed into the beam-line simulation.

The main restriction for additional installations is the very limited space available between magnetic elements. Up to about 80m there is no space for a calorimeter, and there one can only consider the installation of tracking devices, such as Roman Pots or micro-stations [5]. On the other hand, to be able to measure the particle momenta, the tracking devices should be placed after bending magnets.

Therefore the idea is to have tracking devices before and after dipoles, to be able to measure both the integrated particle flow and the particle momenta. The free space after 135 m can be used for a calorimeter.

Taking into account the limited space available for new detectors, the background conditions and magnetic field, the following strategy is proposed (Fig.2):

– the 25 cm space in front of TAS absorber at 19 m from interaction point can be used to install two micro-stations with two half-ring radiation hard silicon or diamond detectors approaching the beam horizontally up to 5-10mm. At this position the particles can not be separated by their momenta, thus the micro-stations will measure the charged particle flow integrated up to energies of $\sim$ 7 TeV in the pseudorapidity range between 7.3 and 9 or 10.5, depending on how close the counter can go to the beam (see Fig. 3). The detector will also provide accurate position measurement which is necessary for linking with roman-pots/micro-stations installed further down the beam line. Combining the position and time-of-flight measurement of these micro-stations with the event vertex measured in central detector will allow to suppress beam-wall background and pile-up events;

– a combination of two horizontal roman-pots/micro-stations can be installed behind the dipole magnets D1 at 85m and 95m. The detectors are the half-rings and approach the beam horizontally from one side up to 10mm. These detectors will cover the pseudorapidity range above 8 units (see Fig. 5). The particles with energies below 2 TeV will escape the detector acceptance, as shown in Fig. 4 and 5.

– a hadronic calorimeter at 135m (in front of TAN iron absorber) with a minimal distance to the beam of 10cm (radius of beam-pipe) will measure the energy in the range 2-5.5 TeV and pseudorapidity between 7 and 11 (see Fig.6). This can be a sandwich type calorimeter with radiation hard sensitive layers, with a transverse size up to $1\times1$ $m^2$ and depth about 7-9 hadronic interaction length. Optionally one can consider to instrument the TAN absorber with sensitive layers.

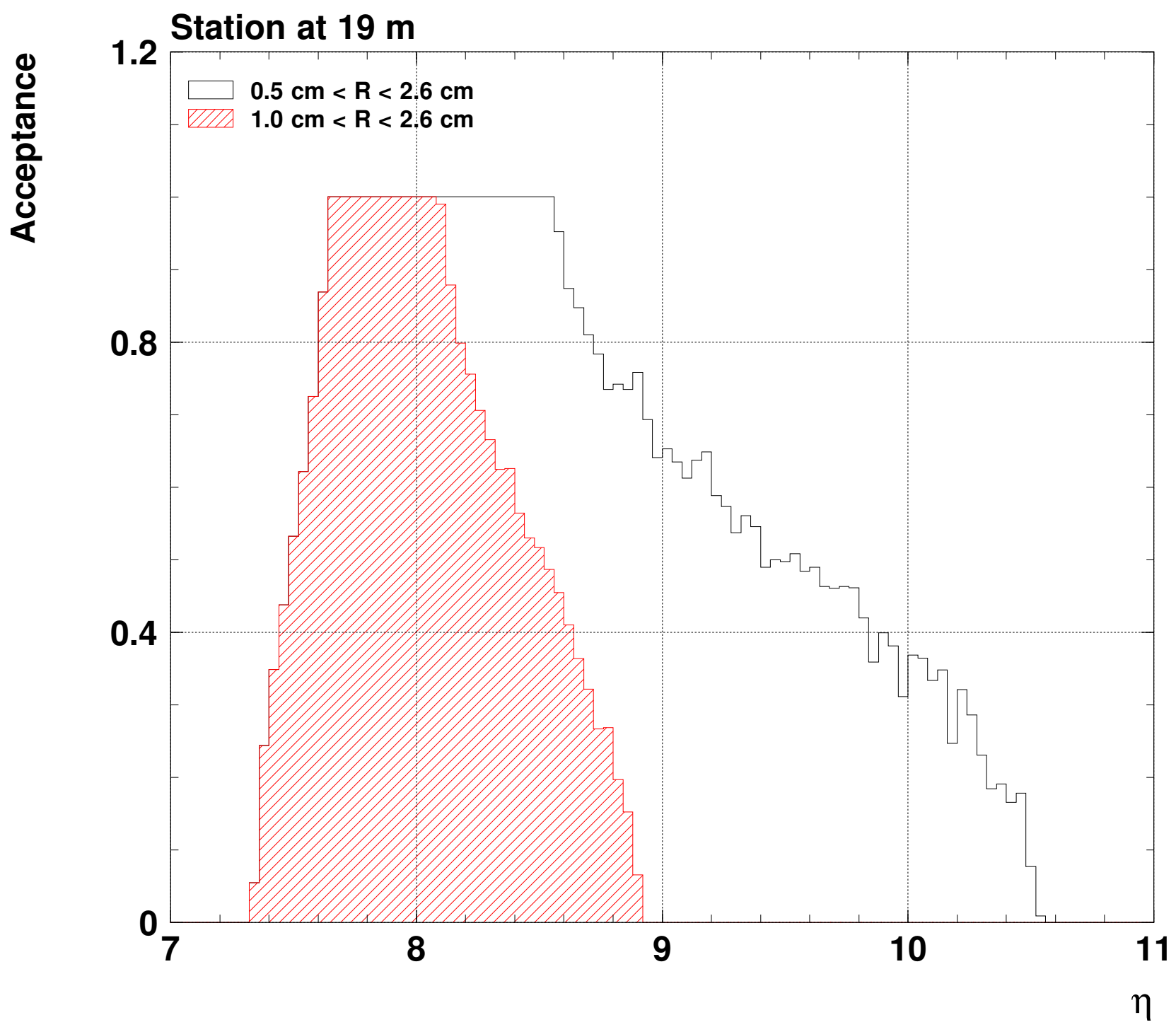

**Fig. 3:** Acceptance of micro-station detector at 19m as a function of pseudorapidity.

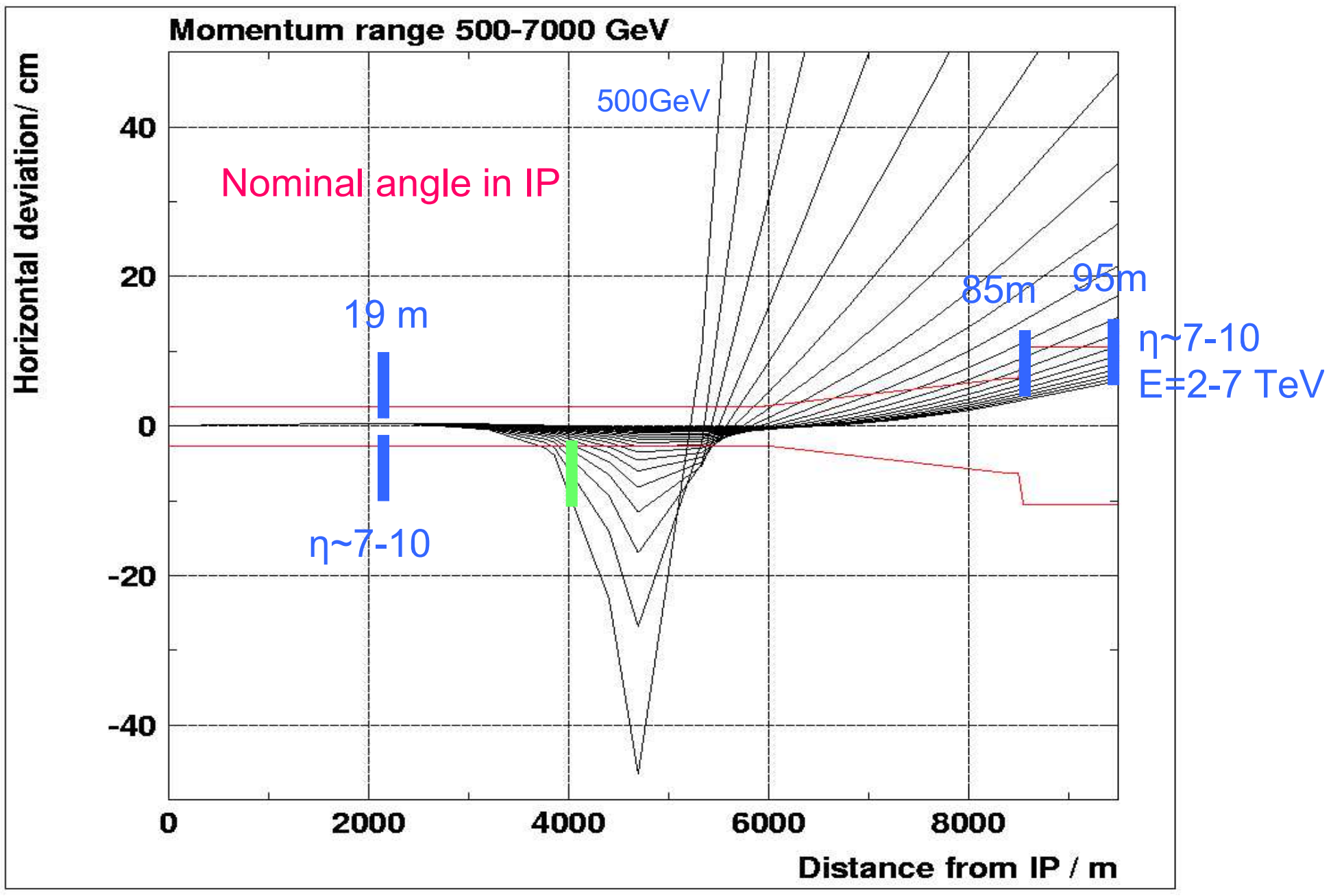

**Fig. 4:** The trajectory of particles in momentum range 500-7000 GeV, scattered from interaction point at 0 deg. The positions of proposed tracking devices at 19m, 85m and 95m are indicated.

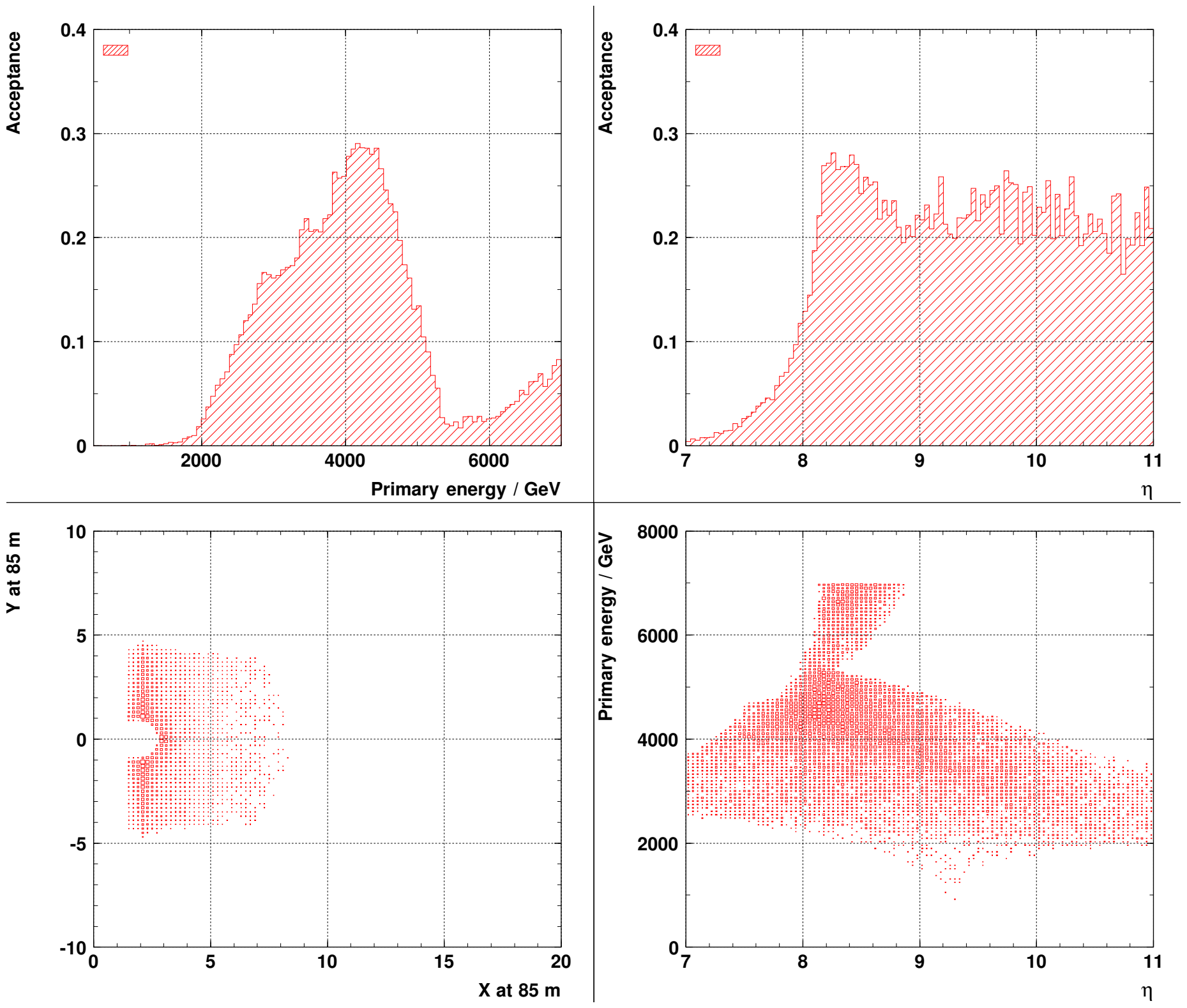

**Fig. 5:** Acceptance of roman pots/micro-station detector at 85 m and 95m as a function of energy and pseudorapidity.

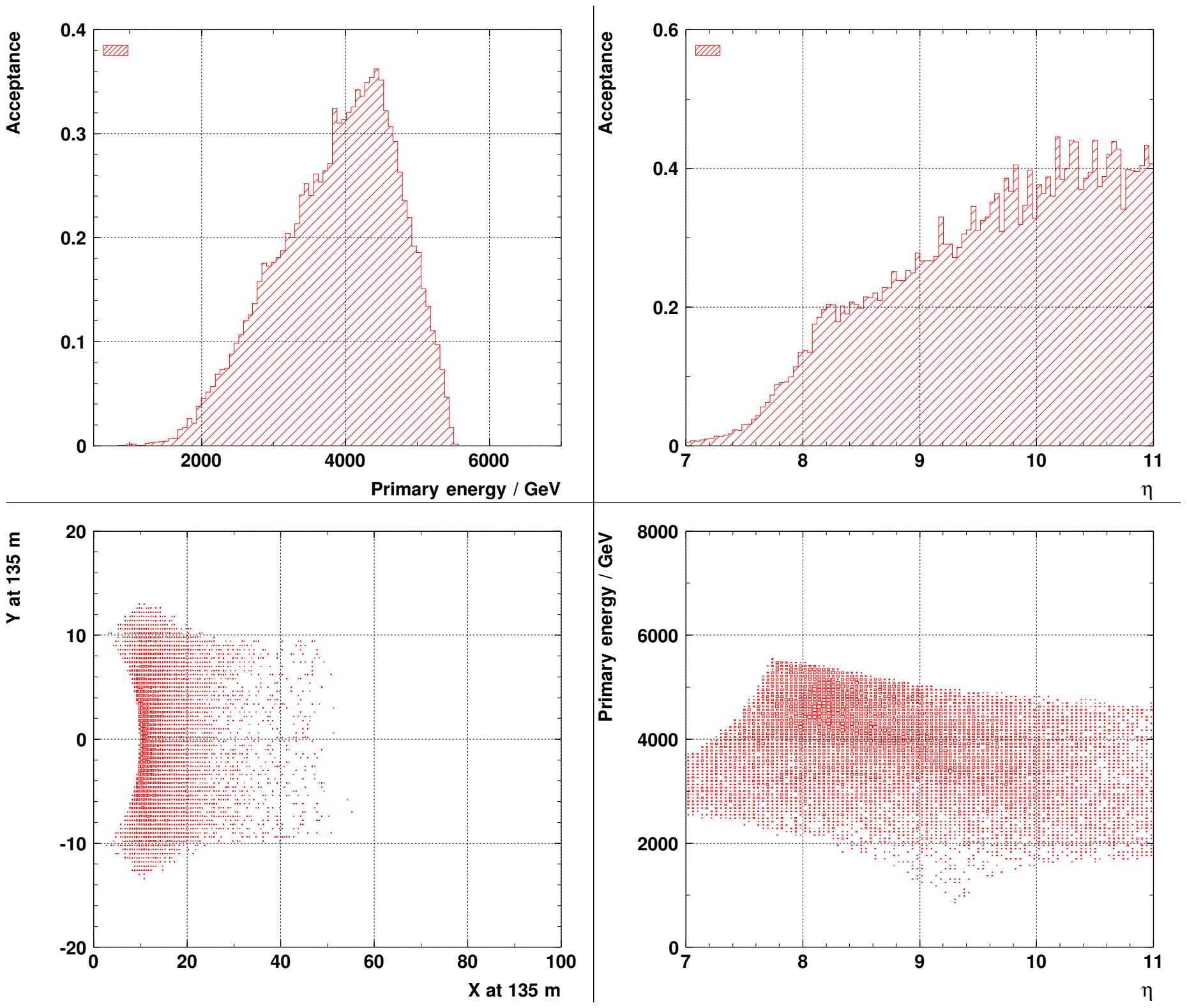

**Fig. 6:** Acceptance of calorimeter at 135 m as a function of energy and pseudorapidity.

The calorimeter covers basically the same kinematic range as the micro-stations at 85 and 95m, but it is needed for energy measurement. In addition it can be used for redundancy, background subtraction and cross calibration.

The acceptances of the proposed detectors as function of energy and pseudorapidity are summarized in Table 1. The result shows that with the proposed installations it will be possible to measure the energy flow in the energy range between 2 and 7 GeV and the pseudorapidity range between 7 and 11.

**Table 1:** Acceptance as a function of $E_p$ and $\eta$

| | 0.5–7 TeV | 2–5.5 TeV |
|---|---|---|
| two roman-pots/micro-stations at 85 and 95m | | |
| $\eta = 7 - 10$ | 11% | 21% |
| $\eta = 7 - 8$ | 10% | 10–20% |
| $\eta = 8 - 9$ | 15-25% | 30–55% |
| $\eta = 9 - 11$ | 20-25% | 55–60% |
| Calorimeter at 135m | | |
| $\eta = 7 - 8$ | 15% | 25% |
| $\eta = 8 - 9$ | 20-25% | 35–55% |
| $\eta = 9 - 11$ | 25-40% | 45–60% |

To be able to measure the particles with momenta below 2 TeV one would need to install detectors in the cold area between the quadrupoles at 40–50 m. This will require essential modifications of cryogenic lines, and can be considered for a future upgrade program.

## 3 Conclusions

We have studied the possibilities of extending the angular acceptance for forward energy measurement at LHC. With additional roman-pots/micro-stations and a calorimeter it will be possible to measure the forward energy in the rapidity range between 7 and 11 units. Such installation will be a valuable addition to the LHC physics program.

# Diffractive Higgs production: theory

*Jeff Forshaw*
Particle Physics Group, School of Physics & Astronomy,
University of Manchester, Manchester, M13 9PL. United Kingdom.

**Abstract**
We review the calculation for Higgs production via the exclusive reaction $pp \to p + H + p$. In the first part we review in some detail the calculation of the Durham group and emphasise the main areas of uncertainty. Afterwards, we comment upon other calculations.

## 1 Introduction

Our aim is to compute the cross-section for the process $pp \to p + H + p$. We shall only be interested in the kinematic situation where all three final state particles are very far apart in rapidity with the Higgs boson the most central. In this "diffractive" situation the scattering protons lose only a very small fraction of their energy, but nevertheless enough to produce the Higgs boson. Consequently, we are in the limit where the incoming protons have energy $E$ much greater than the Higgs mass $m_H$ and so we will always neglect terms suppressed by powers of $m_H/E$. In the diffractive limit cross-sections do not fall as the beam energy increases as a result of gluonic (spin-1) exchanges in the $t$-channel.

Given the possibility of instrumenting the LHC to detect protons scattered through tiny angles with a high resolution [1–4], diffractive production of any central system $X$ via $pp \to p + X + p$ is immediately of interest if the production rate is large enough. Even if $X$ is as routine as a pair of high $p_T$ jets we can learn a great deal about QCD in a new regime [2, 3, 5, 6]. But no doubt the greatest interest arises if $X$ contains "new physics" [7–19]. The possibility arises to measure the new physics in a way that is not possible using the LHC general purpose detectors alone. For example, its invariant mass may be measured most accurately, and the spin and CP properties of the system may be explored in a manner more akin to methods hitherto thought possible only at a future linear collider. Our focus here is on the production of a Standard Model Higgs boson [7, 8, 13, 18, 19]. Since the production of the central system $X$ effectively factorizes, our calculation will be seen to be of more general utility.

Most of the time will be spent presenting what we shall call the "Durham Model" of central exclusive production [7, 8]. It is based in perturbative QCD and is ultimately to be justified a posteriori by checking that there is not a large contribution arising from physics below 1 GeV. A little time will also be spent explaining the non-perturbative model presented by the Saclay group [13] and inspired by the original paper of Bialas and Landshoff [20]. Even less time will be devoted to other approaches which can be viewed, more-or-less, as hybrids of the other two [18, 19].

Apart from the exclusive process we study here, there is also the possibility to produce the new physics in conjunction with other centrally produced particles, e.g. $pp \to p + H + X + p$. This more inclusive channel typically has a much higher rate but does not benefit from the various advantages of exclusive production. Nevertheless, it must be taken into account in any serious phenomenological investigation into the physics potential of central exclusive production [21, 22]

## 2 The Durham Model

The calculation starts from the easier to compute parton level process $qq \to q + H + q$ shown in Figure 1. The Higgs is produced via a top quark loop and a minimum of two gluons need to be exchanged in order that no colour be transferred between the incoming and outgoing quarks. Quark exchange in the $t$-channel leads to contributions which are suppressed by an inverse power of the beam energy and so the diagram in Figure 1 is the lowest order one. Our strategy will be to compute only the imaginary part of the amplitude and we shall make use of the Cutkosky rules to do that – the relevant cut is indicated by the vertical dotted line in Figure 1. There is of course a second relevant diagram corresponding to the Higgs being emitted from the left-hand gluon. We shall assume that the real part of the amplitude is negligible, as it will be in the limit of asymptotically high centre-of-mass energy when the quarks are scattered through small angles and the Higgs is produced centrally.

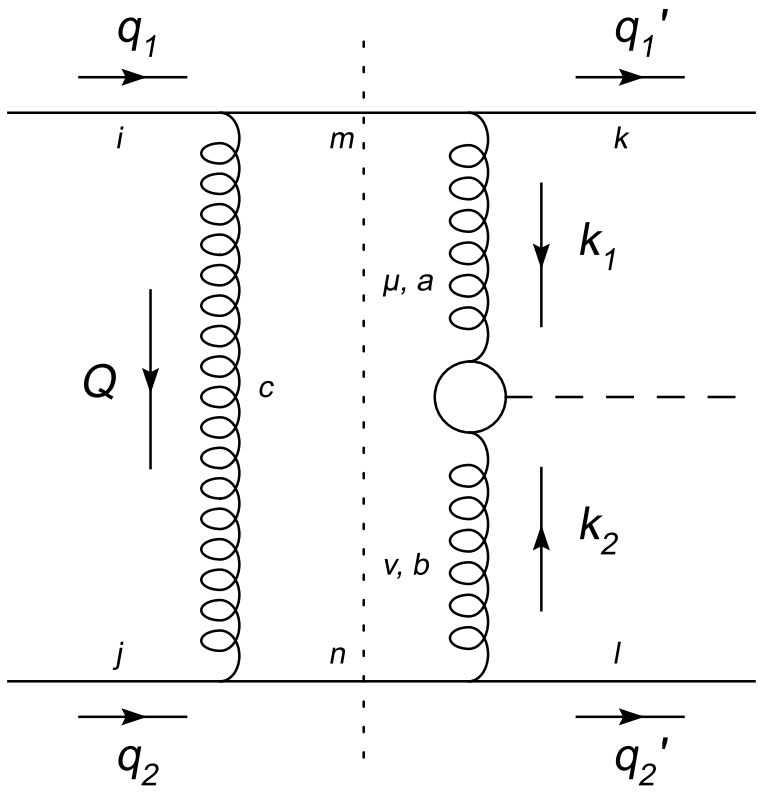

**Fig. 1:** The relevant Feynman graph for $qq \to q + H + q$.

The calculation can be further simplified by making use of the eikonal approximation for those vertices which couple the gluons to the external quarks. The gluons are very soft and so, modulo corrections which are suppressed by the inverse of the beam energy, we can approximate the $qqg$ vertices by $2g\tau^a_{ij} q_{1,2}\delta_{\lambda,\lambda'}$, where $\tau^a$ is a Gell-Mann matrix, $g$ is the QCD coupling and the Kronecker delta tells us that the quark does not change its helicity. The calculation of the amplitude is now pretty straightforward:

$$\begin{aligned} \mathrm{Im}A^{ik}_{jl} &= \frac{1}{2} \times 2 \int d(PS)_2 \, \delta((q_1 - Q)^2)\delta((q_2 + Q)^2) \\ & \quad \frac{2gq_1^\alpha \, 2gq_{2\alpha}}{Q^2} \frac{2gq_1^\mu}{k_1^2} \frac{2gq_2^\nu}{k_2^2} V^{ab}_{\mu\nu} \, \tau^c_{im}\tau^c_{jn}\tau^a_{mk}\tau^b_{nl} \, . \end{aligned} \tag{1}$$

The factor of $1/2$ is from the cutting rules and the factor of 2 takes into account that there are two diagrams. The phase-space factor is

$$d(PS)_2 = \frac{s}{2} \int \frac{d^2\mathbf{Q_T}}{(2\pi)^2} \, d\alpha d\beta \tag{2}$$

where we have introduced the Sudakov variables via $Q = \alpha q_1 + \beta q_2 + Q_T$. The delta functions fix the cut quark lines to be on-shell, which means that $\alpha \approx -\beta \approx \mathbf{Q_T}^2/s \ll 1$ and $Q^2 \approx Q_T^2 \equiv -\mathbf{Q_T}^2$. As always, we are neglecting terms which are energy suppressed such as the product $\alpha\beta$. For the Higgs production vertex we take the Standard Model result:

$$V^{ab}_{\mu\nu} = \delta^{ab} \left( g_{\mu\nu} - \frac{k_{2\mu}k_{1\nu}}{k_1 \cdot k_2} \right) V \tag{3}$$

where $V = m_H^2 \alpha_s/(4\pi v) F(m_H^2/m_t^2)$ and $F \approx 2/3$ provided the Higgs is not too heavy. The Durham group also include a NLO K-factor correction to this vertex. After averaging over colours we have

$$\tau^c_{im}\tau^c_{jn}\tau^a_{mk}\tau^b_{nl} \to \frac{\delta^{ab}}{4N_c^2}.$$

We can compute the contraction $q_1^\mu V_{\mu\nu}^{ab} q_2^\nu$ either directly or by utilising gauge invariance which requires that $k_1^\mu V_{\mu\nu}^{ab} = k_2^\nu V_{\mu\nu}^{ab} = 0$. Writing[1] $k_i = x_i q_i + k_{iT}$ yields

$$q_1^\mu V_{\mu\nu}^{ab} q_2^\nu \approx \frac{k_{1T}^\mu}{x_1} \frac{k_{2T}^\nu}{x_2} V_{\mu\nu}^{ab} \approx \frac{s}{m_H^2} k_{1T}^\mu k_{2T}^\nu V_{\mu\nu}^{ab} \tag{4}$$

since $2k_1 \cdot k_2 \approx x_1 x_2 s \approx m_H^2$. Note that it is as if the gluons which fuse to produce the Higgs are transversely polarized, $\epsilon_i \sim k_{iT}$. Moreover, in the limiting case that the outgoing quarks carry no transverse momentum $Q_T = -k_{1T} = k_{2T}$ and so $\epsilon_1 = -\epsilon_2$. This is an important result; it clearly generalizes to the statement that the centrally produced system should have a vanishing $z$-component of angular momentum in the limit that the protons scatter through zero angle (i.e. $q_{iT}'^2 \ll Q_T^2$). Since we are experimentally interested in very small angle scattering this selection rule is effective. One immediate consequence is that the Higgs decay to $b$-quarks may now be viable. This is because, for massless quarks, the lowest order $q\bar{q}$ background vanishes identically (it does not vanish at NLO). The leading order $b\bar{b}$ background is therefore suppressed by a factor $\sim m_b^2/m_H^2$. Beyond leading order, one also needs to worry about the $b\bar{b}g$ final state.

Returning to the task in hand, we can write the colour averaged amplitude as

$$\frac{\mathrm{Im}A}{s} \approx \frac{N_c^2 - 1}{N_c^2} \times 4\alpha_s^2 \int \frac{d^2\mathbf{Q_T}}{\mathbf{Q_T}^2 \mathbf{k_{1T}}^2 \mathbf{k_{2T}}^2} \frac{-\mathbf{k_{1T}} \cdot \mathbf{k_{2T}}}{m_H^2} V. \tag{5}$$

Using $d^3\mathbf{q_1}' d^3\mathbf{q_2}' d^3\mathbf{q_H} \delta^{(4)}(q_1 + q_2 - q_1' - q_2' - q_H) = d^2\mathbf{q_{1T}}' d^2\mathbf{q_{2T}}' dy\ E_H$ ($y$ is the rapidity of the Higgs) the cross-section is therefore

$$\frac{d\sigma}{d^2\mathbf{q_{1T}}' d^2\mathbf{q_{2T}}' dy} \approx \left(\frac{N_c^2 - 1}{N_c^2}\right)^2 \frac{\alpha_s^6}{(2\pi)^5} \frac{G_F}{\sqrt{2}} \left[\int \frac{d^2\mathbf{Q_T}}{2\pi} \frac{\mathbf{k_{1T}} \cdot \mathbf{k_{2T}}}{\mathbf{Q_T}^2 \mathbf{k_{1T}}^2 \mathbf{k_{2T}}^2} \frac{2}{3}\right]^2 \tag{6}$$

and for simplicity here we have taken the large top mass limit of $V$ (i.e. $m_t \gg m_H$). We are mainly interested in the forward scattering limit whence

$$\frac{\mathbf{k_{1T}} \cdot \mathbf{k_{2T}}}{\mathbf{Q_T}^2 \mathbf{k_{1T}}^2 \mathbf{k_{2T}}^2} \approx -\frac{1}{\mathbf{Q_T}^4}.$$

As it stands, the integral over $Q_T$ diverges. Let us not worry about that for now and instead turn our attention to how to convert this parton level cross-section into the hadron level cross-section we need.[2]

What we really want is the hadronic matrix element which represents the coupling of two gluons into a proton, and this is really an off-diagonal parton distribution function [23]. At present we don't have much knowledge of these distributions, however we do know the diagonal gluon distribution function. Figure 2 illustrates the Durham prescription for coupling the two gluons into a proton rather than a quark. The factor $K$ would equal unity if $x' = x$ and $k_T = 0$ which is the diagonal limit. That we should, in the amplitude, replace a factor of $\alpha_s C_F/\pi$ by $\partial G(x, Q_T)/\partial \ln Q_T^2$ can be easily derived starting from the DGLAP equation for evolution off an initial quark distribution given by $q(x) = \delta(1 - x)$. The Durham approach makes use of a result derived in [24] which states that in the case $x' \ll x$ and $k_T^2 \ll Q_T^2$ the off-diagonality can be approximated by a multiplicative factor, $K$. Assuming a Gaussian form factor suppression for the $k_T$-dependence they estimate that

$$K \approx e^{-bk_T^2/2} \frac{2^{2\lambda+3}}{\sqrt{\pi}} \frac{\Gamma(\lambda + 5/2)}{\Gamma(\lambda + 4)} \tag{7}$$

[1] We can do this because $x_i \sim m_H/\sqrt{s}$ whilst the other Sudakov components are $\sim Q_T^2/s$.

[2] We note that (6) was first derived by Bialas and Landshoff, except that they made a factor of 2 error in the Higgs width to gluons.

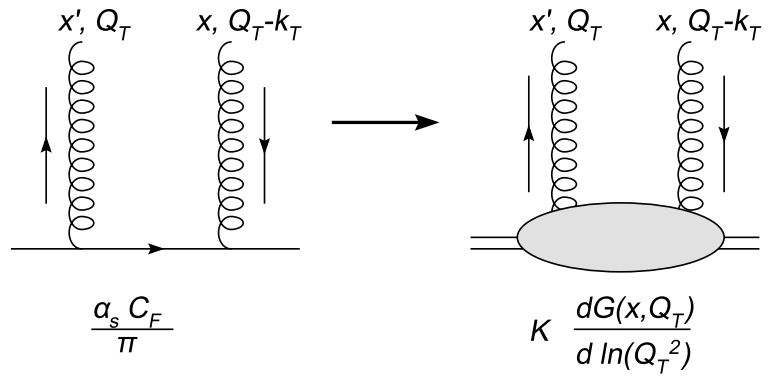

**Fig. 2:** The recipe for replacing the quark line (left) by a proton line (right).

and this result is obtained assuming a simple power-law behaviour of the gluon density, i.e. $G(x,Q) \sim x^{-\lambda}$. For the production of a 120 GeV Higgs boson at the LHC, $K \sim 1.2 \times e^{-bk_T^2/2}$. In the cross-section, the off-diagonality therefore provides an enhancement of $(1.2)^4 \approx 2$. Clearly the current lack of knowledge of the off-diagonal gluon is one source of uncertainty in the calculation. We also do not really know what to take for the slope parameter $b$. It should perhaps have some dependence upon $Q_T$ and for $Q_T \sim 1.5$ GeV, which it will turn out is typical for a 120 GeV scalar Higgs, one might anticipate the same $k_T$-dependence as for diffractive $J/\psi$ production which is well measured, i.e. $b \approx 4$ GeV$^{-2}$.

Thus, after integrating over the transverse momenta of the scattered protons we have

$$\frac{d\sigma}{dy} \approx \frac{1}{256\pi b^2} \frac{\alpha_s G_F \sqrt{2}}{9} \left[ \int \frac{d^2\mathbf{Q_T}}{\mathbf{Q_T}^4} f(x_1, Q_T) f(x_2, Q_T) \right]^2 \tag{8}$$

where $f(x,Q) \equiv \partial G(x,Q)/\partial \ln Q^2$ and we have neglected the exchanged transverse momentum in the integrand. Notice that in determining the total rate we have introduced uncertainty in the normalisation arising from our lack of knowledge of $b$. This uncertainty, as we shall soon see, is somewhat diminished as the result of a similar $b$-dependence in the gap survival factor.

We should about the fact that our integral diverges in the infra-red. Fortunately we have missed some crucial physics. The lowest order diagram is not enough, virtual graphs possess logarithms in the ratio $Q_T/m_H$ which are very important as $Q_T \to 0$; these logarithms need to be summed to all orders. This is Sudakov physics: thinking in terms of real emissions we must be sure to forbid real emissions into the final state. Let's worry about real gluon emission off the two gluons which fuse to make the Higgs. The emission probability for a single gluon is (assuming for the moment a fixed coupling $\alpha_s$)

$$\frac{C_A \alpha_s}{\pi} \int_{Q_T^2}^{m_H^2/4} \frac{dp_T^2}{p_T^2} \int_{p_T}^{m_H/2} \frac{dE}{E} \sim \frac{C_A \alpha_s}{4\pi} \ln^2 \left( \frac{m_H^2}{Q_T^2} \right) .$$

The integration limits are kinematic except for the lower limit on the $p_T$ integral. The fact that emissions below $Q_T$ are forbidden arises because the gluon not involved in producing the Higgs completely screens the colour charge of the fusing gluons if the wavelength of the emitted radiation is long enough, i.e. if $p_T < Q_T$. Now we see how this helps us solve our infra-red problem: as $Q_T \to 0$ so the screening gluon fails to screen and real emission off the fusing gluons cannot be suppressed. To see this argument through to its conclusion we realise that multiple real emissions exponentiate and so we can write the non-emission probability as

$$e^{-S} = \exp \left( -\frac{C_A \alpha_s}{\pi} \int_{Q_T^2}^{m_H^2/4} \frac{dp_T^2}{p_T^2} \int_{p_T}^{m_H/2} \frac{dE}{E} \right) . \tag{9}$$

As $Q_T \to 0$ the exponent diverges and the non-emission probability vanishes faster than any power of $Q_T$. In this way our integral over $Q_T$ becomes (its value is finite):

$$\int \frac{dQ_T^2}{Q_T^4} f(x_1, Q_T) f(x_2, Q_T)\, e^{-S} . \tag{10}$$

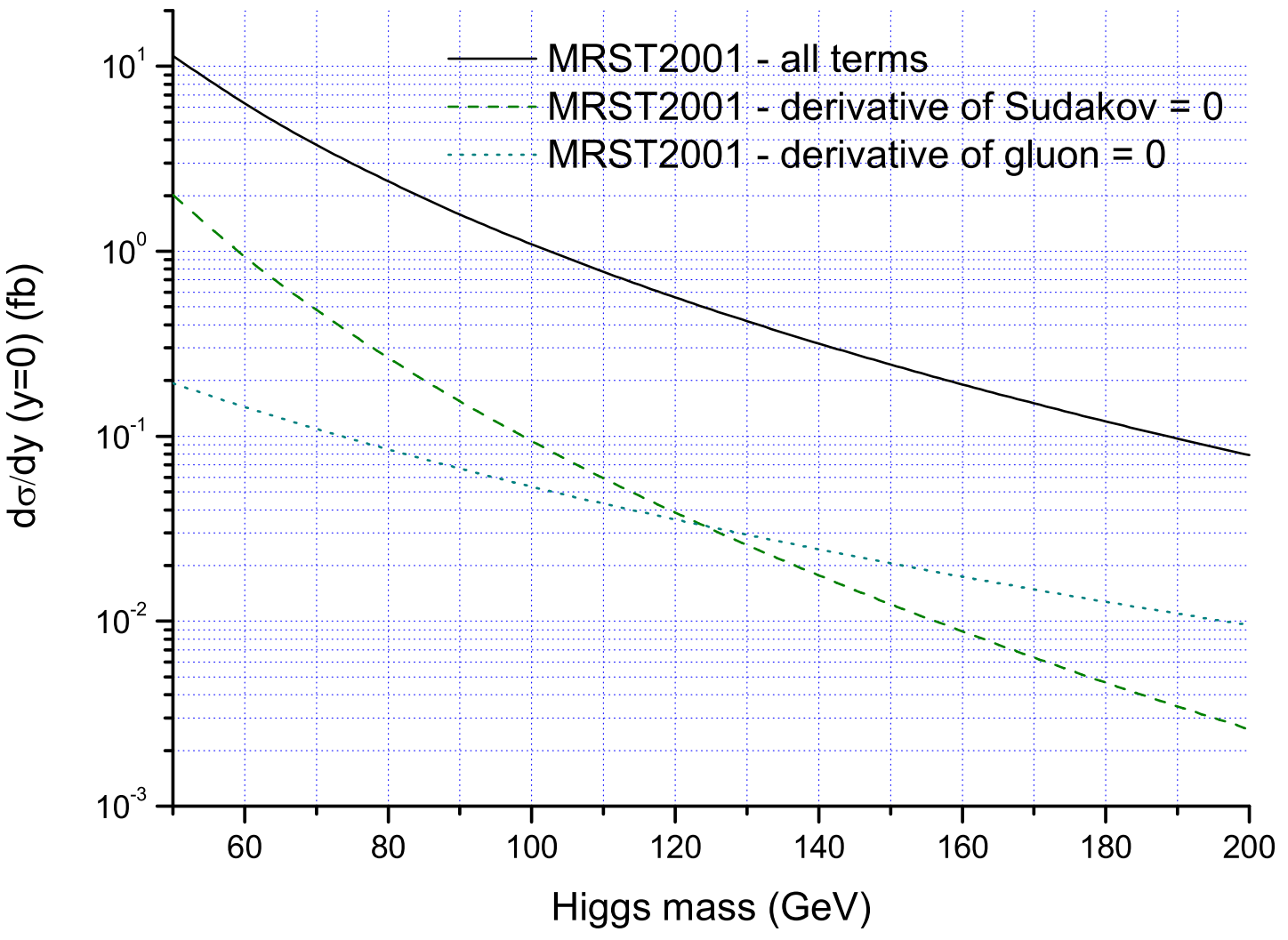

**Fig. 3:** The Higgs cross-section at zero rapidity, and the result obtained if one were to assume that $\partial G(x,Q)/\partial Q = 0$ or that $\partial S/\partial Q = 0$.

There are two loose ends to sort out before moving on. Firstly, note that emission off the screening gluon is less important since there are no associated logarithms in $m_H/Q_T$. Secondly, (9) is correct only so far as the leading double logarithms. It is of considerable practical importance to correctly include also the single logarithms. To do this we must re-instate the running of $\alpha_s$ and allow for the possibility that quarks can be emitted. Including this physics means we ought to use

$$e^{-S} = \exp\left(-\int_{Q_T^2}^{m_H^2/4} \frac{dp_T^2}{p_T^2}\frac{\alpha_s(p_T^2)}{2\pi}\int_0^{1-\Delta} dz\,\left[zP_{gg}(z) + \sum_q P_{qg}(z)\right]\right) \tag{11}$$

where $\Delta = 2p_T/m_H$, and $P_{gg}(z)$ and $P_{qg}(z)$ are the leading order DGLAP splitting functions. To correctly sum all single logarithms requires some care in that what we want is the distribution of gluons in $Q_T$ with no emission up to $m_H$, and this is in fact [25]

$$\tilde{f}(x, Q_T) = \frac{\partial}{\partial \ln Q_T^2}\left(e^{-S/2}\, G(x, Q_T)\right).$$

The integral over $Q_T$ is therefore

$$\int \frac{dQ_T^2}{Q_T^4}\tilde{f}(x_1, Q_T)\tilde{f}(x_2, Q_T) \tag{12}$$

which reduces to (10) in the double logarithmic approximation where the differentiation of the Sudakov factor is subleading.

The numerical effect of correctly including the single logarithms is large. For production of a 120 GeV Higgs at the LHC, there is a factor $\sim 30$ enhancement compared to the double logarithmic approximation, with a large part of this coming from terms involving the derivative of the Sudakov. Figure 3 shows just how important it is to keep those single logarithmic terms coming from differentiation of the Sudakov factor. For the numerical results we used the MRST2001 leading order gluon [26], as included in LHAPDF [27]. Here and elsewhere (unless otherwise stated), we use a NLO QCD K-factor of 1.5 and the one-loop running coupling with $n_f = 4$ and $\Lambda_{\rm QCD} = 160$ MeV. As discussed in the next paragraph, we also formally need an infra-red cut-off $Q_0$ for the $Q_T$-integral; we take $Q_0 = 0.3$ GeV although as we shall see results are insensitive to $Q_0$ provided it is small enough. Finally, all our results include an overall multiplicative "gap survival factor" of 3% (gap survival is discussed shortly).

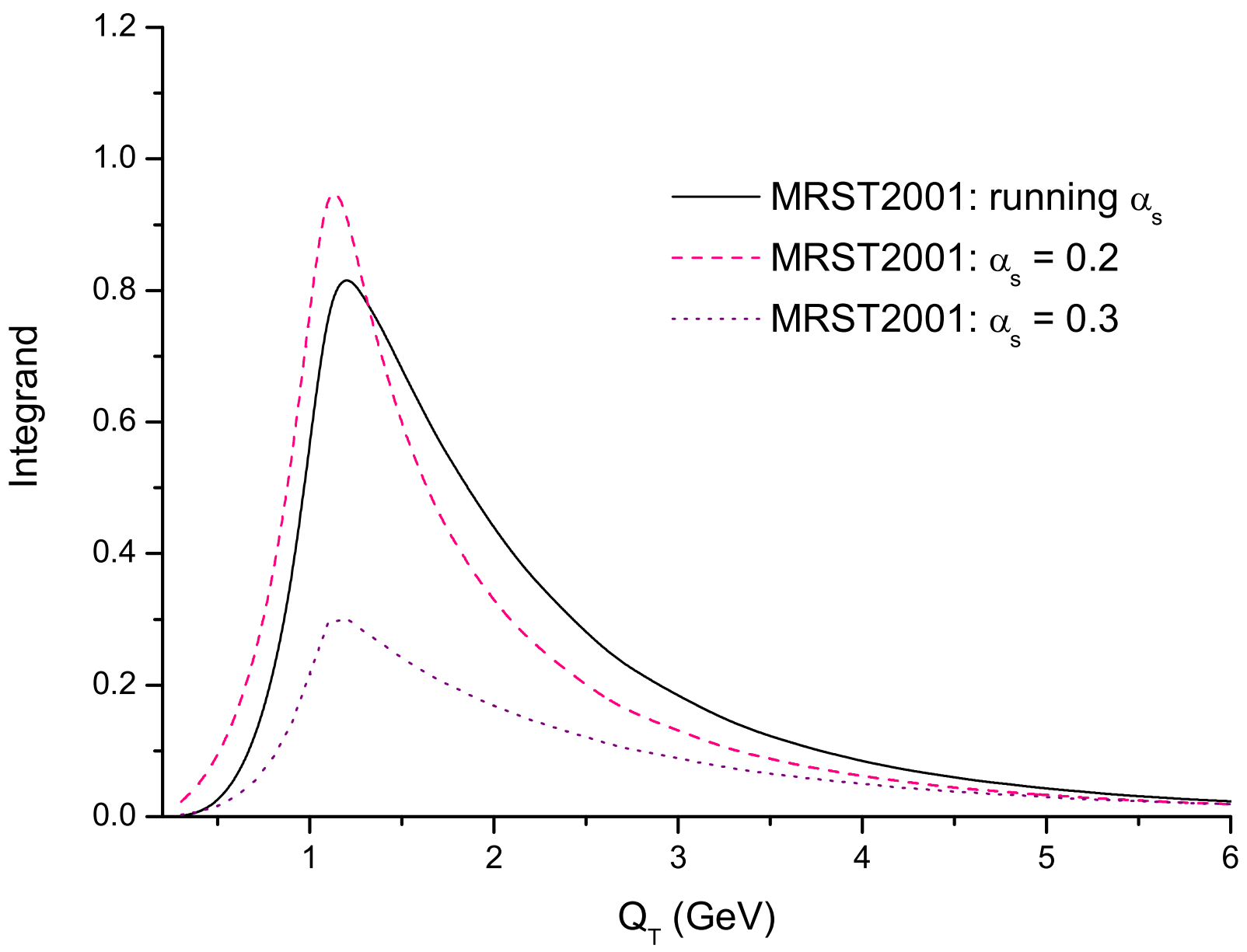

**Fig. 4:** The integrand of the $Q_T$ integral for three different treatments of $\alpha_s$ and $m_H = 120$ GeV.

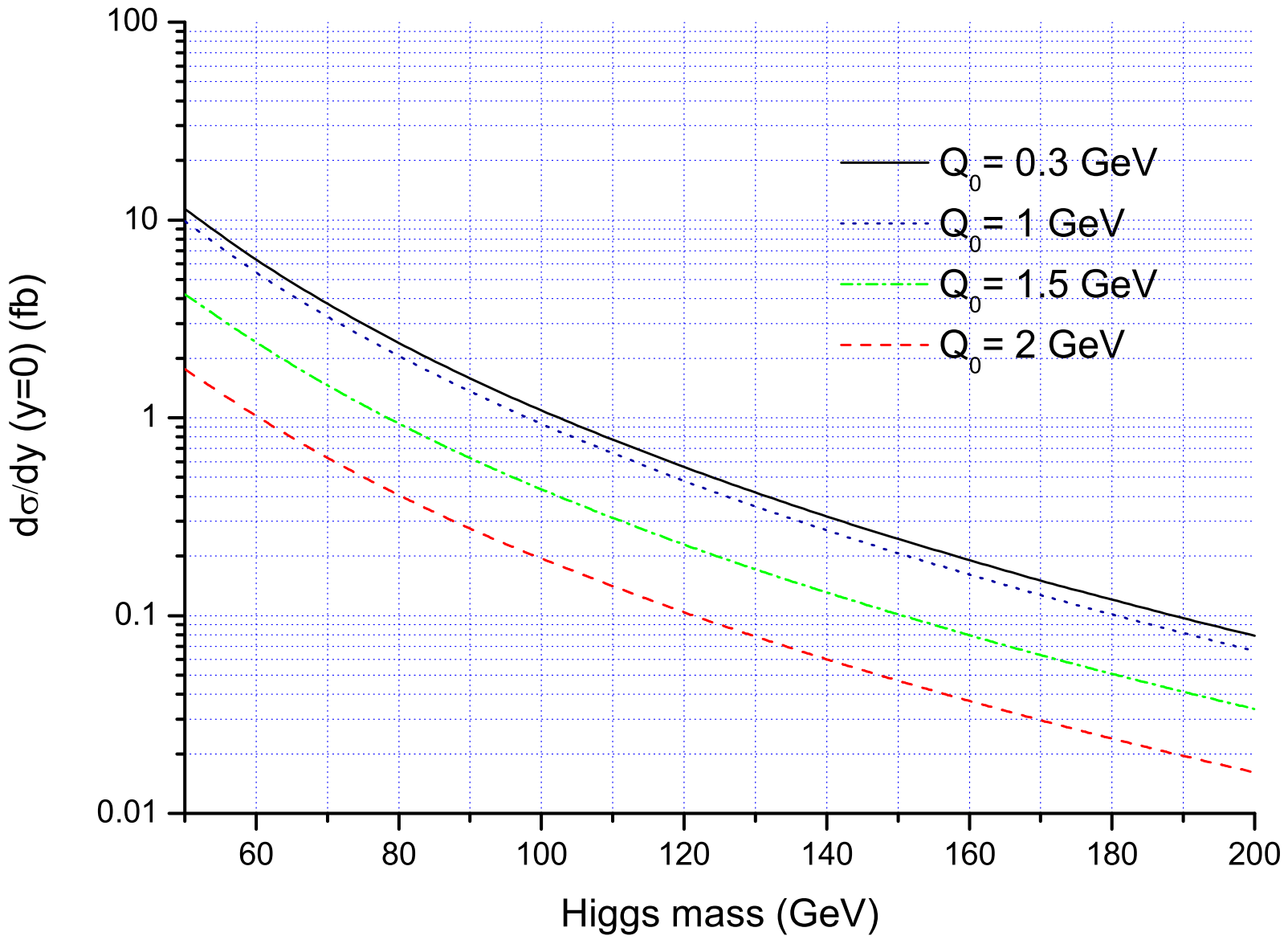

**Fig. 5:** The Higgs cross-section dependence upon the infra-red cutoff $Q_0$.

Formally there is the problem of the pole in the QCD coupling at $p_T = \Lambda_{\rm QCD}$. However, this problem can be side-stepped if the screening gluon has "done its job" sufficiently well and rendered an integrand which is peaked at $Q_T \gg \Lambda_{\rm QCD}$ since an infra-red cutoff on $p_T$ can then safely be introduced. We must be careful to check whether or not this is the case in processes of interest. Indeed, a saddle point estimate of (10) reveals that

$$\exp(\langle \ln Q_T \rangle) \sim \frac{m_H}{2} \exp\left(-\frac{c}{\alpha_s}\right) \tag{13}$$

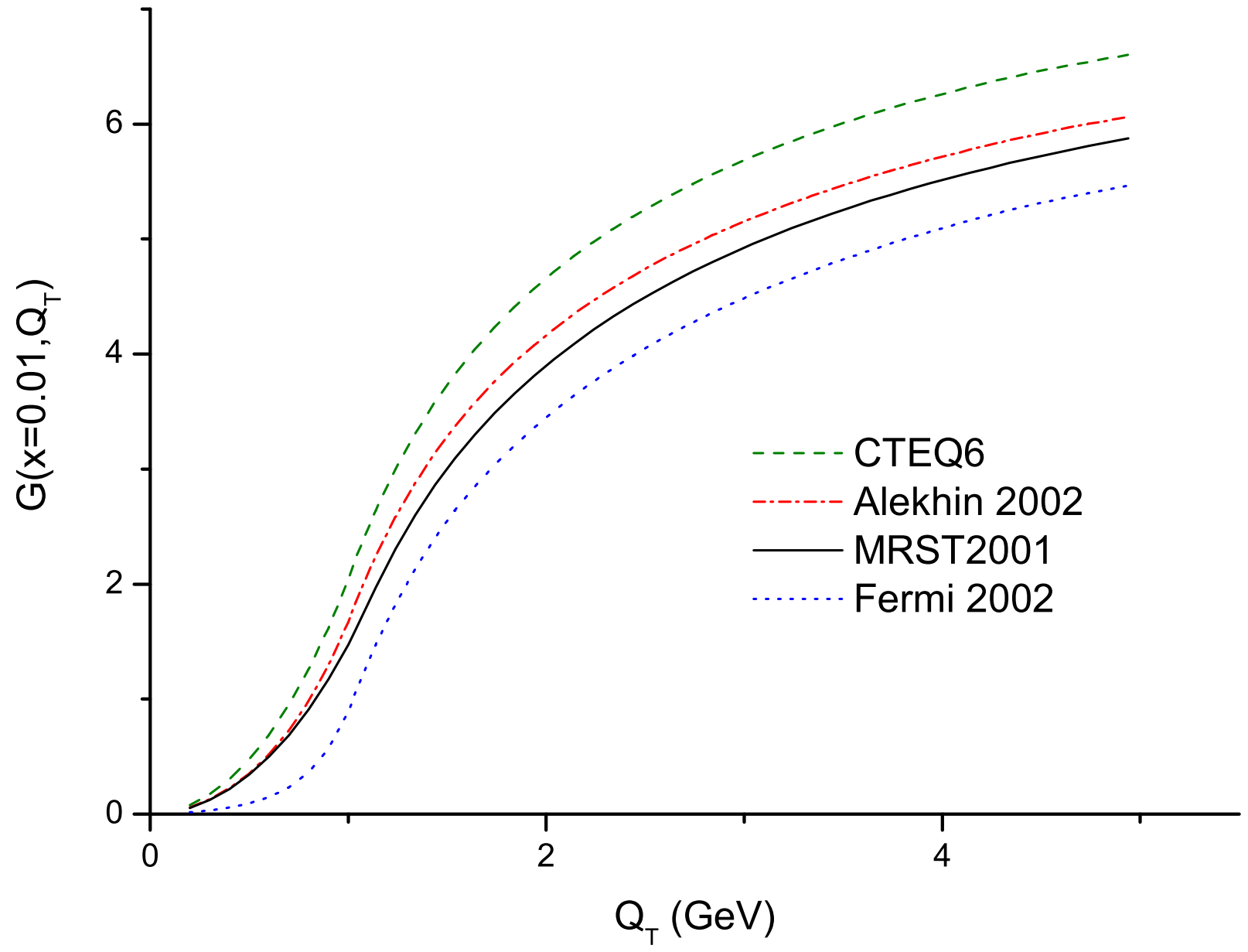

**Fig. 6:** The gluon density function in four different parameterisations.

where $c$ is a constant if the gluon density goes like a power of $Q_T^2$. Clearly there is a tension between the Higgs mass, which encourages a large value of the loop momentum, and the singular behaviour of the $1/Q_T^4$ factor which encourages a low value. Also, as $\alpha_s$ reduces so real emission is less likely and the Sudakov suppression is less effective in steering $Q_T$ away from the infra-red. Putting in the numbers one estimates that $\exp(\langle \ln Q_T^2 \rangle) \approx 4$ GeV$^2$ for the production of a 120 GeV scalar at the LHC which is just about large enough to permit an analysis using perturbative QCD. Figure 4 provides the quantitative support for these statements in the case of a Higgs of mass 120 GeV. The integrand of the $Q_T$ integral in equation (12) is shown for both running and fixed $\alpha_s$. We see that the integrand peaks just above 1 GeV and that the Sudakov factor becomes increasingly effective in suppressing the cross-section as $\alpha_s$ increases. Although it isn't too easy to see on this plot, the peak does move to higher values of $Q_T$ as $\alpha_s$ increases in accord with (13). This plot also illustrates quite nicely that the cross-section is pretty much insensitive to the infra-red cutoff for $Q_0 < 1$ GeV and this is made explicit in Figure 5.

Discussion of the infra-red sensitivity would not be complete without returning to the issue of the unintegrated gluon density. In all our calculations we model the off-diagonality as discussed below equation (7) and we shan't discuss this source of uncertainty any further here.[3] Figure 6 shows the gluon density $G(x, Q)$ as determined in four recent global fits (rather arbitrarily chosen to illustrate the typical variety) [26, 28–30]. Apart from the Fermi2002 fit, they are all leading order fits. Now, none of these parameterisations go down below $Q = 1$ GeV, so what is shown in the figure are the gluons extrapolated down to $Q = 0$. We have extrapolated down assuming that the gluon and its derivative are continuous at $Q = 1$ GeV and that $G(x, Q) \sim Q^2$ at $Q \to 0$.[4] The gluons plotted in Figure 6 are all determined at $x = 0.01$ which would be the value probed in the production of a 120 GeV Higgs at $y = 0$ at the LHC. The key point is to note that it is hard to think of any reasonable parameterisation of the gluon below 1 GeV which could give a substantial contribution to the cross-section. The Sudakov factor suppresses the low $Q^2$ region and also the size of the gluon and its derivative are crucial, and one cannot keep both of these large for $Q < 1$ GeV. Figure 7 shows the integrand of the $Q_T$ integral for different fits to the gluon. In all cases the contribution below 1 GeV is small, although there are clearly important uncertainties in the cross-section. These uncertainties are better seen in Figure 8 which illustrates that one might anticipate a factor of a few uncertainty from this source.

[3] We actually assume a constant enhancement factor of 1.2 per gluon density.

[4] To be precise we extrapolate assuming $G(x, Q) \sim Q^{2+(\gamma-2)Q}$.

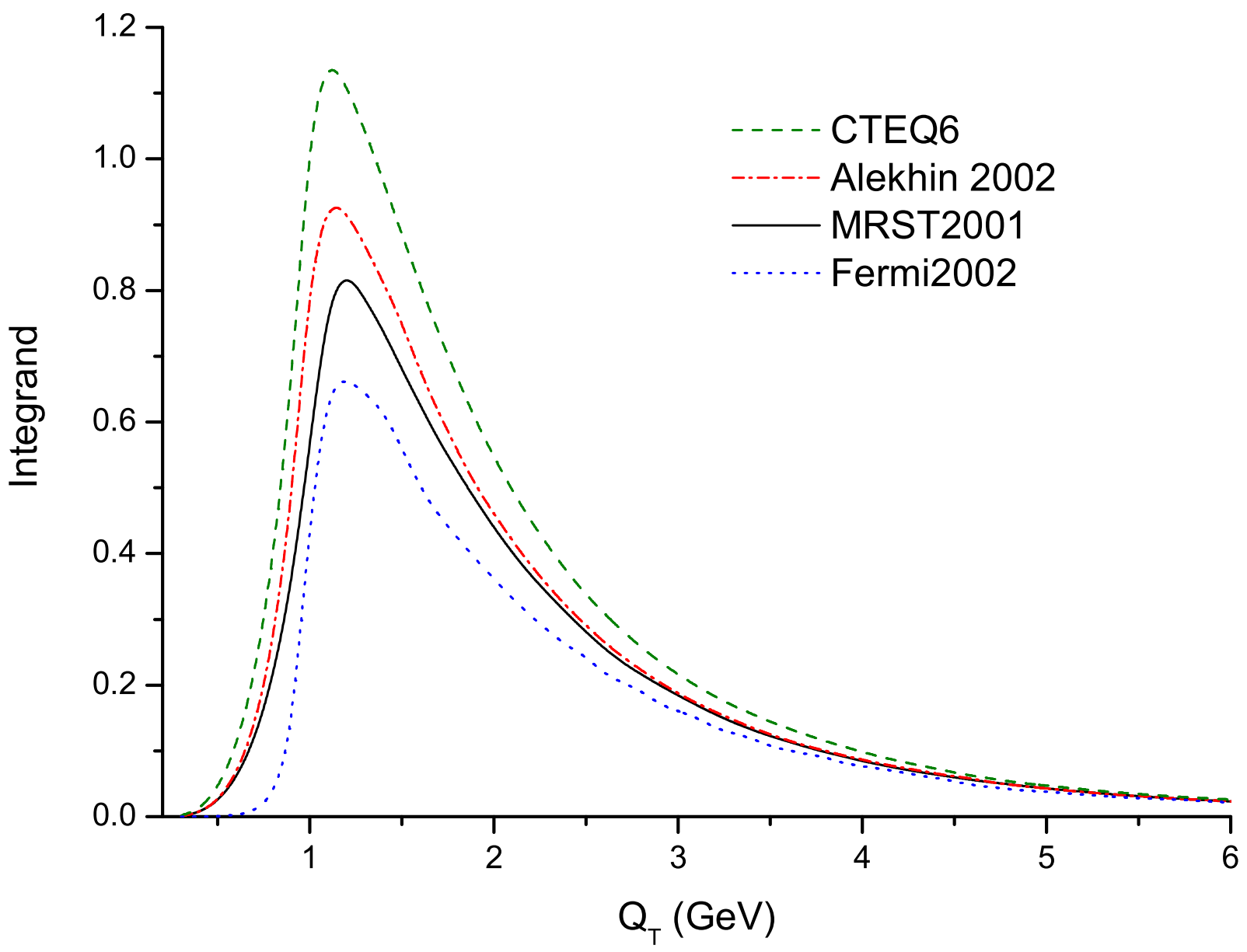

**Fig. 7:** The integrand of the $Q_T$ integral for four recent global fits to the gluon.

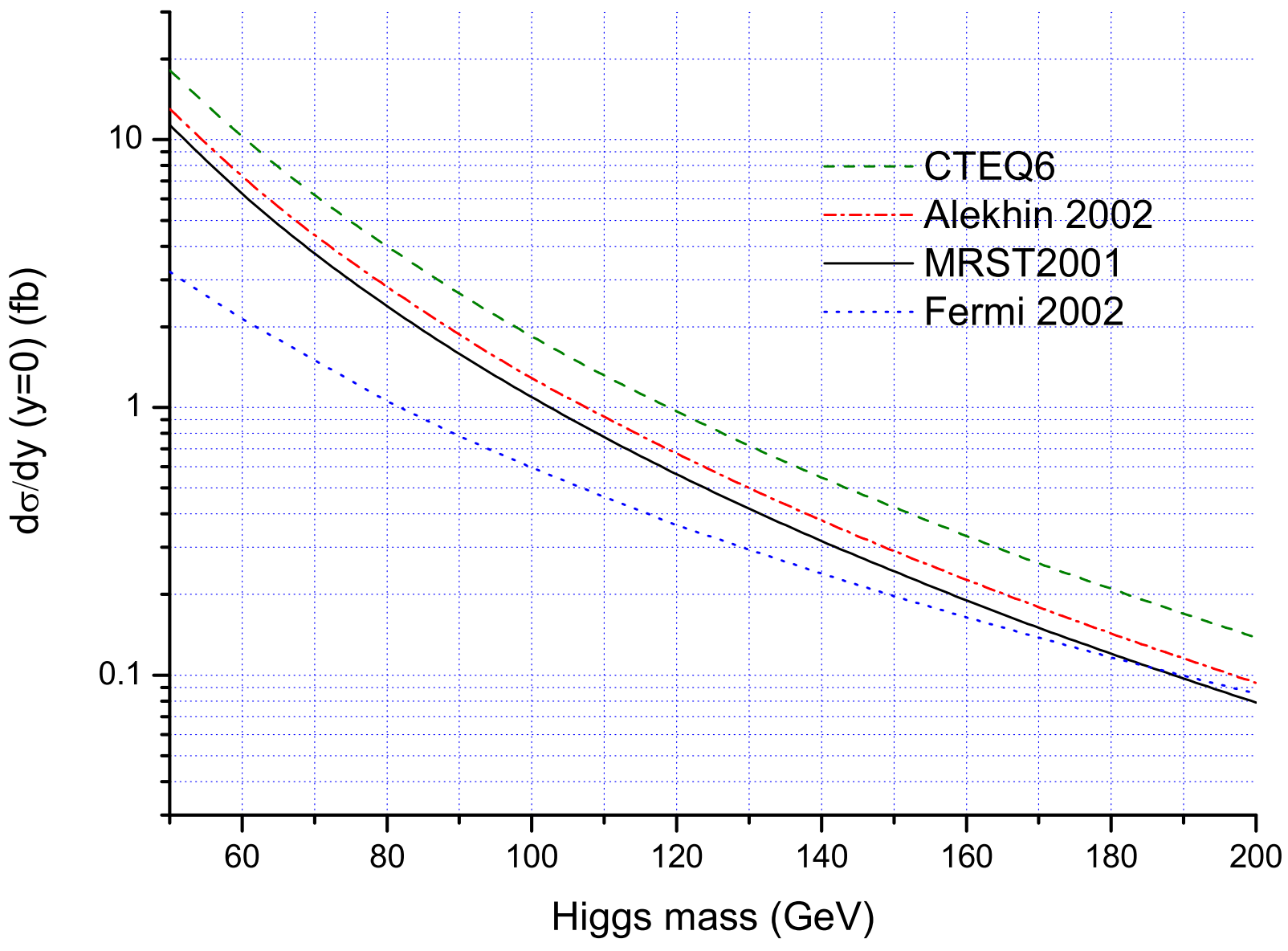

**Fig. 8:** The Higgs cross-section for four recent global fits to the gluon.

We note that although a variety of parameterizations are presented in Figure 8 the way that the actual $Q_T$ dependence of the integrand is obtained is the same in each case. In [31, 32] the uncertainties arising from the way the unintegrated parton densities are obtained from the integrated ones are examined. Here we have followed the prescription presented in [33] which amounts to performing one backward step in a DGLAP parton shower. However, it is known that such showers tend to underestimate the hardness of, for example, the $W/Z$ $p_\perp$ spectra in hadron colliders unless a large intrinsic transverse momentum is added to the perturbative $k_\perp$ distribution of the colliding partons [34, 35]. In [32] it was shown that adding such an intrinsic transverse momentum would harden the $Q_T$ distribution of the integrand in (12) for small $Q_T$ which in turn lowers the cross-section for central exclusive Higgs production by a factor 2 (for a Gaussian intrinsic transverse momentum with $\langle k_\perp^2 \rangle = 2$ GeV$^2$). Investigations into how one could use unintegrated gluon densities obtained by CCFM [36] and LDC [37] evolution for central exclusive Higgs production have also been performed [32]. However, as discussed in more detail elsewhere in these proceedings [23], the available parameterizations, which are all fitted to HERA data only, are not constrained enough to allow for reliable predictions for Higgs production at the LHC.

This is perhaps a good place to mention pseudo-scalar production, as might occur in an extension to the Standard Model. The scalar product, $\mathbf{k_{1T}} \cdot \mathbf{k_{2T}}$, in (6) now becomes $(\mathbf{k_{1T}} \times \mathbf{k_{2T}}) \cdot \mathbf{n}$, where $\mathbf{n}$ is a unit vector along the beam axis. After performing the angular integral the only surviving terms are proportional to the vector product of the outgoing proton transverse momenta, i.e. $\mathbf{q_1}' \times \mathbf{q_2}'$. Notice that this term vanishes, in accord with the spin-0 selection rule, as $\mathbf{q_i}' \to 0$. Notice also that the integrand now goes like $\sim 1/Q_T^6$ (in contrast to the $1/Q_T^4$ in the scalar case). As a result $c$ in (13) is larger (in fact it is linearly proportional to the power of $Q_T$) and the mean value of $Q_T$ smaller. This typically means that pseudo-scalar production is not really accessible to a perturbative analysis.

The Sudakov factor has allowed us to ensure that the exclusive nature of the final state is not spoilt by perturbative emission off the hard process. What about non-perturbative particle production? The protons can in principle interact quite apart from the perturbative process discussed hitherto and this interaction could well lead to the production of additional particles. We need to account for the probability that such emission does not occur. Provided the hard process leading to the production of the Higgs occurs on a short enough timescale, we might suppose that the physics which generates extra particle production factorizes and that its effect can be accounted for via an overall factor multiplying the cross-section we have just calculated. This is the "gap survival factor". Gap survival is discussed in detail elsewhere in these proceedings and so we'll not dwell on it here [38].

The gap survival, $S^2$, is given by

$$d\sigma(p + H + p|\text{no soft emission}) = d\sigma(p + H + p) \times S^2$$

where $d\sigma(p+H+p)$ is the differential cross-section computed above. The task is to estimate $S^2$. Clearly this is not straightforward since we cannot utilize QCD perturbation theory. Let us at this stage remark that data on a variety of processes observed at HERA, the Tevatron and the LHC can help us improve our understanding of "gap survival".

The model presented here provides a good starting point for understanding the more sophisticated treatments [39–41]. Dynamically, one expects that the likelihood of extra particle production will be greater if the incoming protons collide at small transverse separation compared to collisions at larger separations. The simplest model which is capable of capturing this feature is one which additionally assumes that there is a single soft particle production mechanism, let us call it a "re-scattering event", and that re-scattering events are independent of each other for a collision between two protons at transverse separation $r$. In such a model we can use Poisson statistics to model the distribution in the number of re-scattering events per proton-proton interaction:

$$P_n(r) = \frac{\chi(r)^n}{n!} \exp(-\chi(r)) \,. \tag{14}$$

This is the probability of having $n$ re-scattering events where $\chi(r)$ is the mean number of such events for proton-proton collisions at transverse separation $r$. Clearly the important dynamics resides in $\chi(r)$; we expect it to fall monatonically as $r$ increases and that it should be much smaller than unity for $r$ much greater than the QCD radius of the proton. Let us for the moment assume we know $\chi(r)$, then we can determine $S^2$ via

$$S^2 = \frac{\int dr \;\; d\sigma(r) \;\; \exp(-\chi(r))}{\int dr \;\; d\sigma(r)} \tag{15}$$

where $d\sigma(r)$ is the cross-section for the hard process that produces the Higgs expressed in terms of the transverse separation of the protons. Everything except the $r$ dependence of $d\sigma$ cancels when computing $S^2$ and so we need focus only on the dependence of the hard process on the transverse momenta of the scattered protons ($\mathbf{q_i}'$), these being Fourier conjugate to the transverse position of the protons, i.e.

$$\begin{aligned} d\sigma(r) &\propto [(\int d^2\mathbf{q_1}'\, e^{i\mathbf{q_1}'\cdot\mathbf{r}/2}\, \exp(-b\mathbf{q_1}'^2/2)) \times (\int d^2\mathbf{q_2}'\, e^{-i\mathbf{q_2}'\cdot\mathbf{r}/2}\, \exp(-b\mathbf{q_2}'^2/2))]^2 \\ &\propto \exp\left(-\frac{r^2}{2b}\right). \end{aligned} \quad (16)$$

Notice that since the $b$ here is the same as that which enters into the denominator of the expression for the total rate there is the aforementioned reduced sensitivity to $b$ since as $b$ decreases so does $S^2$ (since the collisions are necessarily more central) and what matters is the ratio $S^2/b^2$.

It remains for us to determine the mean multiplicity $\chi(r)$. If there really is only one type of rescattering event[5] independent of the hard scattering, then the inelastic scattering cross-section can be written

$$\sigma_{\text{inelastic}} = \int d^2\mathbf{r}(1 - \exp(-\chi(r))), \quad (17)$$

from which it follows that the elastic and total cross-sections are

$$\sigma_{\text{elastic}} = \int d^2\mathbf{r}(1 - \exp(-\chi(r)/2))^2, \quad (18)$$

$$\sigma_{\text{total}} = 2\int d^2\mathbf{r}(1 - \exp(-\chi(r)/2)). \quad (19)$$

There is an abundance of data which we can use to test this model and we can proceed to perform a parametric fit to $\chi(r)$. This is essentially what is done in the literature, sometimes going beyond a single-channel approach. Suffice to say that this simple approach works rather well. Moreover, it also underpins the models of the underlying event currently implemented in the PYTHIA [42] and HERWIG [43, 44] Monte Carlo event generators which have so far been quite successful in describing many of the features of the underlying event [45–47]. Typically, models of gap survival predict $S^2$ of a few percent at the LHC. Although data support the existing models of gap survival there is considerable room for improvement in testing them further and in so doing gaining greater control of what is perhaps the major theoretical uncertainty in the computation of exclusive Higgs production. In all our plots we took $S^2 = 3\%$ which is typical of the estimates in the literature for Higgs production ath the LHC.

## 3 Other Models

We'll focus in this section mainly on the model presented by what we shall call the Saclay group [13]. The model is a direct implementation of the original Bialas-Landshoff (BL) calculation [20] supplemented with a gap survival factor. It must be emphasised that BL did not claim to have computed for an exclusive process, indeed they were careful to state that "additional...interactions...will generate extra particles...Thus our calculation really is an inclusive one".

Equation (6) is the last equation that is common to both models. BL account for the coupling to the proton in a very simple manner: they multiply the quark level amplitude by a factor of 9 (which corresponds to assuming that there are three quarks in each proton that are able to scatter off each other). Exactly like the Durham group they also include a form factor suppression factor $\exp(-bq_{iT}'^2)$ for each proton at the cross-section level with $b = 4$ GeV$^{-2}$. Since BL are not interested in suppressing radiation, they do have a problem with the infra-red since there is no Sudakov factor. They dealt with this by following the earlier efforts of Landshoff and Nachtmann (LN) in replacing the perturbative gluon propagators with non-perturbative ones [48,49]:

$$\frac{g^2}{k^2} \to A\exp(-k^2/\mu^2).$$

[5]Clearly this is not actually the case, but such a "single channel eikonal" model has the benefit of being simple.

Rather arbitrarily, $g^2 = 4\pi$ was assumed, except for the coupling of the gluons to the top quark loop, where $\alpha_s = 0.1$ was used.

Following LN, $\mu$ and $A$ are determined by assuming that the $p\bar{p}$ elastic scattering cross-section at high energy can be approximated by the exchange of two of these non-perturbative gluons between the $3 \times 3$ constituent quarks: the imaginary part of this amplitude determines the total cross-section for which there are data which can be fitted to. In order to carry out this procedure successfully, one needs to recognize that a two-gluon exchange model is never going to yield the gentle rise with increasing centre-of-mass energy characteristic of the total cross-section. BL therefore also include an additional "reggeization" factor of $s^{\alpha(t)-1}$ in the elastic scattering amplitude where

$$\alpha(t) = 1 + \epsilon + \alpha' \, t$$

is the pomeron trajectory which ensures that a good fit to total cross-section data is possible for $\epsilon = 0.08$ and $\alpha' = 0.25$ GeV$^{-2}$. In this way the two-gluon system is modelling pomeron exchange. They found that $\mu \approx 1$ GeV and $A \approx 30$ GeV$^{-2}$ gave a good fit to the data. Similarly, the amplitude for central Higgs production picks up two reggeization factors.

The inclusive production of a Higgs boson in association with two final state protons is clearly much more infra-red sensitive than the exclusive case where the Sudakov factor saves the day. Having said that, the Saclay model does not include the Sudakov suppression factor. Instead it relies upon the behaviour of the non-perturbative gluon propagators to render the $Q_T$ integral finite. As a result, the typical $Q_T$ is much smaller than in the Durham case. Indeed it may be sufficiently small to make the approximation $Q_T^2 \gg q_{iT}'^2$ invalid which means that the spin-0 selection rule is no longer applicable.

Pulling everything together, the Saclay model of the cross-section for $pp \to p + H + p$ gives

$$\begin{aligned}\frac{d\sigma}{d^2\mathbf{q_{1T}}'d^2\mathbf{q_{2T}}'dy} \approx\; & S^2 \left(\frac{N_c^2-1}{N_c^2}\right)^2 \frac{\alpha_s^2}{(2\pi)^5} \left(\frac{g^2}{4\pi}\right)^4 \frac{G_F}{\sqrt{2}} e^{-bq_{1T}'^2} e^{-bq_{2T}'^2} \\ & x_1^{2-2\alpha(q_{1T}'^2)} x_2^{2-2\alpha(q_{2T}'^2)} \left[9 \int \frac{d^2\mathbf{Q_T}}{2\pi} \mathbf{Q_T}^2 \left(\frac{A}{g^2}\right)^3 \exp(-3\mathbf{Q_T}^2/\mu^2)\frac{2}{3}\right]^2 . \end{aligned} \quad (20)$$

The reggeization factors depend upon the momentum fractions $x_1$ and $x_2$ which satisfy $x_1 x_2 s = m_H^2$ and $y = \frac{1}{2}\ln(x_1/x_2)$. The only difference[6] between this and the original BL result is the factor of $S^2$. Integrating over the final state transverse momenta and simplifying a little gives

$$\frac{d\sigma}{dy} \approx S^2 \frac{\pi}{b + 2\alpha' \ln(1/x_1)} \frac{\pi}{b + 2\alpha' \ln(1/x_2)} \left(\frac{N_c^2-1}{N_c^2}\right)^2 \frac{G_F}{\sqrt{2}} \frac{\alpha_s^2}{(2\pi)^5} \frac{1}{(4\pi)^4} \left(\frac{s}{m_H^2}\right)^{2\epsilon} \frac{1}{g^4} \left[\frac{A^3\mu^4}{3}\right]^2 . \quad (21)$$

Figure 9 shows how the Saclay model typically predicts a rather larger cross-section with a weaker dependence upon $m_H$ than the Durham model. The weaker dependence upon $m_H$ arises because the Saclay model does not contain the Sudakov suppression, which is more pronounced at larger $m_H$, and also because of the choice $\epsilon = 0.08$. A larger value would induce a correspondingly more rapid fall. The Durham use of the gluon density function does indeed translate into an effective value of $\epsilon$ subtantially larger than 0.08. This effect is also to be seen in the dependence of the model predictions upon the centre-of-mass energy as shown in Figure 10. We have once again assumed a constant $S^2 = 3\%$ in this figure despite the fact that one does expect a dependence of the gap survival factor upon the energy.

[6] Apart from the factor 2 error previously mentioned.

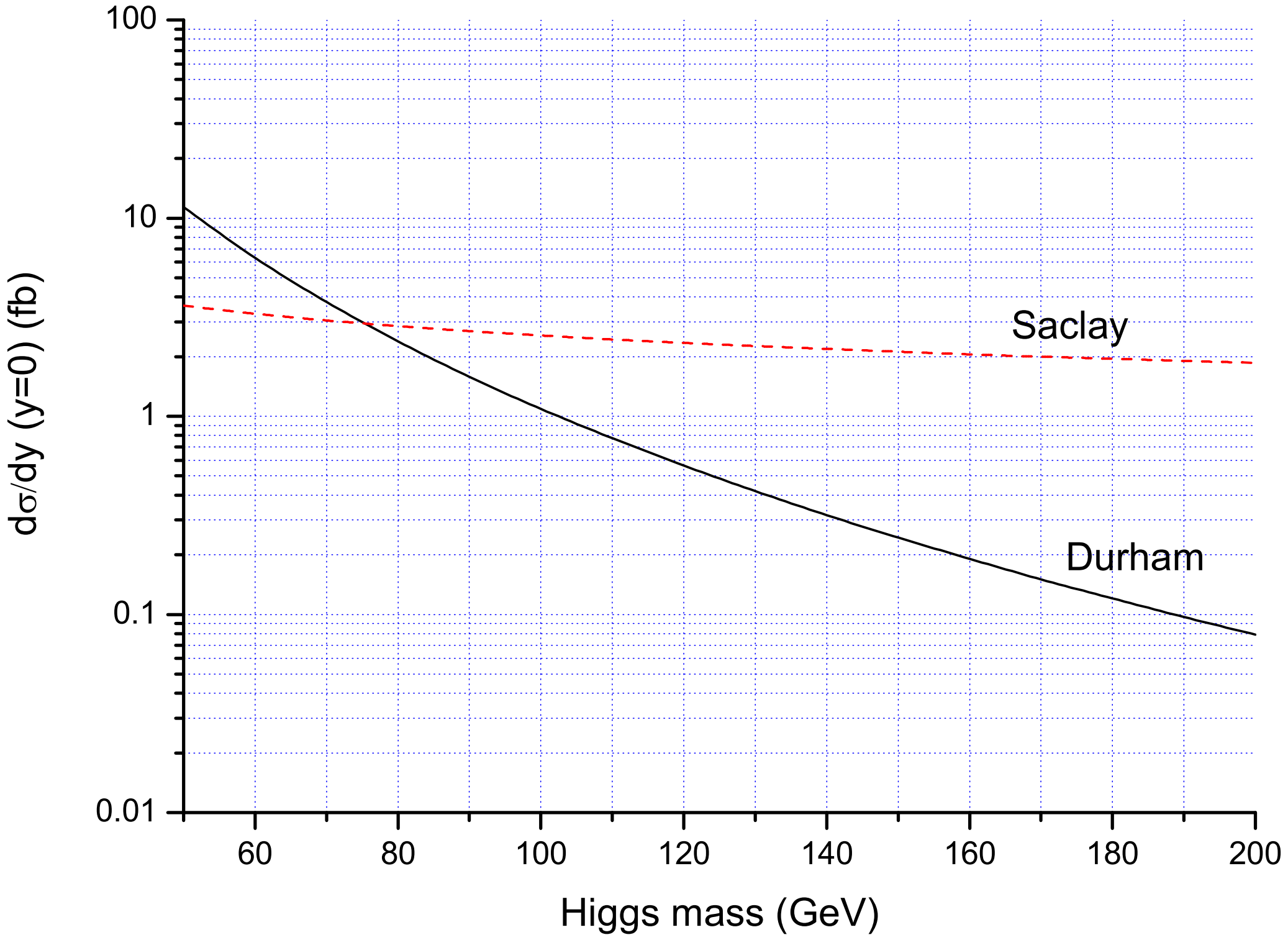

**Fig. 9:** Comparing dependence upon $m_H$ of the Saclay and Durham predictions. $S^2 = 3\%$ in both cases.

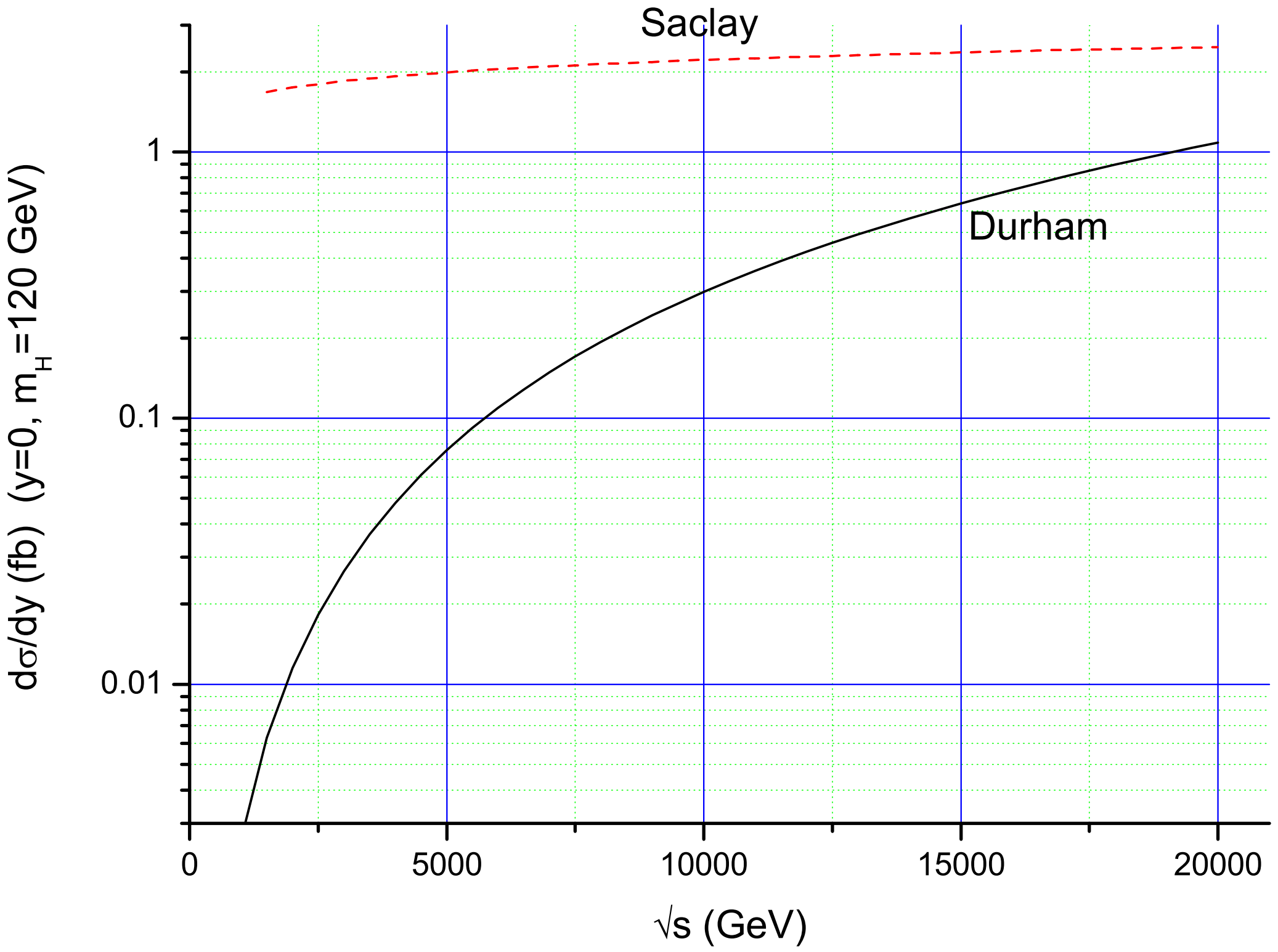

**Fig. 10:** Comparing dependence upon $\sqrt{s}$ of the Saclay and Durham predictions for $m_H = 120$ GeV.

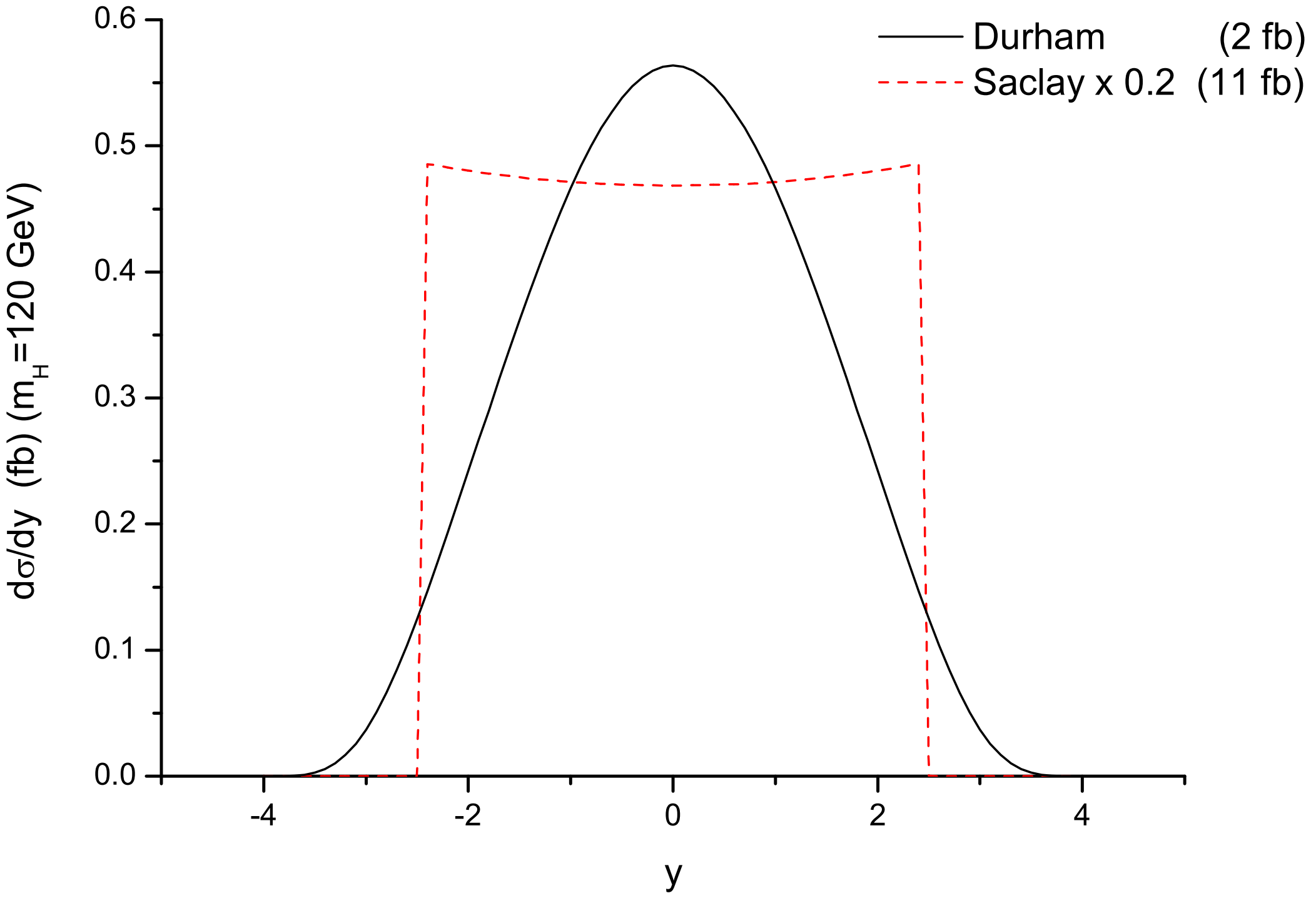

**Fig. 11:** Comparing the $y$ dependence of the Saclay and Durham predictions for $m_H = 120$ GeV. Note that the Saclay prediction has been reduced by a factor 5 to make the plot easier to read. The numbers in parenthesis are the total cross-sections, i.e. integrated over rapidity.

Figure 11 compares the rapidity dependence of the Higgs production cross-section in the two models. The Saclay prediction is almost $y$-independent. Indeed the only $y$-dependence is a consequence of $\alpha' \neq 0$. In both models the calculations are really only meant to be used for centrally produced Higgs bosons, i.e. $|y|$ not too large since otherwise one ought to revisit the approximations implicit in taking the high-energy limit. Nevertheless, the Durham prediction does anticipate a fall as $|y|$ increases, and this is coming because one is probing larger values of $x$ in the gluon density. In contrast, the Saclay prediction does not anticipate this fall and so a cutoff in rapidity needs to be introduced in quoting any cross-section integrated over rapidity. In Figure 11 a cut on $x_{1,2} < 0.1$ is made (which is equivalent to a cut on $|y| < 2.5$) for the Saclay model. After integrating over rapidity, the Durham model predicts a total cross-section of 2 fb for the production of a 120 GeV Higgs boson at the LHC whilst the Saclay model anticipates a cross-section a factor $\sim 5$ larger.

The essentially non-perturbative Saclay prediction clearly has some very substantial uncertainties associated with it. The choice of an exponentially falling gluon propagator means that there is no place for a perturbative component. However, as the Durham calculation shows, there does not seem to be any good reason for neglecting contributions from perturbatively large values of $Q_T$. It also seems entirely reasonable to object on the grounds that one should not neglect the Sudakov suppression factor and that including it would substantially reduce the cross-section.

In [18], the Sudakov factor of equation (11) is included, with the rest of the amplitude computed following Bialas-Landshoff. The perturbative Sudakov factor is also included in the approach of [19], albeit only at the level of the double logarithms. This latter approach uses perturbative gluons throughout the calculation but Regge factors are included to determine the coupling of the gluons into the protons, i.e. rather than the unintegrated partons of the Durham model. In both cases the perturbative Sudakov factor, not suprisingly, is important.

## 4 Concluding remarks

We hope to have provided a detailed introduction to the Durham model for central exclusive Higgs production. The underlying theory has been explained and the various sources of uncertainty highlighted with particular emphasis on the sensitivity of the predictions to gluon dynamics in the infra-red region. We also made some attempt to mention other approaches which can be found in the literature.

The focus has been on the production of a Standard Model Higgs boson but it should be clear that the formalism can readily be applied to the central production of any system $X$ which has a coupling to gluons and invariant mass much smaller than the beam energy. There are many very interesting possibilities for system $X$ which have been explored in the literature and we have not made any attempt to explore them here [2, 3, 8, 11, 15–17]. Nor have we paid any attention to the crucial challenge of separating signal events from background [5, 9]. The inclusion of theoretical models into Monte Carlo event generators and a discussion of the experimental issues relating to central exclusive particle production have not been considered here but can be found in other contributions to these proceedings [50, 51].

It seems that perturbative QCD can be used to compute cross-sections for processes of the type $pp \to p + X + p$. The calculations are uncertain but indicate that rates ought to be high enough to be interesting at the LHC. In the case that the system $X$ is a pair of jets there ought to be the possibility to explore this physics at the Tevatron [52]. Information gained from such an analysis would help pin down theoretical uncertainties, as would information on the rarer but cleaner channel where $X$ is a pair of photons [53]. Of greatest interest is when $X$ contains "new physics" whence this central exclusive production mechanism offers new possibilities for its exploration.

## 5 Acknowledgments

Special thanks to Hannes Jung for all his efforts in making the workshop go so well. Thanks also to Brian Cox, Markus Diehl, Valery Khoze, Peter Landshoff, Leif Lönnblad, James Monk, Leszek Motyka, Andy Pilkington and Misha Ryskin for very helpful discussions.

# Monte Carlo generators for central exclusive diffraction

*Maarten Boonekamp, Creighton Hogg, James Monk, Andrew Pilkington & Marek Tasevsky*

**Abstract**
We review the three Monte Carlo generators that are available for simulating the central exclusive reaction, $pp \to p + X + p$.

## 1 Introduction

The central exclusive mechanism is defined as $pp \to p + X + p$ with no radiation emitted between the intact outgoing beam hadrons and the central system $X$. The study of central exclusive Higgs boson production has been aided with the recent development of Monte Carlo simulations to enable parton, hadron and detector level simulation. The three generators that we shall examine here are DPEMC [1], EDDE [2] and ExHuME [3]. From an experimental perspective, it is important to examine both the similarities and differences between the models in order to assess the physics potential in terms of forward proton tagging at the LHC [4].

Each of the Monte Carlos implements a different model of central exclusive production that is either perturbative or non-perturbative. ExHuME is an implementation of the perturbative calculation of Khoze, Martin and Ryskin [5], the so-called "Durham Model". In this calculation (depicted in fig 1(a)), the two gluons couple perturbatively to the off-diagonal unintegrated gluon distribution in the proton. The Durham approach includes a Sudakov factor to suppress radiation into the rapidity gap between the central system and the outgoing protons and which renders the loop diagram infra-red safe. The bare cross section is suppressed by a soft-survival probability, $\mathcal{S}^2$, that accounts for additional momentum transfer between the proton lines that lead to particle production that could fill in the gap. The current ExHuME default takes $\mathcal{S}^2$ to be 0.03 at the LHC.

In contrast, DPEMC and EDDE treat the proton vertices non-perturbatively. This is acheived in the context of Regge theory, by pomeron exchange from each of the proton lines. DPEMC follows the Bialas-Landshoff approach [6] of parameterising the pomeron flux within the proton. DPEMC also sets the default value of $\mathcal{S}^2$ to 0.03 at the LHC. EDDE uses an improved Regge-eikonal approach [7] to calculate the soft proton vertices and includes a Sudakov suppression factor to prohibit real gluon emission. There is no explicit soft-survival factor present in EDDE: it is assumed that the Regge parameterisation includes the effect of additional interactions between the proton lines. For further details of the calculations underlying both DPEMC and ExHuME please refer to [8].

The connection between the parton level process and the hadronic final state is not the same in the three Monte Carlos. Both ExHuME and EDDE are linked to Pythia [9, 10] for final state parton showering and hadronisation. DPEMC however, overrides the HERWIG [11] internal $\gamma\gamma$ interactions in $e^+e^-$ collisions to simulate double pomeron exchange.

The processes available are similar in each Monte Carlo. Perhaps the most interesting is Higgs boson production with all subsequent decays. In addition, di-jet production is included in all three generators. None of the Monte Carlos yet includes the next-to-leading order 3 jet process, which could be an important, or even the dominant, background to the central exclusive $H \to b\bar{b}$ search channel.

Finally, inclusive double pomeron exchange (shown in figure 1(b)) will also act as a backgound to the exclusive process as there are 2 protons in the final state. These processes are always accompanied by pomeron remnants in the central system and it may be a challenge experimentally to separate these from the system of interest. Two models for these processes are the Cox-Forshaw model (CF), implemented in POMWIG [12], and the Boonekamp-Peschanski-Royon model (BPR) [13] that is included in DPEMC.

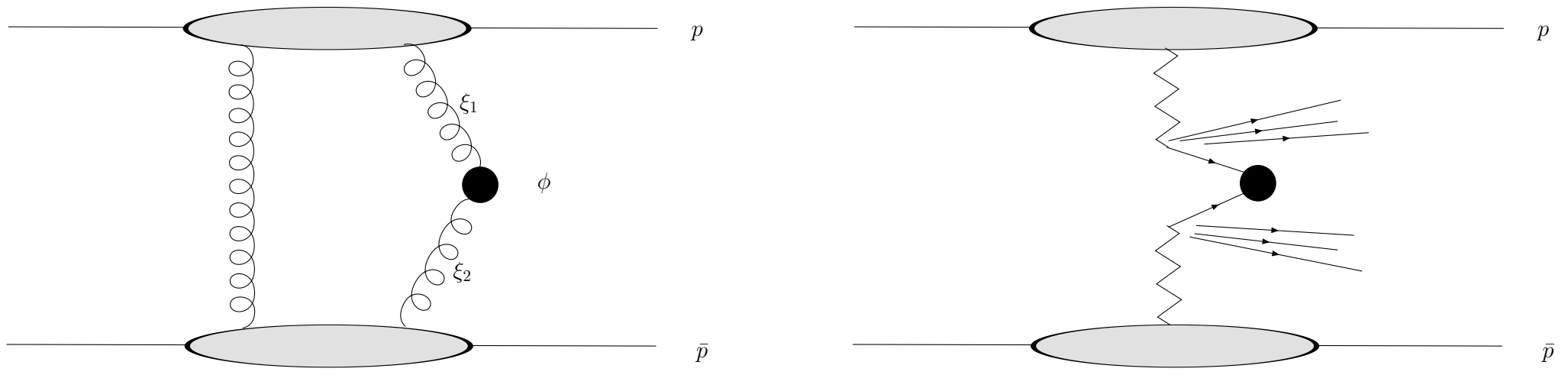

**Fig. 1:** The exclusive production process (a) and the inclusive (double pomeron) production process (b).

## 2 Results

Unless otherwise stated, all plots shown here were produced by using each of the Monte Carlos as they are distributed. constant soft survival factor, $\mathcal{S}^2$, of 0.03 was used in all three generators. In the case of ExHuME, where a parton distribution set must be chosen, the default is the 2002 MRST set, usually supplied via the LHAPDF library.

Using the default settings at the LHC energy of 14 TeV the total cross sections for production of a 120 GeV Higgs boson are 3.0 fb, 1.94 fb and 2.8 fb for DPEMC, EDDE and ExHuME respectively. However, despite these similar cross section predictions, the physics reach of the central exclusive process is predicted to differ significantly between the Monte Carlos. Figure 2(a) shows that ExHuME and EDDE predict that the cross section for exclusive Higgs boson production will fall much faster than DPEMC with an increase in Higgs boson mass. This is a direct effect of the Sudakov suppression factors growing as the available phase space for gluon emission increases with the mass of the central system. The different gluon momentum fraction, $\xi$, dependences lead to the differences in figure 2(b). With a fixed central mass an increase in collision energy is identical to a decrease in $\xi$, and the flatter $\xi$ distributions of DPEMC and EDDE are reflected in the flatter $\sqrt{s}$ dependence compared to ExHuME.

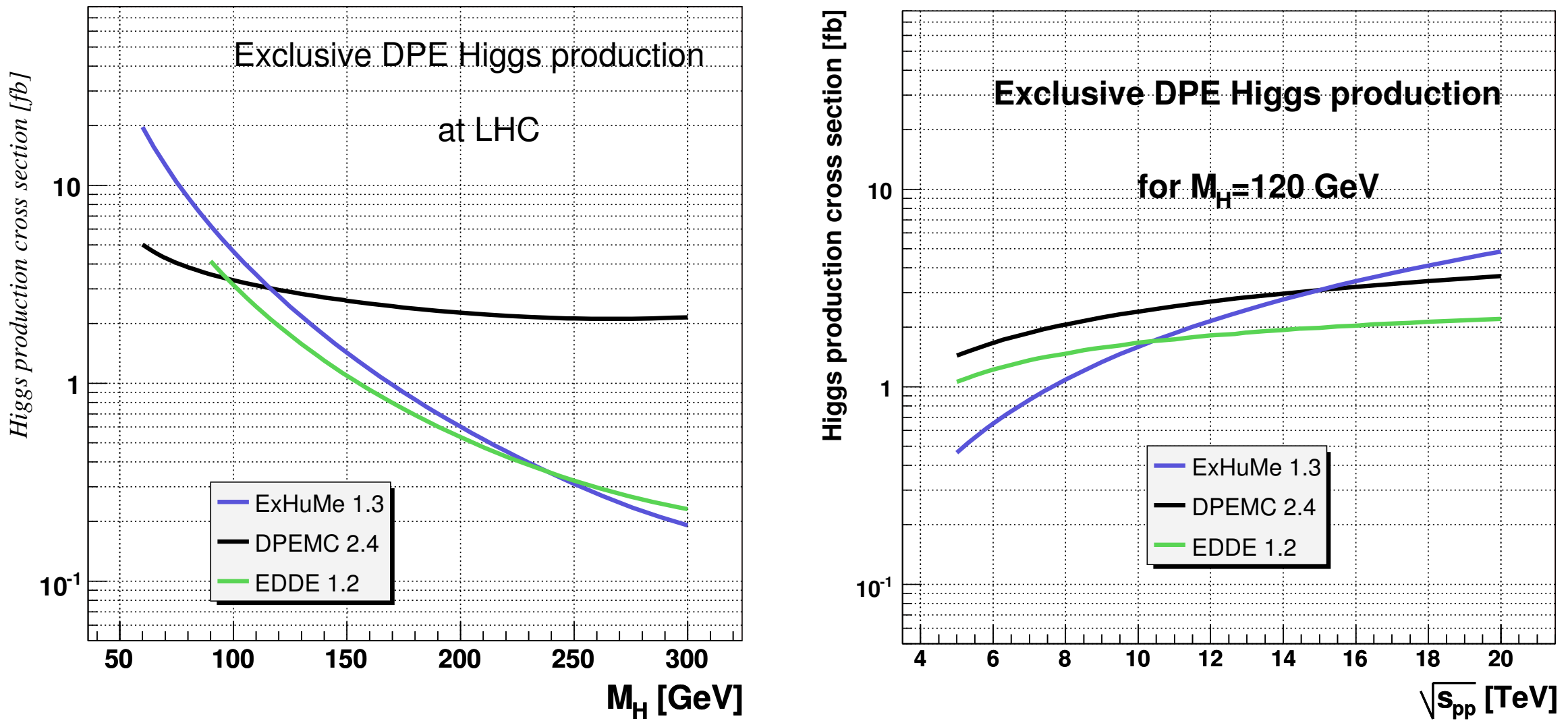

**Fig. 2:** (a) The left hand plot shows the higgs cross section as a function of higgs mass. (b) The right hand plot shows the increase in cross section with the collision energy (fixed gap survival factor).

The physics potential is dependent not only on the total cross section, but also on the rapidity distribution of the central system, which is shown in figure 3(b) together with the $\xi$ distribution for the gluons. The more central rapidity distribution of ExHuME is due to the gluon distributions falling more sharply than the pomeron parameterisation present in DPEMC.

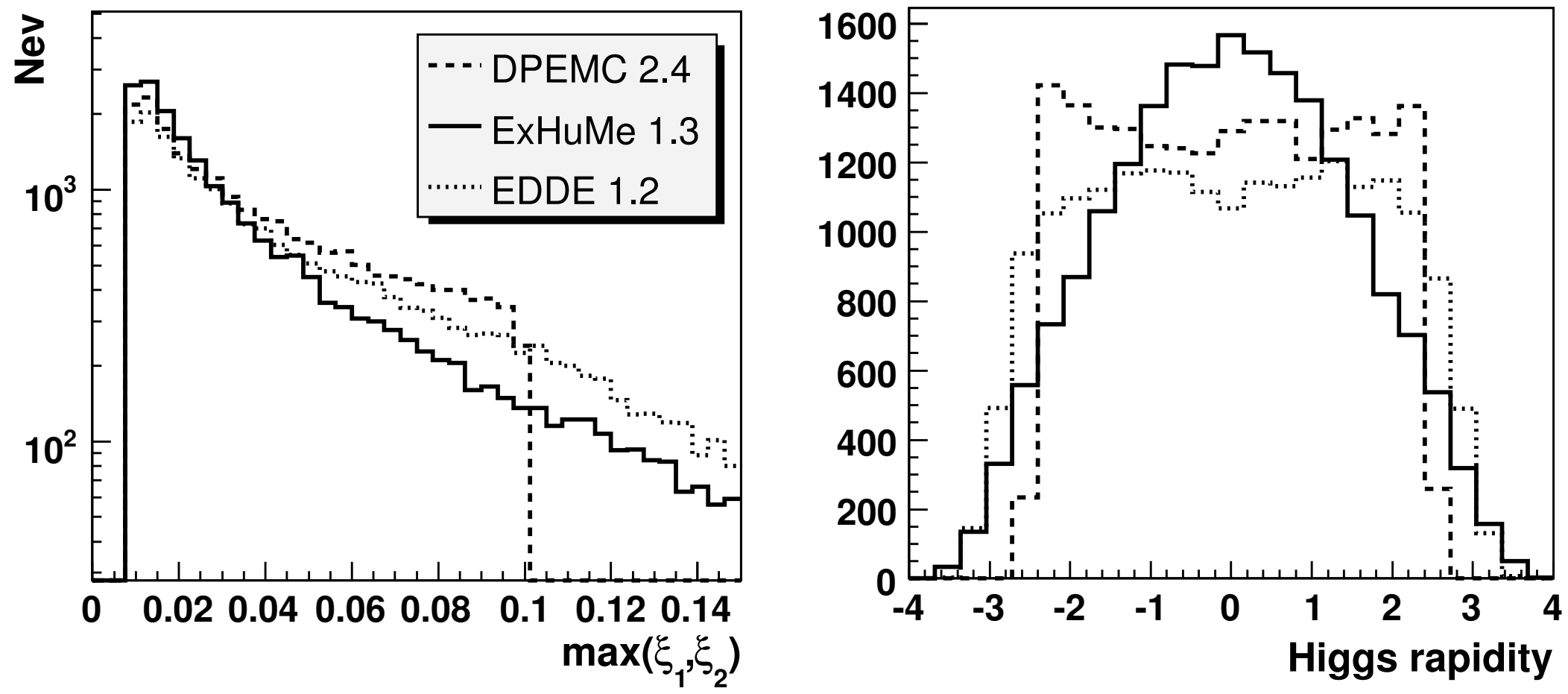

**Fig. 3:** The $max(\xi_1, \xi_2)$(a) and rapidity(b) of the 120 GeV Higgs. ExHuME predicts a steeper fall off in the number of events at high $\xi$ and hence favours a less broad rapidity distribution compared to the soft non-perturbative models. Note that a cut is applied in DPEMC at $\xi = 0.1$, as required by the Bialas-Landshoff approach.

The acceptances of any forward proton taggers that might be installed at the LHC are sensitive to the rapidity distributions of the central system. The differences seen in figure 3(b) are reflected in different acceptance curves shown in figure 4. The predicted acceptances using taggers at 420 and 220 metres as a function of the mass of the central system were obtained using a fast simulation of the CMS detector. The fast simulation includes a parameterisation of the responses of the forward taggers based on a detailed simulation of the detectors [14]. As seen in figure 4, as the mass of the central system increases the combined acceptance using detectors at *both* 220 and 420 metres increases, with the relative difference between the predictions from the three generators decreasing (from about 40% down to 15% for the most extreme relative differences). For a Higgs boson of mass 120 GeV the acceptances are predicted to be 46, 50 and 57% for EDDE, DPEMC and ExHuME respectively.

Changes from the default generator settings can have an effect on all of these distributions. As an example Fig. 5(a) shows the rapidity distribution from ExHuME using the CTEQ6M set compared to the MRST 2002 set of parton distribution functions. The CTEQ pdf has a flatter $\xi$ dependence in the sensitive region of $Q_\perp \simeq 3$ GeV, which leads to a broader peak and sharper fall in the rapidity distribution and a larger cross section of 3.75 fb. This in turn should improve the efficiency of the forward proton taggers because not only are there more events, but there are more events at low rapidity. It is also possible to change the DPEMC code to add a harder $\xi$ dependence of the form $(1-\xi)^\alpha$ to the pomeron flux parameterisation. This would favour a more central rapidity distribution, thus increasing the acceptance in the forward pots.

In di-jet production the di-jet mass fraction, $R_{jj}$ is defined as $R_{jj} = M_{jj}/\sqrt{\hat{s}}$, where $M_{jj}$ is the mass of the di-jet system and $\sqrt{\hat{s}}$ is the total invariant mass of the central system. $R_{jj}$ should be large in a central exclusive event. In current searches for central exclusive di-jet production at the Tevatron [15], the CDF collaboration have experimentally defined exclusive events to be those where $R_{jj} > 0.8$. It should be noted that $M_{jj}$ depends on the particular jet reconstruction algorithm used and the $\sqrt{\hat{s}}$ measurement is dependent either on tagging the outgoing protons or on reconstructing the missing mass using the calorimeter. In figure 6(a) we show the prediction for the $R_{jj}$ fraction in exclusive events at the LHC, whilst in figure 6(b) we show two examples of the inclusive background with pomeron remnants from Pomwig and DPEMC. It is clear that the $R_{jj} > 0.8$ definition for central exclusive production leads to an overlap between the exclusive and inclusive regions for all of the Monte Carlo predictions.

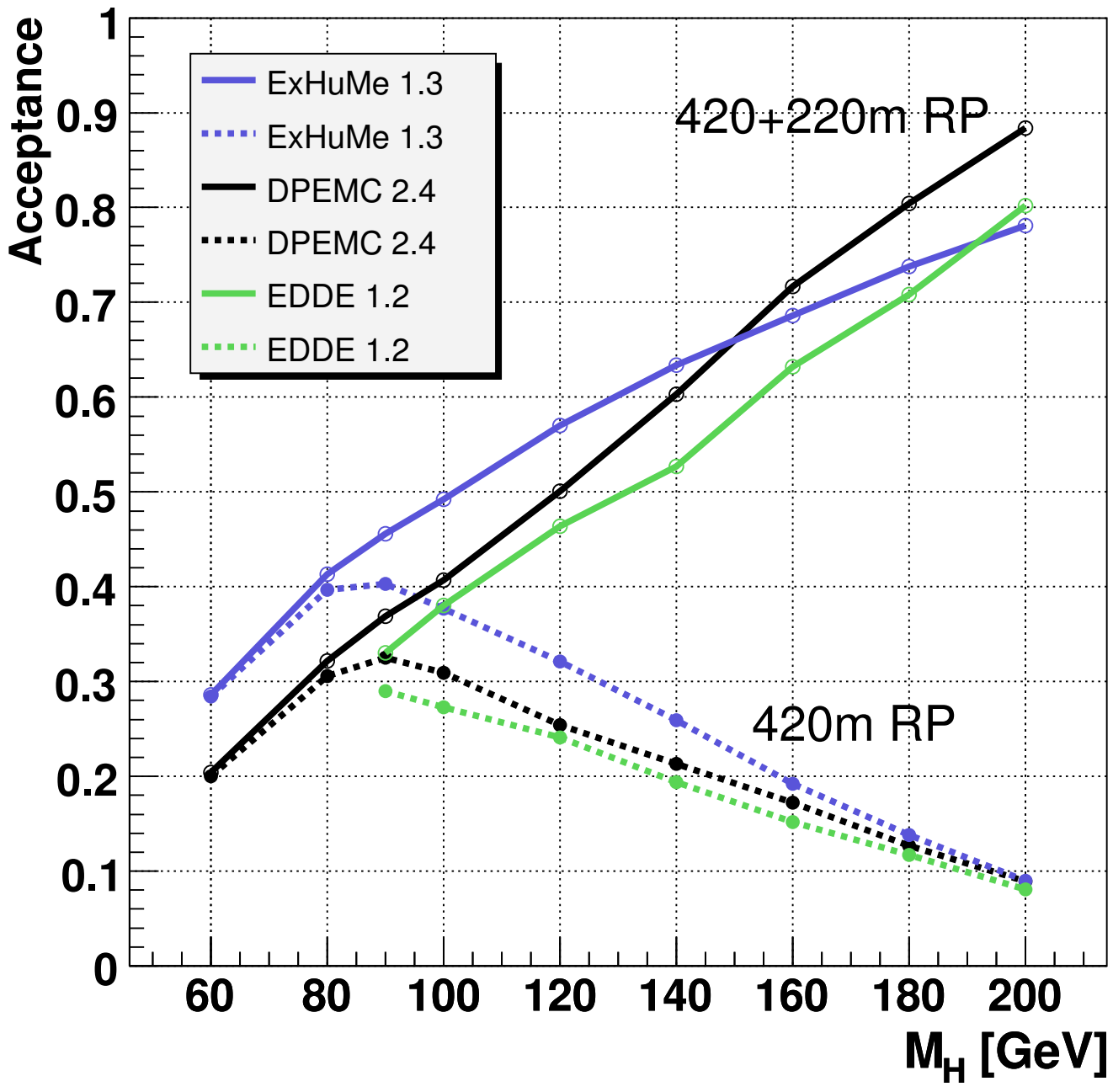

**Fig. 4:** The predicted acceptances for the proposed forward taggers at 420 metres and for a combination of taggers at 220 and 420 metres from the central detector.

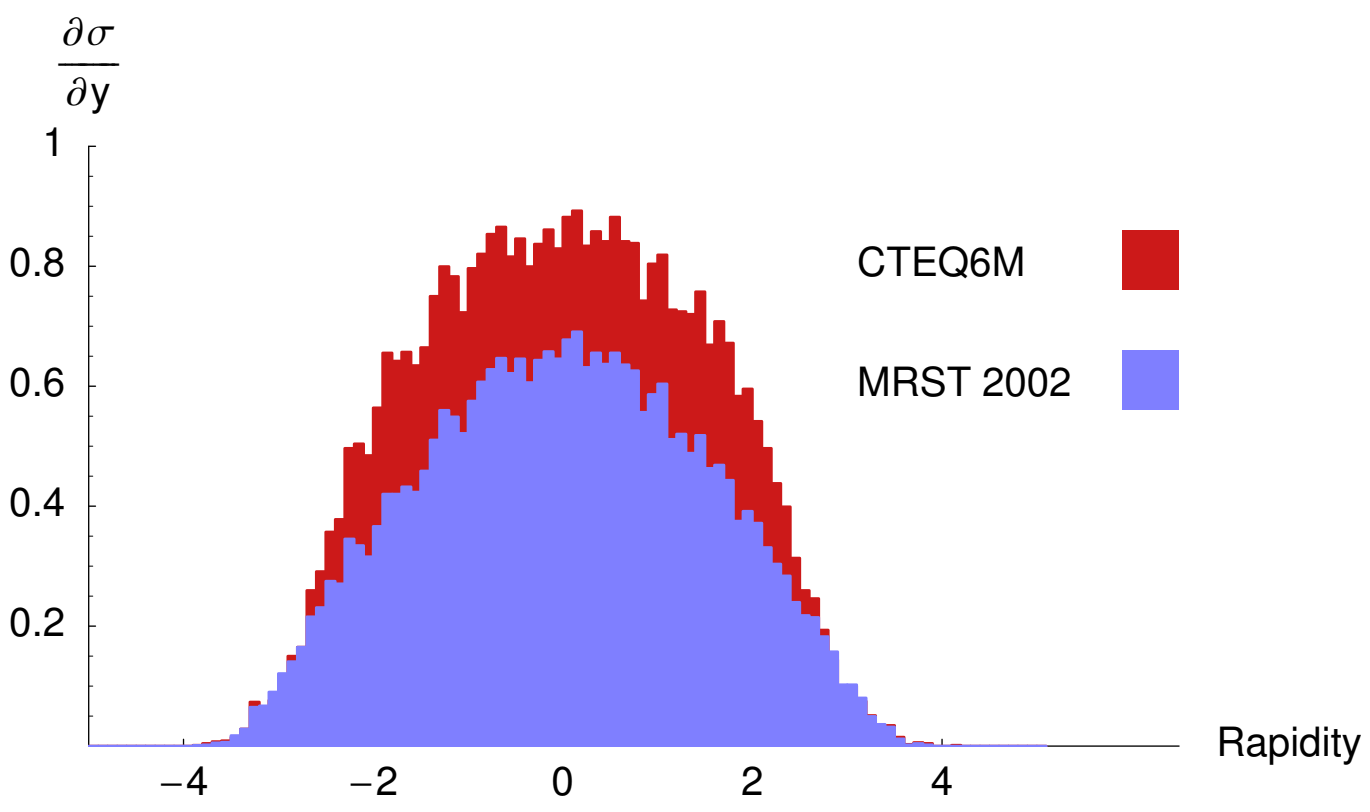

**Fig. 5:** The ExHuME rapidity distributions for production of a 120 GeV Higgs using MRST 2002 pdfs and the CTEQ6M pdfs.

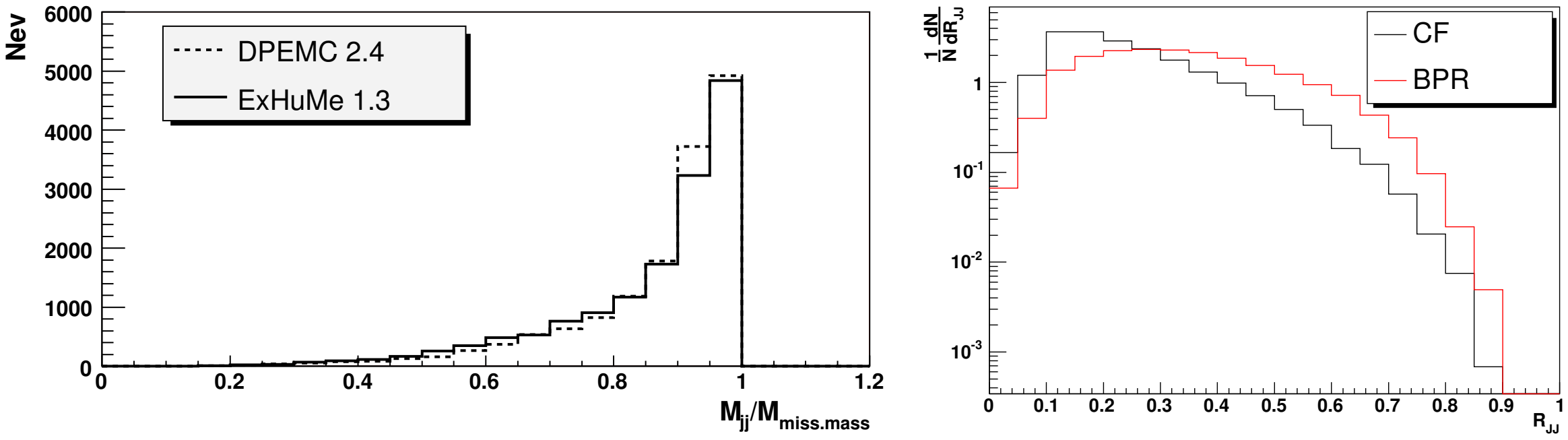

**Fig. 6:** $R_{jj}$ distributions for (a) exclusive di-jet production using DPEMC and ExHuME and (b) the background inclusive di-jet production as predicted by the Cox-Forshaw (CF) and Boonekamp-Peschanski-Royon (BPR) models.

## 3 Summary

At a Higgs boson mass of 120GeV, all three Monte Carlo simulations give similar predictions for the cross section. However, the physics potential decreases for models that include Sudakov suppression, which will limit Higgs boson searches. The differing rapidity distributions of the central system result in different efficiencies for a forward proton tagging programme. In order to fully study the background to the $H \to b\bar{b}$ channel, future additions to the Monte Carlo programs should include the next to leading order three jet process.

### Acknowledgements

We would like to thank Brian Cox, Jeff Forshaw, Valery Khoze, Misha Ryskin and Roman Ryutin for discussion throughout the proceedings.

# Diffractive Structure Functions and Diffractive PDFs

*Editors: M. Arneodo[a] and P. van Mechelen[b]*
[a]Università del Piemonte Orientale, Novara, Italy
[b]University of Antwerpen, Antwerpen, Belgium

## Abstract

This section of the Proceedings contains papers summarising the current status of the $F_2^D$ measurements at HERA, the extraction of the diffractive parton distribution functions and the relevance of a direct measurement of $F_L^D$.

The selection of a pure sample of inclusive diffractive events, $ep \to eXp$, is a challenging task. Three alternative approaches have been used so far by the H1 and ZEUS collaborations at HERA:

1. a fast proton in the final state is required;
2. a rapidity gap in the forward direction is required;
3. the different shape of the $M_X$ distribution for diffractive and non-diffractive events is exploited.

The results obtained with these approaches exhibit a level of agreement which varies from tolerable to poor. This is not surprising since different final states are selected, in which the reaction $ep \to eXp$ appears with different degrees of purity. The paper by Newman and Schilling presents a systematic comparison of the results available, quantifies the differences and discusses their origins, when understood.

NLO QCD fits to the diffractive structure function $F_2^D$ are used to extract the diffractive parton distribution functions (dPDFs) in the proton. They can be interpreted as conditional probabilities to find a parton in the proton when the final state of the process contains a fast proton of given four-momentum. They are essential to determine the cross sections of less inclusive processes in $ep$ diffractive scattering, such as dijet or charm production. They are also a non-negotiable ingredient for the prediction of the cross sections for inclusive diffractive processes at the LHC.

Several groups have so far performed such fits to the available data. The results of these fits are presented in the papers by Newman and Schilling, Groys et al. and Watt et al. All fits give diffractive PDFs largely dominated by gluons. However, significant differences are apparent, reflecting the differences in the data, but also in the fitting procedure. Newman and Schilling and Groys et al. assume the so-called Regge factorisation hypothesis, i.e. take $F_2^D = f_{I\!P}(x_{I\!P}, t) \cdot F_2^{I\!P}(\beta, Q^2)$. This assumption has no basis in QCD and is critically discussed by Groys et al. and by Watt et al. The latter also argue that the leading-twist formula used by Newman and Schilling and by Groys et al. is inadequate in large parts of the measured kinematics, and use a modified expression which includes an estimate of power-suppressed effects.

The parametrisations of the dPDFs discussed in these three papers are available in a code library discussed in the paper by Schilling.

Finally, the paper by Newman addreses the importance of measuring the longitudinal diffractive structure function $F_L^D$. A measurement of $F_L^D$ to even modest precision would provide a very powerful independent tool to verify our understanding of diffraction and to test the gluon density extracted indirectly in QCD fits from the scaling violations of $F_2^D$.

# HERA Diffractive Structure Function Data and Parton Distributions

*Paul Newman*[a], *Frank-Peter Schilling*[b]
[a] School of Physics and Astronomy, University of Birmingham, B15 2TT, United Kingdom
[b] CERN/PH, CH-1211 Geneva 23, Switzerland

**Abstract**
Recent diffractive structure function measurements by the H1 and ZEUS experiments at HERA are reviewed. Various data sets, obtained using systematically different selection and reconstruction methods, are compared. NLO DGLAP QCD fits are performed to the most precise H1 and ZEUS data and diffractive parton densities are obtained in each case. Differences between the $Q^2$ dependences of the H1 and ZEUS data are reflected as differences between the diffractive gluon densities.

## 1 Introduction

In recent years, several new measurements of the semi-inclusive 'diffractive' deep inelastic scattering (DIS) cross section for the process $ep \to eXY$ at HERA have been released by the H1 and ZEUS experiments [1–6]. The data are often presented in the form of a $t$-integrated reduced diffractive neutral current cross section $\sigma_r^{D(3)}$, defined through[1]

$$\frac{d^3\sigma^{ep\to eXY}}{\mathrm{d}x_{I\!P}\,\mathrm{d}x\,\mathrm{d}Q^2} = \frac{4\pi\alpha^2}{xQ^4}\left(1 - y + \frac{y^2}{2}\right)\sigma_r^{D(3)}(x_{I\!P}, x, Q^2)\ , \qquad (1)$$

or in terms of a diffractive structure function $F_2^{D(3)}(x_{I\!P}, \beta, Q^2)$. Neglecting any contributions from $Z^0$ exchange,

$$\sigma_r^{D(3)} = F_2^{D(3)} - \frac{y^2}{1 + (1-y)^2} F_L^{D(3)}\ , \qquad (2)$$

such that $\sigma_r^{D(3)} = F_2^{D(3)}$ is a good approximation except at very large $y$. The new data span a wide kinematic range, covering several orders of magnitude in $Q^2$, $\beta$ and $x_{I\!P}$.

Within the framework of QCD hard scattering collinear factorisation in diffractive DIS [7], these data provide important constraints on the diffractive parton distribution functions (dpdf's) of the proton. These dpdf's are a crucial input for calculations of the cross sections for less inclusive diffractive processes in DIS, such as dijet or charm production [8, 9]. In contrast to the case of inclusive scattering, the dpdf's extracted in DIS are not expected to be directly applicable to hadron-hadron scattering [7, 10–12]. Indeed, diffractive factorisation breaks down spectacularly when HERA dpdf's are applied to diffractive proton-proton interactions at the TEVATRON [13]. It may, however, be possible to recover good agreement by applying an additional 'rapidity gap survival probability' factor to account for secondary scattering between the beam remnants [14–17]. The HERA dpdf's thus remain an essential ingredient in the prediction of diffractive cross sections at the LHC, notably the diffractive Higgs cross section [18]. Although the poorly known rapidity gap survival probability leads to the largest uncertainty in such calculations, the uncertainty due to the input dpdf's also plays a significant role. In [3], the H1 collaboration made a first attempt to assess the uncertainty from this source, propagating the experimental errors from the data points to the 'H1 2002 NLO fit' parton densities and assessing the theoretical uncertainties from various sources.

In this contribution, we investigate the compatibility between various different measurements of $F_2^D$ by H1 and ZEUS. We also apply the techniques developed in [3] to ZEUS data in order to explore the consequences of differences between the H1 and ZEUS measurements in terms of dpdf's.

[1] For a full definition of all terms and variables used, see for example [3].

## 2 Diffractive Selection Methods and Data Sets Considered

One of the biggest challenges in measuring diffractive cross sections, and often the source of large systematic uncertainties, is the separation of diffractive events in which the proton remains intact from non-diffractive events and from proton-dissociation processes in which the proton is excited to form a system with a large mass, $M_{\rm Y}$. Three distinct methods have been employed by the HERA experiments, which select diffractive events of the type $ep \rightarrow eXY$, where $Y$ is a proton or at worst a low mass proton excitation. These methods are complimentary in that their systematics due to the rejection of proton dissociative and non diffractive contributions are almost independent of one another. They are explained in detail below.

- **Roman Pot Spectrometer Method.** Protons scattered through very small angles are detected directly in detectors housed in 'Roman Pot' insertions to the beampipe well downstream the interaction point. The proton 4-momentum at the interaction point is reconstructed from the position and slope of the tracks in these detectors, given a knowledge of the beam optics in the intervening region. The Roman Pot devices are known as the Leading Proton Spectrometer (LPS) in the case of ZEUS and the Forward Proton Spectrometer (FPS) in H1. The Roman pot method provides the cleanest separation between elastic, proton dissociative and non-diffractive events. However, acceptances are rather poor, such that statistical uncertainties are large in the data sets obtained so far.
- **Rapidity Gap Method.** This method is used by H1 for diffractive structure function measurements and by both H1 and ZEUS for the investigation of final state observables. The outgoing proton is not observed, but the diffractive nature of the event is inferred from the presence of a large gap in the rapidity distribution of the final state hadrons, separating the $X$ system from the unobserved $Y$ system. The diffractive kinematics are reconstructed from the mass of the $X$ system, which is well measured in the main detector components. The rapidity gap must span the acceptance regions of various forward[2] detector components. For the H1 data presented here, these detectors efficiently identify activity in the pseudorapidity range $3.3 < \eta \lesssim 7.5$. The presence of a gap extending to such large pseudorapidities is sufficient to ensure that $M_{\rm Y} \lesssim 1.6$ GeV. In light of the poor knowledge of the $M_{\rm Y}$ spectrum at low masses, no attempt is made to correct the data for the small remaining proton dissociation contribution, but rather the cross sections are quoted integrated over $M_{\rm Y} < 1.6$ GeV.
- **$M_{\rm X}$ Method.** Again the outgoing proton is not observed, but rather than requiring a large rapidity gap, diffractive events are selected on the basis of the inclusive $\ln M_{\rm X}^2$ distribution. Diffractive events are responsible for a plateau in this distribution at low $\ln M_{\rm X}^2$, such that they can be selected cleanly for the lowest $M_{\rm X}$ values. At intermediate $M_{\rm X}$, non-diffractive contributions are subtracted on the basis of a two component fit in which the non-diffractive component rises exponentially. This method is used for diffractive structure function measurements by ZEUS. It does not discriminate between elastic and low $M_{\rm Y}$ proton-dissociative contributions. Results are quoted for $M_{\rm Y} < 2.3$ GeV.

Four recent data sets are considered, for which full details of luminosities and kinematic ranges can be found in Table 1.

- Published data from ZEUS taken in 1998 and 1999, using the $M_{\rm X}$ method and taking advantage of the increased forward acceptance offered by a new plug calorimeter ('ZEUS-$M_{\rm X}$') [1].
- Published ZEUS data obtained with the LPS using data taken in 1997 ('ZEUS-LPS') [2].
- Preliminary H1 data obtained using the rapidity gap method, combining three measurements using different data sets from the period 1997-2000 for different regions in $Q^2$ ('H1-LRG') [3–5].
- Preliminary H1 data obtained using the FPS, based on data taken in 1999 and 2000 ('H1-FPS') [6].

[2] The forward hemisphere is that of the outgoing proton beam, where the pseudorapidity $\eta = -\ln\tan\theta/2$ is positive.

**Table 1:** Overview of the data sets discussed here. The quoted kinematic ranges in $Q^2$, $\beta$ and $x_{I\!P}$ correspond to the bin centres.

| Label | Ref. | Reconstruction Method | Lumi $\mathcal{L}[\mathrm{pb}^{-1}]$ | Kinematic range $M_{\mathrm{Y}}$[GeV] | $Q^2[\mathrm{GeV}^2]$ | $\beta$ | $x_{I\!P}$ |
|---|---|---|---|---|---|---|---|
| ZEUS-$M_{\mathrm{X}}$ | [1] | $M_{\mathrm{X}}$ method | 4.2 | $< 2.3$ | 2.7..55 | 0.003..0.975 | 0.0001..0.03 |
| ZEUS-LPS | [2] | Roman Pot | 12.8 | $M_p$ | 2.4..39 | 0.007..0.48 | 0.0005..0.06 |
| H1-LRG | [3–5] | Rapidity Gap | 3.4..63 | $< 1.6$ | 1.5..1600 | 0.01..0.9 | 0.0001..0.05 |
| H1-FPS | [6] | Roman Pot | 25 | $M_p$ | 2.6..20 | 0.01..0.7 | 0.002..0.05 |

## 3 Comparisons between Data Sets

In this section, the $x_{I\!P}$ dependences of the data from the different measurements are compared at fixed values of $Q^2$ and $\beta$. Since the various measurements are generally presented at different $Q^2$ and $\beta$ values, it is necessary to transport the data to the same values. The $\beta$ and $Q^2$ values of the H1-LRG data are chosen as the reference points. The factors applied to data points from the other measurements are evaluated using two different parameterisations, corresponding to the results of QCD fits to 1994 H1 data [19] and to a subset of the present H1-LRG data at intermediate $Q^2$ [3] (see also section 4). In order to avoid any significant bias arising from this procedure, data points are only considered further here if the correction applied is smaller than $50\%$ in total and if the correction factors obtained from the two parameterisations are in agreement to better than $25\%$. In practice, these criteria only lead to the rejection of data points in the ZEUS-$M_{\mathrm{X}}$ data set at $Q^2 = 55$ GeV$^2$ and $\beta = 0.975$, where the poorly known high $\beta$ dependence of the diffractive cross section implies a large uncertainty on the factors required to transport them to $\beta = 0.9$. Elsewhere, there is reasonable agreement between the factors obtained from the two parameterisations and no additional uncertainties are assigned as a consequence of this procedure.

Since the various data sets correspond to different ranges in the outgoing proton system mass, $M_{\mathrm{Y}}$, additional factors are required before comparisons can be made. For all data and fit comparisons, all data are transported to the H1 measurement range of $M_{\mathrm{Y}} < 1.6$ GeV and $|t| < 1$ GeV$^2$. The leading proton data are scaled by a factor 1.1 [20] to correspond to the range $M_{\mathrm{Y}} < 1.6$ GeV and the ZEUS-$M_{\mathrm{X}}$ data are scaled to the same range by a further factor of 0.7 [1], such that the overall factor is 0.77. The uncertainties on these factors are large, giving rise to normalisation uncertainties of perhaps $15\%$ between the different data sets.

The ZEUS-LPS and H1-FPS data are compared in figure 1. Within the experimental uncertainties, the two data sets are in good agreement. Both data sets are also consistent with a parameterisation of the H1-LRG data [3] based on the H1 2002 NLO QCD fit, which is also shown. This good agreement between the H1-LRG and the Roman Pot data is also shown explicitly in figure 3.

In figure 2, a comparison is made between the H1-LRG and the ZEUS-$M_{\mathrm{X}}$ data after all factors have been applied. For much of the kinematic range, there is tolerable agreement between the two data sets. However, there are clear regions of disagreement. One is at the largest $\beta$ (smallest $M_{\mathrm{X}}$), where the H1 data lie significantly above the ZEUS data for $Q^2 \lesssim 20$ GeV$^2$. Another is at intermediate and low $\beta$, where the two data sets show significantly different dependences on $Q^2$. With the factor of 0.77 applied to the ZEUS data, there is good agreement at low $Q^2$, but the ZEUS data lie below the H1 data at large $Q^2$. If the factor of 0.77 is replaced with a value closer to unity, the agreement improves at large $Q^2$, but the H1 data lie above the ZEUS data at low $Q^2$. These inconsistencies between the different data sets are discussed further in section 4.

For completeness, figure 3 shows a comparison between all four data sets considered.

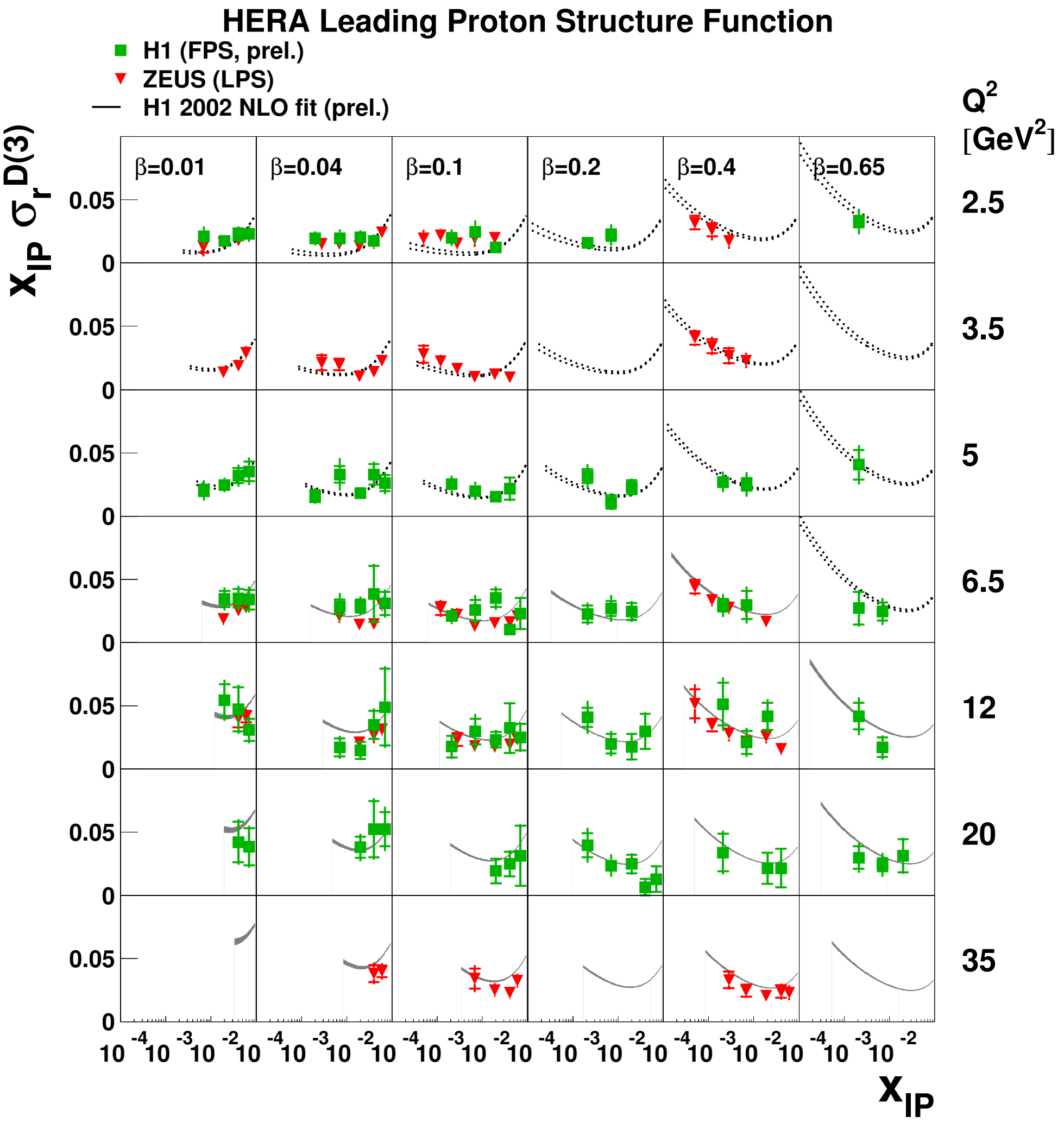

**Fig. 1:** Comparison of the Roman Pot data from H1 and ZEUS, scaled by a factor 1.1 such that they correspond to $M_{\rm Y} < 1.6$ GeV. The $Q^2$ and $\beta$ values have been shifted to the H1-LRG bin centres using small translation factors. The upper and lower curves form an error band on the predictions from the H1 2002 NLO QCD fit to the H1-LRG data (experimental errors only). Dotted lines are used for kinematic regions which were not included in the fit. Normalisation uncertainties of $^{+12\%}_{-10\%}$ on the ZEUS LPS data and $15\%$ on the factor applied to shift the datasets to $M_{\rm Y} < 1.6$ GeV are not shown.

## 4 Diffractive Parton Distributions

### 4.1 Theoretical Framework and Fit to H1-LRG Data

In this contribution, we adopt the fitting procedure used by H1 in [3], where next-to-leading order (NLO) QCD fits are performed to diffractive reduced cross section, $\sigma_r^{D(3)}$, data [3, 21] with $6.5 \leq Q^2 \leq 800$ GeV$^2$ and the $\beta$ and $x_{I\!P}$ ranges given in table 1.

The proof that QCD hard scattering collinear factorisation can be applied to diffractive DIS [7] implies that in the leading $\log(Q^2)$ approximation, the cross section for the diffractive process $ep \to eXY$ can be written in terms of convolutions of universal partonic cross sections $\hat{\sigma}^{ei}$ with diffractive parton distribution functions (dpdf's) $f_i^D$ [11, 22, 23], representing probability distributions for a parton $i$ in the proton under the constraint that the proton is scattered with a particular 4 momentum. Thus, at

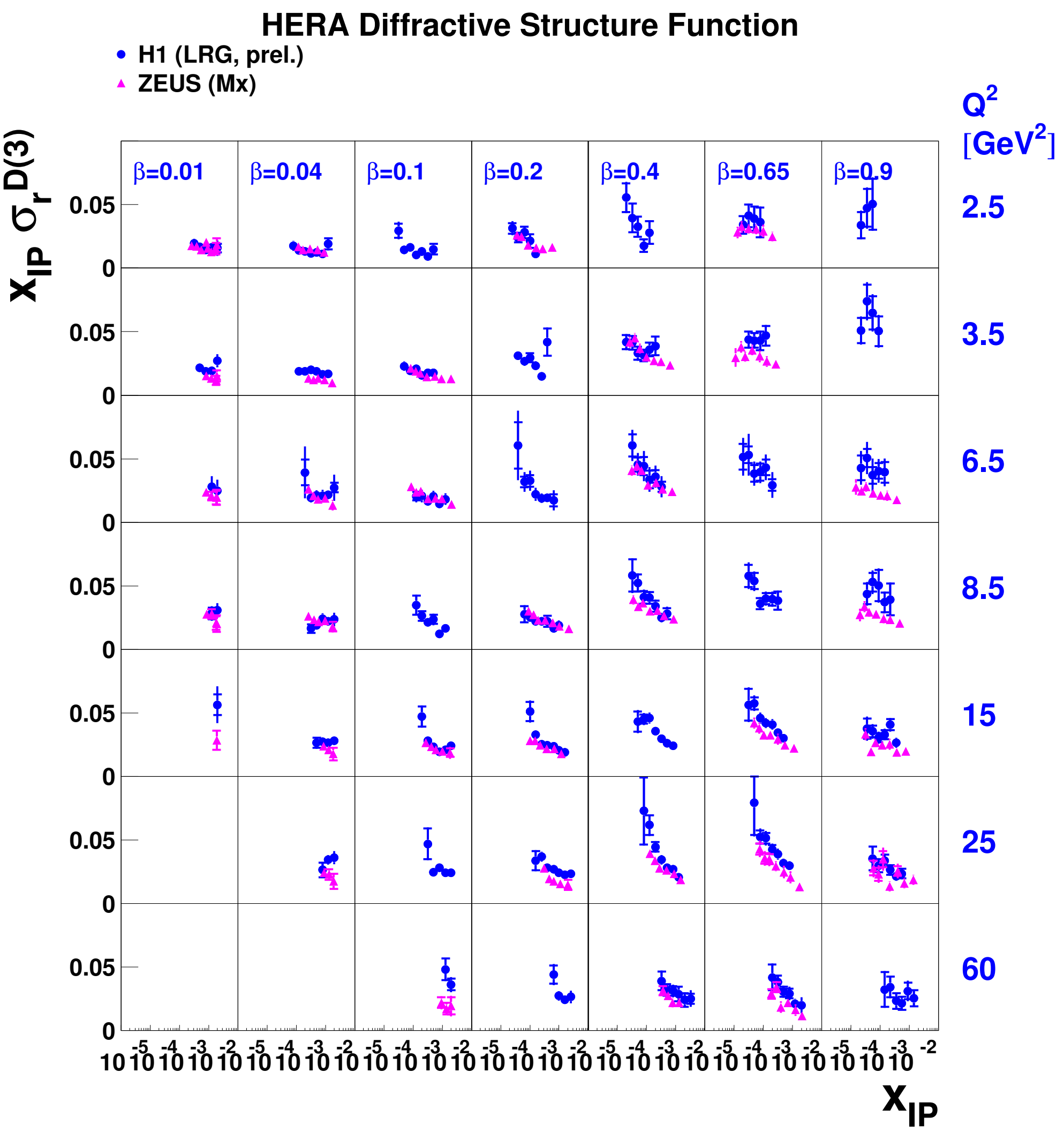

**Fig. 2:** Comparison of the ZEUS-$M_{\rm X}$ data with a subset of the H1-LRG data. The $Q^2$ and $\beta$ values of the ZEUS data have been shifted to the H1 bin centres using small translation factors. The ZEUS data have also been multiplied by a universal factor of 0.77, such that both data sets correspond to $M_{\rm Y} < 1.6$ GeV. Normalisation uncertainties of $15\%$ on this factor and of $\pm 6.7\%$ on the H1 data are not shown.

leading twist,[3]

$$\frac{{\rm d}^2\sigma(x,Q^2,x_{I\!P},t)^{ep\to eXp'}}{{\rm d}x_{I\!P}\,{\rm d}t} = \sum_i \int_x^{x_{I\!P}} {\rm d}\xi\ \hat{\sigma}^{ei}(x,Q^2,\xi)\ f_i^D(\xi,Q^2,x_{I\!P},t)\ . \tag{3}$$

This factorisation formula is valid for sufficiently large $Q^2$ and fixed $x_{I\!P}$ and $t$. It also applies to the case of proton dissociation into a system of fixed mass $M_{\rm Y}$ and thus to any cross section which is integrated over a fixed range in $M_{\rm Y}$. The partonic cross sections $\hat{\sigma}^{ei}$ are the same as those for inclusive DIS and the dpdf's $f_i^D$, which are not known from first principles, should obey the DGLAP evolution equations [25].

In addition to the rigorous theoretical prescription represented by equation (3), an additional assumption is necessary for the H1 fits in [3], that the shape of the dpdf's is independent of $x_{I\!P}$ and $t$ and that their normalisation is controlled by Regge asymptotics [26]. Although this assumption has no solid basis in QCD, it is compatible with the data fitted. The diffractive parton distributions can then be factorised into a term depending only on $x_{I\!P}$ and $t$ and a term depending only on $x$ (or $\beta$) and $Q^2$:

$$f_i^D(x_{I\!P}, t, x, Q^2) = f_{I\!P/p}(x_{I\!P}, t) \cdot f_i^{I\!P}(\beta = x/x_{I\!P}, Q^2) \; . \tag{4}$$

[3] A framework also exists to include higher order operators [24].

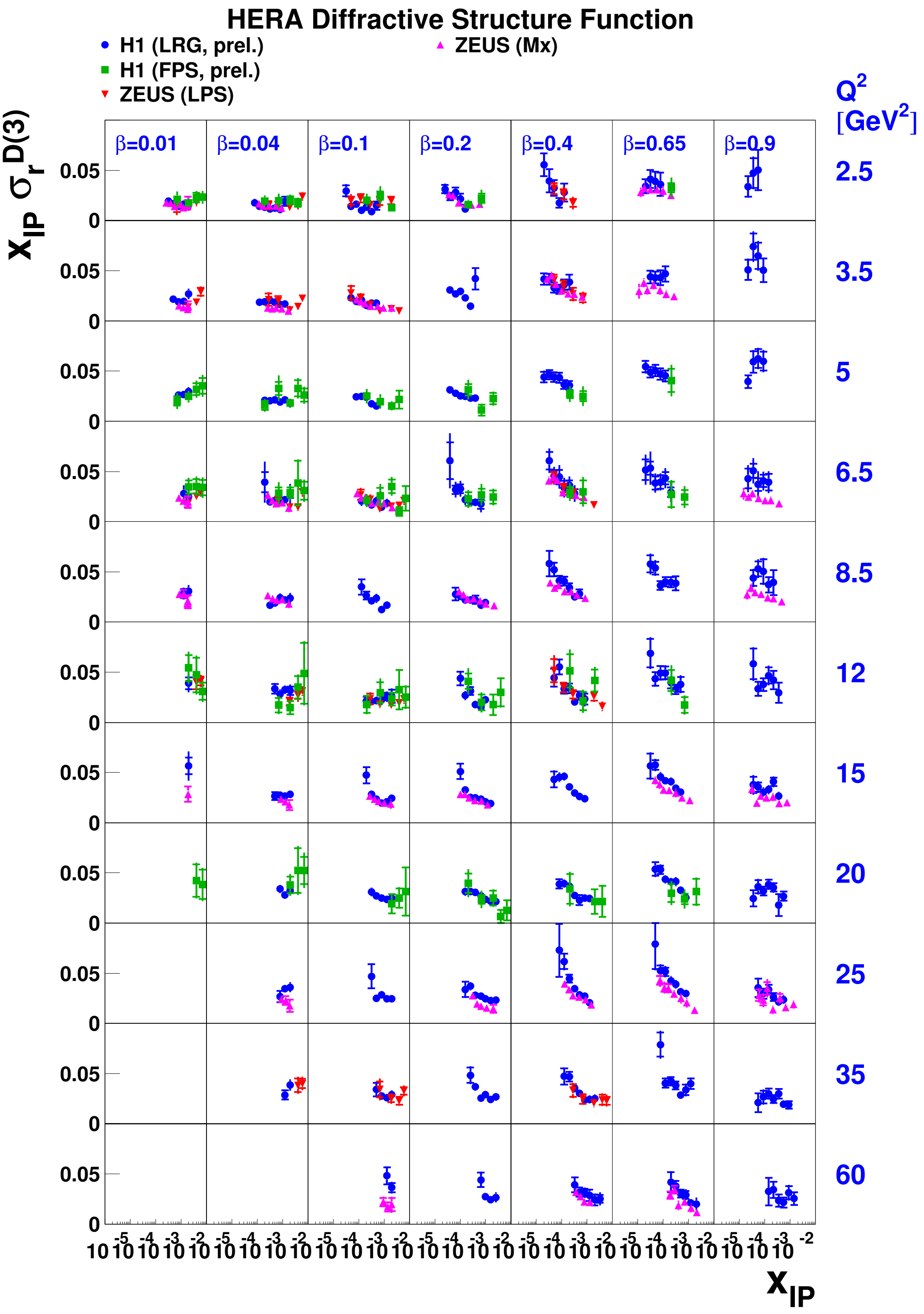

**Fig. 3:** Summary plot of all diffractive DIS data sets considered here. Additional H1-LRG data with $Q^2 < 2.5$ GeV$^2$, $Q^2 = 45$ GeV$^2$ and $Q^2 > 60$ GeV$^2$ are not shown. The $Q^2$ and $\beta$ values for all data sets have been shifted to the H1 bin centres using small translation factors. The ZEUS data have been multiplied by a universal factor of 0.77 and the LPS and FPS data by factors of 1.1, such that all data sets correspond to $M_Y < 1.6$ GeV. Relative normalisation uncertainties of 15% due to these factors and further normalisation uncertainties of $\pm 6.7\%$ (H1-LRG) and $^{+12\%}_{-10\%}$ (ZEUS-LPS) data are not shown.

Under this 'Regge' factorisation assumption, the diffractive exchange can be treated as an object (a 'pomeron', $I\!P$) with a partonic structure given by parton distributions $f_i^{I\!P}(\beta, Q^2)$. The variable $\beta$ then corresponds to the fraction of the pomeron longitudinal momentum carried by the struck parton. The 'pomeron flux factor' $f_{I\!P/p}(x_{I\!P}, t)$ represents the probability that a pomeron with particular values of $x_{I\!P}$ and $t$ couples to the proton.

In the fit, the $x_{I\!P}$ dependence is parameterised using a Regge flux factor

$$f_{I\!P/p}(x_{I\!P}, t) = A \cdot \int_{t_{cut}}^{t_{min}} \frac{e^{B_{I\!P} t}}{x_{I\!P}^{2\alpha_{I\!P}(t)-1}} \, \mathrm{d}t \; , \tag{5}$$

where $t_{cut} = -1.0 \text{ GeV}^2$, $|t_{min}|$ is the minimum kinematically allowed value of $|t|$ and the pomeron trajectory is assumed to be linear, $\alpha_{I\!P}(t) = \alpha_{I\!P}(0) + \alpha'_{I\!P} t$. The parameters $B_{I\!P}$ and $\alpha'$ and their uncertainties are fixed as described in [3]. The value of $A$ is chosen such that the flux factor is normalised to unity at $x_{I\!P} = 0.003$. The pomeron intercept is then obtained from the $x_{I\!P}$ dependence of the data and takes the value $\alpha_{I\!P}(0) = 1.173 \; \pm 0.018 \text{ (stat.)} \; \pm 0.017 \text{ (syst.)} \; ^{+0.063}_{-0.035} \text{ (model)}$.

The description of the data is improved with the inclusion of an additional separately factorisable sub-leading exchange with a trajectory intercept of $\alpha_{I\!R}(0) = 0.50$ and parton densities taken from a parameterisation of the pion [27]. This exchange contributes significantly only at low $\beta$ and large $x_{I\!P}$.

The dpdf's are modelled in terms of a light flavour singlet

$$\Sigma(z) = u(z) + d(z) + s(z) + \bar{u}(z) + \bar{d}(z) + \bar{s}(z) \; , \tag{6}$$

with $u = d = s = \bar{u} = \bar{d} = \bar{s}$ and a gluon distribution $g(z)$ at a starting scale $Q_0^2 = 3 \text{ GeV}^2$. Here, $z$ is the momentum fraction of the parton entering the hard sub-process with respect to the diffractive exchange, such that $z = \beta$ for the lowest-order quark parton model process, whereas $0 < \beta < z$ for higher order processes. The singlet quark and gluon distributions are parameterised using the form

$$z p_i(z, Q_0^2) = \left[ \sum_{j=1}^{n} C_j^i P_j(2z - 1) \right]^2 e^{\frac{0.01}{z-1}} \; , \tag{7}$$

where $P_j(\xi)$ is the $j^{\text{th}}$ member of a set of Chebychev polynomials[4]. The series is squared to ensure positivity. The exponential term is added to guarantee that the dpdf's tend to zero in the limit of $z \to 1$. It has negligible influence on the extracted partons at low to moderate $z$. The numbers of terms in the polynomial parameterisations are optimised to the precision of the data, with the first three terms in the series used for both the quark singlet and the gluon distributions, yielding 3 free parameters ($C_j^{\Sigma}$ and $C_j^g$) for each. The normalisation of the sub-leading exchange contribution at high $x_{I\!P}$ is also determined by the fit such that the total number of free parameters is 7. The data used in the fit are restricted to $M_{\mathrm{X}} > 2$ GeV to suppress non-leading twist contributions. The effects of $F_L^D$ are considered through its relation to the NLO gluon density, such that no explicit cut on $y$ is required.

The NLO DGLAP equations are used to evolve the dpdfs to $Q^2 > Q_0^2$ using the method of [28], extended for diffraction. No momentum sum rule is imposed. Charm quarks are treated in the massive scheme (appearing via boson gluon fusion processes) with $m_c = 1.5 \pm 0.1$ GeV. The strong coupling is set via[5] $\Lambda_{\mathrm{QCD}}^{\overline{\mathrm{MS}}} = 200 \pm 30$ MeV. The statistical and experimental systematic errors on the data points and their correlations are propagated to obtain error bands for the resulting dpdfs, which correspond to increases in the $\chi^2$ by one unit [29]. A theoretical error on the dpdfs is estimated by variations of $\Lambda_{\rm QCD}$, $m_c$ and the parameterisation of the $x_{I\!P}$ dependences as described in [3]. No theoretical uncertainty is assigned for the choice of parton parameterisation, though the results are consistent within the quoted uncertainties if alternative approaches [30] are used. No inhomogeneous term of the type included in [31] is considered here. The presence of such a term would lead to a reduction in the gluon density extracted.

[4] $P_1 = 1$, $P_2 = \xi$ and $P_{j+1}(\xi) = 2\xi P_j(\xi) - P_{j-1}(\xi)$.

[5] Although this value is rather different from the world average, we retain it here for consistency with previous H1 preliminary results, where it has been used consistently for QCD fits [3] and final state comparisons [8, 9].

The central fit gives a good description of the data, with a $\chi^2$ of $308.7$ for 306 degrees of freedom. The resulting diffractive quark singlet and gluon distributions are shown in figure 4. Both extend to large fractional momenta $z$. Whereas the singlet distribution is well constrained by the fit, there is a substantial uncertainty in the gluon distribution, particularly for $z \gtrsim 0.5$. The fraction of the exchanged momentum carried by gluons integrated over the range $0.01 < z < 1$ is $75 \pm 15\%$ (total error), confirming the conclusion from earlier work [19] that diffraction is a gluon-induced phenomenon. These dpdf's have been astonishingly successful in describing diffractive final state data in DIS such as charm [9] and jet [8] production, which, being induced by boson-gluon fusion-type processes, are roughly proportional to the diffractive gluon density.

### 4.2 Fit to ZEUS Data

A very similar fit to that described in section 4.1 is performed to the ZEUS-$M_{\rm X}$ data and the implications of the differences between the data sets to the dpdf's are investigated. The data are fitted in their original binning scheme, but are scaled to $M_{\rm Y} < 1.6$ GeV using the factor of 0.77. As for the fit to the H1 data, the first 3 terms are included in the polynomial expansions for the quark and gluon densities at the starting scale for QCD evolution. The same fit program, prescription and parameters are used as was the case for the H1 2002 NLO fit, with the following exceptions.

- ZEUS-$M_{\rm X}$ data with $Q^2 > 4$ GeV$^2$ are included in the fit, whereas only H1 data with $Q^2 > 6.5$ GeV$^2$ are included. It has been checked that the result for ZEUS is not altered significantly if the minimum $Q^2$ value is increased to 6 GeV$^2$.
- The quadratic sum of the statistical and systematic error is considered, i.e. there is no treatment of correlations between the data points through the systematics.
- No sub-leading Reggeon exchange component is included in the parameterisation. Including one does not improve or alter the fit significantly.
- The Pomeron intercept is fitted together with the dpdf's, in contrast to the two stage process of [3]. This does not influence the results significantly, though it does decrease the uncertainty on $\alpha_{\rm I\!P}(0)$.

The fit describes the ZEUS-$M_{\rm X}$ data well ($\chi^2 = 90$ for 131 degrees of freedom) and yields a value for the Pomeron intercept of $\alpha_{I\!P}(0) = 1.132 \pm 0.006$ (experimental error only). This value is in agreement with the H1 result if the full experimental and theoretical errors are taken into account. A good fit is thus obtained without any variation of $\alpha_{I\!P}(0)$ with $Q^2$ or other deviation from Regge factorisation.

The diffractive parton densities from the fit to the ZEUS-$M_{\rm X}$ data are compared with the results from H1 in figure 4. The differences observed between the H1 and the ZEUS data are directly reflected in the parton densities. The quark singlet densities are closely related to the measurements of $F_2^D$ themselves. They are similar at low $Q^2$ where the H1 and ZEUS data are in good agreement, but become different at larger $Q^2$, where discrepancies between the two data sets are observed. This difference between the $Q^2$ dependences of the H1 and ZEUS data is further reflected in a difference of around a factor of 2 between the gluon densities, which are roughly proportional to the logarithmic $Q^2$ derivative $\partial F_2^D / \partial \ln Q^2$ [32].

The H1-LRG and ZEUS-$M_{\rm X}$ data are shown together with the results from both QCD fits in figure 5. Both fits give good descriptions of the data from which they are obtained. The differences between the two data sets are clearly reflected in the fit predictions, most notably in the $Q^2$ dependence.

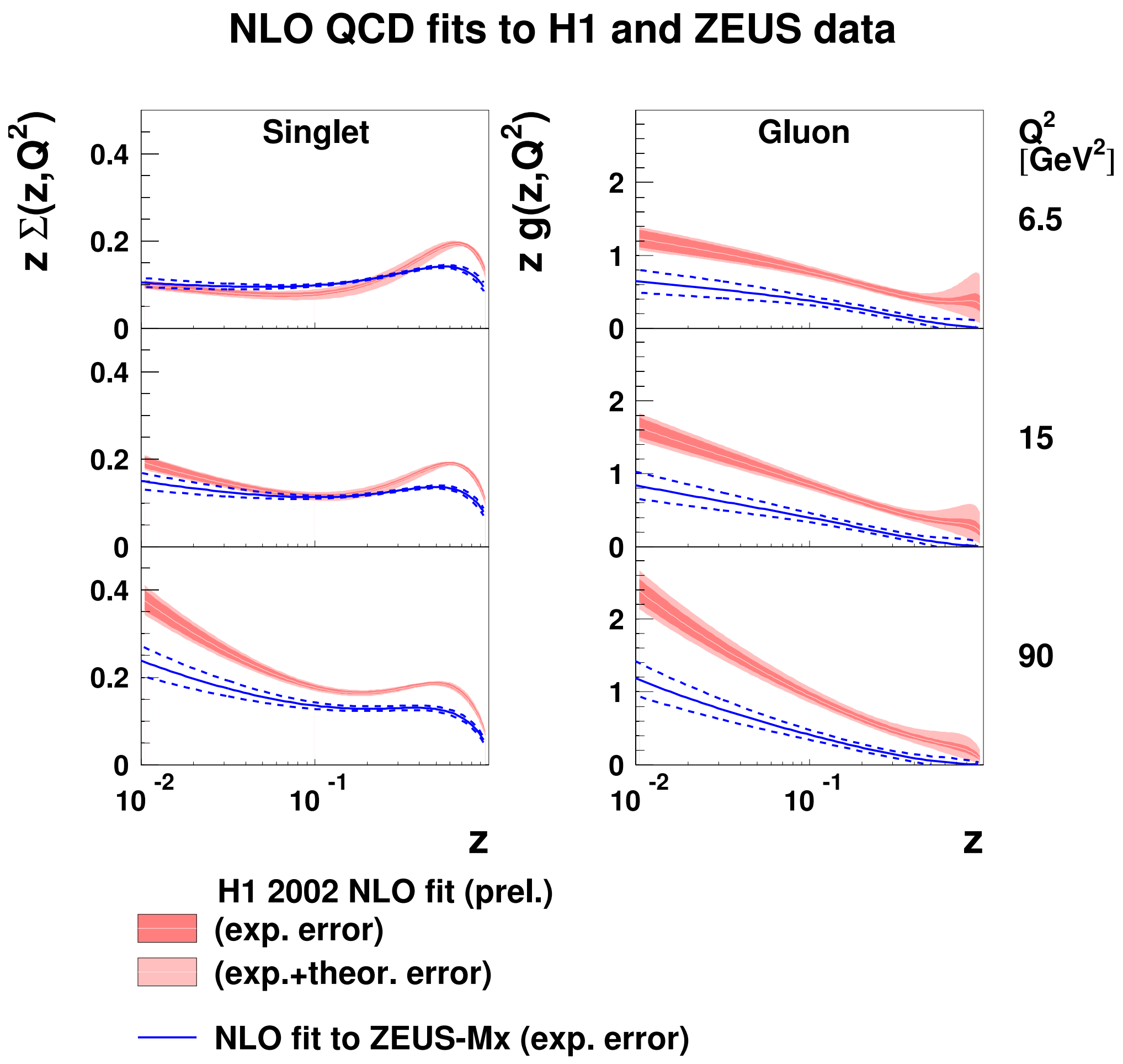

**Fig. 4:** Diffractive quark singlet and gluon pdf's for various $Q^2$ values, as obtained from the NLO DGLAP fits to the H1-LRG and ZEUS-$M_X$ data. The bands around the H1 result indicate the experimental and theoretical uncertainties. The dotted lines around the result for ZEUS indicate the experimental uncertainty. The ZEUS data used in the fit are scaled by a normalisation factor of 0.77 to match the H1-LRG range of $M_Y < 1.6$ GeV. This factor is reflected in the normalisations of the quark and gluon densities. An uncertainty of $15\%$ on this factor is not included in the error bands shown.

## 5 Summary

Recent diffractive structure function data from H1 and ZEUS have been compared directly. The leading proton data from both experiments (H1-FPS and ZEUS-LPS) are in good agreement with one other and with the H1 large rapidity gap data (H1-LRG). There is reasonable agreement between the H1-LRG and the ZEUS-$M_X$ data over much of the kinematic range. However, differences are observed at the highest $\beta$ (smallest $M_X$) and the $Q^2$ dependence at intermediate to low $\beta$ is weaker for the ZEUS-$M_X$ data than is the case for the H1-LRG data.

An NLO DGLAP QCD fit has been performed to the ZEUS-$M_X$ data, using the same theoretical framework, assumptions and parameterisations as have been employed previously for the H1-2002-prelim NLO QCD fit to a subset of the H1-LRG data. As a consequence of the differences between the $Q^2$ dependences of the H1-LRG and ZEUS-$M_X$ data, the gluon density obtained from the ZEUS data is significantly smaller than that for H1.

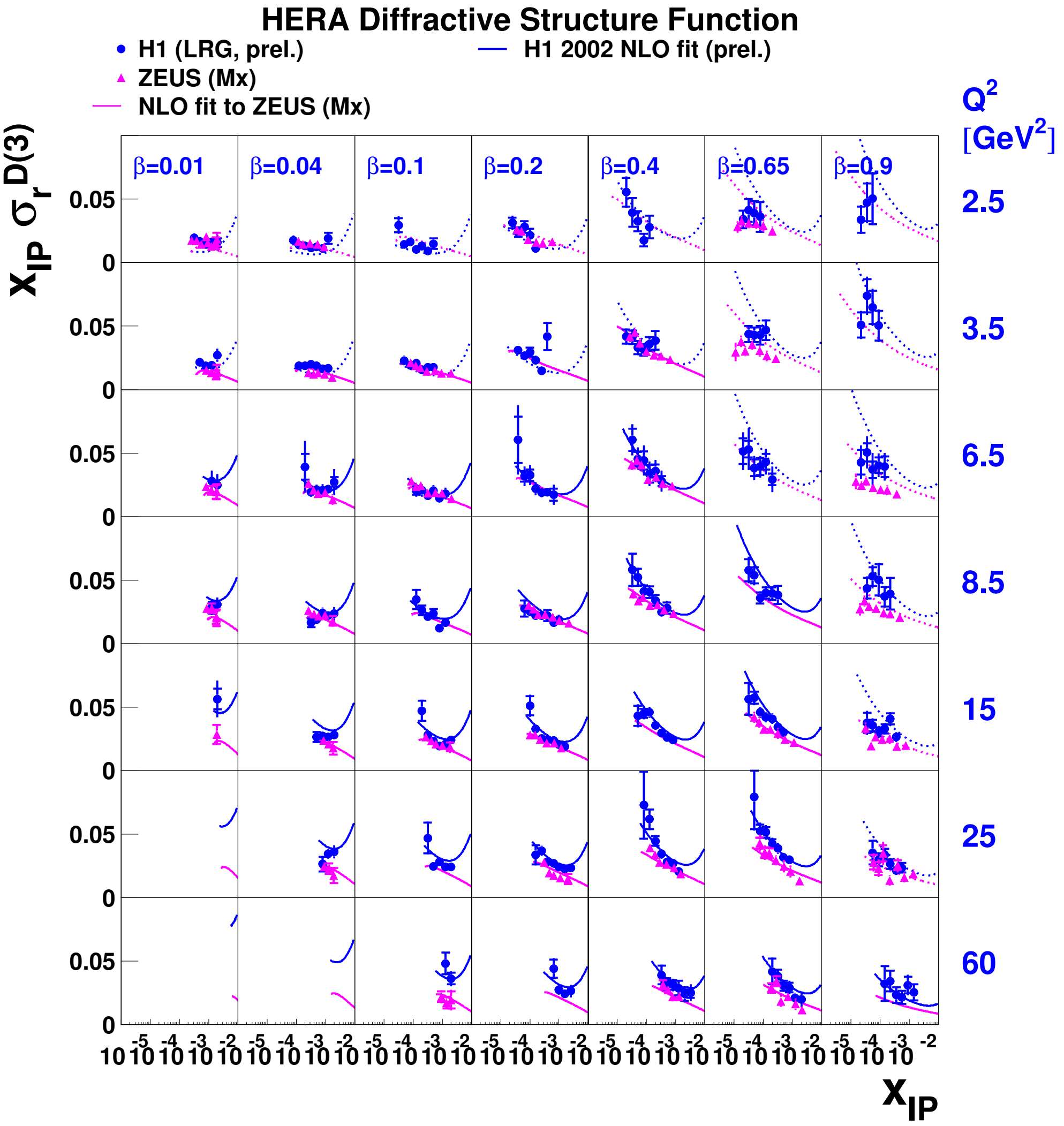

**Fig. 5:** As figure 2, but also showing the predictions using the NLO QCD fits to the H1-LRG and ZEUS-$M_X$ data (uncertainties not shown).

# Diffractive parton distributions from the HERA data

*Michael Groys*[a], *Aharon Levy*[a] *and Alexander Proskuryakov*[b]
[a]Raymond and Beverly Sackler Faculty of Exact Sciences, School of Physics and Astronomy, Tel Aviv University, Tel Aviv, Israel
[b]Institute of Nuclear Physics, Moscow State University, Moscow, Russia

**Abstract**

Measurements of the diffractive structure function, $F_2^D$, of the proton at HERA are used to extract the partonic structure of the Pomeron. Regge Factorization is tested and is found to describe well the existing data within the selected kinematic range. The analysis is based on the next to leading order QCD evolution equations. The results obtained from various data sets are compared.

## 1 Introduction

In the last 10 years a large amount of diffractive data was accumulated at the HERA collider [1–3]. There are three methods used at HERA to select diffractive events. One uses the Leading Proton Spectrometer (LPS) [3] to detect the scattered proton and by choosing the kinematic region where the scattered proton looses very little of its initial longitudinal energy, it ensures that the event was diffractive. A second method [2] simply requests a large rapidity gap (LRG) in the event and fits the data to contributions coming from Pomeron and Reggeon exchange. The third method [1] relies on the distribution of the mass of the hadronic system seen in the detector, $M_X$, to isolate diffractive events and makes use of the Forward Plug Calorimeter (FPC) to maximize the phase space coverage. We will refer to these three as ZEUS LPS, H1 and ZEUS FPC methods.

The experiments [4–6] provide sets of results for inclusive diffractive structure function, $x_{I\!P}F_2^{D(3)}$, in different regions of phase space. In extracting the initial Pomeron parton distribution functions (pdfs), the data are fitted assuming the validity of Regge factorization.

In the present study, Regge factorization is tested. New fits, based on a NLO QCD analysis, are provided and include the contribution of the longitudinal structure function. The obtained PDFs are systematically analyzed. A comparison of the different experimental data sets is provided. Additional quantities derived from the fit results are also presented.

In order to make sure that diffractive processes are selected, a cut of $x_{I\!P} < 0.01$ was performed, where $x_{I\!P}$ is the fraction of the proton momentum carried by the Pomeron. It was shown [7] that this cuts ensures the dominance of Pomeron exchange. In addition, a cut of $Q^2 > 3$ GeV$^2$ was performed on the exchanged photon virtuality for applying the NLO analysis. Finally, a cut on $M_X > 2$ GeV was used so as to exclude the light vector meson production.

## 2 Regge factorization

The Regge Factorization assumption can be reduced to the following,

$$F_2^{D(4)}(x_{I\!P}, t, \beta, Q^2) = f(x_{I\!P}, t) \cdot F(\beta, Q^2), \quad (1)$$

where $f(x_{I\!P}, t)$ represents the Pomeron flux which is assumed to be independent of $\beta$ and $Q^2$ and $F(\beta, Q^2)$ represents the Pomeron structure and is $\beta$ and $Q^2$ dependent. In order to test this assumption, we check whether the flux $f(x_{I\!P}, t)$ is indeed independent of $\beta$ and $Q^2$ on the basis of the available experimental data.

The flux is assumed to have a form $\sim x_{I\!P}^{-A}$ (after integrating over $t$ which is not measured in the data) . A fit of this form to the data was performed in different $Q^2$ intervals, for the whole $\beta$ range, and for different $\beta$ intervals for the whole $Q^2$ range.

Figure 1 shows the $Q^2$ dependence of the exponent $A$ for all three data sets, with the $x_{I\!P}$ and $M_X$ cuts as described in the introduction. The H1 and the LPS data show no $Q^2$ dependence. The ZEUS FPC data show a small increase in $A$ at the higher $Q^2$ region. It should be noted that while for the H1 and LPS data, releasing the $x_{I\!P}$ cut to 0.03 seems to have no effect, the deviation of the ZEUS FPC data from a flat dependence increases from a 2.4 standard deviation (s.d.) to a 4.2 s.d. effect (not shown).

The $\beta$ dependence of $A$ is shown in figure 2. All three data sets seem to show no $\beta$ dependence, within the errors of the data. Note however, that by releasing the $x_{I\!P}$ cut to higher values, a strong dependence of the flux on $\beta$ is observed (not shown).

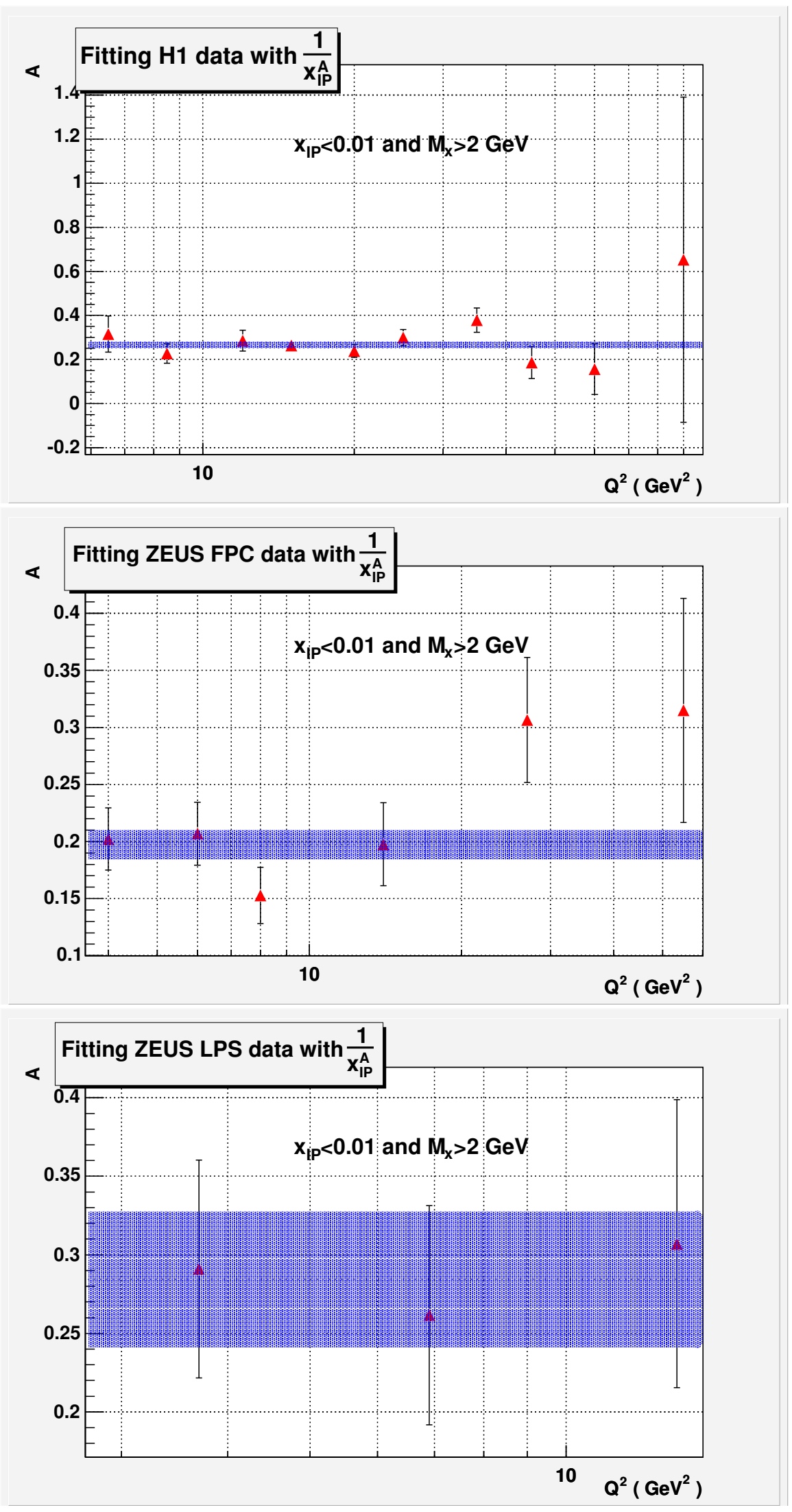

**Fig. 1:** $A$ as a function of $Q^2$ for $x_{I\!P} < 0.01$ and $M_X > 2$ GeV, for the three data sets, as indicated in the figure. The line corresponds to a fit over the whole $Q^2$ region

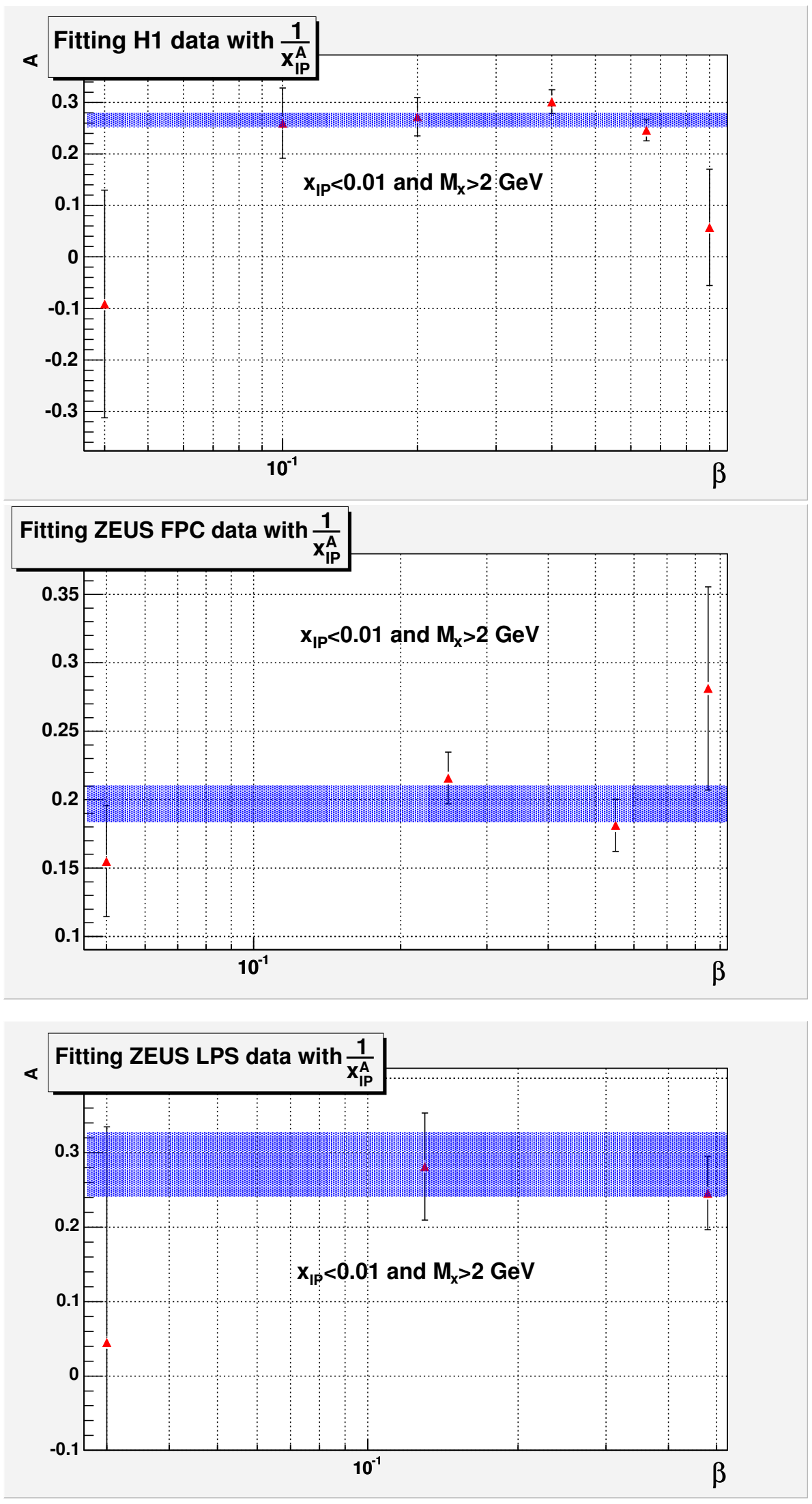

**Fig. 2:** $A$ as a function of $\beta$ for $x_{I\!P} < 0.01$ and $M_X > 2$ GeV, for the three data sets, as indicated in the figure. The line corresponds to a fit over the whole $\beta$ region

We thus conclude that for $x_{I\!P} < 0.01$, the Pomeron flux seems to be independent of $Q^2$ and of $\beta$ and thus the Regge factorization hypothesis holds.

## 3 NLO QCD fits

We parameterized the parton distribution functions of the Pomeron at $Q_0^2$ = 3 GeV$^2$ in a simple form of $Ax^b(1-x)^c$ for $u$ and $d$ quarks (and anti-quarks) and set all other quarks to zero at the initial scale. The gluon distribution was also assumed to have the same mathematical form. We thus had 3 parameters for quarks, 3 for gluons and an additional parameter for the flux, expressed in terms of the Pomeron intercept $\alpha_{I\!P}(0)$. Each data set was fitted to 7 parameters and a good fit was achieved for each. The H1 and ZEUS FPC had $\chi^2$/df $\approx$ 1, while for the LPS data, the obtained value was 0.5. The data together with the results of the fits are shown in figure 3. The following values were obtained for $\alpha_{I\!P}(0)$, for each of the three data sets: $\alpha_{I\!P}(0) = 1.138 \pm 0.011$, for the ZEUS FPC data, $\alpha_{I\!P}(0) = 1.189 \pm 0.020$, for the ZEUS LPS data, $\alpha_{I\!P}(0) = 1.178 \pm 0.007$, for the H1 data.

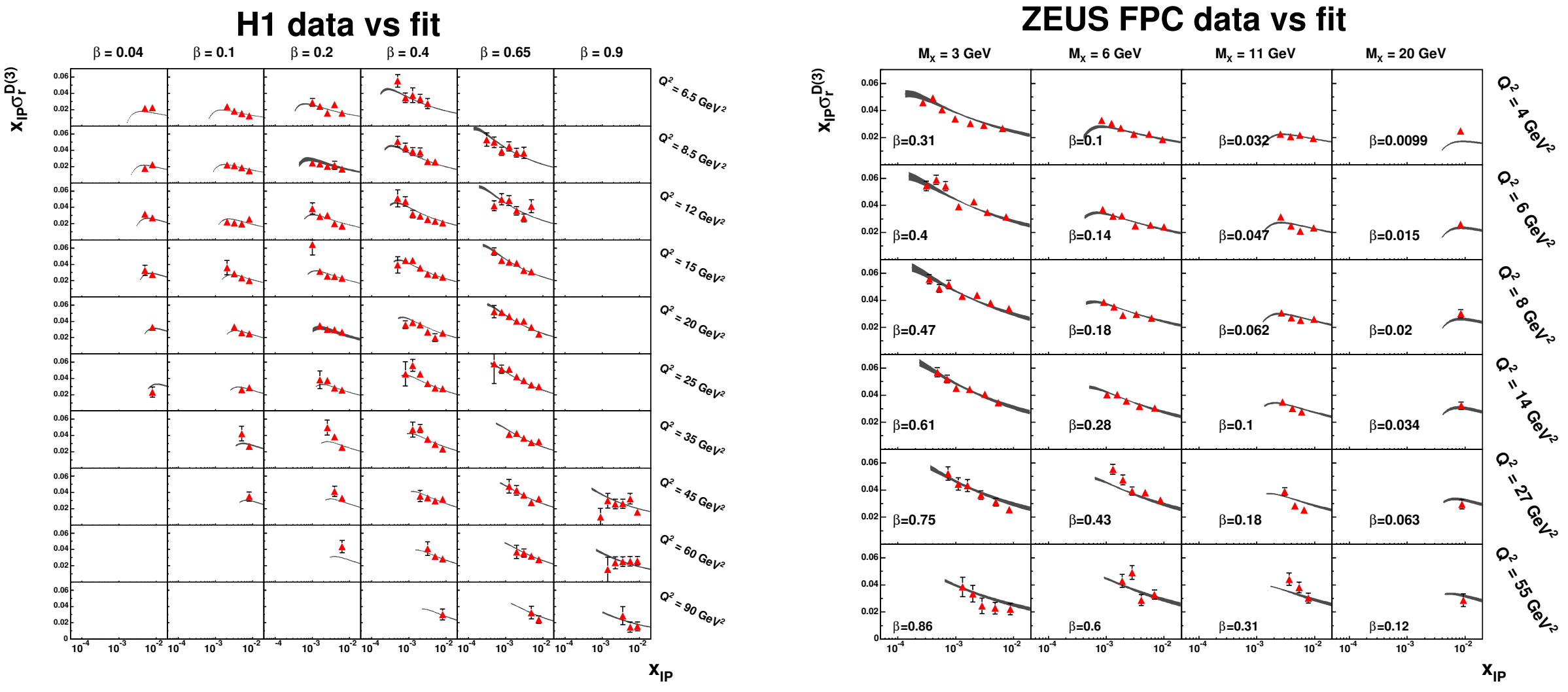

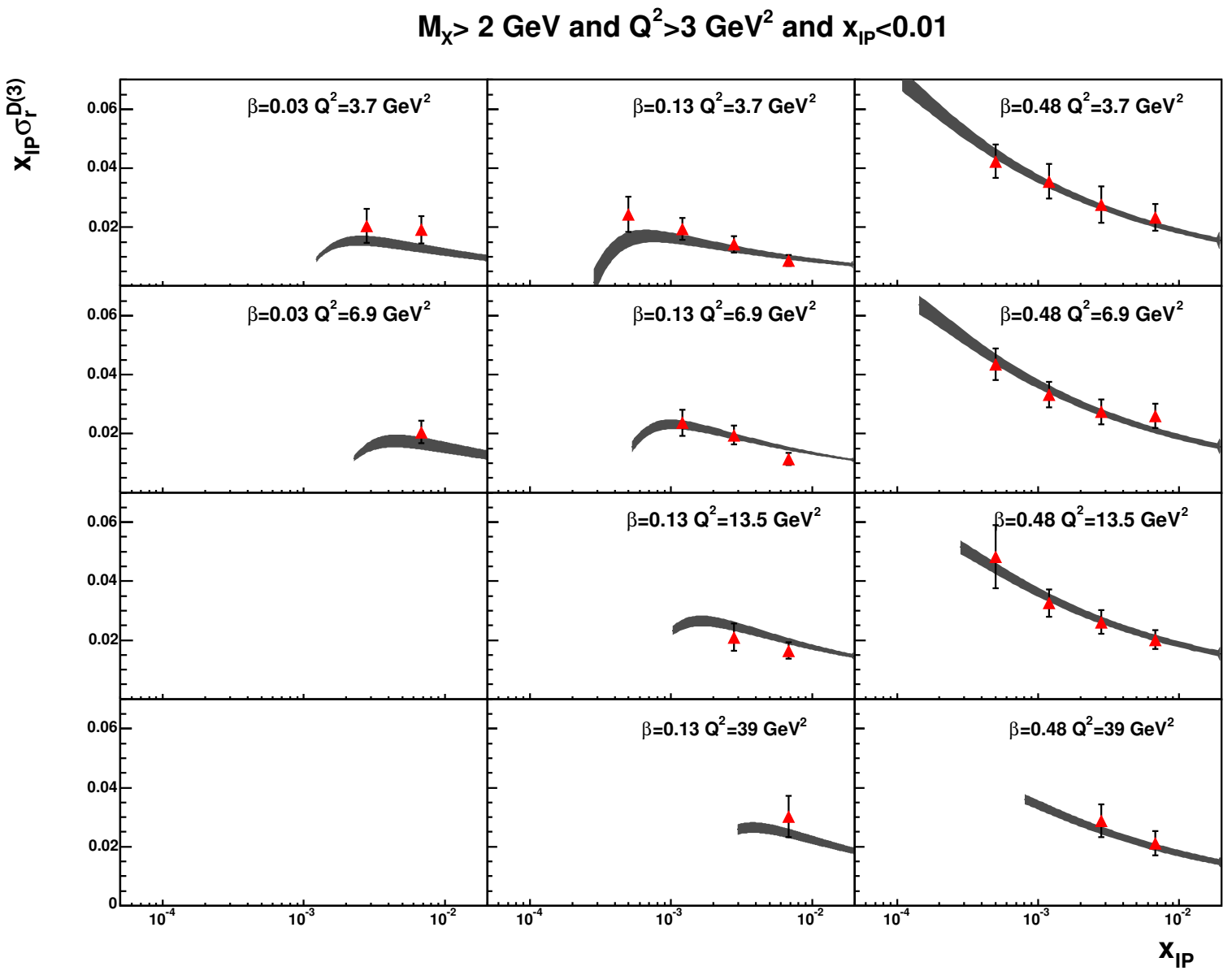

**Fig. 3:** The diffractive reduced cross section of the proton multiplies by $x_{I\!P}$, as a function of $x_{I\!P}$ for the different data sets (the most right plot is for the LPS data) in different bins of $Q^2$ and $\beta$, as indicated in the figure. The bands are the results of the fits including uncertainties.

The parton distribution functions are shown in figure 4 for the H1 and the ZEUS FPC data points. Because of the limited $\beta$ range covered by the LPS data, the resulting pdfs uncertainties are large and are not shown here. In fact one gets two solutions; one where the gluon contribution is dominant and another one where the gluons and the quarks contribute about equally. Note however that once the diffractive charm structure function data [8] are included in the fit, the gluon dominant solution is chosen (see below). For the H1 fit one sees the dominance of the gluons in all the $\beta$ range. For the ZEUS FPC data, the quark constituent of the Pomeron dominates at high $\beta$ while gluons dominate at low $\beta$. We can quantify this by calculating the Pomeron momentum carried by the gluons. Using the fit results one gets for the H1 data 80-90%. while for the ZEUS FPC data, 55-65%.

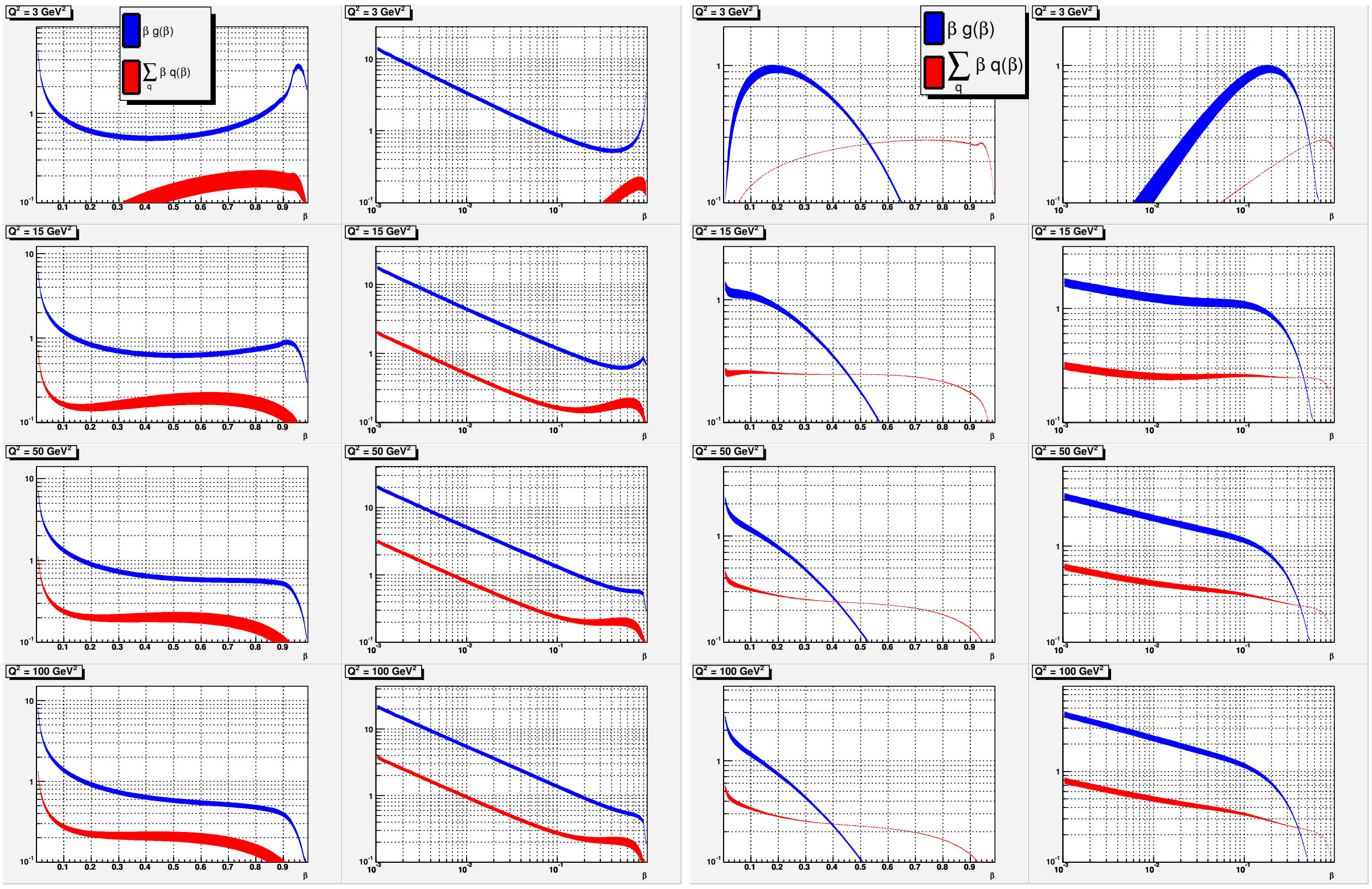

**Fig. 4:** Quark and gluon pdfs of the Pomeron as obtained from the H1 data fit (left two figures) and from the ZEUS FPC data fit (two rightmost figures) as a function of $\beta$, at different values of $Q^2$.

## 4 Comparison of the data sets

One way of checking the compatibility of all three data sets is to make an overall fit for the whole data sample. Since the coverage of the $\beta$ range in the LPS data is limited, we compare only the H1 and the ZEUS FPC data. A fit with a relative overall scaling factor of the two data sets failed. Using the fit results of one data sets superimposed on the other shows that the fit can describe some kinematic regions, while failing in other bins. This leads to the conclusion that there seems to be some incompatibility between the two data sets.

## 5 Comparison to $F_2^{D(3)}$(charm)

The ZEUS collaboration measured the diffractive charm structure function, $F_2^{D(3),c\bar{c}}$ [8] and these data were used together with the LPS data for a combined fit [6]. The charm data are shown in figure 5 as function of $\beta$. The full line shown the resulting best fit, where the contribution from charm was calculated as photon-gluon fusion. In the same figure one sees the prediction from the NLO QCD fit to the ZEUS FPC data (dashed line). Clearly, the gluons from the ZEUS FPC fit can not describe the diffractive charm data.

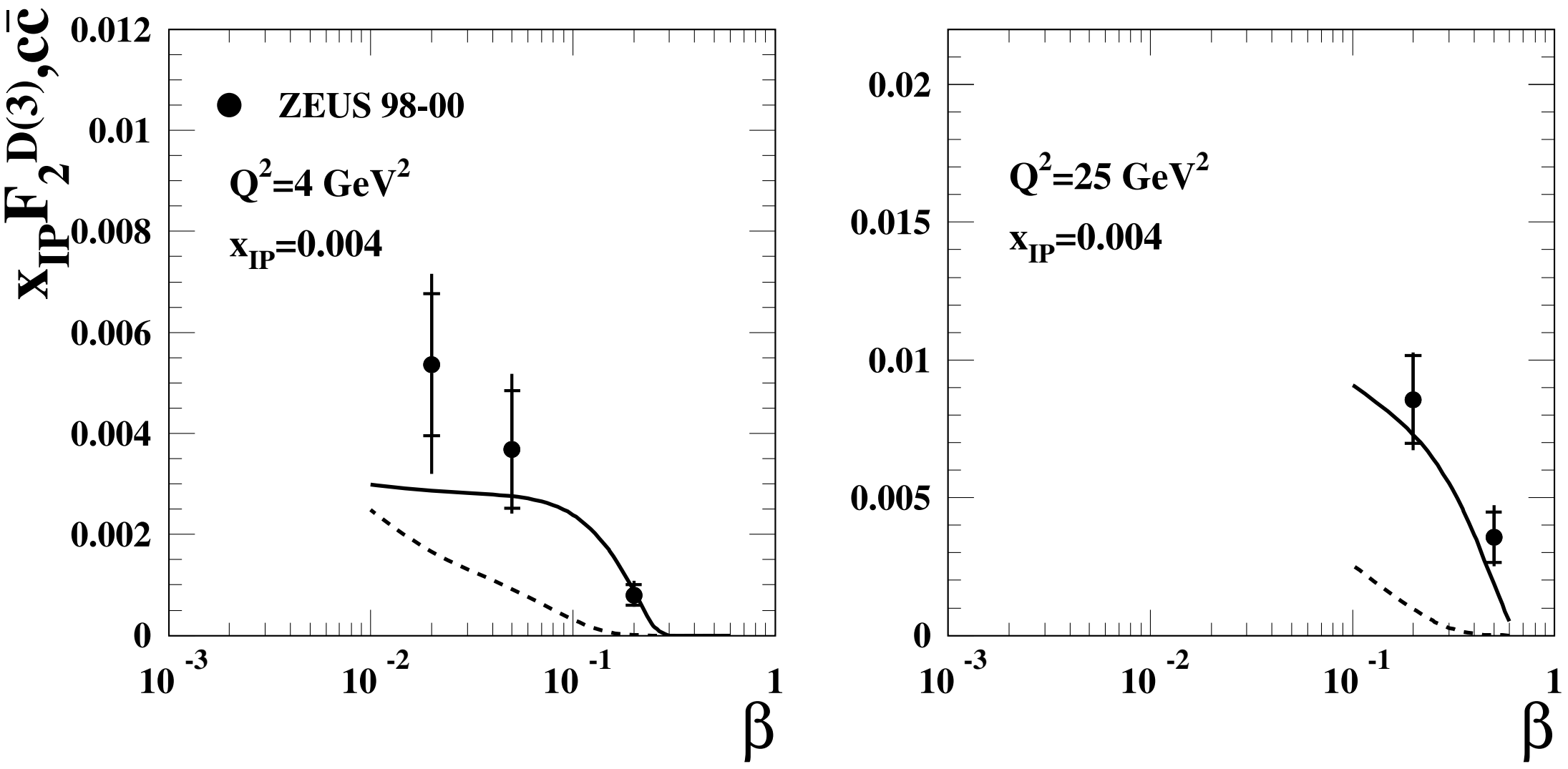

**Fig. 5:** Diffractive charm structure function, $F_2^{D(3),c\bar{c}}$, as a function of $\beta$ for values of $Q^2$ and $x_{I\!P}$ as indicated in the figure. The full line is the result of a combined fit to the LPS and the diffractive charm data. The dashed line is the prediction using the gluons from the ZEUS FPC fit.

## Acknowledgements

We would like to thank Prof. Halina Abramowicz for her useful and clarifying comments during this analysis. We would also like to thank Prof. John Collins for providing the program to calculate the NLO QCD equations for the diffractive data. This work was supported in part by the Israel Science Foundation (ISF).

# Diffractive parton distributions

G. Watt[a], A.D. Martin[b], M.G. Ryskin[b,c]
[a] Deutsches Elektronen-Synchrotron DESY, 22607 Hamburg, Germany
[b] Institute for Particle Physics Phenomenology, University of Durham, DH1 3LE, UK
[c] Petersburg Nuclear Physics Institute, Gatchina, St. Petersburg, 188300, Russia

**Abstract**
We discuss the perturbative QCD description of diffractive deep-inelastic scattering, and extract diffractive parton distributions from recent HERA data. The asymptotic collinear factorisation theorem has important modifications in the sub-asymptotic HERA regime. In addition to the usual *resolved* Pomeron contribution, the *direct* interaction of the Pomeron must also be accounted for. The diffractive parton distributions are shown to satisfy an *inhomogeneous* evolution equation, analogous to the parton distributions of the photon.

## 1 Diffractive parton distributions from Regge factorisation

It is conventional to extract diffractive parton distribution functions (DPDFs) from diffractive deep-inelastic scattering (DDIS) data using two levels of factorisation. Firstly, collinear factorisation means that the diffractive structure function can be written as [1]

$$F_2^{\rm D(3)}(x_{I\!P},\beta,Q^2) = \sum_{a=q,g} C_{2,a} \otimes a^{\rm D}, \tag{1}$$

where the DPDFs $a^{\rm D} = zq^{\rm D}$ or $zg^{\rm D}$ satisfy DGLAP evolution:

$$\frac{\partial a^{\rm D}}{\partial \ln Q^2} = \sum_{a'=q,g} P_{aa'} \otimes a'^{\rm D}, \tag{2}$$

and where $C_{2,a}$ and $P_{aa'}$ are the *same* hard-scattering coefficients and splitting functions as in inclusive DIS. The factorisation theorem (1) applies when $Q$ is made large, therefore it is correct up to power-suppressed corrections. It says nothing about the mechanism for diffraction. What *is* the exchanged object with vacuum quantum numbers ('Pomeron') which *causes* the large rapidity gap (LRG) characterising diffractive interactions?

In a second stage [2] Regge factorisation is usually assumed, such that

$$a^{\rm D}(x_{I\!P},z,Q^2) = f_{I\!P}(x_{I\!P})\, a^{I\!P}(z,Q^2), \tag{3}$$

where the Pomeron PDFs $a^{I\!P} = zq^{I\!P}$ or $zg^{I\!P}$. The Pomeron flux factor $f_{I\!P}$ is taken from Regge phenomenology,

$$f_{I\!P}(x_{I\!P}) = \int_{t_{\rm cut}}^{t_{\rm min}} {\rm d}t \quad {\rm e}^{B_{I\!P}\,t}\, x_{I\!P}^{1-2\alpha_{I\!P}(t)}. \tag{4}$$

Here, $\alpha_{I\!P}(t) = \alpha_{I\!P}(0) + \alpha'_{I\!P}\,t$, and the parameters $B_{I\!P}$, $\alpha_{I\!P}(0)$, and $\alpha'_{I\!P}$ should be taken from fits to soft hadron data. Although the first fits to use this approach assumed a 'soft' Pomeron, $\alpha_{I\!P}(0) \simeq 1.08$ [3], all recent fits require a substantially higher value to describe the data. In addition, a secondary Reggeon contribution is needed to describe the data for $x_{I\!P} \gtrsim 0.01$. This approach is illustrated in Fig. 1(a), where the virtualities of the $t$-channel partons are strongly ordered as required by DGLAP evolution. The Pomeron PDFs $a^{I\!P}$ are parameterised at some arbitrary low scale $Q_0^2$, then evolved up to the factorisation scale, usually taken to be the photon virtuality $Q^2$.

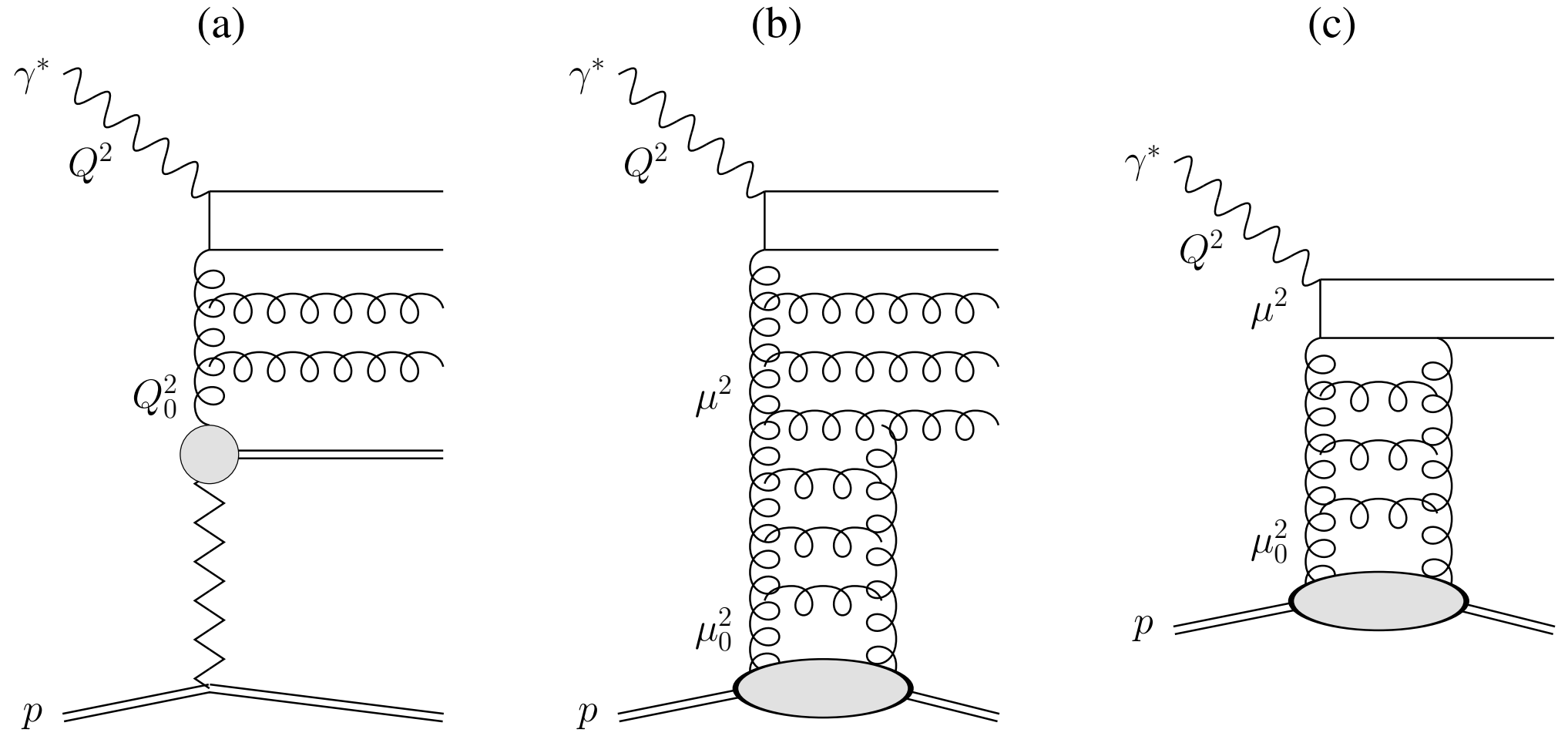

**Fig. 1:** (a) Resolved Pomeron contribution in the 'Regge factorisation' approach. (b) Resolved Pomeron contribution in the 'perturbative QCD' approach. (c) Direct Pomeron contribution in the 'perturbative QCD' approach.

Although this approach has been found to give a good description[1] of the DDIS data [4–7], it has little theoretical justification. The 'Regge factorisation' of (3) is merely a simple way of parameterising the $x_{I\!P}$ dependence of the DPDFs. Note, however, that the effective Pomeron intercept $\alpha_{I\!P}(0)$ has been observed to depend on $Q^2$ [8]. The fact that the required $\alpha_{I\!P}(0)$ is greater than the 'soft' value indicates that there is a significant perturbative QCD (pQCD) contribution to DDIS.

## 2 Diffractive parton distributions from perturbative QCD

In pQCD, Pomeron exchange can be described by two-gluon exchange, two gluons being the minimum number needed to reproduce the quantum numbers of the vacuum. Two-gluon exchange calculations are the basis for the colour dipole model description of DDIS, in which the photon dissociates into $q\bar{q}$ or $q\bar{q}g$ final states. Such calculations have successfully been used to describe HERA data. The crucial question, therefore, is how to reconcile two-gluon exchange with collinear factorisation as given by (1) and (2). Are these two approaches compatible?

Generalising the $q\bar{q}$ or $q\bar{q}g$ final states to an arbitrary number of parton emissions from the photon dissociation, and replacing two-gluon exchange by exchange of a parton ladder, we have diagrams like that shown in Fig. 1(b) [9–12]. Again, the virtualities of the $t$-channel partons are strongly ordered: $\mu_0^2 \ll \ldots \ll \mu^2 \ll \ldots \ll Q^2$. The scale $\mu^2$ at which the Pomeron-to-parton splitting occurs can vary between $\mu_0^2 \sim 1$ GeV$^2$ and the factorisation scale $Q^2$. Therefore, to calculate the inclusive diffractive structure function, $F_2^{\mathrm{D}(3)}$, we need to integrate over $\mu^2$:

$$F_2^{\mathrm{D}(3)}(x_{I\!P}, \beta, Q^2) = \int_{\mu_0^2}^{Q^2} \frac{\mathrm{d}\mu^2}{\mu^2}\, f_{I\!P}(x_{I\!P}; \mu^2)\, F_2^{I\!P}(\beta, Q^2; \mu^2). \tag{5}$$

Here, the perturbative Pomeron flux factor can be shown to be [12]

$$f_{I\!P}(x_{I\!P}; \mu^2) = \frac{1}{x_{I\!P} B_D} \left[ R_g \frac{\alpha_S(\mu^2)}{\mu}\, x_{I\!P} g(x_{I\!P}, \mu^2) \right]^2. \tag{6}$$

The diffractive slope parameter $B_D$ comes from the $t$-integration, while the factor $R_g$ accounts for the skewedness of the proton gluon distribution [13]. There are similar contributions from sea quarks, where $g(x_{I\!P}, \mu^2)$ in (6) is replaced by $S(x_{I\!P}, \mu^2)$, together with an interference term. In the fits presented here, we use the MRST2001 NLO gluon and sea-quark distributions of the proton [14]. The Pomeron structure function in (5), $F_2^{I\!P}(\beta, Q^2; \mu^2)$, is calculated from Pomeron PDFs, $a^{I\!P}(z, Q^2; \mu^2)$, evolved using NLO DGLAP from a starting scale $\mu^2$ up to $Q^2$, taking the input distributions to be LO Pomeron-to-parton splitting functions, $a^{I\!P}(z, \mu^2; \mu^2) = P_{aI\!P}(z)$ [11,12]. At first glance, it would appear that the perturbative Pomeron flux factor (6) behaves as $f_{I\!P}(x_{I\!P}; \mu^2) \sim 1/\mu^2$, so that contributions from large $\mu^2$ are strongly suppressed. However, at large $\mu^2$, the gluon distribution of the proton behaves as $x_{I\!P} g(x_{I\!P}, \mu^2) \sim (\mu^2)^\gamma$, where $\gamma$ is the anomalous dimension. In the BFKL limit of $x_{I\!P} \to 0$, $\gamma \simeq 0.5$, so $f_{I\!P}(x_{I\!P}; \mu^2)$ would be approximately independent of $\mu^2$. The HERA domain is in an intermediate region: $\gamma$ is not small, but is less than 0.5. It is interesting to plot the integrand of (5) as a function of $\mu^2$, as shown in Fig. 2. Notice that there is a large contribution from $\mu^2 > 3$ GeV$^2$, which is the value of the input scale $Q_0^2$ typically used in the 'Regge factorisation' fits of Sect. 1. Recall that fits using 'Regge factorisation' include contributions from $\mu^2 \leq Q_0^2$ in the input distributions, but neglect all contributions from $\mu^2 > Q_0^2$; from Fig. 2 this is clearly an unreasonable assumption.

[1] Note that the H1 2002 NLO fit [4] uses the 2-loop $\alpha_S$ with $\Lambda_{\mathrm{QCD}} = 200$ MeV for 4 flavours. This gives $\alpha_S$ values much smaller than the world average, meaning that the H1 2002 diffractive gluon density is artificially enhanced.

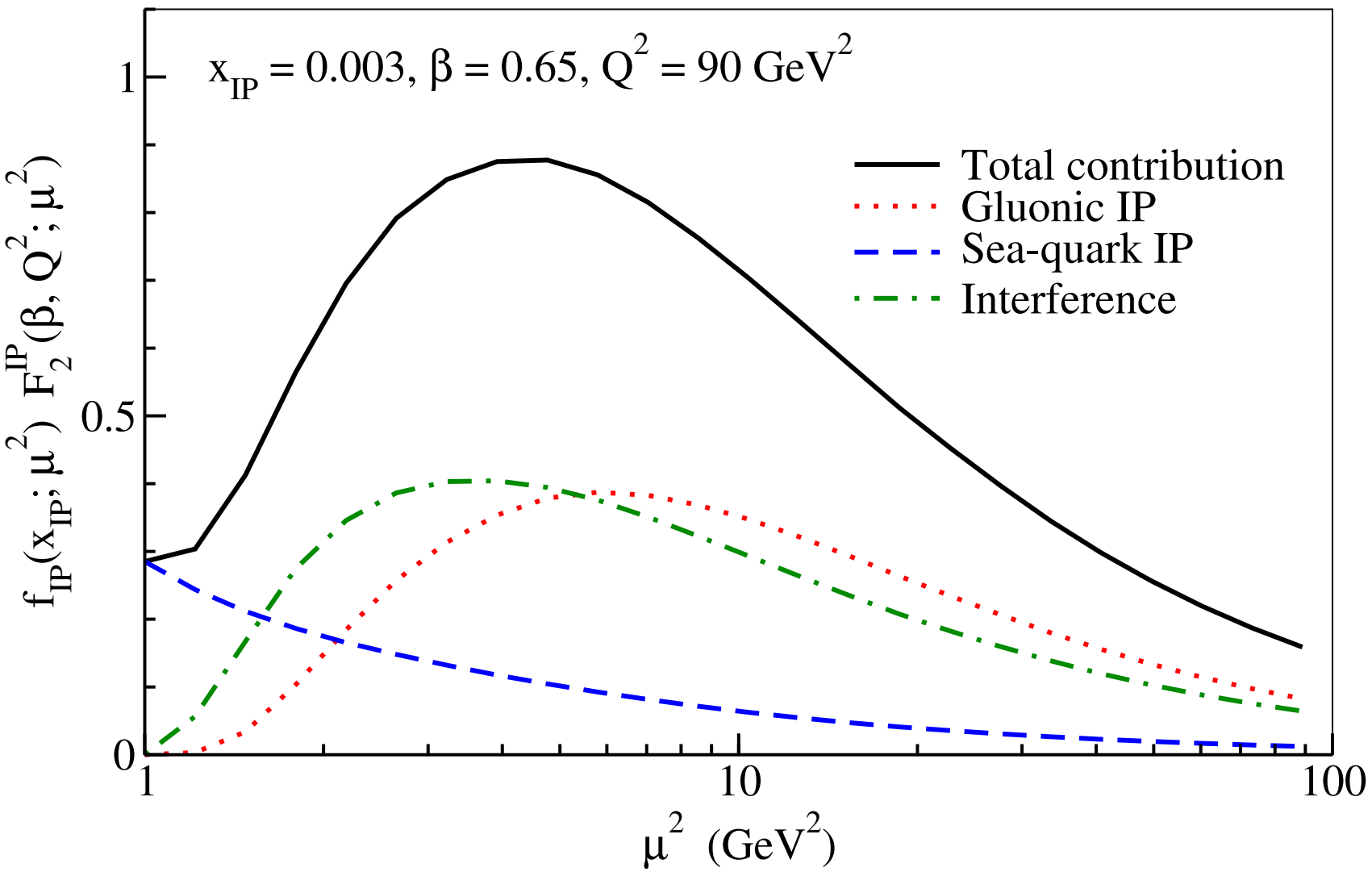

**Fig. 2:** Contributions to $F_2^{\mathrm{D}(3)}$ as a function of $\mu^2$.

As well as the *resolved* Pomeron contribution of Fig. 1(b), we must also account for the *direct* interaction of the Pomeron in the hard subprocess, Fig. 1(c), where there is no DGLAP evolution in the upper part of the diagram. Therefore, the diffractive structure function can be written as

$$F_2^{\mathrm{D}(3)} = \underbrace{\sum_{a=q,g} C_{2,a} \otimes a^{\mathrm{D}}}_{\text{Resolved Pomeron}} + \underbrace{C_{2,I\!P}}_{\text{Direct Pomeron}} ; \tag{7}$$

cf. (1) where there is no direct Pomeron contribution. The direct Pomeron coefficient function, $C_{2,I\!P}$, calculated from Fig. 1(c), will again depend on $f_{I\!P}(x_{I\!P}; \mu^2)$ given by (6). Therefore, it is formally suppressed by a factor $1/\mu^2$, but in practice does not behave as such; see Fig. 2.

The contribution to the DPDFs from scales $\mu > \mu_0$ is

$$a^{\mathrm{D}}(x_{I\!P}, z, Q^2) = \int_{\mu_0^2}^{Q^2} \frac{\mathrm{d}\mu^2}{\mu^2}\, f_{I\!P}(x_{I\!P}; \mu^2)\, a^{I\!P}(z, Q^2; \mu^2). \tag{8}$$

Differentiating (8), we see that the evolution equations for the DPDFs are [12]

$$\frac{\partial a^{\mathrm{D}}}{\partial \ln Q^2} = \sum_{a'=q,g} P_{aa'} \otimes a'^{\mathrm{D}} + P_{aI\!P}(z)\, f_{I\!P}(x_{I\!P}; Q^2); \tag{9}$$

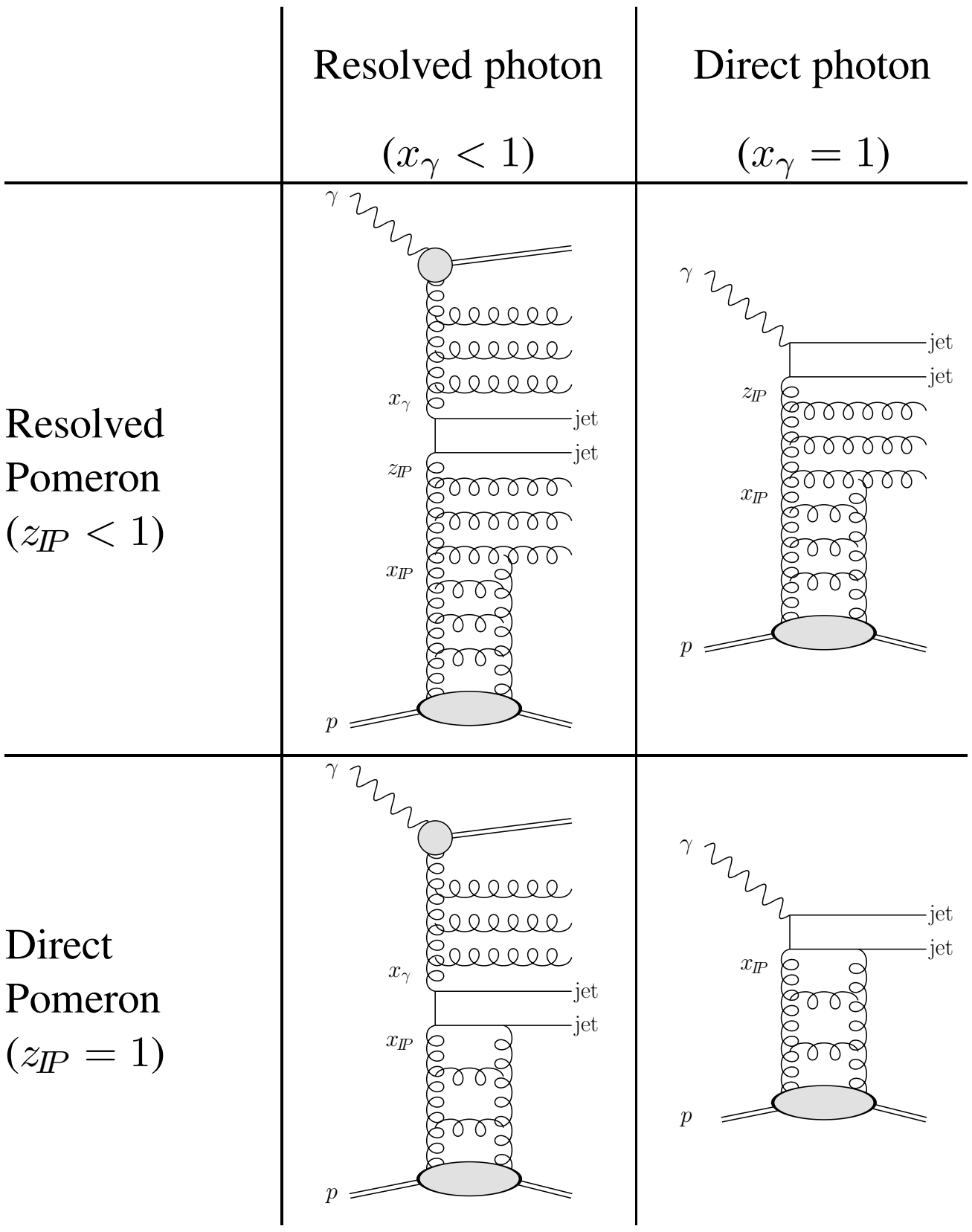

**Fig. 3:** The four classes of contributions to diffractive dijet photoproduction at LO. Both the photon and the Pomeron can be either 'resolved' or 'direct'.

cf. (2) where the second term of (9) is absent. That is, the DPDFs satisfy an *inhomogeneous* evolution equation [10, 12], with the extra inhomogeneous term in (9) leading to more rapid evolution than in the 'Regge factorisation' fits described in Sect. 1. Note that the inhomogeneous term will change the $x_{I\!P}$ dependence evolving upwards in $Q^2$, in accordance with the data, and unlike the 'Regge factorisation' assumption (3). Again, the inhomogeneous term in (9) is formally suppressed by a factor $1/Q^2$, but in practice does not behave as such; see Fig. 2.

Therefore, the diffractive structure function is analogous to the photon structure function, where there are both resolved and direct components and the photon PDFs satisfy an inhomogeneous evolution equation, where at LO the inhomogeneous term accounts for the splitting of the point-like photon into a $q\bar{q}$ pair. If we consider, for example, diffractive dijet photoproduction, there are four classes of contributions; see Fig. 3. The relative importance of each contribution will depend on the values of $x_\gamma$, the fraction of the photon's momentum carried by the parton entering the hard subprocess, and $z_{I\!P}$, the fraction of the Pomeron's momentum carried by the parton entering the hard subprocess.

## 3 Description of DDIS data

A NLO analysis of DDIS data is not yet possible. The direct Pomeron coefficient functions, $C_{2,I\!P}$, and Pomeron-to-parton splitting functions, $P_{aI\!P}$, need to be calculated at NLO within a given factorisation scheme (for example, $\overline{\mathrm{MS}}$). Here, we perform a simplified analysis where the usual coefficient functions $C_{2,a}$ and splitting functions $P_{aa'}$ ($a, a' = q, g$) are taken at NLO, but $C_{2,I\!P}$ and $P_{aI\!P}$ are taken at LO [12]. We work in the fixed flavour number scheme, where there is no charm DPDF. Charm quarks are produced via $\gamma^* g^{I\!P} \to c\bar{c}$ at NLO [15] and $\gamma^* I\!P \to c\bar{c}$ at LO [16]. For light quarks, we include the direct Pomeron process $\gamma_L^* I\!P \to q\bar{q}$ at LO [12], which is higher-twist and known to be important at large $\beta$.

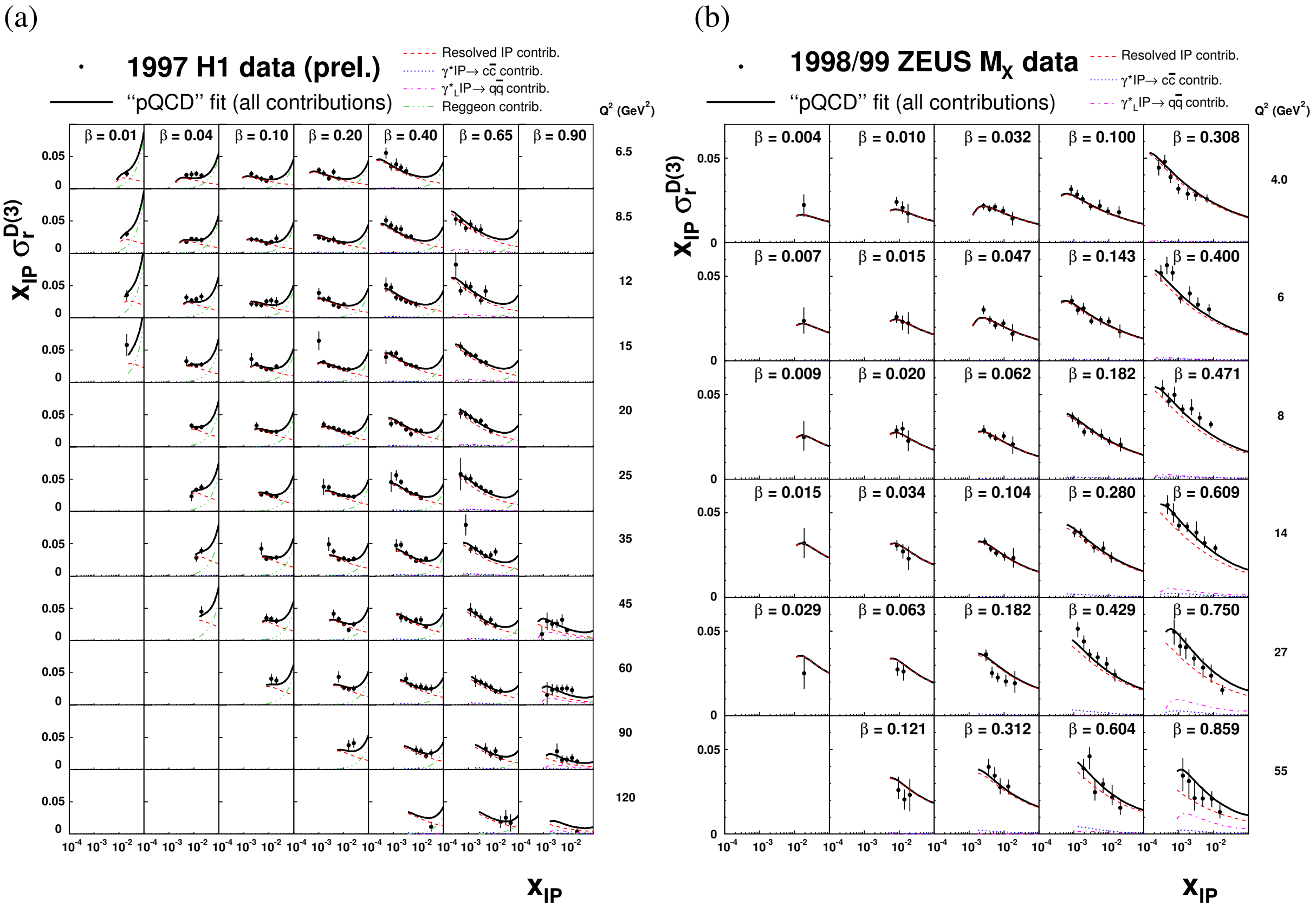

**Fig. 4:** "pQCD" fits to (a) H1 LRG and (b) ZEUS $M_X$ data.

To see the effect of the direct Pomeron contribution and the inhomogeneous evolution, we make two types of fits:

**"Regge"** : The 'Regge factorisation' approach discussed in Sect. 1, where there is no direct Pomeron contribution and no inhomogeneous term in the evolution equation.

**"pQCD"** : The 'perturbative QCD' approach discussed in Sect. 2, where these effects are included.

We make separate fits to the recent H1 LRG (prel.) [4] and ZEUS $M_X$ [8] $\sigma_r^{\mathrm{D}(3)}$ data, applying cuts $Q^2 \geq 3$ GeV$^2$ and $M_X \geq 2$ GeV, and allowing for overall normalisation factors of 1.10 and 1.43 to account for proton dissociation up to masses of 1.6 GeV and 2.3 GeV respectively. Statistical and systematic experimental errors are added in quadrature. The strong coupling is set via $\alpha_S(M_Z) = 0.1190$. We take the input forms of the DPDFs at a scale $Q_0^2 = 3$ GeV$^2$ to be

$$z\Sigma^{\mathrm{D}}(x_{I\!P}, z, Q_0^2) = f_{I\!P}(x_{I\!P})\; C_q\, z^{A_q}(1-z)^{B_q}, \tag{10}$$

$$zg^{\mathrm{D}}(x_{I\!P}, z, Q_0^2) = f_{I\!P}(x_{I\!P})\; C_g\, z^{A_g}(1-z)^{B_g}, \tag{11}$$

where $f_{I\!P}(x_{I\!P})$ is given by (4), and where $\alpha_{I\!P}(0)$, $C_a$, $A_a$, and $B_a$ ($a = q, g$) are free parameters. The secondary Reggeon contribution to the H1 data is treated in a similar way as in the H1 2002 fit [4], using the GRV pionic parton distributions [17]. Good fits are obtained in all cases, with $\chi^2/\mathrm{d.o.f.} = 0.75$, 0.71, 0.76, and 0.84 for the "Regge" fit to H1 data, "pQCD" fit to H1 data, "Regge" fit to ZEUS $M_X$ data, and "pQCD" fit to ZEUS $M_X$ data respectively. The "pQCD" fits are shown in Fig. 4, including a breakdown of the different contributions. The DPDFs are shown in Fig. 5. Note that the "pQCD" DPDFs are smaller than the corresponding "Regge" DPDFs at large $z$ due to the inclusion of the higher-twist $\gamma_L^* I\!P \to q\bar{q}$ contribution. Also note that the "pQCD" DPDFs have slightly more rapid evolution than the "Regge" DPDFs due to the extra inhomogeneous term in the evolution equation (9). There is a large difference between the DPDFs obtained from the H1 LRG and ZEUS $M_X$ data due to the different $Q^2$ dependence of these data sets; see also [6, 7].

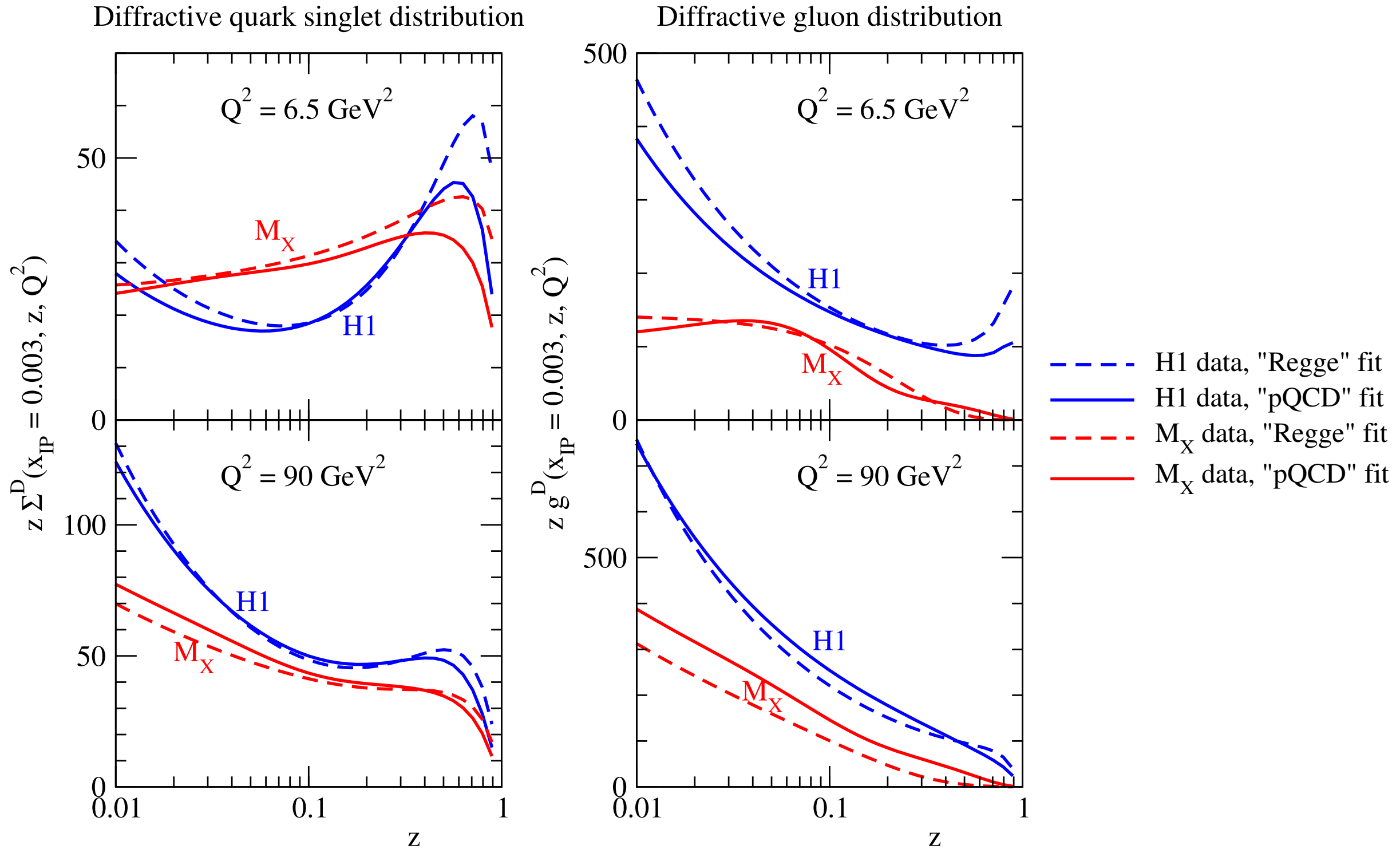

**Fig. 5:** DPDFs obtained from separate fits to H1 LRG and ZEUS $M_X$ data using the "Regge" and "pQCD" approaches.

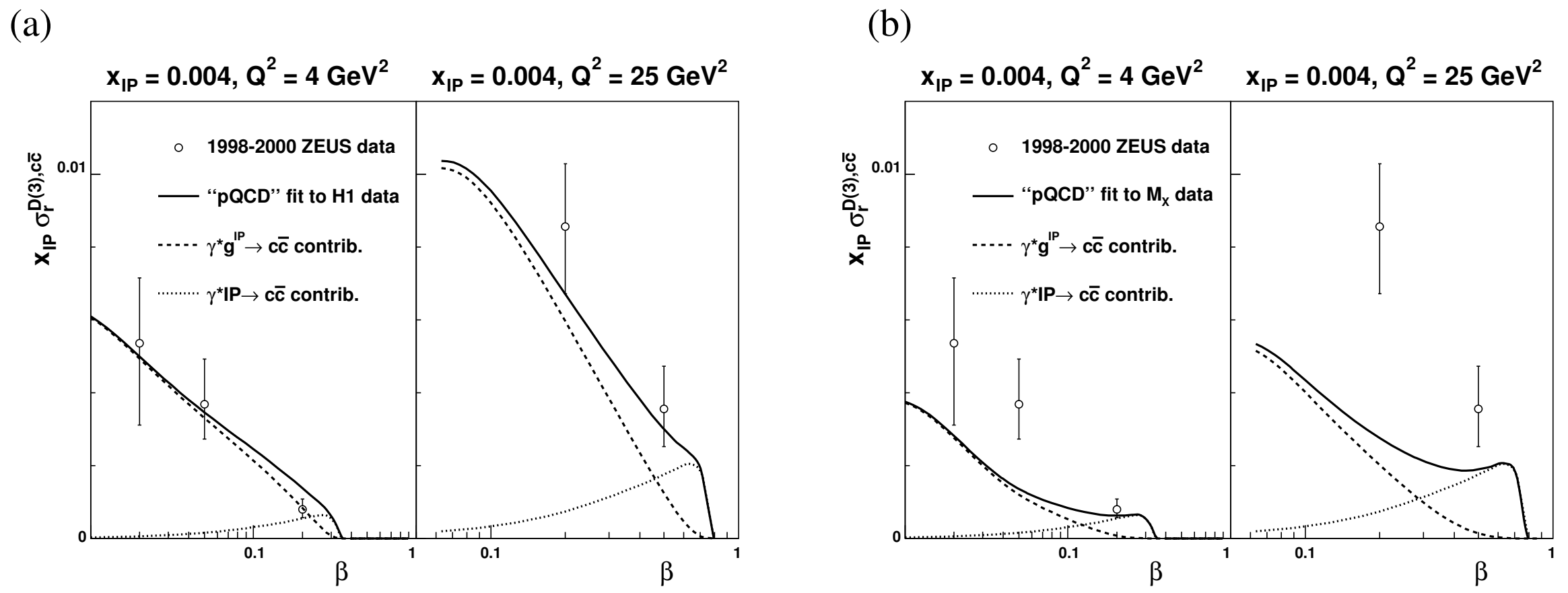

**Fig. 6:** Predictions for ZEUS LRG diffractive charm production data using DPDFs from the "pQCD" fits to (a) H1 LRG and (b) ZEUS $M_X$ data. Note the large direct Pomeron ($\gamma^* I\!P \to c\bar{c}$) contribution at moderate $\beta$.

The predictions from the two "pQCD" fits for the charm contribution to the diffractive structure function as measured by ZEUS using the LRG method [18] are shown in Fig. 6. Our H1 LRG fit gives a good description, while our ZEUS $M_X$ fit is too small at low $\beta$. Note that the direct Pomeron contribution is significant at moderate $\beta$. These charm data points were included in the determination of DPDFs from ZEUS LPS data [5], but only the resolved Pomeron ($\gamma^* g^{I\!P} \to c\bar{c}$) contribution was included and not the direct Pomeron ($\gamma^* I\!P \to c\bar{c}$) contribution. Therefore, the diffractive gluon distribution from the ZEUS LPS fit [5] needed to be artificially large to fit the charm data at moderate $\beta$.

## 4 Conclusions and outlook

To summarise, diffractive DIS is more complicated than inclusive DIS. Collinear factorisation holds, but we need to account for the direct Pomeron coupling, leading to an inhomogeneous evolution equation (9). Therefore, the treatment of DPDFs has more in common with photon PDFs than with proton PDFs. The H1 LRG and ZEUS $M_X$ data seem to have a different $Q^2$ dependence, leading to different DPDFs. This issue needs further attention.[2] For a NLO analysis of DDIS data, the direct Pomeron coefficient functions, $C_{2,I\!P}$, and Pomeron-to-parton splitting functions, $P_{aI\!P}$, need to be calculated at NLO. There are indications [16] that there are large $\pi^2$-enhanced virtual loop corrections ('K-factors') similar to those found in the Drell–Yan process. As with all PDF determinations, the sensitivity to the form of the input parameterisation, (10) and (11), and input scale $Q_0^2$ needs to be studied.[3] The inclusion of jet and heavy quark DDIS data, and possibly $F_L^{\rm D(3)}$ if it is measured [19], would help to constrain the DPDFs further. The extraction of DPDFs from HERA data will provide an important input for calculations of diffraction at the LHC.

[2] In particular, the main assumption of the $M_X$ method, that the diffractive contribution to the $\ln M_X^2$ distribution is constant, while the non-diffractive contribution rises exponentially, is motivated by Regge theory in the limit that $t = 0$, $\alpha_{I\!P}(0) \equiv 1$, and $Q^2 \ll M_X^2$, and it is not clear that this should be true in general.

[3] The ZEUS LPS fit [5], where the data have large statistical uncertainties, found that the shape of the DPDFs had a significant dependence on the functional form of the initial parameterisation. Even for the other data sets, where the statistical uncertainties are smaller, there seems to be a fairly strong dependence on the value of the input scale $Q_0^2$.

# DPDF: A Library for Diffractive Parton Distributions

*Frank-Peter Schilling*
CERN/PH, CH-1211 Geneva 23, Switzerland

**Abstract**
A code library is presented which provides a common interface to available parameterizations of diffractive parton distribution functions determined from QCD fits to HERA diffractive structure function data.

## 1 Introduction

In recent years, various precise measurements of the diffractive reduced cross section[1] $\sigma_r^{D(3)}(x_{I\!P}, \beta, Q^2)$ have been made by the HERA experiments H1 and ZEUS. Within the framework of QCD factorization in diffractive DIS [1], several sets of *diffractive parton distributions* (dpdf's) have been obtained from leading or next-to-leading order DGLAP QCD fits to these data[2]. The extracted dpdf's are a crucial input for the calculations of the cross sections of less inclusive diffractive processes such as diffractive jet, heavy quark or even Higgs production.

Since these diffractive pdf's are used in many different Monte-Carlo generators as well as in fixed order QCD calculations, it is desirable to provide them through a common software interface, similar in spirit to the common PDFLIB [2] and LHAPDF [3] packages for non-diffractive pdf's. To achieve this, the DPDF library has been developed. When a new dpdf set becomes available, it then needs to be implemented only in one place. Furthermore, additional features such as custom QCD evolution, structure function calculation and error information are available. Thus, the DPDF library provides a useful way to make the knowledge from HERA available to the TEVATRON, LHC and theory communities.

## 2 Theoretical Framework

The concept of QCD factorization in diffractive DIS implies that the diffractive $\gamma^* p$ cross section can be expressed as a convolution of universal diffractive parton distributions $f_i^D$ with process-dependent coefficient functions:

$$\frac{\mathrm{d}^2\sigma(x, Q^2, x_{I\!P}, t)^{\gamma^* p \to p' X}}{\mathrm{d}x_{I\!P}\, \mathrm{d}t} = \sum_i \int_x^{x_{I\!P}} \mathrm{d}\xi\; \hat{\sigma}^{\gamma^* i}(x, Q^2, \xi)\; f_i^D(x_{I\!P}, t, \xi, Q^2) \; . \quad (1)$$

The diffractive pdf's $f_i^D(x_{I\!P}, t, \beta, Q^2)$ can be extracted from a DGLAP QCD analysis of the diffractive reduced cross section $\sigma_r^D$.

For many (but not all) of the included parameterizations the $(x_{I\!P}, t)$ dependence factorizes ("Regge factorization") so that a *flux factor* $f_{I\!P/p}(x_{I\!P}, t)$ and dpdf's $f_i^{I\!P}(\beta, Q^2)$ are defined separately:

$$f_i^D(x_{I\!P}, t, \beta, Q^2) = f_{I\!P/p}(x_{I\!P}, t) \cdot f_i^{I\!P}(\beta, Q^2) \; . \quad (2)$$

For those parameterizations which include a secondary Reggeon exchange contribution (often using a pion structure function) in order to describe the data at high $x_{I\!P}$, such a possibility is also included. The dpdf's are typically parameterized in terms of a light quark flavor singlet and a gluon distribution, which are evolved using the (N)LO DGLAP equations[3].

[1] The reduced cross section $\sigma_r^D$ corresponds to the structure function $F_2^D$ if contributions from $F_L^D$ and $xF_3^D$ are neglected.
[2] In some cases, final state data were used in addition in order to better constrain the diffractive gluon density.
[3] For details of the parameterizations, see the original publications.

**Table 1:** Overview of the diffractive pdf sets implemented in the DPDF package. The $Q^2$, $\beta$ and $x_{I\!P}$ ranges correspond to the approximate kinematic range of the data used in the fit.

| Set | Fit | Var | Name | Ref. | Order | $Q^2$ (GeV$^2$) | $\beta$ | $x_{I\!P}$ |
|---|---|---|---|---|---|---|---|---|
| 1 | 4 | – | H1-1997-LO-Fit-1 | H1 Coll. [5] | LO | 4.5..75 | 0.04..0.9 | $< 0.05$ |
| 1 | 5 | – | H1-1997-LO-Fit-2 | | LO | | | |
| 1 | 6 | – | H1-1997-LO-Fit-3 | | LO | | | |
| 2 | 1 | – | H1-2002-NLO | H1 Coll. (prel.) [6] | NLO | 6.5..800 | 0.01..0.9 | $< 0.05$ |
| 2 | 2 | – | H1-2002-LO | | LO | | | |
| 3 | 1 | 1..3 | ACTW-NLO-A | Alvero et al. [7] | NLO | 6.0..75 | 0.20..0.7 | $< 0.01$ |
| 3 | 2 | 1..3 | ACTW-NLO-B | | NLO | | | |
| 3 | 3 | 1..3 | ACTW-NLO-C | | NLO | | | |
| 3 | 4 | 1..3 | ACTW-NLO-D | | NLO | | | |
| 3 | 5 | 1..3 | ACTW-NLO-SG | | NLO | | | |
| 4 | – | – | BGH-LO | Buchmueller et al. [8] | LO | 4.5..75 | 0.04..0.9 | $< 0.01$ |
| 5 | – | – | HS-NLO | Hautmann and Soper [9] | NLO | | | |
| 6 | – | – | ZEUS-LPS | ZEUS Coll. [11, 12] | NLO | 2.4..39 | 0.01..0.5 | $< 0.01$ |
| 7 | 1 | – | MRW-NLO-Lambda | Martin et al. [10] | NLO | 2.4..90 | 0.01..0.9 | $< 0.05$ |
| 7 | 2 | – | MRW-NLO-MRST | | NLO | | | |
| 8 | – | – | ZEUS-MX | Groys et al. [13] | NLO | 4.0..55 | 0.01..0.9 | $< 0.01$ |

## 3 Implementation

DPDF is a FORTRAN 77 package. A C++ wrapper will be provided. There is an external dependency on the QCDNUM [4] package, which can be disabled.

### 3.1 Available Parameterizations

Currently the following dpdf sets are implemented: the fits performed by the H1 collaboration in [5], the preliminary H1 fits presented in [6], the fits by Alvero et al. [7], a parameterization of the semi-classical model by Buchmueller et al. [8], the fits by Hautmann and Soper [9] and by Martin et al. [10], the ZEUS fit from [11, 12] and a fit to recent ZEUS data presented at this workshop [12, 13].

Details of the available dpdf sets are presented in table 1, including the kinematic ranges of the data which were included in the fits. This information can be used as a guideline for the range of validity of the fits. Note in particular that typically only data for $x_{I\!P} < 0.05$ or $< 0.01$ are included in the fits, which introduces an additional uncertainty when these fits are used for comparisons with experimental data at higher $x_{I\!P}$.

### 3.2 Interface to QCDNUM

DPDF provides an interface to the NLO DGLAP QCD evolution package QCDNUM [4]. It is possible to perform a QCD evolution of the given pdf set from its starting scale $Q_0^2$ using either the original evolution scheme and parameters, or by providing modified parameters. The benefits are:

- QCDNUM calculates the full (N)LO structure functions $F_2$ and $F_L$ for light and heavy quarks, which can be used for consistent comparisons with experimental data;
- The QCD evolution parameters such as $\alpha_s$ can be varied for systematic studies;

- The dpdf's can be evolved to $Q^2$ or $\beta$ values beyond the grid on which the original parameterization is provided, which is particularly interesting for LHC applications.

### 3.3 Usage

The DPDF package can be obtained from [14]. In the following the principal user subroutines of the library are listed.

- The package is initialized for a given dpdf by calling `dpdf_init(iset,ifit,ivar)` where `iset`, `ifit` and `ivar` are the parameters as given in table 1.
- The diffractive proton pdf's for either Pomeron or sub-leading Reggeon exchange or their sum (if provided) are returned at given values of $(x_{I\!P}, t, \beta, Q^2)$ in an array `xpq(-6:6)` using `dpdf_ppdf`. The result may also be integrated over $t$.
- If provided, the flux factors $f_{I\!P}(x_{I\!P}, t)$ and the parton densities $f_i(\beta, Q^2)$ can be obtained separately from `dpdf_flux` and `dpdf_pdf`.
- The diffractive structure function can be obtained from `dpdf_f2d`.
- QCD evolution using QCDNUM can be performed using default parameters for the given set with `dpdf_evolve_std` and using modified evolution parameters with `dpdf_evolve`.

Note that the details of the user interface may change in the future. For details refer to the user manual available from [14].

## 4 Outlook

It is planned to update DPDF if new dpdf sets become available. Additional features such as the possibility of error dpdf's (as for LHAPDF) are foreseen. The code and manual are available from [14].

# Prospects for $F_L^D$ Measurements at HERA-II

*Paul Newman*
School of Physics and Astronomy, University of Birmingham, B15 2TT, United Kingdom

**Abstract**
The theoretical interest in the longitudinal diffractive structure function $F_L^D$ is briefly motivated and possible measurement methods are surveyed. A simulation based on realistic scenarios with a reduced proton beam energy at HERA-II using the H1 apparatus shows that measurements are possible with up to $4\sigma$ significance, limited by systematic errors.

## 1 Introduction

In order to understand inclusive diffraction fully, it is necessary to separate out the contributions from transversely and longitudinally polarised exchange photons. Here, the formalism of [1] is adopted, where by analogy with inclusive scattering and neglecting weak interactions, a reduced cross section $\sigma_r^D$ is defined,[1] related to the experimentally measured cross section by

$$\frac{d^3\sigma^{ep\to eXY}}{\mathrm{d}x_{I\!P}\,\mathrm{d}\beta\,\mathrm{d}Q^2} = \frac{2\pi\alpha^2}{\beta\,Q^4}\cdot Y_+\cdot\sigma_r^D(x_{I\!P},\beta,Q^2)\,,\qquad \text{where}\qquad \sigma_r^D = F_2^D - \frac{y^2}{Y_+}F_L^D \tag{1}$$

and $Y_+ = 1 + (1-y)^2$. The structure function $F_L^D$, is closely related to the longitudinal photon contribution, whereas the more familiar $F_2^D$ contains information on the sum of transverse and longitudinal photon contributions.

It is generally understood [2] that at high $\beta$ and low-to-moderate $Q^2$, $\sigma_r^D$ receives a significant, perhaps dominant, higher twist contribution due to longitudinally polarised photons. Definite predictions [3] exist for this contribution, obtained by assuming 2-gluon exchange, with a similar phenomenology to that successfully applied to vector meson cross sections at HERA. The dominant role played by gluons in the diffractive parton densities [1] implies that the leading twist $F_L^D$ must also be relatively large. Assuming the validity of QCD hard scattering collinear factorisation [4], this gluon dominance results in a leading twist $F_L^D$ which is approximately proportional to the diffractive gluon density. A measurement of $F_L^D$ to even modest precision would provide a very powerful independent tool to verify our understanding of the underlying dynamics and to test the gluon density extracted indirectly in QCD fits from the scaling violations of $F_2^D$. This is particularly important at the lowest $x$ values, where direct information on the gluon density cannot be obtained from jet or $D^*$ data due to kinematic limitations and where novel effects such as parton saturation or non-DGLAP dynamics are most likely to become important.

Several different methods have been proposed to extract information on $F_L^D$. It is possible in principle to follow the procedure adopted by H1 in the inclusive case [5, 6], exploiting the decrease in $\sigma_r^D$ at large $y$ relative to expectations for $F_2^D$ alone (see equation 1). This method may yield significant results if sufficient precision and $y$ range can be achieved [7], though assumptions are required on the $x_{I\!P}$ dependence of $F_2^D$, which is currently not well constrained by theory. An alternative method, exploiting the azimuthal decorrelation between the proton and electron scattering planes caused by interference between the transverse and longitudinal photon contributions [8], has already been used with the scattered proton measured in the ZEUS LPS [9]. However, due to the relatively poor statistical precision achievable with Roman pots at HERA-I, the current results are consistent with zero. If the potential of the H1 VFPS is fully realised, this method may yet yield significant results in the HERA-II data [10]. However, if the necessary data are taken, the most promising possibility is to extract $F_L^D$ by comparing data at the same $Q^2$, $\beta$ and $x_{I\!P}$, but from different centre of mass energies $\sqrt{s}$ and hence from different $y$ values. The longitudinal structure function can then be extracted directly and model-independently from the measured data using equation 1. In this contribution, one possible scenario is investigated, based on modified beam energies and luminosities which are currently under discussion as a possible part of the HERA-II programme.

[1] It is assumed here that all results are integrated over $t$. The superscript $(3)$ usually included for $F_2^{D(3)}$ and other quantities is dropped for convenience.

## 2 Simulated $F_L^D$ Measurement

Given the need to obtain a large integrated luminosity at the highest possible beam energy for the remainder of the HERA programme and the fixed end-point in mid 2007, it is likely that only a relatively small amount of data can be taken with reduced beam energies. A possible scenario is investigated here in which 10 $\mathrm{pb}^{-1}$ are taken at just one reduced proton beam energy of $E_p = 400$ GeV, the electron beam energy being unchanged at 27.5 GeV. Since the maximum achievable instantaneous luminosity at HERA scales like the proton beam energy squared [11], this data sample could be obtained in around 2-3 months at the current level of HERA performance. It is assumed that a larger data volume of 100 $\mathrm{pb}^{-1}$ is available at $E_p = 920$ GeV, which allows for downscaling of high rate low $Q^2$ inclusive triggers.[2] The results presented here can be used to infer those from other scenarios given that the statistical uncertainty scales like $\sigma_r^{D\,400}/\sqrt{\mathcal{L}_{400}} + \sigma_r^{D\,920}/\sqrt{\mathcal{L}_{920}}$, where $\sigma_r^{D\,E_p}$ and $\mathcal{L}_{E_p}$ are the reduced cross section and the luminosity at a proton beam energy of $E_p$, respectively.

The longitudinal structure function can be extracted from the data at the two beam energies using

$$F_L^D = \frac{Y_+^{400}\,Y_+^{920}}{y_{400}^2 Y_+^{920} - y_{920}^2 Y_+^{400}} \left(\sigma_r^{D\,920} - \sigma_r^{D\,400}\right) , \qquad (2)$$

where $y_{E_p}$ and $Y_+^{E_p}$ denote $y$ and $Y_+$ at a beam energy $E_p$. It is clear from equation 2 that the best sensitivity to $F_L^D$ requires the maximum difference between the reduced cross sections at the two beam energies, which (equation 1) implies the maximum possible $y$ at $E_p = 400$ GeV. By measuring scattered electrons with energies $E'_e$ as low as 3 GeV [5], the H1 collaboration has obtained data at $y = 0.9$. This is possible with the use of the SPACAL calorimeter in combination with a measurement of the electron track in either the backward silicon tracker (BST) or the central jet chamber (CJC). For HERA-II running, the corresponding available range of scattered electron polar angle is $155^\circ < \theta'_e < 173^\circ$, which is used in the current study.[3] Three intervals in $y$ are considered, corresponding at $E_p = 400$ GeV to $0.5 < y_{400} < 0.7$, $0.7 < y_{400} < 0.8$ and $0.8 < y_{400} < 0.9$. It is ensured that identical ranges in $\beta$, $x_{I\!P}$ and $Q^2$ are studied at $E_p = 920$ GeV by choosing the bin edges such that $y_{920} = y_{400} \cdot 400/920$. Since the highest possible precision is required in this measurement, the restriction $x_{I\!P} < 0.02$ is imposed, which leads to negligible acceptance losses with a typical cut on the forwardmost extent of the diffractive system $\eta_{\max} < 3.3$. The kinematic restrictions on $E'_e$, $\theta'_e$ and $x_{I\!P}$ lead to almost no change in the mean $Q^2$, $M_X^2$ or $\beta \simeq Q^2/(Q^2 + M_X^2)$ as either $y$ or $E_p$ are varied. In contrast, $x_{I\!P} = Q^2/(s\,y\,\beta)$ varies approximately as $1/y$. As is shown in Fig. 1, at the average $\beta = 0.23$, there is at least partial acceptance for all $y$ bins in the range $7 < Q^2 < 30$ GeV$^2$, which is chosen for this study, leading to an average value of $Q^2$ close to 12 GeV$^2$.

The simulation is performed using the RAPGAP [13] Monte Carlo generator to extract the number of events per unit luminosity in each bin at each centre of mass energy. The values of $F_2^D$ and $F_L^D$, and hence $\sigma_r^{D\,920}$ and $\sigma_r^{D\,400}$ are obtained using an updated version of the preliminary H1 2002 NLO QCD fit [1].

[2] Alternative scenarios in which a smaller data volume at large $E_p$ is taken in a short, dedicated run, could potentially lead to better controlled systematics at the expense of increased statistical errors.

[3] One interesting alternative running scenario [12] is to obtain data at $E_p = 920$ GeV with the vertex shifted by 20 cm in the outgoing proton direction, which would allow measurements up to $\theta'_e = 175^\circ$, giving a low $Q^2$ acceptance range which closely matches that for the $E_p = 400$ GeV data at the normal vertex position.

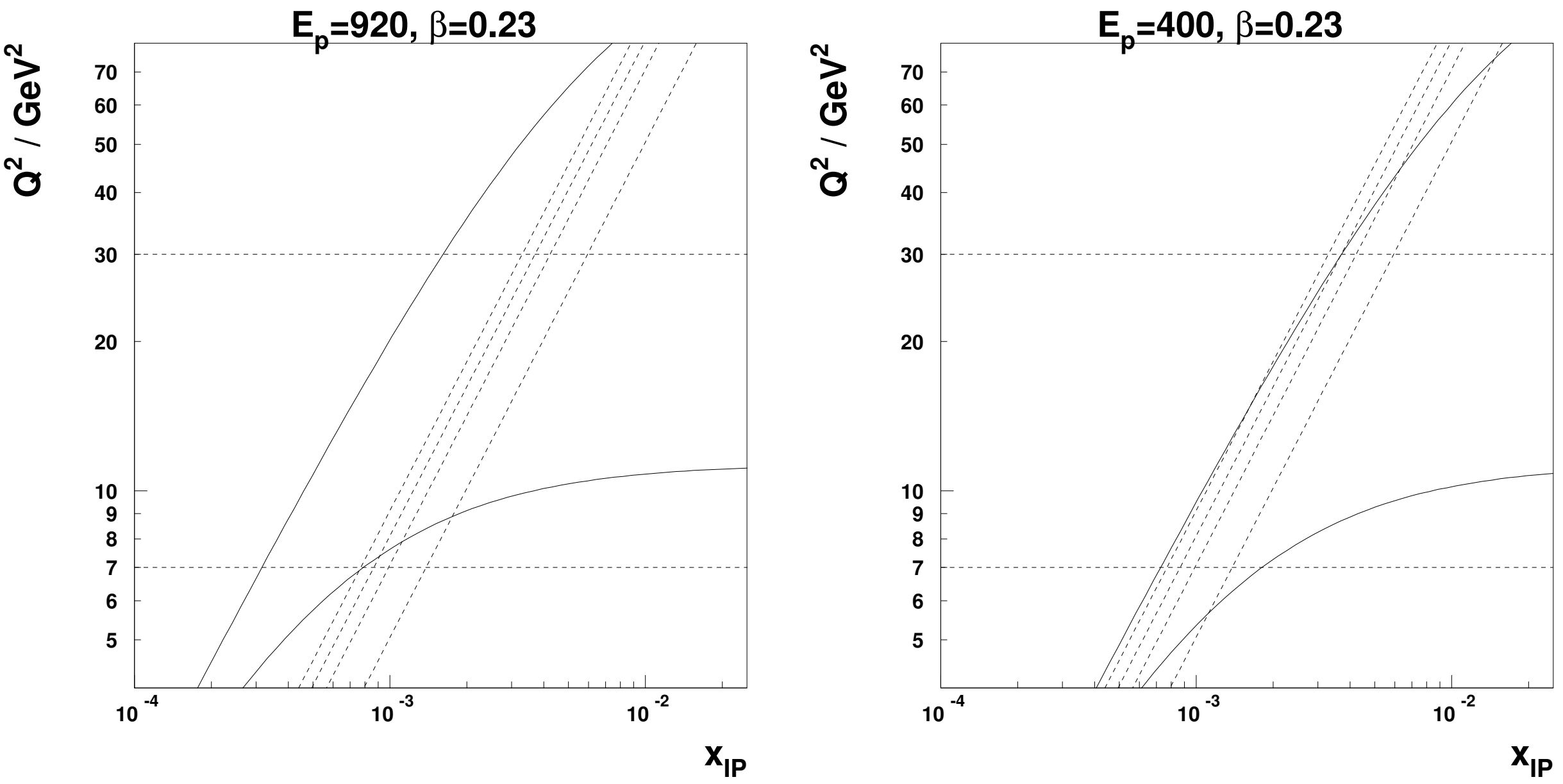

**Fig. 1:** Illustration of the kinematic plane in $Q^2$ and $x_{I\!P}$ at proton energies of 920 GeV and 400 GeV, with fixed $\beta = x/x_{I\!P} = 0.23$. The solid lines illustrate the experimental limits of $155^\circ < \theta'_e < 173^\circ$. The horizontal dashed lines illustrate the $Q^2$ range used for the simulation. The diagonal dashed lines illustrate the binning in $y$, corresponding at $E_p = 400$ GeV to $y = 0.9$ (leftmost line), $y = 0.8$, $y = 0.7$ and $y = 0.5$ (rightmost line).

The expected precision on $F_L^D$ is obtained by error propagation through equation 2. The systematic uncertainties are estimated on the basis of previous experience with the H1 detector [1, 5]. At the large $y$ values involved, the kinematic variables are most accurately reconstructed using the electron energy and angle alone. The systematic uncertainties on the measurements of these quantities are assumed to be correlated between the two beam energies. With the use of the BST and CJC, the possible bias in the measurement of $\theta'_e$ is at the level of $0.2\ \mathrm{mrad}$. The energy scale of the SPACAL calorimeter is known with a precision varying linearly from 2% at $E'_e = 3$ GeV to 0.2% at $E'_e = 27.5$ GeV. Other uncertainties which are correlated between the two beam energies arise from the photoproduction background subtraction (important at large $y$ and assumed to be known with a precision of $25\%$) and the energy scale for the hadronic final state used in the reconstruction of $M_X$ and hence $x_{I\!P}$ (taken to be known to 4%, as currently). Sources of uncertainty which are assumed to be uncorrelated between the low and high $E_p$ measurements are the luminosity measurement (taken to be $\pm 1\%$), the trigger and electron track efficiencies ($\pm 1\%$ combined) and the acceptance corrections, obtained using RAPGAP ($\pm 2\%$). The combined uncorrelated error is thus $2.4\%$. Finally, a normalisation uncertainty of $\pm 6\%$ due to corrections for proton dissociation contributions is taken to act simultaneously in the two measurements. Other sources of uncertainty currently considered in H1 measurements of diffraction are negligible in the kinematic region studied here.

Full details of the simulated uncertainties on the $F_L^D$ measurements are given in Table 1. An illustration of the corresponding expected measurement, based on the $F_L^D$ from the H1 2002 fit is shown in Fig. 2. The most precise measurement is obtained at the highest $y$, where $F_L^D$ would be determined to be unambiguously different from its maximum value of $F_2^D$ and to be non-zero at the $4\sigma$ level. Two further measurements are obtained at lower $y$ values. The dominant errors arise from statistical uncertainties and from uncertainties which are uncorrelated between the two beam energies. Minimising the latter is a major experimental challenge to be addressed in the coming years.

**Table 1:** Summary of the simulation at $Q^2 = 12$ GeV and $\beta = 0.23$. The first three columns contain the $y$ ranges used at $E_p = 400$ GeV and $E_p = 920$ GeV and the $x_{I\!P}$ values. The next two columns contain the values of the diffractive structure functions. These are followed by the uncorrelated ($\delta_{\rm unc}$) and proton dissociation ($\delta_{\rm norm}$) uncertainties and the correlated systematics due to the electron energy ($\delta E_e'$) and angle ($\delta\theta_e'$) measurements, the hadronic energy scale ($\delta M_X$) and the photoproduction background ($\delta\gamma p$), all in percent. The last three columns summarise the systematic, statistical and total uncertainties.

| $y_{400}$ | $y_{920}$ | $x_{I\!P}$ | $F_2^D$ | $F_L^D$ | $\delta_{\rm unc}$ | $\delta_{\rm norm}$ | $\delta E_e'$ | $\delta\theta_e'$ | $\delta M_X$ | $\delta\gamma p$ | $\delta_{\rm syst}$ | $\delta_{\rm stat}$ | $\delta_{\rm tot}$ |
|---|---|---|---|---|---|---|---|---|---|---|---|---|---|
| 0.5 – 0.7 | 0.217 – 0.304 | 0.0020 | 15.72 | 3.94 | 34 | 6 | 8 | 2 | 7 | 0 | 36 | 20 | 41 |
| 0.7 – 0.8 | 0.304 – 0.348 | 0.0016 | 20.87 | 5.25 | 19 | 6 | 3 | 2 | 5 | 6 | 22 | 17 | 28 |
| 0.8 – 0.9 | 0.348 – 0.391 | 0.0014 | 24.47 | 6.16 | 14 | 6 | 6 | 1 | 2 | 13 | 21 | 13 | 25 |

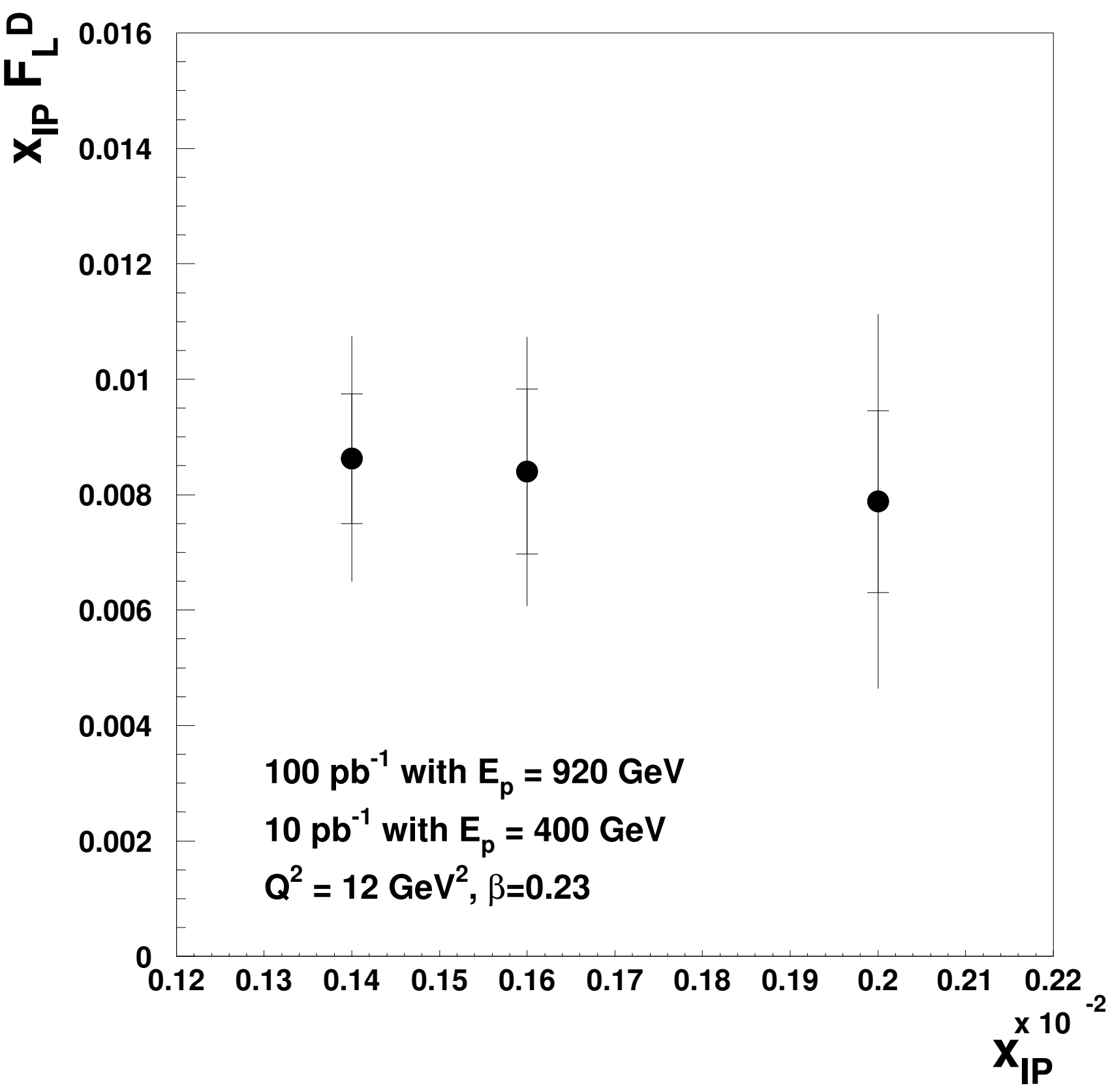

**Fig. 2:** Illustration of the simulated result for $F_L^D$, showing the three data points with statistical (inner bars) and total (outer bars) errors.

Only one possible scenario has been investigated here, leading to a highly encouraging result at relatively low $\beta$, which would provide a very good test of the leading twist $F_L^D$ and thus of the gluon density extracted in QCD fits to $F_2^D$. It may also be possible to obtain results at high $\beta$, giving information on the higher twist contributions in that region, for example by restricting the analysis to lower $x_{I\!P}$.

## Acknowledgements

For comments, corrections, cross-checks and code, thanks to Markus Diehl, Joel Feltesse, Max Klein and Frank-Peter Schilling!

# Diffractive Dijet Production at HERA

*A. Bruni*[1], *M. Klasen*[2,3], *G. Kramer*[3] *and S. Schätzel*[4]
[1] INFN Bologna, Via Irnerio 46, 40156 Bologna, Italy
[2] Laboratoire de Physique Subatomique et de Cosmologie, Université Joseph Fourier/CNRS-IN2P3, 53 Avenue des Martyrs, 38026 Grenoble, France
[3] II. Inst. für Theoret. Physik, Universität Hamburg, Luruper Chaussee 149, 22761 Hamburg, Germany
[4] DESY FLC, Notkestr. 85, 22607 Hamburg, Germany

**Abstract**
We present recent experimental data from the H1 and ZEUS Collaborations at HERA for diffractive dijet production in deep-inelastic scattering (DIS) and photoproduction and compare them with next-to-leading order (NLO) QCD predictions using diffractive parton densities. While good agreement is found for DIS, the dijet photoproduction data are overestimated by the NLO theory, showing that factorization breaking occurs at this order. While this is expected theoretically for resolved photoproduction, the fact that the data are better described by a global suppression of direct *and* resolved contribution by about a factor of two comes as a surprise. We therefore discuss in some detail the factorization scheme and scale dependence between direct and resolved contributions and propose a new factorization scheme for diffractive dijet photoproduction.

## 1 Introduction

It is well known that in high-energy deep-inelastic $ep$-collisions a large fraction of the observed events are diffractive. These events are defined experimentally by the presence of a forward-going system $Y$ with four-momentum $p_Y$, low mass $M_Y$ (in most cases a single proton and/or low-lying nucleon resonances), small momentum transfer squared $t = (p - p_Y)^2$, and small longitudinal momentum transfer fraction $x_{I\!P} = q(p - p_Y)/qp$ from the incoming proton with four-momentum $p$ to the system $X$ (see Fig. 1).

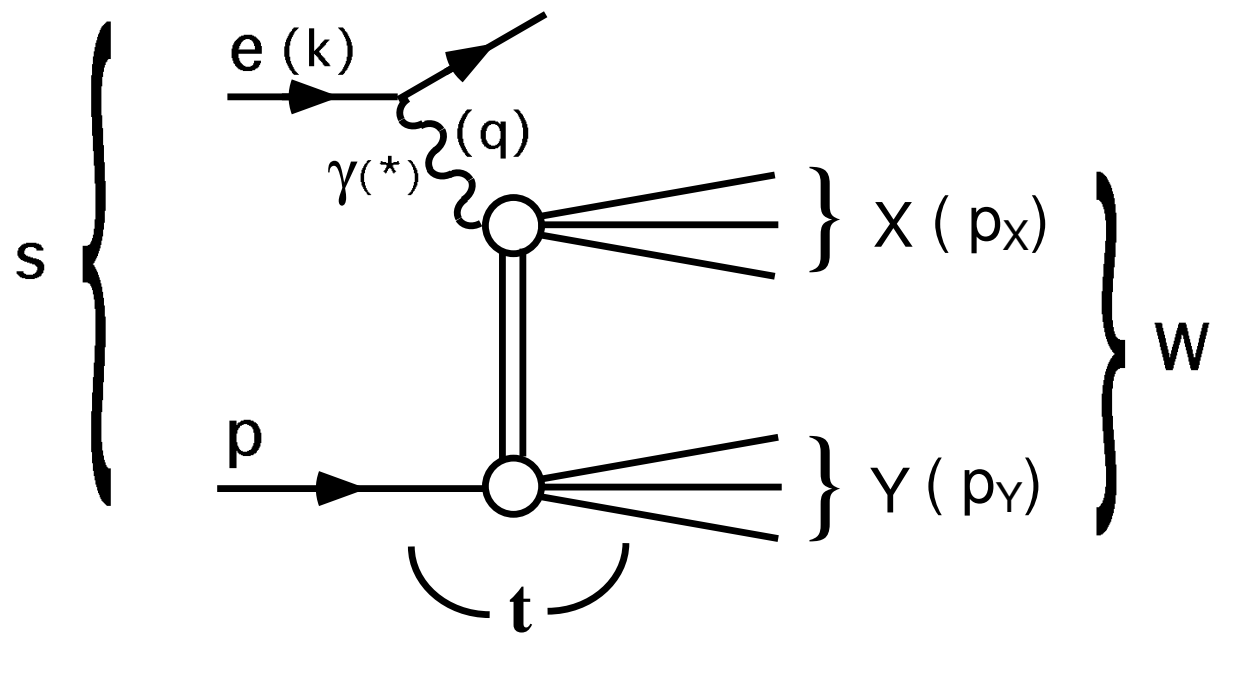

**Fig. 1:** Diffractive scattering process $ep \to eXY$, where the hadronic systems $X$ and $Y$ are separated by the largest rapidity gap in the final state.

The presence of a hard scale, as for example the photon virtuality $Q^2 = -q^2$ in deep-inelastic scattering (DIS) or the large transverse jet momentum $p_T^*$ in the photon-proton centre-of-momentum frame, should then allow for calculations of the production cross section for the central system $X$ with the known methods of perturbative QCD. Under this assumption, the cross section for the inclusive production of two jets, $e + p \to e + 2\,\text{jets} + X' + Y$, can be predicted from the well-known formulæ for jet production in non-diffractive $ep$ collisions, where in the convolution of the partonic cross section with the parton distribution functions (PDFs) of the proton the latter ones are replaced by the diffractive PDFs. In the simplest approximation, they are described by the exchange of a single, factorizable pomeron/Regge-pole.

The diffractive PDFs have been determined by the H1 Collaboration at HERA from high-precision inclusive measurements of the DIS process $ep \to eXY$ using the usual DGLAP evolution equations in leading order (LO) and next-to-leading order (NLO) and the well-known formula for the inclusive cross section as a convolution of the inclusive parton-level cross section with the diffractive PDFs [1]. For a similar analysis of the inclusive measurements of the ZEUS Collaboration see [2,3]. A longer discussion of the extraction of diffractive PDFs can also be found in these proceedings [4] and in [5]. For inclusive diffractive DIS it has been proven by Collins that the formula referred to above is applicable without additional corrections and that the inclusive jet production cross section for large $Q^2$ can be calculated in terms of the same diffractive PDFs [6]. The proof of this factorization formula, usually referred to as the validity of QCD factorization in hard diffraction, may be expected to hold for the direct part of photoproduction ($Q^2 \simeq 0$) or low-$Q^2$ electroproduction of jets [6]. However, factorization does not hold for hard processes in diffractive hadron-hadron scattering. The problem is that soft interactions between the ingoing two hadrons and their remnants occur in both the initial and final state. This agrees with experimental measurements at the Tevatron [7]. Predictions of diffractive dijet cross sections for $p\bar{p}$ collisions as measured by CDF using the same PDFs as determined by H1 [1] overestimate the measured cross section by up to an order of magnitude [7]. This suppression of the CDF cross section can be explained by considering the rescattering of the two incoming hadron beams which, by creating additional hadrons, destroy the rapidity gap [8].

Processes with real photons ($Q^2 \simeq 0$) or virtual photons with fixed, but low $Q^2$ involve direct interactions of the photon with quarks from the proton as well as resolved photon contributions, leading to parton-parton interactions and an additional remnant jet coming from the photon (for a review see [9]). As already said, factorization should be valid for direct interactions as in the case of DIS, whereas it is expected to fail for the resolved process similar as in the hadron-hadron scattering process. In a two-channel eikonal model similar to the one used to calculate the suppression factor in hadron-hadron processes [8], introducing vector-meson dominated photon fluctuations, a suppression by about a factor of three for resolved photoproduction at HERA is predicted [10]. Such a suppression factor has recently been applied to diffractive dijet photoproduction [11,12] and compared to preliminary data from H1 [13] and ZEUS [14]. While at LO no suppression of the resolved contribution seemed to be necessary, the NLO corrections increase the cross section significantly, showing that factorization breaking occurs at this order at least for resolved photoproduction and that a suppression factor $R$ must be applied to give a reasonable description of the experimental data.

As already mentioned elsewhere [11,12], describing the factorization breaking in hard photoproduction as well as in electroproduction at very low $Q^2$ [15] by suppressing the resolved contribution only may be problematic. An indication for this is the fact that the separation between the direct and the resolved process is uniquely defined only in LO. In NLO these two processes are related. The separation depends on the factorization scheme and the factorization scale $M_\gamma$. The sum of both cross sections is the only physically relevant cross section, which is approximately independent of the factorization scheme and scale [16]. As demonstrated in Refs. [11, 12] multiplying the resolved cross section with the suppression factor $R = 0.34$ destroys the correlation of the $M_\gamma$-dependence between the direct and resolved part, and the sum of both parts has a stronger $M_\gamma$-dependence than for the unsuppressed case ($R = 1$), where the $M_\gamma$-dependence of the NLO direct cross section is compensated to a high degree against the $M_\gamma$-dependence of the LO resolved part.

In the second Section of this contribution, we present the current experimental data from the H1 and ZEUS Collaborations on diffractive dijet production in DIS and photoproduction and compare these data to theoretical predictions at NLO for two different scenarios: suppression of only the resolved part by a factor $R = 0.34$ as expected from LO theory and proposed in [8], and equal suppression of all direct and resolved contributions by a factor $R = 0.5$, which appears to describe the data better phenomenologically. This motivates us to investigate in the third Section the question whether certain parts of the direct contribution might break factorization as well and therefore need a suppression factor.

The introduction of the resolved cross section is dictated by perturbation theory. At NLO, collinear singularities arise from the photon initial state, which are absorbed at the factorization scale into the photon PDFs. This way the photon PDFs become $M_\gamma$-dependent. The equivalent $M_\gamma$-dependence, just with the opposite sign, is left in the NLO corrections to the direct contribution. With this knowledge, it is obvious that we can obtain a physical cross section at NLO, *i.e.* the superposition of the NLO direct and LO resolved cross section, with a suppression factor $R < 1$ and no $M_\gamma$-dependence left, if we also multiply the $\ln M_\gamma$-dependent term of the NLO correction to the direct contribution with the same suppression factor as the resolved cross section. We are thus led to the theoretical conclusion that, contrary to what one may expect, not *all* parts of the direct contribution factorize. Instead, the *initial state* singular part appearing beyond LO breaks factorization even in direct photoproduction, presumably through soft gluon attachments between the proton and the collinear quark-antiquark pair emerging from the photon splitting. This would be in agreement with the general remarks about initial state singularities in Ref. [6].

In the third Section of this contribution, we present the special form of the $\ln M_\gamma$-term in the NLO direct contribution and demonstrate that the $M_\gamma$-dependence of the physical cross section cancels to a large extent in the same way as in the unsuppressed case ($R = 1$). These studies can be done for photoproduction ($Q^2 \simeq 0$) as well as for electroproduction with fixed, small $Q^2$. Since in electroproduction the initial-state singularity in the limit $Q^2 \to 0$ is more directly apparent than for the photoproduction case, we shall consider in this contribution the low-$Q^2$ electroproduction case just for demonstration. This diffractive dijet cross section has been calculated recently [15]. A consistent factorization scheme for low-$Q^2$ virtual photoproduction has been defined and the full (direct and resolved) NLO corrections for inclusive dijet production have been calculated in [17]. In this work we adapt this inclusive NLO calculational framework to diffractive dijet production at low-$Q^2$ in the same way as in [15], except that we multiply the $\ln M_\gamma$-dependent terms as well as the resolved contributions with the same suppression factor $R = 0.34$, as an example, as in our earlier work [11, 12, 15]. The exact value of this suppression factor may change in the future, when better data for photoproduction and low-$Q^2$ electroproduction have been analyzed. We present the $\ln M_\gamma$-dependence of the partly suppressed NLO direct and the fully suppressed NLO resolved cross section $\mathrm{d}\sigma/\mathrm{d}Q^2$ and their sum for the lowest $Q^2$ bin, before we give a short summary in section 4.

## 2 Comparison of H1 and ZEUS Data with NLO Theory Predictions

In this Section, diffractive PDFs [1–3] extracted from diffractive structure function data are used in NLO calculations to test factorisation in diffractive dijet production. Dijet production is directly sensitive to the diffractive gluon (Fig. 2) whereas in inclusive measurements the gluon is determined from scaling violations.

### 2.1 Diffractive Dijet Production in DIS

H1 has measured the cross sections for dijet production [13] in the kinematic range $Q^2 > 4$ GeV$^2$, $165 < W < 242$ GeV (photon-proton centre-of-mass energy) and $x_{I\!P} < 0.03$. Jets are identified using the inclusive $k_T$ cluster algorithm and selected by requiring $E_T^{*,\mathrm{jet}}(1,2) > 5, 4$ GeV and $-3 < \eta^*_{\mathrm{jet}} < 0$.[1] NLO predictions have been obtained by interfacing the H1 diffractive PDFs with the DISENT program [18]. The renormalisation and factorisation scales were set to the transverse energy of the leading parton jet. The NLO parton jet cross sections have been corrected for hadronisation effects using the leading order (LO) generator RAPGAP [19] with parton showers and the Lund string fragmentation model. Comparisons of the DISENT and RAPGAP predictions with the measured cross section differential in $z_{I\!P}^{\rm jets}$, an estimator for the fraction of the momentum of the diffractive exchange entering the hard scatter, are shown in Fig. 3a. The inner band around the NLO calculation indicates the $\approx 20\%$ uncertainty resulting from a variation of the renormalisation scale by factors 0.5 and 2. The uncertainty in the diffractive PDFs is not shown. Within this additional uncertainty, which is large at high $z_{I\!P}^{\rm jets}$, the cross section is well described. The cross section differential in $\log_{10}(x_{I\!P})$, $p_T^{\rm jet1}$, and $Q^2$ is shown in Figs. 3b and 4. All distributions are well described and QCD factorisation is therefore in good agreement with dijet production in diffractive DIS.

[1]The '$*$' denotes variables in the photon-proton centre-of-mass system.

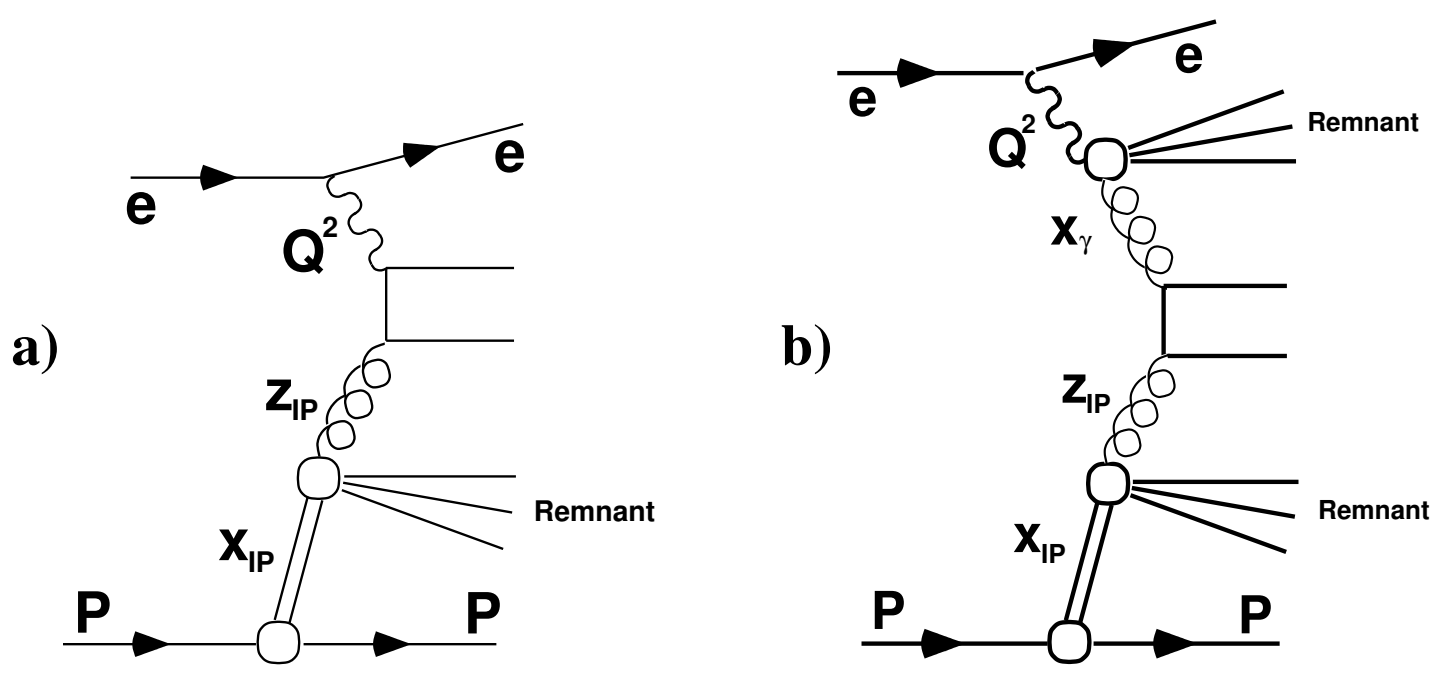

**Fig. 2:** Example processes for a) direct photon and b) resolved photon interactions.

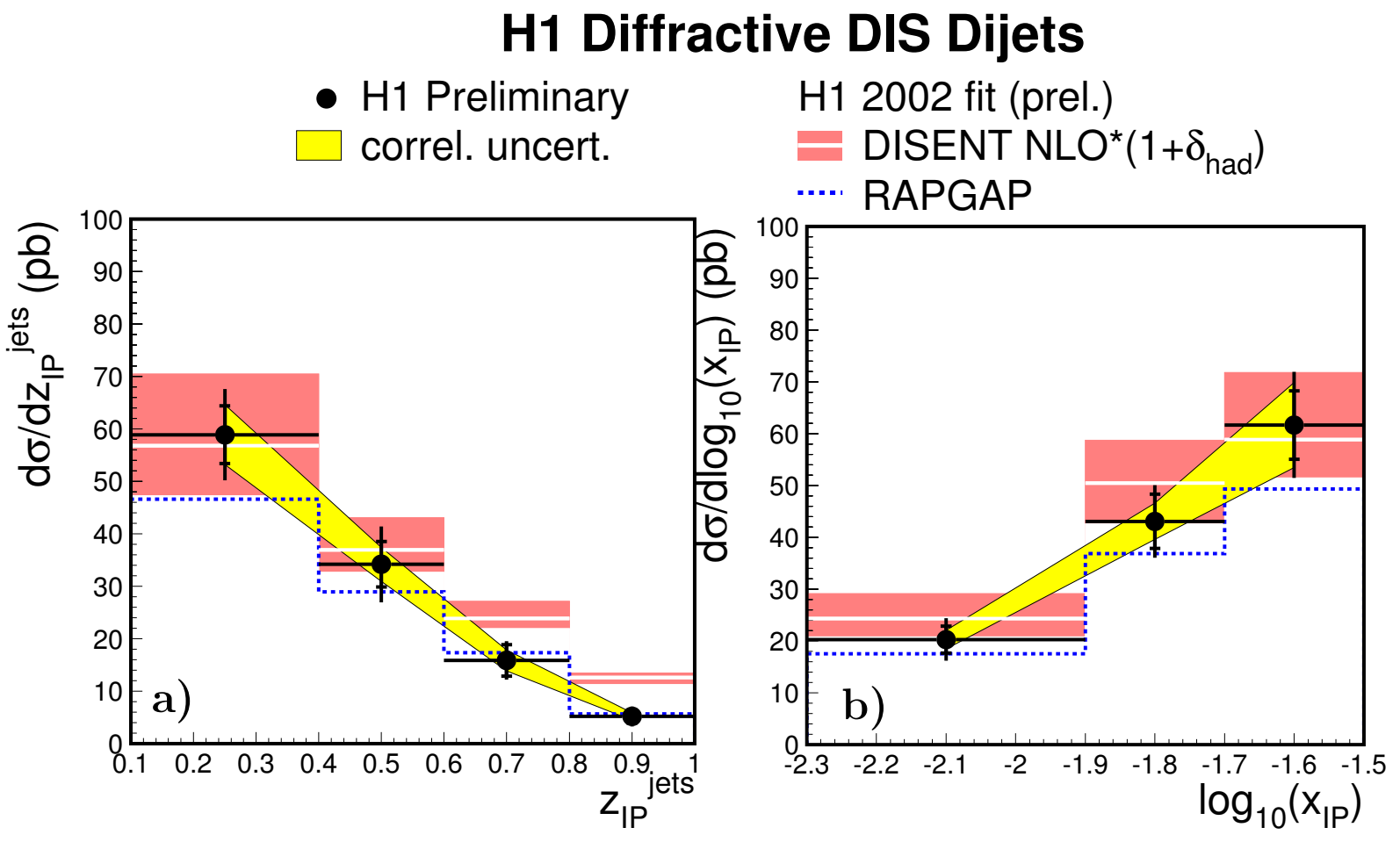

**Fig. 3:** Diffractive DIS dijet cross sections compared with a NLO prediction based on diffractive PDFs and with RAPGAP.

Similar results are presented by ZEUS [20]; the dijet cross sections have been measured in the kinematic range $5 < Q^2 < 100$ GeV$^2$, $100 < W < 200$ GeV, $x_{I\!P} < 0.03$. The jets were identified using the inclusive $k_T$ algorithm in the $\gamma p$ frame and required to satisfy $E_T^{*,\rm jet}(1,2) > 5,4$ GeV and $-3.5 < \eta^*_{\rm jet} < 0.0$. NLO predictions have been obtained with the DISENT program interfaced to three different sets of diffractive PDFs: from fit to H1 data [1], from fit to the ZEUS $M_X$ data (GLP) [3] and from fit to ZEUS LPS and $F_2^{D,charm}$ data [2]. Comparisons of the DISENT predictions with the measured cross section differential in $E_T^{*,\rm jet}$, $\eta^*_{\rm jet}$, $z_{I\!P}^{\rm jets}$ and $x_\gamma^{obs}$ are shown in Fig. 5. The $20 - 30\%$ uncertainty in the NLO calculations resulting from a variation of the renormalisation and factorisation scales is not shown. Within the experimental and QCD scale uncertainties, the predictions based on the H1 and ZEUS-LPS PDFs give a good description of the dijet cross section. The normalisation of the prediction using the GLP fit is substantially lower than those from the other two sets of PDFs. For ZEUS, the difference observed between the three sets may be interpreted as an estimate of the uncertainty associated with the diffractive PDFs and with the definition of the diffractive region. The dijet data could be included in future fits in order to better constrain the diffractive gluon distribution.

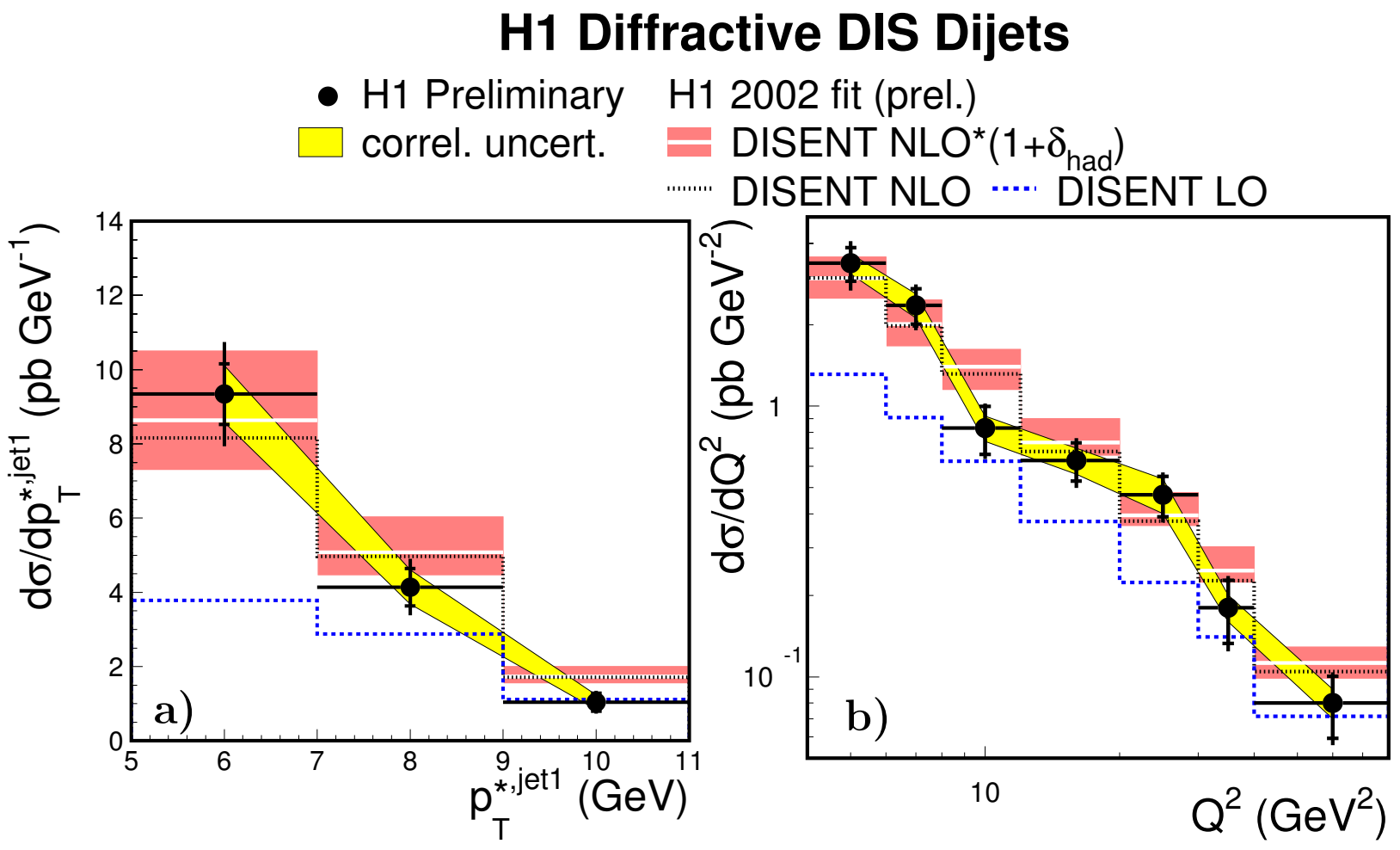

**Fig. 4:** Diffractive DIS dijet cross sections compared with a NLO prediction based on diffractive PDFs.

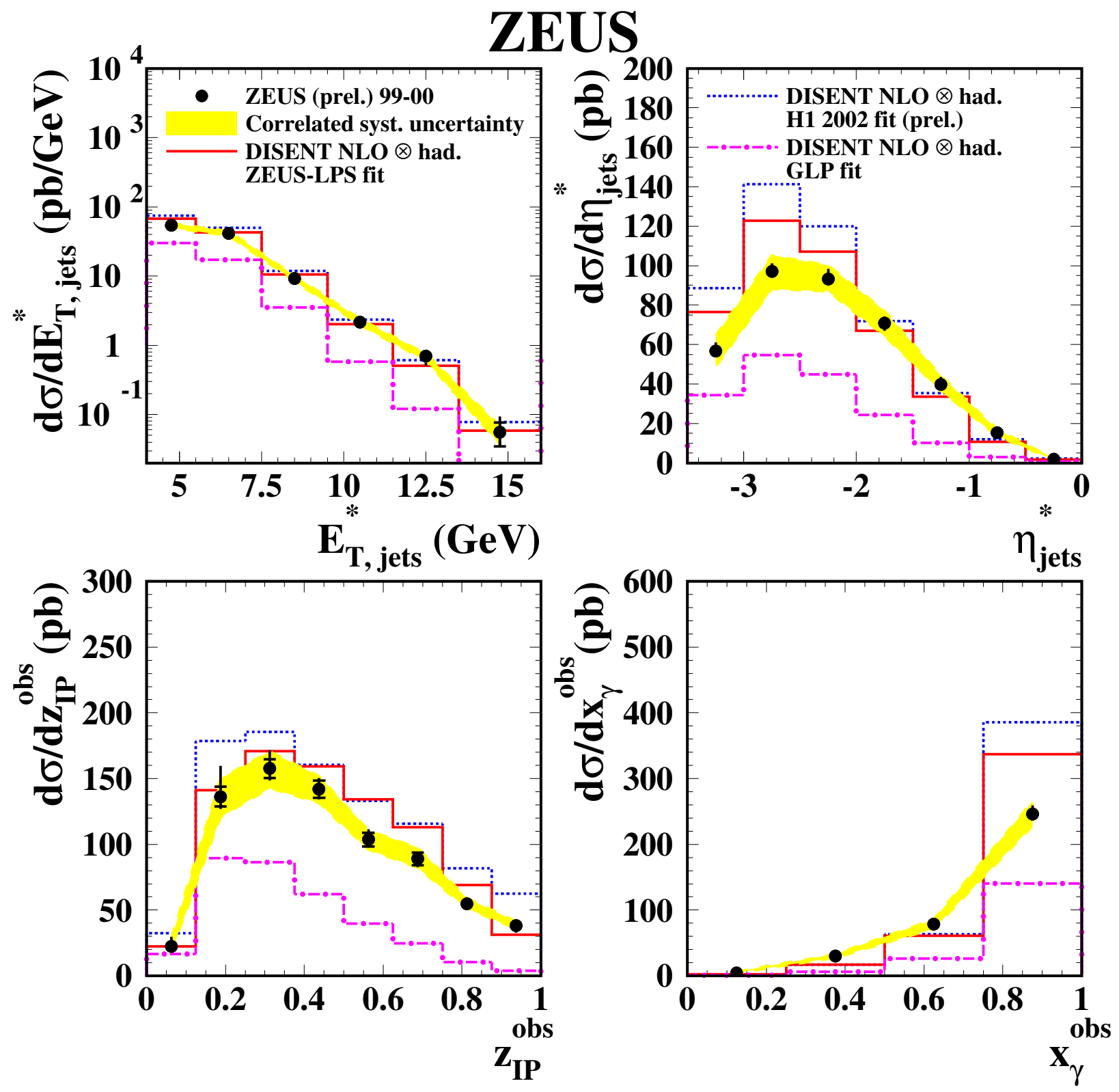

**Fig. 5:** Diffractive DIS dijet cross sections compared with NLO predictions based on three sets of diffractive PDFs.

Within the experimental and theoretical uncertainties and assuming the H1 diffractive PDFs, factorisation is in good agreement with diffractive $D^*$ production [21, 22] in the DIS kinematic region.

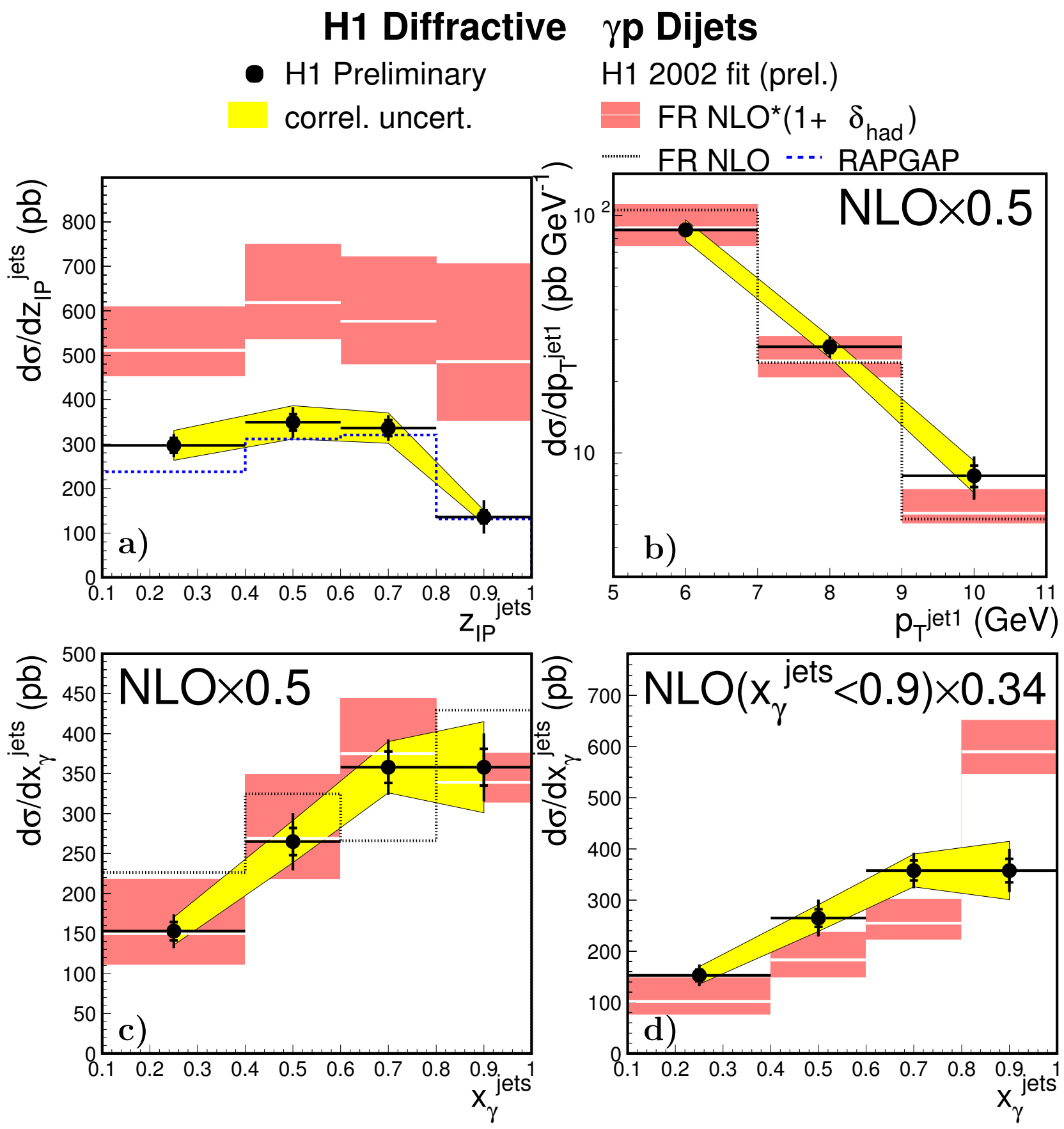

**Fig. 6:** a) Diffractive dijet photoproduction cross section differential in $z_{I\!P}^{\text{jets}}$ compared with a NLO prediction based on diffractive PDFs and RAPGAP. b)-d): Cross section differential in $p_T^{\text{jet1}}$and $x_\gamma^{\text{jets}}$, compared with the NLO prediction modified as follows: in b) and c) the calculation is scaled by a global factor 0.5 whereas in d) only the "resolved" part is scaled by 0.34.

## 2.2 Diffractive Photoproduction of Dijets

In photoproduction, a sizeable contribution to the cross section is given by resolved photon processes (Fig. 2b) in which only a fraction $x_\gamma < 1$ of the photon momentum enters the hard scatter. The photoproduction dijet cross section measured by H1 ($Q^2 < 0.01$ GeV$^2$, $165 < W < 242$ GeV, $x_{I\!P} < 0.03$, $E_T^{\text{jet}}(1,2) > 5, 4$ GeV, $-1 < \eta_{\text{jet}} < 2$, inclusive $k_T$ algorithm) is shown in Fig. 6 [13]. NLO predictions have been obtained with the Frixione *et al.* program [23] interfaced to the H1 diffractive PDFs. The parton jet calculation is corrected for hadronisation effects using RAPGAP. The cross section differential in $z_{I\!P}^{\text{jets}}$ is shown in Fig. 6a. The calculation lies a factor $\approx 2$ above the data. Fig. 6b and 6c show the cross section as a function of $p_T^{\text{jet1}}$ and $x_\gamma^{\text{jets}}$ and the NLO predictions have been scaled down by a factor 0.5. Good agreement is obtained for this global suppression. In Fig. 6d, only the "resolved" part for which $x_\gamma^{\text{jets}} < 0.9$ at the parton level is scaled by a factor 0.34. This factor was proposed by Kaidalov *et al.* [10] for the suppression of the resolved part in LO calculations. The calculation for $x_\gamma^{\text{jets}} > 0.9$ is left unscaled. This approach is clearly disfavoured.

The ZEUS measurement [24] ($Q^2 < 0.01$ GeV$^2$, $x_{I\!P} < 0.025$, $0.2 < y < 0.85$, $E_T^{\text{jet}}(1,2) > 7.5, 6.5$ GeV, $-1.5 < \eta < 1.5$, inclusive $k_T$ algorithm) is shown in Figs. 7 and 8 separately for samples enriched in "direct" ($x_\gamma^{\text{jets}} > 0.75$) and "resolved" ($x_\gamma^{\text{jets}} < 0.75$) processes, respectively. The NLO [12] prediction using the H1 diffractive PDFs is also presented corrected for hadronization effects and with the "resolved" part scaled by the factor 0.34. No evidence is observed for a suppression of resolved photon processes relative to direct photon processes in any particular kinematic region.

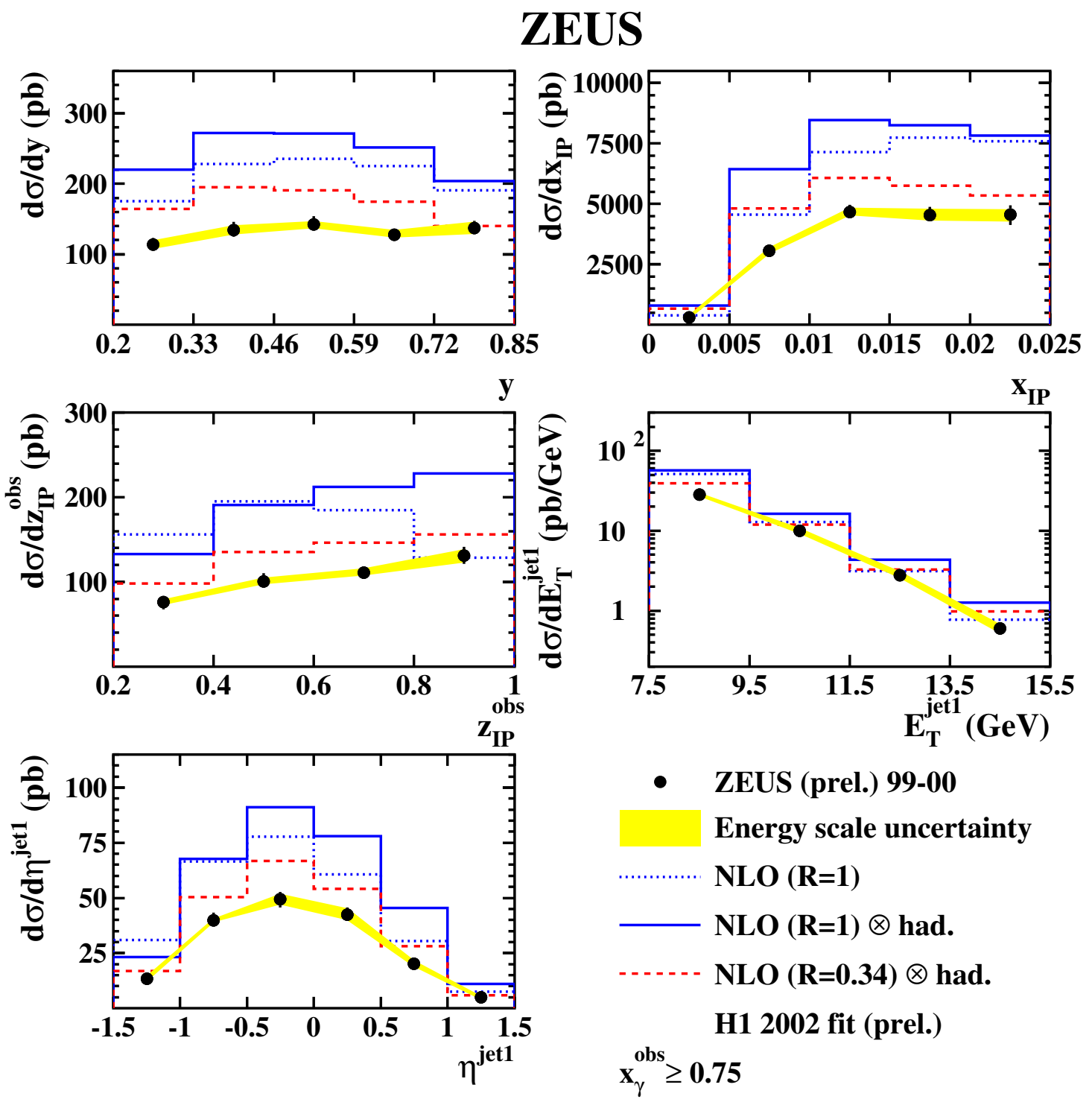

**Fig. 7:** Direct enriched photoproduction. Diffractive dijet photoproduction cross section differential in $y$, $x_{I\!P}$, $z_{I\!P}^{\rm jets}$, $E_T^{\rm jet^1}$ and $\eta_{\rm jet_1}$ compared with a NLO prediction based on diffractive PDFs. The NLO prediction is also presented corrected for hadronization effects and with the "resolved" part scaled by 0.34.

Diffractive dijet photoproduction is overestimated by calculations based on PDFs which give a good description of the diffractive DIS data. Factorisation is broken in photoproduction relative to DIS by a factor $\approx 0.5$ with no observed dependence on $x_\gamma$ or other kinematic variables.

## 3 Factorization and its Breaking in Diffractive Dijet Production

The fact that equal suppression of direct *and* resolved photoproduction by a factor $R = 0.5$ appears to describe the H1 and ZEUS data better phenomenologically motivates us to investigate in some detail the question whether certain parts of the direct contribution might break factorization as well and therefore need a suppression factor. These studies can be done for photoproduction ($Q^2 \simeq 0$) as well as for electroproduction with fixed, small $Q^2$. Since in electroproduction the initial-state singularity in the limit $Q^2 \to 0$ is more directly apparent than for the photoproduction case, we shall consider in this contribution the low-$Q^2$ electroproduction case just for demonstration.

A factorization scheme for virtual photoproduction has been defined and the full NLO corrections for inclusive dijet production have been calculated in [17]. They have been implemented in the NLO Monte Carlo program JETVIP [25] and adapted to diffractive dijet production in [15]. The subtraction term, which is absorbed into the PDFs of the virtual photon $f_{a/\gamma}(x_\gamma, M_\gamma)$, can be found in [26]. The main term is proportional to $\ln(M_\gamma^2/Q^2)$ times the splitting function

$$P_{q_i \leftarrow \gamma}(z) = 2N_c Q_i^2 \frac{z^2 + (1-z)^2}{2}, \tag{1}$$

where $z = p_1 p_2 / p_0 q \in [x; 1]$ and $Q_i$ is the fractional charge of the quark $q_i$. $p_1$ and $p_2$ are the momenta of the two outgoing jets, and $p_0$ and $q$ are the momenta of the ingoing parton and virtual photon, respec-

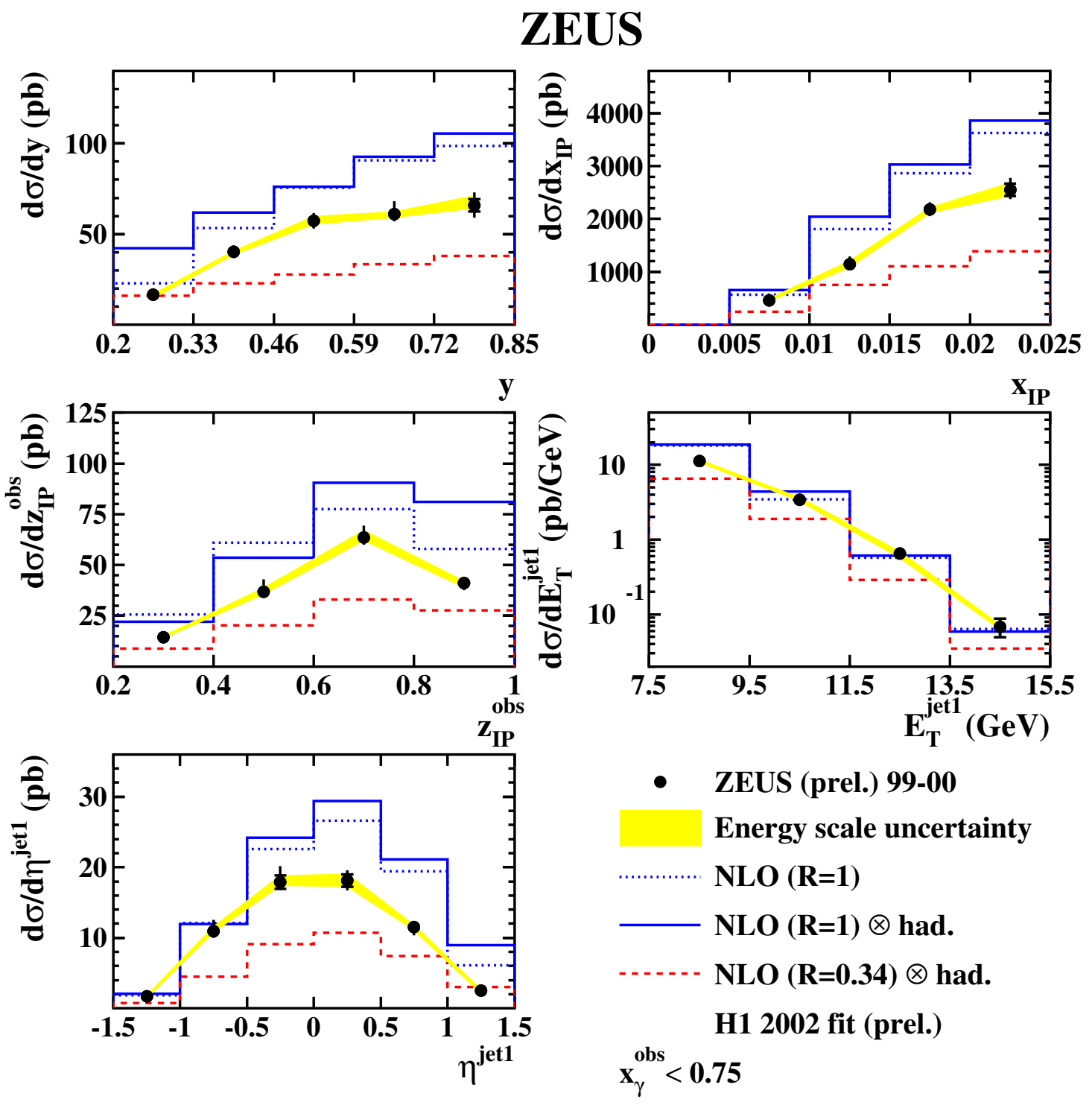

**Fig. 8:** Resolved enriched photoproduction. Diffractive dijet photoproduction cross section differential in $y$, $x_{I\!P}$, $z_{I\!P}^{\rm jets}$, $E_T^{\rm jet^1}$ and $\eta_{\rm jet_1}$ compared with a NLO prediction based on diffractive PDFs. The NLO prediction is also presented corrected for hadronization effects and with the "resolved" part scaled by 0.34.

tively. Since $Q^2 = -q^2 \ll M_\gamma^2$, the subtraction term is large and is therefore resummed by the DGLAP evolution equations for the virtual photon PDFs. After this subtraction, the finite term $M(Q^2)_{\overline{\rm MS}}$, which remains in the matrix element for the NLO correction to the direct process [17], has the same $M_\gamma$-dependence as the subtraction term, *i.e.* $\ln M_\gamma$ is multiplied with the same factor. As already mentioned, this yields the $M_\gamma$-dependence before the evolution is turned on. In the usual non-diffractive dijet photoproduction these two $M_\gamma$-dependences cancel, when the NLO correction to the direct part is added to the LO resolved cross section [16]. Then it is obvious that the approximate $M_\gamma$-independence is destroyed, if the resolved cross section is multiplied by a suppression factor $R$ to account for the factorization breaking in the experimental data. To remedy this deficiency, we propose to multiply the $\ln M_\gamma$-dependent term in $M(Q^2)_{\overline{\rm MS}}$ with the same suppression factor as the resolved cross section. This is done in the following way: we split $M(Q^2)_{\overline{\rm MS}}$ into two terms using the scale $p_T^*$ in such a way that the term containing the slicing parameter $y_s$, which was used to separate the initial-state singular contribution, remains unsuppressed. In particular, we replace the finite term after the subtraction by

$$
\begin{aligned}
M(Q^2,R)_{\overline{\rm MS}} &= \left[-\frac{1}{2N_c}P_{q_i\leftarrow\gamma}(z)\ln\left(\frac{M_\gamma^2 z}{p_T^{*2}(1-z)}\right)+\frac{Q_i^2}{2}\right]R \\
&\quad -\frac{1}{2N_c}P_{q_i\leftarrow\gamma}(z)\ln\left(\frac{p_T^{*2}}{zQ^2+y_s s}\right), \qquad (2)
\end{aligned}
$$

where $R$ is the suppression factor. This expression coincides with the finite term after subtraction (see Ref. [26]) for $R = 1$, as it should, and leaves the second term in Eq. (2) unsuppressed. In Eq. (2) we have suppressed in addition to $\ln(M_\gamma^2/p_T^{*2})$ also the $z$-dependent term $\ln(z/(1-z))$, which is specific to the $\overline{\rm MS}$ subtraction scheme as defined in [17]. The second term in Eq. (2) must be left in its original form,

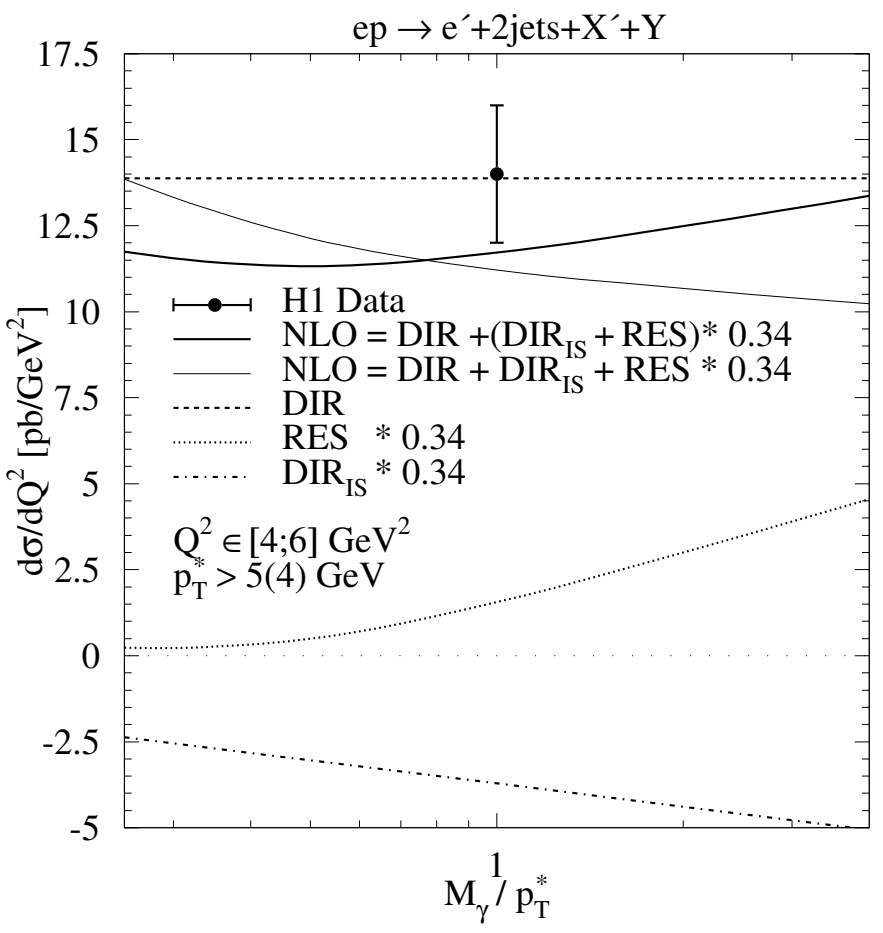

**Fig. 9:** Photon factorization scale dependence of resolved and direct contributions to $\mathrm{d}\sigma/\mathrm{d}Q^2$ together with their weighted sums for (i) suppression of the resolved cross section and for (ii) additional suppression of DIR$_{\mathrm{IS}}$, using SaS1D virtual photon PDFs [30].

*i.e.* being unsuppressed, in order to achieve the cancellation of the slicing parameter ($y_s$) dependence of the complete NLO correction in the limit of very small $Q^2$ or equivalently very large $s$. It is clear that the suppression of this part of the NLO correction to the direct cross section will change the full cross section only very little as long as we choose $M_\gamma \simeq p_T^*$. The first term in Eq. (2), which has the suppression factor $R$, will be denoted by $\mathrm{DIR}_{\mathrm{IS}}$ in the following.

To study the left-over $M_\gamma$-dependence of the physical cross section, we have calculated the diffractive dijet cross section with the same kinematic constraints as in the H1 experiment [27]. Jets are defined by the CDF cone algorithm with jet radius equal to one and asymmetric cuts for the transverse momenta of the two jets required for infrared stable comparisons with the NLO calculations [28]. The original H1 analysis actually used a symmetric cut of 4 GeV on the transverse momenta of both jets [29]. The data have, however, been reanalyzed for asymmetric cuts [27].

For the NLO resolved virtual photon predictions, we have used the PDFs SaS1D [30] and transformed them from the DIS$_\gamma$ to the $\overline{\mathrm{MS}}$ scheme as in Ref. [17]. If not stated otherwise, the renormalization and factorization scales at the pomeron and the photon vertex are equal and fixed to $p_T^* = p_{T,jet1}^*$. We include four flavors, *i.e.* $n_f = 4$ in the formula for $\alpha_s$ and in the PDFs of the pomeron and the photon. With these assumptions we have calculated the same cross section as in our previous work [15]. First we investigated how the cross section $\mathrm{d}\sigma/\mathrm{d}Q^2$ depends on the factorization scheme of the PDFs for the virtual photon, *i.e.* $\mathrm{d}\sigma/\mathrm{d}Q^2$ is calculated for the choice SaS1D and SaS1M. Here $\mathrm{d}\sigma/\mathrm{d}Q^2$ is the full cross section (sum of direct and resolved) integrated over the momentum and rapidity ranges as in the H1 analysis. The results, shown in Fig. 2 of Ref. [26], demonstrate that the choice of the factorization scheme of the virtual photon PDFs has negligible influence on $\mathrm{d}\sigma/\mathrm{d}Q^2$ for all considered $Q^2$. The predictions agree reasonably well with the preliminary H1 data [27].

We now turn to the $M_\gamma$-dependence of the cross section with a suppression factor for DIR$_{\mathrm{IS}}$. To show this dependence for the two suppression mechanisms, (i) suppression of the resolved cross section only and (ii) additional suppression of the DIR$_{\mathrm{IS}}$ term as defined in Eq. (2) in the NLO correction of the direct cross section, we consider $\mathrm{d}\sigma/\mathrm{d}Q^2$ for the lowest $Q^2$-bin, $Q^2 \in [4,6]$ GeV$^2$. In Fig. 9, this cross section is plotted as a function of $\xi = M_\gamma/p_T^*$ in the range $\xi \in [0.25;4]$ for the cases (i) (light full curve) and (ii) (full curve). We see that the cross section for case (i) has an appreciable $\xi$-dependence in the considered $\xi$ range of the order of $40\%$, which is caused by the suppression of the resolved contribution only. With the additional suppression of the DIR$_{\mathrm{IS}}$ term in the direct NLO correction, the $\xi$-dependence of $\mathrm{d}\sigma/\mathrm{d}Q^2$ is reduced to approximately less than $20\%$, if we compare the maximal and the minimal value of $\mathrm{d}\sigma/\mathrm{d}Q^2$ in the considered $\xi$ range. The remaining $\xi$-dependence is caused by the NLO corrections to the suppressed resolved cross section and the evolution of the virtual photon PDFs. How the compensation of the $M_\gamma$-dependence between the suppressed resolved contribution and the suppressed direct NLO term works in detail is exhibited by the dotted and dashed-dotted curves in Fig. 9. The suppressed resolved term increases and the suppressed direct NLO term decreases by approximately the same amount with increasing $\xi$. In addition we show also $\mathrm{d}\sigma/\mathrm{d}Q^2$ in the DIS theory, *i.e.* without subtraction of any $\ln Q^2$ terms (dashed line). Of course, this cross section must be independent of $\xi$. This prediction agrees very well with the experimental point, whereas the result for the subtracted and suppressed theory (full curve) lies slightly below. We notice, that for $M_\gamma = p_T^*$ the additional suppression of $\mathrm{DIR_{IS}}$ has only a small effect. It increases $\mathrm{d}\sigma/\mathrm{d}Q^2$ by $5\%$ only.

## 4 Summary

Experimental data from the H1 and ZEUS Collaborations at HERA for diffractive dijet production in DIS and photoproduction have been compared with NLO QCD predictions using diffractive parton densities from H1 and ZEUS. While good agreement was found for DIS assuming the H1 diffractive PDFs, the dijet photoproduction data are overestimated by the NLO theory, showing that factorization breaking occurs at this order. While this is expected theoretically for resolved photoproduction, the fact that the data are better described by a global suppression of direct *and* resolved contribution by about a factor of two has come as a surprise. We have therefore discussed in some detail the factorization scheme and scale dependence between direct and resolved contributions and proposed a new factorization scheme for diffractive dijet photoproduction.

## Acknowledgements

M.K. thanks the II. Institute for Theoretical Physics at the University of Hamburg for hospitality while this work was being finalized.

# Effect of absorptive corrections on inclusive parton distributions

G. Watt[a], A.D. Martin[b], M.G. Ryskin[b,c]
[a] Deutsches Elektronen-Synchrotron DESY, 22607 Hamburg, Germany
[b] Institute for Particle Physics Phenomenology, University of Durham, DH1 3LE, UK
[c] Petersburg Nuclear Physics Institute, Gatchina, St. Petersburg, 188300, Russia

**Abstract**
We study the effect of absorptive corrections due to parton recombination on the parton distributions of the proton. A more precise version of the GLRMQ equations, which account for non-linear corrections to DGLAP evolution, is derived. An analysis of HERA $F_2$ data shows that the small-$x$ gluon distribution is enhanced at low scales when the absorptive effects are included, such that there is much less need for a negative gluon distribution at 1 GeV.

## 1 Parton recombination at small $x$

At very small values of $x$ it is expected that the number density of partons within the proton becomes so large that they begin to recombine with each other. This phenomenon of parton recombination is also referred to as absorptive corrections, non-linear effects, screening, shadowing, or unitarity corrections, all leading to saturation. The first perturbative QCD (pQCD) calculations describing the fusion of two Pomeron ladders into one were made by Gribov-Levin-Ryskin (GLR) [1] and by Mueller-Qiu (MQ) [2]. The GLRMQ equations add an extra non-linear term, quadratic in the gluon density, to the usual DGLAP equations for the gluon and sea-quark evolution. The evolution of the gluon distribution is then given by

$$\frac{\partial xg(x,Q^2)}{\partial \ln Q^2} = \frac{\alpha_S}{2\pi} \sum_{a'=q,g} P_{ga'} \otimes a' \; - \; \frac{9}{2} \frac{\alpha_S^2(Q^2)}{R^2\, Q^2} \int_x^1 \frac{\mathrm{d}x'}{x'} \left[x' g(x',Q^2)\right]^2 , \qquad (1)$$

where $R \sim 1$ fm is of the order of the proton radius. The GLRMQ equations account for all 'fan' diagrams, that is, all possible $2 \to 1$ ladder recombinations, in the double leading logarithmic approximation (DLLA) which resums all powers of the parameter $\alpha_S \ln(1/x) \ln(Q^2/Q_0^2)$.

There has been much recent theoretical activity in deriving (and studying) more precise non-linear evolution equations, such as the Balitsky-Kovchegov (BK) and Jalilian-Marian–Iancu–McLerran–Weigert–Leonidov–Kovner (JIMWLK) equations (see [3] for a review). Note that the BK and JIMWLK equations are both based on BFKL evolution. However, for the most relevant studies in the HERA and LHC domain ($x \gtrsim 10^{-4}$), the predominant theoretical framework is collinear factorisation with DGLAP-evolved parton distribution functions (PDFs). At very small values of $x$ it might be expected that the DGLAP approximation would break down, since large $\alpha_S \ln(1/x)$ (BFKL) terms would appear in the perturbation series in addition to the $\alpha_S \ln(Q^2/Q_0^2)$ terms resummed by DGLAP evolution. However, it turns out that the resummed NLL BFKL calculations of the gluon splitting function $P_{gg}$ [4] and the gluon transverse momentum distribution [5] are rather close to the DGLAP calculations. Moreover, the convolution $P_{gg} \otimes g(x, Q^2)$ coincides with the NNLO DGLAP result and is close to the NLO DGLAP result for $x \gtrsim 10^{-4}$ [6]. Hence, in the analysis of current data, it is reasonable to ignore BFKL effects.

If recombination effects are significant, it is therefore important that they be incorporated into the global DGLAP parton analyses which determine the PDFs from deep-inelastic scattering (DIS) and related hard-scattering data. Such a programme, based on GLRMQ evolution (which accounts for gluon-induced screening only), was implemented some years ago [7], before the advent of HERA. The input gluon and sea-quark distributions were *assumed* to have a small-$x$ behaviour of the form $xg, xS \sim x^{-0.5}$ at an input scale of $Q_0^2 = 4$ GeV$^2$. The inclusion of shadowing effects, both in the form of the input PDFs and in the GLRMQ evolution, was found to significantly *decrease* the size of the small-$x$ gluon distribution in comparison with the result with no absorptive corrections. A crucial observation is that, at that time (1990), $F_2$ data were only available for $x_B \geq 0.07$, and so these results were largely dependent on the theoretical assumptions made for the starting distributions. However, with HERA, we now have $F_2$ data down to $x_B \sim 10^{-4}$ or less, and so the PDFs at small $x$ can be determined directly from the HERA data.

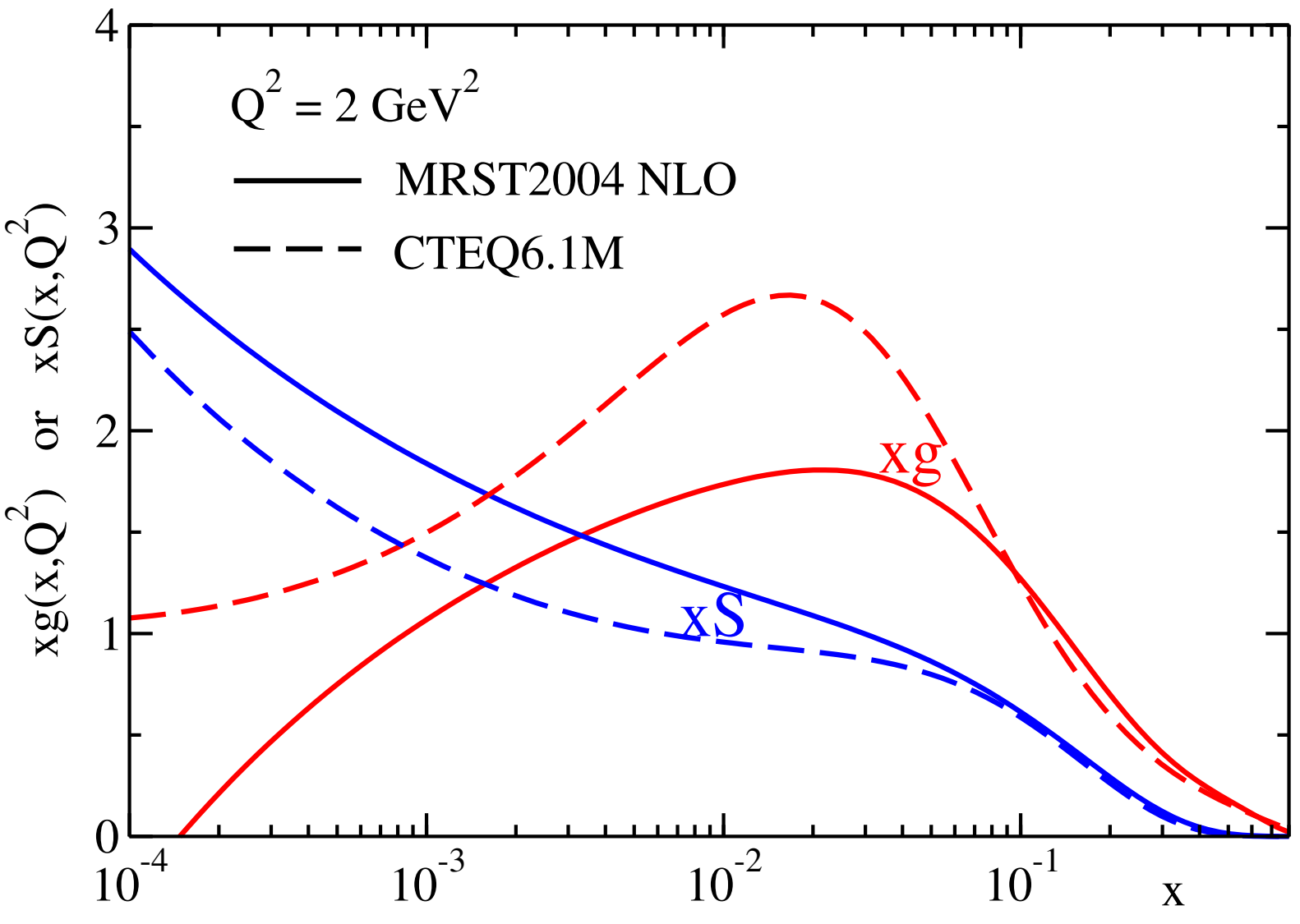

**Fig. 1:** The behaviour of the gluon and sea-quark distributions at $Q^2 = 2$ GeV$^2$ found in the MRST2004 NLO and CTEQ6.1M global analyses. The valence-like behaviour of the gluon is evident.

In fact, the advent of HERA data has led to a puzzling behaviour of the small-$x$ gluon and sea-quark PDFs at low scales $Q^2$. If we write $xg \sim x^{-\lambda_g}$ and $xS \sim x^{-\lambda_S}$, then the expectation of Regge theory is that $\lambda_g = \lambda_S = \lambda_{\rm soft}$ for low scales $Q \lesssim Q_0 \sim 1$ GeV, where $\lambda_{\rm soft} \simeq 0.08$ [8] is the power of $s$ obtained from fitting soft hadron data. At higher $Q \gtrsim 1$ GeV, QCD evolution should take over, increasing the powers $\lambda_g$ and $\lambda_S$. However, the current MRST2004 NLO [9] and CTEQ6.1M [10] PDF sets exhibit a very different behaviour at low scales from that theoretically expected; see Fig. 1. In fact, the MRST group has found that a negative input gluon distribution at $Q_0 = 1$ GeV is required in all their NLO DGLAP fits since MRST2001 [11]. The CTEQ group, who take a slightly higher input scale of $Q_0 = 1.3$ GeV, also find a negative gluon distribution when evolving backwards to 1 GeV.

Since data at small $x_B$ now exist, the introduction of the absorptive corrections is expected to *increase* the size of the input gluon distribution at small $x$ to maintain a satisfactory fit to the data. To understand this, note that the negative non-linear term in the GLRMQ equation (1) slows down the evolution. Therefore, it is necessary to start with a *larger* small-$x$ gluon distribution at low scales $Q \sim Q_0$ to achieve the *same* PDFs at larger scales required to describe the data. If the non-linear term is neglected, the input small-$x$ gluon distribution is forced to be artificially small in order to *mimic* the neglected screening corrections.

We have anticipated that the introduction of absorptive corrections will *enhance*[1] the small-$x$ gluon at low scales, and hence could possibly avoid what appears to be anomalous behaviour at small $x$. Thus, here, we perform such a study using an abridged version of the MRST2001 NLO analysis [11], improving on our previous analysis [13]. First, we derive a more precise form of the GLRMQ equations.

[1]Eskola *et al.* [12] have found that taking input gluon and sea-quark distributions at $Q^2 = 1.4$ GeV$^2$, then evolving upwards with the GLRMQ equations based on LO DGLAP evolution, improves the agreement with $F_2$ data at small $x_B$ and low $Q^2$ compared to the standard CTEQ sets, and leads to an enhanced small-$x$ gluon distribution for $Q^2 \lesssim 10$ GeV$^2$. Note, however, that there is a large NLO correction to the splitting function $P_{qg}$ which changes completely the relationship between the quark and gluon distributions, and so weakens the conclusion of Ref. [12].

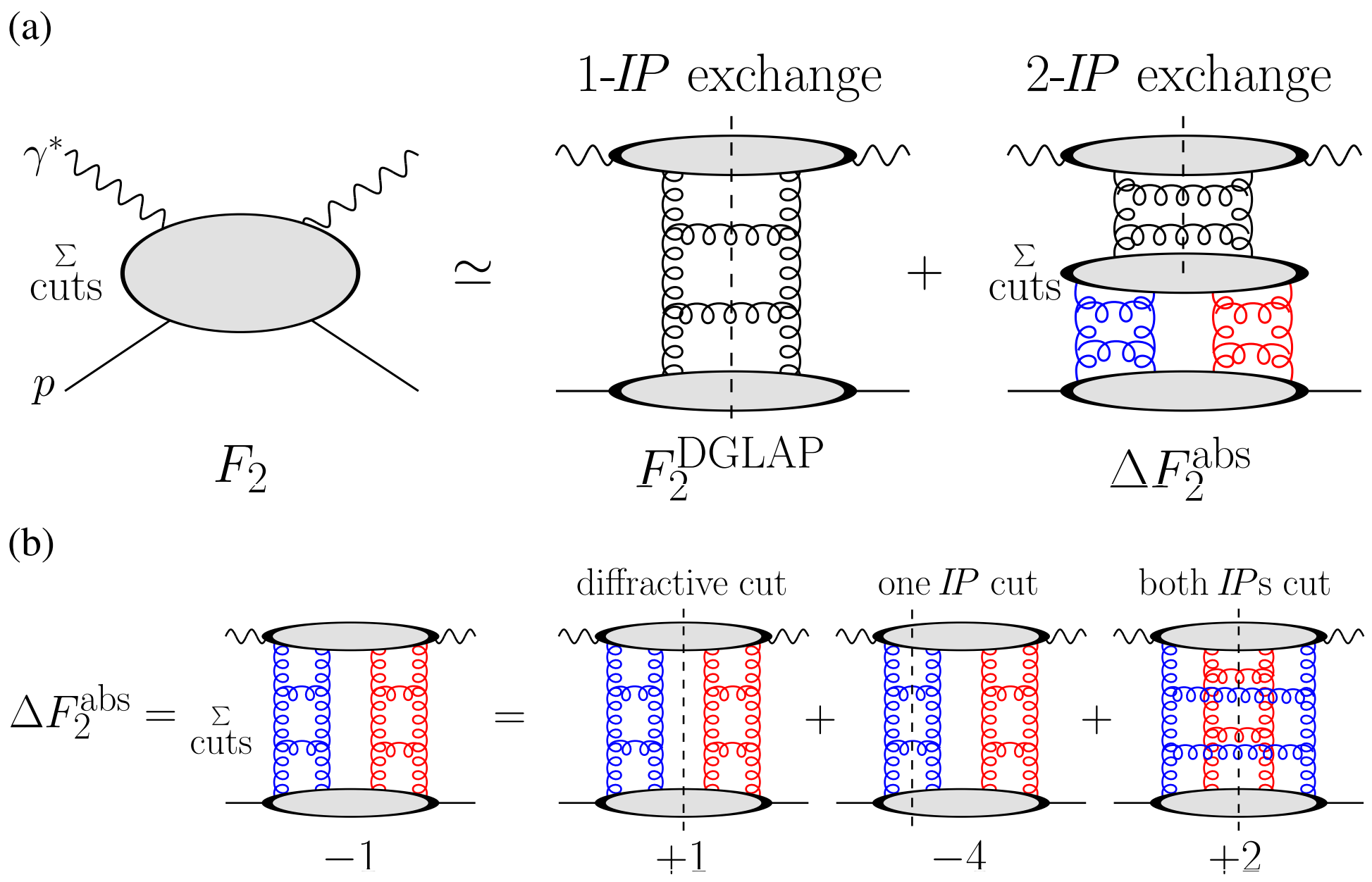

**Fig. 2:** (a) Absorptive corrections to $F_2$ due to the $2 \to 1$ Pomeron contribution. (b) Application of the AGK cutting rules. For simplicity, the upper parton ladder, shown in the right-hand diagram of (a), is hidden inside the upper blob in each diagram of (b).

## 2 Non-linear evolution from diffractive DIS

The inclusive proton structure function, $F_2(x_B, Q^2)$, as measured by experiment, can be approximately written as a sum of the single Pomeron exchange (DGLAP) contribution and absorptive corrections due to a $2 \to 1$ Pomeron merging; see Fig. 2(a). That is,

$$F_2(x_B, Q^2) = F_2^{\rm DGLAP}(x_B, Q^2) + \Delta F_2^{\rm abs}(x_B, Q^2). \tag{2}$$

In computing $\Delta F_2^{\rm abs}$ we need to sum over all possible cuts. The Abramovsky-Gribov-Kancheli (AGK) cutting rules [14] were originally formulated in Reggeon field theory but have been shown to also hold in pQCD [15]. Application of the AGK rules gives the result that relative contributions of $+1$, $-4$, and $+2$ are obtained according to whether neither Pomeron, one Pomeron, or both Pomerons are cut; see Fig. 2(b). Therefore, the sum over cuts is equal to *minus* the diffractive cut and so the absorptive corrections can be computed from a calculation of the $t$-integrated diffractive structure function $F_2^{\rm D(3)}(x_{I\!P}, \beta, Q^2)$, where $\beta \equiv x_B/x_{I\!P}$ and $x_{I\!P}$ is the fraction of the proton's momentum transferred through the rapidity gap.

The pQCD description of $F_2^{\rm D(3)}$ is described in [16, 17], and in a separate contribution to these proceedings. Working in the fixed flavour number scheme (FFNS), it can be written as

$$F_2^{\rm D(3)} = \underbrace{F_{2,\rm non-pert.}^{\rm D(3)}}_{\text{soft Pomeron}} + \underbrace{F_{2,\rm pert.}^{\rm D(3)} + F_{2,\rm direct}^{{\rm D(3)},c\bar{c}} + F_{L,\rm tw.4}^{\rm D(3)}}_{\text{QCD Pomeron}}, \tag{3}$$

apart from the secondary Reggeon contribution. The separation between the soft Pomeron and QCD Pomeron is provided by a scale $\mu_0 \sim 1$ GeV. For simplicity, we take $\mu_0$ to be the same as the scale $Q_0$ at which the input PDFs are taken in the analysis of $F_2$ data, so $\mu_0 = Q_0 = 1$ GeV, the value used in the MRST2001 NLO analysis [11]. The contribution to the absorptive corrections arising from the soft Pomeron contribution of (3) is already included in the input PDFs, therefore

$$\Delta F_2^{\rm abs} = -\frac{1}{1 - f_{\rm p.diss.}} \int_{x_B}^{1} {\rm d}x_{I\!P}\; \left[F_{2,\rm pert.}^{\rm D(3)} + F_{2,\rm direct}^{{\rm D(3)},c\bar{c}} + F_{L,\rm tw.4}^{\rm D(3)}\right], \tag{4}$$

where $f_{\text{p.diss.}}$ is the fraction of diffractive events in which the proton dissociates. In practice, we take $f_{\text{p.diss.}} = 0.5$ and take an upper limit of 0.1 instead of 1 for $x_{I\!P}$ in (4).[2]

First consider the contribution to (4) from the $F_{2,\text{pert.}}^{\text{D(3)}}$ term.[3] It corresponds to a $2 \to 1$ Pomeron merging with a cut between the two Pomeron ladders and can be written as

$$F_{2,\text{pert.}}^{\text{D(3)}}(x_{I\!P}, \beta, Q^2) = \sum_{a=q,g} C_{2,a} \otimes a_{\text{pert.}}^{\text{D}}, \tag{5}$$

where $C_{2,a}$ are the *same* coefficient functions as in inclusive DIS. The diffractive PDFs, $a^{\text{D}} = zq^{\text{D}}$ or $zg^{\text{D}}$, where $z \equiv x/x_{I\!P}$, satisfy an *inhomogeneous* evolution equation [17]:

$$a_{\text{pert.}}^{\text{D}}(x_{I\!P}, z, Q^2) = \int_{\mu_0^2}^{Q^2} \frac{\mathrm{d}\mu^2}{\mu^2}\, f_{I\!P}(x_{I\!P}; \mu^2)\, a^{I\!P}(z, Q^2; \mu^2) \tag{6}$$

$$\Longrightarrow \frac{\partial a_{\text{pert.}}^{\text{D}}}{\partial \ln Q^2} = \frac{\alpha_S}{2\pi} \sum_{a'=q,g} P_{aa'} \otimes a'^{\text{D}}_{\text{pert.}} \;+\; P_{aI\!P}(z)\, f_{I\!P}(x_{I\!P}; Q^2). \tag{7}$$

Here, $f_{I\!P}(x_{I\!P}; Q^2)$ is the perturbative Pomeron flux factor,

$$f_{I\!P}(x_{I\!P}; \mu^2) = \frac{1}{x_{I\!P} B_D} \left[ R_g \frac{\alpha_S(\mu^2)}{\mu}\, x_{I\!P} g(x_{I\!P}, \mu^2) \right]^2. \tag{8}$$

The diffractive slope parameter $B_D$ comes from the $t$-integration, while the factor $R_g$ accounts for the skewedness of the proton gluon distribution [19]. There are similar contributions from (light) sea quarks, where $g$ in (8) is replaced by $S \equiv 2(\bar{u} + \bar{d} + \bar{s})$, together with an interference term. A sum over all three contributions is implied in (6) and in the second term of (7). The Pomeron PDFs in (6), $a^{I\!P}(z, Q^2; \mu^2)$, are evolved using NLO DGLAP from a starting scale $\mu^2$ up to $Q^2$, taking the input distributions to be LO Pomeron-to-parton splitting functions, $a^{I\!P}(z, \mu^2; \mu^2) = P_{aI\!P}(z)$ [17].

From (2),

$$a(x, Q^2) = a^{\text{DGLAP}}(x, Q^2) + \Delta a^{\text{abs}}(x, Q^2), \tag{9}$$

where $a(x, Q^2) = xg(x, Q^2)$ or $xS(x, Q^2)$, and

$$\Delta a^{\text{abs}}(x, Q^2) = -\frac{1}{1 - f_{\text{p.diss.}}} \int_x^1 \mathrm{d}x_{I\!P}\; a_{\text{pert.}}^{\text{D}}(x_{I\!P}, x/x_{I\!P}, Q^2). \tag{10}$$

Differentiating (9) with respect to $Q^2$ gives the evolution equations for the (inclusive) gluon and sea-quark PDFs:

$$\boxed{\frac{\partial a(x, Q^2)}{\partial \ln Q^2} = \frac{\alpha_S}{2\pi} \sum_{a'=q,g} P_{aa'} \otimes a' \;-\; \frac{1}{1 - f_{\text{p.diss.}}} \int_x^1 \mathrm{d}x_{I\!P}\; P_{aI\!P}(x/x_{I\!P})\, f_{I\!P}(x_{I\!P}; Q^2).} \tag{11}$$

Thus (11) is a more precise version of the GLRMQ equations (1), which goes beyond the DLLA and accounts for sea-quark recombination as well as gluon recombination. Consider the recombination of gluons into gluons, for example, in the DLLA where $x \ll x_{I\!P}$, then $P_{gI\!P} = 9/16$ [17]. Taking $R_g = 1$ and $f_{\text{p.diss.}} = 0$, then (11) becomes

$$\frac{\partial xg(x, Q^2)}{\partial \ln Q^2} = \frac{\alpha_S}{2\pi} \sum_{a'=q,g} P_{ga'} \otimes a' \;-\; \frac{9}{16} \frac{\alpha_S^2(Q^2)}{B_D\, Q^2} \int_x^1 \frac{\mathrm{d}x_{I\!P}}{x_{I\!P}} \left[ x_{I\!P} g(x_{I\!P}, Q^2) \right]^2. \tag{12}$$

[2]The value of $f_{\text{p.diss.}} = 0.5$ is justified by a ZEUS comparison [18] of proton-tagged diffractive DIS data with data which allowed proton dissociation up to masses of 6 GeV, where $f_{\text{p.diss.}} = 0.46 \pm 0.11$ was obtained.

[3]The other two contributions to (4) are described after (13).

Comparing to (1) this is simply the GLRMQ equation with $R^2 = 8B_D$. For numerical results we take $B_D = 6$ (4) GeV$^{-2}$ for light (charm) quarks, which would correspond to $R = \sqrt{8B_D} = 1.4$ (1.1) fm.

The procedure for incorporating absorptive corrections into a (NLO) global parton analysis (in the FFNS) is as follows:

1. Parameterise the $x$ dependence of the input PDFs at a scale $Q_0 \sim 1$ GeV.
2. Evolve the PDFs $xg(x,Q^2)$ and $xS(x,Q^2)$ using the non-linear evolution equation (11). (The non-singlet distributions are evolved using the usual linear DGLAP equations.)
3. Compute

$$F_2(x_B,Q^2) = \sum_{a=q,g} C_{2,a} \otimes a \; - \; \frac{1}{1-f_{\rm p.diss.}} \int_{x_B}^1 {\rm d}x_{I\!P} \; \left[ F^{{\rm D}(3),c\bar{c}}_{2,{\rm direct}} + F^{{\rm D}(3)}_{L,{\rm tw.4}} \right], \tag{13}$$

and compare to data. Here, the two terms inside the square brackets are beyond collinear factorisation, that is, they cannot be written as a convolution of coefficient functions with the PDFs. The first term inside the square brackets corresponds to the process $\gamma^* I\!P \to c\bar{c}$. The second term corresponds to the process $\gamma^* I\!P \to q\bar{q}$, for light quarks with a longitudinally polarised photon. These contributions are calculated as described in Ref. [17].

As usual, these three steps should be repeated with the parameters of the input PDFs adjusted until an optimal fit is obtained. This procedure is our recommended way of accounting for absorptive corrections in a global parton analysis. However, in practice, available NLO DGLAP evolution codes, such as the QCDNUM [20] program, are often regarded as a 'black box', and it is not trivial to modify the usual linear DGLAP evolution to the non-linear evolution of (11). Therefore, we adopt an alternative iterative procedure which avoids the explicit implementation of non-linear evolution, but which is equivalent to the above procedure.

## 3 Effect of absorptive corrections on inclusive PDFs

We model our analysis of HERA $F_2$ data [21] on the MRST2001 NLO analysis [11], which was the first in which a negative gluon distribution was required at the input scale of $Q_0 = 1$ GeV. (The more recent MRST sets have not changed substantially at small $x$.) We apply cuts $x_B \leq 0.01$, $Q^2 \geq 2$ GeV$^2$, and $W^2 \geq 12.5$ GeV$^2$, leaving 280 data points. The input gluon and sea-quark distributions are taken to be

$$xg(x,Q_0^2) \;=\; A_g\, x^{-\lambda_g}(1-x)^{3.70}(1+\epsilon_g\sqrt{x}+\gamma_g x) \;-\; A_-\, x^{-\delta_-}(1-x)^{10}, \tag{14}$$

$$xS(x,Q_0^2) \;=\; A_S\, x^{-\lambda_S}(1-x)^{7.10}(1+\epsilon_S\sqrt{x}+\gamma_S x), \tag{15}$$

where the powers of the $(1-x)$ factors are taken from [11], together with the valence-quark distributions, $u_V$ and $d_V$, and $\Delta \equiv \bar{d} - \bar{u}$. The $A_g$ parameter is fixed by the momentum sum rule, while the other nine parameters are allowed to go free. Since we do not fit to DIS data with $x_B > 0.01$, we constrain the input gluon and sea-quark distributions, and their derivatives with respect to $x$, to agree with the MRST2001 NLO parton set [11] at $x = 0.2$. This is done by including the value of these MRST PDFs at $x = 0.2$, and their derivatives, as data points in the fit, with an error of 10% on both the value of the MRST PDFs and their derivatives. Therefore, the PDFs we obtain are not precisely constrained at large $x$, but this paper is primarily concerned with the small-$x$ behaviour of the PDFs.

The procedure we adopt is as follows:

(i) Start by performing a standard NLO DGLAP fit to $F_2$ data with no absorptive corrections.

(ii) Tabulate $\Delta F_2^{\rm abs}$, given by (4), and $\Delta a^{\rm abs}$, given by (10), using PDFs $g(x_{I\!P},\mu^2)$ and $S(x_{I\!P},\mu^2)$ obtained from the previous fit.

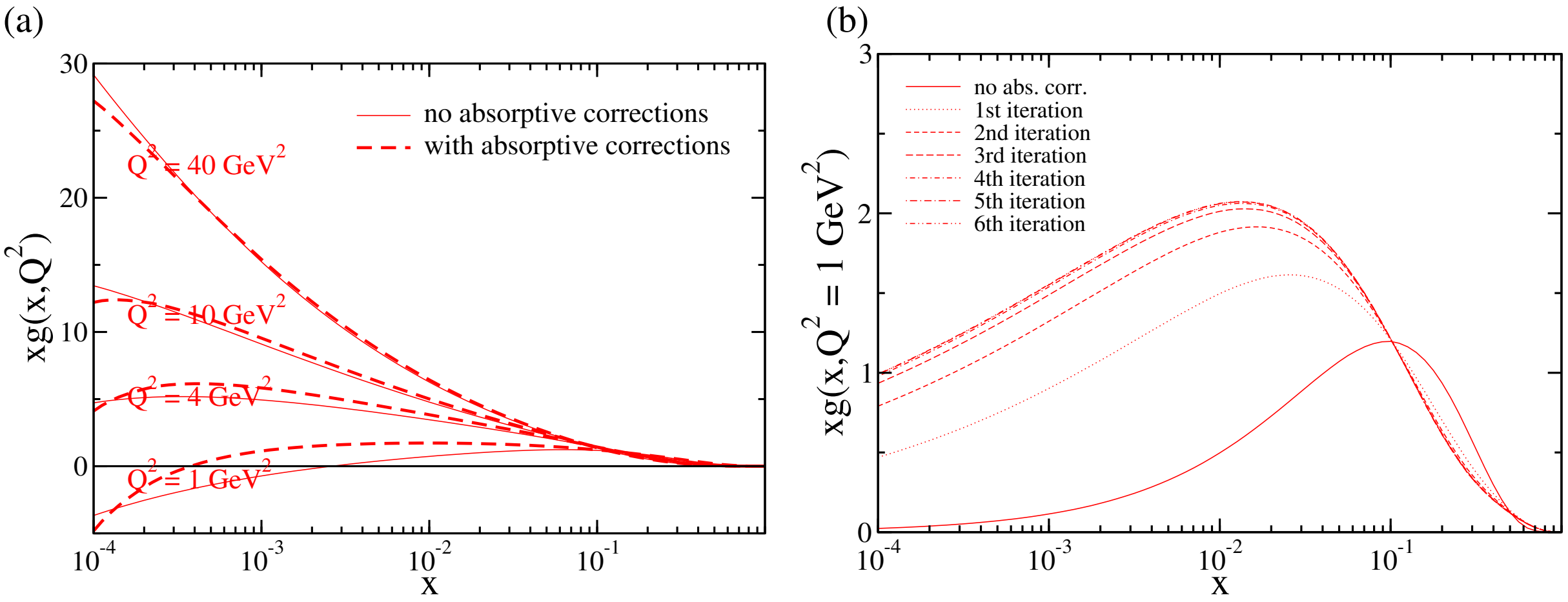

**Fig. 3:** (a) The gluon distribution obtained from fits to $F_2$ data, before and after absorptive corrections have been included. (b) The effect of successive iterations on the gluon distribution obtained from fits to $F_2$, taking a positive definite input gluon at 1 GeV. Each iteration introduces another level of $2 \to 1$ Pomeron mergings.

(iii) Perform a standard NLO DGLAP fit to 'corrected' data, $F_2^{\rm DGLAP} = F_2 - \Delta F_2^{\rm abs}$, to obtain PDFs $a^{\rm DGLAP}$. Then correct these PDFs to obtain $a = a^{\rm DGLAP} + \Delta a^{\rm abs}$. These latter PDFs $a$ then satisfy the non-linear evolution equations (11).

(iv) Go to (ii).

Each successive iteration of steps (ii) and (iii) introduces another level of $2 \to 1$ Pomeron mergings, so that eventually all the 'fan' diagrams are included, achieving the same effect as the procedure described at the end of Section 2.

Note that the correction to the PDFs, $a = a^{\rm DGLAP} + \Delta a^{\rm abs}$, in each step (iii), was omitted in our previous analysis [13]. Consequently, the effect of the absorptive corrections on the PDFs at large scales was overestimated. Also in [13], the known LO $P_{a\mathbb{P}}(z)$ were multiplied by free parameters ('K-factors'), determined from separate fits to diffractive DIS data, in an attempt to account for higher-order pQCD corrections to the LO Pomeron-to-parton splitting functions. However, since these K-factors took unreasonable values, with some going to zero, here we have chosen to fix them to 1. Therefore, the updated analysis, presented here, does not require a simultaneous fit to the diffractive DIS data.

In Fig. 3(a) we show the gluon distribution at scales $Q^2 = 1$, 4, 10, and 40 GeV$^2$ obtained from fits before and after absorptive corrections have been included. Both fits are almost equally good with $\chi^2$/d.o.f. values of 0.86 and 0.87 for the fits without and with absorptive corrections respectively. At low $Q^2$ the absorptive corrections give an increased gluon distribution at small $x$, apart from at $x \lesssim 10^{-4}$ where there are only a few data points and where additional absorptive effects (Pomeron loops) may become important. The non-linear term of (11) slows down the evolution, so that by 40 GeV$^2$ the two gluon distributions are roughly equal; see Fig. 3(a).

We repeated the fits without the negative term in the input gluon distribution, that is, without the second term in (14). When absorptive corrections were included, almost the same quality of fit was obtained ($\chi^2/\text{d.o.f.} = 0.90$), while without absorptive corrections the fit was slightly worse ($\chi^2/\text{d.o.f.} = 0.95$). We conclude that absorptive corrections lessen the need for a negative gluon distribution at $Q^2 = 1$ GeV$^2$. The gluon distributions obtained from six successive iterations of steps (ii) and (iii) above are shown in Fig. 3(b). The convergence is fairly rapid, with only the first three iterations having a significant effect, that is, the 'fan' diagrams which include $8 \to 4 \to 2 \to 1$ Pomeron mergings.

Although we have seen that the inclusion of absorptive corrections has reduced the need for a *negative* gluon, it has not solved the problem of the *valence-like* gluon. That is, the gluon distribution at low scales still decreases with decreasing $x$, whereas from Regge theory it is expected to behave as $xg \sim x^{-\lambda_{\rm soft}}$ with $\lambda_{\rm soft} \simeq 0.08$. We have studied several possibilities of obtaining a satisfactory fit with this behaviour [13]. The only modification which appears consistent with the data (and with the desired $\lambda_g = \lambda_S$ equality) is the inclusion of power-like corrections, specifically, a global shift in all scales by about 1 GeV$^2$. (Note that a similar shift in the scale is required in the dipole saturation model [22].) However, we do not have a solid theoretical justification for this shift. Therefore, a more detailed, and more theoretically-motivated, investigation of the effect of power corrections in DIS is called for.

# Multiple Scattering at HERA and at LHC - Remarks on the AGK Rules

*J. Bartels*
II. Institut für Theoretische Physik, Universität Hamburg
Luruper Chaussee 149, D-22761 Hamburg, Germany

**Abstract**
We summarize the present status of the AGK cutting rules in perturbative QCD. Particular attention is given to the application of the AGK analysis to multiple scattering in DIS at HERA and in $pp$ collisions at the LHC

## 1 Introduction

Multiple parton interactions play an important role both in electron proton scattering at HERA and in high energy proton proton collisions at the LHC. At HERA, the linear QCD evolution equations provides, for not too small $Q^2$, a good description of the $F_2$ data (and of the total $\gamma^* p$ cross section, $\sigma_{tot}^{\gamma^* p}$). This description corresponds to the emission of partons from a single chain (Fig.1a). However, at low $Q^2$ where the transition to nonperturbative strong interaction physics starts, this simple picture has to supplemented with corrections. First, there exists a class of models [1] which successfully describe this transition region; these models are based upon the idea of parton saturation: they assume the existence of multiple parton chains (Fig.1b) which interact with each other, and they naturally explain the observed scaling behavior, $F_2(Q^2, x) \approx F_2(Q^2/Q_s^2(x))$ with $Q_s^2(x) = Q_0^2(1/x)^\lambda$. Next, in the photoproduction region, $Q^2 \approx 0$, direct evidence for the presence of multiple interactions also comes from the analysis of final states [2]. A further strong hint at the presence of multi-chain configurations comes from the observation of a large fraction of diffractive final states in deep inelastic scattering at HERA. In the final states analysis of the linear QCD evolution equations, it is expected that the produced partons are not likely to come with large rapidity intervals between them. In the momentum-ordered single chain picture (Fig.1a), therefore, diffractive final states should be part of the initial conditions (inside the lower blob in Fig.1a), i.e. they should lie below the scale $Q_0^2$ which seperates the parton description from the nonperturbative strong interactions. This assignment of diffractive final states, however, cannot be complete. First, data have shown that the Pomeron which generates the rapidity gap in DIS diffraction is harder than in hadron - hadron scattering; furthermore, there are specific diffractive final states with momentum scales larger than $Q_0^2$, e.g. vector mesons built from heavy quarks and diffractive dijets (illustrated in Fig.2): the presence of such final states naturally requires corrections to the single chain picture (Fig.2b). From a $t$-channel point of view, both Fig.1b and Fig.2b belong to the same class of corrections, characterized by four gluon states in the $t$-channel.

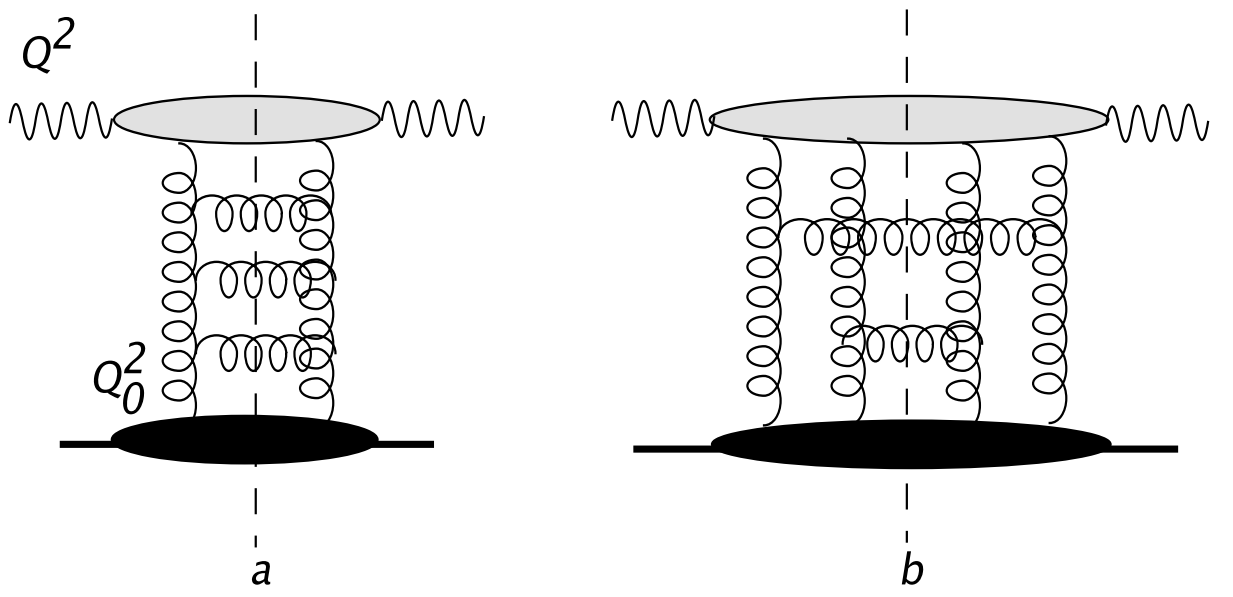

**Fig. 1:** Contributions to the total cross section $\sigma_{tot}^{\gamma^* p}$: (a) the single chain representing the linear QCD evolution equations; (b) gluon production from two different gluon chains.

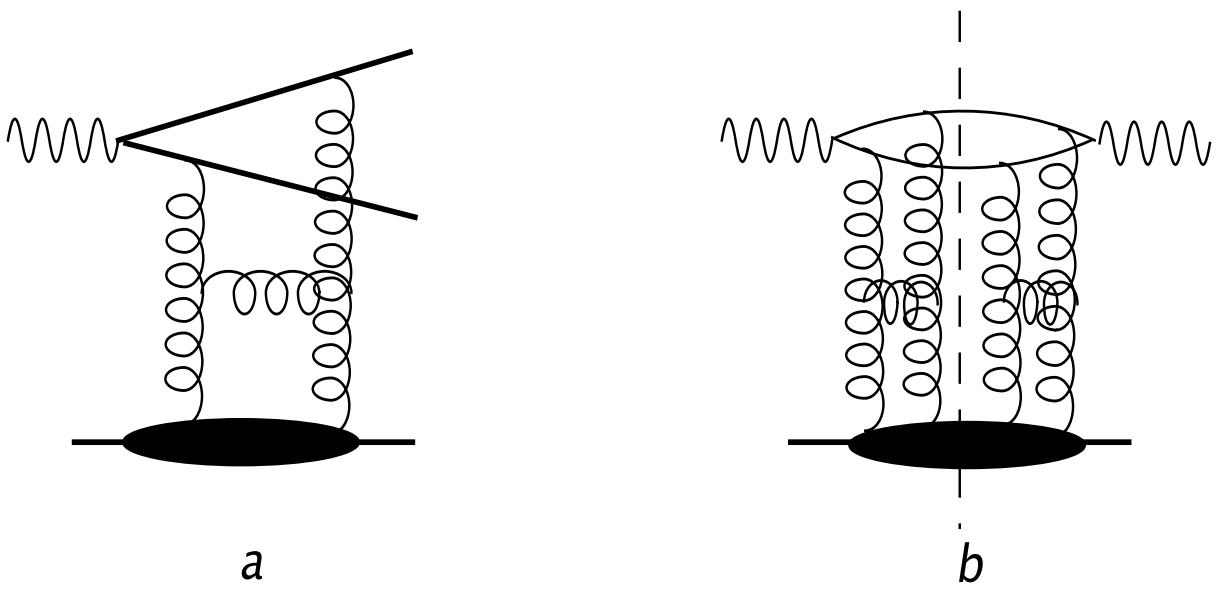

**Fig. 2:** Hard diffractive final states.(a) dijet production; (b) the diffractive cross section as $s$-channel discontinuity of a two-ladder diagram.

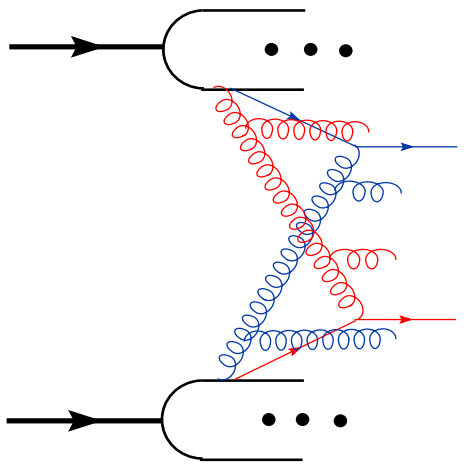

**Fig. 3:** Jet production in $pp$ collisions from two different parton chains

In proton-proton collisions corrections due to multiple interactions should be important in those kinematic regions where parton densities for small momentum fractions values and for not too large momentum scales are being probed, e.g. jet production near the forward direction. Another place could be the production of multijet final states (Fig.3): multiple jets may come from different parton chains, and these contributions may very well affect the background to new physics beyond the standard model. Moreover, the modelling of multijet configurations will be necessary for understanding the underlying event structure in $pp$ collidions [3].

From the point of view of collinear factorization, multiple interactions with momentum ordered parton chains are higher-twist effects, i.e they are suppressed by powers of the hard momentum scale. At small $x$, however, this suppression is compensated by powers of the large logarithms, $\ln 1/x$: multiple interactions, therefore, are mainly part of small-$x$ physics. In this kinematic region the Abramovsky-Gribov-Kanchelli (AGK) [4] rules can be applied to the analysis of multi-gluon chains, and it is the aim of this article to present a brief overview about the current status of the AGK rules in pQCD.

As we will discuss below, in the analysis of multiple parton chains the couplings of $n$ gluons to the proton play an essential role. Regge factorization suggests that these coupling should be universal, i.e. the couplings in $\gamma^* p$ collisions at HERA are the same as those in $pp$ scattering at the LHC. Therefore, a thorough analysis of the role of multiple interactions in deep inelastic electron-proton scattering at HERA should be useful for a solid understanding of the structure of events at the LHC.

## 2 Basics of the AGK cutting rules

The original AGK paper [4], which was written before the advent QCD, addresses the question how, in the optical theorem,

$$\sigma_{tot}^{pp} = \frac{1}{s} Im\, T_{2\to 2} = \sum_f \int d\Omega_f |T_{i\to f}|^2 \tag{1}$$

the presence of multi-Pomeron exchanges (Fig.4) in the total hadron-hadron cross section leads to observable effects in the final states (rhs of eq.(1)). Based upon a few model-independent assumptions on the couplings of multi-Pomeron exchanges to the proton, the authors derived simple 'cutting rules': different contributions to the imaginary part belong to different cuts across the multi-Pomeron diagrams, and each cut has its own, quite distinct, final state characteristics. As a result, the authors found counting rules for final states with different particle multiplicities, and they proved cancellations among rescattering corrections to single-particle and double-particle inclusive cross sections.

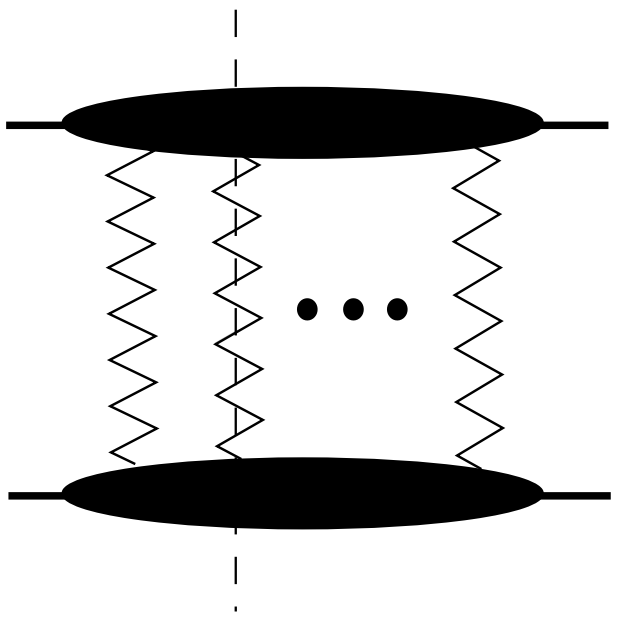

**Fig. 4:** $s$-cut through a multi-Pomeron exchange: the zig-zag lines stand for nonperturbative Pomerons.

In the QCD description of hard (or semihard) final states a close analogy appears between (color singlet) gluon ladders and the nonperturbative Pomeron: multiple parton chains (for example, the two chains in Fig.1b) can be viewed as cuts through two perturbative BFKL Pomerons. In the same way as in the original AGK paper, the question arises how different cuts through a QCD multi-ladder diagram can be related to each other. In the following we briefly describe how AGK cutting rules can be derived in pQCD [5,6]. In the subsequent section we will present a few new results which come out from pQCD calculations, going beyond the original AGK rules.

One of the few assumptions made in the original AGK paper states that the coupling of the Pomerons to the external particle are (i) symmetric under the exchange of the Pomerons (Bose symmetry), and (ii) that they remain unchanged if some of the Pomerons are beeing cut. These properties also hold in pQCD, but they have to be reformulated: (i') the coupling of (reggeized) gluons to external particles is symmetric under the exchange of reggeized gluons, and (ii') it remains unchanged if we introduce cutting lines between the gluons. In QCD, however, the color degree of freedom also allows for another possibility: inside the n-gluon state (with total color zero), a subsystem of two gluons can form an antisymmetric color octet state: in this case the two gluons form a bound state of a reggeized gluon (bootstrap property). For the case of $\gamma^*\gamma^*$ scattering, explicit calculations [7] have shown that the coupling of $n$ gluons to virtual photons can be written as a sum of several pieces: the fully symmetric ('irreducible') one which satisfies (i') and (ii'), and other pieces which, by using the bootstrap property, can be reduced to symmetric couplings of a smaller number of gluons ('cut reggeons'). This decomposition is illustrated in Fig.5.

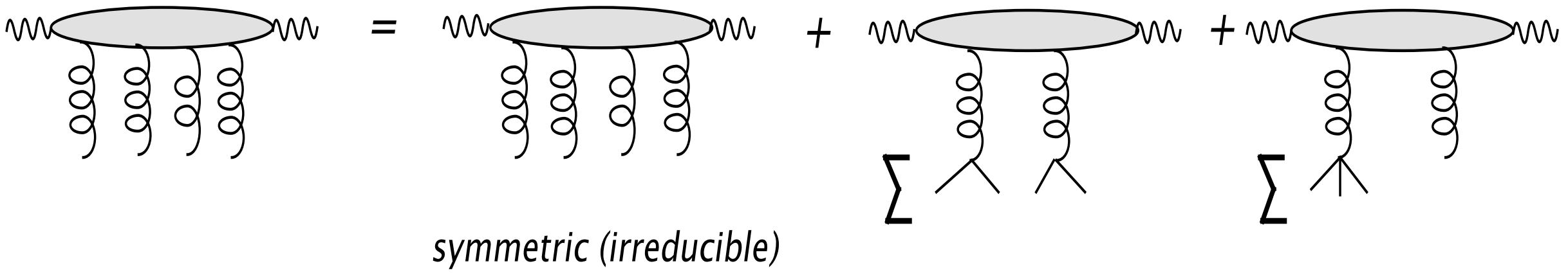

Fig.5 Decomposition of the coupling of four gluons to a virtual photon. In the last two terms on the rhs it is understood that we have to sum over different pairings of gluons at the lower end.

Since the bootstrap property is related to the regeization of the gluon and, therefore, is expected to be valid to all orders perturbation theory, also these properties of the couplings of multi-gluon states to external particles should be of general validity. In this short review we will mainly concentrate on the symmetric couplings.

As an illustrative example, we consider the coupling of four gluons to a proton. The simplest model of a symmetric coupling is a sum of three pieces, each of which contains only the simplest color structure:

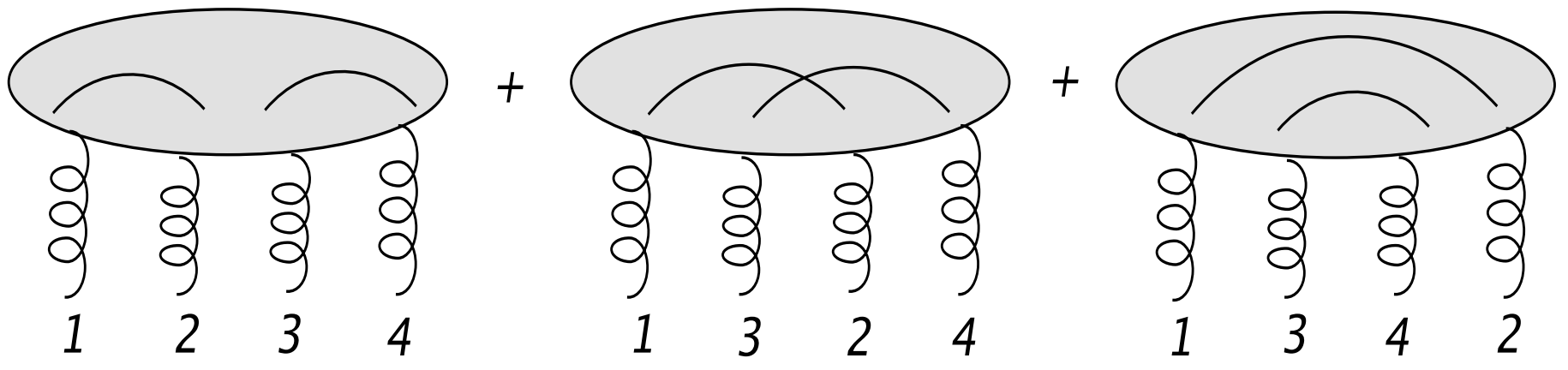

Fig.6 The symmetric coupling of four gluons to an external particle. The lines inside the blob denote the color connection, e.g. the first term has the color structure $\delta_{a_1a_2}\delta_{a_3a_4}$.

The best-known cutting rule for the four gluon exchange which follows [5,6] from this symmetry requirement is the ratio between the three different pairings of lines:

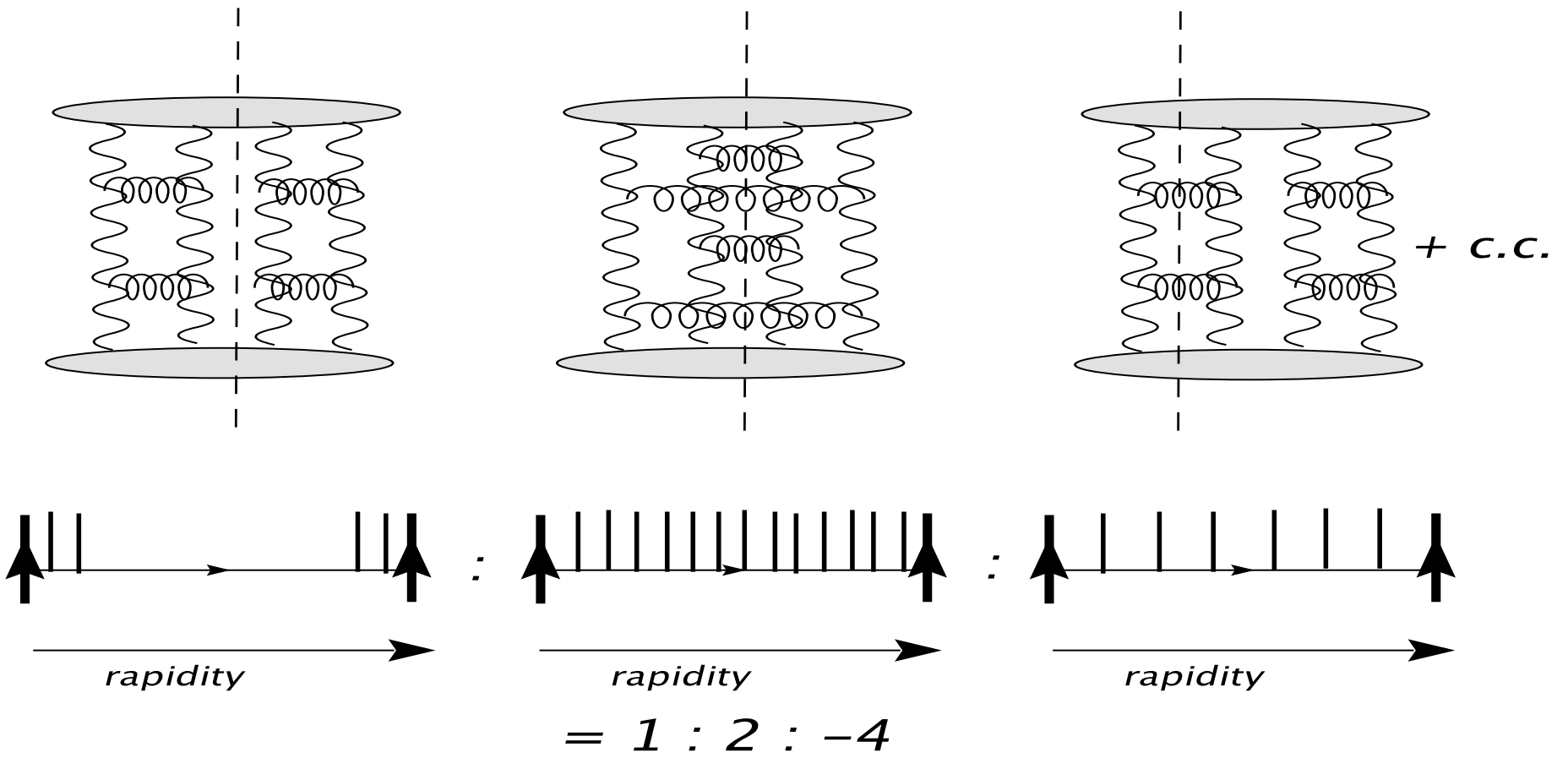

Fig 7: different cutting lines in the four-gluon exchange.

Each term, on the partonic level, corresponds to a certain multiplicity structure of the final state: a rapidity gap ('zero multiplicity'), double multiplicity, and single multiplicity. Simple combinatorics then leads to the ratio

$$1 : 2 : -4. \tag{2}$$

In order to be able to generalize and to sum over an arbitrary number of gluon chains, it is convenient to use an eikonal ansatz:

$$N_{2n}^{A}(\boldsymbol{k}_1, a_1; \ldots; \boldsymbol{k}_{2n}, a_{2n}; \omega) =$$

$$\frac{1}{\sqrt{(N_c^2-1)^n}} \left( \sum_{Pairings} \phi^A(\boldsymbol{k}_1, \boldsymbol{k}_2; \omega_{12})\delta_{a_1a_2} \cdot \ldots \cdot \phi^A(\boldsymbol{k}_{2n-1}, \boldsymbol{k}_{2n}; \omega_{2n-1,2n})\delta_{a_{2n-1}a_{2n}} \right). \tag{3}$$

Inserting this ansatz into the hadron - hadron scattering amplitude, using the large-$N_c$ approximation, and switching to the impact parameter representation, one obtains, for the contribution of $k$ cut gluon ladders, the well-known formula:

$$ImA_k = 4s \int d^2\boldsymbol{b} e^{i\boldsymbol{qb}} P(s, \boldsymbol{b}) \tag{4}$$

where

$$P(s, \boldsymbol{b}) \;=\; \frac{[\Omega(s,\boldsymbol{b})]^k}{k!} e^{-\Omega(s,\boldsymbol{b})}, \tag{5}$$

and $\Omega$ stands for the (cut) two-gluon ladder.

Another result [6] which follows from the symmetry properties of the n gluon-particle coupling is the cancellation of rescattering effects in single and double inclusive cross sections. In analogy with the AGK results on the rescattering of soft Pomerons, it can be shown that the sum over multi-chain contributions and rescattering corrections cancels (Fig.8), leaving only the single-chain contribution (in agreement with the factorization obtained in the collinear analysis). This statement, however, holds only for rescattering between the two projectile: it does not affect the multiple exchanges between the tagged jet and the projectile (Fig.9) which require a seperate discussion (see below).

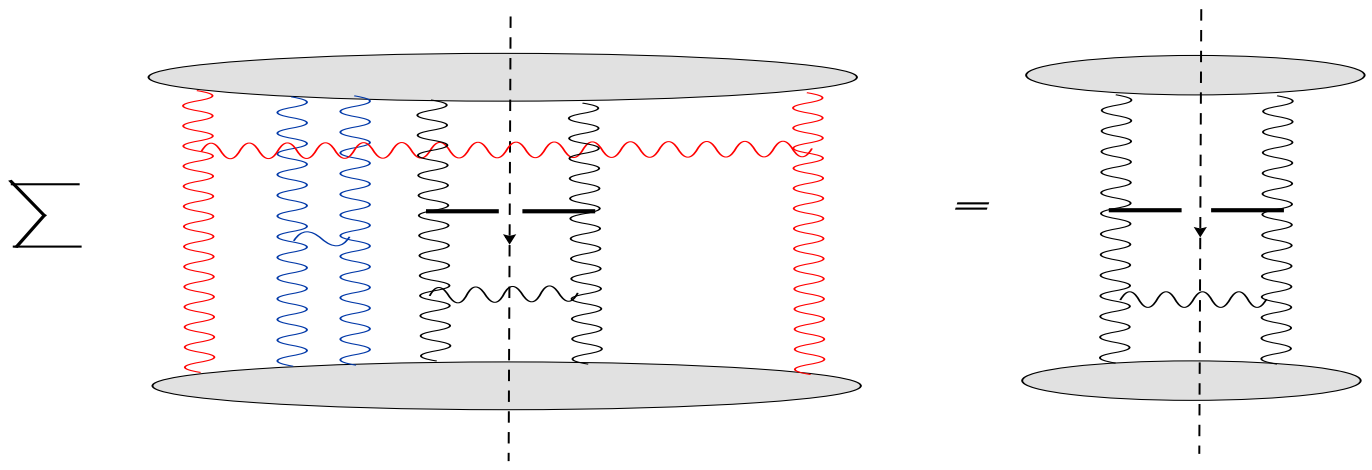

Fig 8: AGK cancellations in the one-jet inclusive cross section.

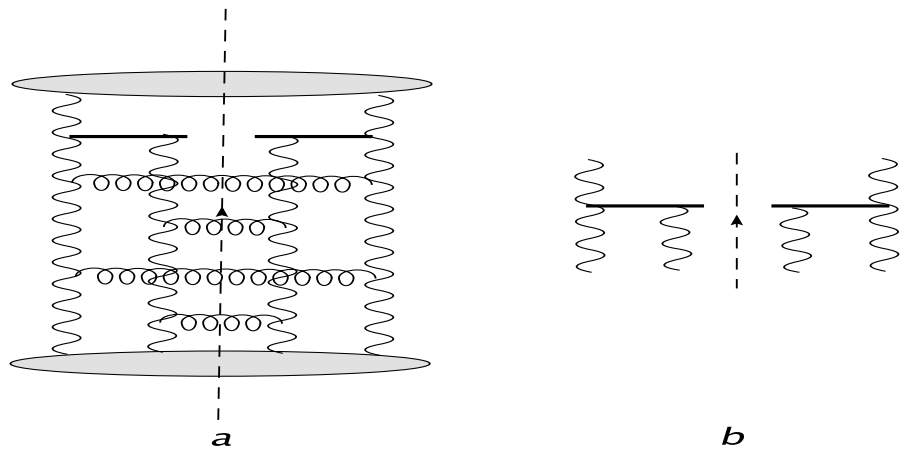

Fig 9: (a) Nonvanishing rescattering corrections in the one-jet inclusive cross section; (b) a new vertex: $g\,+\,2g \rightarrow jet$.

All these results can be generalized to include also the soft Pomeron: all one needs to assume is that the couplings of soft Pomerons and reggeized gluons are symmetric under interchanges, and they are not altered if cutting lines are introduced.

## 3 New results

Explicit calculations in QCD lead to futher results on multiple interactions. First, in the four gluon exchange there are other configurations than those shown in Fig.7; one example is depicted in Fig.10. Here the pairing of gluon chains switches from $(14)(23)$ in the upper part (= left rapidity interval) to $(12)(34)$ in the lower part (= right rapidity interval).

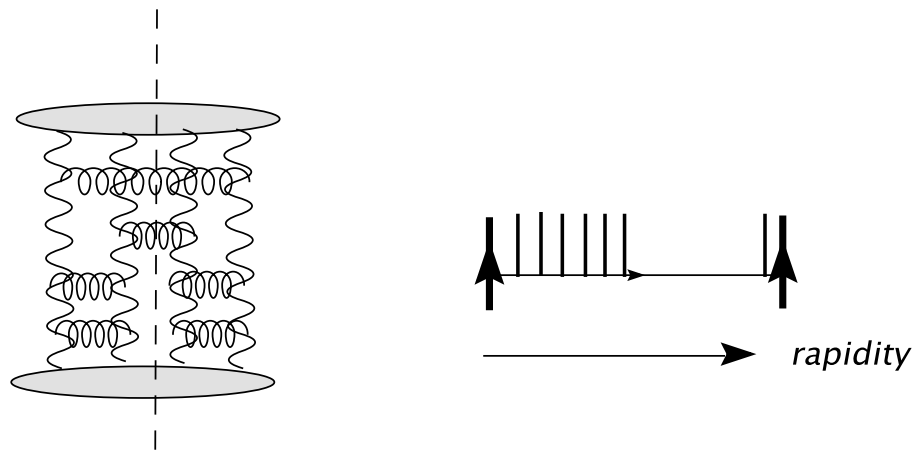

Fig 10: Decomposition into two rapidity intervals: the upper (left) interval has double multiplicity, the lower (right) one corresponds to a rapidity gap.

One can show that the ratio $1 : 2 : -4$ holds for each rapidity interval. In [6] this has been generalized to an arbitrary number of exchanged gluon lines.

Another remark applies to the applicability of the cutting rules to rescattering corrections in the single jet inclusive cross section (Fig.9). Below the jet vertex we, again, have an exchange of four gluon lines, similar to the diagram in the middle of Fig.7. As to the cutting rules, however, there is an important difference between the two situations. In Fig.7, the blob above the four gluons is totally inclusive, i.e. it contains an unrestricted sum over $s$-channel intermediate states, whereas in Fig.9 the part above the four gluon state is semi-inclusive , i.e. it contains the tagged jet. This 'semi-inclusive' nature destroys the symmetry above the four gluon states, and the cutting rules have to be modified [8, 9]. In particular, eqs.(3) - (4) are not applicable to the rescattering corrections between the jet and projectile. A further investigation of these questions is in progress [10].

Finally a few comments on reggeization and cut reggeons. Clearly there are more complicated configurations than those which we have discussed so far; an example appears in $\gamma^* p$ scattering (deep inelastic electron proton scattering). In contrast to $pp$ scattering, the coupling of multi-gluon chains to the virtual photon can be computed in pQCD, and the LO results, for the case of $n = 4$ gluons, are illustrated in Fig.11.

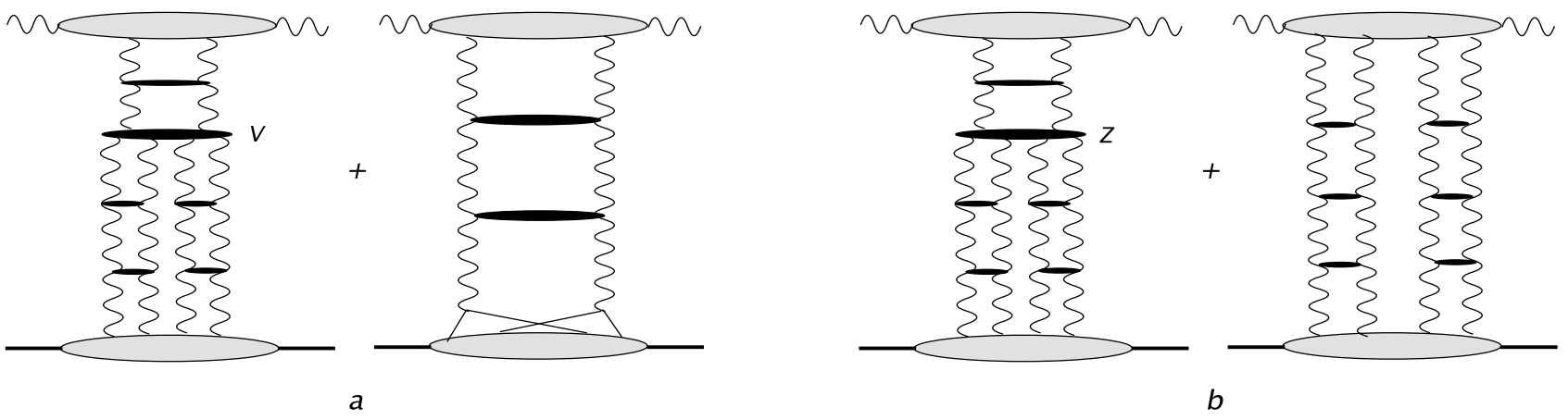

Fig.11: Four-gluon contributions to $\gamma^* p$ proton scattering: two equivalent ways of summing over all contributions. (a) the decomposition of Fig.5 with the pQCD triple Pomeron vertex. (b) an alternative way of summation which explicitly shows the coupling of two Pomerons to the photon vertex and which leads to a new vertex $Z$.

It turns out that we have two alternative possibilities: in the completely inclusive case (total cross section), it is convenient to chose Fig.11a, i.e. the sum of all contributions can be decomposed into two sets of diagrams. In the first set, at the top of the diagram two gluons couple to the quark-antiquark pair, and the subsequent transition to the four-gluon state goes via the pQCD triple Pomeron vertex. This vertex, as a function of the 4 gluons below, has the symmetry properties described above. As a result, we can apply the cutting rules to the four gluon state, as discussed before. However, there is also the second term in Fig.11a, which consists of a two gluon state only: this is the reggeizing contribution we have mentioned before. As indicated in the figure, the splitting of the reggized gluons at the bottom amounts to a change in the (nonperturbative) coupling. We want to stress that, because of the inclusive nature of this set of diagrams, the triple Pomeron vertex $V$ in Fig.11a, similar to the BFKL kernel, contains both real and virtual contributions. For this reason, the decomposition in Fig.11a is applicable to inclusive cross sections, and it is not convenient for investigating specific final states such as, for example, diffractive final states with a fixed number of quarks and gluons in the final state.

There exists an alternative way of summing all contributions (Fig.11b) which is completely equivalent to Fig.11a but allows to keep track of diffractive $q\bar{q}$, $q\bar{q}g$,... final states: this form is illustrated in Fig.11b. One recognizes the 'elastic intermediate state' which was not visible in Fig.11a, and the new triple Pomeron vertex $Z$ which contains only real gluon production. This vertex $Z$, as discussed in [11] is no longer symmetric under permutations of the gluons at the lower end; consequently, we cannot apply the AGK cutting rules to the four gluon states below.. These findings for multiple scattering effects in DIS imply, strictly speaking, that cross sections for diffractive $q\bar{q}$ or $q\bar{q}g$ states cannot directly be inserted into the counting rules (2).

Also $pp$ scattering will contain corrections due to multiple interactions which are more complex. There are, for example, graphs which contain the $2 \to 4$ gluon vertex $V$, leading to a change of the number of gluon lines ( Fig.12).

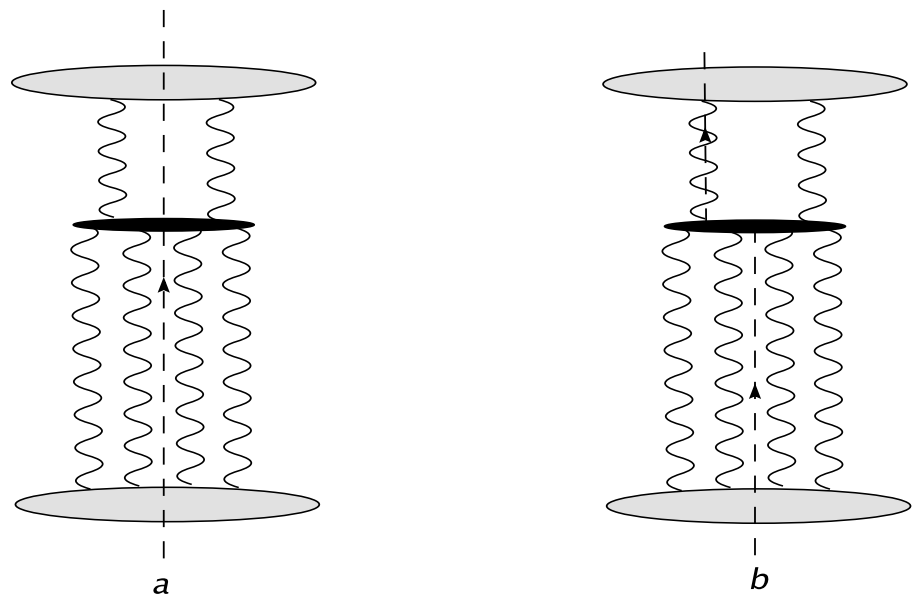

Fig 12: A correction in which the number of lines changes.
The black vertex denotes the $2 \to 4$ gluon vertex.

Since this $2 \to 4$ gluon vertex, as a function of the four gluons below the vertex, satisfies the symmetry requirements listed above, we can apply our previous analysis to the cutting lines below the vertex. In addition, however, one can ask how the lines continue above the $2 \to 4$ gluon vertex: we show two examples, one of them containing a cut (reggeized) gluon. Concentrating on this two-gluon state (i.e. we imagine that we have already summed over all possible cutting lines below the vertex $V$), the counting rules are quite different: in contrast to the even-signature Pomeron, the gluon is a odd-signature reggeon. Consequently, the cut gluon is suppressed w.r.t. the uncut gluon by one power in $\alpha_s$, and this suppression leads to the following hierarchy of cutting lines: the cut between the gluons belongs to leading order, the cut through one of the two reggeized gluons is supressed by one power in $\alpha_s$, the cut through both reggeized gluons is double suppressed (order $\alpha_s^2$). A closer analysis of this question is under investigation [10].

## 4 Conclusions

Corrections due to multiple interactions seem to be important in DIS at small $x$ and low $Q^2$; they are expected to play a significant role also in multijet production in *pp* scattering. The study of the AGK rules to pQCD provides help in understanding the systematics of multiple gluon chains. Results described in this review represent the beginning of a systematic analysis. We have listed a few questions which require further work.

As an immediate application, we believe that a quantitative analysis of multiple scattering at HERA will provide a useful input to the modelling of final states at the LHC.

A question of practical importance which we have not addressed at all is the hadronization of partonic final states. All statements on ratios of 'particle densities in the final states' made in this paper refer to the parton (gluon) level. However, the hadronization of events which, for example, belong to a double-cut ladder configuration may be quite different from the one obtained by applying just the normal single-chain hadronization to each chain seperately. The answer to this question [1] goes beyond the AGK analysis discussed in this paper.

[1]I thank G.Gustafson for a very useful discussion on this point.

# Multiple Interactions in DIS

*Henri Kowalski*
Deutsches Elektronen Synchrotron DESY, 22603 Hamburg

**Abstract**
The abundance of diffractive reactions observed at HERA indicates the presence of multiple interactions in DIS. These interactions are analysed, first in a qualitative way, in terms of QCD Feynman diagrams. Then a quantitative evaluation of diffractive and multiple interaction is performed with the help of the AGK cutting rules applied within an Impact Parameter Dipole Saturation Model. The cross-sections for multiple and diffractive interactions are found to be of the same order of magnitude and to exhibit a similar $Q^2$ dependence.

## 1 Introduction

One of the most important observations of HERA experiments is the rapid rise of the structure function $F_2$ with decreasing $x$ indicating the presence of abundant gluon radiation processes [1]. The observation of a substantial diffractive component in DIS processes, which is also quickly rising with decreasing $x$, is equally important. The diffractive contribution at HERA is of a leading-twist type, i.e. the fraction of diffractive events remains constant or decreases only logarithmically with increasing $Q^2$. The presence of a substantial diffractive component suggests that, in addition to the usual partonic single ladder contribution, also multi-ladder processes should be present.

In this talk I will first discuss the general role of multi-ladder contributions in DIS scattering, called for historical reasons multi-Pomeron processes. The concept of a Pomeron is very useful in the discussion of high energy scattering processes since it relates, by the AGK cutting rules [2], seemingly different reactions like inclusive, diffractive and multiple scattering. I will present a numerical estimate of the magnitude of diffractive and of multi-Pomeron contributions, using AGK cutting rules within a dipole model which has been shown to provide a good description of HERA DIS data [3].

## 2 General Analysis

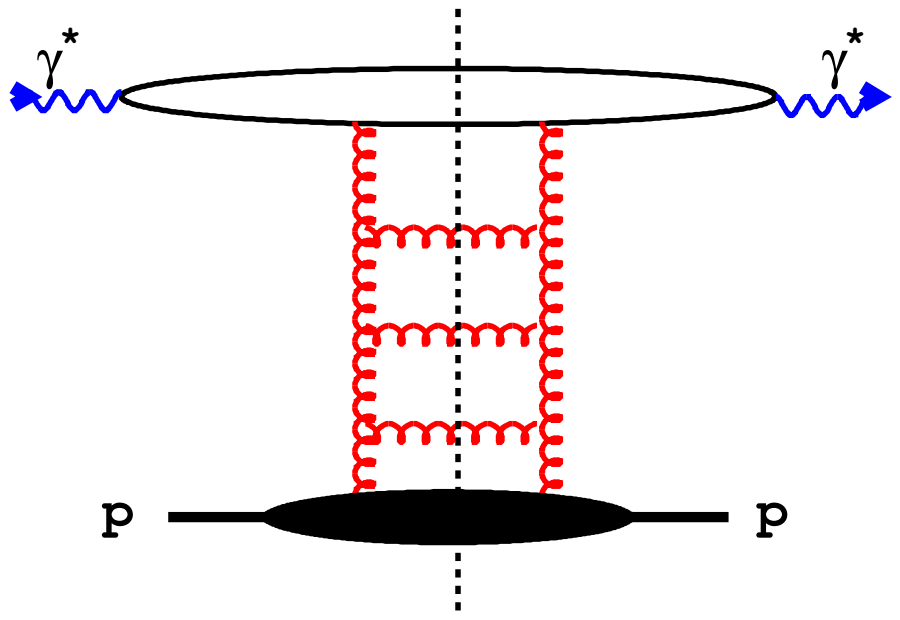

**Fig. 1:** The single gluon-ladder contribution to the total $\gamma^*p$ cross section. The blob at the lower end of the diagrams contains the physics below the scale $Q_0^2$ which seperates hard from soft physics, whereas the blob at the upper end contains hard physics that can be described by pQCD. The dashed line denotes the cut.

Let us first recall that the main properties of HERA interactions can be related to the properties of the elastic amplitude, $A_{\gamma^*p\to\gamma^*p}$, which, by the optical theorem, is directly related to the total $\gamma^*p$ cross-section:

$$\sigma_{\gamma^*p} = \frac{1}{W^2} Im\, A_{\gamma^*p\to\gamma^*p}(W^2, t=0). \tag{1}$$

Here $W$ denotes the $\gamma^* p$ CMS energy and $t$ the 4-momentum transfer of the elastically scattered proton. At not too small $Q^2$, the total cross section is dominated by the single ladder exchange shown in Fig. 1; the ladder structure also illustrates the linear DGLAP evolution equations that are used to describe the $F_2$ data. In the region of small $x$, gluonic ladders are expected to dominate over quark ladders. The cut lines in Fig. 1 mark the final states produced in a DIS event: a cut parton (gluon) hadronizes and leads to jets or particles seen in the detector. It is generally expected that partons produced from a single chain are unlikely to generate large rapidity gaps between them, since large gaps are exponentially suppressed as a function of the gap size. Therefore, in the single ladder contribution of Fig. 1, diffractive final states only reside inside the blob at the lower end, i.e. lie below the initial scale $Q_0^2$.

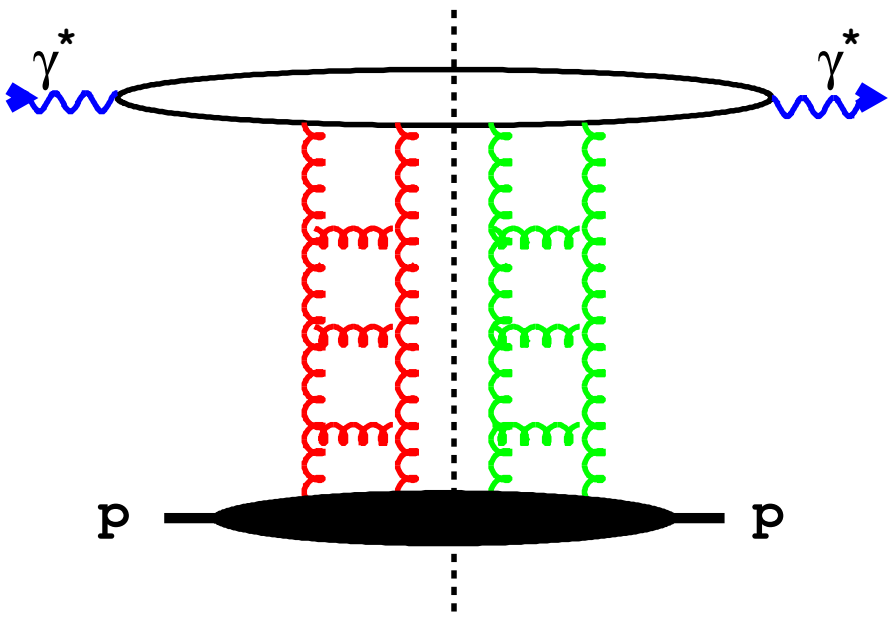

**Fig. 2:** The double-gluon ladder contribution to the inclusive diffractive $\gamma^* p$ cross section

The properties of diffractive reactions at HERA, however, give clear indications that significant contributions from multi-ladder exchanges should be present: not all diffractive final states are soft, in particular the diffractive production of jets and charm was observed [4, 5]. In addition, the inclusive diffractive cross-section is rising as quickly as the total cross-section with increasing $W$ [6] and the exclusive diffractive production of $J/\Psi$ and $\Upsilon$ vector meson exhibits a rise with energy which is about twice as fast [7]. In short, the Pomeron exchanged in inclusive diffractive DIS is harder than the hadronic soft Pomeron and therefore, one should expect that the majority of the observed diffractive final states cannot be absorbed into the blob of soft physics of Fig. 1. Instead, double ladder exchange, Fig. 2, provides a potential source for these harder diffractive states: the cut blob at the upper end may contain $q\bar{q}$ and $q\bar{q}g$ states which hadronize into harder jets or particles. Further evidence for the presence of multi-ladder contribution comes from saturation models which have been shown to successfully describe HERA $F_2$ data in the transition region at low $Q^2$ and small $x$: these models are explicitly built on the idea of summing over multiple exchanges of single ladders (or gluon densities).

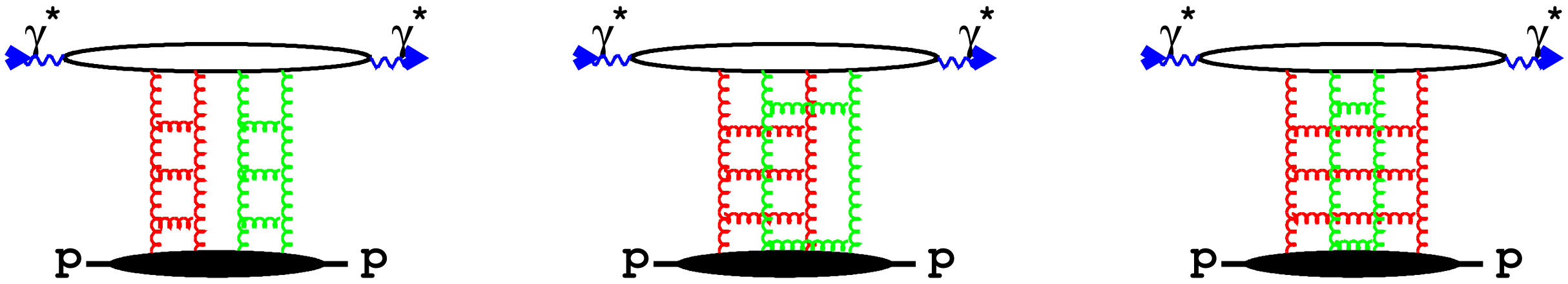

**Fig. 3:** The double-gluon ladder contribution to the elastic $\gamma^* p$ amplitude

Let us analyse the content of a double ladder exchange contribution (for a more detailed analysis see Ref. [8]). It is easiest to begin with the elastic $\gamma^* p$ scattering amplitude, Fig. 3: from a $t$-channel point of view, the two gluon ladders form a four gluon intermediate state which has to be symmetric under permutations of the gluon lines (Bose symmetry). Therefore, on the amplitude level one cannot distinguish between different diagrams of Fig. 3. Invoking now the optical theorem, (1), different contributions to the total cross section correspond to different cuts through the two-ladder diagrams: they are shown in Fig. 4, ordered w.r.t. the density of cut gluons. In Fig. 4a, the cut runs between the two ladders: on the both sides of the cut there is a color singlet ladder, and we expect a rapidity gap between the upper blob (containing, for example, a diffractive $q\bar{q}$ final state) and the proton remnants inside the lower blob. Similarly, the diagram of Fig. 4b describes a single cut ladder with a final state similar to the one ladder contribution in Fig. 1; this contribution simply represents a correction to the one ladder contribution. Finally, the diagram of Fig. 4c belongs to final states with double density of cut partons. As outlined in [9], the correct counting of statistic factors and combinatorics leads to the result that the contributions shown in Fig. 4 a - c are identical, up to the overall counting factors $1 : -4 : 2$.

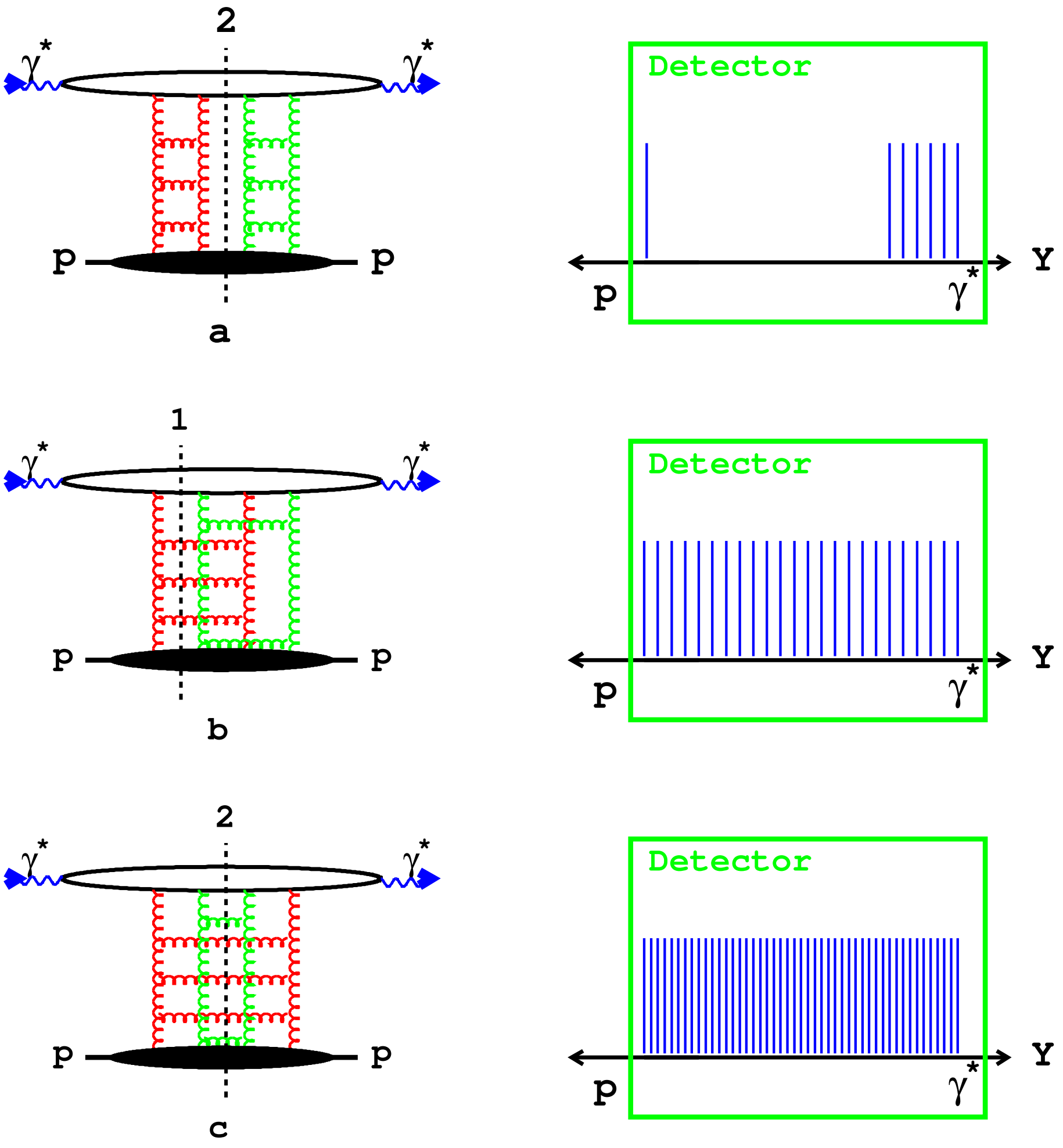

**Fig. 4:** Three examples of 2-ladder contributions (lhs), with the corresponding, schematical, detector signatures (rhs). *Top row*: the diagram (a) with the cut positions (2) describes diffractive scattering. *Middle row*: the diagram (b) with the cut position (1) describes inclusive final states with single densityof cut partons. *Bottom row*: the diagram (c) with the cut position (2) describes inclusive final states with increased multiplicity.

Experimentally it is easy to differentiate between diffractive and *single* or *multiple* inclusive final states since diffractive states exhibit large rapidity gaps. The *multiple* inclusive final states should also be distinct from the *single* inclusive ones since, at least naively, we would expect that in the *multiple* case the particle multiplicity should be considerably higher. At low $x$, however, the relation between the number of virtual states excited in the interaction (as measured by $F_2$) and the final particle multiplicity cannot be straight-forward since the growth of $F_2$ with decreasing $x$ is faster than the multiplicity increase. This may indicate that the hadronization mechanism may be different from the string picture commonly used in the hadronization procedure of single chain parton showers. The influence of multiple scattering on the particle multiplicity of the final states should also be damped by the energy conservation. The cut through several Pomerons leads clearly to more gluons produced in the final state, but the available energy to produce particles in the hadronization phase remains the same. A detailed Monte Carlo program is therefore necessary to evaluate this effect.

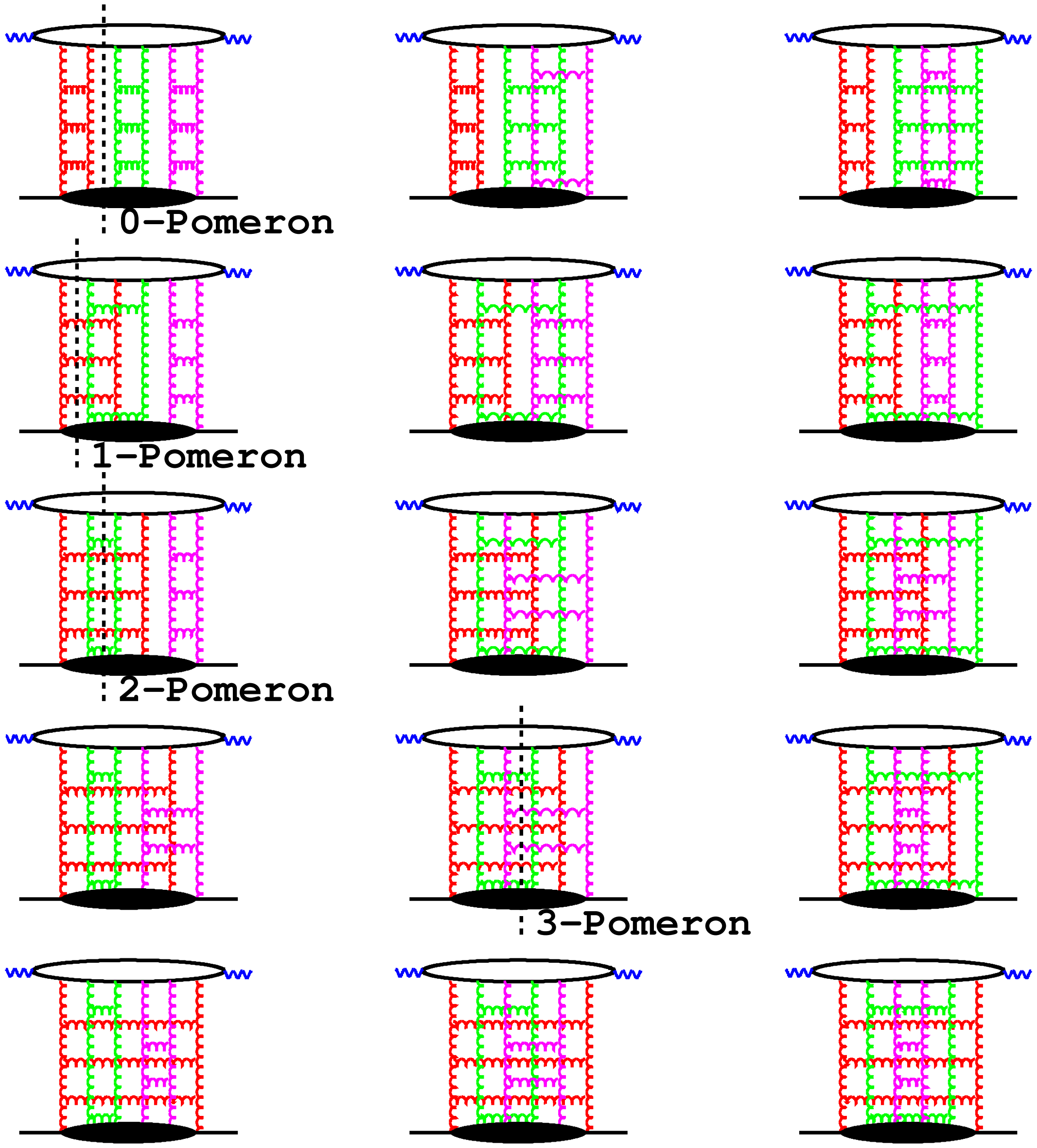

**Fig. 5:** 3-Pomeron contributions to the elastic $\gamma^* p$ amplitude. All 15 possible diagrams are shown with some examples of Pomeron cuts.

The number of diagrams contributing to the reaction amplitude increases very quickly with the number of Pomerons. For the 3-Pomeron amplitude the gluons can be paired in 15 possible ways, shown in Fig. 5 with the examples of 0-Pomeron, 1-Pomeron, 2-Pomeron and 3-Pomeron cuts. For $m$-Pomerons the number of possible gluon pairs and also diagrams is:

$$(2m-1)(2m-3)(2m-5).... = (2m-1)!/(2^{m-1}(m-1)!).$$

Assuming that all the diagrams for a given multi-Pomeron exchange amplitude contribute in the same way, the above analysis suggests that the probability for different cuts to contribute should be given by the combinatorial factors. This is the content of the AGK rules which were obtained from the analysis of field theoretical diagrams well before QCD was established [2] and which relate the cross-section, $\sigma_k$, for observing a final state with $k$-cut Pomerons with the amplitudes for exchange of $m$ Pomerons, $F^{(m)}$:

$$\sigma_k = \sum_{m=k}^{\infty} (-1)^{m-k}\, 2^m \, \frac{m!}{k!(m-k)!} F^{(m)}. \tag{2}$$

The same result is also obtained from a detailed analysis of the Feynman diagram contributions in QCD with the oversimplified assumption that only the symmetric part of the two-gluon couplings contributes [9].

## 3 Multiple Interactions in the Dipole Model

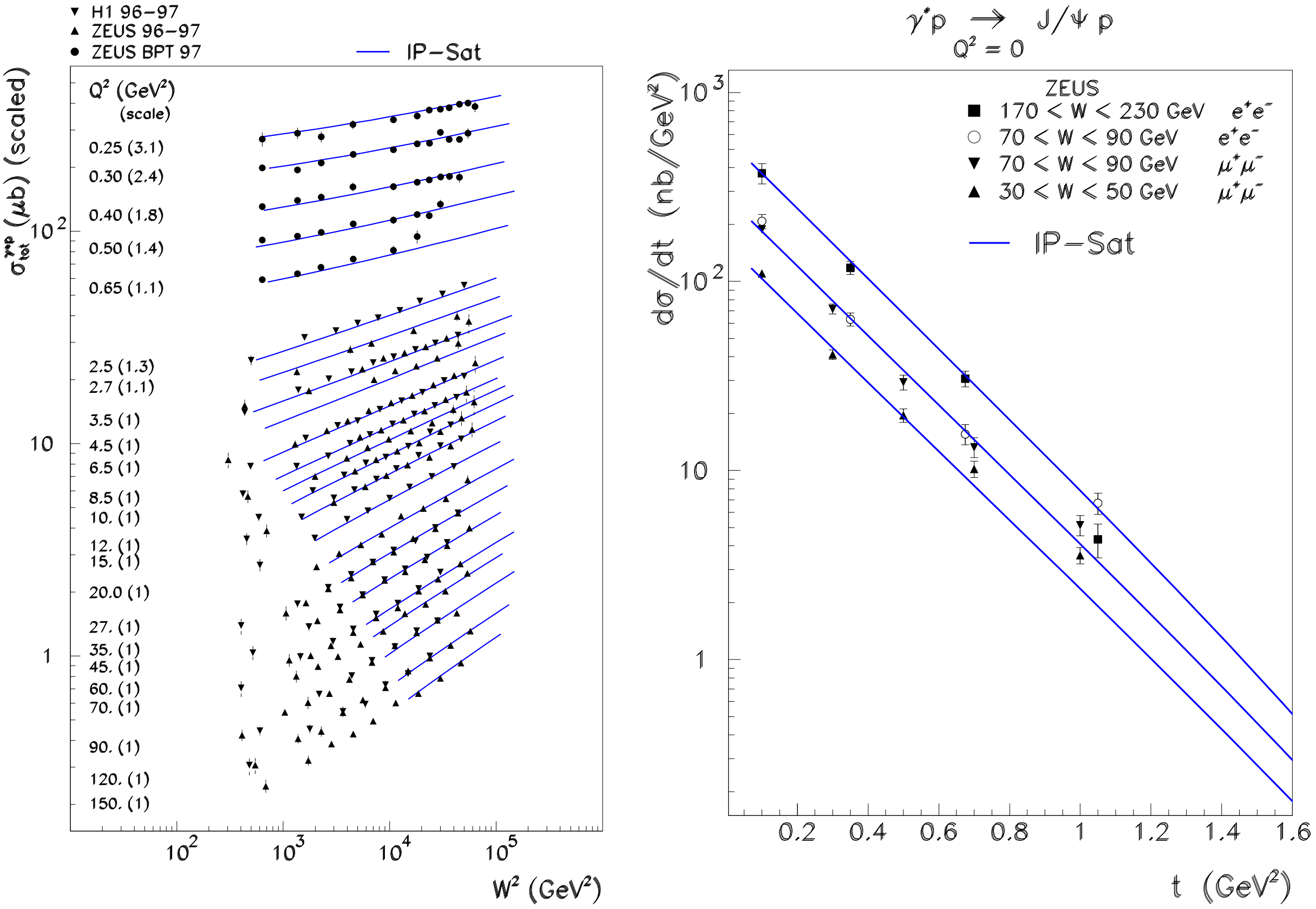

**Fig. 6:** LHS: The $\gamma^* p$ cross-section as a function of $W^2$. RHS: The differential cross section for exclusive diffractive $J/\Psi$ production as a function of the four-momentum transfer $t$. The solid line shows a fit by the IP saturation model.

The properties of the multi-Pomeron amplitude and of the cut Pomeron cross-sections can be quantitatively studied in a dipole model. Let us first recall the main properties of the dipole picture, see Ref. [10, 11] and [3]. In the model the $\gamma^* p$ interaction proceeds in three stages: first the incoming vitual photon fluctuates into a quark-antiquark pair, then the $q\bar{q}$ pair elastically scatters on the proton, and finally the $q\bar{q}$ pair recombines to form a virtual photon. The total cross-section for $\gamma^* p$ scattering, or equivalently $F_2$, is obtained by averaging the dipole cross-sections with the photon wave functions, $\psi(r, z)$, and integrating over the impact parameter, $b$:

$$F_2 = \frac{Q^2}{4\pi^2 \alpha_{em}} \int d^2 r \int \frac{dz}{4\pi} \psi^* \psi \int d^2 b \frac{d\sigma_{qq}}{d^2 b} \,. \tag{3}$$

Here $\psi^*\psi$ denotes the probability for a virtual photon to fluctuate into a $q\bar{q}$ pair, summed over all flavors and helicity states. The dipole cross-section is assumed to be a function of the opacity $\Omega$:

$$\frac{d\sigma_{qq}}{d^2b} = 2\left(1 - \exp(-\frac{\Omega}{2})\right) . \tag{4}$$

At small-$x$ the opacity $\Omega$ can be directly related to the gluon density, $xg(x, \mu^2)$, and the transverse profile of the proton, $T(b)$:

$$\Omega = \frac{\pi^2}{N_C} r^2 \alpha_s(\mu^2)\, xg(x, \mu^2)\, T(b) . \tag{5}$$

The parameters of the gluon density are determined from the fit to the total inclusive DIS cross-section, as shown in Fig. 6 [3]. The transverse profile was determined from the exclusive diffractive $J/\Psi$ cross-sections shown in the same figure. The opacity function $\Omega$ determined in this way has predictive properties; it allows to describe other measured reactions, e.g. charm structure function or elastic diffractive $J/\Psi$ production shown in Fig.7.

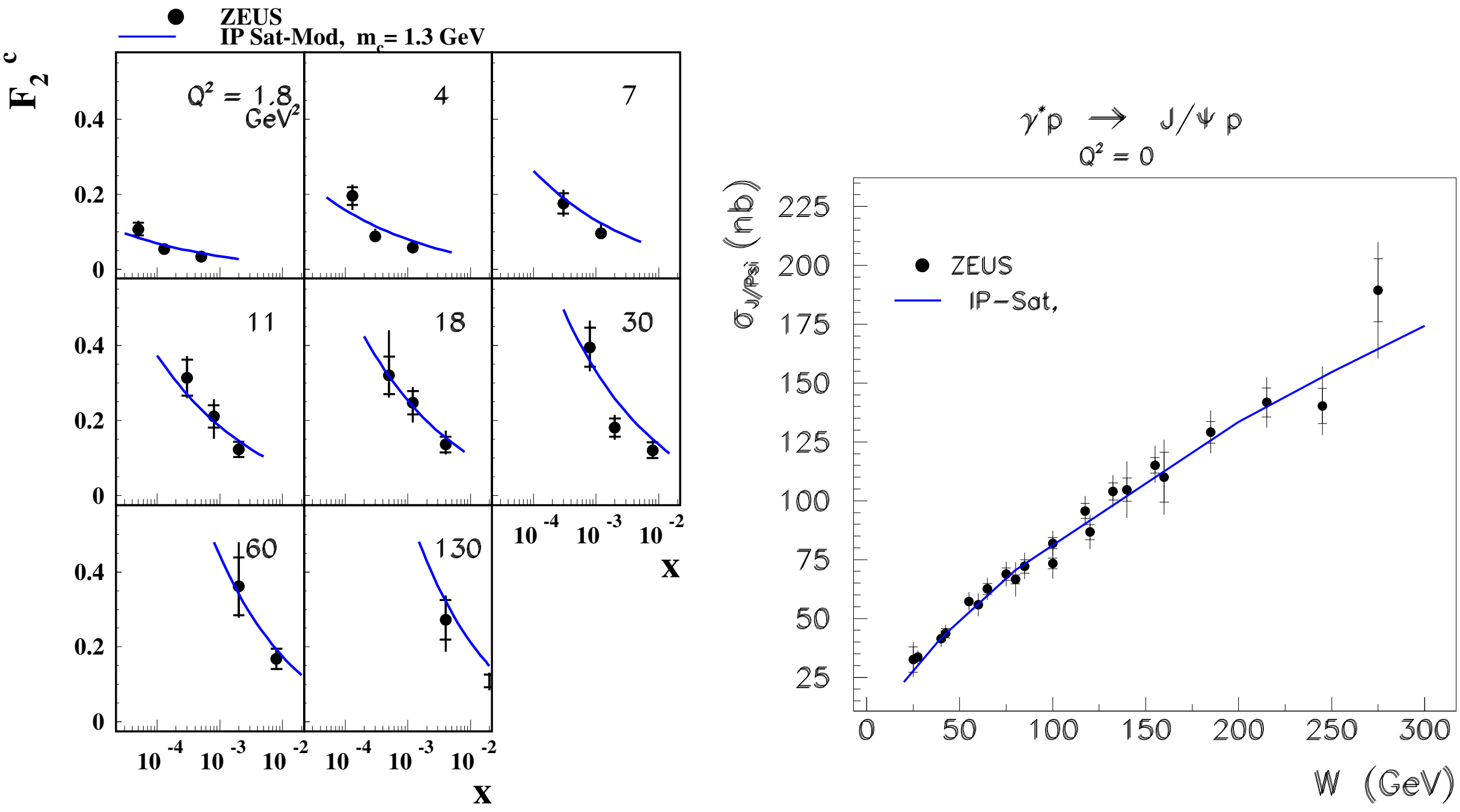

**Fig. 7:** LHS: Charm structure function, $F_2^c$. RHS: Total elastic $J/\Psi$ cross-section. The solid line shows the reswult of the IP saturation model.

For a small value of $\Omega$ the dipole cross-section, Eq (4), is equal to $\Omega$ and therefore proportional to the gluon density. This allows to identify the opacity with the single Pomeron exchange amplitude of Fig. 1. The multi-Pomeron amplitude is determined from the expansion:

$$\frac{d\sigma_{qq}}{d^2b} = 2\left(1 - \exp(-\frac{\Omega}{2})\right) = 2\sum_{m=1}^{\infty}(-1)^{m-1}\left(\frac{\Omega}{2}\right)^m \frac{1}{m!} \tag{6}$$

as

$$F^{(m)} = \left(\frac{\Omega}{2}\right)^m \frac{1}{m!}, \tag{7}$$

since the dipole cross-section can be expressed as a sum of multi-Pomeron amplitudes [12] in the following way:

$$\frac{d\sigma_{qq}}{d^2b} = 2\sum_{m=1}^{\infty}(-1)^{m-1}\,F^{(m)}\,. \tag{8}$$

The cross-section for $k$ cut Pomerons is then obtained from the AGK rules, eq. 2, and from the multi-Pomeron amplitude, Eq. (7), as:

$$\frac{d\sigma_k}{d^2b} = \sum_{m=k}^{\infty}(-1)^{m-k}\,2^m\,\frac{m!}{k!(m-k)!}\left(\frac{\Omega}{2}\right)^m\frac{1}{m!} = \frac{\Omega^k}{k!}\sum_{m=k}^{\infty}(-1)^{m-k}\,\frac{\Omega^{m-k}}{(m-k)!} \tag{9}$$

which leads to a simple expression:

$$\frac{d\sigma_k}{d^2b} = \frac{\Omega^k}{k!}\exp(-\Omega)\,. \tag{10}$$

The diffractive cross-section is given by the difference between the total and the sum over all cut cross-sections:

$$\frac{d\sigma_{diff}}{d^2b} = \frac{d\sigma_{tot}}{d^2b} - \sum_{k=1}^{\infty}\frac{d\sigma_k}{d^2b} = 2\left(1-\exp\left(-\frac{\Omega}{2}\right)\right) - (1-\exp(-\Omega)) = \left(1-\exp\left(-\frac{\Omega}{2}\right)\right)^2\,. \tag{11}$$

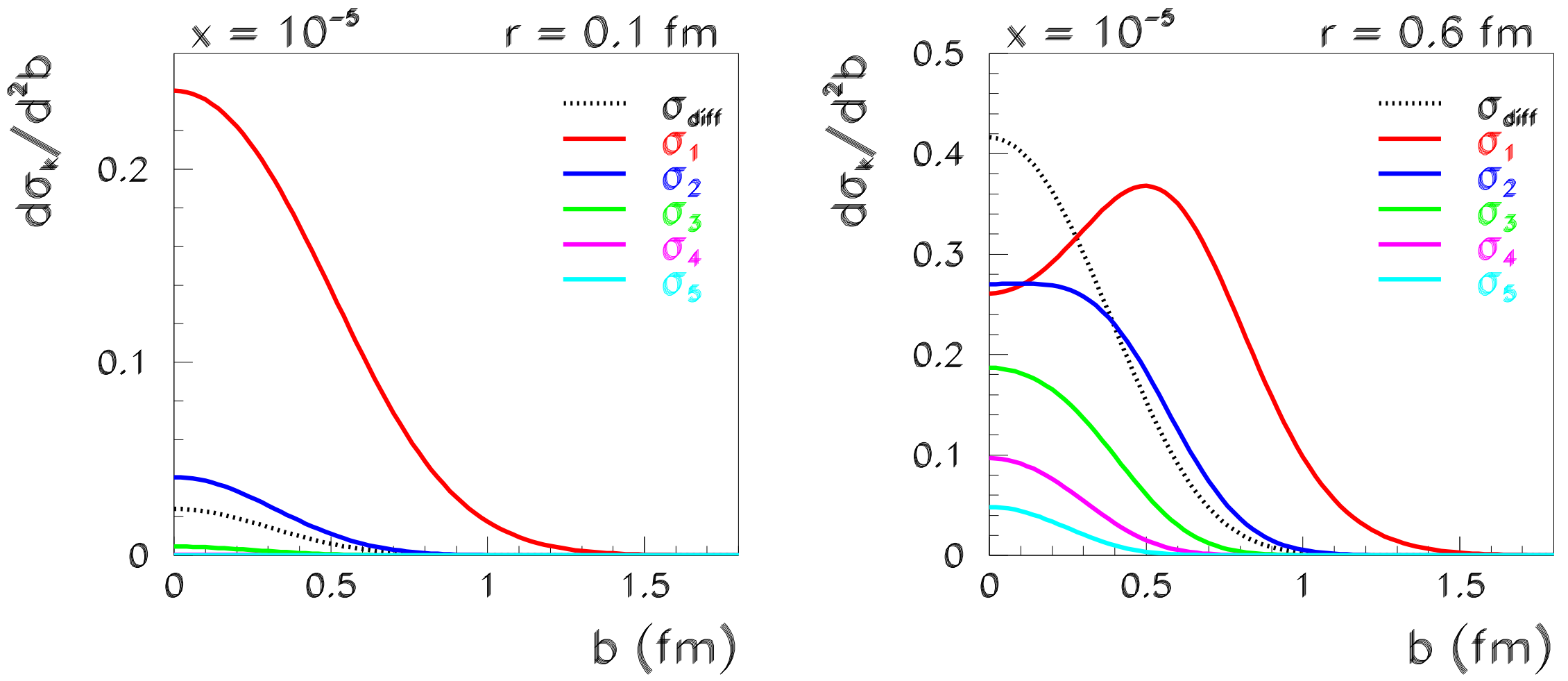

**Fig. 8:** Examples of $b$ dependence of various cut dipole and diffractive cross-sections.

The cut cross-sections determined in the dipole model analysis of HERA data have several interesting properties shown in Fig. 8: for small dipoles ($r = 0.1$ fm) the opacity $\Omega$ is also small, so the single cut cross-section, $\sigma_1$, dominates. This leads to particle production emerging only from the one-cut pomeron, which should correspond, in the context of e.g. the LUND model, to a fragmentation of only one string. For larger dipoles ($r = 0.6$ fm) the dipole cross-section starts to be damped in the middle of the proton (at $b \approx 0$) by saturation effects. Therefore, the single cut cross-section is suppressed in the middle while the multiple cut cross-sections, $\sigma_2$, $\sigma_3$, etc, become substantial and increasingly concentrated in the proton center. These, fairly straight-forward properties of dipoles indicate that in the central scattering events the multiple scattering probability will be enhanced, which may lead at the LHC to substantial effects in a surrounding event multiplicity.

The contribution to $F_2$ from the $k$-cut Pomeron exchanges are computed in the analogous way to $F_2$:

$$F_2^k = \frac{Q^2}{4\pi^2\alpha_{em}}\int d^2r\int\frac{dz}{4\pi}\psi^*\psi\int d^2b\frac{d\sigma_k}{d^2b}. \tag{12}$$

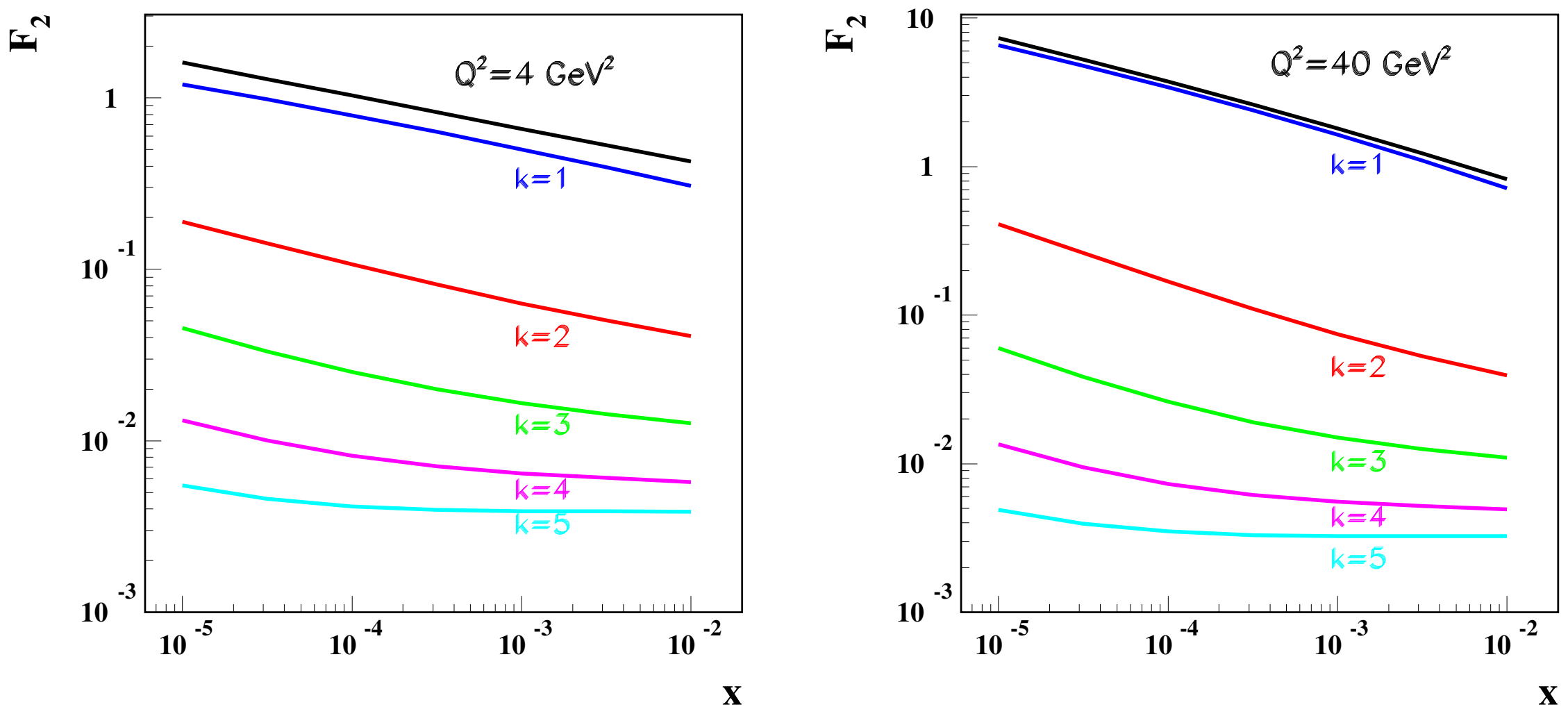

**Fig. 9:** $F_2$ and the contributions of k-cut Pomeron processes, $F_2^k$.

These contributions are shown, together with $F_2$, as a function of $x$ for two representative $Q^2$ values in Fig. 9. One finds that multiple interaction contributions, i.e. $k \geq 2$, in the perturbative region, at $Q^2 = 4$ GeV$^2$, are substantial. In the typical HERA range of $x \approx 10^{-3} - 10^{-4}$, the $k = 2$ contribution is around $10\%$ of $F_2$ and the contributions of higher cuts are also non-negligible. For example, the contribution of the 5-cut Pomeron exchanges is still around 0.5%, which means that at HERA, many thousand events may come from this type of process. Figure 10 shows the fraction of the multpile interaction processes, $F_2^{MI} = F_2^{k=2} + F_2^{k=3} + F_2^{k=4} + F_2^{k=5}$ in $F_2$, at the same $Q^2$ values. At $Q^2 = 4$ GeV$^2$ the fraction of multiple scattering events is around $14\%$ and at $Q^2 = 40$ GeV$^2$ around $6\%$, in the HERA $x$ region, which indicates that the decrease of multiple scattering with increasing $Q^2$ is only logarithmic. The fraction of diffractive processes, shown for comparison, is of the same order, and drops also logarithmically with $Q^2$. The logarithmic drop of the diffractive contribution expected in the dipole model is confirmed by the data [6].

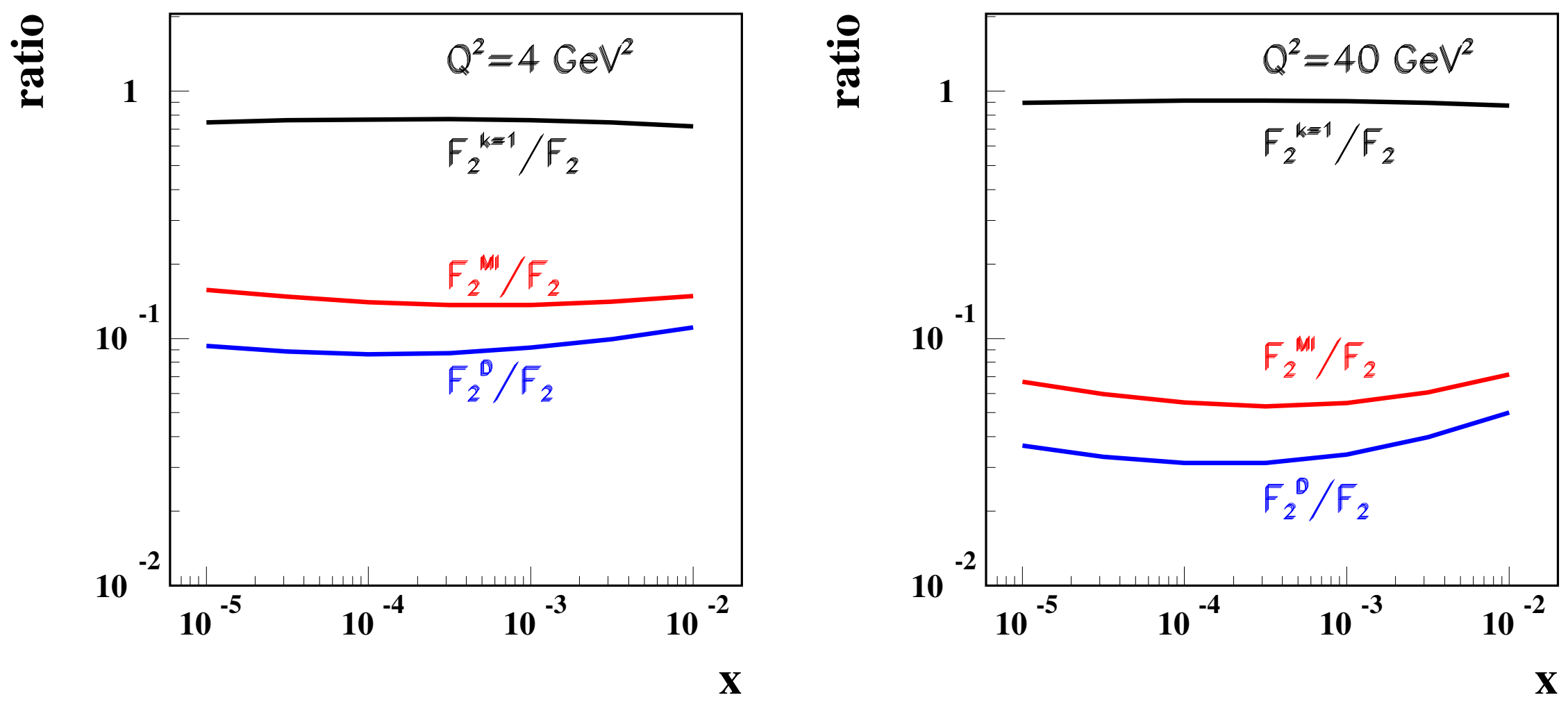

**Fig. 10:** Fractions of single (k=1), multiple interaction (MI) and diffraction (D) in DIS

The dipole model provides a straight-forward extrapolation to the region of low $Q^2$, which is partly perturbative and partly non-perturbative. Figure 11 shows the contribution to $F_2$ of $k$-cut Pomeron processes and the fractions of multiple interactions and diffractive processes at $Q^2 = 0.4$ GeV$^2$.

Note also that, as a byproduct of this investigation, the ratio of diffractive and inclusive cross-sections, $F_2^D/F_2$ is found to be almost independent of $x$, in agreement with the data and also other dipole model predictions [6, 13, 14]. The absolute amount of diffractive effects is underestimated, since the evaluation of diffraction through AGK rules is oversimplified. It is well known [14], that a proper evaluation of diffraction should also take into account the $q\bar{q}g$ contribution which is missing in the simple AGK schema.

In conclusion, we find that the impact parameter dependent dipole saturation model [3] reproduces well the main properties of the data and leads to the prediction that multiple interaction effects at HERA should be of the order of diffractive effects, which are known to be substantial. The multiple interaction effects should decrease slowly (logarithmically) with increasing $Q^2$, similarly to the diffractive contribution.

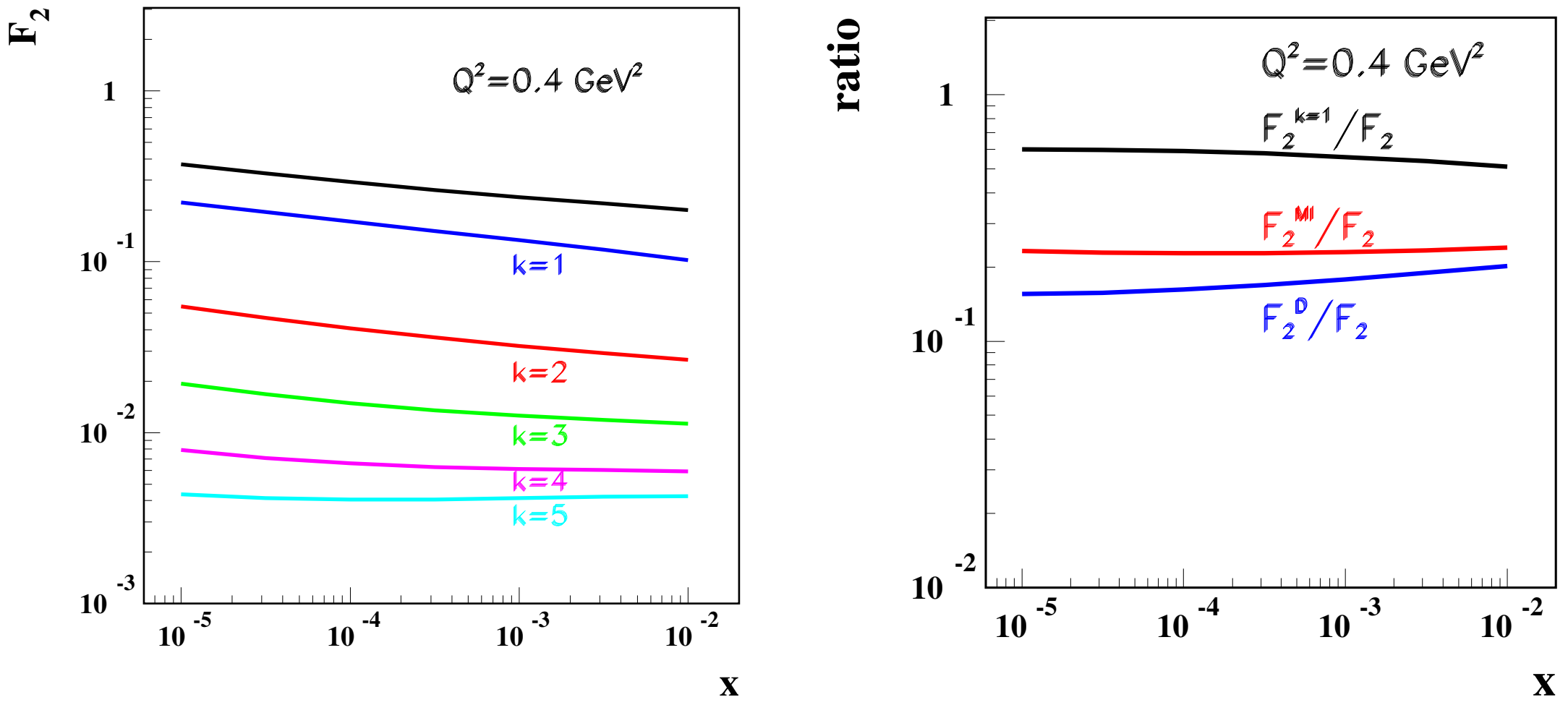

**Fig. 11:** Left: $F_2$ and the contributions of $k$-cut Pomeron processes. Right: Fractions of single (k=1), multiple interaction (MI) and diffraction (D) in DIS at $Q^2 = 0.4$ GeV$^2$.

## 4 Acknowledgements

I would like to thank Jochen Bartels for many illuminating discussions about physics of the multi-ladder QCD diagrams.

# From HERA to LHC through the Color Glass Condensate

*Raju Venugopalan*
Physics Department
Brookhaven National Laboratory
Upton, NY 11973, USA.

**Abstract**

A classical effective field theory, the Color Glass Condensate (CGC), provides a unified treatment of high parton density effects in both DIS and hadron-hadron collisions at very high energies. The validity and limitations of $k_\perp$ factorization can be studied in this effective theory. Multi-parton correlations in the effective theory are described by universal dipole and multipole operators. The evolution of these operators with energy provides a sensitive test of multi-parton dynamics in QCD at high energies.

## 1 Introduction

In the Bjorken limit of QCD, $Q^2 \to \infty$, $s \to \infty$, $x_{\rm Bj} \approx Q^2/s = $ fixed, we have a powerful framework to compute a large number of processes to high accuracy. Underlying this machinery is the Operator Product Expansion (OPE), where cross-sections are identified as a convolution of short distance "coefficient functions" which are process dependent and long distance parton distribution functions which are universal. The evolution of the parton distribution functions with $x$ and $Q^2$ is described by splitting functions, which determine the probability of "parent" partons to split into a pair of "daughter" partons. Both coefficient functions and splitting functions for DIS inclusive cross-sections are now available to Next-Next-Leading-Order (NNLO) accuracy [1].

While this is a tremendous achievement, the contribution of high $Q^2$ processes to the total cross-section is very small. The bulk of the cross-section can perhaps be better understood in the Regge asymptotic limit: $x_{\rm Bj} \to 0$, $s \to \infty$, $Q^2 = $ fixed. The BFKL renormalization group equation [3] describes the leading $\alpha_S \ln(1/x)$ behavior of gluon distributions in this limit. The solutions of the BFKL equation predict that gluon distributions grow very rapidly with decreasing $x$. In the Regge asymptotics, since the transverse size of the partons is fixed, this growth of distributions will lead to the overlapping of partons in the transverse plane of the hadron. In this regime, contributions that were power suppressed in the BFKL scheme become important. These are recombination and screening effects which slow down the growth of gluon distributions leading ultimately to a saturation of these distributions [4, 5]. Such effects must appear at small $x$ because the occupation number [1] of partons in QCD be at most of order $1/\alpha_S$.

Thus qualitatively, the competition between Bremsstrahlung and recombination/screening effects becomes of the same order when

$$\frac{1}{2\,(N_c^2-1)}\,\frac{x\,G(x,Q^2)}{\pi R^2 Q^2} \approx \frac{1}{\alpha_S(Q^2)}\,, \tag{1}$$

where $R$ is the radius of the target. This relation is solved self-consistently when $Q \equiv Q_s(x)$. The scale $Q_s(x)$ is termed the saturation scale and it grows as one goes to smaller values of $x$. When $Q^2 \leq Q_s^2$, higher twist effects are important; at sufficiently small $x$, $Q_s^2 \gg \Lambda^2_{\rm QCD}$, which makes feasible a weak coupling analysis of these effects. At HERA, reasonable fits of small $x$ inclusive and diffractive data for $x \leq 10^{-2}$ are obtained in saturation models with $Q_s^2(x) \approx Q_0^2\,(x_0/x)^\lambda$, with $Q_0^2 = 1$ GeV$^2$ and $x_0 = 3 \cdot 10^{-4}$. Detailed estimates suggest that the saturation scale for gluons is $Q_s(x) \approx 1.4$ GeV at $x \approx 10^{-4}$ [7]. The applicability of weak coupling techniques at these scales is dubious. Nevertheless, they cannot be ruled out since the effective scale at which the coupling runs can be larger than the estimate. Leading twist evolution of "shadowed" distributions at the saturation scale can extend out to significantly large values of $x$. A hint of this possibility is suggested by the fact that geometrical scaling- the dependence of cross-sections on the dimensionless ratio $Q^2/Q_s^2$ alone-extends out to $Q^2 \approx 450$ GeV$^2$ at HERA [8].

[1] This corresponds to the number of partons per unit transverse area, per unit transverse momentum, per unit rapidity, in light cone gauge. This condition has its gauge invariant counterpart in the requirement that the field strength squared not exceed $1/\alpha_S$.

The possibility that weak coupling may apply at high energies is good news. Some of the remarkable regularities in high energy scattering data may be understood in a systematic way. The OPE, for instance, is no longer a good organizing principle since its usefulness is predicated on the twist expansion. In the next section, we will discuss an effective field theory approach which may provide a more efficient organizing principle at high parton densities.

## 2 The Color Glass Condensate

The physics of high parton densities can be formulated as a classical effective theory [6] because there is a Born-Oppenheimer separation between large x and small x modes [9] which are respectively the slow and fast modes in the effective theory. Large x partons are static sources of color charge for the dynamical wee (small x) parton fields. The generating functional of wee partons has the form

$$\mathcal{Z}[j] = \int [d\rho]\, W_{\Lambda^+}[\rho] \left\{ \frac{\int^{\Lambda^+} [dA]\delta(A^+) e^{iS[A,\rho] - j\cdot A}}{\int^{\Lambda^+} [dA]\delta(A^+) e^{iS[A,\rho]}} \right\} \tag{2}$$

where the wee parton action has the form

$$S[A,\rho] = \frac{-1}{4} \int d^4x\, F_{\mu\nu}^2 + \frac{i}{N_c} \int d^2x_\perp dx^- \delta(x^-)\, \mathrm{Tr}\left(\rho(x_\perp) U_{-\infty,\infty}[A^-]\right)\,. \tag{3}$$

In Eq. (2), $\rho$ is a two dimensional classical color charge density and $W[\rho]$ is a weight functional of sources (which sits at momenta $k^+ > \Lambda^+$: note, $x = k^+/P^+_{\rm hadron}$). The sources are coupled to the dynamical wee gluon fields (which in turn sit at $k^+ < \Lambda^+$) via the gauge invariant term which is the second term on the RHS of Eq. (3). Here $U_{-\infty,\infty}$ denotes a path ordered exponential of the gauge field $A^-$ in the $x^+$ direction. The first term in Eq. (3) is the QCD field strength tensor squared — thus the wee gluons are treated in full generality in this effective theory, which is formulated in the light cone gauge $A^+ = 0$. The source $j$ is an external source — derivatives taken with respect to this source (with the source then put to zero) generate correlation functions in the usual fashion.

The argument for why the sources are classical is subtle and follows from a coarse graining of the effective action. The weight functional for a large nucleus is a Gaussian in the source density [6, 11], with a small correction for SU($N_c$) coming from the $N_c - 2$ higher Casimir operators [10]. The variance of the Gaussian, the color charge squared per unit area $\mu_A^2$, proportional to $A^{1/3}$, is a large scale — and is the only scale in the effective action [2]. Thus for $\mu_A^2 \gg \Lambda_{\rm QCD}^2$, $\alpha_S(\mu_A^2) \ll 1$, and one can compute the properties of the theory in Eq. (2) in weak coupling.

The saddle point of the action in Eq. (3) gives the classical distribution of gluons in the nucleus. The Yang-Mills equations can be solved analytically to obtain the classical field of the nucleus as a function of $\rho$: $A_{\rm cl.}(\rho)$ [6, 11, 12]. One can determine, for Gaussian sources, the occupation number $\phi = dN/\pi R^2/dk_\perp^2 dy$ (the number of partons per unit transverse momentum, per unit rapidity y, where $y = \ln(1/x)$) of wee partons in the classical field of the nucleus. One finds for $k_\perp \gg Q_s^2$, the Weizsäcker-Williams spectrum $\phi \sim Q_s^2/k_\perp^2$; for $k_\perp \leq Q_s$, one obtains a complete resummation to all orders in $k_\perp$,

[2] $\mu_A^2$ is simply related in the classical theory to the saturation scale $Q_s^2$ via the relation $Q_s^2 = \alpha_S N_c \mu_A^2 \ln(Q_s^2/\Lambda_{\rm QCD}^2)$

which gives $\phi \sim \frac{1}{\alpha_S}\ln(Q_s/k_\perp)$. (The behavior at low $k_\perp$ can, more accurately, be represented as $\frac{1}{\alpha_S}\Gamma(0,z)$ where $\Gamma$ is the incomplete Gamma function and $z = k_\perp^2/Q_s^2$ [13]).

A high energy hadron is a Color Glass Condensate for the following reasons [2]. The 'color' is obvious since the parton degrees of freedom are colored. It is a glass because the sources, static on time scales much larger than time scales characteristic of the system, induce a stochastic (space-time dependent) coupling between the partons under quantum evolution — this is analogous to a spin glass. Finally, the matter is a condensate because the wee partons have large occupation numbers (of order $1/\alpha_S$) and have momenta peaked about $Q_s$. These properties are enhanced by quantum evolution in $x$. The classical field retains its structure — while the saturation scale grows: $Q_s(x') > Q_s(x)$ for $x' < x$.

Small fluctuations about the effective action in Eq. (3) give large corrections of order $\alpha_S\ln(1/x)$ (see Ref. [14]). The Gaussian weight functional is thus fragile under quantum evolution of the sources. A Wilsonian renormalization group (RG) approach systematically treats these corrections [15]. In particular, the change of the weight functional $W[\rho]$ with $x$ is described by the JIMWLK- non-linear RG equations [15]. These equations form an infinite hierarchy of ordinary differential equations for the gluon correlators $\langle A_1 A_2 \cdots A_n\rangle_Y$, where $Y = \ln(1/x)$ is the rapidity. The JIMWLK equation for an arbitrary operator $\langle O\rangle$ is

$$\frac{\partial\langle\mathcal{O}[\alpha]\rangle_Y}{\partial Y} = \left\langle \frac{1}{2}\int_{x_\perp,y_\perp} \frac{\delta}{\delta\alpha_Y^a(x_\perp)}\,\chi^{ab}_{x_\perp,y_\perp}[\alpha]\,\frac{\delta}{\delta\alpha_Y^b(y_\perp)}\mathcal{O}[\alpha]\right\rangle_Y , \tag{4}$$

where $\alpha = (\nabla_\perp^2)^{-1}\rho$. Here $\chi$ is a non-local object expressed in terms of path ordered (in rapidity) Wilson lines of $\alpha$ [2]. This equation is analogous to a (generalized) functional Fokker-Planck equation, where $Y$ is the "time" and $\chi$ is a generalized diffusion coefficient. It illustrates the stochastic properties of operators in the space of gauge fields at high energies. For the gluon density, which is proportional to a two-point function $\langle\alpha^a(x_\perp)\alpha^b(y_\perp)\rangle$, one recovers the BFKL equation in the limit of low parton densities.

## 3 Dipoles in the CGC

In the limit of large $N_c$ and large $A$ ($\alpha_S^2 A^{1/3} \gg 1$), the JIMWLK hierarchy closes for the two point correlator of Wilson lines because the expectation value of the product of traces of Wilson lines factorizes into the product of the expectation values of the traces:

$$\langle \mathrm{Tr}(V_x V_z^\dagger)\mathrm{Tr}(V_z V_y^\dagger)\rangle \longrightarrow \langle \mathrm{Tr}(V_x V_z^\dagger)\rangle\,\langle \mathrm{Tr}(V_z V_y^\dagger)\rangle\,, \tag{5}$$

where $V_x = \mathcal{P}\exp\left(\int dz^-\alpha^a(z^-,x_\perp)T^a\right)$. Here $\mathcal{P}$ denotes path ordering in $x^-$ and $T^a$ is an adjoint SU(3) generator. In Mueller's dipole picture, the cross-section for a dipole scattering off a target can be expressed in terms of these 2-point dipole operators as [16, 17]

$$\sigma_{q\bar{q}N}(x,r_\perp) = 2\int d^2b\;\mathcal{N}_Y(x,r_\perp,b)\,, \tag{6}$$

where $\mathcal{N}_Y = 1 - \frac{1}{N_c}\langle\mathrm{Tr}(V_x V_y^\dagger)\rangle_Y$, the imaginary part of the forward scattering amplitude. Note that the size of the dipole, $\vec{r}_\perp = \vec{x}_\perp - \vec{y}_\perp$, and the impact parameter, $\vec{b} = (\vec{x}_\perp + \vec{y}_\perp)/2$. The JIMWLK equation for the two point Wilson correlator is identical in the large $A$, large $N_c$ mean field limit to an equation derived independently by Balitsky and Kovchegov — the Balitsky-Kovchegov equation [18], which has the operator form

$$\frac{\partial\mathcal{N}_Y}{\partial Y} = \frac{\alpha_S N_c}{\pi}\,\mathcal{K}_{\mathrm{BFKL}}\otimes\left\{\mathcal{N}_Y - \mathcal{N}_Y^2\right\}\,. \tag{7}$$

Here $\mathcal{K}_{\mathrm{BFKL}}$ is the well known BFKL kernel. When $\mathcal{N} \ll 1$, the quadratic term is negligible and one has BFKL growth of the number of dipoles; when $\mathcal{N}$ is close to unity, the growth saturates. The approach to unity can be computed analytically [19]. The B-K equation is the simplest equation including both the Bremsstrahlung responsible for the rapid growth of amplitudes at small $x$ as well as the repulsive many body effects that lead to a saturation of this growth.

A saturation condition which fixes the amplitude at which this change in behavior is significant, say $\mathcal{N} = 1/2$, determines the saturation scale. One obtains $Q_s^2 = Q_0^2 \exp(\lambda Y)$, where $\lambda = c\alpha_S$ with $c \approx 4.8$. The saturation condition affects the overall normalization of this scale but does not affect the power $\lambda$. In fixed coupling, the power $\lambda$ is large and there are large pre-asymptotic corrections to this relation-which die off only slowly as a function of $Y$. BFKL running coupling effects change the behavior of the saturation scale completely–one goes smoothly at large $Y$ to $Q_s^2 = Q_0^2 \exp(\sqrt{2b_0 c(Y+Y_0)})$ where $b_0$ is the coefficient of the one-loop QCD $\beta$-function. The state of the art computation of $Q_s$ is the work of Triantafyllopoulos, who obtained $Q_s$ by solving NLO-resummed BFKL in the presence of an absorptive boundary (which corresponds to the CGC) [20]. The pre-asymptotic effects are much smaller in this case and the coefficient $\lambda \approx 0.25$ is very close to the value extracted from saturation model fits to the HERA data [21]. Fits of CGC inspired models to the HERA data have been discussed elsewhere [22] and will not be discussed here.

## 4 Hadronic scattering and $k_\perp$ factorization in the CGC

Collinear factorization is the pQCD mechanism to compute hard scattering. At collider energies, a new window opens up where $\Lambda_{\rm QCD}^2 \ll M^2 \ll s$, where $M$ is the invariant mass of the final state. In principle, cross-sections in this window can be computed in the collinear factorization language–however, one needs to sum up large logarithmic corrections in $s/M^2$. An alternative formalism is that of $k_\perp$-factorization [23, 24], where one has a convolution of $k_\perp$ dependent "un-integrated" gluon distributions from the two hadrons with the hard scattering matrix. In this case, the in-coming partons from the wavefunctions have non-zero $k_\perp$. Levin et al. [25] suggested that at high energies the typical $k_\perp$ is the saturation scale $Q_s$. The rapidity dependence of the unintegrated distributions is given by the BFKL or BK equations. However, unlike the structure functions, it has not been proven that these unintegrated distributions are universal functions.

At small $x$, both the collinear factorization and $k_\perp$ factorization limits can be understood in a systematic way in the framework of the Color Glass Condensate. The expectation value of an operator $\mathcal{O}$ can be computed as

$$\langle \mathcal{O} \rangle_Y = \int [d\rho_1]\,[d\rho_2]\, W_{x_1}[\rho_1]\, W_{x_2}[\rho_2]\, \mathcal{O}(\rho_1, \rho_2)\,, \tag{8}$$

where $Y = \ln(1/x_F)$ and $x_F = x_1 - x_2$. Quantum information, to leading logarithms in $x$, is contained in the source functionals $W_{x_1(x_2)}[\rho_1(\rho_2)]$ of the two hadrons. The operator $\mathcal{O}$ corresponding to the final state is expressed in terms of gauge fields $A^\mu[\rho_1, \rho_2](x)$. Inclusive gluon production in the CGC is computed by solving the Yang-Mills equations $[D_\mu, F^{\mu\nu}]^a = J^{\nu,a}$ for $A^\mu[\rho_1, \rho_2]$, where the current is given by $J^\nu = \rho_1\, \delta(x^-)\delta^{\nu+} + \rho_2\, \delta(x^+)\delta^{\nu-}$ with initial conditions determined by the Yang-Mills fields of the two hadrons before the collision. These are obtained self-consistently by matching the solutions of the Yang-Mills equations on the light cone [26]. Since we have argued in Section 2 that we can *compute* the Yang-Mills fields in the nuclei before the collision, the classical problem is in principle completely solvable. Quantum corrections not enhanced by powers of $\alpha_S \ln(1/x)$ can be computed systematically. Those terms enhanced by powers of $\alpha_S \ln(1/x)$ are absorbed into the weight functionals $W[\rho_{1,2}]$.

Hadronic scattering in the CGC can therefore be studied through a systematic power counting in the density of sources in powers of $\rho_{1,2}/k_{\perp;1,2}^2$. This power counting is more relevant at high energies than whether the incoming projectile is a hadron or a nucleus. In addition, one can study the applicability of collinear and $k_\perp$ factorization at small $x$ in this approach.

The power counting is applicable as well to a proton at small $x$. The relevant quantity here is $Q_s$, which, as one may recall, is enhanced both for large $A$ and small $x$. As long as $k_\perp \gg Q_s \gg \Lambda_{\rm QCD}$, one can consider the proton or nucleus as being dilute. To lowest order in $\rho_{p1}/k_\perp^2$ and $\rho_{p2}/k_\perp^2$, one can compute inclusive gluon production analytically [26]. At large transverse momenta, $Q_s \ll k_\perp$, the scattering can be expressed in a $k_\perp$-factorized form. The inclusive cross-section is expressed as the product of two unintegrated ($k_\perp$ dependent) distributions times the matrix element for the scattering. The comparison of this result to the collinear pQCD $gg \to gg$ process and the $k_\perp$ factorized $gg \to g$ was performed in Ref. [27]. At this order, the result is equivalent to the pQCD result first derived by Gunion and Bertsch [28]. This result for gluon production is substantially modified, as we shall discuss shortly, by high parton density effects either because the target is a large nucleus or because small values of $x$ are being probed in the hadron (as in forward $pp$ scattering).

$k_\perp$ factorization is a good assumption at large momenta for quark pair-production. This was worked out in the CGC approach by François Gelis and myself [29]. The result for inclusive quark pair production can be expressed in $k_\perp$ factorized form as

$$\frac{d\sigma_1}{dy_p dy_q d^2p_\perp d^2q_\perp} \propto \int \frac{d^2k_{1\perp}}{(2\pi)^2}\frac{d^2k_{2\perp}}{(2\pi)^2}\delta(k_{1\perp}+k_{2\perp}-p_\perp-q_\perp) \times\phi_1(k_{1\perp})\phi_2(k_{2\perp})\frac{\mathrm{Tr}\left(\left|m_{ab}^{-+}(k_1,k_2;q,p)\right|^2\right)}{k_{1\perp}^2 k_{2\perp}^2}\,, \tag{9}$$

where $\phi_1$ and $\phi_2$ are the unintegrated gluon distributions in the projectile and target respectively (with the gluon distribution defined as $xG(x,Q^2) = \int_0^{Q^2} d(k_\perp^2)\,\phi(x,k_\perp)$).

The matrix element $\mathrm{Tr}\left(\left|m_{ab}^{-+}(k_1,k_2;q,p)\right|^2\right)$ is identical to the result derived in the $k_\perp$–factorization approach [23, 24]. In the limit $|\vec{k_{1\perp}}|, |\vec{k_{2\perp}}| \to 0$, $\mathrm{Tr}\left(\left|m_{ab}^{-+}(k_1,k_2;q,p)\right|^2\right)\Big/(k_{1\perp}^2 k_{2\perp}^2)$ is well defined–after integration over the azimuthal angles in Eq. (9), one obtains the usual matrix element $|\mathcal{M}|^2_{gg\to q\bar{q}}$, recovering the lowest order pQCD collinear factorization result.

### 4.1 Gluon and quark production in forward $pp$ and $pA$ collisions

Many analytical results are available when one of the hadrons is dilute and the other is dense. This may correspond to either $pA$ collisions or forward $pp$ collisions. One solves the Yang–Mills equations $[D_\mu, F^{\mu\nu}] = J^\nu$ with the light cone sources $J^{\nu,a} = \delta^{\nu+}\,\delta(x^-)\,\rho_p^a(x_\perp) + \delta^{\nu-}\,\delta(x^+)\,\rho_A^a(x_\perp)$, to determine the gluon field produced-to lowest order in the source density of one projectile ($\rho_p/k_\perp^2 \ll 1$)and to all orders ($\rho_A/k_\perp^2 \sim 1$) in the source density of the other. The inclusive gluon production cross-section, in this framework, was first computed by Kovchegov and Mueller [30] and shown to be $k_\perp$ factorizable in Ref. [31, 34]. The "unintegrated" gluon distribution in the dense system however is here replaced by the gluon "dipole" distribution $\mathcal{N}_Y$ we discussed previously. It is no longer a leading twist object but includes all twists enhanced by high parton density effects. The well known "Cronin" effect observed in Deuteron-Gold collisions at RHIC is obtained in this formalism and can be simply understood in terms of the multiple scattering of a parton from the projectile with those in the target. The energy evolution of the dipole distribution is given by the BK equation, leading to a suppression of the Cronin effect at high densities due to the shadowing of nuclear distributions. This prediction appears to be confirmed by the RHIC data. The "dipole" operators extracted from DIS can therefore be used to predict inclusive hadron production in $pp$ and $pA$ collisions. One can similarly compute Drell-Yan and photon production in forward $pp$ and $pA$ collisions [33, 35].

Unlike gluon production, neither quark pair-production nor single quark production is strictly $k_\perp$ factorizable. The pair production cross-section can however still be written in $k_\perp$ factorized form as a product of the unintegrated gluon distribution in the proton times a sum of terms with three unintegrated distributions, $\phi_{g,g}$, $\phi_{q\bar{q},g}$ and $\phi_{q\bar{q},q\bar{q}}$. These are respectively proportional to 2-point (dipole), 3-point and 4-point correlators of the Wilson lines we discussed previously. Again, these operators include all twist contributions. For instance, the distribution $\phi_{q\bar{q},g}$ is the product of fundamental Wilson lines coupled to a $q\bar{q}$ pair in the amplitude and adjoint Wilson lines coupled to a gluon in the complex conjugate amplitude. For large transverse momenta or large-mass pairs, the 3-point and 4-point distributions collapse to the unintegrated gluon distribution, and we recover the previously discussed $k_\perp$-factorized result for pair production in the dilute/$pp$-limit. Single quark distributions are straightforwardly obtained and depend only on the 2-point quark and gluon correlators and the 3-point correlators. For Gaussian sources, as in the McLerran-Venugopalan-model, these 2-,3- and 4-point functions can be computed exactly as discussed in Ref. [32].

The situation gets complicated when one enters a regime where both projectiles are dense–as defined in our power counting. $k_\perp$ factorization breaks down decisively and analytical approaches are likely not possible. Nevertheless, numerical techniques have been developed, which allow the computation of final states, at least to leading logs in $x$ [38].

The results for gluon and quark production in forward $pp$ and $pA$ or $dA$ collisions (for a review, see Ref. [37]), coupled with the previous results for inclusive and diffractive [33–36] distributions in DIS, suggest an important new paradigm. *At small $x$ in DIS and hadron colliders, previously interesting observables such as quark and gluon densities are no longer the only observables to capture the relevant physics. Instead, they should be complemented by dipole and multipole correlators of Wilson lines that seem ubiquitous in all high energy processes and are similarly gauge invariant and process independent. The renormalization group running of these operators may be a powerful and sensitive harbinger of new physics.*

# Vector Boson Fusion at CMS

*N. Amapane[1,a], M. Arneodo[2], R. Bellan[1], S.Bolognesi[1], G. Cerminara[1], C. Mariotti[3]*
1) Torino University and INFN Torino, 2) University of Eastern Piedmont, Novara and INFN Torino,
3) INFN Torino, a) now at CERN

**Abstract**
The processes of boson-boson scattering and of Higgs production in boson-boson fusion hold the key to electroweak symmetry breaking. A preliminary study has been performed using a fast simulation of the CMS detector. The results are encouraging and suggest that, after few years of data taking at LHC, the region above 1 TeV can be explored, which is interesting if the Higgs is not found.

## 1 Vector Boson Fusion at CMS

### 1.1 Introduction

The Standard Model predicts that, without the Higgs boson, the scattering amplitude of the *longitudinally polarized vector boson* ($V_L$) fusion process violates unitarity at about 1-1.5 TeV. The longitudinal polarization of the *V* arises from the *V* getting massive, i.e. when the symmetry breaks spontaneously. The cross section as a function of the $V_LV_L$ invariant mass will show a resonance at M($V_LV_L$)=M(H) if the Higgs is there; otherwise, the cross section will deviate from the Standard Model prediction at high values of M(VV). Therefore, *VV* scattering can probe the Electroweak Symmetry Breaking with or without the assumption the Higgs mechanism.

### 1.2 The Signal Selection

Two channels have been studied using *Pythia* [1] and the *CMS Fast Simulation* [2]:

- $pp \to \mu\mu jjjj$ [3] through the processes:
  - $pp \to V_LV_Ljj \to Z_LZ_Ljj \to \mu\mu jjjj$,
  - $pp \to Z_LW_Ljj \to Z_LW_Ljj \to \mu\mu jjjj$.
- $pp \to \mu\nu jjjj$ [4] through the process:
  - $pp \to V_LV_Ljj \to W_LW_Ljj \to \mu\nu jjjj$.

The study has been done for high Higgs masses: $m_H = 500$ GeV and $m_H = 1000$ GeV, and for the no-Higgs scenario. The latter has been simulated in Pythia by setting $m_H = 10000$ GeV (the Higgs exchange diagram is suppressed by a $m_H^2$ term in the denominator of the Higgs propagator). The cross sections of the signal processes are shown in Table 1.

**Table 1:** Signal cross section (in fb) for different Higgs masses.

| **Processes** | $m_H = 500$ GeV | $m_H = 1000$ GeV | $m_H = 10\,000$ GeV |
|---|---|---|---|
| $pp \to Z_LZ_Ljj \to \mu\mu jjjj$ | 9.1 | 3.0 | 1.7 |
| $pp \to Z_LW_Ljj \to \mu\mu jjjj$ | 0.7 | 1.0 | 1.5 |
| $pp \to W_LW_Ljj \to \mu\nu jjjj$ | 64.4 | 26.9 | 19.7 |

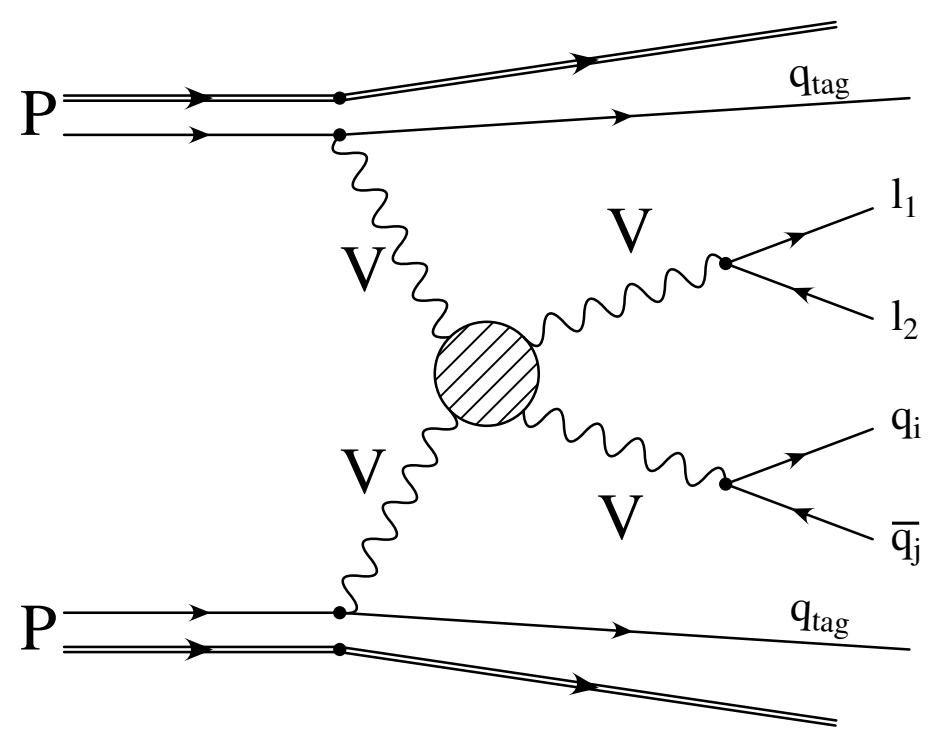

**Fig. 1:** The signal topology. $l_1$ and $l_2$ can be $\mu^{\pm}$ or $\mu$ and $\nu$.

The signal has a well defined topology (see Figure 1):

- one $\mu^+$ and one $\mu^-$ (or one $\mu$ and one $\nu$) in the final state, with high $p_T$ and low $\eta$ coming from the *Z* (*W*) boson;
- two jets with high $p_T$ and low $\eta$, coming from the vector boson decay;
- two energetic jets with high $p_T$, in the forward-backward regions (large $\eta$ and $\Delta\eta$).

The aim of the work is to reconstruct the invariant mass of the *VV*-fusion system in both the channels and estimate its resolution. We also attempted a first estimate of the signal to background ratio assuming that the main background processes are:

- $t\bar{t}$ *background*: a six fermion final state, like the signal, but the jets are mainly in the central region; therefore, by requiring two jets at high $\eta$ and with a large $\Delta\eta$ between them this kind of background can be rejected.
- *VV associated production*: a four fermion final state; it needs however to be kept under control in the case in which one boson decays leptonically since there are several jets from gluon radiation in the final state. The most effective variables to distinguish this background from the signal are the transverse momenta of the jets and of the leptons.
- *V plus one and two hard jets*: it is simple to reject this background because it has a topology not very similar to that of the signal and the additional jets have a very low $p_T$ (since they are generated by the parton shower). However it is fundamental to keep it under control since it has a very large cross section.

The cross section of the background processes are shown in Table 2.

**Table 2:** Background cross section (in fb).

| **Background** | Cross Section [fb] | **Background** | Cross Section [fb] |
|---|---|---|---|
| $t\bar{t}$, 1 $\mu$ | $622 \cdot 10^3$ | $t\bar{t}$, 1 $\mu^-$ and 1 $\mu^+$ | $620 \cdot 10^3$ |
| $ZZ \to \mu^-\mu^+jj$ | 653 | $ZW \to \mu^-\mu^+jj$ | 663 |
| $WW \to \mu\nu^+jj$ | $11 \cdot 10^3$ | $W + jj \to \mu\nu jj$ | $77 \cdot 10^3$ |
| $Z + jet \to \mu^-\mu^+j$ | $13 \cdot 10^6$ | $W + j \to \mu\nu j$ | $184 \cdot 10^6$ |

### 1.3 The Results

A set of cuts has been applied to enhance the signal with respect to the background. A good resolution (estimated using MC info) on the most important observables has been achieved. In particular:

- $Z \to \mu\mu$ invariant mass: $R_z \sim 1.5\%$;
- $V \to jj$ invariant mass, $\mu\mu jjjj$ channel: $R_v \sim 13\%$;
- $V \to jj$ invariant mass, $\mu\nu jjjj$ channel: $R_v \sim 10\%$.

The difference between the two latter resolutions reflects the fact that for the $jjjj\mu\mu$ channel the pile-up has been considered whereas in the $jjjj\mu\nu$ it was not. The resolution on the energy scale of the process ($M_{inv}(VV)$) is:

- $4\%$ for the $pp \to \mu\mu jjjj$ channel;
- $8\%$ for the $pp \to \mu\nu jjjj$ channel.

The difference is due to the worse resolution on the neutrino $p_T$ and $p_z$ reconstruction. The resulting background efficiency is lower than one percent while the signal efficiency reaches $30\%$ for the $jjjj\mu\mu$ channel and $50\%$ for the $jjjj\mu\nu$ channel. A high significance ($S/\sqrt{B}$) has been achieved for an integrated luminosity of $100 fb^{-1}$: for the $\mu\mu jjjj$ samples it is about 8 in the interval $M_{inv}^{VV} \in [0,1]$ TeV for the Higgs mass set to 500 GeV and about 10 for $M_{inv}^{VV} > 1$ TeV for the no-Higgs scenario. Similar values have been obtained for the $\mu\nu jjjj$ channel: a significance of about 5, in the interval $M_{inv}^{VV} \in [0,1]$ TeV, for the Higgs mass set to 500 GeV and about 2.4 in the interval $M_{inv}^{VV} > 1$ TeV for the no-Higgs scenario. In Figs. 2 (no-Higgs scenario) and 3 ($m_H = 500$ GeV) the number of reconstructed events and the selection efficiency as a function of the *VV* invariant mass are shown.

### 1.4 Future Plans

Further studies are in progress, since for those presented here the Pythia generator was used, which only simulates a subset of the relavant diagrams, and cannot simulate the full set of background processes (notably not the scattering of transversely polarised vector bosons). To better describe the signal (and the background as well) a *Matrix Element Monte Carlo* must be used. *Phase* [5] is the best candidate, since it simulates all processes that lead to a six fermion final state, at order $\alpha_{QED}^6$. Up to now only the channel $pp \to \mu\nu jjjj$ has been computed; therefore for the the $\mu\mu jjjj$ final state the *MadGraph* [6] event generator was used. This can simulate the $2l4j$ final state through the production (in Narrow Width Approximation) of intermediate vector bosons and their subsequent semileptonic decay.

Moreover it is crucial to redo the analysis, processing the events through the *Full Simulation* [7] of the CMS detector in order to properly take into account the detector resolution.

### 1.5 Summary

In conclusion, Electroweak Symmetry Breaking can be probed through the fusion of longitudinally polarized vector bosons with the CMS detector at LHC. The signal reconstruction and the background rejection algorithms have been successfully tested with the Fast Simulation. In the near future the study will be repeated with the Full Simulation of the detector and with dedicated generators.

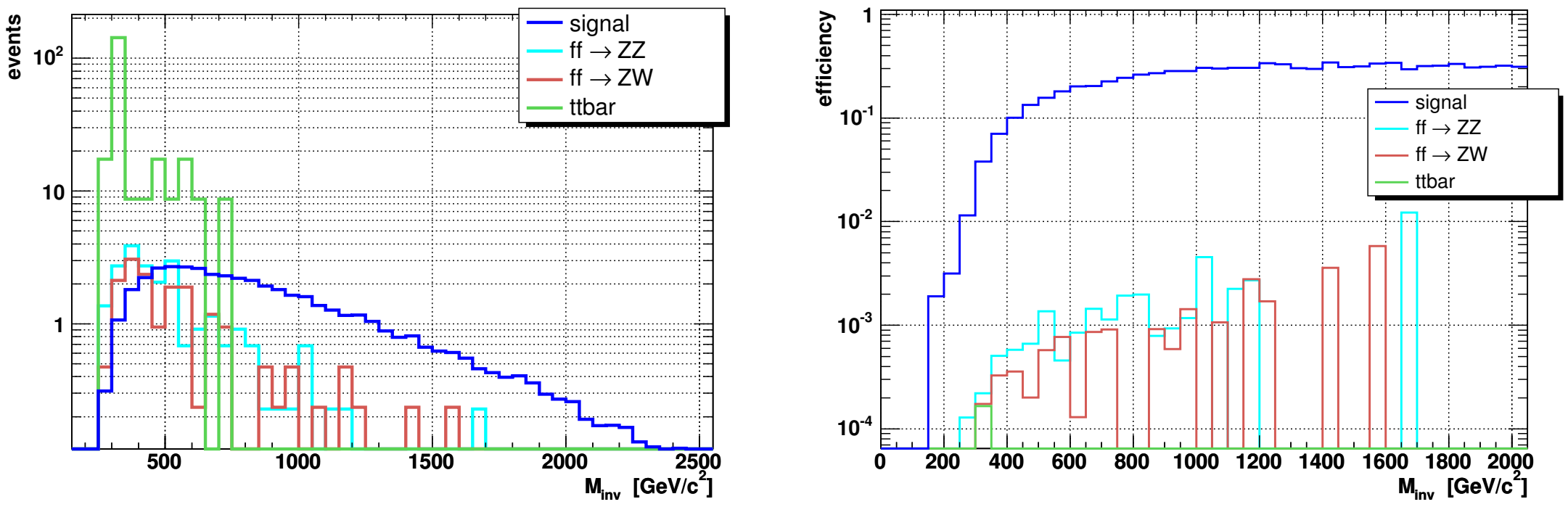

**Fig. 2:** (Left) Number of reconstructed events as a function of the *VV* invariant mass and (Right) the selection efficiency as a function of the invariant mass of the VV-fusion process; both for the $\mu\mu jjjj$ final state in the no-Higgs scenario and an integrated luminosity of 100 fb$^{-1}$.

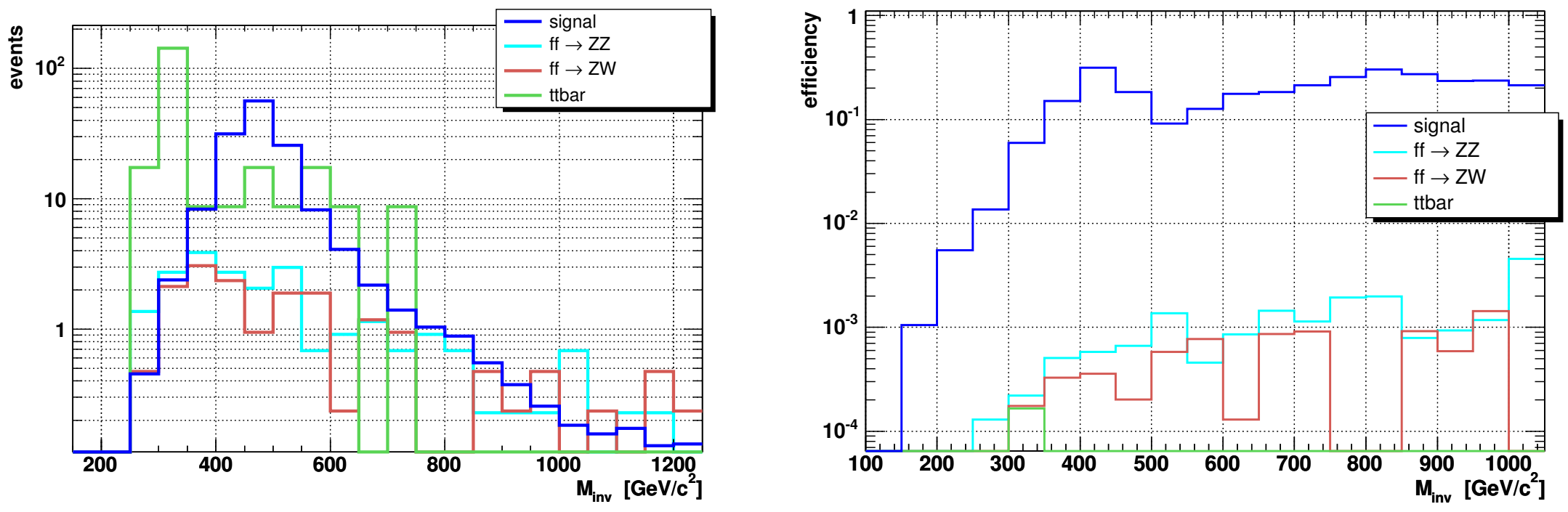

**Fig. 3:** (Left) Number of reconstructed events as a function of the *VV* invariant mass and (Right) the selection efficiency as a function of the invariant mass of the VV-fusion process; both for the $\mu\mu jjjj$ final state for $m_H = 500$ GeV and an integrated luminosity of 100 fb$^{-1}$.

# Part VI

# Working Group 5: Monte Carlo Tools

## List of participants in the working group

V. Lendermann, S. Nikitenko, E. Richter-Was, P. Robbe, M.Seymour

P. Bartalini, O. Behnke, D. Beneckestein, D. Bourilkov, G. Bruni, A. Buckley, A. Bunyatyan, C. Buttar, J. Butterworth, S. Butterworth, S. Caron, S. Chekanov, J. Collins, M. Corradi, B. Cox, Z. Czyczula, A. Dainese, G. Davatz, R. Field, A. Geiser, S. Gieseke, T. Gleisberg, P. Golonka, G. Grindhammer, R. Craig Group, O. Gutsche, S. Höche, G. Iacobucci, S. Jadach, A. W. Jung, H. Jung, B. Kersevan, F. Krauss, S. Lausberg, N. Lavesson, K. Lohwasser, L. Lönnblad, M. Mangano, S. Maxfield, J. Monk, A.Moraes, C.Pascaud, K.Rabbertz, P.Richardson, L.Rinaldi, E.Rodrigues, M.Ruspa, A.Schälicke, T. Schörner-Sadenius, S. Schumann, T. Sjöstrand, M. Skrzypek, A. Sobol, J. Stirling, P. Szczypka, P. Thompson, N. Tuning, J. Turnau, B. Ward, Z. Was, B. Waugh, M. Whalley, J. Winter

# Introduction to Monte Carlo Tools

*V. Lendermann* [1], *A. Nikitenko* [2], *E. Richter-Was* [3], *P. Robbe* [4], *M. H. Seymour* [5]
[1] Kirchhoff-Institut für Physik, Universität Heidelberg, Im Neuenheimer Feld 227, 69120 Heidelberg, Germany;
[2] Imperial College, London, UK; on leave from ITEP, Moscow, Russia;
[3] Institute of Physics, JU/IFJ-PAN, 30-059 Krakow, ul. Reymonta 4, Poland;
[4] Laboratoire de l'Accélérateur Linéaire, Université Paris-Sud, 91898 Orsay, France;
[5] School of Physics & Astronomy, University of Manchester, UK, and Theoretical Physics Group, CERN.

**Abstract**
The activities of working group 5 'Monte Carlo Tools' of the HERA–LHC Workshop are summarized. The group concerned itself with the developments and tunings of Monte Carlo models in light of the HERA–LHC connection, interfaces and libraries for parton density functions, Monte Carlo running, validation and tuning frameworks as well as some data analysis tools.

## 1 Introduction

The goals of working group 5 were

- to review existing and developing Monte Carlo (MC) models used for studies at HERA and the LHC;
- to examine and possibly improve MC models for the LHC physics using HERA data;
- to prioritize possible measurements at HERA which would allow tuning of these MC models;
- to pursue the development of frameworks for running, validating and tuning of MC and analysis programs;
- to improve and further develop common interfaces and libraries used with MC event generators;
- to review data analysis tools developed by the HERA collaborations which can be useful for studies at the LHC.

Both theorists and experimentalists from the HERA and LHC communities came together, in order to share their experience, identify crucial issues, and discuss the future developments of the programs, libraries and frameworks.

The physics topics discussed in the group have overlapped with those of all the other working groups in this workshop. Therefore many presentations were given in common sessions with other groups, most notably with working group 2 'Multijet Final States and Energy Flows'. These presentations covered the models of multiparton interactions, new developments in parton shower models, matrix element / parton shower (ME+PS) matching and simulations of multijet final states. The contributions to the present proceedings reviewing these studies are published in the chapter of working group 2. Further contributions are summarized below.

## 2 Libraries for Parton Density Functions

In the past, the PDFLIB library [1] was the standard package containing parametrizations of the proton, photon and pion parton density functions (PDFs). The LHC studies have necessitated the development of a new library which should include not only the central values of PDFs but also the error sets. The PDFLIB interface appeared not well suited to meet the new requirements. Therefore a new library, LHAPDF (Les Houches Accord PDF library [2]) was created following the Les Houches meeting in 2001. During this workshop the library was extensively developed by *M. Whalley* and *D. Bourilkov* [3]. Several recent PDF sets were included, both those by the leading theory groups and by the H1 and ZEUS collaborations. The fits are important for better estimations of PDF uncertainties. A particularly interesting cross check would be, for example, the comparison of the fits provided by the TeVatron collaborations with the independent predictions obtained by the DGLAP evolution of the HERA PDFs to the TeVatron region.

The photon and pion PDFs were also included in the library, thus allowing its use for all HEP analyses, in particular, for HERA studies. The library has thus developed to the level at which it can fully replace PDFLIB. Several tests of the applicability and performance of the library were made within the H1 collaboration (*V. Lendermann*).

Another topic discussed is the creation of a collection of diffractive PDF parametrizations. The project was presented by *F. P. Schilling* in a meeting of working group 4 'Diffraction'.

## 3 Monte Carlo event generators

The status and plans for major leading order (LO) and next-to-leading order (NLO) QCD programs, as well as generators with $k_T$ factorisation were discussed.

### 3.1 Leading order Monte Carlo programs

Currently, the major FORTRAN MC event generators, PYTHIA [4] and HERWIG [5], are undergoing the transition to object-oriented software technologies. The C++ versions of both generators, PYTHIA7 [6] and HERWIG++ [7], are built in the common framework THEPEG [8] which is based on the CLHEP class library [9].

PYTHIA7/THEPEG includes some basic $2 \rightarrow 2$ matrix elements (ME), several built-in PDF parametrizations, remnant handling, initial- and final-state parton showers, Lund string fragmentation and particle decays. There have been plans to rework the fragmentation model, in order to include junction strings, and to implement multiple interactions (*L. Lönnblad*). However, *T. Sjöstrand* recently started a completely new C++ implementation, PYTHIA8 [10].

In parallel to the work on the C++ PYTHIA versions, the development of the FORTRAN PYTHIA6 continues. It remains the main platform for new physical concepts. Version 6.3 [11, 12] includes a completely new framework for simulation of parton showers and multiple interactions by *T. Sjöstrand* and *P. Skands*. Currently this version works for $pp$ interactions only.

The development of HERWIG continues mainly in C++ (*S. Gieseke*, *A. Ribon*, *P. Richardson*, *M. H. Seymour*, *P. Stephens* and *B.R. Webber*). The current FORTRAN version 6.5 is foreseen as the final FORTRAN version of HERWIG (apart from possible bug fixes). It is interfaced to the JIMMY generator for multiple interactions (*J. M. Butterworth*, *J. R. Forshaw*, *M. H. Seymour*, and *R. Walker*).

HERWIG++ includes a new parton shower algorithm and an improved model of cluster fragmentation. The $e^+e^-$ event generation is implemented in HERWIG++ 1.0. The next version including hadronic interactions is in progress. The plans for the near future are to fully implement the matrix element–parton shower matching according to the Catani–Krauss–Kuhn–Webber (CKKW) scheme [13], as well as multiple interactions. A new framework for accessing particle data and simulations of particle decays is currently being constructed by *P. Richardson*. The treatment of hadronic decays will include spin correlations.

Further physics models can be incorporated into the same THEPEG framework. In particular, it is planned to make a C++ version of ARIADNE [14] based on THEPEG (*L. Lönnblad*). ARIADNE implements the Dipole Cascade Model (DCM) [15, 16] as an alternative to the DGLAP-based shower models used in PYTHIA or HERWIG.

Despite the great success of ARIADNE in modelling hadronic final state observables, as measured at LEP and HERA, additional work is necessary to make ARIADNE fully suitable for modelling inter-actions at the LHC. This was, in particular, shown in a study presented by *Z. Czyczula* on the impact of parton shower models on the generation of $bbH$, $H \to \tau\tau$ at the LHC [17]. The planned features include a remodelling of initial-state $g \to q\bar{q}$ splittings as well as the introduction of the $q \to g^{\star}q$ process.

Another study was presented by *N. Lavesson* of ME+PS matching in ARIADNE on the example of $W$+jet production at the TeVatron [18]. It is also planned to include the matching to fixed order tree-level matrix elements *à la* CKKW [13, 18, 19] for the most common subprocesses at the LHC. When these plans are realised (we hope during 2006), it should be safe to use ARIADNE for LHC predictions.

An alternative C++ implementation is performed in the SHERPA program (*T. Gleisberg*, *S. Höche*, *F. Krauss*, *A. Schälicke*, *S. Schumann*, *J. Winter*) [20] which is capable of simulating lepton–lepton, lepton–photon, photon–photon and fully hadronic collisions, such as proton–proton reactions. In its current version SHERPA includes the ME generator AMEGIC++ providing the matrix elements for hard processes and decays in the SM, MSSM and the ADD model, the parton shower module APACIC++ containing virtuality-ordered initial- and final-state parton showers, ME+PS matching using the CKKW algorithm, the AMISIC++ module for a simple hard underlying event model taken from PYTHIA and an interface to the PYTHIA string fragmentation and hadron decays. Studies were presented on ME+PS matching considering $W/Z$+jet production at the TeVatron and at the LHC (*S. Schumann*) [18], and on the underlying event simulations (*S. Höche*) [12].

None of the above C++ programs is available for $ep$ interactions yet, and so no applications and tests of these programs at HERA have been possible.

Further talks were given on the RAPGAP event generator [21] by *H. Jung* and on the ACERMC event generator [22] by *B. Kersevan* and *E. Richter-Was*. RAPGAP is one of the major generators used at HERA. It includes leading-order QCD matrix elements, LEPTO [23] and ARIADNE parton cascade models, as well as simulations of hard diffraction. Both $ep$ and $pp$ versions are available. Recently, the Les Houches Accord interface for fragmentation models was included; this allows the choice between the PYTHIA and HERWIG fragmentation models. This feature may allow better estimations of measurement uncertainties accounting for the transition from the parton to the hadron level of final states. It is planned to include double-diffractive scattering for $pp$ collisions to allow simulation of diffractive Higgs production.

The ACERMC event generator simulates the Standard Model backgrounds to the Higgs production in $pp$ collisions. It includes LO QCD matrix elements produced by MADGRAPH/HELAS [24], as well as both PYTHIA and HERWIG parton shower and fragmentation models via the Les Houches Accord interface. During this workshop, the ARIADNE parton shower model and the LHAPDF library were implemented. The program is also interfaced to TAUOLA [25] for precise treatment of $\tau$ decays and to PHOTOS [25, 26] for simulations of QED radiative decays. The study on the impact of parton shower models on generation of $bbH$, $H \to \tau\tau$ at the LHC [17], mentioned above, was performed using ACERMC.

The program can be linked with the ACERDET package [27] which provides a fast and simplified simulation of the expected ATLAS detector effects (energy smearing, acceptance corrections) as well as the usual analysis steps (jet reconstruction algorithms, isolation criteria, etc.). This allows a quick estimation of the feasibility of measurements in an LHC experiment, not necessarily by the members of the experimental collaboration.

### 3.2 NLO Monte Carlo programs

NLO QCD calculations are required to make theoretical predictions at the level of precision currently reached in particle scattering experiments. However, writing a hadron level MC program implementing an NLO model is a very complicated task, which has currently been solved only for a few $pp$ reactions [28]. An important step forward would be an $ep$ version of MC@NLO. It would be a major benefit for HERA studies of heavy quark and multijet production and would also allow an extensive validation of the NLO QCD calculations with HERA data. The development of the program started recently [29].

### 3.3 Monte Carlo programs with $k_T$ factorization

The CASCADE event generator presented by *H. Jung* [30] provides an implementation of the CCFM model for parton cascades [31]. The program was very successful in describing hadronic final states at low $x$ at HERA. First applications of CASCADE for the studies at the LHC were presented by *G. Davatz* [17]. The plans include an implementation of quark lines into CCFM cascades (currently, only gluon lines are implemented), as well as a new model for multiparton interactions based on the AGK cutting rules [32].

A reformulation of the CCFM model into the link dipole chain (LDC) model [33] provides a simplified formalism, which has been incorporated into the LDCMC program by *H. Kharraziha* and *L. Lönnblad* [14, 34]. An LDCMC version for deep inelastic *ep* scattering is available within the framework of the ARIADNE event generator. A *pp* version is planned.

In conjunction with these models, special sessions of working group 2 were dedicated to possibilities of determining the unintegrated parton distributions in the proton [35].

## 4 Comparisons of MC models with data

Models for particular subprocesses and their tuning using HERA data were reviewed in the corresponding working groups. As mentioned above, many discussions were carried out in the common sessions of working group 5 with the other groups.

One topic considered in WG5 is a comparison of leading proton data with several MC models (*G. Bruni*, *G. Iacobucci*, *L. Rinaldi*, *M. Ruspa*) [36]. The *ep* data from ZEUS and *pp* data from ISR and fixed-target experiments were confronted to the HERWIG (together with POMWIG [37] and SANG to simulate diffraction), LEPTO, ARIADNE and PYTHIA simulations. This exercise revealed that the simulation of the leading-proton momenta, both longitudinal and transverse to the beams, does not reproduce the properties of the data.

This study can be especially important for the understanding of diffractive processes and backgrounds for them at the LHC.

## 5 MC running, tuning and validation frameworks

During this workshop much progress was made in developing common frameworks that provide a convenient handling of MC and analytical programs and allow quick comparisons of MC simulations and analytical calculations with the results of HERA and other HEP experiments. The developments of HZTOOL/JETWEB, RUNMC and NLOLIB packages were presented and actively discussed.

The HZTOOL [38, 39] library provides a comprehensive collection of FORTRAN routines to produce various distributions using Monte Carlo event generators. The routines allow easy reproduction of the experimental distributions by modelling programs and give access to published data from the EMC, SPS, LEP, HERA and TeVatron experiments. A number of studies for the LHC and the future linear collider are also included. The library can be linked with all major FORTRAN MC event generators, and with a number of NLO QCD programs from the NLOLIB package (see below). The development of the library started within the workshop ‘Future Physics at HERA’ and steadily continued in the last few years.

In the current workshop, the emphasis was put on the HERA results relevant for the LHC [38]. Several measurements by H1 and ZEUS were implemented which allow tuning of multiparton interaction models in MC event generators (work by *D. Beneckenstein, A. Bunyatyan, J. M. Butterworth, H. Jung, S. Lausberg, K. Lohwasser, V. Lendermann, B. M. Waugh*). Common tunings of multiple interaction models based on the TeVatron and HERA results may constitute in the future an interesting outcome of the current efforts. Recent H1 and ZEUS results on heavy quark production in *ep* collisions were also added (*A. W. Jung, A. Geiser, O. Gutsche, P. Thompson*). In addition, calculations of benchmark cross-sections for heavy flavour production were included [40].

Based on HZTOOL, JETWEB [38,41] is a facility for tuning and validating Monte Carlo models through a World Wide Web interface. A relational database of reaction data and predictions from different models is accessed through a Java servlet, enabling a user to find out how well a given model agrees with a range of data and to request predictions using parameter sets that are not yet in the database.

The transition of experimental analysis frameworks and Monte Carlo generators to object-oriented software technologies necessitates a proper development of the MC running, tuning and validation frameworks. For this reason, HZTOOL/JETWEB is currently subject to a major redesign within the CEDAR ('Combined e-Science Data Analysis Resource for high energy physics') project [42]. CEDAR should comprise:

- an extensive archive of data from particle scattering experiments, based on the Durham HEP database [43];
- validation and tuning of Monte Carlo programs, parton distribution functions and other high-energy physics calculation programs, building on JETWEB;
- access to well-defined versions of these programs and code management support for developers;
- a standardized set of data formats for specifying HEP measurements as used in HepData and Monte Carlo event generator configurations as used in JETWEB;
- Grid compatibility for distribution of JETWEB Monte Carlo submissions and to enable secure addition of experimental data to the HepData catalogue by experimental collaborations.

A particularly important step in building CEDAR will be designing a C++ equivalent for HZTOOL, as well as providing an interface to the new C++ MC event generators.

A complementary approach using the object-oriented software design was realised in the RUNMC framework [44] by *S. Chekanov*. While JETWEB is a Web server system, RUNMC is a desktop application written in C++ and Java. It provides a unified approach to generate MC events and to analyse different MC models. All major FORTRAN MC event generators can be run via RUNMC. The output of FORTRAN MC programs is converted to C++ classes for further analysis and for graphical representation (histograms). The graphical user interface of RUNMC allows an initialization of MC models and histograms in a unified manner, and provides monitoring of the event generation. The program provides an interface to HZTOOL. It also allows loading of 'project files' which can contain external calculations, MC tunings, histogram definitions, etc. In particular, these files can include C++ data analysis code, similar to the HZTOOL FORTRAN analysis routines.

A further project, discussed in working group 5, is a common framework for the NLO QCD programs, NLOLIB [45], which was initiated within the workshop 'Monte Carlo Generators for HERA Physics'. Since hadron level Monte Carlo programs implementing QCD NLO calculations are not (yet) available for many processes, parton level NLO calculations are extensively used. NLOLIB is aimed at becoming a container for virtually all NLO QCD programs. It provides:

- a set-up for compiling and linking the programs on diverse UNIX platforms;
- a unified access to the NLO event records;
- a unified steering for parameters and settings;
- a unified access to PDF libraries;
- an interface to HZTOOL, thus allowing easy comparisons with experimental results;
- examples of the analysis code which can be linked with the library.

During the workshop the structure of the framework was further developed by *K. Rabbertz*. In addition to already implemented programs for $ep$ (DISENT [46], DISASTER++ [47], MEPJET [48]) and $e^+e^-$ (RACOONWW [49]) physics, an effort was made to integrate further $ep$ programs: NLOJET++ [50] (*K. Rabbertz*) and JETVIP [51,52] (*T. Schörner-Sadenius*). The integration of the NLO programs for $pp$ physics is surely possible, but requires additional effort.

## 6 Data analysis tools

Apart from MC related topics, general analysis tools, aimed at searches for specific final states, were presented.

One such tool is SBUMPS [53], currently being developed by *S. Chekanov*, which performs automatic searches for resonance peaks in invariant-mass distributions of two or more tracks. The program can be useful for searches of new states as well as for the reconstruction of known resonances.

Recently, interest in hadron spectroscopy at HERA increased sharply with the observations of narrow peaks in inclusive invariant-mass distributions which can be interpreted as pentaquarks [54]. These studies have inspired the development of the automated peak searching tool, which can be used in data analyses at any particle scattering experiment.

A general strategy for searches for new physics was presented by *S. Caron*. The approach was developed and used by the H1 Collaboration for searches of new phenomena at HERA [55]. It involves a statistical algorithm to search for deviations from the Standard Model in the distributions of the scalar sum of transverse momenta or invariant mass of final-state particles and to quantify their significance.

## 7 Conclusions

A number of interesting developments of MC models, programs, libraries and frameworks were presented in working group 5. The general status and prospects for major established and currently developed MC generators were reviewed. In common sessions with working group 2, the models of multiparton interactions, new developments in parton shower models, matrix element/parton shower matching and simulations of multijet final states were extensively discussed. Direct communication between theoreticians and experimentalists from the HERA and LHC communities allowed the pursuit of several developments and studies. In particular the recent advances in the development of the LHAPDF library, HZTOOL, RUNMC and NLOLIB frameworks were inspired by discussions within working group 5. It is hoped that this will help further studies on validation and tuning of the MC models for multiparton interactions, parton showers, and heavy flavour production.

# The Les Houches Accord PDFs (LHAPDF) and LHAGLUE

*M R Whalley*[†], *D Bourilkov*[‡] *and R C Group*[‡]
[†]University of Durham, Durham, DH1 3LE, UK
[‡]University of Florida, Gainesville, FL 32611, USA

**Abstract**
We describe the development of the LHAPDF library from its initial implementation following the Les Houches meeting in 2001 to its present state as a functional replacement for PDFLIB. Brief details are given of how to install and use the library together with the PDF sets available. We also describe LHAGLUE, an add-on PDFLIB look-a-like interface to LHAPDF, which facilitates using LHAPDF with existing Monte Carlo generators such as PYTHIA and HERWIG.

## 1 LHAPDF – Introduction

Parton Density Functions (PDFs), which describe the partonic content of hadrons, need to be well understood and of sufficiently high precision if theoretical predictions are to match the experimental accuracies expected from future LHC data. These PDFs, which are produced by several different groups (e.g. MRST, CTEQ, Alekhin and more recently ZEUS and H1), are derived from fitting deep inelastic and related hard scattering data using parameterisations at low $Q_0^2$ ($\approx$ 1–7 (GeV/c)$^2$) and evolving these to higher $Q^2$. These PDFs are typically presented as grids in $x$-$Q^2$ with suitable interpolation codes provided by the PDF authors. The CERN PDFLIB library [1] has to date provided a widely used standard FORTRAN interface to these PDFs with the interpolation grids built into the PDFLIB code itself. However, it is realised that PDFLIB would be increasingly unable to meet the needs of the new generation of PDFs which often involve large numbers of sets ($\approx$20–40) describing the uncertainties on the individual partons from variations in the fitted parameters. As a consequence of this, at the Les Houches meeting in 2001 [2], the beginnings of a new interface were conceived — the so-call "Les Houches Accord PDF"— LHAPDF. This has further been developed over the course of the HERA-LHC workshop incorporating many new features to enable it to replace PDFLIB as the standard tool to use. The development is briefly described in this writeup together with LHAGLUE, an interface to LHAPDF, which provides PDF access using almost identical calling routines as PDFLIB.

## 2 LHAPDF – Development during the Workshop

In its initial incarnation (Version 1), LHAPDF had two important features which distinguished it from the methods used by PDFLIB in handling PDFs.

Firstly the PDFs are defined by the analytical formulae used in the original fitting procedures, with external files of parameters, which describe the momentum $x$ distributions of the partons at the relevant $Q_0^2$. Evolution codes within LHAPDF then produce the PDF at any desired $Q^2$ at the users request. At present LHAPDF provides access to two evolution codes, EVLCTEQ for the CTEQ distributions and QCDNUM 16.12 [3] for the other PDF sets. This represents a radical difference from the existing methods used by the PDF authors to present their distributions where large grid files and interpolation routines are the norm. In PDFLIB these interpolation codes and grids are essentially compiled into a single FORTRAN library. The advantage of the LHAPDF method is that the compiled code is separate from the parameter files, which are typically small. Thus to add new PDF sets does not necessarily need the code to be recompiled and the library rebuilt.

Secondly, the concept is introduced of a "set" being a related collection of PDFs (e.g. an error set) all of which are accessible to the program after initialisation of that set. This allows LHAPDF to handle the multi-set "error" PDFs produced in recent years which give predicted uncertainties to the PDF values. All the PDFs in a set are initialised together and are therefore available to the user.

V1 was written by Walter Giele of Fermilab who in 2002 released a working version which could be downloaded from a web-site together with the parameter files for a limited number of PDF sets. There was also a manual and example files. One of the present authors MRW became involved and took over maintenance and development of LHAPDF in March 2003. The limitations of the idealised situation in V1 with respect to making LHAPDF a replacement tool for PDFLIB soon became apparent.

The primary problem was that V1 contained only a limited number of PDF sets and, since the method was reliant on the $x$ parameterisations at $Q_0^2$ being available, it would be virtually impossible to include many of the older sets which are still needed for comparisons. A second and serious problem is the compute time taken in the initialisation phase of the individual members of a PDF set (i.e. calling the routine InitPDF described later). This can take in the region of 2 seconds per call on a 1GHz machine and is therefore unacceptable in the situation of a program which makes repeated use of the different members. [1]

A solution introduced in LHAPDF Version 2, which helps to solve the above problems, was to include the option to make the original grid files and interpolation codes available in LHAPDF **in addition** to the V1 method of parameter files and "on-the-fly" evolution. For some PDF sets both methods would be available and for others only the latter. The operation of the program was made identical for both methods with the content of the input file (with extension ".LHpdf" for the former and ".LHgrid" for the latter) dictating which is used. Not only does this allow all the older PDF sets to be included but also there is no time penalty in changing between members of the same set since all are loaded in the initialisation phase. LHAPDF V2 was released in March 2004 including many of the older PDF sets as well as some new ones.

LHAPDF Version 3 was released in September 2004 and, as well incorporating more older and some new PDF sets (e.g. ZEUS and H1), it also included the code for LHAGLUE, a newly developing add-on interface to LHAPDF which provides PDFLIB look-alike access. In addition to having subroutine calls identical to those in PDFLIB it also incorporates a PDF numbering scheme to simplify usage. It should be noted however that, because of the greatly increased number of new PDF sets, it was not possible to follow the original numbering scheme of PDFLIB and a new one was devised. This is described in more detail in Section 5.

The major feature of Version 4, which was released in March 2005, was the incorporation of the photon and pion PDFs. All the photon and pion PDFs that were implemented in PDFLIB were put into LHAPDF using identical code and using the ".LHgrid" method. The LHAGLUE numbering scheme in these cases more closely resembles that of PDFLIB than it does for the protons.

In addition in V4 there were new proton PDFs (MRST2004 and an updated Alekhin's a02m), a new simpler file structure with all the source files being in a single "src" directory, some code changes to incorporate access to $\Lambda_{4/5}^{QCD}$ and a more rigorous implementation of the $\alpha_s$ evolution as being exactly that used by the PDF author.

All the LHAPDF and LHAGLUE data and code, in addition to being made available on the new web site (http://hepforge.cedar.ac.uk/lhapdf/), is also included in the GENSER subproject of the LHC Computing Grid.

## 3 LHAPDF – Development after the Workshop

Since the last HERA-LHC meeting there has been one minor release of LHAPDF (Version 4.1 in August 2005). In this version the installation method has changed to be more standard with the "configure; make;

[1] A third problem reported at the workshop concerning small differences (up to $\approx 0.5\%$) between the PDFs produced by LHAPDF for MRST and the authors' code directly is now believed to be due to slight mismatches of grid boundaries at the heavy quark thresholds and will be corrected in future MRST grids.

make install" sequence familiar to many and also a small amount of code has been altered to be more compliant with proprietary FORTRAN 95 compilers. As mentioned in the previous section the web site for public access to LHAPDF from which the source code can be obtained has changed. Since this is the current and most recent version we assume V4.1 in the following referring to earlier versions where necessary.

## 4 Using LHAPDF

Once the code and PDF data sets have been downloaded from the relevant web site and installed following the instructions given therein, using LHAPDF is simply a matter of linking the compiled FORTRAN library **libLHAPDF.a** to the users program. Table 1 lists the LHAPDF routines available to the user, which are of three types:

- Initialisation (selecting the required PDF set and its member)
- Evolution (producing the momentum density functions (*f*) for the partons at selected *x* and *Q*)
- Information (displaying for example $\alpha_s$, descriptions, etc.)

**Table 1:** LHAPDF commands

| Command | Description |
|---|---|
| **call InitPDFset(***name* **)** | Initialises the PDF set to use. |
| **call InitPDF(***member* **)** | Selects the member from the above PDF set. |
| **call evolvePDF(***x,Q,f* **)** | Returns the momentum density function, $f(x, Q)$, for protons or pions. |
| **call evolvePDFp(***x,Q,P2,ip2,f* **)** | Returns the momentum density function for photons[2]. |
| **call numberPDF(***num* **)** | Returns the number (*num*) of PDF members in the set. |
| **call GetDesc( )** | Prints a description of the PDF set. |
| **alphasPDF(***Q* **)** | Function giving the value of $\alpha_s$ at *Q* GeV. |
| **call GetLam4(***mem,qcdl4* **)** | Returns the value of $\Lambda_4^{QCD}$ for the specific member. |
| **call GetLam5(***mem,qcdl5* **)** | Returns the value of $\Lambda_5^{QCD}$ for the specific member. |
| **call GetOrderPDF(***order* **)** | Returns the order of the PDF evolution. |
| **call GetOrderAs(***order* **)** | Returns the order of the evolution of $\alpha_s$ . |
| **call GetRenFac(***muf* **)** | Returns the renormalisation factor. |
| **call GetQmass(***nf,mass* **)** | Returns the mass of the parton of flavour *nf*. |
| **call GetThreshold(***nf,Q* **)** | Returns the threshold value for parton of flavour *nf*. |
| **call GetNf(***nfmax* **)** | Returns the number of flavours. |

The evolution commands utilise a double precision array *f(-6:6)* where the arguments range from -6 to +6 for the different (anti)partons as shown in Table 2 below.

**Table 2:** The flavour enumeration scheme used for *f(n)* in LHAPDF

| parton | $\bar{t}$ | $\bar{b}$ | $\bar{c}$ | $\bar{d}$ | $\bar{u}$ | $\bar{d}$ | $g$ | $d$ | $u$ | $s$ | $c$ | $b$ | $t$ |
|---|---|---|---|---|---|---|---|---|---|---|---|---|---|
| $n$ | $-6$ | $-5$ | $-4$ | $-3$ | $-2$ | $-1$ | 0 | 1 | 2 | 3 | 4 | 5 | 6 |

Specifying the location of the PDF sets in the code should be especially mentioned at this point. The argument (*name*) in **InitPDFset** should specify the complete path (or at least to a symbolic link to this path). From version 4.1 onwards, however, a new routine **InitPDFsetByName** can be used in which only the name of the PDF set need by specified. This works in conjunction with the script '`lhapdf-config`' which is generated at the configure stage of the installation which provides the correct path to the PDF sets. The location of this script must therefore be in the users execution path. Tables 3 and 4 list the complete range of PDF set available. The equivalent numbers to use in LHAGLUE, as described in the next section, are also listed in these tables.

[2]In **evolvePDFp** $P2$ is the vitruality of the photon in GeV $^2$, which should by 0 for an on-shell photon, and *ip2* is the parameter to evaluate the off-shell anomalous component. See the PDFLIB manual [1] for details.

**Table 3:** The Proton PDF sets available in LHAPDF.

| Ref | Prefix | Suffix (number of sets) | type | LHAGLUE numbers |
|---|---|---|---|---|
| [4] | alekhin_ | 100 (100), 1000 (1000) | p | **40100-200**, **41000-1999** |
| [5] | a02m_ | lo (17), nlo (17), nnlo (17) ) ) | g | 40350-67, 40450-67, 40550-67 |
| [6] | botje_ | 100 (100),1000 (1000) | p | **50100-200**, **51000-1999** |
| [7] | cteq | 61 (41) | p,g | **10100-40**, 10150-90 |
| [8] | cteq | 6 (41) | p,g | **10000-40**, 10050-90 |
| | cteq | 6m, 6l, 6ll | p | **10000, 10041, 10042** |
| [9] | cteq | 5m, 5m1, 5d, 5l | g | 19050, 19051, 19060, 19070 |
| [10] | cteq | 4m, 4d, 4l | g | 19150, 19160, 19170 |
| [11] | fermi2002_ | 100 (100), 1000 (1000) | p | **30100-200**, **31000-2000** |
| [12] | GRV98 | lo, nlo(2) | g | 80060, 80050-1 |
| [13] | H12000 | msE (21), disE (21), loE (21) | g | 70050-70, 70150-70, 70250-70 |
| [14] | MRST2004 | nlo | p,g | **20400**, 20450 |
| | MRST2004 | nnlo | g | 20470 |
| [15] | MRST2003 | cnlo | p,g | **20300**, 20350 |
| | MRST2003 | cnnlo | g | 20370 |
| [16] | MRST2002 | nlo (2) | p,g | **20200**, 20250 |
| | MRST2002 | nnlo | g | 20270 |
| | MRST2001 | E (31) | p,g | **20100-130**, 20150-180 |
| [17] | MRST2001 | nlo(4) | p,g | **20000-4**, 20500-4 |
| | MRST2001 | lo, nnlo | g | 20060, 20070 |
| [18] | MRST98 | (3) | p | **29000-3** |
| | MRST98 | lo (5), nlo (5) dis (5), ht | g | 29040-5, 29050-5,29060-5,29070-5 |
| [19] | ZEUS2002_ | TR (23), FF (23), ZM (23) | p | **60000-22**, **60100-22**, **60200-22** |
| [20] | ZEUS2005_ | ZJ (23) | p | **60300-22** |

**Notes:**

LHAPDF → PrefixSuffix.LHpdf (type p), filename → PrefixSuffix.LHgrid (type g).
Where both p and g are present (p,g) the user has the choice of either.
LHAGLUE numbers in **bold** are the type p (.LHpdf) sets.

## 5 LHAGLUE

The LHAGLUE interface [21] to LHAPDF is designed along the lines of the existing interface from PYTHIA to PDFLIB. [3]. For both HERWIG and PYTHIA the existing 'hooks' for PDFLIB have been utilised for the LHAGLUE interface. This makes it possible to link it exactly like PDFLIB with no further changes to PYTHIA's or HERWIG's source code needing to be implemented.

[3]DB would like to thank T. Sjöstrand and S. Mrenna for discussions on this topic.

**Table 4:** The Pion and Photon PDF sets available in LHAPDF.

| Prefix | Suffix | LHAGLUE numbers | Prefix | Suffix | LHAGLUE numbers |
|---|---|---|---|---|---|
| **Pion PDFs** | | | **Photon PDFs** | | |
| OWPI | (2) | 211-12 | DOG | 0, 1 | 311, 312 |
| SMRSPI | (3) | 231-3 | DGG | (4) | 321-4 |
| GRVPI | 0, 1 | 251, 252 | LACG | (4) | 331-4 |
| ABFKWPI | (3) | 261-3 | GSG | 0 (2), 1 | 341-2, 343 |
| All filenames are PrefixSuffix.LHgrid | | | GSG96 | 0, 1 | 344, 345 |
| The nomenclature used here is | | | GRVG | 0 (2), 1 (2) | 351-2, 353-4 |
| essentially the same as in PDFLIB and | | | ACFGPG | (3) | 361-3 |
| the relevant publication references | | | WHITG | (6) | 381-6 |
| can be found in the PDFLIB manual [1]. | | | SASG | (8) | 391-8 |

The interface contains three subroutines (similar to PDFLIB) and can be used seamlessly by Monte Carlo generators interfaced to PDFLIB or in standalone mode. These are described in Table 5. In addition any of the LHAPDF routines, **except** the initialisation routines **InitPDFset** and **InitPDF**, described in Table 1, can also be used, for example to return the value of the strong coupling constant $\alpha_s$ (**alphasPDF**($Q$)), or to print the file description (**call GetDesc**()).

There are also several CONTROL switches specified through the 20 element character array LHAPARM and COMMON blocks which determine how the interface operates.

- Location of the LHAPDF library of PDFs (pathname):
  From version LHAPDF v4.1 onwards, and the LHAGLUE routines distributed with it, the location of the PDFsets data files is set automatically using the ”lhapdf-config” script as described in the previous section, provided that the prescribed installation instructions have been used.
  For previous versions (4.0 and earlier) the common block **COMMON/LHAPDFC/LHAPATH** is used where **LHAPATH** is a **character*132** variable containing the full path to the PDF sets. The default path is subdir ’PDFsets’ of the current directory.
- Statistics on under/over-flow requests for PDFs outside their validity ranges in $x$ and $Q^2$.
  a) **LHAPARM(16) .EQ. ‘NOSTAT’** → No statistics (faster)
  b) **LHAPARM(16) .NE. ‘NOSTAT’** → Default: collect statistics
  c) **call PDFSTA** at the end to print out statistics.
- Option to use the values for $\alpha_s$ as computed by LHAPDF in the Monte Carlo generator as well in order to ensure uniform $\alpha_s$ values throughout a run
  a) **LHAPARM(17) .EQ. ‘LHAPDF’** → Use $\alpha_s$ from LHAPDF
  b) **LHAPARM(17) .NE. ‘LHAPDF’**→ Default (same as LHAPDF V1/V3)
- Extrapolation of PDFs outside the LHAPDF validity range given by $x_{min/max}$ and $Q^2_{min/max}$.
  a) Default → PDFs “frozen” at the boundaries.
  b) **LHAPARM(18) .EQ. ‘EXTRAPOLATE’** → Extrapolate PDFs at own risk
- Printout of initialisation information in PDFSET (by default)
  a) **LHAPARM(19) .EQ. ‘SILENT’** → No printout (silent mode).
  b) **LHAPARM(19) .EQ. ‘LOWKEY’** → Print 5 times (almost silent).
- Double Precision values of $\Lambda^{QCD}_{4/5}$ applicable to the selected PDF are available (as read-only) in the COMMON block: **COMMON/W50512/QCDL4,QCDL5** → as in PDFLIB.

**Table 5:** LHAGLUE commands

| Command | Description |
|---|---|
| **call PDFSET(***parm,value)* | For initialisation (called once) where PARM and VALUE are LOCAL arrays in the calling program specified as **CHARACTER*20 PARM(20)** **DOUBLE PRECISION VALUE(20)** |
| **call STRUCTM(***X,Q,UPV,DNV,USEA,DSEA,STR,CHM,BOT,TOP,GLU)* | For the proton (and pion) PDFs: where X and Q are the input kinematic variables and the rest are the output PDF of the valence and sea quarks and the gluon. |
| **call STRUCTP(***X,Q2,P2,IP2,UPV,DNV,USEA,DSEA,STR,CHM,BOT,TOP,GLU)* | For the photon PDFs: as above with the additional input variables P2 and IP2 [2]. |

The LHAGLUE interface can be invoked in one of 3 ways, Standalone, PYTHIA or HERWIG, depending on the value of *parm(1)* when calling **PDFSET(***parm,value).*

– **Standalone** mode
  PARM(1)= 'DEFAULT'
  VALUE(1) = "*PDF number*"
– **PYTHIA mode**
  PARM(1) = 'NPTYPE' ← set automatically in PYTHIA
  In this case the user must supply MSTP(51) and MSTP(52) in the PYTHIA common block
  **COMMON/PYPARS/MSTP(200),PARP(200),MSTI(200).PARI(200)**
  MSTP(52) = 2 ← to use an external PDF library
  MSTP(51)= "*PDF number*"
– **HERWIG mode**
  PARM(1) = 'HWLHAPDF'← set by the user.
  In this case one sets for the beam and target particles separately
  AUTPDF(1) = 'HWLHAPDF'
  AUTPDF(2) = 'HWLHAPDF'
  MODPDF(1) = "*PDF number*"
  MODPDF(2) = "*PDF number*"
  Note that HERWIG specifies the"*PDF number*" for each of the colliding particles separately and care should be taken that the same PDF members are used when appropriate.

The user then simply links their own standalone code, or the HERWIG/PYTHIA main program and the HERWIG/PYTHIA code [4], with the LHAPDF library **libLHAPDF.a** making sure the 'PDFsets' directory is specified as described above.

The LHAGLUE interface has been tested extensively at TEVATRON and LHC energies for the proton PDFs and with HERA examples for the photon PDFs. Results with new and legacy PDF sets, using LHAPDF, PDFLIB or internal implementations in the Monte Carlo generators, and comparing cross sections produced with PYTHIA and HERWIG, give us confidence in the consistency of the LHAGLUE interface and the underlying LHAPDF library [22].

[4]It is important when starting with a fresh PYTHIA or HERWIG download the user must first rename the 'dummy' subroutines **STRUCTM**, **STRUCTP** and **PDFSET** in the PYTHIA/ HERWIG source codes exactly as if one were to link to PDFLIB.

## 6 Summary and Future Development

Both LHAPDF and the interface LHAGLUE have been developed over the period of the Workshop to a point where they can now be used as a serious replacement for PDFLIB. Indeed, except for the PDF authors' own code, they are the only place to obtain the latest PDFs. There is however still considerable development in progress and the latest PDF sets will be incorporated as and when they become available. One major development area is to include the possibility of having more than one PDF set initialised concurrently. This may be necessary in interactions between different beam and target particles types and also including photon and pion PDFs. This will be the aim of the next LHAPDF release.

## Acknowledgements

MRW wishes to thank the UK PPARC for support from grant PP/B500590/1. DB wishes to thank the USA National Science Foundation for support under grants NSF ITR-0086044 and NSF PHY-0122557.

# THEPEG
# Toolkit for High Energy Physics Event Generation

*Leif Lönnblad*
Department of Theoretical Physics, Lund University, Sweden

**Abstract**
I present the status of the THEPEG project for creating a common platform for implementing C++ event generators. I also describe briefly the status of the new versions of PYTHIA, HERWIG and ARIADNE which are implemented using this framework.

## 1 Introduction

Monte Carlo Event Generators (EGs) have developed into essential tools in High Energy Physics. Without them it is questionable if it at all would be possible to embark on large scale experiments such as the LHC. Although the current EGs work satisfactorily, the next generation of experiments will substantially increase the demands both on the physics models implemented in the EGs and on the underlying software technology.

The current EGs are typically written in Fortran and their basic structure was designed almost two decades ago. Meanwhile there has been a change in programming paradigm, towards object oriented methodology in general and C++ in particular. This applies to almost all areas of high-energy physics, but in particular for the LHC program, where all detector simulation and analysis is based on C++. When designing the next generation of EGs it is therefore natural to use C++. Below is a brief description of the THEPEG [1] project for designing a general framework in C++ for implementing EG models, and also the PYTHIA7 and ARIADNE programs which uses THEPEG to implement their respective physics models. Also HERWIG++ is implemented in the THEPEG framework, but this program is described elsewhere in these proceedings [2]

## 2 Basic structure

THEPEG is a general platform written in C++ for implementing models for event generation. It is made up from the basic model-independent parts of PYTHIA7 [3, 4], the original project of rewriting the Lund family of EGs in C++. When the corresponding rewrite of the HERWIG program [5] started it was decided to use the same basic infrastructure as PYTHIA7 and therefore the THEPEG was factorized out of PYTHIA7 and is now the base of both PYTHIA7 and HERWIG++ [6]. Also the coming C++ version of ARIADNE [7] is using THEPEG.

THEPEG uses CLHEP [8] and adds on a number of general utilities such as smart pointers, extended type information, persistent I/O, dynamic loading and some extra utilities for kinematics, phase space generation etc.

The actual event generation is then performed by calling different *handler* classes for hard partonic sub-processes, parton densities, QCD cascades, hadronization etc. To implement a new model to be used by THEPEG, the procedure is then to write a new C++ class inheriting from a corresponding handler class and implement a number of pre-defined virtual functions. Eg. a class for implementing a new hadronization model would inherit from the abstract `HandronizationHandler` class, and a new parton density parameterization would inherit from the `PDFBase` class.

To generate events with THEPEG one first runs a setup program where an `EventGenerator` object is set up to use different models for different steps of the generation procedure. All objects to be chosen from are stored in a *repository*, within which it is also possible to modify switches and parameters of the implemented models in a standardized fashion, using so called *interface* objects. Typically the user would choose from a number of pre-defined `EventGenerator` objects and only make minor changes for the specific simulation to be made. When an `EventGenerator` is properly set up it is saved persistently to a file which can then be read into a special run program to perform the generation, in which case special `AnalysisHandler` objects may be specified to analyze the resulting events. Alternatively it can be read into eg. a detector simulation program or a user supplied analysis program, where it can be used to generate events.

## 3 Status

### 3.1 THEPEG

THEPEG version 1.0$\alpha$ is available [1] and is working. As explained above, it contains the basic infrastructure for implementing and running event generation models. It also contains some simple physics models, such as some $2 \rightarrow 2$ matrix elements, a few parton density parameterizations and a near-complete set of particle decays. However, these are mainly in place for testing purposes, and to generate realistic events, the PYTHIA7 and/or HERWIG++ programs are needed.

Currently the program only works under Linux using the `gcc` compiler. This is mainly due to the use of dynamic linking of shared object files, which is inherently platform-dependent. Recently, the build procedure has been redesigned using the `libtool` facility [9], which should allow for easy porting to other platforms in the future.

Although THEPEG includes a general structure for implementing basic fixed-order matrix element generation to produce the initial hard subprocesses in the event generation, a general procedure for reading such parton level events from external programs using the Les Houches accord [10] has been developed and will be included in the next release[1].

The documentation of THEPEG is currently quite poor. Recently the actual code documentation was converted to Doxygen format [11], which will hopefully facilitate the documentation process. The lack of documentation means that there is currently a fairly high threshold for a beginner to start using and/or developing physics modules for THEPEG. The situation is further complicated since the user interface is currently quite primitive. THEPEG has a well worked through low-level interface to be able to set parameter and switches, etc. in classes introduced to the structure from the outside. However, the current external user interface is a simple command-line facility which is not very user-friendly. A Java interface is being worked on, but is not expected to be released until next year.

### 3.2 PYTHIA 7 (and PYTHIA8)

PYTHIA7 version 1.0$\alpha$ is available [4] and is working. It contains a reimplementation of the parton shower and string fragmentation models currently available in the 6.1 version of PYTHIA [12]. In an unfortunate turn of events, the principal PYTHIA author, Torbjörn Sjöstrand, has decided to leave the THEPEG collaboration and is currently developing a new `C++` version of PYTHIA (called PYTHIA8 [13]) on his own. This means that the development of PYTHIA7 is stopped, but hopefully it will be possible to interface the different modules in PYTHIA8 so that they can be used within the general framework of THEPEG.

[1] A snapshot of the current development version is available from [1]

### 3.3 ARIADNE

The reimplementation of the ARIADNE [7, 14] program using the framework of THEPEG has just started and is, hence, not publically available yet. Although this is mainly a pure rewrite of the fortran version of ARIADNE, it will contain some improvements, such as the CKKW matching [15, 16] also planned for HERWIG++. In addition, an improved version of the LDCMC [17] is planned.

## 4 Conclusions

THEPEG was intended to be *the* standard platform for event generation for the LHC era of collider physics. Unfortunately, this does not seem to become a reality. Besides the recent split between PYTHIA and THEPEG, there will also be other separate programs such as SHERPA [18, 19]. This is, of course, not an optimal situation, especially not for the LHC experiments, which surely would have preferred a uniform interface to different event generator models.

# PYTHIA

*T. Sjöstrand*
CERN, Geneva, Switzerland, and
Department of Theoretical Physics, Lund University, Sweden

The PYTHIA program is a standard tool for the generation of high-energy collisions, containing a realistic description of the full story, from a hard interaction involving a few partons to an observable hadronic final state of hundreds of particles. The current PYTHIA 6.3 version is described in detail in the manual [1], with the most recent update notes to be found on the PYTHIA webpage
`http://www.thep.lu.se/~torbjorn/Pythia.html` ,
together with the code itself, sample main programs and some further material. The latest published version is [2] and a recent brief review is found in [3]. The 6.3 version includes new transverse-momentum-ordered showers and a new multiple-interactions and beam-remnant scenario [4], described elsewhere in these proceedings.

From the onset, all PYTHIA code has been written in Fortran 77. For the LHC era, the experimental community has made the decision to move heavy computing completely to C++. Hence the main future development line is PYTHIA 8, which is a re-implementation in C++. Many obsolete options will be removed and various aspects modernized in the process.

With the rise of automatic matrix-element code generation and phase-space sampling, input of process-level events via the Les Houches Accord (LHA) [5] reduces the need to have extensive process libraries inside PYTHIA itself. Thus emphasis is on providing a good description of subsequent steps of the story, involving elements such as initial- and final-state parton showers, multiple parton–parton interactions, string fragmentation, and decays. All the latter components now exist as C++ code, even if in a preliminary form, with finer details to be added, and still to be better integrated and tuned. At the current stage, however, there is not even the beginning of a PYTHIA 8 process library; instead a temporary interface is provided to PYTHIA 6, so that all hard processes available there can be generated and sent on to PYTHIA 8, transparent to the user.

PYTHIA 8 is intended to be a standalone program, i.e. does not require any external libraries. However, in addition to the LHA interface, hooks also exist for external parton distribution functions, particle decays and random numbers, and more may be added.

This project was started in September 2004, and so is still at an early stage. A first public version, PYTHIA 8.040, can be found on the PYTHIA webpage (look under the “Future” link). This should be viewed as a development snapshot, to allow early feedback from the LHC experimental community, and cannot be used for any serious physics studies. It is intended/hoped that a first realistic version, PYTHIA 8.100, could be ready by early 2007, but even this version will be clearly limited in its capabilities, and strongly focused on LHC applications. It is therefore to be expected that PYTHIA 6 and PYTHIA 8 will co-exist for several years.

# HERWIG

*Michael H. Seymour*
School of Physics & Astronomy, University of Manchester, and Theoretical Physics Group, CERN

**Abstract**
I review the status of the current fortran version of HERWIG. Progress towards its replacement, Herwig++, is reviewed elsewhere in these proceedings.

## 1 Introduction

HERWIG [1] is a Monte Carlo event generator for simulation of hadronic final states in lepton–lepton, lepton–hadron and hadron–hadron collisions. It incorporates important colour coherence effects in the final state [2] and initial state [3] parton showers, as well as in heavy quark processes [4] and the hard process generation [5]. It uses the cluster [6] hadronization model and a cluster-based simulation of the underlying event [7]. While earlier versions [8] concentrated on QCD and a few other SM processes, recent versions contain a vast library of MSSM [9] and other BSM processes. A review of current Monte Carlo event generators including HERWIG can be found in [10].

We are currently in a period of intense activity, finalizing the HERWIG program and writing a completely new event generator, HERWIG++. In this very short contribution, I can do little more than mention the areas of progress and provide references to sources of more details.

## 2 HERWIG version 6.5

HERWIG version 6.5 was released [11] in October 2002. Its main new features were an interface to the Les Houches Accord event format [12], the hooks needed by the MC@NLO package [13] and various bug fixes and minor improvements. It was advertised as the final fortran version of HERWIG before work switched to HERWIG++.

Despite this, the period since then has seen intense development with several new subversion releases and new features, most notably version 6.505, which featured an improved interface to the Jimmy generator for multiparton interactions, which I will discuss in more detail shortly. The most recent version is 6.507, which can be obtained from the HERWIG web site [14].

Development of fortran HERWIG is now slowing, and the only new feature still foreseen is the implementation of matrix element corrections to the production of Higgs bosons, both SM and MSSM, preliminary versions of which have been discussed in [15]. Beyond this, the HERWIG collaboration has made a commitment to all running (and ceased) experiments to support their use of HERWIG throughout their lifetimes. Due to lack of manpower, making the same promise to the LHC experiments would divert too much effort away from support of HERWIG++, and we will only support their use of HERWIG until we believe that HERWIG++ is a stable alternative for production running.

## 3 Jimmy

Early versions of the Jimmy model [16] generated jet events in photoproduction using a multiparton interaction picture. The recent update [17] enables it to work efficiently as a generator of underlying events in high $E_T$ jet events and other hard processes in hadron–hadron collisions for the first time. For a given pdf set, the main adjustable parameters are `PTJIM`, the minimum transverse momentum of partonic scattering, and `JMRAD(73)`, related to the effective proton radius. Varying these one is able to get a good description of the CDF data [18] and other data held in the JetWeb database [19] that are sensitive to underlying event effects in hard process events. However, a poor description of minimum bias data in which there is no hard scale is still obtained. This is probably due to the fact that `PTJIM` is a hard cutoff and there is no soft component below it; preliminary attempts to rectify this are encouraging [20]. It is interesting to note that with tunings that give equally good descriptions of current data, Jimmy predicts twice as much underlying event activity as PYTHIA at the LHC.

# Herwig++

*Stefan Gieseke*
Institut für Theoretische Physik, Universität Karlsruhe, 76128 Karlsruhe, Germany

**Abstract**
I briefly review the status of the Herwig++ event generator. Current achievements are highlighted and a brief summary of future plans is given.

## 1 Introduction

Herwig++ [1] is a new Monte Carlo event generator for simulating collider physics, written in the object oriented programming language C++. The idea is to rewrite the well-established multi–purpose event generator HERWIG [2] and to improve it where necessary [3]. The Lund event generators PYTHIA [4] and ARIADNE [5] are also being rewritten at the moment. Herwig++ and ARIADNE will both be based on a common event generation framework, called ThePEG [6] which will make it possible to exchange single modules of the event generation and allows us to have a common, or at least a very similar user interface. PYTHIA8, the rewrite of PYTHIA (6.3) will be written independently of this project but may become integrated into the structure later [7]. A further object oriented event generator, SHERPA [8], is established as an independent project.

## 2 Event Generation

In its present version (1.0) Herwig++ is capable of simulating $e^+e^-$ annihilation events. The physics simulation consists of several steps, going from small (perturbative) to large (non–perturbative) distance scales. First, the effective CM energy of the annihilating $e^+e^-$ pair is selected according to some model structure function of the electron, thereby radiating photons that carry some fraction of the original energy. Next, we set up the $q\bar{q}$ final state and a hard matrix element correction is applied [9]. In the next step, parton showers are radiated from the coloured final state particles. These effectively resum large soft and collinear logarithms. The parton shower is modelled in terms of new evolution variables with respect to the FORTRAN program [10]. This, and the use of splitting functions for massive particles allow us to simulate the suppression of soft and collinear radiation from heavy particles dynamically (dead cone effect) which has previously only been modelled in a crude way. Parton showers from initial state particles in a hard scattering and from decays of heavy particles (particularly $t$–quarks) have been formulated for various situations in [10]. The next stage of the simulation is the hadronization of the outgoing coloured particles. First, remaining gluons are split into non–perturbative $q\bar{q}$–pairs. Colour connected particles are paired into colourless clusters. The invariant mass spectrum of these clusters contains a long high–mass tail that still contains a large scale. These heavy clusters are further split into pairs of lighter clusters. Once all clusters are below a certain mass threshold they decay into pairs of hadrons. The hadron species are selected only according to a handful of parameters. It is this stage where it has been observed in previous versions of the FORTRAN program that the meson/baryon number ratio in $e^+e^-$ annihilation events was difficult to obtain when a large number of highly excited mesons is available in the program [11]. In the current version the hadron selection is reorganised and we obtain more stable results. Finally, the produced hadrons decay into stable hadrons according to some models. In version 1.0 the hadronic decays were modelled similarly to the decays in the FORTRAN version.

We have tested the simulation of $e^+e^-$ annihilation events in very great detail [1]. We considered event shape variables, jet rates, hadron yields and many more observables. One observable of special interest has been the $b$–quark fragmentation function that we found to be well–described on the basis of the parton shower only. This is a result of the new shower algorithm for heavy quarks. The overall result was that we are capable of simulating $e^+e^-$ events at least as well good as with the FORTRAN version.

## 3 Current and Future Developments

Many new features are currently being implemented for the event simulation at hadron colliders. The list of hard matrix elements will be slightly extended in the next version in order to cover some basic processes. In principle we can also rely on parton level event generators and read in event files that follow the Les Houches Accord [12]. The parton shower will include initial state radiation and gluon radiation in the perturbative decay of heavy particles. Some related aspects of estimating uncertainties from initial state parton showers were addressed in [13]. A large effort went into remodelling and updating the secondary hadronic decays. A future version should also be able to simulate hard jets in deep inelastic scattering. Exhaustive tests of our generator output against current data from the experiments at HERA and the Tevatron will be made in order to validate and understand our program. In the long–term we plan to include a larger number of simple processes, mainly $2 \to 2$ and some $2 \to 3$, both Standard Model processes and some BSM processes as well. The modelling of the underlying event will at first only be on the basis of the simple so–called UA5 model that is also available in the FORTRAN version. A refinement towards a more sophisticated multiple interaction model [14, 15] is planned.

# The Event Generator SHERPA

*T. Gleisberg, S. Höche, F. Krauss, A. Schälicke, S. Schumann, J. Winter*
Institut für theoretische Physik, TU Dresden, D-01062 Dresden, Germany
E-mail: steffen@theory.phy.tu-dresden.de

**Abstract**
In this contribution the multi-purpose event generation framework SHERPA is presented and the development status of its physics modules is reviewed. In its present version, SHERPA is capable of simulating lepton-lepton, lepton-photon, photon-photon and fully hadronic collisions, such as proton-proton reactions.

SHERPA [1] is an independent approach for a framework for event generation at high energy collider experiments. The program is entirely written in the object-oriented programming language C++. This is reflected in particular in the structure of the program. In SHERPA, the various tasks related to the generation of events are encapsulated in a number of specific modules. These physics modules are initialized and steered by the SHERPA framework. This structure facilitates a high modularity of the actual event generator and allows for the easy replacement/modification of entire physics models, e.g. the parton shower or the fragmentation model. The current version SHERPA-1.0.6 incorporates the following physics modules:

- ATOOLS-2.0.6: This is the toolbox for all other modules. ATOOLS contain classes with mathematical tools like vectors and matrices, organization tools such as read-in or write-out devices, and physics tools like particle data or classes for the event record.
- BEAM-1.0.6: This module manages the treatment of the initial beam spectra for different colliders. At the moment two options are implemented for the beams: they can either be monochromatic, and therefore require no extra treatment, or, for the case of an electron collider, laser backscattering off the electrons is supported leading to photonic initial states.
- PDF-1.0.6: In this module the handling of initial state radiation (ISR) is located. It provides interfaces to various proton and photon parton density functions, and to the LHAPDFv3 interface. In addition, an analytical electron structure function is supplied.
- MODEL-1.0.6: This module comprises the basic physics parameters (like masses, mixing angles, etc.) of the simulation run. Thus it specifies the corresponding physics model. Currently three different physics models are supported: the Standard Model (SM), its Minimal Supersymmetric extension (MSSM) and the ADD model of large extra dimensions. For the input of MSSM spectra a run-time interface to the program Isasusy 7.67 [2] is provided. The next release of SHERPA will in addition support the SLHA format of spectrum files [3].
- EXTRA_XS-1.0.6: In this module a collection of analytic expressions for simple $2 \to 2$ processes within the SM and the corresponding classes embedding them into the SHERPA framework are provided. This includes methods used for the definition of the parton shower evolution, such as color connections and the hard scale of the process. The classes for phase space integration, which are common with AMEGIC, are located in a special module called PHASIC.
- AMEGIC++-2.0.6: AMEGIC [4] is SHERPA's own matrix element generator. It works as a generator-generator: during the initialization run the matrix elements for a set of given processes within the SM, the MSSM or the ADD model, as well as their specific phase space mappings are created by AMEGIC and stored in library files. In the initialization of the production run, these libraries are linked to the program. They are used to calculate cross sections and to generate single events based on them.

- PHASIC++-1.0.6: Here all classes dealing with the Monte Carlo phase space integration are located. As default the adaptive multi-channel method of [5] together with a Vegas optimization [6] for the single channels is used for the evaluation of the initial state and final state integrals.
- APACIC++-2.0.6: APACIC [7] contains classes for the simulation of both the initial and the final state parton shower. The sequence of parton emissions in the shower evolution is organized in terms of the parton's virtual mass as ordering parameter. Coherence effects are accounted for by explicit ordering of the opening angles in subsequent branchings. This treatment is similar to the Pythia [8] parton shower approach. All features for a consistent merging with matrix elements [9] are included.
- AMISIC++-1.0.6: AMISIC contains classes for the simulation of multiple parton interactions according to [10]. SHERPA extends this treatment of multiple interactions by allowing for the simultaneous evolution of an independent parton shower in each of the subsequent (semi-)hard collisions. This shower evolution is done by the APACIC module.
- SHERPA-1.0.6: Finally, SHERPA is the steering module that initializes, controls and evaluates the different phases in the entire process of event generation. Furthermore, all necessary routines for combining the parton showers and matrix elements, which are independent of the specific parton shower are found in this module. In addition, this subpackage provides an interface to the Lund String Fragmentation of Pythia 6.214 including its hadron decay routines.

SHERPA is publicly available from http://www.sherpa-mc.de. It has successfully been tested for various processes of great relevance at future colliders [11]. Present activities of developing SHERPA cover the modeling of the underlying event and an alternative fragmentation model [12].

# ARIADNE at HERA and at the LHC

*Leif Lönnblad*
Department of Theoretical Physics, Lund University, Sweden

**Abstract**
I describe briefly the status of the ARIADNE program implementing the Dipole Cascade Model and comment both on its performance at HERA, and the uncertainties relating to the extrapolation to LHC energies.

## 1 Introduction

ARIADNE [1] is a Fortran subroutine library to be used with the PYTHIA event generator [2]. By simply adding a few lines to a PYTHIA steering routine, the PYTHIA parton shower is replaced by the dipole cascade in ARIADNE. For lepton–hadron DIS it can also be used together with the LEPTO [3] generator in a similar fashion. However, even if it thus simple to use ARIADNE also for the LHC, there are a few caveats of which the user should be aware. In this brief presentation of the program, I will first go through the main points of the final-state dipole shower relevant for $\mathrm{e^+e^-}$-annihilation, then I will present the extention of the model to lepton–hadron DIS, and finally describe how the model works for hadron–hadron collisions.

## 2 The Basic Dipole Model

In the Dipole Cascade Model (DCM) [4, 5], the bremsstrahlung of gluons is described in terms of radiation from colour dipoles between gluons and quarks. Thus, in an $\mathrm{e^+e^-} \to q\bar{q}$ event, a gluon, $g_1$ may be emitted from the colour-dipole between the $q$ and $\bar{q}$. In this emission the initial dipole is replaced by two new ones, one between $q$ and $g_1$ and one between $g_1$ and $\bar{q}$. These may then continue radiating independently in a cascade where each step is a $2 \to 3$ partonic splitting or, equivalently, a splitting of a dipole into two. The splittings are ordered in a transverse momentum variable, $\mathrm{p}_\perp$, defined in a Lorentz-invariant fashion, which also defines the scale in $\alpha_S$.

There are several advantages of this model. One is that the coherence effects approximated by angular ordering [6] in eg. the HERWIG [7] parton cascade, are automatically included. Another is that the first order $\mathrm{e^+e^-} \to qg\bar{q}$ matrix element correction is in some sense built-in. A major *dis*advantage is that the $g \to q\bar{q}$ splitting does not enter naturally in this formalism. Final-state $g \to q\bar{q}$ splittings are, however, easy to add [8] and for final-state cascades in $\mathrm{e^+e^-}$-annihilation the description is complete. Ariadne is generally considered to be the program which best reproduces event shapes and other hadronic final-state observables at LEP (see eg. [9]).

## 3 ARIADNE at HERA

While for $\mathrm{e^+e^-}$-annihilation, the DCM is formally equivalent to conventional angular ordered parton showers to modified leading logarithmic accuracy, the situation for collisions with incoming hadrons is quite different. In a conventional shower the struck quark in eg. lepton–hadron DIS is evolved backwards with an initial-state cascade according to DGLAP [10–13] evolution. In contrast, the DCM model for DIS [14] describes all gluon emissions in terms of final-state radiation from colour-dipoles, in a similar way as in $\mathrm{e^+e^-}$-annihilation, with the initial dipole now being between the struck quark and the hadron remnant. Contrary to $\mathrm{e^+e^-}$-annihilation, the remnant must now be treated as an extended object and, since radiation of small wavelengths from an extended antenna is suppressed, the emission of high-$p_\perp$ gluons in the DCM is suppressed in the remnant direction.

Despite this suppression, which is modeled semi-classically, the net result is that gluon emissions are allowed in a much larger phase space region than in a conventional parton shower, especially for limited $Q^2$ values. Although the emissions are ordered in $p_\perp$, they are not ordered in rapidity (or $x$). Hence, if tracing the emissions in rapidity, they will be unordered in $p_\perp$, and there are therefore qualitative similarities between the DCM and BFKL evolution [15–17]. This is in contrast to conventional showers which are purely DGLAP-based and where the emissions are ordered both in scale and in $x$. One of the striking phenomenological consequences of this is that ARIADNE is one of the few programs which are able to describe the high rate of forward (in the proton direction) jets measured in small-$x$ DIS at HERA [18–20], an observable which conventional parton showers completely fail to reproduce. In fact, ARIADNE is in general considered to be the program which best describe hadronic final-state observables at HERA [21].

This does not mean that the DCM is perfect in any way. Most notably, the initial-state $g \to q\bar{q}$ and $q \to g^\star q$ (where the $q$ is emitted into the final-state) splittings are not easily included. While the former process has been included as an explicit initial-state splitting step [22], the latter is currently absent in the the ARIADNE program. In addition, the treatment of the initial-state $g \to q\bar{q}$ splitting has been found to be somewhat incomplete, as it by construction imposes ordering in both $p_\perp$ and rapidity, thus excluding certain regions of the allowed phase space. At HERA, the incomplete treatment of the $g \to q\bar{q}$ and $q \to g^\star q$ splittings can be shown to be a small effect. However, this is not always the case at the LHC.

## 4 ARIADNE at LHC

Given the great success of ARIADNE at LEP and HERA, it is natural to assume that it also would do a good job at the Tevatron and the LHC. In principle, the extention of the DCM to hadron–hadron collisions is trivial, and indeed it is simple to run ARIADNE together with PYTHIA for hadron–hadron collisions. Whichever hard sub-process, PYTHIA generates, the relevant dipoles between hard partons and hadron remnants are constructed and are allowed to radiate. In addition, the initial-state $g \to q\bar{q}$ splittings are included from both sides. However, for many processes there are modifications needed.

The most obvious processes are Drell-Yan and vector boson production, where a quark from one hadron annihilates with an anti-quark from the other. The gluon radiation is then initiated by the dipole between the two remnants, and we have a suppression in both directions. However, it is unphysical to give the remnants a large transverse momentum from the recoil of the gluon emission. In DIS, this is resolved by introducing so-called recoil gluons [14], but here it is clear that the recoil should be taken by the vector boson or the Drell-Yan lepton pair. Such a procedure was introduced in [23], and together with a correction where the first emission is matched to the $qg \to qZ$ and $q\bar{q} \to gZ$ matrix elements, it describes well eg. the $Z^0$ $p_\perp$ spectrum measured at the Tevatron [24,25]. There are still some differences wrt. conventional parton showers. Eg. the rapidity correlation between the vector boson and the hardest jet is more flat in ARIADNE due to the increased phase space for emissions [26]. Although $W$ and $Z^0$ production at the Tevatron is not a small-$x$ process, the effect is related to higher rate of forward jets in ARIADNE for DIS. Such correlations have not yet been measured at the Tevatron, but another related effect is the somewhat harder $p_\perp$-spectrum of the $Z^0$ for low $p_\perp$ in ARIADNE, which is compatible with Tevatron measurements [26]. For a conventional cascade to be able to describe the low-$p_\perp$ spectrum, a quite substantial "non-perturbative" intrinsic transverse momentum must be added to the incoming quarks [27,28].

Going from the Tevatron to the LHC, there is a substantial increase in phase space for QCD radiation, and it can be argued that $W$ and $Z^0$ production at the LHC is a small-$x$ process with $x \propto m_Z/\sqrt{S} < 0.01$. Indeed ARIADNE predicts a harder $p_\perp$-spectrum for the $W$ at the LHC as compared to conventional showers [29].

Also Higgs production can be argued to be almost a small-$x$ process at the LHC, if the Higgs is found with a mass around the "most likely" value of $\approx$ 120 GeV. However, Higgs production is a gluon-initiated process, and the absence of the $q \to g^{\star}q$ splitting is a serious deficiency giving a much softer $p_{\perp}$-spectrum for the Higgs in ARIADNE as compared to conventional showers [30]. Hence the predictions from ARIADNE for this and similar processes can currently not be trusted. Furthermore, the increased phase-space at the LHC means that predictions also for quark-initiated processes may become affected by the deficiencies in the treatment of initial-state $g \to q\bar{q}$ mentioned above.

## 5 Conclusion

The success of the DCM as implemented in ARIADNE in describing hadronic final-state observables as measured at LEP and HERA makes it tempting to use it also to make predictions for the LHC. The temptation is even more difficult to resist as it is so simple to run ARIADNE together with PYTHIA for any LHC process. Currently, this must be done with great care. As explained above, it is possible to obtain reasonable predictions for vector boson production. Also standard jet-production should be fairly safe. However, for Higgs production, one of the most interesting processes at LHC, ARIADNE in its current state turns out to be quite useless.

ARIADNE is currently being rewritten in C++ within the framework of THEPEG [31, 32]. The planned features includes a remodeling of initial-state $g \to q\bar{q}$ splittings as well as the introduction of the $q \to g^{\star}q$ process. In addition the matching to fixed-order tree-level matrix elements à la CKKW [26, 33–35] will be implemented for the most common sub-processes. When this version is released, hopefully during 2006, it should therefore be safe to use ARIADNE to produce LHC predictions.

# The Monte Carlo Event Generator AcerMC and package AcerDET

*B. Kersevan*[1,2], *E. Richter-Was*[3,4,*]
[1] Faculty of Mathematics and Physics, University of Ljubljana, Jadranska 19, SI-1000, Slovenia
[2] Experimental Particle Physics Department, Jozef Stefan Institute, P.P. 3000, Jamova 39, SI-1000 Ljubljana, Slovenia.
[3] Institute of Physics, Jagiellonian University, 30-059 Krakow, ul. Reymonta 4, Poland.
[4] Institute of Nuclear Physics PAN, 31-342 Krakow, ul. Radzikowskiego 152, Poland.

**Abstract**

The **AcerMC** Monte Carlo Event Generator is dedicated for the generation of Standard Model background processes at *pp* LHC collisions. The program itself provides a library of the massive matrix elements and phase space modules for generation of selected processes. The hard process event, generated with one of these modules, can be completed by the initial and final state radiation, hadronisation and decays, simulated with either `PYTHIA`, `ARIADNE` or `HERWIG` Monte Carlo event generator and (optionally) with `TAUOLA` and `PHOTOS`. Interfaces to all these packages are provided in the distribution version. The matrix element code has been derived with the help of the `MADGRAPH` package. The phase-space generation is based on the multi-channel self-optimising approach using the modified Kajantie-Byckling formalism for phase space construction and further smoothing of the phase space was obtained by using a modified `ac-VEGAS` algorithm.

## 1 Introduction

The physics programme of the general purpose LHC experiments, ATLAS [1] and CMS [2], focuses on the searches for the *New Physics* with the distinctive signatures indicating production of the Higgs boson, SUSY particles, exotic particles, etc. The expected environment will in most cases be very difficult, with the signal to background ratio being quite low, on the level of a few percent after final selection in the signal window.

Efficient and reliable Monte Carlo generators, which enable one to understand and predict background contributions, are becoming key elements in the discovery perspective. As the cross-section for signal events is rather low, even rare Standard Model processes might become the overwhelming backgrounds in such searches. In several cases, generation of such processes is not implemented in the general purpose Monte Carlo generators, when the complicated phase space behaviour requires dedicated (and often rather complex) pre-sampling, whilst the general purpose Monte Carlo generators due to a large number of implemented processes tend to use simpler (albeit more generic) phase space sampling algorithms. In addition, the matrix element expressions for these processes are often quite lengthy and thus require complicated calculations. Only recently, with the appearance of modern techniques for automatic computations, their availability *on demand* became feasible for the tree-type processes. With the computation power becoming more and more easily available even very complicated formulas can now be calculated within a reasonable time frame.

## 2 The Monte Carlo Event Generator AcerMC

The physics processes implemented in **AcerMC** library [3–5] represent such a set of cases. They are all being key background processes for the discovery in the channels characterised by the presence of the heavy flavour jets and/or multiple isolated leptons. For the Higgs boson searches, the $t\bar{t}H$, $ZH, WH$ with $H \to b\bar{b}$, the $gg \to H$ with $H \to ZZ^* \to 4\ell$, the $b\bar{b}h/H/A$ with $h/H/A \to \tau\tau, \mu\mu$ are the most obvious examples of such channels.

*Supported in part by the Polish Government grant KBN 1 P03 091 27 (years 2004-2006) and by the EC FP5 Centre of Excellence "COPIRA" under the contract No. IST-2001-37259.

Let us shortly discuss the motivation for these few Standard Model background processes which are implemented in the **AcerMC 2.x** library.

**The $t\bar{t}b\bar{b}$ production** is a dominant irreducible background for the Standard Model (SM) and Minimal Supersymmetric Standard Model (MSSM) Higgs boson search in the associated production, $t\bar{t}H$, followed by the decay $H \to b\bar{b}$. Proposed analysis [1] requires identifying four b-jets, reconstruction of both top-quarks in the hadronic and leptonic mode and visibility of the peak in the invariant mass distribution of the remaining b-jets. The irreducible $t\bar{t}b\bar{b}$ background contributes about 60-70% of the total background from the $t\bar{t}$ events ($t\bar{t}b\bar{b}$, $t\bar{t}bj$, $t\bar{t}jj$).

**The $Wb\bar{b}$ production** is recognised as a substantial irreducible background for the Standard Model (SM) and Minimal Supersymmetric Standard Model (MSSM) Higgs boson search in the associated production, $WH$, followed by the decay $H \to b\bar{b}$.

**The $Wt\bar{t}$ production** is of interest because it contributes an overwhelming background [7] for the measurement of the Standard Model Higgs self-couplings at LHC in the most promising channel $pp \to HH \to WWWW$.

**The $Z/\gamma^*(\to \ell\ell)b\bar{b}$ production** has since several years been recognised as one of the most substantial irreducible (or reducible) backgrounds for the several Standard Model (SM) and Minimal Supersymmetric Standard Model (MSSM) Higgs boson decay modes as well as for observability of the SUSY particles. There is a rather wide spectrum of *regions of interest* for this background. In all cases the leptonic $Z/\gamma^*$ decay is asked for, but events with di-lepton invariant mass around the mass of the Z-boson mass or with the masses above or below the resonance peak could be of interest. The presented process enters an analysis either by the accompanying b-quarks being tagged as b-jets, or by the presence of leptons from the b-quark semi-leptonic decays in these events, in both cases thus contributing to the respective backgrounds.

**The $Z/\gamma^*(\to \ell\ell, \nu\nu, b\bar{b})t\bar{t}$ production** is an irreducible background to the Higgs search in the invisible decay mode (case of $Z \to \nu\nu$) in the production with association to the top-quark pair [8]. With the $Z/\gamma^*(\to b\bar{b})$ it is also an irreducible resonant background to the Higgs search in the $t\bar{t}H$ production channel but with the Higgs boson decaying to the b-quark pair.

The complete **EW production** of the $gg, q\bar{q} \to (Z/W/\gamma^* \to)b\bar{b}t\bar{t}$ final state is also provided. It can be considered as a benchmark for the previous process, where only the diagrams with resonant $gg, q\bar{q} \to (Z/\gamma^* \to)b\bar{b}t\bar{t}$ are included. It thus allows the verification of the question, whether the EW resonant contribution is sufficient in case of studying the $t\bar{t}b\bar{b}$ background away from the Z-boson peak, like for the $t\bar{t}H$ with Higgs-boson mass of 120 GeV.

**The $gg, q\bar{q} \to t\bar{t}t\bar{t}$ production**, interesting process per se, is a background to the possible Higgs self-coupling measurement in the $gg \to HH \to WWWW$ decay, [7].

**The $gg, q\bar{q} \to (WWbb \to)f\bar{f}f\bar{f}b\bar{b}$ and $gg, q\bar{q} \to (t\bar{t} \to)f\bar{f}bf\bar{f}\bar{b}$** processes give possibility to study spin correlations in the top-quark pair production and decays as well as the effect from the off-shell production. Those are important for the selection optimisation eg. in the $gg \to H \to WW$ channel, see the discussion in [9]. As an example, Fig. 1 illustrates spin correlation effects in the top-pair production and decays, namely asymmetry in the correlations between lepton and antilepton direction in the rest frame of top-quark, for events generated with $2 \to 6$ matrix element. Such correlation is absent if only $2 \to 2$ matrix element is used for events generation, followed by the independent decays of each top-quark.

A set of **control channels**, i.e. the $q\bar{q} \to Z/\gamma^* \to \ell\ell$, $gg, q\bar{q} \to t\bar{t}$, $q\bar{q} \to W \to \ell\nu$ and $gg \to (t\bar{t} \to)WbW\bar{b}$ processes, have been added to **AcerMC** in order to provide a means of consistency and cross-check studies.

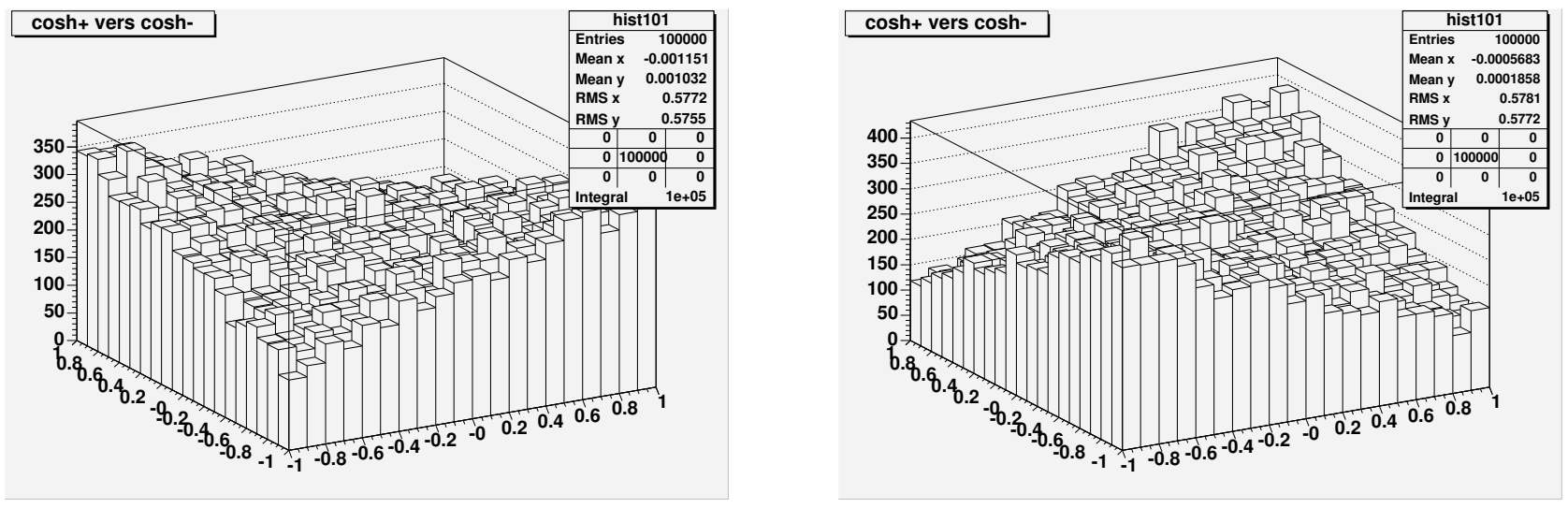

**Fig. 1:** The correlations between $\cos\Theta$ (azimuthal angle) of lepton and antilepton from $t\bar{t} \to \ell\bar{\nu}b\bar{\ell}\nu\bar{b}$ decays measured in the rest frame of the top-quark with respect to the anti-top quark direction. Left plot is for $gg \to (WWb\bar{b} \to)f\bar{f}f\bar{f}b\bar{b}$ process, right plot for $q\bar{q} \to (WWb\bar{b} \to)f\bar{f}f\bar{f}b\bar{b}$ process.

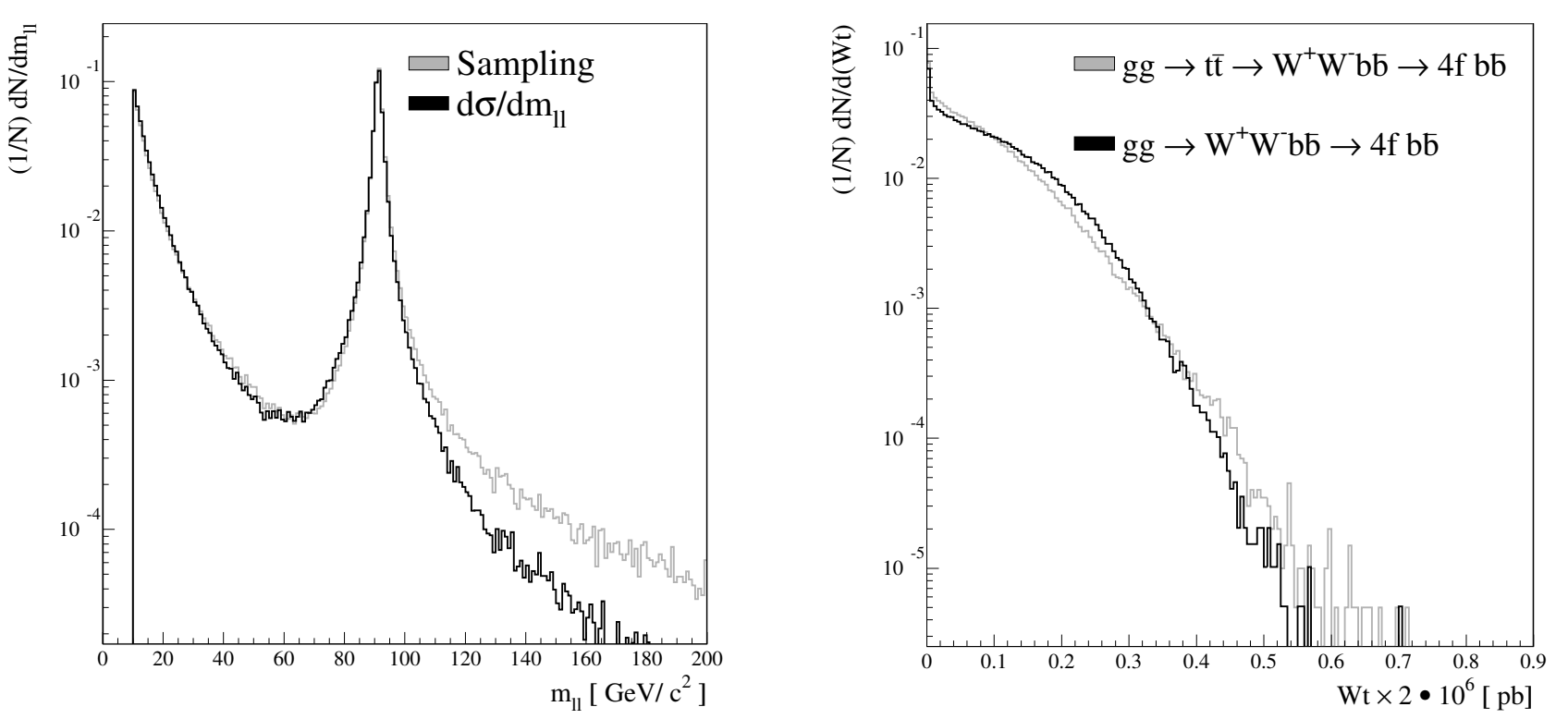

**Fig. 2:** Left: A representative invariant mass distribution comparisons between the (normalised) sampling functions and the normalised differential cross-section for the $\ell\bar{\ell}$ pair in the process $gg \to Z^0/\gamma^* b\bar{b} \to \ell\bar{\ell}b\bar{b}$ process. Right: The weight distribution of the sampled events for the $gg \to t\bar{t} \to b\bar{b}W^+W^- \to b\bar{b}\ell\bar{\nu}_\ell\bar{\ell}\nu_\ell$ (light gray histogram) and $gg \to b\bar{b}W^+W^- \to b\bar{b}\ell\bar{\nu}_\ell\bar{\ell}\nu_\ell$ (black histogram) processes. One can observe the well defined weight range for the two processes; as it turns out the weight distribution is even marginally better for the (more complex) second process, possibly because the higher number of sampling channels manage to cover the event topologies in phase space to a better extent.

This completes the list of the native **AcerMC** processes implemented so far. The hard process event, generated with one of these modules, can be completed by the initial and final state radiation, hadronisation and decays, simulated with either `PYTHIA`, `ARIADNE` or `HERWIG` Monte Carlo event generator and (optionally) with `TAUOLA` and `PHOTOS`. Interfaces to all these packages are provided in the distribution releases. The matrix element code has been derived with the help of the `MADGRAPH` package. The phase-space generation is based on the multi-channel self-optimising approach [3] using the modified Kajantie-Byckling formalism for phase space construction and further smoothing of the phase space was obtained by using a modified `ac-VEGAS` algorithm.

The improved and automated phase space handling provided the means to include the $2 \to 6$ processes like e.g. $gg \to t\bar{t} \to b\bar{b}W^+W^- \to b\bar{b}\ell\bar{\nu}_\ell q_1\bar{q}_2$ which would with the very complicated phase space topologies prove to be too much work to be handled manually. The studies show that the overall unweighting efficiency which can be reached in the $2 \to 6$ processes by using the recommended phase space structuring is on the order of 10 percent. An example of the implemented sampling functions and the actual differential distributions for the $gg \to Z^0/\gamma^* b\bar{b} \to \ell\bar{\ell}b\bar{b}$ process and of the weight distribution for the $gg \to t\bar{t} \to b\bar{b}W^+W^- \to b\bar{b}\ell\bar{\nu}_\ell q_1\bar{q}_2$ process are shown in Fig.2

In its latest version, the **AcerMC-2.4** package is interfaced also to `ARIADNE 4.1` [12] parton shower model and the LHAPDF structure functions [13].

It is not always the case that the matrix element calculations in the lowest order for a given topology represent the total expected background of a given type. This particularly concerns the heavy flavour content of the event. The heavy flavour in a given event might occur in the hard process of a much simpler topology, as the effect of including higher order QCD corrections (eg. in the shower mechanism). This is the case for the b-quarks present in the inclusive Z-boson or W-boson production, which has a total cross-section orders of magnitude higher than the discussed matrix-element-based $Wb\bar{b}$ or $Zb\bar{b}$ production. Nevertheless, the matrix-element-based calculation is a very good reference point to compare with parton shower approaches in different fragmentation/hadronisation models. It also helps to study matching procedures between calculations in a fixed $\alpha_{QCD}$ order and parton shower approaches. For very exclusive hard topologies matrix-element-based calculations represent a much more conservative approximation than the parton shower ones [6].

## 3 The AcerDET package

The package **AcerDET** [14] is designed to complete the **AcerMC** generator framework with the easy-to-use simulation and reconstruction algorithms for phenomenological studies on high $p_T$ physics at LHC The package provides, starting from list of particles in the event, the list of reconstructed jets, isolated electrons, muons and photons and reconstructed missing transverse energy. The **AcerDET** represents a simplified version of the package called `ATLFAST` [15], used since several years within ATLAS Collaboration. In the **AcerDET** version some functionalities of the former one have been removed, only the most crucial detector effects are implemented and the parametrisations are largely simplified. Therefore it is not representing in details neither ATLAS nor CMS detectors. Nevertheless, we believe that the package can be well adequate for some feasibility studies of the high $p_T$ physics at LHC.

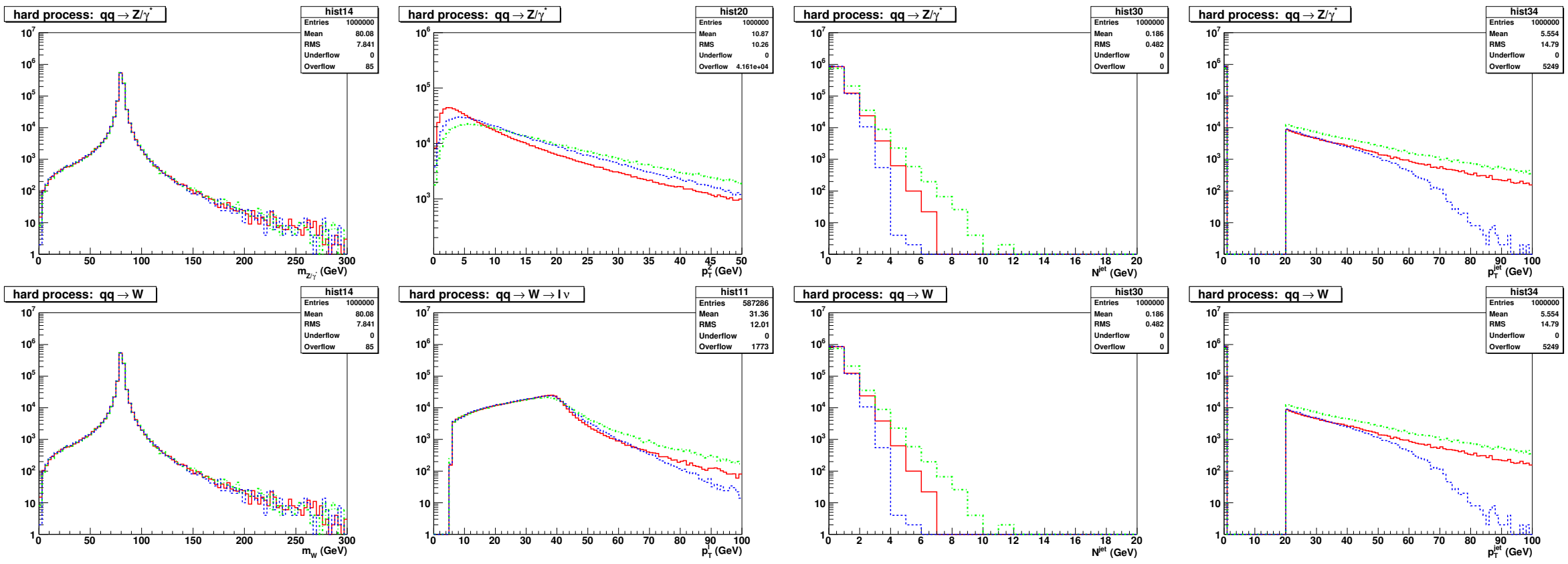

**Fig. 3:** A few examples of theoretical systematic uncertainties from parton shower model on experimentally observable distributions from Drell-Yan W and Z boson production at LHC (see text).

Fig. 3 shows possible application of the **AcerMC** control processes and **AcerDET** package for studying theoretical systematic uncertainties on the experimentally observed distributions from the parton shower model. The control channels $q\bar{q} \to Z/\gamma^* \to \ell\ell$, and $q\bar{q} \to W \to \ell\nu$ were processed with parton shower model as implemented in `PYTHIA` (red), `HERWIG` (blue) and `ARIADNE` (green). The comparison includes the distributions of the invariant mass of the di-lepton or lepton-neutrino system, transverse momenta of the Z boson, transverse mass of the W, multiplicity of jets from ISR parton shower and transverse momenta of the hardest jets reconstructed with **AcerDET** package. Perfect agreement on the most left plots confirms consistent starting point for the evolution of the ISR QCD parton shower. The differences observed on remaining plots should be attributed to the systematic theoretical uncertainties of the parton shower models.

# RAPGAP

*Hannes Jung*
Deutsches Elektronen-Synchroton DESY Hamburg, FRG

**Abstract**
RAPGAP, originally developed to describe rapidity gap event in $ep$ collisions, has evolved into a multi-purpose Monte Carlo event generator for diffractive and non-diffractive processes at $ep$ colliders both for high $Q^2$ and in the photoproduction regime ($Q^2 \sim 0$) as well as hard (single diffractive and non-diffractive) processes for $pp$ and $p\bar{p}$ colliders. A detailed description of the program as well as the source code can be found under [1]. In the following only new developments are described.

## 1 NLO and Order $\alpha_s$ matrix elements

The $\mathcal{O}(\alpha_s)$ matrix elements for light quarks are divergent for $p_T^2 \to 0$, and usually a $p_T^2$ cutoff is applied. The $\overline{MS}$ factorization scheme provides a description which finite parts of the matrix elements are treated explicitely and which parts are included in the parton distribution functions. A consistent implementation of the NLO formalism for $F_2$ in DIS including initial state parton showering is described in detail in [2]. The LO ($\alpha_s^0$) and the NLO ($\alpha_s$) part are treated according the $\overline{MS}$ subtraction scheme, reformulated such that it properly can be used together with initial and final state parton showers, avoiding any double counting [3]. When using this scheme, the NLO parton densities calculated in the $\overline{MS}$ scheme should be selected. The program then transforms the parton densities from the $\overline{MS}$ to the $BS$ scheme for parton showers. However, at present only the BGF part is implemented.

## 2 Les Houches interface

A generic format for the transfer of parton level event configurations from matrix element event generators (MEG) to showering and hadronization event generators (SHG) [4] is provided by the *Les Houches interface*. RAPGAP gives the possibility to write the full parton level events to the file `rapgap.gen`, which can be read in directly by the PYTHIA and HERWIG programs to perform the hadronization. This option is best suited to estimate the uncertainty coming from hadronization correction.

## 3 Proton dissociation ala DIFFVM

Dissociation of the proton according to the model in DIFFVM [5] can be included for diffractive events. The proton dissociation part of the cross section is given by

$$\frac{d\sigma}{dM_Y^2\, dt\, dx_{I\!P}} \sim \frac{1}{M_Y^{2(1+\epsilon_Y)}} \exp\left(-B_{diss}|t|\right)$$

with $\epsilon_Y$ describing the dependence on the dissociation mass $M_Y$ and $B_{diss}$ the $t$-dependence. The dissociative system $Y$ is split into a $quark - gluon - diquark$ system for masses $M_Y > 2$ GeV whereas for masses $0.939 < M_Y < 2$ GeV the system is fragmented according to the nucleon resonances as implemented in DIFFVM [5].

## 4 Future Plans

In the next future it is planned to include double diffractive scattering for *pp* collisions to allow simulation of diffractive Higgs production.

# CASCADE

*Hannes Jung*
Deutsches Elektronen-Synchroton (DESY), Hamburg, FRG

**Abstract**
CASCADE is a full hadron-level Monte Carlo event generator for $ep$, $\gamma p$, $pp$ and $p\bar{p}$ processes.

CASCADE uses the unintegrated parton distribution functions convoluted with off-mass shell matrix elements for the hard scattering. The CCFM [1] evolution equation is an appropriate description valid for both small and moderate $x$ which describes parton emission in the initial state in an angular ordered region of phase space. For inclusive quantities it is equivalent to the BFKL and DGLAP evolution in the appropriate asymptotic limits. The angular ordering of the CCFM description makes it directly applicable for Monte Carlo implementation. The following processes are available: $\gamma^* g^* \to q\bar{q}(Q\bar{Q})$ $\gamma g^* \to J/\psi g$, $g^* g^* \to q\bar{q}(Q\bar{Q})$ and $g^* g^* \to h^0$.

A detailed description of CASCADE, the source code and manual can be found under [2]. A discussion of different unintegrated gluon densities can be found in [3–5].

The unintegrated gluon density $x\mathcal{A}_0(x, k_\perp, \bar{q})$ is a function of the longitudinal momentum fraction $x$ the transverse momentum of the gluon $k_\perp$and the scale (related to the angle of the gluon) $\bar{q}$. Given this distribution, the generation of a full hadronic event is separated into three steps:

- The hard scattering process is generated,

$$\sigma = \int dk_{t\,1}^2 dk_{t\,2}^2 dx_1 dx_2 \mathcal{A}(x_1, k_{t\,1}, \bar{q}) \mathcal{A}(x_2, k_{t\,2}, \bar{q}) \sigma(g_1^* g_2^* \to X)\,, \quad (1)$$

  with $X$ being $q\bar{q}$, $Q\bar{Q}$, $J/\psi g$ or $h^0$ states. The hard cross section is calculated using the off-shell matrix elements given in [6] for $q\bar{q}$ and $Q\bar{Q}$, $\gamma g^* \to J/\psi g$ in [7] and for Higgs production $g^* G^* \to h^0$ in [8]. The gluon momentum is given in Sudakov representation:

$$k = x_g p_p + \bar{x}_g p_e + k_t \simeq x_g p_p + k_t\,. \quad (2)$$

  where the last expression comes from the high energy approximation ($x_g \ll 1$), which then gives $-k^2 \simeq k_t^2$.
- The initial state cascade is generated according to CCFM in a backward evolution approach.
- The hadronization is performed using the Lund string fragmentation implemented in PYTHIA [9].

The backward evolution there faces one difficulty: The gluon virtuality enters in the hard scattering process and also influences the kinematics of the produced quarks and therefore the maximum angle allowed for any further emission in the initial state cascade. This virtuality is only known after the whole cascade has been generated, since it depends on the history of the gluon evolution (as $\bar{x}_g$ in eq.( 2) may not be neglected for exact kinematics). In the evolution equations itself it does not enter, since there only the longitudinal energy fractions $z$ and the transverse momenta are involved. This problem can only approximately be overcome by using $k^2 = k_t^2/(1 - x_g)$ for the virtuality which is correct in the case of no further gluon emission in the initial state. This problem is further discussed in [5, 10]

The CCFM evolution equations have been solved numerically [11] using a Monte Carlo method. Several sets of un-integrated gluon densities are available which have the input parameters were fitted to describe the structure function $F_2(x, Q^2)$ in the range $x < 5 \cdot 10^{-3}$ and $Q^2 > 4.5$ GeV$^2$ as measured at H1 [12] and ZEUS [13].

Also the unintegrated gluon densities described in [5] including non-linear terms [14] are available within CASCADE.

# Leading proton production in $ep$ and $pp$ experiments: how well do high-energy physics Monte Carlo generators reproduce the data?

*G. Bruni, G. Iacobucci, L. Rinaldi, M. Ruspa*
INFN Bologna and University of Eastern Piedmont

**Abstract**
The simulation of leading-proton production at high-energy colliders as obtained by the HERWIG, LEPTO and PYTHIA Monte Carlo generators is analysed and compared to the measurements of HERA and fixed-target experiments. The discrepancies found between real data and Monte Carlo events could be responsible for inaccurate simulation of particle multiplicities and hadronic final states, which could eventually generate problems in computing the Standard-Model backgrounds to new physics at the LHC collider.

## 1 Introduction

The production of final state baryons carrying a large fraction of the available energy but a small transverse momentum (leading baryons) is crucial for a deep understanding of strong interactions beyond the perturbative expansion of QCD. Indeed, in high-energy collisions, the QCD-hardness scale decreases from the central, large $p_T$ region, to the soft, non-perturbative hadronic scale of the target-fragmentation region. Therefore, the measurement of leading baryons in the final state of high-energy collisions allows to gather information on the non-perturbative side of strong interactions.

Another reason of interest in leading-baryon production comes from the fact that the energy carried away by the leading baryon(s) produced in a high-energy collision is not available for the production of the central-hadronic system. Therefore, the leading-baryon spectra should be well simulated for a proper accounting of the hadronic multiplicities and energies, e.g. at the LHC collider where an appropriate simulation of these quantities will be the ground for a reliable calculation of the Standard-Model backgrounds to new physics.

Here we will review the data on the production of leading protons and compare them to the most popular Monte Carlo generators available.

## 2 The data and the Monte Carlo generators used for the comparison

### 2.1 The proton-proton data

Although the experimental data on leading-proton production are scarce, a few measurements in a large $x_L$ range are available, where $x_L$ represents the fractional longitudinal momentum of the proton. In proton-proton collisions, leading-proton production has been studied both at the ISR [2, 3] and in fixed-target experiments [4–6]. The $x_L$ spectra measured in fixed-target experiments are shown in Fig. 1a-c,e.

### 2.2 The $ep$ data

Cross sections for the production of leading protons were also measured at the HERA collider [7–9]. More recently, the ZEUS Collaboration made a new measurement [10] of the cross-section for the semi-inclusive reaction $ep \rightarrow eXp$ in deep-inelastic scattering using $12.8\,\text{pb}^{-1}$ of data collected during 1997. The single-differential cross sections, $d\sigma_{ep\rightarrow eXp}/dx_L$ and $d\sigma_{ep\rightarrow eXp}/p_T^2$, and the double-differential cross section, $d^2\sigma_{ep\rightarrow eXp}/dx_L dp_T^2$, were measured in the kinematic range $Q^2 > 3\,\text{GeV}^2$ and $45 < W < 225$ GeV, where $W$ is the total mass of the hadronic system. The protons were measured using the leading-proton spectrometer (LPS) [11] in the range $x_L > 0.56$ and $p_T^2 < 0.5$ GeV$^2$, where $p_T$ is the scattered-proton transverse momentum.

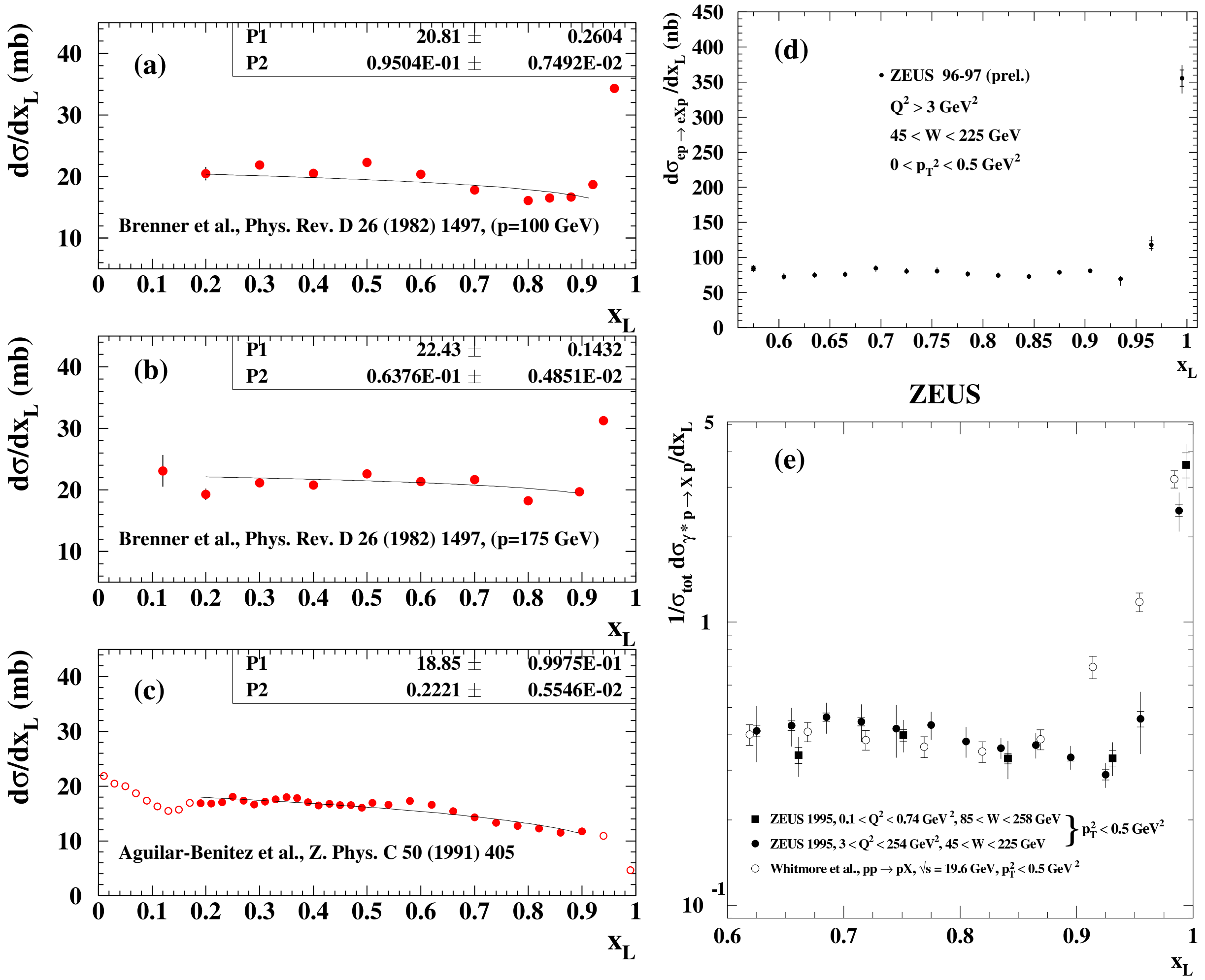

**Fig. 1:** Differential cross sections $d\sigma/dx_L$ measured in fixed-target experiments (a, b, c and e) and by the ZEUS Collaboration [9, 10] (d, e).

### 2.3 The Monte Carlo generators

Large samples of Monte Carlo $ep$ events were generated to be compared to the data. The LEPTO generator was used either with the MEPS or the ARIADNE packages; in the latter case the diffractive component of the cross section was simulated using the Soft Color Interaction model. Events were also generated with HERWIG. Since this Monte Carlo does not simulate diffractive events, the POMWIG generator was used to account for the single diffractive events, and the SANG generator to account for the diffractive events in which the scattered-proton dissociates in a higher-mass hadronic system.

Proton–proton events at the LHC center of mass energy (14 TeV) were generated with PYTHIA.

## 3 Discussion

### 3.1 The $x_L$ spectrum

Figure 1a, b and c show the $d\sigma/dx_L$ obtained by the fixed target experiments [4–6] which measured leading protons in a wide range of $x_L$. The cross section for such events shows a peak for values of the final-state proton momentum close to the maximum kinematically allowed value, the so-called diffractive peak. Below the diffractive peak the cross section is lower and consistent with a flat one. In this region, under the assumption of Regge factorisation [12], the fraction of events with a leading proton is expected to be approximately independent of the energy and type of the incoming hadron. The lines superimposed

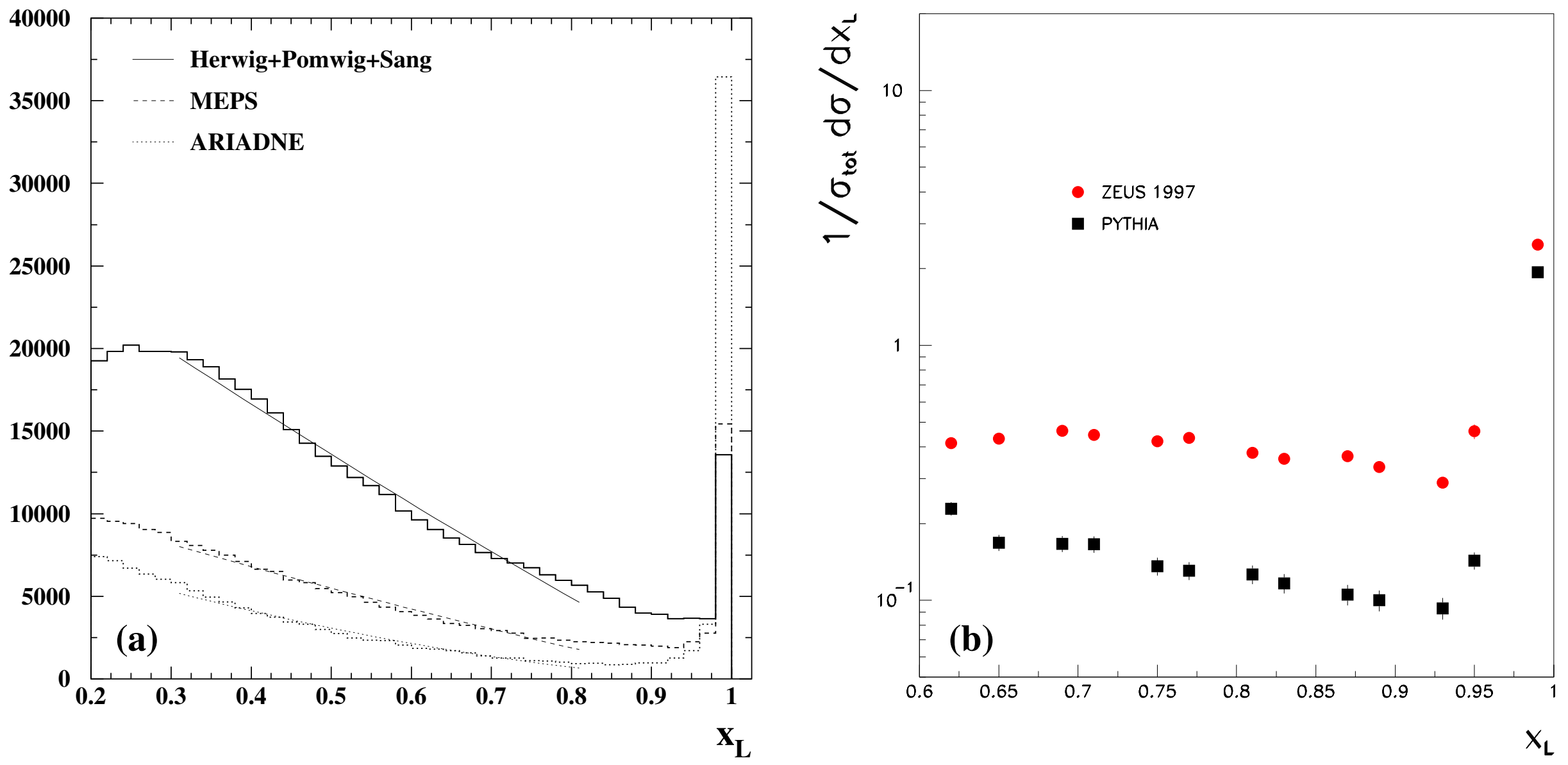

**Fig. 2:** (a) Differential cross section $d\sigma_{ep\to eXp}/dx_L$ generated with LEPTO, with MEPS or ARIADNE, and with HERWIG+POMWIG+SANG Monte Carlos; (b) Comparison between the normalised differential cross section $1/\sigma_{tot}\ d\sigma_{pp\to Xp}/dx_L$ simulated with PYTHIA and $1/\sigma_{tot}\ d\sigma_{ep\to eXp}/dx_L$ measured by the ZEUS Collaboration [9].

to the data are the results of fits in the range $0.1 < x_L < 0.9$ to the function $(1 - x_L)^\alpha$ that is commonly used to characterise the longitudinal distributions of leading particles. The values of $\alpha$ obtained are 0.1, 0.06 and 0.22 respectively for Fig. 1a, b and c. Fig. 1d shows the preliminary $d\sigma/dx_L$ obtained by ZEUS. Below the diffractive peak the cross section is again consistent with a flat one, i.e. $\alpha \sim 0$. A comparison between the normalised cross section $1/\sigma_{tot}\ d\sigma/dx_L$ obtained by the fixed-target data [6] and by the $ep$ data is shown in Fig. 1e. For $x_L < 0.9$ the fraction of events with a leading proton is indeed consistent for the $pp$ and the $ep$ data set.

The $x_L$ distributions of the simulation of the HERA events are shown in Fig. 2a. Already at first glance, the difference w.r.t. the data is evident, since the spectra are much more populated at low $x_L$ in the Monte Carlo than they are in the data. Indeed, the fits to the same functional form as for the data give $\alpha$ = 1.0 for LEPTO-MEPS, $\alpha$ = 1.4 for LEPTO-ARIADNE and $\alpha$ = 1.0 for HERWIG+POMWIG+SANG.

In Fig. 2b the $x_L$ distribution obtained from the simulation with PYTHIA of $pp$ events at the LHC center of mass energy (14 TeV) is compared to the ZEUS data. As discussed previously, according to the vertex factorisation hypothesis, the fraction of events with a leading proton is expected to be consistent in the $ep$ and in the $pp$ case. The simulation appears to approximately agree with the data in the diffractive peak region but is not able to describe the data neither in shape nor in normalisation in the region outside the diffractive peak.

In general we conclude that the fraction of beam energy carried away by the leading proton in the Monte Carlo is on average much smaller than in the data, with the consequence that the energy available in the simulation for the production of the central-hadronic system is correspondingly larger than in nature.

### 3.2 The $p_T^2$ spectrum

Although the $p_T^2$ distribution of the leading proton is less important for the hadronic final states than the $x_L$ distribution, it is interesting to investigate how well the generators can reproduce it.

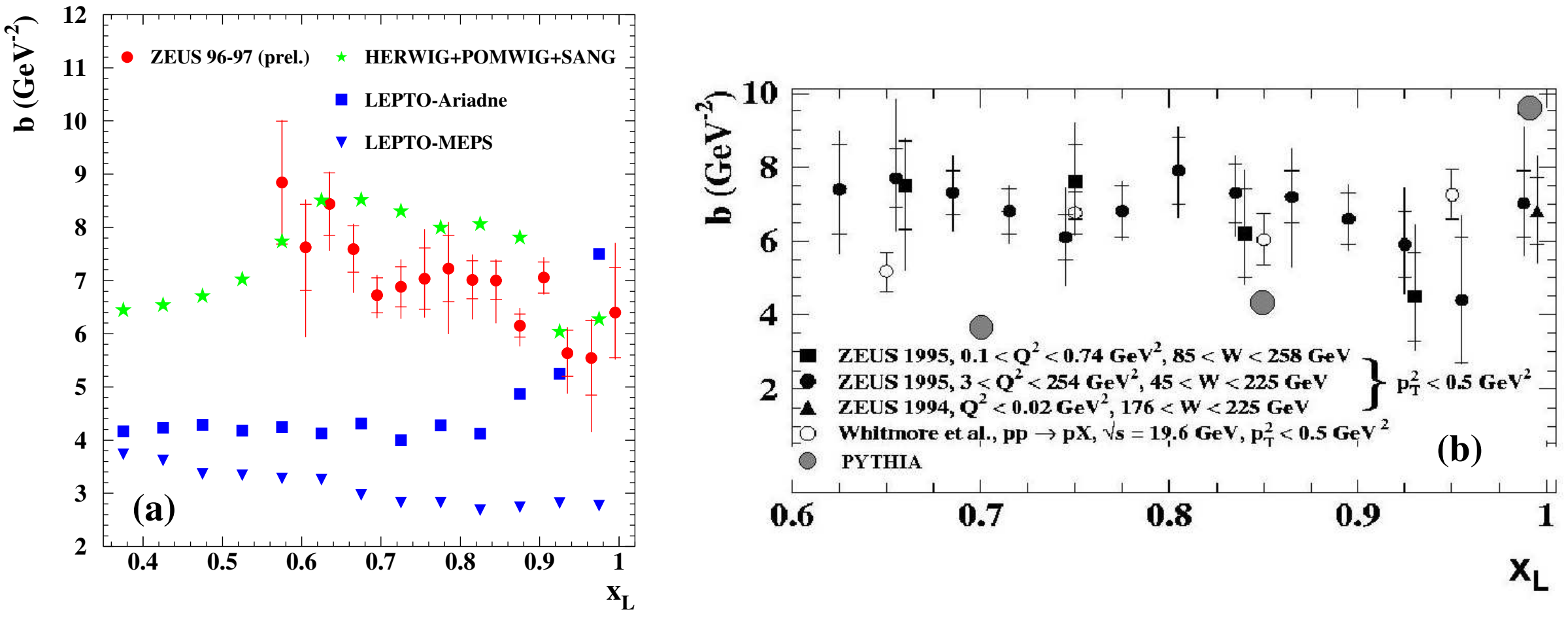

**Fig. 3:** The slope-parameters b obtained from a single-exponential fit to the differential cross section $d\sigma_{ep \to eXp}/dp_T^2$ ($d\sigma_{pp \to Xp}/dp_T^2$ for PYTHIA $pp$ sample) to the function $A \cdot e^{-b \cdot p_T^2}$ in each of the $x_L$ bins shown. (a) The red dots are the ZEUS preliminary data [10], while the other symbols represent the results of the Monte Carlos described in the picture; (b) The grey dots are the $pp$ events simulated with PYTHIA, while the other symbols are the data described in the picture.

In Fig. 3a the red dots show the values of the slope-parameter b obtained from a single-exponential fit to the function $e^{-b \cdot p_T^2}$ in each of the $x_L$ bins of the ZEUS $d\sigma/dp_T^2$ measurement. The b slopes obtained by a similar fit performed on the simulated events in the $ep$ case and in the $pp$ case are also reported in Fig. 3a and Fig. 3b, respectively. In the $pp$ case the extracted b-slopes have been corrected for the expected shrinkage of the diffractive peak.

The b-slopes values resulting from the fit to the $p_T^2$ distribution of the HERWIG+POMWIG+SANG sample appear to be in the right ball park.

The LEPTO generator shows too small b-slope values, smaller than those of the data by approximately 3 GeV$^{-2}$. In the case the matrix-element parton showers are used to generate the events, since the dependence of b on $x_L$ is similar to that of the data, it is conceivable to fix the difference by tuning the primordial $k_T$ of the generation. If the ARIADNE package is used instead, it seems quite difficult to improve the situation in a similar way, since the generated b values increase with $x_L$, a feature that is not seen in the data.

The b-slopes values resulting from the fit to the $p_T^2$ distribution of the PYTHIA $pp$ sample are approximately consistent with the ZEUS data in the diffractive peak region, but lower than the data again by approximately 3 GeV$^{-2}$ in the region outside the peak.

### 3.3 Reweighting of the PYTHIA leading proton spectrum

A sample of $pp$ proton events generated with PYTHIA has been used to simulate the many interactions per bunch crossing (pile-up events) occurring at the LHC luminosities in a recent study on the diffractive production of a Higgs boson at the LHC [13]. The simulated leading proton spectrum has been reweighted both in $x_L$ and in $p_T^2$ with the following function, calculated for each $x_L$ bin of Fig. 2b:

$$f(x_L) = [-37.22 + 135.1 \cdot x_L - 148.5 \cdot x_L^2 + 54.3 \cdot x_L^3] \cdot \frac{e^{-b_{\text{ZEUS}} t}}{e^{-b_{\text{PYTHIA}} t}} \cdot \frac{b_{\text{ZEUS}}}{b_{\text{PYTHIA}}},$$

where $b_{ZEUS} = 7$ GeV$^{-2}$ and $b_{PYTHIA} = 4.4$ GeV$^{-2}$. The polynomial form in $x_L$ is the result of a fit to the ratio ZEUS/PYTHIA of the differential cross sections $1/\sigma_{tot}\ d\sigma/dx_L$ of Fig. 2b; thus $f(x_L)$ provides the number of simulated leading protons to be consistent with the ZEUS data.

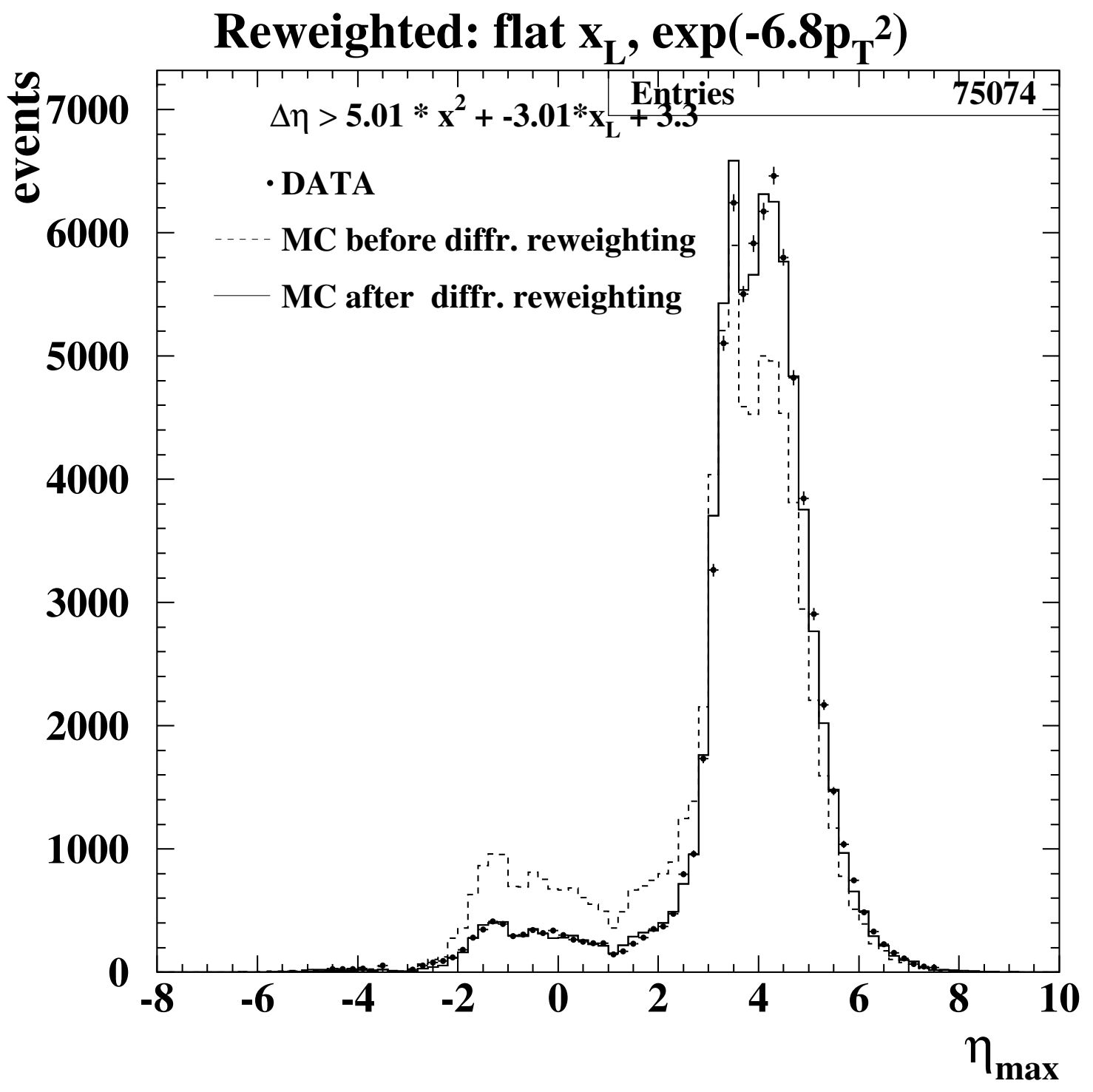

**Fig. 4:** The $\eta_{max}$ distribution of the ZEUS data [10] (dots) and of the LEPTO-MEPS Monte Carlo obtained with the default $x_L$ and $p_T^2$ distributions (dashed-line histogram). The MC distribution after reweighting to a flat $x_L$ and to $e^{-6.8p_T^2}$ is shown by the full-line histogram.

### 3.4 The fraction of diffractive — large-rapidity gap – - events w.r.t. the total

One way to identify a diffractive event produced in $ep$ or $pp$ interactions is to search for a large-rapidity gap (LRG) in the pseudorapidity distribution of the particles produced. In ZEUS, a diffractive-LRG event was tagged by $\eta_{max} < 2$, where $\eta_{max}$ corresponds to the pseudorapidity of the most forward (i.e. proton direction) energy deposit in the calorimeter exceeding 400 MeV. The $\eta_{max}$ distribution for ZEUS DIS events with $Q^2 > 3$ GeV$^2$ is shown by the dots in Fig. 4. The two regions of non-diffractive events (with $\eta_{max}$ between 2 and 8) and of diffractive events (which distribute at $\eta_{max}$ values below 2) are clearly distinguishable.

The LEPTO-MEPS events were passed through the standard simulation of the ZEUS trigger and detector, and through the same reconstruction and analysis programs as the data. The $\eta_{max}$ distribution of the MC events after such processing is shown by the dashed histogram in Fig. 4. We note that the diffractive events with $\eta_{max} < 2$ generated with the Soft Color Interaction algorithm in LEPTO-MEPS are more than twice those found in the data. Therefore, the Soft Color Interaction algorithm, as implemented in the LEPTO generator, fails to describe the data in the range of the ZEUS-LPS detector, i.e. $x_L > 0.56$.

## 4 Summary

The data on leading-proton production in $ep$ and $pp$ scattering have been compared to the most popular Monte Carlo generators available to simulate high-energy physics events. This exercise has revealed that the simulation of the leading-proton momenta, both longitudinal and transverse to the beams, does not reproduce the properties of the data. In particular, the $x_L$ distribution of the leading protons would be made more close to that of the data if a proper accounting of the energy available for the hard-scattering process could be achieved.

Although the HERWIG generator has been successful in simulating many features of high-energy physics final states, it does not contain the diffractive component of the cross section, and the $x_L$ spectrum it produces is far from being almost flat, as seen in the data.

The Soft Color Interaction model in its standard implementation in LEPTO is producing twice the fraction of diffractive-LRG events seen in the ZEUS data at $x_L > 0.56$, therefore distorting in a significant way the multiplicities and hadronic energies present in real events.

The PYTHIA Monte Carlo has been used to simulate $pp$ events at the LHC center of mass energy (14 TeV), and then also compared to $ep$ data. The generator has been shown to reproduce the longitudinal and transverse momentum of the data in the diffractive peak region; however, it underestimates both the cross section and the $p_T^2$-slopes at lower values of the scattered proton momentum, contradicting the hypothesis of vertex factorisation, which is supported by the data.

All the above arguments generate some concern that the hadronic multiplicities of the MC generators taken into account here have been tuned consistently and that they can produce an accurate simulation of the final states of the Standard-Model processes at the LHC energies.

## 5 Acknowledgements

We would like to thank Torbjörn Sjöstrand for useful discussions.

# NLOLIB - A Common Interface for Fixed-Order Calculations

*Klaus Rabbertz*[†] *and Thomas Schörner-Sadenius*[o]
[†]University of Karlsruhe, EKP, Postfach 6980, D-76128 Karlsruhe, *klaus.rabbertz@cern.ch*
[o]University of Hamburg, IExpPh, Luruper Chaussee 149, D-22761 Hamburg, *schorner@mail.desy.de*

**Abstract**
We present the current status of the NLOLIB framework which provides an interface to various higher order perturbative calculations, thus allowing for simple comparisons of these calculations with each other and with measured distributions. We show, as a newly included example of the NLOLIB abilities, a comparison of calculations for jet production in deeply inelastic ep scattering.

## 1 Introduction

Progress in particle physics relies, to a large extent, on the comparison of data to theoretical predictions. Most commonly, the theoretical calculations are available to the experiments in the form of Monte Carlo event generators producing event records that contain all generated particles, their four-momenta, the decay trees etc. in a commonly adopted format. The events that represent the outcome of such a prediction can be used directly by experiment-specific detector simulation and analysis software in order to perform detailed comparisons with experimental data.

Due to complications in the involved mathematical techniques most programs providing higher order electroweak or QCD calculations, however, require large numbers of events to be generated with positive and negative weights. These can not be easily used in simulations because the meaning of the cancellation of positive and negative weights in combination with detector influences is not overly clear. In addition, producing the necessary numbers of events is extremely time-consuming. Nevertheless the results are very useful for comparisons to data and they are used in a variety of experiments to perform, for example, precision measurements of standard model parameters like the strong coupling parameter, $\alpha_s$.

Although many such programs have been developed for a variety of physical processes, and sometimes even more than one program exist for the same purpose, the usage and the presentation of the results has not been unified so far. This leads to a number of unnecessary technical problems for the user who wants to compare the predictions of more than one program to some measured distribution. To be more specific, the user has to learn, for each single program he or she wants to use, how to install and compile the program, how to implement the specific distributions or processes of interest, and how to extract the results. It should be noted that the programs are also implemented in different programming languages, mostly *FORTRAN* and *C++*.

NLOLIB seeks to simplify the physicists' life by unifying the steering and providing a common framework to implement new quantities and to extract the results. A first version of NLOLIB was developed in the workshop on 'Monte Carlo Generators for HERA Physics' 1998/99 and aimed to integrate and compare three different programs for next-to-leading order calculations in electron-proton scattering [1]. Now this scope will be extended to include also proton-(anti)proton and electron-positron collisions.

## 2 Implemented programs

**RacoonWW** provides tree-level cross-sections to all processes where an electron-positron collision yields four fermions in the final state. In addition, it contains a number of higher order corrections, see [2] for details.

**DISENT** [3] is a next-to-leading order calculation for the production of one or two jets (processes up to $\mathcal{O}(\alpha_s^2)$) for deep-inelastic ep scattering. It uses a subtraction scheme [4] for the cancellation of divergencies. DISENT is the standard program for DIS jet production at HERA and has been used in a variety of analyses.

**DISASTER++** [5] offers more possibilities to separate different terms in the derivation of the cross sections; otherwise it provides a functionality similar to DISENT. It also employs the subtraction method.

**JetViP** [6, 7] is a next-to-leading order calculation for jet production in deep-inelastic ep scattering and $\mathrm{e}^+\mathrm{e}^-$ collisions that contains processes up to $\mathcal{O}(\alpha_s^2)$ and implements both direct and resolved contributions to the cross-section. The cancellation of divergencies is performed using the phase-space slicing method [8] which leads to dependencies of the resolved contribution on the unphysical slicing parameter, $y_{cut}$. The direct predictions of JetViP have been shown to be compatible [7,9] with those of DISENT and DISASTER++. So far, only the implementation into NLOLIB of the direct photon ep scattering part of JetViP has been thoroughly tested; the resolved photon part is implemented in principal but needs more testing. The implementation of the $\mathrm{e}^+\mathrm{e}^-$ part into NLOLIB has just started.

**MEPJET** [10] was the first complete next-to-leading order calculation available for deep-inelastic ep scattering and is based on the phase-space slicing method in combination with the technique of crossing functions [11]. The predictions of MEPJET show some discrepancies with respect to the results of DISASTER++, DISENT and JetViP [9].

**NLOJET++** [12] incorporates next-to-leading order predictions for ep, pp and e+e- scattering using the subtraction method. Due to its very different way of allowing users to implement their favourite quantities, an integration into NLOLIB on the same footing as for the other programs seems not to be feasible. However, it will be tried to achieve an approach as similar as possible. Currently, only the original version of NLOJET++ in its unchanged form is included.

**FMNR** [13]: FMNR is a program for the calculation of next-To-leading order photoproduction jet cross-sections with heavy quarks in the final state. The implementation of FMNR into NLOLIB has only just begun.

## 3 Getting started

Once, a new release is finished, a compressed tar archive will be made available like it is done already now on `http://www.desy.de/∼nlolib`.

Since the way in which NLOLIB is installed has changed considerably in the course of this workshop from the original version [1], some short instructions on how to get started are given here.

Originally, the `make` tool together with a set of `perl` scripts containing hardware-specific settings were used. However, this procedure was not easily maintainable, so it was decided to employ the GNU `autotools` [14]. For this to work `automake` versions 1.7 or higher and `autoconf` versions 2.57 or higher are needed. In addition, the CERN libraries including a version of PDFLIB [15] and the HzTool [16] libraries as available from our web page are required.

When these conditions are met, the following scheme should be followed for installing NLOLIB:

- Retrieve the NLOLIB source code from the web page at DESY (at a later stage it will also be downloadable from the CVS server in Karlsruhe) and copy it to your working directory (assumed to be `∼/nlolib` in the following).
- In `nlolib.sh` (for c-type shells in `nlolib.csh`) set the correct paths to the PDFLIB and the CERNLIB libraries (`libpdflib.a` or `libpdflib804.a`, `libkernlib.a`, `libpacklib.a`), the HzTool libraries (`libhztool.a`, `libmyhztool.a`) and the directories where to put the NLOLIB binaries and libraries, typically `∼/nlolib/bin` and `∼/nlolib/lib`.
- Go to the working directory and source `nlolib.sh` (or `nlolib.csh`):
  $\sim \backslash nlolib >$ `source nlolib.sh`

- Usually, the following three steps can be skipped. But in case the `configure` script below fails, it has to be recreated by doing:
  - $\sim \backslash nlolib >$ `aclocal`
  - $\sim \backslash nlolib >$ `automake`
  - $\sim \backslash nlolib >$ `autoconf`
- $\sim \backslash nlolib >$ `./configure`
- This step can be skipped if a complete recompile is not necessary/wanted:
  $\sim \backslash nlolib >$ `make clean`
- $\sim \backslash nlolib >$ `make`
- $\sim \backslash nlolib >$ `make install`

Since the running of NLOLIB depends on the simulation program required by the user no general rules can be given here concerning the NLOLIB execution.

## 4 Jet cross-sections in ep NC DIS

As a new check of the NLOLIB framework we tested the predictions of the DISENT and JetViP programs for deep-inelastic ep scattering single-inclusive jet production against the stand-alone versions of the programs and against data published by the H1 collaboration [17]. The phase-space of the measurement is determined by two requirements on the scattered electron: the energy of the scattered electron $E'$ must be larger than 10 GeV, and its polar angle must be larger than $156^o$. In addition, two kinematic cuts are applied to select well-reconstructed low-$Q^2$ DIS events: $5 < Q^2 < 100$ GeV$^2$ and $0.2 < y < 0.6$.

Jet reconstruction for the selected events is performed in the Breit reference frame with the longitudinally invariant $k_\perp$ algorithm [18] in the inclusive mode [19]. Jets are selected by requiring their transverse energy in the Breit frame to be larger than 5 GeV, $E_T^{breit} > 5$ GeV, and their pseudorapidity in the laboratory frame, $\eta^{lab}$, to be between $-1$ and $2.5$.

### 4.1 Comparison of calculations and data

We first present a comparison of event and jet quantities between the stand-alone versions and the versions implemented in NLOLIB of DISENT and JetViP. For this purpose 1 million events have been generated with each of the four programs using CTEQ4M as proton PDFs and $Q^2$ as renormalization and factorization scale. Figure 1 shows the differential cross-sections as functions of $Q^2$, $y$, $E_e$ and $\theta_e$ for the four predictions. A very good agreement between the predictions is observed.

Also the comparison of the various predictions for the jet pseudorapidities in the Breit and laboratory reference frames, $\eta^{Breit}$ and $\eta^{lab}$, and for the jet transverse energy in the Breit frame, $E_T^{Breit}$, shows satisfactory agreement. There are, however, small discrepancies between the two JetViP and the two DISENT predictions in the $\eta^{Breit}$ distribution and a rather large discrepancy between the stand-alone JetViP prediction on the one hand side and the other three calculations on the other hand side for $\eta^{lab}$, see Fig. 2.

Figure 3 finally compares the four predictions to the published H1 data which are presented as inclusive jet cross-sections as functions of $E_T^{Breit}$ in different ranges of $\eta^{lab}$. Also for these published observables the agreement of the various predictions is reasonable.

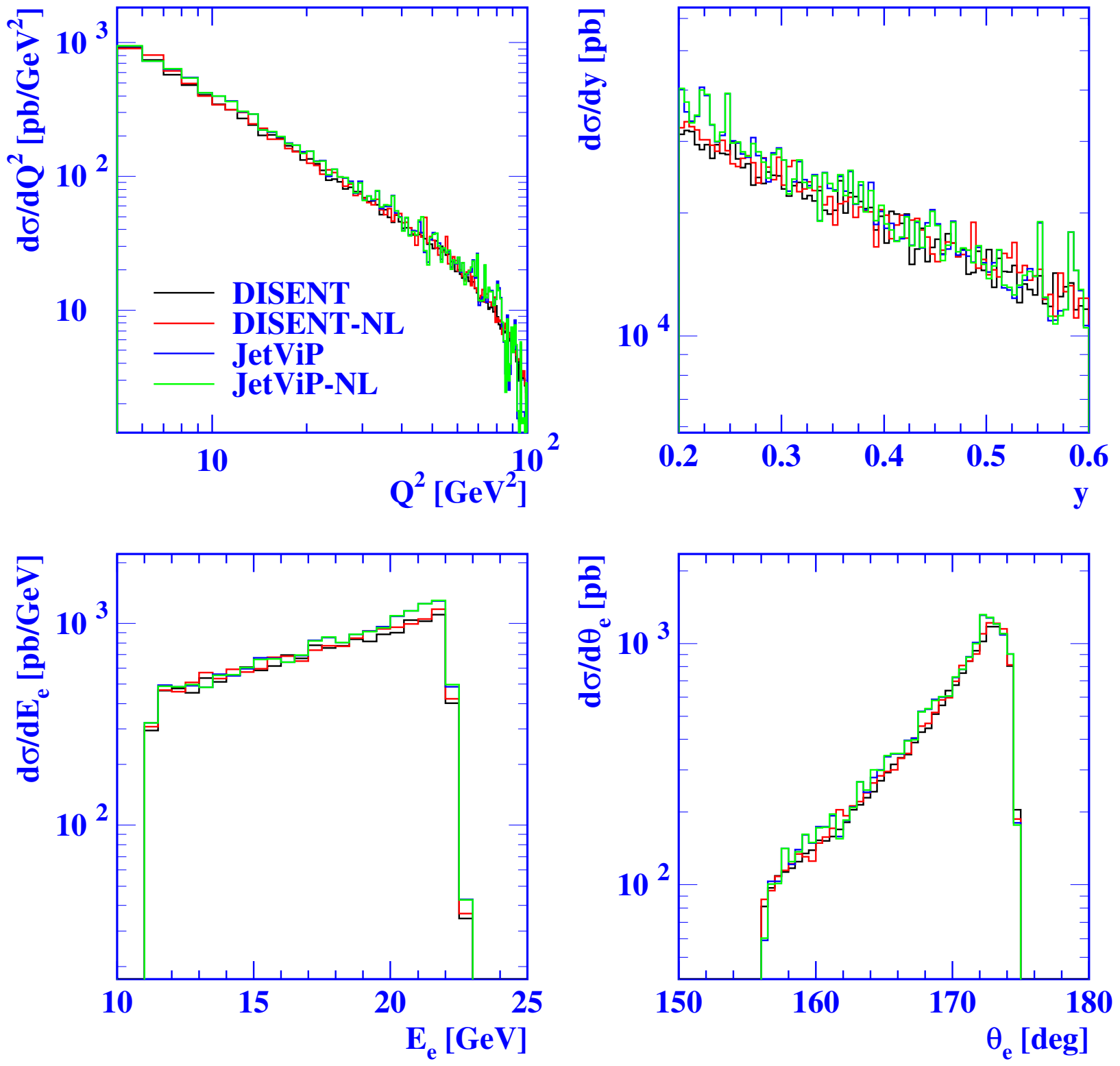

**Fig. 1:** Cross-sections as functions of $Q^2$, $y$, $E_e$ and $\theta_e$ for the different DISENT and JetViP predictions.

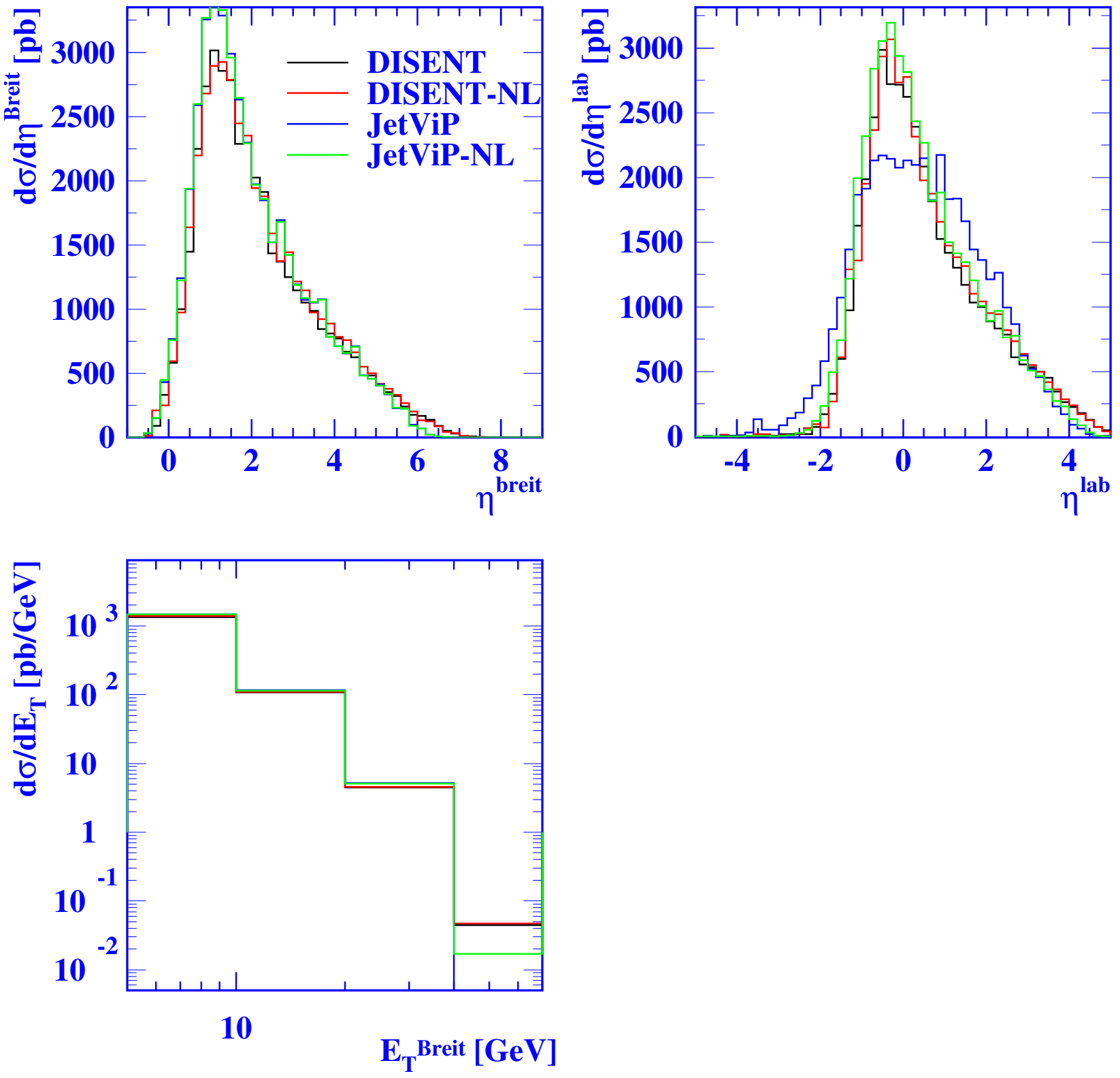

**Fig. 2:** Inclusive jet cross-sections as functions of $\eta^{Breit}$, $\eta^{lab}$ and $E_T^{Breit}$ for the different DISENT and JetViP predictions.

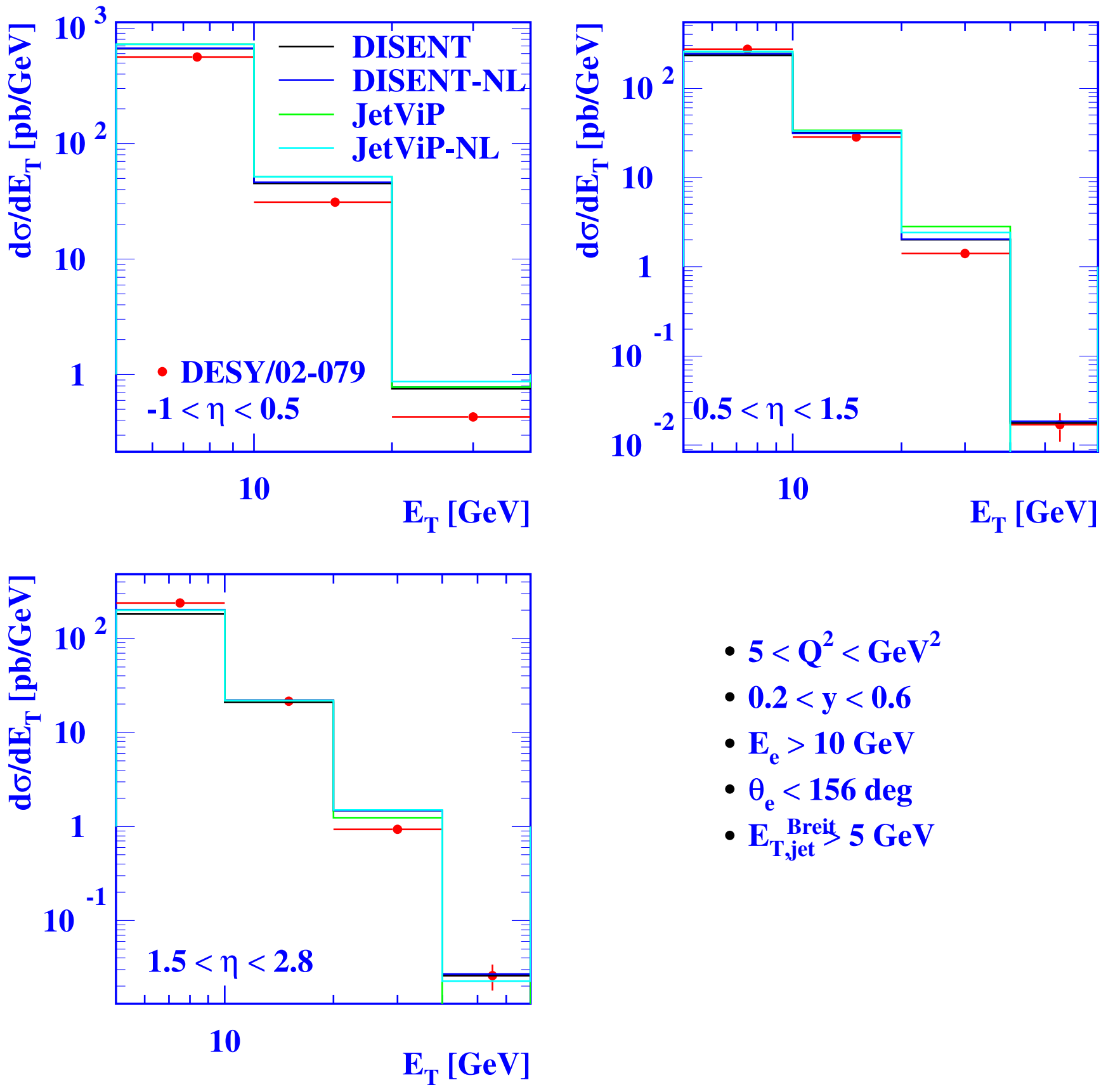

**Fig. 3:** Inclusive jet cross-sections as functions of $E_T^{Breit}$ in different ranges of $\eta^{lab}$. The H1 data are compared to the different DISENT and JetViP predictions.

### 4.2 How to obtain the theoretical distributions

The NLOLIB calculations shown in this section have been obtained by running DISENT and JetViP via the HzTool interface in NLOLIB and using the HzTool routine for the H1 data analysis, `hz02079.f`. The steering files for the DISENT and JetViP job can be found in `nlolib/steering` and are called `dis02079.t` for DISENT and `jv02079.t` for JetViP. The command to run the DISENT job is thus (assuming the command is issued in the `nlolib` directory)

$\sim \backslash nlolib >$ `bin/hzttol < steering/dis02079.t`

for JetViP the command is

$\sim \backslash nlolib >$ `bin/hzttol < steering/jv02079.t`

In both cases a HBOOK file `test.hbook` is created that contains, in subdirectory 02079, the results of the calculation and the published H1 data points. A *PAW* macro `epjets.kumac` that creates the plots shown here will soon be available.

## 5 Summary

Some results with recently implemented higher order calculations have been shown, but clearly many items on our agenda unfortunately are still to be done.

For the implementation of JetViP into NLOLIB first, and most importantly, the $e^+e^-$ mode has to be implemented and tested — so far only the ep mode has been done. Secondly, the resolved photon contribution has to be tested more thoroughly and the discrepancies between DISENT and JetViP in the pseudorapidity distributions need to be sorted out.

Concerning NLOJET++ a similar approach like for the other programs has to be set up and thoroughly tested in an all-program comparison, for example of event shapes or jet cross-sections in deep-inelastic scattering. Then, jet cross-sections in hadron-hadron collisions can be derived with NLOJET++. In addition, the use of PDFLIB will be replaced by LHAPDF [20].

Finally, the work on the implementation of more programs, for example FMNR or further proton–(anti)proton programs, needs to be followed up.

# HZTool

*J. M. Butterworth* [1], *H. Jung* [2], *V. Lendermann* [3], *B. M. Waugh* [1]
[1] Department of Physics and Astronomy, University College London, Gower Street, London, WC1E 6BT, England
[2] Deutsches Elektronen-Synchrotron, Notkestr. 85, 22607 Hamburg, Germany;
[3] Kirchhoff-Institut für Physik, Universität Heidelberg, Im Neuenheimer Feld 227, 69120 Heidelberg, Germany

**Abstract**
HZTOOL is a library of Fortran routines used for tuning and validating of Monte Carlo and analytical models of high energy particle collisions. The library includes an extensive collection of routines reproducing $ep$, $\gamma p$, $p\bar{p}$ and $\gamma\gamma$ data published by the HEP experiments, as well as calculations as they were performed in the experimental data analyses. During this workshop the library was further developed to include HERA and TeVatron results relevant for the studies at the LHC.

## 1 Introduction

Data from high-energy physics experiments have seen the triumph of the Standard Model both in precision electroweak measurements and in the verification of QCD to a reasonable degree of precision. However, a number of aspects of high energy collisions remain poorly understood due to technical difficulties in the calculation. This is particularly the case for measurements of the hadronic final state in high energy collisions, where the specific event shapes variables, jet algorithms and kinematic cuts may be rather complex.

Accurate models of the final state are often needed to design new experiments and to interpret the data from them. Simulation programs employing fits to existing data address these problems. However, consistent tuning of the parameters of these programs, and examination of the physics assumptions they contain, is non-trivial due to the wide variety of colliding beams, regions of phase space, and complex observables. Comparing a new calculation to a sensible set of relevant data is in practice extremely time consuming and prone to error.

HZTOOL [1] is created to improve this situation. It is a library of Fortran routines allowing reproduction of the experimental distributions and easy access to the published data. Basically, each subroutine corresponds to a published paper. If supplied with the final states of a set of simulated collisions, these routines will perform the analysis of the final state exactly as it was performed in the paper, providing simulated data points which may be compared to the measurement. HZTOOL currently contains measurements from $ep$, $\gamma p$, $\gamma\gamma$ and $p\bar{p}$ collisions. Others may easily be added.

While it is designed to be used as simply as possible as a standalone library, HZTOOL is also a key component of JETWEB [2]. JETWEB is a Web-based facility for tuning and validating Monte Carlo models. A relational database of reaction data and predictions from different models is accessed via the Web, enabling a user to find out how well a given model agrees with a range of data and to request predictions using parameter sets that are not yet in the database.

HZTOOL can also be used together with RUNMC [3], which is an object-oriented frond-end for Monte Carlo programs incorporating a sophisticated graphical user interface.

The library was initially developed within the workshop "Future Physics at HERA" [4] but has since expanded to become a more general toolkit. The package was managed by *T. Carli* for quite some time and many people have contributed routines and general development to HZTOOL since it first appeared. Nowadays, HZTOOL and JETWEB are further developed within the CEDAR project [5]. The current maintainers are *J. Butterworth*, *H. Jung*, *E. Nurse* and *B. Waugh*[1]. It is planned to design a C++ equivalent, in order to provide a native interface to the new C++ Monte Carlo programs, to enable a straightforward implementation of new HEP data analyses performed in C++.[2]

## 2 HZTOOL Usage

Each analysis subroutine books, fills and outputs two sets of histograms: one reproducing the published data and another one filled by the chosen simulation program. The routine names relate to the publication. The preferred convention is:

`HZHyymmnnn` where yymmnnn is the arXiv:hep–ex preprint number.

Alternative naming schemes, used for older routines or when a hep–ex number is not available, are:

`HZDyynnn` where yynnn is the DESY preprint number.
`HZCyynnn` where yynnn is the CERN preprint number.
`HZFyynnnE` where yynnn is the FNAL preprint number.
`HZyynnn` where yynnn is the DESY preprint number[3].

If none of the above number schemes exist, the routine name is generally derived from the journal publication (e.g. `HZPRT154247`). Occasionally a single publication contains results taken under more than one set of beam conditions, in which case there will be a routine for each beam condition, distinguished by appending a letter to the expected name (e.g. `HZC88172A`, `HZC88172B`).

HZTOOL is a library, and the main program, which is usually a Monte Carlo event generator, must be provided by the user. The code of HZTOOL routines is basically independent of the program used to simulate the collisions. The Monte Carlo generators currently supported are: ARIADNE [6], CASCADE [7], HERWIG [8] (including JIMMY [9]), LEPTO [10], PYTHIA [11], PHOJET [12], QCDINS [13], RAPGAP [14], RIDI [15], and DJANGOH [19]. The production versions of these programs are all currently written in Fortran. The library can also be accessed within the NLOLIB framework for running NLO QCD programs [16]. Besides the HZTOOL library it is also necessary to link in the CERNLIB library routines and possibly PDFLIB [17] or LHAPDF [18]. Examples of the main programs can be found in the HZSTEER package [20] which provides the executable programs for JETWEB to submit from its backend.

To ease the implementation of the analysis code, HZTOOL provides the relevant jet finders (the cluster and cone algorithms with various options), as well as a number of utilities to calculate event shape variables, to perform Lorentz boosts etc.

## 3 Recent Developments

Within this workshop, an effort was made to include all results from HERA and other HEP experiments which can be helpful for tuning of MC models used for event simulations at the LHC. In particular, the models for multiple interactions and for heavy flavour production were considered.

The publications which may be relevant for the tuning of multiple interaction models and which are available in HZTOOL are listed in [21]. The names for the corresponding HZTOOL subroutines are also specified. From this list, the newly written routines are: `HZH9505001` (*J. M. Butterworth*, *B. M. Waugh*), `HZH9810020` (*S. Lausberg*, *V. Lendermann*), `HZH0006017` (*D. Beneckenstein*, *V. Lendermann*) and `HZH0302034` (*K. Lohwasser*, *V. Lendermann*). The routine `HZ95219` was extended by *A. Buniatian* to include the results from the corresponding H1 paper which are especially sensitive to underlying events in the photoproduction of jets. The models of multiple interactions and efforts of their tuning are reviewed in [22].

[1] The maintainers can be contacted at `hztool@cedar.ac.uk`. To receive announcements of new releases, send an e-mail to `majordomo@cedar.ac.uk` with `subscribe hztool-announce` in the body of the e-mail.

[2] More details may be found at `http://hepforge.cedar.ac.uk/rivet`.

[3] This naming should not be used for new routines; the HZD prefix is preferred.

As for heavy flavour production, a number of HERA measurements of open charm and beauty production are included in the library [23]. From those, the newly written routines are: HZH0108047 (*P. D. Thompson*), HZH0312057 (*O. Gutsche*), HZH0408149 (*A. W. Jung*). New publications [24] are to be implemented. The following routines for the TeVatron results [25] were also recently provided: HZH9905024 (*O. Gutsche*), HZH0307080 (*H. Jung*), HZH0412071 (*H. Jung*, *K. Peters*). Furthermore a set of *Benchmark* cross sections have been defined for easy comparison of different calculations: HZDISCC, HZDISBB for charm and beauty production in DIS, HZHERAC, HZHERAB for photoproduction of charm and beauty and HZLHCC, HZLHCBB for charm and beauty production at the LHC [26].

# CEDAR

*A Buckley [1], J M Butterworth [2], S Butterworth [2], L Lönnblad [3], W J Stirling [1], M Whalley [1], B M Waugh [2]*
[1] Institute for Particle Physics Phenomenology, University of Durham, DH1 3LE, UK
[2] Department of Physics and Astronomy, UCL, Gower Street, London, WC1E 6BT
[3] Department of Theoretical Physics, Sölvegatan 14A, S-223 62 Lund, Sweden

**Abstract**
The CEDAR collaboration is developing a set of tools for tuning and validating theoretical models by comparing the predictions of event generators with data from particle physics experiments. CEDAR is also constructing resources to provide access to well defined versions of high-energy physics software and support for software developers. Here we give an overview of the CEDAR project and its status and plans.

## 1 Introduction

Despite the success of the Standard Model in accurately describing a wide range of phenomena in high-energy particle physics, there are aspects of high-energy collisions where technical difficulties in the relevant calculations make it hard to attain a good understanding This is particularly true where non-perturbative QCD is involved, as in the description of hadronic collisions, where the final state is influenced by the parton distribution functions (PDFs) of the colliding beams, by multiple soft interactions leading to an "underlying event" and by hadronisation of the outoing partons.

These theoretical uncertainties can limit the precision of new measurements, as well as hindering the planning of future experiments. Building accurate models of hadronic processes is important for these reasons as well as for the insight they may offer into the fundamental physics involved. However, the models that are constructed typically have a number of parameters that can be varied, constrained only by how well the resulting predictions agree with experiment.

Tuning these free parameters and testing the models against experimental data is a difficult task because the data are so varied, involving different beam particles, different regions of phase space and complex observables. Changing a single parameter in a model can affect the predictions for different measurements in very different ways, and tuning to a limited set of data may result in a contradiction with other data not taken into account.

It is thus important to compare models simultaneously with as wide a range as possible of experimental results, and the aim of CEDAR [1] is to simplify this task.

The rest of this contribution will describe in turn the projects making up CEDAR.

## 2 HZTool and Rivet

The first requirement is for a library of routines to enable, for each experimental measurement of interest, a comparable prediction to be produced from any given Monte Carlo generator. This role is currently filled by the Fortran library HZTool, described in more detail in another contribution to these proceedings [2]. The HZTool library is being maintained by CEDAR, with subroutines for various measurements contributed by a number of authors within and outside the CEDAR collaboration. A number of HERA routines were written within this workshop.

Work is underway to build a replacement for HZTool, to be called Rivet (Robust Independent Validation of Experiment and Theory). This will use an object-oriented design, implemented in C++, together with standard interfaces (such as HepMC [3] and AIDA [4]) to make the new framework more flexible and extensible than the Fortran HZTool. For example, it will be easier to incorporate new Monte Carlo generators into Rivet than into HZTool.

## 3 JetWeb

JetWeb [5] provides a web interface to HZTool, along with a relational database of both experimental data and model predictions generated using HZTool. The core of JetWeb is a set of Java servlets that manipulate an object model representing data and predictions. A user can use a web form to specify a model and choice of parameters, and the data they wish to compare to this model. If this model and parameter set are already in the database, a set of comparison plots and statistics is returned. Otherwise the user may request a set of Monte Carlo jobs to be run using their specified model, and the results will be added to the database.

The existing JetWeb database has been frozen, although it can still be searched, while the design and functionality are improved. JetWeb is being adapted to use the data already stored in HEPDATA rather than duplicating this in its own database, and to make the addition of further Monte Carlo models (beyond the currently supported HERWIG [6] and PYTHIA [7]) easier.

## 4 HEPDATA

HEPDATA [8] is a well established and widely used source of scattering data from HEP experiments. As part of CEDAR it has been converted from the existing hierarchical structure to a relational database using MySQL. The next steps in this part of the CEDAR project will provide front ends so that the data in the relational database can be accessed through a searchable web interface and also directly by JetWeb and other users.

## 5 HepML

In order to simplify the transfer of data between different parts of the CEDAR project and other software frameworks, an XML schema [9] is being developed to specify particle reactions and experimental results (as provided by HEPDATA) as well as generator programs and parameters.

This schema is separate from the HepML developed within the MCDB project [10], which is designed primarily as a format for event records. It may be that some parts of the schemas can be unified later or become parts of a more general schema.

## 6 HepCode

HepCode [11] aims to provide access to well defined versions of Monte Carlo programs, parton distribution functions and other high-energy physics calculation programs. Currently it is simply a list of codes with details for each of the processes calculated, the order of the calculation, the authors and the programming language used, along with a link to further information where this is available.

HepCode will eventually feature a search facility so that users can find a set of available programs simply by entering the details of a particular scattering process. It may also be possible to have links from matching data records in HEPDATA or from papers in bibliographic databases such as SPIRES.

## 7 HepForge

CEDAR also provides a development environment, HepForge, for authors of HEP software, including Monte Carlo generators. In addition to the core CEDAR projects (HZTool, HZSteer, JetWeb, HepML) other projects using HepForge are fastNLO [12], Herwig++ [13], Jimmy [14], KtJet [15], LHAPDF [16], RunMC [17] and ThePEG [18].

Facilities provided to developers include a code repository (using CVS or Subversion), a bug tracker (using Trac), a wiki for documentation and communication between project contributors, and mailing lists for project discussions, queries and announcements.

# RunMC: an object-oriented analysis framework to generate Monte Carlo events for current and future HEP experiments

*S. Chekanov*
HEP division, Argonne National Laboratory, 9700 S.Cass Avenue, Argonne, IL 60439, USA
E-mail: `chekanov@mail.desy.de`

**Abstract**
RunMC is a C++ object-oriented framework aimed to generate and to analyze high-energy collisions using Monte Carlo simulations. The package provides a common interface to different Monte Carlo models using modern physics libraries developed for the LHC and NLC experiments. Physics calculations can easily be loaded and saved as external project modules. This simplifies the development of complicated physics calculations for high-energy physics in large collaborations.

## 1 Introduction

Monte Carlo models (MC) written in FORTRAN 77 are widely used in many high-energy physics laboratories worldwide. These models are known to be fast, robust and well tested. However, the main choice for future high-energy experiments is an object-oriented programming language, either C++ (the LHC experiments at CERN) or Java (the NLC project). Some steps towards converting the FORTRAN MC models to the C++ programming language have already been undertaken [1]. However, such models written in C++ will require a thoughtful verification to insure that their predictions are consistent with the original FORTRAN-based MC programs, as well as with the physics results obtained in the past. Such verifications will go over certain time, and a tool which allows to perform such comparisons is urgently needed.

A program which allows running of both FORTRAN-coded and C++ MC models using a common C++ programming environment should be valuable. This is important not only for comparisons and verifications of these MC models. Such a C++ framework can also extend the lifetime of FORTRAN-based models especially for the LHC, NLC and TEVATRON communities, and can provide compatibility of most popular MC models with the new software to be used for current and future HEP experiments.

The RunMC package [2] is a common C++ frond-end of Monte Carlo models which provides a unified approach to generate and analyze very different MC models independent of their native codes. In this approach, the MC output (typically the HEPEVT record) is converted to C++ classes for further analysis or graphical representation (histograms). The graphical user interface (GUI) of this program helps to initialize MC models and histograms, as well as to monitor the event generation.

The RunMC program fully complies with the change in the programming paradigm of data analysis, and meets the requirements of future high-energy experiments. Instead of FORTRAN-based analysis tools, such as PAW [3] and HBOOK [4], it uses the modern CERN C++ analysis packages, CLHEP [5] and ROOT [6].

In this respect, the RunMC program is similar to the JetWeb server [7], which also provides the ability to compare the existing MC models and to confirm the physics assumptions they contain. However, in contrast to JetWeb, the RunMC program was designed as a stand-alone desktop application. Therefore, the user has full access to his calculations and to the program itself.

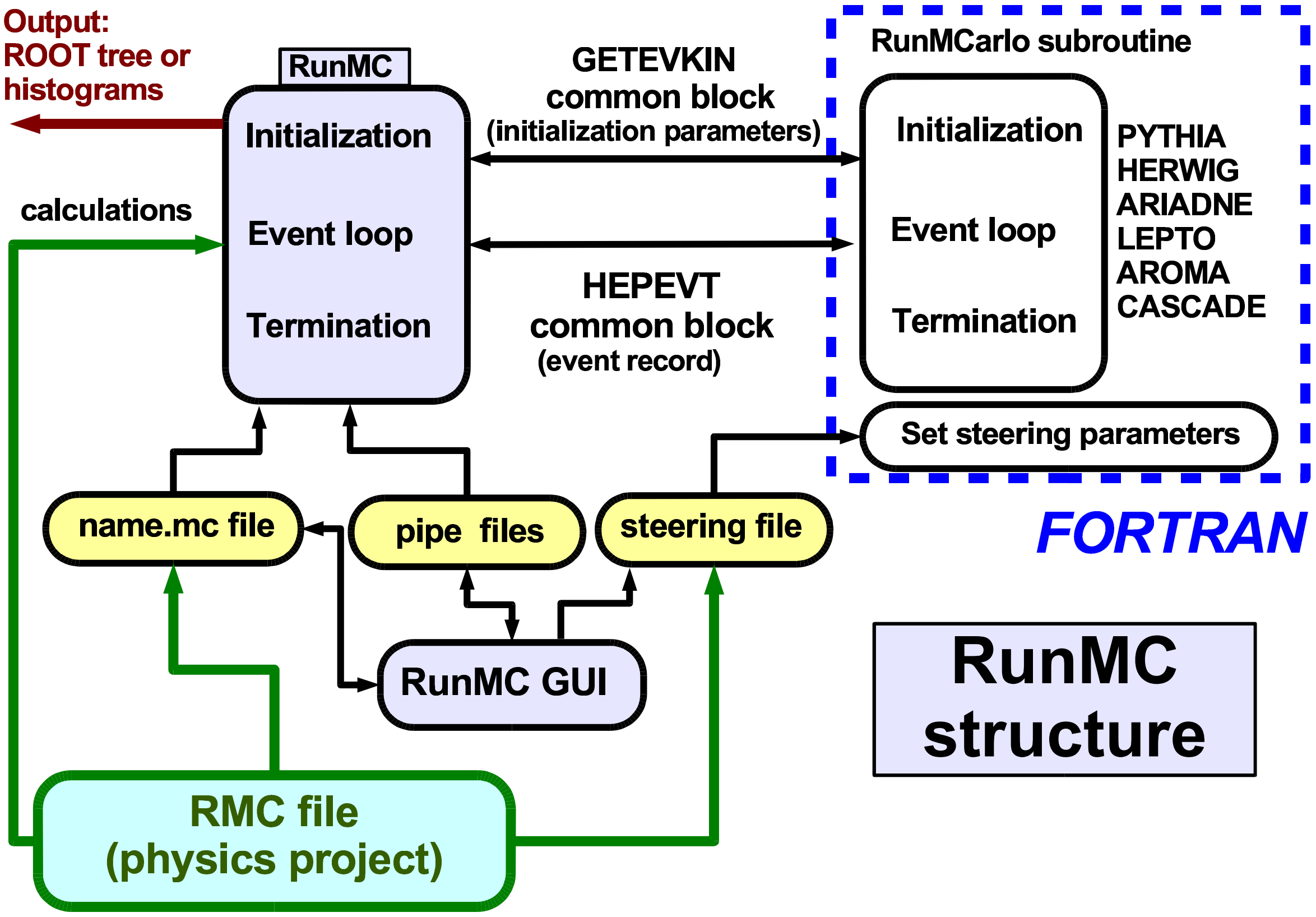

**Fig. 1:** The diagram shows the structure of the RunMC program.

The RunMC program also provides an interface to the popular HZTOOL package [8], thus many physics calculations from HERA, LEP and TEVATRAN can easily be accessed. In addition, within the RunMC approach, the concept of project modules was introduced (in fact, the HZTOOL package is one of such modules). A project file, which can contain external calculations, MC tunings, histogram definitions, etc., can be loaded to RunMC with the same ease as a document can be opened in the Microsoft Word program. The project files are small and platform independent, therefore, it is fairly simple to share complicated physics calculations between scientists in large collaborations.

## 2 Program structure

RunMC consists of a GUI and several RunMC MC programs. There are two implementations of the RunMC GUI: one is written using the Wide Studio C++ classes [9] and an alternative GUI is based on Java. Due to complete independence of RunMC GUI from RunMC MC programs, one can run jobs in the background without any GUI or pop-up window.

The RunMC programs integrate the C++ ROOT and CLHEP packages with native implementations of MC models. A schematic structure of RunMC is illustrated in Fig. 1. The following MC models are included to RunMC (version 3.3): PYTHIA 6.2 [10], HERWIG 6.5 [11], ARIADNE 4.12 [12], LEPTO 6.5 [13], AROMA 2.2 [14], CASCADE 1.2 [15], PHOJET 1.05 [16], RAPGAP 3.1 [17]. There are several executable RunMC MC programs corresponding to each MC model.

RunMC GUI communicates with the RunMC MC programs using pipe files located in the directory “$RUNMC/pipes”. Here, “$RUNMC” denotes the installation directory which has to be defined by the user. All the directories to be discussed below are assumed to be located in this directory.

### 2.1 RunMC GUI

RunMC GUI allows an interaction between the user and RunMC MC programs. At present, two types of RunMC GUI are available: a user interface based on C++ (can be executed with the command "runmc") and that based on Java (the command "jrunmc"). Below we will describe only the C++ RunMC GUI.

The task of RunMC GUI is to generate the output file "project.mc", where "project" is a user-defined name of the current calculation. This file contains the most important information for MC running, such as the type of MC model, the number of output events, the type of colliding particles, their energies, RunMC output (histograms or ntuples) etc.

RunMC GUI adopts the following strategy to define the histograms: The window "Variables" contains the names of the variables (with some additional comments) defined for a certain physics project. The user should select the appropriate variable and copy it to the window "Histograms" by clicking on the corresponding variable name. If two one-dimensional histograms are defined, a two-dimensional histogram can be build from these two histograms using RunMC GUI.

The variable names are divided into the three categories: event-based variables (characterizing the event as whole), single-particle densities (filled for each particle/jet; the variable name starts with "@") and two-particle densities (filled for each particle/jet pair; the name starts with "@@"). The histograms can also be filled in the user-defined subroutine "user-run.cpp"; in this case the naming convention for the variables discussed above is unnecessary.

During the event generation, the ROOT canvas can display the output histograms (up to eight in total) with filled events. The output log from RunMC MC is written to ".analmc.log" (a symbolic link to the "project.log" file). Possible errors are redirected to the file "project.err", which is constantly monitored by RunMC GUI.

The ROOT histograms are automatically modified at the end of the fill if they are required to be normalized to the total number of events or converted to differential cross sections. Note that there is no need to wait until the end of the current run: once the histogram statistics is sufficient, one can terminate the run by clicking "Stop" on the GUI window. Histograms should be saved in the ROOT file "project.root" for further studies. The style of the histograms can further be modified using the ROOT canvas editor.

### 2.2 The RunMC MC programs

The RunMC MC programs integrating Monte Carlo models with ROOT C++ classes have the generic names "analmc.MCname", where "MCname" denotes the MC name. The main C++ function of RunMC MC is located in the file "analmc.cpp" (in the directory "main/src"). The C++ code accesses the HEPEVT common block of a given MC program via a C-like structure. The RunMC MC program receives the initial parameters via the symbolic link ".analmc.ln" pointing to the file "project.mc".

Each MC model has its own FORTRAN subroutine "runmcarlo" which provides an interface to the native MC code. This interface program (in the file "RUNMC-MCname.f") is located in the directory "main/mcarlo/MCname". The task of the subroutine "runmcarlo" is to fill the HEPEVT common block. In addition, some initial settings are done by accessing a C/C++ structure with the initial parameters defined in the "project.mc" file. The main function in "analmc.cpp" calls this interface subroutine and fills the C/C++ structure which represents the complete HEPEVT event record. The output is copied to the class "HEPLIST" which can be accessed by external calculations. The HEPLIST class consists of several vectors based on the LorentzVector vector class (from the CLHEP library) which represents four-momentum of a particle (or a jet). The definition of the HEPLIST class, as well as other include files, can be found in the "main/inc" directory. Note that the user still can access more elementary event records (such as FORTRAN HEPEVT common block) which can be used to transform them to other event classes and physics calculations.

There are several physics packages available inside RunMC MC to transform the original four-momentum vector of particles/jets to the required observable:

– the transformations provided by the physics vector class “LorentzVector” from CLHEP can be used, since a particle or a jet is represented as a general four-vector based on this class;
– the event-shape calculations are available using the package developed by M. Iwasaki [18];
– the longitudinally-invariant $k_T$ algorithm as implemented in C++ [19] can be used for the jet reconstruction. In addition to this package, the JADE and Durham jet algorithms are implemented according to M. Iwasaki [18];
– the Breit frame calculations were implemented for $ep$ deep inelastic scattering.

The physics packages and their documentation are located in the directory “main/physics”.

## 3 User calculations

For a new physics calculation, the directory “proj” should be modified. This user directory can contain external calculations, steering cards for MC initializations, as well as the standard RunMC functions which are necessary to initialize and fill the histograms.

The user directory should always contain the file “project.mc” created by RunMC GUI. This file can be edited manually without the RunMC GUI program using any text editor. On Linux/Unix, one can load this file to RunMC GUI by executing the command “runmc project.mc” from the shell prompt (or using the option “Projects→read MC” of RunMC GUI).

The directory “proj” can contain steering files “MCname.cards” to redefine initial MC parameters. Such files can be created via RunMC GUI (“MC settings” option). For more flexibility, the MC initialization parameters can also be overwritten by FORTRAN-coded subroutines located in the directory “proj/ini”. If this is not done, the default MC parameters will be selected according to the RunMC option.

To define histograms, user-defined variables should be calculated in the file “proj/user_afill.cpp”. The output of this function is a pointer. The output variable name should always be associated with this pointer. The variable names should be specified in the file “user-name.txt”. It includes the variable names to the list “Variables” accessed by RunMC GUI. Finally, to compile the source codes and to rebuild all RunMC MC programs to take into account changes made in the project source files, one should type “make” in the “proj” directory. All MC programs will be recompiled and RunMC GUI will be updated with new histograms. Then, the command “runmc” (or “jrunmc”) should be executed from the directory “proj” to start RunMC GUI. The main advantage of this approach is that once a necessary variable is defined, new histogram definitions do not require the MC recompilation.

RunMC histograms can also be filled using the conventional method, i.e. in the function located in “user-run.cpp”. In this case, the initialization of histograms is not required, as long as the file “project.mc” defines which histograms should be filled and what presentation style should be used to fill the histograms. The histograms can be initialized in the file “user-init.cpp” using the standard ROOT procedure.

## 4 Physics calculations as external RMC projects

In order to share complicated analysis calculations or to store them for future use, the directory “proj” can be packed into an external RMC file with the extention “rmc”. For example, “project.rmc” is the RMC file which has the user-defined name “project”. The “proj” directory inside of it has a file “project.mc” with RunMC GUI settings, user-defined external functions, libraries, make files MC steering files, etc.

RunMC GUI can automatically load and recompile such project RMC files (see details in [2]). The user can also save his/her calculations into a RMC file for future analysis. As it was mentioned, the project files are compact and platform independent, therefore, it is fairly simple to share physics calculations between the users, as long as the RunMC package is installed.

At present, several RunMC project files are available on the Web [2] (they are also included in the directory "archive" of RunMC):

- the default project. Only pre-installed variables can be included in the calculations.
- HERA kinematic variables ($Q^2$, $x$, etc.);
- jets at HERA and LHC using the longitudinally-invariant $k_T$ algorithm in the Breit frame. In addition, the ratio of jet cross sections at the parton and hadron levels are calculated (the so-called hadronisation corrections);
- $D^*$ cross sections in $ep$ collisions at HERA;
- cross sections for strange-particle production in $ep$ collisions at HERA;
- the HZTOOL package [8];
- the event-shape variables in $e^+e^-$ at NLC energies;
- several examples of how to visualize tracks and $k_T$ jets in 3D for a single MC event ($e^+e^-$, $ep$, $pp$ collisions).

The RMC project files discussed above only illustrate how to set up and to develop new physics calculations in the RunMC framework. For practical applications, these examples should be modified.

## 5 RunMC ROOT tree analyzer

In addition to the standard functionality of the MC event simulation, RunMC GUI can also use ROOT trees as the input for physics calculations.

The ROOT tree can be generated by selecting the option "HEPEVT" or "RUNMC", in addition or instead of the ROOT histogram option. Then, the MC events should be generated as usual, but this time a ROOT tree with the extension ".rtup" or ".htup" will be created. Then, RunMC can run over this ROOT tree if, instead of the MC model, the option "RUNMC" or "HEPEVT" is selected. Several ROOT trees can automatically be included in the analysis, as long as they are in the same directory. The analysis of the ROOT trees is very similar to the standard run over MC events. External RMC files can be used to include new calculations, variables and histograms.

The main advantage of the RunMC ROOT tree analyzer is that physics calculations can be validated significantly faster than when RunMC is used to generate events and to fill histograms at the same time. In case of the ROOT trees, RunMC can fill histograms by a factor of $\sim$10–15 faster, thus the RMC project files can be validated and analyzed more efficiently.

With this additional functionality, the RunMC program can also be used to analyze experimental data if the event record is converted to the appropriate ROOT tree. The data analysis can be performed using exactly the same RMC project files as for the usual MC simulation runs.

# A C++ framework for automatic search and identification of resonances

*S. Chekanov*
HEP division, Argonne National Laboratory, 9700 S.Cass Avenue, Argonne, IL 60439, USA
E-mail: `chekanov@mail.desy.de`

**Abstract**
This paper describes the first proof-of-concept version of a C++ program designed for peak searches in invariant-mass distributions. The program can be useful for searches of new states as well as for the reconstruction of known resonances.

Presently, there is a considerable progress in understanding the rich spectrum of hadron resonances. However, even in case of known baryonic resonances, there are still many open questions. The Particle Data Group (PDG) quotes more than 100 baryonic states, but only half of them are reasonably well established. Recently, the revitalized interest to baryon spectroscopy was triggered by observation of narrow peaks in the $K^{\pm}n$ and $K^0_S p$ invariant-mass distributions which can be interpreted as pentaquarks. There is also longstanding interest to other exotic multiquark states, glueballs, hybrids, baryonia, etc. The HERA experiments have active program for searching such states [1]. At the LHC this physics program will continue.

In hadron spectroscopy, searches for new resonances and measurements of known states are based on the reconstruction of invariant-mass distributions of two or more tracks in order to find the rest mass of the originally decaying particle. The usual procedure for such studies is to assign certain masses to tracks, and then to combine their four-momenta to form the invariant-mass distributions. From the observed peaks, one can determine resonance masses, widths and cross sections.

Obviously, a particle identification is highly desirable for any experiment in order to reduce combinatorial background. Without the ability to identify tracks, the combinatorial background rises as $\sim 0.5n^2$, where $n$ is the number of produced particles used in the searches. Thus high-energy experiments have to deal with a significant background for searches in fully inclusive events. For the reconstruction of three- and four-body decays, the combinatorial background is even higher than for the two-body decays.

Particle identification can be achieved using various approaches, such as the energy loss per unit length ($dE/dx$), silicon-strip detectors, Cherenkov counters etc.. In many cases, the particle identification cannot be perfect. For example, when the $dE/dx$ method is used for pentaquark searches, it is difficult to obtain a high-purity sample of kaons or protons for tracks with large momenta ($p > 1$ GeV) due to significant overlap of the $dE/dx$ bands for different particle species in this momentum range. In this case, several mass assumptions for invariant-mass distributions are needed to be checked in order to exclude peaks due to possible misidentification.

Searches for new resonances using various mass assumptions remain to be a tedious task since no much progress has been made so far to develop a tool which can automate this procedure. For example, the reconstruction of two-, three-, four-body decays involving only three mass assumptions leads to 36 possible non-identical invariant-mass distributions (6 distributions for two-body decays, 10 - for three-particle decays and 20 - for four-particle decays). All these invariant-mass distributions should be reconstructed, analyzed and possible reflections from known PDG states, when different mass assumptions are used, should be disregarded. Obviously, this time consuming work can be simplified and automated.

The program called "SBumps", which is still at the early stage of the development, attempts to accomplish this task. It rather represents the first proof-of-concept version of a program which helps to perform automatic searches of peaks in invariant-mass distributions. To perform the reconstruction of invariant-mass distributions, the user should specify:

- the event record, i.e. a list of tracks and (optionally) the probabilities that tracks belong to certain particle species. Such probabilities can be obtained using various particle-identification technics;
- the names of particle species used during the mass assignments;
- the statistical significance of expected peaks;
- instrumental invariant-mass resolution;
- how many tracks should be combined to the invariant-mass distributions.

Once the initial conditions are specified, the program runs over the event list and reconstructs all possible invariant-mass distributions with the mass assumptions as specified by the user. SBumps can reconstruct at the same time $2-$, $3-$ and $4-$ body decays in different combinations. At the end of the run, the program saves all the created invariant-mass distributions to the ROOT histograms [2].

At the second stage, the program analyses the created distributions and attempts to find statistically significant peaks. SBumps uses the build-in ROOT fast peak finder (from the TSpectrum class), which uses a fast deconvolution method [3] based on a Markov approach for peak searching in presence of a background and statistical noise. This algorithm was mainly developed for narrow, high-amplitude peaks which are characteristic for $\gamma$-ray physics. Therefore, this algorithm is not completely appropriate for peak searches in the invariant-mass spectra without tunings of the initial parameters of this algorithm. In case of the SBumps package, the deconvolution techniques is only used to identify the so-called seeds, i.e. bins with positions of possible peaks above a smooth background. The number of seeds can be rather large, and not all of them correspond to statistically significant peaks. At this stage, several adjustable parameters are available, such as the resolution of neighboring peaks, the peak sensitivity and the peak thresholds. Such parameters can be set using the steering file before each run.

Next, the program evaluates each seed by calculating the statistical significance, $N(S)/\Delta N$, where $\Delta N = \sqrt{N(B) + N(S)}$, with $N(S)(N(B))$ being the number of the signal (background) events. The calculation was done by comparing the values at the seed positions with the background level, which is determined from the neighboring bins. Then the peaks, which have the statistical significance above the level specified by the user, are considered for further analysis. This part of the program can be further improved, introducing more complicated algorithms. For example, sufficiently broad resonances might be overlooked by the present prescription.

The two-step procedure described above simplifies the peak search since the user normally does not need to deal with tunings of the initial parameters for the fast deconvolution method. At the same time, the method is sufficiently fast and can possibly be extended by introducing other algorithms to identify the peaks. For example, algorithms based on a fitting procedure can also be used; in this case, the seed positions can be used as the initial parameters of the fit functions.

At the third stage, the program attempts to identify the found peaks by comparing them with the masses of known PDG states. For this, a look-up table containing the information on established PDG resonances is used. Since the errors on the reconstructed peaks and the errors on the masses of known resonances taken from the PDG look-up table are known, the program matches the peaks using a simple criteria: $\mid L \mid /\Delta L < S$, where $S$ is a free parameter given by the user, $L$ is the distance between the PDG mass and the peak position and $\Delta L$ is the error on the reconstructed peak position combined with the error on the mass of known PDG resonances. For this matching procedure, the information on charge of decaying resonance is properly taken into account. In future, the information on specific decay channel can also be taken into account in order to reduce misidentification in the matching procedure.

Figure 1 shows the output of the SBumps program for events generated with PYTHIA Monte Carlo model. This example shows one histogram for two-body decays with two automatically identified peaks corresponding to known states. The required statistical significance for the final peaks was $4\sigma$.

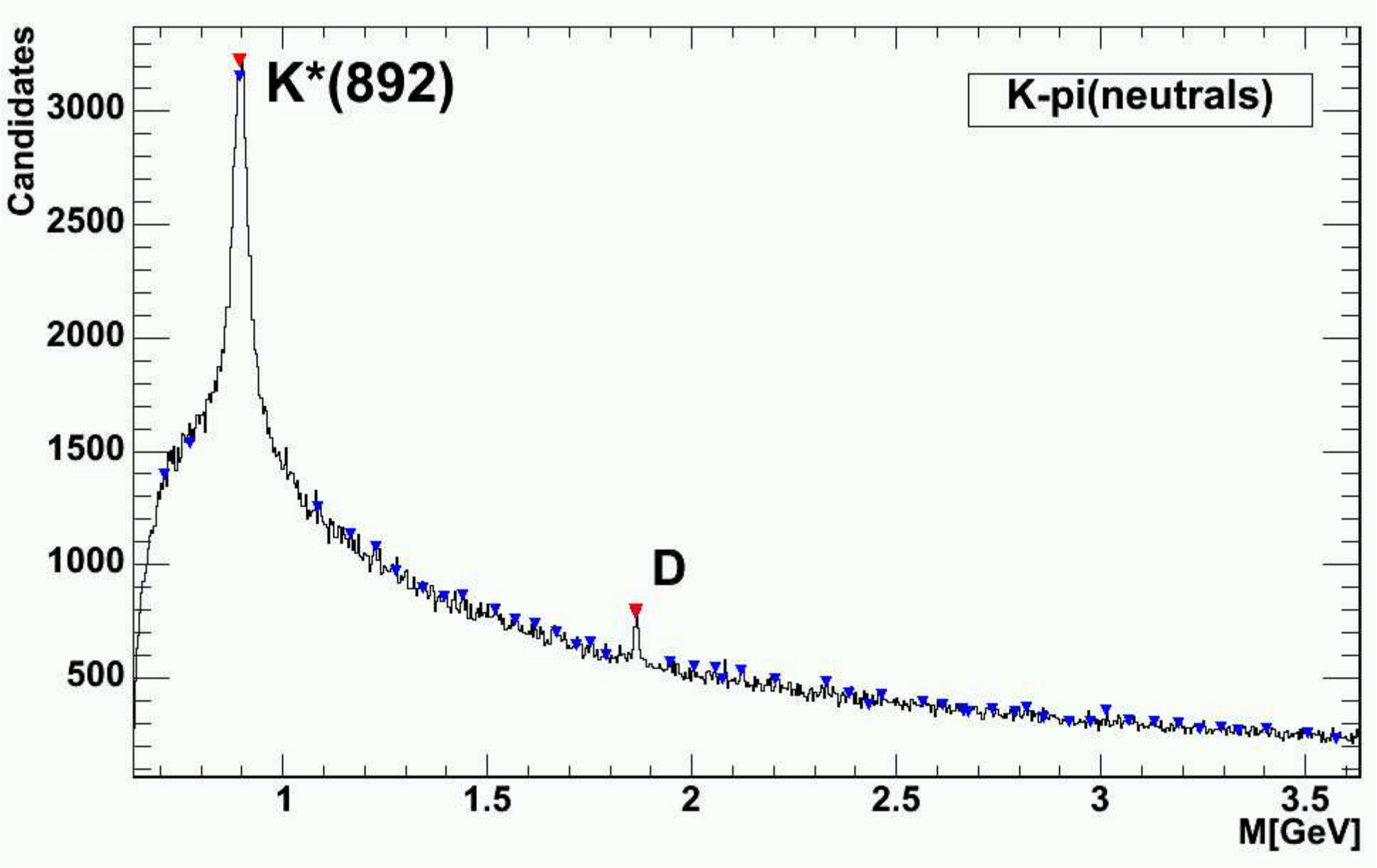

**Fig. 1:** The two-particle invariant-mass distribution calculated using the events generated with the PYTHIA Monte Carlo model ($pp$ collisions at 1.4 TeV), when two mass hypothesis are used ($K^{\pm}$ and $\pi$ mesons). The blue triangle symbols show the seeds used for the calculations of statistical significance. The seeds were found by using the fast deconvolution method. Most of the seed peaks were disregarded after the final calculation of the statistical significance. The program correctly identifies the known PDG states, $K^*$ and $D^0$, after comparing the reconstructed peak positions with the PDG look-up table.

This example shows that the program can easily find and identify rather broad resonances ($K^*$), as well as narrow peaks which do not have high statistical significance.

The proof-of-concept version of the SBumps program is available as a loadable RMC file of the RunMC analysis framework [4]. After loading the “sbumps.rmc” module, RunMC first creates a ROOT tree with Monte Carlo events. The initial conditions are given in the file “sbumps.cards”. The executable file “sbumps.exe” can be used to run over the events. At the end of the calculations, the ROOT browser should display the reconstructed histograms with invariant-mass distributions. The peaks which have the statistical significance above that specified by the user should be labeled.